%% file: EVLI.TEX
\pdfoutput=1     
\documentclass[12pt,titlepage,makeidx]{book}
 \usepackage{emlines}
 \usepackage{emlines2}
 \usepackage{verbatim}
\usepackage{latexsym}
\usepackage{amsmath}
\usepackage{fancybox} 
\usepackage{graphicx} 

 \usepackage[cmtip,arrow]{xy} 
 \usepackage{pb-diagram,pb-xy} 

\usepackage{tikz}
\usetikzlibrary{calc, shadows, backgrounds}

\renewcommand{\baselinestretch}{1.1}
\newcommand{\mai}[2]{{#1}_{#2}} 
\newcommand{\maj}[2]{{#1}^{#2}} 
\newcommand{\maij}[3]{{#1}_{#2}^{#3}} 
\newcommand{\map}[2]{{#1}\{{ #2}\}}

\newcommand{\mapij}[4]{{#1}\{{ #2}\}_{#3}^{#4}} 
\newcommand{\mapq}[3]{{#1}\{_{#2}^{#3}\}} 
 
\newcommand{\mapqj}[4]{{#1}\{_{#2}^{#3}\}^{#4}} 
\newcommand{\mapqij}[5]{{#1}\{_{#2}^{#3}\}_{#4}^{#5}} 
\def\ma1{\!+\!1} 
\def\me1{\!-\!1}

\def\til{\widetilde} 
\def\overli{\overline} 
\def\overha{\widehat} 

\def\haat{\widehat}

\def\argmin \mathop{\rm argmin}
\def\Re{I\kern -0.37 em R}
\def\Na{I\kern -0.37 em N}
\def\Qe{I\kern -0.37 em Q}

\def\aalpha{a} 
\def\bbeta{b} 
\def\fto{\colon\!} 
\def\g{\,\vert \,}
\def\ate{\,\mbox{:}\,}
\def\atee{\,\mbox{::}\,} 
\def\tria{\,\triangle \,}

\def\del{\partial\,} 
\def\E{\,\mbox{E}} 
\def\Var{\,\mbox{Var}} 
\def\Cov{\,\mbox{Cov}} 
\def\Vec{\,\mbox{Vec}} 
 
\def\kron{\otimes} 
\def\tr{\mbox{tr}} 
\def\arg{\mbox{arg}} 
\def\diag{\mbox{diag}} 
 
\def\ev{\,\mbox{ev}\,} 
\def\evb{\overline{\ev}} 
\def\sev{\mbox{sev}} 
\def\sevb{\overline{\sev}} 
\def\uno{\mbox{\bf 1}} 
\def\zero{\mbox{\bf 0}} 
 
\def\Bin{\mbox{Bi}} 
\def\Mno{\mbox{Mn}} 
\def\Dir{\mbox{Di}} 
\def\Poi{\mbox{Ps}} 
\def\Gam{\mbox{Ga}} 
 
\def\Bne{\mbox{NB}} 
\def\Bbi{\mbox{BB}} 
 
\def\Hip{\mbox{Hy}} 
\def\DM{\mbox{DM}} 
\def\olP{T} 
\def\ch{\mbox{ch}} 
 
\def\pa{\mbox{pa}}

\def\If{\mbox{ if }} 
\def\Then{\mbox{ then }} 
 
\def\And{\mbox{ and }} 
\def\Or{\mbox{ or }} 
\def\Where{\mbox{ where }} 
\def\For{\mbox{ for }} 

\def\leqk{\leq_k}
\def\leqt{\leq_t}
\def\leqc{\leq_c}
\def\leqd{\leq_d}
\def\sqcapc{\sqcap_c}
\def\sqcapd{\sqcap_d}
\def\sqcapk{\sqcap_k}
\def\sqcapt{\sqcap_t}
\def\sqcupc{\sqcup_c}
\def\sqcupd{\sqcup_d}
\def\sqcupk{\sqcup_k}
\def\sqcupt{\sqcup_t}
\def\bi{\mbox{BI}}
\def\bt{\mbox{BT}}

\def\mb{{mb}} 
\def\pa{{pa}} 
\def\ch{{ch}} 
\def\sp{{f}} 
 
\def\diag{\mbox{diag}} 
\def\nze{\mbox{nze}} 
\def\Cov{\mbox{Cov}} 
\def\Var{\mbox{Var}} 
\def\div{\mbox{div}} 
\def\grad{\mbox{grad}}

\def\Chi2{\mbox{Q}} 
\def\pprod{\prod\nolimits} 
\def\ssum{\sum\nolimits}

\def\bbigvee{\bigvee\nolimits} 
\def\bbigwedge{\bigwedge\nolimits} 
\def\g{\,\vert\,}
\def\del{\partial\,}
\def\Pr{\mbox{Pr}}
\def\tr{\mbox{tr}}
\def\arg{\mbox{arg}}

\def\etab{\overli{\eta}}
\def\mb{\overli{m}}

\def\Tb{\overli{T}}
\def\Vb{\overli{V}}
\def\Wb{\overli{W}}
\def\sh{\overha{s}}
\def\tho{\overha{\theta}}

\def\uno{\mbox{\bf 1}}
\def\zero{\mbox{\bf 0}}

\def\aand{\mbox{ and }}
\def\oor{\mbox{ or }}
\def\wwhere{\mbox{ where }}

\def\uu{{\mathcal{U}}}
\def\Re{{R}}

\def\hy{{}^{1}_{1}\mbox{H}} 
\def\hequ{{}^{4}_{2}\mbox{He}} 
\def\cado{{}^{12}_{\;\, 6}\mbox{C}} 
\def\catr{{}^{13}_{\;\, 6}\mbox{C}}  
 
\def\nitr{{}^{13}_{\;\, 7}\mbox{N}} 
\def\niqa{{}^{14}_{\;\, 7}\mbox{N}} 
\def\niqi{{}^{15}_{\;\, 7}\mbox{N}} 
\def\oxqi{{}^{15}_{\;\, 8}\mbox{O}}

 \def\rr{\noindent - }

\renewcommand{\baselinestretch}{1.1}
\makeindex

\begin{document}

 \title{Cognitive Constructivism and \\ 
        the Epistemic Significance of \\ 
        Sharp Statistical Hypotheses \\ 
        in Natural Sciences }

 \author{Julio Michael Stern}

 \date{IME-USP \\  
 Institute of Mathematics and Statistics \\ 
 of the University of S\~{a}o Paulo \\ 
 \mbox{} \\ \mbox{} \\ 
 Version 2.31, May 01, 2013.} 

 \maketitle 

 \cleardoublepage  
 
 \mbox{} \vspace{3cm} \mbox{}  

 {\flushright {\it 
 \`{A} Marisa e a nossos filhos, \ \mbox{} \\ 
 Rafael, Ana Carolina, e Deborah. \ \mbox{} \\  } 
 \mbox{} }  

 \vfill 

 \mbox{} 

 \pagebreak 

 \mbox{} 

 \pagebreak 

 \mbox{} \vspace{3cm} \mbox{}  

 \begin{flushright}

 {\it ``Remanso de rio largo, \  viola da solid\~{a}o: \ \ \mbox{} \\ 
 Quando vou p'ra dar batalha, \  convido meu cora\c{c}\~{a}o.''} \ \mbox{} \\ 
 Gentle backwater of wide river, fiddle to solitude:  \ \mbox{} \\ 
 When going to do battle, I invite my heart.    \ \mbox{} \\ 

 \mbox{} 

 Jo\~{a}o Guimar\~{a}es Rosa (1908-1967). \ \mbox{} \\ 
 Grande Sert\~{a}o, Veredas. \ \mbox{} \\

 \mbox{} \\ \mbox{}

 {\it ``Sert\~{a}o \'{e} onde o homem tem de ter 
   a dura nuca e a m\~{a}o quadrada. \ \ \mbox{} \\ 
   (Onde quem manda \'{e} forte, com astucia e com cilada.) \ \ \mbox{} \\ 
   Mas onde \'{e} bobice a qualquer resposta, \ \ \mbox{} \\ 
   \'{e} a\'{\i} que a pergunta se pergunta.''} \ \mbox{} \\ 

 {\it ``A gente vive repetido, o repetido... \ \ \mbox{} \\ 
 Digo: o real n\~{a}o est\'{a} na sa\'{\i}da nem na chegada: \ \ \mbox{} \\  
 ele se disp\~{o}em para a gente \'{e} no meio da travessia.''} \ \ \mbox{} \\ 

 \mbox{} 

 Sertao is where a man's might must prevail,  \ \mbox{} \\ 
 where he has to be strong, smart and wise.    \ \mbox{} \\ 
 But there, where any answer is wrong,    \ \mbox{} \\ 
 there is where the question asks itself.    \ \mbox{} \\ 

 We live repeating the reapeated... \ \mbox{} \\ 
 I say: the real is neither at the departure nor at the arrival:  \ \mbox{} \\ 
 It presents itself to us at the middle of the journey.  \ \mbox{} \\

 \end{flushright} 

 \mbox{} 

 \vfill 

 \mbox{} 

 \pagebreak 

 \mbox{} 



\tableofcontents


\setcounter{chapter}{0}

 \clearpage \input{CAPE0.TEX} \clearpage 
 \clearpage \input{CAPE1.TEX} \clearpage 
 \clearpage \input{CAPE2.TEX} \clearpage 
 \clearpage \input{CAPE3.TEX} \clearpage 
 \clearpage \input{CAPE4.TEX} \clearpage 
 \clearpage

\input{CAPE5.TEX}  \clearpage 
 \clearpage \input{CAPE6.TEX}  \clearpage 
 \clearpage \input{CAPEPI.TEX}  \clearpage 
 \clearpage \input{CAPREF.TEX}  \clearpage 

\setcounter{chapter}{0}
\appendix 

 \clearpage \input{CAPFBST.TEX}  \clearpage 
 \clearpage \input{CAPBIN.TEX}  \clearpage 
 \clearpage \input{CAPMIX.TEX}  \clearpage 
 \clearpage

\input{CAPDOPT.TEX}  \clearpage 
 \clearpage

\input{CAPENT.TEX}  \clearpage 
 \clearpage \input{CAPFAC.TEX} \clearpage 
 \clearpage

\input{CAPMM.TEX}  \clearpage 
 \clearpage \input{CAPSOPT.TEX}  \clearpage 
 \clearpage \input{CAPRES.TEX}   \clearpage 
 \clearpage \input{CAPARTA.TEX}  \clearpage 

\end{document}

%% file: CAPE0.TEX
\chapter*{Preface}
\addcontentsline{toc}{chapter}{Preface}        
\markboth{PREFACE}{PREFACE}


 \begin{flushright} 
 {\it ``Life is like riding a bicycle. \ \mbox{} \\ 
 To keep your balance you must keep moving.''} \\   
 Albert Einstein. \ \mbox{} 
 \end{flushright} 

 \mbox{}

 The main goals of this book are to develop an epistemological framework
based on Cognitive Constructivism, and to provide a general introduction
to the Full Bayesian Significance Test (FBST). 
 The FBST was first presented  in Pereira and Stern (1999)  
 as a coherent Bayesian method for accessing the statistical 
 significance of sharp or precise statistical hypotheses. 
 A review of the FBST is given in the appendices, including:   
 \\ \mbox{\ } 
 a) Some examples of its practical application;   
 \\ \mbox{\ } 
 b) The basic computational techniques used in its implementation; 
 \\ \mbox{\ }
 c) Its statistical properties; 
 \\ \mbox{\ }
 d) Its logical or formal algebraic properties; 
 \\  
 The items above have already been explored in
 previous presentations and courses given. 
 In this book we shall focus on presenting 
 \\ \mbox{\ }
 e) A coherent epistemological framework for precise statistical hypotheses.

 The FBST grew out of the necessity of testing sharp statistical
hypothesis in several instances of the consulting practice of its
authors. By the end of the year 2003, various interesting applications
of this new formalism had been published by members of the Bayesian
research group at IME-USP, some of which outperformed previously
published solutions based on alternative methodologies, see for example
Stern and Zacks (2002). In some applications, the FBST offered simple,
elegant and complete solutions whereas alternative methodologies offered
only partial solutions and / or required convoluted problem
manipulations, see for example Lauretto et al. (2003).

 The FBST measures the significance of a sharp hypothesis in a way that
differs completely from that of Bayes Factors, the method of choice of
orthodox Bayesian statistics. 
 These methodological differences fired interesting debates that
motivated us to investigate more thoroughly the logical and algebraic
properties of the new formalism. These investigations also gave us the
opportunity to interact  with people in communities that were interested
in more general belief calculi, mostly from the areas of Logic and
Artificial Intelligence, see for example Stern (2003, 2004) and Borges
and Stern (2007).

 However, as both Orthodox Bayesian Statistics and Frequentist
Statistics have their own well established epistemological frameworks,
namely, Decision Theory and Popperian Falsificationism, respectively,
there was still one major gap to be filled: the establishment of an
epistemological framework for the FBST formalism. Despite the fact that
the daily practice of Statistics rarely leads to epistemological
questions, the distinct formal properties of the FBST repeatedly brought
forward such considerations. Consequently, defining an epistemological
framework fully compatible with the FBST became an unavoidable task, as
part of our effort to answer the many interesting questions posed by our
colleagues.

 Besides compatibility with the FBST logical properties, this new
epistemological framework was also required to fully support sharp
(precise or lower dimensional) statistical hypothesis. In fact,
contrasting with the decision theoretic epistemology of the orthodox
Bayesian school, which is usually hostile or at least unsympathetic to
this kind of hypothesis, this new epistemological framework actualy
puts, as we will see in the following chapters, sharp hypothesis at the
center stage of the philosophy of science.

\section*{Cognitive Constructivism}
\addcontentsline{toc}{section}{Cognitive Constructivism}

 The epistemological framework chosen to the aforementioned task was
Cognitive Constructivism, as presented in chapters 1 to 4, and
constitute the core lectures of this course. 
 The central epistemological concept supporting the notion of a sharp
statistical hypothesis is that of a systemic eigen-solution. According
to Heinz von Foerster, the four essential attributes of such
eigen-solutions are: discreteness (sharpness), stability, separability
(decoupling) and composability. Systemic eigen-solutions correspond to
the ``objects'' of knowledge, which may, in turn, be represented by
sharp hypotheses in appropriate statistical  models. These are the main
topics discussed of chapter 1.

 Within the FBST setup, the {\it e-value} of a hypothesis, $H$, defines
the measure of its {\it Epistemic Value} or the {\it Value of the
Evidence} in support of $H$, provided by the observations. This measure
corresponds, in turn, to the  ``reality'' of the object described by the
statistical hypothesis. The FBST formalism is reviewed in Appendix A.

 In chapter 2 we delve into this epistemological framework from a 
broader perspective, linking it to the philosophical schools of
Objective Idealism and Pragmatism. The general approach of this chapter
can be summarized by the ``wire walking''  metaphor, according to which
one strives to keep in balance at a center of equilibrium, to avoid the
dangers of extreme positions that are faraway from it, see Figure J.1.
In this context, such extreme positions are related to the
epistemological positions of Dogmatic Realism and Solipsistic
Subjectivism.

 Chapters 3 and 4 relate to another allegory, namely, the Bicycle
Metaphor: In a bike, it is very hard to achieve a static equilibrium,
that is, to keep one's balance by standing still. Fortunately, it is
easy to achieve a dynamic equilibrium, that is, to ride the bike running
forward. In order to keep the bike running, one has to push the left and
right pedals alternately, which will inevitably result in a gentile
oscillation. Hence a double (first and second order) paradox: In order
to stay in equilibrium one has to move forward, and in order to move
forward one has to push left and right of the center. Overcoming the
fear generated by this double paradox is a big part of learning to ride
a bike.

 Chapters 3 and 4 illustrate realistic and idealistic metaphorical
pushes in the basic cycle of the constructivist epistemic ride. They
work like atrial and ventricular systoles of a hart in the life of the
scientific system. From an individual point of view, these realistic and
idealistic pushes also correspond to an impersonal, extrospective or
objective perspective versus a personal, introspective or subjective
perspective in science making.

 Chapter 5 explores the stochastic evolution of complex systems and is
somewhat independent of chapters 1 to 4. In this chapter, the evolution
of scientific theories is analyzed within the basic epistemological
framework built in chapters 1 to 4. Also, while in chapters 1 to 4 many
of the examples used to illustrate the topics under discussion come from
statistical modeling, in chapter 5, many of the examples come from
stochastic optimization.     

 Chapter 6 how some misperceptions in science or misleading
interpretations can lead to ill-posed problems, paradoxical situations
and even misconceived philosophical dilemmas.
 It also (re)presents some of the key concepts of Cog-Con using simple 
and intuitive examples. 
 Hence, this last chapter may actually be the first one to read.

 Figures J.2, J.3 and J.4 illustrate the bicycle metaphor. The first is
a cartoon, by K.Przibram, of Ludwig Boltzmann, the second a photography
of Albert Einstein, and the third a photography of Niels Bohr. They are
all riding their bikes, an activity that appears to be highly beneficial
to theoretical Probability. Boltzmann advocated for an
atomistic and probabilistic interpretation of thermodynamics, that is,
viewing thermodynamics as a limit approximation of Statistical
Mechanics. His position was thoroughly rejected by the scientific
establishment of his time. One of the main criticisms to his work was
the introduction of ``metaphysical'', that is, non ``empirical'' or non
``directly observable'' entities. In 1905, annus mirabilis, Einstein
published his paper on Brownian motion, providing a rigorous
mathematical description of observable macroscopic fluctuation phenomena
that could only be explained in the context of Statistical Mechanics.
Sadly, Boltzmann died the next year, before his theories were fully
appreciated. Discretization and probability are also basic concepts in
Quantum Mechanics. The famous philosophical debates between Bohr and
Einstein, involving these two concepts among others, greatly contributed
to the understanding of the new theory.

\section*{Basic Tools for the (Home) Works}
\addcontentsline{toc}{section}{Basic Tools for the (Home) Works}

 The fact that focus of this summer course will be on epistemological
questions should not be taken as an excuse for working not so hardly
with statistical modeling, data analysis, computer implementation, and
the like. After all, this course will give to successful students 4 full
credits in the IME-USP graduate programs!

  In the core lectures we will illustrate the topics under discussion
with several `concrete' mathematical and statistical models. We have
made a conscious effort to choose illustration models ivolving only
mathematical concepts already familiar to our prospective students.
Actually, most of these models are entail mathematical techniques that
are used in the analysis and the computational implementation of the
FBST, or that are closely related to them.
 Appendices A through K should help the students with their homeworks.
We point out, however, that the presentation quality of these appendices
is very heterogeneous. Some are (I hope) didactic and well prepared,
some are only snapshots from slide presentations, and finally, some are
just commented computer codes.

\section*{Acknowledgements and Final Remarks}
\addcontentsline{toc}{section}{Acknowledgements and Final Remarks}

 The main goal of this book is to explore the FBST formalism and
Bayesian statistics from a constructivist epistemological perspective.
In order to accomplish this, ideas from many great masters, including
philosophers like Peirce, Maturana, von Foerster, and Luhmann, 
statisticians like Peirce, Fisher, de Finetti, Savage, Good, Kemthorne,
Jaynes, Jeffreys and Basu, ans physicists like Boltzmann, Planck, 
de Broglie, Bohr, Heisenberg, and Born have been used. 
 I hope it is clear from the text how much I admire and feel I owe to
these giants, even when my attitude is less then reverential. By that I
mean that I always felt free to borough from the many ideas I like, and
was also unashamed to reject the few I do not. The progress of science
has always relied on the free and open discussion of ideas, in contrast
to rigid cults of personality. I only hope to receive from the reader
the same treatment and that, among the ideas presented in this work, he
or she finds some that will be considered interesting and worthy of be
kept in mind.

  Chapters 1 to 4, released as Stern (2005a) and the Technical Reports  
Stern (2006a-c), have been used in January-February of 2007 (and again
for 2008) in the IME-USP Summer Program for the discipline MAE-5747
Comparative Statistical Inference. 
 Chapter 5, released as the Technical Report by Stern (2007c), has also
been used in the second semester of 2007 in the discipline MAP-427 -
Nonlinear Programming. 
 A short ``no-math'' article based on part of the material in Chapter 1
has been published (in Portuguese) in the journal 
 {\it Scientiae Studia}. 
 Revised and corrected versions of articles based on the material
presented at Chapters 1, 2 and 3 have also been either published or
accepted for publication in the journal 
 {\it Cybernetics \& Human Knowing}. 
 In the main text and the appendices I have used several results
concerning the FBST formalism, some of its applications, and also other
statistical and optimization models and techniques, developed in
collaboration with or by other researchers. Appropriate acknowledgements
and references are given in the text. 

  The author has benefited from the support of FAPESP, CNPq, BIOINFO,
the Institute of Mathematics and Statistics of the University of S\~{a}o
Paulo, Brazil, and the Mathematical Sciences Department at
SUNY-Binghamton, USA. The author is grateful to many people for helpful
discussions, most specially, Wagner Borges, Soren Brier, Carlos Humes,
Joseph Kadane, Luis Gustavo Esteves, Marcelo Lauretto, Fabio Nakano,
Osvaldo Pessoa, Rafael Bassi Stern, Sergio Wechsler, and  Shelemyahu
Zacks. The author also received interesting comments and suggestions
from the participants of FIS-2005, the Third Conference on the
Foundations of Information Science, and several anonymous referees. 
 The alchemical transmutation of my original drafts into proper English
text is a non-trivial operation, in which I had the help of Wagner
Borges and several referees.  

  But first and foremost I want to thank Professor Carlos Alberto de
Bragan\c{c}a Pereira (Carlinhos). I use to say that he teached me much
of the (Bayesian) Statistics I know, the easy part, after un-teaching me
much of the (frequentist) Statistics I thought I knew, the hard part.
Carlinhos is a lover of the scientific debate, based on the critical
examination of concepts and ideas, always poking and probing established
habits and frozen ideas with challenging questions. This is an attitude,
we are told, he shared with his Ph.D. advisor, the late Prof. Debabrata
Basu.  
 
 Just as an example, one of Carlinhos favorit questions is: Why do we
(ever) randomize? 
 I hope that some of the ideas presented in chapter 3 can contribute to
the discussion of this fundamental issue.  Carlinhos extensive
consulting practice for the medical community makes him (some times,
painfully) aware of the need of tempering randomization procedures with
sophisticated protocols that take into account the patients' need of
receiving proper care.
 
 This work has its focus on epistemological aspects. The topics under
discussion are, however, surprisingly close to, and have many times been
directly motivated by, our consulting practice in statistical modeling
and operations research. The very definition of the FBST was originally
inspired by some juridical consulting projects, see Stern (2003). This
does not mean that many of these interrelated issues tend to be ignored
in everyday practice, like the proverbial bird that ignores the air
which supports it, or the fish that ignores the water in which it swims.
  
 The author can be reached at \ \ {\tt jmstern@hotmail.com\ }. 

 \mbox{} 

 \begin{flushright} 
 Julio Michael Stern  \ \ \ \ \ \ \ \ \ \ \ \ \ \ \  \mbox{}  \\  
 S\~{a}o Paulo, 20/12/2007.  \ \ \ \ \ \ \ \ \ \ \ \  \mbox{} 
 \end{flushright}

 \pagebreak

 \subsubsection{Version Control} 

  \noindent 
 - Version 1.0 - December 20, 2007.

  \noindent 
  - Version 1.1 - April 9, 2008.  
  Several minor corrections to the main text and 
  some bibliographic updates. 
  The appendices have been reorganized as follows: 
  Appendix A presents a short review of the FBST, including its
definition and main statistical and logical properties; 
 Appendix B fully reviews the  distribution theory used to build
Multinomial-Dirichlet statistical models;
 Appendix C summarizes several statistical models
used to illustrate the core lectures; 
 Appendix D (previously a separate handout) gives a short introduction 
to deterministic optimization; 
 Appendix E reviews some important concepts related to the
Maximum Entropy formalism and asymptotic convergence; 
 Appendix F, on sparse factorizations, provides some technical details
related to the discussions on decoupling procedures in chapter 3;
 Appendix G presents a technical miscellanea on Monte Carlo Methods;
 Appendix H provides a short derivation of some stochastic optimization
algorithms and evolution models; 
 Appendix I lists some open research programs;  
 Appendix J contains all bitmap figures and, finally, 
 Appendix K brings to bear pieces of difficult to get reading material.
They will be posted at my web page, subject to the censorship of our
network administrator and his understanding of Brazilian copyright laws
and regulations.  
  All computer code was removed from text   
 and is now available at my web page, 
 {\tt www.ime.usp.br/$\sim$jstern} \ . 

 This version has been used for a tutorial at 
 MaxEnt-2008, the  
 28th International Workshop on Bayesian Inference  
 and Maximum Entropy Methods in Science and Engineering,   
 held on July 6-11, at Borac\'{e}ia, S\~{a}o Paulo, Brazil.

  \noindent 
  - Version 1.2 - December 10, 2008.  
  Minor corrections to the main text and appendices, 
  and some bibliographic updates. 
  New section F.1 on dense matrix factorizations.  
  This section also defines the matrix notation now used 
  consistently throughout the book. 

  \noindent 
  - Version 2.0 - December 19, 2009.  
  New section 4.5 and chapter 6, presented at the conference MBR'09 - 
  Model Based Reasoning in Science and Technology - 
  held at Campinas, Brazil. 
  Most of the figures at exhibition in the 
  art gallery are now in the separate file,  \\  
   \verb#www.ime.usp.br/~jstern/pub/gallery2.pdf# .

  \noindent 
  - Version 2.3 - November 02, 2012. 
  New sections D.3.1 on Quadratic and Linear Complementarity Problems 
  and E.6 on Reaction Networks and Kirchhok's Laws. 
  Updated References. Minor corrections throughout the text.

 \vfill

 \mbox{} 

%% file: CAPE1.TEX
 
 \chapter{Eigen-Solutions and Sharp Statistical Hypotheses}

 {\flushright {\it 
 ``Eigenvalues have been found ontologically to be \\ 
 discrete, stable, separable and composable ...'' \\ }  
 Heinz von Foerster (1911 - 2002), \mbox{} \\ 
 Objects: Tokens for Eigen-Behaviours. \mbox{} \\ }  


 \section{Introduction}
 \markboth{CHAPTER 1: EIGEN-SOLUTIONS AND STATISTICAL HYPOTHESES}
  {1.1 \ INTRODUCTION}

 In this chapter, a few epistemological, ontological and sociological
questions concerning the statistical significance of sharp hypotheses 
in the scientific context are investigated within the framework provided
by Cognitive Constructivism, or the Constructivist Theory (ConsTh)  
as presented in  Maturana and Varela (1980), Foerster (2003) 
and Luhmann (1989, 1990, 1995). 
 Several conclusions of the study, however, remain valid, mutatis
mutandis, within various other organizations and systems,  
 see for example Bakken and Hernes (2002), Christis (2001), 
 Mingers (2000) and Rasch (1998).
 
The author's interest in this research topic emerged from his
involvement in the development of the Full Bayesian Significance
Test (FBST), a novel Bayesian solution to the statistical problem of
measuring the support of sharp hypotheses, first presented in  Pereira
and Stern (1999).  
 The problem of measuring the support of sharp hypotheses poses several
conceptual and methodological difficulties for traditional statistical
analysis under both the frequentist (classical) and the orthodox
Bayesian approaches. 
 The solution provided by the FBST has significant advantages over
traditional alternatives, in terms of its statistical and logical
properties. Since these properties have already been thoroughly analyzed
in previous  papers, see references, the focus herein is directed
exclusively to epistemological and ontological questions.

 Despite the fact that the FBST is fully compatible with  Decision
Theory (DecTh), as shown in Madruga et al. (2001), which, in turn, 
provides a strong and coherent  epistemological framework to orthodox
Bayesian Statistics, its logical properties open the possibility of
using and  benefiting from alternative epistemological settings.  
 In this chapter, the epistemological framework of  ConsTh is
counterposed to that of DecTh. The contrast, however,  is limited in
scope by our interest in statistics and is carried out  in a rather
exploratory an non exhaustive form. The epistemological  framework of
ConsTh is also counterposed to that of Falsificationism,  the
epistemological framework within which classical  frequentist  
statistical test of hypotheses are often presented, 
as shown in Boyd (1991) and Popper (1959, 1963). 

 In section 2, the fundamental notions of Autopoiesis and
Eigen-Solutions in autopoietic systems are reviewed. 
 In section 3, the same is  done with the notions of Social Systems and
Functional Differentiation and in section 4, a ConsTh view of science is
presented. 
 In section 5, the material presented in sections 2, 3 and 4 is related
to the statistical significance of sharp scientific hypotheses and the
findings therein are counterposed to traditional interpretations such as
those of DecTh. 
 In section 6, a few sociological analyses for differentiation
phenomena are reviewed. 
 In sections 7 and 8, the final conclusions are established. 

 In sections 2, 3, 4, and 6, well established concepts of the ConsTh 
are presented. However, in order to overcome an unfortunately common 
scenario, an attempt is made to make them accessible to a scientist or
statistician who is somewhat familiar with traditional frequentist, and
decision-theoretic statistical interpretations, but unfamiliar with the
constructivist approach to epistemology. 
 Rephrasing these concepts (once again) is also avoided.     
 Instead, quoting the primary sources is preferred  whenever it can be
clearly (in our context) and synthetically done. 
 The contributions in sections 5, 7 and 8, relate mostly to the analysis
of the role of  quantitative methods specifically  designed to measure
the statistical support of sharp hypotheses. 
 A short review of the FBST is presented in Appendix A.


 \section{Autopoiesis and Eigen-Solutions}
 \markboth{CHAPTER 1: EIGEN-SOLUTIONS AND STATISTICAL HYPOTHESES}
  {1.2 \ AUTOPOIESIS}

 The concept of autopoiesis tries to capture an essential characteristic
of living organisms (auto=self, poiesis=production). Its purpose and
definition are stated in Maturana and Varela (1980, p.84 and 78-79):

 \begin{quote} 
 {\it ``Our aim was to propose the characterization of living systems that
explains the generation of all the phenomena proper to them. We have
done this by pointing at Autopoiesis in the physical space as a
necessary and sufficient  condition for a system to be a living one.''}  

 {\it ``An autopoietic system is organized (defined as a unity) as a 
network of processes of production (transformation and destruction) 
of components that produces the components which:
  
 (i) through their interactions and transformations continuously
regenerate and realize the network of processes (relations) that
produced them; and
  
 (ii) constitute it (the machine) as a concrete unity in the space in
which they (the components) exist by specifying the topological domain
of its realization as such a network.''} 
  \end{quote} 

 Autopietic systems are non-equilibrium (dissipative) dynamical systems
exhibiting (meta) stable structures, whose organization remains 
invariant over (long periods of) time, despite the frequent substitution
of their   components. 
 Moreover, these components are produced by the same structures they 
regenerate. 
 For example, the macromolecular population of a single cell can be
renewed thousands of times during its lifetime, see Bertalanffy (1969).
 The investigation of these regeneration processes in the autopoietic 
system production network leads to the definition of cognitive domain,   
 Maturana and Varela (1980, p.10):  


  \begin{quote} 
 {\it 
 ``The circularity of their
organization continuously brings them back to the same internal state
(same with respect to the cyclic process). Each internal state requires
that certain conditions (interactions with the environment) be satisfied
in order to proceed to the next state. Thus the circular organization
implies the prediction that an interaction that took place once will
take place again.  
   If this does not happen the system maintains its integrity (identity
  with respect to the observer) and enters into a new prediction. In a
  continuously changing environment these predictions can only be
  successful if the environment does no change in that which is predicted.
  Accordingly,   the predictions implied in the organization of the living
  system are not predictions of particular events, but of classes of
  inter-actions. 
 Every interaction is a particular interaction, but every prediction is
a prediction of a class of interactions that is defined by those
features of its elements that will allow the living system to retain its
circular organization after the interaction, and thus, to interact
again. This makes living systems inferential systems, and their domain
of interactions a cognitive domain.''} 
  \end{quote} 

 The characteristics of this circular (cyclic or recursive) regenerative
processes and their eigen (auto, equilibrium, fixed, homeostatic,
invariant, recurrent, recursive) -states, both in concrete and abstract
autopoietic systems, are further investigated in 
 Foerster (2003) and Segal (2001):  

  \begin{quote} 
 {\it ``The meaning of recursion is to run through one's own path again. 
 One of its results is that under certain conditions there exist indeed
solutions which, when reentered into the formalism, produce again the
same solution. These are called ``eigen-values'', ``eigen-functions'',
``eigen-behaviors'', etc., depending on which domain this formation is
applied - in the domain of numbers, in functions, in behaviors, etc.''} 
 Segal (2001, p.145).  
 \end{quote} 

 The concept of eigen-solution for an autopoietic system is the key to
distinguish specific objects in a cognitive domain. von Foerster also
establishes four essential attributes of eigen-solutions that will
support the analyses conducted in this chapter and conclusions 
established herein.  

  \begin{quote} 
 {\it ``Objects are tokens for eigen-behaviors. Tokens stand for
something else. In exchange for money (a token itself for gold held by
one's government, but unfortunately no longer redeemable), tokens are
used to gain admittance to the subway or to play pinball machines. In
the cognitive realm, objects are the token names we give to our
eigen-behavior.    

 This is the constructivist's insight into what takes place when we 
talk about our experience with objects.''} 
 Segal (2001, p.127). 

 {\it ``Eigenvalues have been found ontologically to be discrete, stable,
separable and composable, while ontogenetically to arise as equilibria
that determine themselves through circular processes. Ontologically,
Eigenvalues and objects, and likewise, ontogenetically, stable behavior
and the manifestation of a subject's ``grasp'' of an object cannot be
distinguished.''} 
 Foerster (2003, p.266). 
   \end{quote} 

 The arguments used in this study rely heavily on two 
qualitative properties of eigen-solutions, refered by von Foerster by
the terms ``Discrete'' and ``Equilibria''. 
 In what follows, the meaning of these qualifiers, as they are   
understood by von Foerster and used herein, are examined: 

 a- Discrete (or sharp):    

    \begin{quote} 
 {\it ``There is an additional point I want to make, an important point.
 Out of an infinite continuum of possibilities, recursive operations
carve out a precise set of discrete solutions. Eigen-behavior generates
discrete, identifiable entities. Producing discreteness out of infinite
variety has incredibly important consequences. It permits us to begin
naming things. Language is the possibility of carving out of an infinite
number of possible experiences those experiences which allow stable
interactions of your-self with yourself.''} 
 Segal (2001, p.128). 
 \end{quote}

 It is important to realize that, in the sequel, the term ``discrete'',
used by von Foerster to qualify eigen-solutions in general,  should be
replaced, depending on the specific context, by terms such as 
lower-dimensional, precise, sharp, singular etc. 
 Even in the familiar case of linear algebra, if we define the
eigen-vectors corresponding to a singular eigen-value $c$ of a linear
transformation $T(\ )$ only by its essential property of directional
invariance, $T(x)=cx$, we obtain one dimensional sub-manifolds which, in
this case, are subspaces or lines trough the origin. Only if we add the
usual (but non essential) normalization condition, $||x||=1$, do we get
discrete eigen-vectors.

 b- Equilibria (or stable):  

  A stable eigen-solution of the operator $Op(\ )$, defined by the
 fixed-point or invariance equation, $x_{inv}=Op(x_{inv})$, 
 can be found, built or computed as the limit, $x_\infty$,  
 of the sequence $\{ x_n \}$,  defined by recursive
 application of the operator, $x_{n+1}=Op(x_n)$.  
  Under appropriate conditions, such as within a domain of attraction, 
 the process convergence and its limit eigen-solution 
 will not depend on the starting point, $x_0$. 
  In the linear algebra example, using almost any staring point, 
 the sequence generated by the recursive relation 
 $x_{n+1} = T(x_n) / ||T(x_n)||\ ,$ 
 i.e. the application of $T$ followed by normalization, 
 converges to the unitary eigen-vector corresponding to the 
 largest eigen-value. 


 In sections 4 and 5 it is shown, for statistical analysis in a 
scientific context, how the property of sharpness indicates that many, 
and perhaps some of the most relevant, scientific hypotheses are sharp,
and how  the property of stability, indicates  that considering these
hypotheses  is natural and reasonable. The statistical consequences of
these findings  will be discussed in sections 7 and 8. 
 Before that, however, a few other ConsTh concepts must be introduced
in sections 3 and 6.

  Autopoiesis found its name in the work of Maturana and Varela (1980), 
 together with a simple, powerful and elegant formulation using the 
 modern language of system's theory. 
  Nevertheless, some of the basic theoretical concepts, 
 such as those of self-organization and autonomy of living organisms, 
 have long historical grounds that some authors trace back to Kant.  
  As seen in Kant (1790, sec. 65) for example, a (self-organized)   
 {\it ``Organism''} is characterized as an entity in which,  

    \begin{quote} 
 {\it ``... every part is thought as `owing' its presence to the 
 `agency' of all the remaining parts, and also as existing 
 `for the sake of the others' and of the whole, that is as an 
 instrument, or organ.''}   

  {\it ``Its parts must in their collective unity reciprocally 
  produce one another alike as to form and combination, 
  and thus by their own causality produce a whole, 
  the conception of which, conversely, 
  -in a being possessing the causality according to 
   conceptions that is adequate for such a product- 
  could in turn be the cause of the whole according to a principle, 
  so that, consequently,  
  the nexus of `efficient causes' 
  (progressive causation, nexus effectivus) 
  might be no less estimated as an 
  `operation brought about by final causes'       
  (regressive causation, nexus finalis).'' } 
    \end{quote} 

  Baruch Spinoza, in his Ethics (1677, Part III, Propositions 6, 7 and 8), 
 defines the {\it Conatus} (effort, endeavour, impetus) of 
 self-preservation as the true essence of a being. 
  This concept has also been regarded as a remote precursor 
 of autopoiesis.         
  \begin{quote} 
  Prop. III-6: Everything, in so far as it is in itself, 
 endeavours to persist in its own being. 
    
  Prop. III-7: The endeavour, wherewith everything endeavours to
 persist in its own being, is nothing else but the actual 
 essence of the thing in question. 

  Prop. III-8: The endeavour,  whereby  a thing  endeavours to persist 
 in its being, involves  no  finite  time,  but  an indefinite time.  
  \end{quote}

   For further historical comments we refer the reader to 
 Zelleny (1980).

 \section{Functional Differentiation}
 \markboth{CHAPTER 1: EIGEN-SOLUTIONS AND STATISTICAL HYPOTHESES}
  {1.3 \ FUNCTIONAL DIFFERENTIATION}

In order to give appropriate answers to environmental complexities,
autopoietic systems can be hierarchically organized as Higher Order 
Autopoietic Systems. As in Maturana and Varela (1980, p.107,108,109), 
this notion is defined via the concept of Coupling: 

    \begin{quote} 
 {\it ``Whenever the conduct of two or more units is such that there
is a domain in which the conduct of each one is a function of the
conduct of the others, it is said that they are coupled in that
domain.''}

 {\it ``Such a composite system will necessarily be defined as a unity 
by the coupling relations of its component autopoietic systems in the 
space that the nature of the coupling specifies, and will remain as a 
unity as long as the component systems retain their autopoiesis which 
allows them to enter into those coupling relations.''}

 {\it ``An autopoietic system whose autopoiesis entails the autopoiesis of
the coupled autopoietic units which realize it, is an autopoietic
system of higher order.''}  
    \end{quote} 

 A typical example of a hierarchical system is a Beehive, a third order 
autopoietic system, formed by the coupling of individual Bees, the
second  order systems, which, in turn, are formed by the coupling of
individual Cells,  the first order systems.  

 The philosopher and sociologist Niklas Luhmann applied this notion to
the study of modern human societies and its systems. Luhmann's basic
abstraction is to look at social systems only at its higher 
hierarchical level, in which it is seen as an autopoietic communications
network. 
 In Luhmann's terminology, a communication event consists of:    
 Utterance, the form of transmission; 
 Information, the specific content; and 
 Understanding, the relation to future events in the network, such as
the  activation or suppression of future communications. 

    \begin{quote} 
{\it ``Social systems use communication as their particular mode of 
autopoietic (re)production. Their elements are communications that are
recursively produced and reproduced by a network of communications that
are not living units, they are not conscious units, they are not
actions. Their unity requires a synthesis of three selections, namely  
information, utterance and understanding (including misunderstanding).''} 
 Luhmann (1990b, p.3). 
    \end{quote}

 For Luhmann, society's best strategy to deal with increasing complexity
is the same as one observes in most biological organisms, namely,
differentiation. Biological organisms differentiate in specialized
systems, such as organs and tissues of a pluricellular life form 
(non-autopoietic or allopoietic systems), or specialized individuals in
an insect colony (autopoietic system). 
 In fact, societies and organisms can be characterized  by the way in
which they differentiate into systems. 
 For Luhmann, modern societies are characterized by a vertical 
differentiation into autopoietic functional systems, where  
each system is characterized by its code, program and (generalized)
media. The code gives a bipolar reference to the system, of what is
positive, accepted, favored or valid, versus what is negative, rejected,
disfavored or invalid. The program gives a specific context where the
code is applied, and the media is the space in which the system operates.

 Standard examples of social systems are:

 - Science: with a true/false code, working in a program set by a
 scientific theory, and having articles in 
 journals and proceedings as its media;   

 - Judicial: with a legal/illegal code, working in a program set by
existing laws and regulations, and having certified legal documents 
as its media;  

 - Religion: with a good/evil code, working in a program set by sacred
and hermeneutic texts, and having study, prayer and good deeds as its media;  

 - Economy: with a property/lack thereof code, working in a program set
 by economic planning scenarios and pricing methods, and having
 money and money-like financial assets as its media.  

 Before ending this section, a notion related to the
break-down of autopoiesis is introduced: Dedifferentiation 
(Entdifferenzierung) is the degradation of the system's internal
coherence, through adulteration, disruption, or dissolution of its own
autopoietic relations. One form of dedifferentiation (in either
biological or social systems) is the system's penetration by external
agents who try to use system's resources in a way that is not
compatible with the system's autonomy. In Lumann's conception of modern
society each system may be aware of events in other systems, 
that is, be cognitively open, but is required to maintain its 
differentiation, that is, be operationally closed. 
 In Luhmann's (1989, p.109) words:   

    \begin{quote} 
 {\it ``With functional differentiation... 
  Extreme elasticity is purchased at the cost of the peculiar rigidity 
 of its contextual conditions. 
 Every binary code claims universal validity, but only for its own 
 perspective. Everything, for example, can be either true of false, 
 but only true or false according to the specific theoretical programs 
 of the scientific system. Above all, this means that no function system 
 can step in for any other. None can replace or even relieve any other. 
 Politics can not be substituted for economy, nor economy for science, 
 nor science for law or religion, nor religion for politics, etc., 
 in any conceivable intersystem relations.''} 
    \end{quote}

 \section{Eigensolutions and Scientific Hypotheses}
 \markboth{CHAPTER 1: EIGEN-SOLUTIONS AND STATISTICAL HYPOTHESES}
  {1.4 \ SCIENTIFIC HYPOTHESES}

 The interpretation of scientific knowledge as an eigensolution of a
research process is part of a constructive approach to epistemology.
 Figure 1 presents an idealized  structure and dynamics of knowledge
production. 
 This diagram represents, on the Experiment side (left column) the  
laboratory or field operations of an empirical science, where experiments
are designed and built, observable effects are generated and
measured, and the experimental data bank is assembled. 
 On the Theory side (right column), the diagram represents the
theoretical work of statistical analysis, interpretation and (hopefully)
understanding according to accepted patterns. If necessary, new
hypotheses (including whole new theories) are formulated, motivating the
design of new experiments. 
 Theory and experiment constitute a double feed-back cycle making it clear
that the design of experiments is guided by the existing theory and its
interpretation, which, in turn, must be constantly checked, adapted or
modified in order to cope with the observed experiments. The whole
system constitutes an autopoietic unit, as seen in 
Krohn and K\"{u}ppers (1990, p.214): 

     \begin{quote} 
 {\it ``The idea of knowledge as an eigensolution of an operationally
closed combination between argumentative and experimental activities
attempts to answer the initially posed question of how the construction
of knowledge binds itself to its construction in a new way. The
coherence of an  eigensolution does not refer to an objectively given
reality but follows from the operational closure of the construction.
Still, different decisions on the selection of couplings may lead to
different, equally valid eigensolutions. Between such different
solutions no reasonable choice is possible unless a new operation of
knowledge is constructed exactly upon the differences of the given
solutions. But again, this frame of reference for explicitly relating
different solutions to each other introduces new choices with respect to
the coupling of operations and explanations. It does not reduce but
enhances the dependence of knowledge on decisions. On the other hand, 
the internal restrictions imposed by each of the chosen couplings do not
allow for any arbitrary construction of results. Only few are suitable
to mutually serve as inputs in a circular operation of knowledge.''}  
  \end{quote}

 
 \section{Sharp Statistical Hypotheses}
 \markboth{CHAPTER 1: EIGEN-SOLUTIONS AND STATISTICAL HYPOTHESES}
  {1.5 \ SHARP HYPOTHESES}

 Statistical science is concerned with inference and application of
probabilistic models. From what has been presented in the preceding
sections,  it becomes clear what the role of Statistics in scientific
research is, at least in the ConsTh view of scientific research:
 Statistics has a dual task,  to be performed both in the Theory and
the Experiment sides of the diagram in  Figure 1:


 \mbox{}  \\ 

 \begin{table}[h] 
 \begin{center}
 \begin{tabular}{c c c c c} 

 Experiment & &     & & Theory \\  \\   

 Operation-  & $\Leftarrow$ & Experiment  

                       &  $\Leftarrow$ &  Hypotheses   \\ 
 alization & & design   & &  formulation                \\ 
   $\Downarrow$   & &               & & $\Uparrow$        \\  
 Effects       &  \multicolumn{3}{c}{True/False} & Creative  \\ 
 observation    &  \multicolumn{3}{c}{eigen-solution} & interpretation  \\     

   $\Downarrow$   & &               & & $\Uparrow$         \\ 
 Data    & & Mnemetic & &  Statistical   \\ 
 acquisition  & $\Rightarrow$ & explanation  
 & $\Rightarrow$ & analysis \\  \\ 
 
 \multicolumn{2}{l}{Sample space} & & 
 \multicolumn{2}{r}{Parameter space} 
 \end{tabular} 
 \mbox{} \\ \mbox{} \\ 
 \mbox{} \\ \mbox{} \\ 
 \centerline{Figure 1: Scientific production diagram.}
 \end{center}    
 \end{table}

 - At the Experiment side of the diagram, the task of statistics is to
make probabilistic statements about the occurrence of pertinent events,
i.e. describe probabilistic distributions for 
 what, where, when or which 
 events can occur. If the events are to occur in the future,
these descriptions are called predictions, as is often the case in the
natural sciences. It is also possible (more often in social sciences) to
deal with observations related to past events, that may or may not be
experimentally generated or repeated, imposing limitations to the
quantity and/or quality of the available data. Even so, the habit of
calling this type of statement {``predictive probabilities''} will be
maintained. 

 - At the Theory side of the diagram, the role of statistics is to
measure the statistical support of hypotheses, i.e. to measure,
quantitatively, the hypotheses plausibility or possibility in the
theoretical framework where they were formulated, given the observed data.
 From the material presented in the preceding sections, it is also clear
that,  in this role, statistics is primarily concerned with measuring
the statistical support of sharp hypotheses, for hypotheses sharpness
(precision or discreteness) is an essential attribute of
eigen-solutions.

 Let us now examine how well the traditional statistical paradigms, and
in contrast the FBST, are able to take care of this dual task. In
order to examine this question, the first step is to distinguish what
kind of  probabilistic statements can be made. We make use of tree
statement categories: Frequentist, Epistemic and Bayesian:  

 Frequentist probabilistic statements are made exclusively on the 
basis of the frequency of occurrence of an event in a (potentially)
 infinite sequence of observations generated by 
 a random variable.

 Epistemic probabilistic statements are made on the basis of the
epistemic status (degree of belief, likelihood, truthfulness, validity)
of an event from the possible outcomes generated by a random variable.
This generation may be actual or potential, that is, may have been
realized or not, may be observable or not, may be repeated an infinite
or finite number of times. 

 Bayesian probabilistic statements are epistemic probabilistic statements
generated by the (in practice, always finite) recursive use of Bayes
formula:  
 \[ p_{n}(\theta) \propto p_{n-1}(\theta) p(x_n | \theta) \ . \]

 In standard models, the parameter $\theta$, a non observed random variable,
and the sample $x$, an observed random variable, are related through their
joint probability distribution, $p(x,\theta)$. 
 The prior distribution, $p_0(\theta)$, is the starting point 
for the Bayesian recursion operation. 
 It represents the initial available information about $\theta$. 
 In particular, the prior may represent 
 no available information, like distributions obtained via the 
 maximum entropy principle, see Dugdale (1996) and Kapur (1989). 
 The posterior distribution, $p_n(\theta)$, represents the available
information on the parameter after the n-th ``learning step'', in which
Bayes formula is used to incorporate the information carried by 
 observation $x_n$. 
 Because of the recursive nature of the procedure, the posterior
distribution in a given step is used as prior 
 in the next step.    

 Frequentist statistics dogmatically demands that all probabilistic
statements be frequentist. Therefore, any direct probabilistic
statement on the parameter space is categorically forbidden. Scientific
hypotheses are epistemic statements about the parameters of a
statistical model. 
 Hence, frequentist statistics can not make any direct statement about
the statistical significance (truthfulness) of hypotheses. 
 Strictly speaking it can only make statements at the Experiment side of
the diagram. The frequentist way of dealing with questions on Theory
side of the diagram, is to embed them some how into the Experiment side.
 One way of doing this is by using a construction in which the whole data
acquisition process is viewed as a single outcome of an imaginary infinite
meta random process, and then make a frequentist statement, on the meta
process, about the frequency of unsatisfactory outcomes of some
incompatibility measure of the observed data bank with the hypothesis.
 This is the classic (and often forgotten) rationale used when stating a
p-value. So we should always speak of the p-value of the data bank
(not of the hypothesis). The resulting conceptual confusion and
frustration (for most working scientists) with this kind of convoluted
reasoning is captured by a wonderful parody of Galileo's
dialogues in Rouanet et al. (1998). 

 A p-value is the probability of getting a sample that is more extreme
than the one we got. We should therefore specify which criterion is used
to define what we mean by more extreme, i.e., how do we order the sample
space, and usually there are several possible criteria to do that, 
 for examples, see Pereira and Wechsler (1993).

 \addtocounter{figure}{1}

 \begin{figure}[htb]
 \centerline{\includegraphics[height=5.5in, width=6.0in]{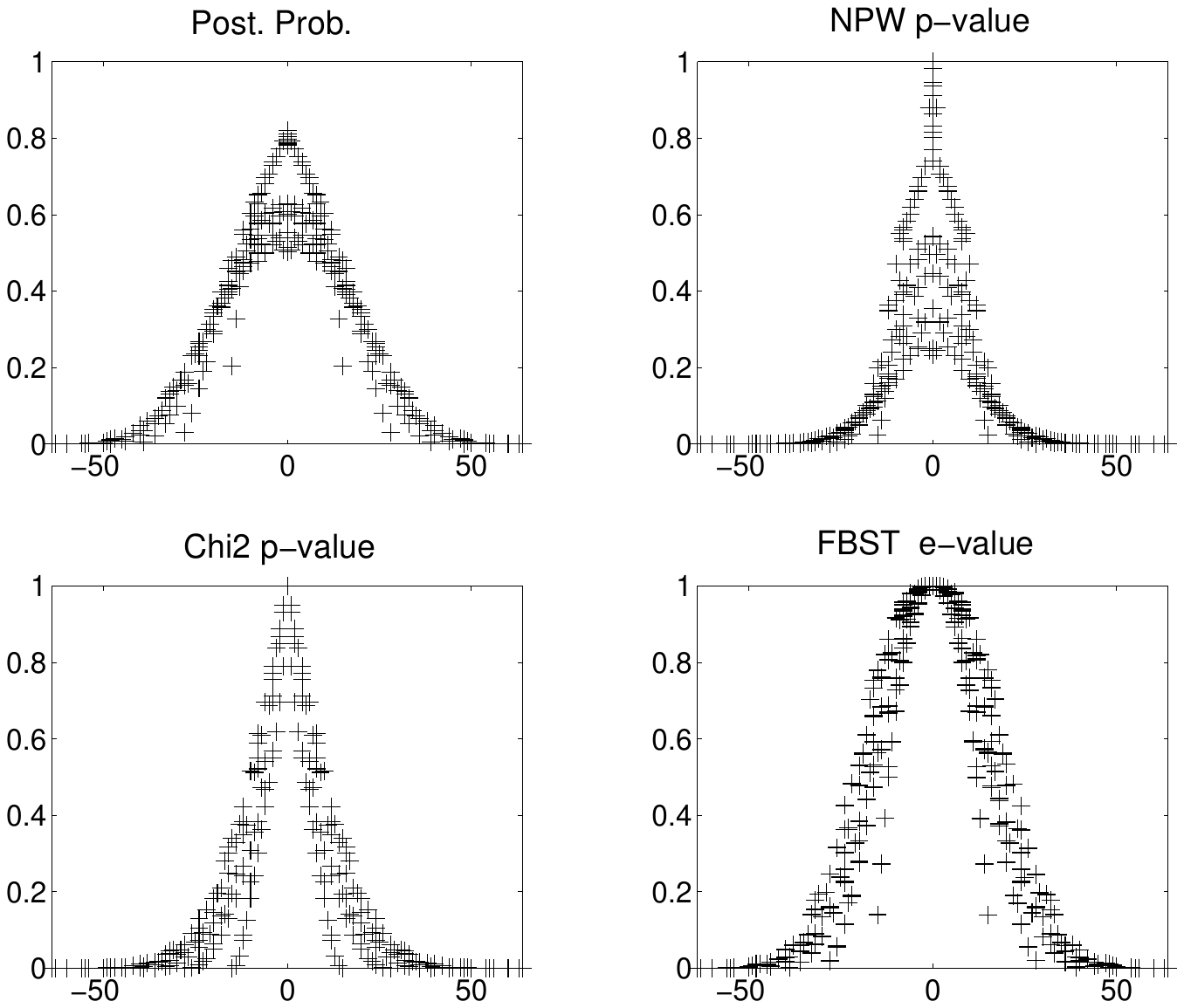}} 
 \vspace{-0.5cm} 
 \centerline{Figure 2: Independence Hypothesis, n=16.} 
 \end{figure}

 Figure 2 compares four statistics, namely,
 orthodox Bayesian posterior probabilities, 
 Neyman-Pearson-Wald (NPW) p-values, 
 Chi-square approximate p-values, and the FBST 
 evidence value in favor of $H$.
 In this example $H$ is the independence hypothesis in a $2\times 2$ 
 contingency table, for sample size $n=16$, see section A1 and B1. 
  The horizontal axis shows the ``diagonal asymmetry'' statistics  
 (difference between the diagonal products). 
  The statistics $D$ is an estimator of an unormalized version of 
 Person's correlation coefficient, $\rho$. 
  For detailed explanations, see Irony et al. (1995, 2000), 
 Stern and Zacks (2002) and Madruga, Pereira and Stern (2003).  
 \[ 
    D= x_{1,1}x_{2,2} - x_{1,2}x_{2,1} \ \ , \ \ \ 
    \rho= \frac{\sigma_{1,2}}{\sqrt{\sigma_{1,1}\sigma_{2,2}}}= 
    \frac{\theta_{1,1}\theta_{2,2} -\theta_{1,2}\theta_{2,1}} 
    {\sqrt{\theta_{1,1}\theta_{1,2}\theta_{2,1}\theta_{2,2}}} \ . 
 \]   

 Samples that are ``perfectly compatible with the hypothesis'',   
 that is, having no asymmetry, are near the center of the plot, 
 with increasingly incompatible samples to the sides.  
  The envelope curve for the resulting FBST e-values, to be commented 
 later in  this section, is smooth 
 and therefore level at its maximum,  where it reaches the value 1.    
  
 In contrast the envelope curves for the p-values take the form of a cusp, 
 i.e. a pointed curve, that is broken (non differentiable) 
 at its maximum, where it also reaches the value one.   
 The acuteness of the cusp also increases with increasing sample size. 
  In the case of NPW p-values we see, at the top of the cusp, a ``ladder'' 
 or ``spike'', with several samples with no asymmetry, but having
 different outcome probabilities, ``competing'' for the higher p-value.

 This is a typical collateral effect of the artifice that converts a 
question  about the significance of $H$, asking for a probability in 
the parameter space as an answer, into a question, conditional on $H$ 
being truth, about the outcome probability of the observed sample, 
offering a probability in the sample space as an answer. 
 This qualitative analysis of the p-value methodology gives us
an insight on typical abuses of  the expression  
{\it ``increase sample size to reject''}. 
 In the words of I.J. Good (1983, p.135): 

   \begin{quote} 
 {\it ``Very often the statistician doesn't bother to make it quite
clear whether his null hypothesis is intended to be sharp or only
approximately sharp....

 It is hardly surprising then that many
Fisherians (and Popperians) say that - you can't get (much) evidence in
favor of the null hypothesis but can only refute it.''} 
   \end{quote}

 In Bayesian statistics we are allowed to make probabilistic statements
on the parameter space, and also, of course, in the sample space. Thus it
seems that Bayesian statistics is the right tool for the job, and so it
is! Nevertheless, we must first examine the role played by DecTh 
in orthodox Bayesian statistics. 
  Since the pioneering work of de Finetti, Savage and many others,
orthodox Bayesian Statistics has developed strong and coherent
foundations grounded on DecTh, where many basic 
questions could be successfully analyzed and solved. 

 This foundations can be stratified in two layers:  

 - In the first layer, DecTh provides a coherence system for the use of 
probability statements, in the sense of Finetti (1974, 1981). 
 In this context, the FBST use of probability theory is fully compatible 
with DecTh, as shown in Madruga et al. (2001).  

 - In the second layer, DecTh provides an epistemological framework 
 for the interpretation of statistical procedures.  
 The FBST logical properties open the possibility of using and
benefiting   from alternative epistemological settings such as ConsTh.  
 Hence, DecTh does not have to be ``the tool for all trades''.  

 We claim that, in the specific case of  statistical procedures for
measuring the support (significance tests) for sharp scientific
hypotheses,  ConsTh provides a more adequate epistemological framework
than DecTh. 
 This point is as important as it is subtle. In order to understand it
let us first remember the orthodox paradigm, as it is concisely stated
in Dubins and Savage (1965, 12.8, p.229,230).   
 In a second quote, from Savage (1954, 16.3, p.254)  
 we find that sharp hypotheses, even if important, make
little sense in this paradigm, a position that is accepted throughout
decision theoretic Bayesian statistics,  
 as can also be seen in Levi (1974) and Maher et al. (1993).

    \begin{quote} 
 {\it ``Gambling problems in which the distributions of various
quantities are prominent in the description of the gambler's fortune
seem to embrace the whole of theoretical statistics according to one
view (which might be called the decision-theoretic Bayesian view) of
the subject.  

 ...From the point of view of decision-theoretic statistics, the gambler
in this problem is a person who must ultimately act in one of two ways
(the two guesses), one of which would be appropriate under one
hypothesis ($H_0$) and the other under its negation ($H_1$). 

 ...Many problems, of which this one is an instance, are roughly of the
following type. A person's opinion about unknown parameters is described
by a probability distribution; he is allowed successively to purchase
bits of information about the parameters, at prices that may depend
(perhaps randomly) upon the unknown parameters themselves, until he
finally chooses a terminal action for which he receives an award that
depends upon the action and parameters.'' 

  ``I turn now to a different and, at least for me, delicate topic in
connection with applications of the theory of testing. Much attention is
given in the literature of statistics to what purport to be tests of
hypotheses, in which the null hypothesis is such that it would not
really be accepted by anyone. 
 ... extreme (sharp) hypotheses, as I shall call them...  

 ...The unacceptability of extreme (sharp) null hypotheses is perfectly
well known; it is closely related to the often heard maxim that science
disproves, but never proves, hypotheses, The role of extreme (sharp)
hypotheses in science and other statistical activities seems to be
important but obscure. In particular, though I, like everyone who
practice statistics, have often ``tested'' extreme (sharp) hypotheses, I
cannot give a very satisfactory analysis of the process, nor say clearly
how it is related to testing as defined in this chapter and other
theoretical discussions.''}   
   \end{quote}

 As it is clearly seen, in the DecTh framework we speak about the
betting odds for ``the hypothesis wining on a gamble taking place in the
parameter space''. But since sharp hypotheses are zero (Lebesgue) measure
sets, our betting odds must be null, i.e. sharp hypotheses must be 
(almost surely) false. If we accept the ConsTh view that an important
class of  hypotheses concern the identification of eigen-solutions, and
that those are ontologically sharp, we have a paradox!  

 From these considerations it is not surprising that frequentist and
DecTh orthodoxy consider sharp hypotheses, at best as anomalous crude
approximations used when the scientist is incapable of correctly
specifying error bounds, cost, loss or utility functions, etc., 
or then just consider them to be {\it ``just plain silly''}. 
 In the words of  D.Williams (2002, p.234):

   \begin{quote} 
 {\it ``Bayesian significance of sharp hypothesis: a plea for sanity:
 ...It astonishes me therefore that some Bayesian now assign non-zero
prior probability  that a sharp hypothesis is exactly true to obtain
results which seem to support strongly null hypotheses which
frequentists would very definitely reject. (Of course, it is blindingly
obvious that such results must follow).''}  
   \end{quote}

 But no matter how many times statisticians reprehend scientist for their
sloppiness and incompetence, they keep formulating sharp hypotheses, as
if they where magnetically attracted to them. From the ConsTh plus FBST 
perspective they are, of course, just doing the right thing!   

 Decision theoretic statistics has also developed methods to deal with
sharp hypotheses, posting sometimes a scary caveat emptor for those
willing to use them.  
 The best known of such methods are Jeffreys' tests, based on Bayes
Factors that assign a positive prior probability mass to the sharp
hypothesis. 
 This positive prior mass is supposed to work like a handicap  system
designed to balance the starting odds and make the game ``fair''.  
 Out of that we only get new paradoxes, like the well documented Lindley's
paradox. In opposition to its frequentist counterpart, this is an 
 ``increase sample size to accept'' effect, see Shafer (1982). 


 The FBST e-value or evidence value supporting the hypothesis, $\ev(H)$,  
was specially designed to effectively evaluate the support for
a sharp  hypothesis, $H$. 
 This support function is based on the  posterior probability 
measure of a set called the tangential set, $\Tb(H)$, which is a non
zero measure set  (so no null probability paradoxes), see Pereira and
Stern (1999),  Madruga et al. (2003) and subsection A1 of the appendix. 

 Although $\ev(H)$ is a probability in the parameter space, it is
also a possibilistic support function. The word {\it possibilistic} 
carries a heavy load, implying that $\ev(H)$ complies with a very
specific logic (or algebraic) structure, 
 as seen in Darwishe and Ginsberg (1992), Stern (2003, 2004),  
 and subsection A3 of the appendix. 
 Furthermore the e-value has many necessary or desirable properties 
for a statistical support function, such as:

 1- Give an intuitive and simple measure of significance for the 
hypothesis in test, ideally, a {\it probability} defined directly
in the  original or {\it natural parameter space}. 

 2- Have an intrinsically geometric definition, independent of any
non-geometric aspect, like the particular parameterization of the
(manifold representing the) null hypothesis being tested, or the
particular coordinate system chosen for the parameter space, i.e., be an
 {\it invariant} procedure.

 3- Give a measure of significance that is smooth, i.e. 
 {\it continuous and differentiable}, on the hypothesis parameters and 
 sample statistics, under appropriate regularity conditions of the model.

 4- Obey the {\it likelihood principle} , i.e., the information
gathered from observations should be represented by, and only by, the
likelihood function.

 5- Require {\it no ad hoc artifice} like assigning a positive prior
probability to zero measure sets, or setting an arbitrary initial
belief ratio between hypotheses. 

 6- Be a {\it possibilistic} support function. 

 7- Be able to provide a {\it consistent} 
 test for a given sharp hypothesis. 

 8- Be able to provide {\it compositionality} operations in  
 complex models. 

 9- Be an {\it exact} procedure, not requiring ``large sample'' 
asymptotic approximations. 

 10- Allow the incorporation of previous experience or expert's opinion via
 (subjective) {\it prior distributions}.

 For a careful and detailed explanation of the FBST definition, its
computational implementation, statistical and logical properties, and
several already developed applications, the reader is invited to consult
some of the articles in the reference list. Appendix A provides a short 
review of the FBST, including its definition and main properties.


 \section{Semantic Degradation}
 \markboth{CHAPTER 1: EIGEN-SOLUTIONS AND STATISTICAL HYPOTHESES}
  {1.6 \ SEMANTIC DEGRADATION}

 In this section some constructivist analyses of 
dedifferentiation phenomena in social systems are reviewed. 
 If the conclusions in the last section are correct, it is
surprising how many times DecTh, sometimes with a very narrow 
 pseudo-economic interpretation, was misused in 
scientific statistical analysis. 
 The difficulties of testing sharp hypotheses in the traditional
statistical  paradigms are well documented, and extensively discussed in
the literature, see for example the articles in Harlow et al. (1997).   
 We hope the material in this section can help us understand these
difficulties as symptoms of problems with much deeper roots. 
  By no means the author is the first to point out the 
danger of analyses carried out by blind transplantation of categories
between heterogeneous systems. In particular, regarding the abuse of
economical analyses, Luhmann (1989, p.164) states:    

   \begin{quote} 
 {\it ``In this sense, it is meaningless to speak of ``non-economic''
costs. This is only a metaphorical way of speaking that transfers the
specificity of the economic mode of thinking indiscriminately to other
social systems.''} 
    \end{quote}

 For a sociological analysis of this phenomenon in the context of
 science, see for example 
 Fuchs (1996, p.310) and DiMaggio and Powell (1991, p.63):     
 
    \begin{quote} 
 {\it ``...higher-status sciences may, more or less aggressively,
colonize lower-status fields in an attempt at reducing them to their own
First Principles.  For particle physics, all is quarks and the four
forces. For neurophysiology, consciousness is the aggregate outcome of
the behavior of neural networks. For sociobiology, philosophy is done by
ants and rats with unusual large brains that utter metaphysical nonsense
according to acquired reflexes. In short, successful and credible chains
or reductionism usually move from the top to the bottom of disciplinary
prestige hierarchies.''}     

 {\it ``This may explain the popularity of giving an ``economical 
understanding'' to processes in functionally distinct areas even if (or
perhaps because) this  semantics is often hidden by statistical theory
and methods based on decision theoretic analysis.  This also may explain
why some areas, like ecology, sociology or psychology, are (or where)
far more prone to suffer this kind of  dedifferentiation by semantic
degradation than others, like physics.''}   
    \end{quote}

 Once the forces pushing towards systemic degradation are clearly
exposed, we hope one can understand the following corollary of von
Foerster  famous ethical and aesthetical imperatives: \\    
 - Theoretical imperative: Preserve systemic autopoiesis and semantic
integrity, for de-differentiation is in-sanity itself. \\   
 - Operational imperative: Chose the right tool for each job: 
 ``If you only have a hammer,  everything looks like a nail''.

 
 \section{Competing Sharp Hypotheses}
 \markboth{CHAPTER 1: EIGEN-SOLUTIONS AND STATISTICAL HYPOTHESES}
  {1.7 \ COMPETING SHARP HYPOTHESES}

 In this section we examine the concept of {\it Competing Sharp
Hypotheses.}  This concept has several variants, but the basic idea is
 that a good scientist should never test a single sharp hypothesis, for
it would be an unfair faith of the poor sharp hypothesis standing all
alone against everything else in the world. Instead, a good scientist
should  always confront a sharp hypothesis with a competing sharp
hypotheses, making the test a fair game. 
 As seen in Good (1983, p.167,135,126): 

   \begin{quote} 
 {\it 
 ``Since I regard refutation and corroboration as both valid criteria
for this demarcation it is convenient to use another term, Checkability,
to embrace both processes. I regard checkability as a measure to which a
theory is scientific, where checking is to be taken in both its positive
and negative senses, confirming and disconfirming.''  

 ``...If by the truth of Newtonian mechanics we mean that it is
approximately true in some appropriate well defined sense we could
obtain strong evidence that it is true; but if we mean by its truth that
it is exactly true then it has already been refuted.''   

``...I think that the initial probability is positive for every
self-consistent scientific theory with consequences verifiable in a
probabilistic sense. No contradiction can be inferred from this 
assumption since the number of statable theories is at most countably
infinite (enumerable).''  

 ``...It is very difficult to decide on numerical values for the
probabilities, but it is not quite so difficult to judge the ratio of
the subjective initial probabilities of two theories by comparing their
complexities. This is one reason why the history of science is
scientifically  important.''  
 }
    \end{quote}

 The competing sharp hypotheses argument does not directly contradict 
the epistemological framework presented in this chapter, and it may be 
appropriate under certain circumstances. It may also mitigate or
partially  remediate the paradoxes pointed out in the previous sections
when testing sharp  hypotheses in the traditional frequentist or
orthodox Bayesian settings.  However, the author does not believe that
having competing sharp hypotheses is neither a necessary condition for
good science practice,  nor an accurate description  of science history.

 Just to stay with Good's example, let us quickly examine the very first
major incident in the tumultuous  debacle of Newtonian mechanics. 
 This incident was Michelson's experiment on the effect of 
 ``aethereal wind'' over the speed of light, see   
 Michelson and Morley (1887) and  Lorentz et al. (1952).  
 A clear and lively  historical account to this experiment can be 
found in  Jaffe (1960). 
 Actually Michelson found no such effect, i.e. he found the speed of
light to be constant, invariant with the relative speed of the observer.  
 This result, a contradiction in Newtonian mechanics, is easily
explained by  Einstein's special theory of relativity. The fundamental
difference between the  two theories is their symmetry or invariance
groups: Galileo's group for  Newtonian mechanics, Lorentz' group for
special relativity. 
 A fundamental result of physics, Noether's Theorem, states that for 
every continuous symmetry in a physical theory, there must exist 
 an invariant quantity or conservation law. 
 For detail the reader is refered to 
 Byron  and Fuller (1969, V-I, Sec. 2.7), 
 Doncel et al. (1987), Gruber et al. (1980-98), Houtappel et al. (1965), 
 French (1968), Landau and Lifchitz (1966), Noether (1918), 
 Wigner (1970), Weyl (1952).
 Conservation laws are sharp hypotheses ideally suited for experimental 
checking. Hence, it seems that we are exactly in the situation of
competing sharp  hypotheses, and so we are today, from a far away
historical perspective. 
 But this is a post-mortem analysis of Newtonian mechanics.  
 At the time of the experiment there was no competing theory.  
 Instead of confirming an effect, specified only within an order of
magnitude,  Michelson found, for his and everybody else's astonishment, 
an, up to the experiment's precision, null effect.

 Complex experiments like Michelson's require a careful analysis of
experimental errors, identifying all significant source of 
measurement noise and fluctuation. 
 This kind of analysis is usual in experimental physics, and    
motivates  a brief comment on a secondary source of criticism on the
use of sharp hypotheses. 
 In the past, one often had to work with over simplified statistical
models. This situation was usually imposed by limitations such as the
lack of better or more realistic  models, or the unavailability of the
necessary  numerical algorithms or the computer power to use them. 
 Under these limitations, one often had to use minimalist statistical
models or approximation techniques, even when  these models or
techniques were not recommended. 
 These models or techniques were instrumental to provide feasible tools 
for statistical analysis, but made it very difficult to work 
(or proved very ineffective) with complex systems, scarce observations, 
very large data sets, etc.  
 The need to work with complex models, and other difficult situations
requiring the use of sophisticated statistical methods and techniques,
is very common (and many times inescapable) in research areas dealing
with complex systems like biology, medicine, social sciences,
psychology, and many other fields, some of them distinguished with the
mysterious appellation of ``soft'' science. A colleague once put it to
me like this: ``It seems that physics got all the easy problems...''. 
 
 If there is one area where the computational techniques of Bayesian
statistics have made dramatic contributions in the last decades, that is
the analysis of complex models. The development of advanced statistical 
computational techniques like Markov Chain Monte Carlo (MCMC) methods,
Bayesian and neural networks, random fields models, and many others,
make us hope that most of the problems related to the use of over
simplified models can now  be overcome. 
 Today good statistical practice requires all statistically significant
influences to be incorporated into the model, and one seldom finds an
acceptable excuse not to do so; see also Pereira and Stern (2001).

 \section{Final Remarks}
 \markboth{CHAPTER 1: EIGEN-SOLUTIONS AND STATISTICAL HYPOTHESES}
  {1.8 \ FINAL REMARKS}

 It should once more be stressed that most of the material presented
in sections 2, 3, 4, and 6 is not new in ConsTh. 
 Unfortunately ConsTh has had a minor impact in statistics, and 
sometimes provoked a hostile reaction from the ill-informed. 
 One possible explanation of this state of affairs may be found in the
historical development of ConsTh. 
 The constructivist reaction to a dogmatic  
realism prevalent in hard sciences, specially in the XIX and the
beginning of the XX century, raised a very outspoken rhetoric
intended to make explicitly clear how naive and fragile the
foundations of this  over simplistic realism were.  
 This rhetoric was extremely successful, quickly awakening and forever
changing the minds of those directly interested in the fields of history
and philosophy of science, and spread rapidly into many other areas.
 Unfortunately the same rhetoric could, in a superficial reading, make  
ConsTh be perceived as either hostile or intrinsically incompatible with
the use of quantitative and statistical methods, or leading to an
extreme forms of subjectivism.  

 In ConsTh, or (objective) Idealism as presented in this chapter,
neither does one claim to have access to a ``thing in itself'' or ``Ding
an sich'' in the external  environment, see Caygill (1995), as do
dogmatic forms of realism, nor does one surrender to solipsism, as
do skeptic forms of
 subjectivism,   including some representatives of the subjectivist
school of probability and statistics, 
 as seen in Finetti (1974, 1.11, 7.5.7).  
 In fact, it is the role of the external constraints imposed by the
environment, together with the internal autopoietic relations of the
system, to guide the convergence of the learning process to precise
eigen-solutions, these being at the end, the ultimate or real objects of
 scientific knowledge. 
  As stated by Luhmann (1990a, 1995): 

    \begin{quote} 
  {\it ``...constructivism maintains nothing more than the 
  unapproachability of the external world ``in itself'' and the closure 
  of knowing - without yielding, at any rate, to the old skeptical or 
  ``solipsistic'' doubt that an external world exists at all-...''}  
   Luhmann (1990a, p.65). 

 {\it ``...at least in systems theory, they (statements) refer to 
 the real world. Thus the concept of system refers to something that 
 in reality is a system and thereby incurs the responsibility of 
 testing its statements against reality.''} 
  Luhmann (1995, p.12). 

 {\it ``...both subjectivist and objectivist theories of knowledge 
 have to be replaced by the system / environment distinction, which then
 makes the distinction subject / object irrelevant.'' }    
  Luhmann (1990a, p.66). 
    \end{quote}


 The author hopes to have shown that ConsTh not only gives a balanced
and effective view of the theoretical / experimental aspects of
scientific research but also that it is well suited (or even better
suited) to give the necessary epistemological foundations for the use of
quantitative methods of statistical analysis needed in the practice of
science. 
 It should also be stressed, according to author's interpretation
of ConsTh, the importance of measuring the statistical support for sharp
hypotheses. 
 In this setting, the author believes that, due to its statistical and
logical characteristics, the FBST is the right tool for the job, 
and hopes to have motivated the reader to find more about the
FBST definition, theoretical properties, efficient computational
implementation, and several of the already developed applications, in
some of the articles in the reference list.        
 This perspective opens interesting areas for further research.  
 Among them, we mention the following two.

  \subsection{Noether and de Finetti Theorems}

  The first area for further research has to do with some
similarities between Noether theorems in physics, and de Finetti type
theorems in statistics. 
 Nother theorems provide invariant physical quantities or conservation
laws from symmetry transformation groups of the physical theory, and
conservation laws are sharp hypotheses by excellence. 
 In a similar way, de Finetti type  theorems provide invariant
distributions from symmetry transformation groups of the statistical 
model. Those invariant distributions can in turn provide  prototypical
sharp hypotheses in many application areas. 
 Physics has its own heavy apparatus to deal with the all important 
issues of invariance and symmetry. 
 Statistics, via de Finetti theorems, can provide  such an apparatus for
other areas, even in situations that are not naturally embedded in a
heavy mathematical formalism,  see Feller (1968, ch.7) 
 and also Diaconis (1987, 1988), Eaton (1989), Nachbin (1965) 
Renyi (1970) and Ressel (1987).

 \subsection{Compositionality}

  The second area for further research has to do with one of the
properties of eigen-solutions  mentioned by von Foerster that has not
been directly explored in this chapter, namely that eigen-solutions are
``composable'',  see Borges and Stern (2005) and section A4. 
 Compositionality properties concern the relationship between the
credibility,  or truth value, of a  complex hypothesis, $H$, and those
of  its elementary constituents, $H^j$, $j=1\ldots k$.  
 Compositionality questions play a central role in analytical
philosophy.

 According to Wittgenstein (2001, 2.0201, 5.0, 5.32):  

 - Every complex statement can be analyzed from its 
    elementary constituents. 
 
 - Truth values of elementary statement are the results of 
 those statements' truth-functions (Wahrheitsfunktionen).

 - All truth-function are results of successive applications to 
 elementary constituents of a finite number of
 truth-operations (Wahrheitsoperationen).

 Compositionality questions also play a central role in far more 
 concrete contexts, like that of 
 reliability engineering, see Birnbaum et al. (1961, 1.4):     
  
    \begin{quote} 
  {\it ``One of the main purposes of a mathematical theory of
reliability is  to develop means by which one can evaluate the
reliability of a structure  when the reliability of its components are
known. The present study will be concerned with this kind of
mathematical development. It will be necessary for this purpose to
rephrase our intuitive concepts of structure, component, reliability,
etc. in more formal language, to restate carefully our assumptions, and
to introduce an appropriate mathematical apparatus.''}  
    \end{quote}

 In Luhmann (1989, p.79) we find the following remark on the evolution
of science that  directly hints the importance of this property:  

    \begin{quote} 
 {\it ``After the (science) system worked for several centuries under
these conditions it became clear where it was leading. This is something
that idealization, mathematization, abstraction, etc. do not describe
adequately. It concerns the increase in the capacity of decomposition
and recombination, a new formulation of knowledge as the product of
analysis and synthesis. In this case analysis is what is most important
because the further decomposition of the visible world into still
further decomposable molecules and atoms, into genetic structures of
life or even into the sequence human/role/action/ action-components as
elementary units of systems uncovers an enormous potential for
recombination.''} 
    \end{quote}

 In the author's view, the composition (or re-combination) of scientific
knowledge and its use, so relevant in  technology development and
engineering, can give us a different perspective (perhaps a, bottom-up,
as opposed to the top-down perspective in this chapter) on the
importance of sharp hypotheses in science and technology practice. It
can also provide some insight on the valid forms of iteration of science
with other social systems or, in Luhmann's terminology, how science does
(or should) ``resonate'' in human society.

%% file: CAPE2.TEX
 

 \chapter{Language and the Self-Reference Paradox}

 \mbox{} 
 
 {\flushright {\it 
 ``If the string is too tight it will snap, \\   
 but if it is too loose it will not play.''}     \\ 
 Siddhartha Gautama. \mbox{} \\ } 

 \mbox{} 

  {\flushright {\it \small  
  ``The most beautiful thing we can experience is the mysterious. \\ 
  It is the source of all true art and all science. \ He to whom  \\ 
  this emotion is a stranger, who can no longer pause to wonder  \\ 
  and  stand rapt in awe, is as good as dead:  His eyes are closed.''} \\ 
  Albert Einstein (1879 - 1955). \mbox{} \\ } 
   
  \mbox{}

 \section{Introduction}
 \markboth{CHAPTER 2: LANGUAGE AND SELF-REFERENCE}
  {2.1 \ INTRODUCTION}

 In Chapter 1 it is shown how the eigen-solutions found in the
practice of science are naturally represented by statistical sharp
hypotheses.  
 Statistical sharp hypotheses are routinely stated as natural ``laws'', 
conservation ``principles'' or invariant ``transforms'', 
and most often take the form of functional equations, like $h(x)=c$. 
 Chapter 1 also discusses why the eigen-solutions' essential attributes 
of  discreteness (sharpness), stability, and composability, indicate
that considering such hypotheses in the practice of science is natural
and reasonable. 
 Surprisingly, the two standard statistical theories for testing 
hypotheses, classical (frequentist p-values) and orthodox Bayesian 
(Bayes factors), have well known and documented problems for handling
or interpreting sharp  hypotheses.  
 These problems are thoroughly reviewed, from   statistical,
methodological, systemic and epistemological perspectives.   

 Chapter 1 and appendix A present the FBST, or Full Bayesian 
Significance Test, an unorthodox Bayesian significance test specifically
designed for this task. 
 The mathematical and statistical properties of the FBST
are carefully analyzed. In particular, it is shown how the FBST fully
supports the test and identification of eigen-solutions in the practice
of science, using procedures that take into account all the essential
attributes pointed by von Foerster. 
 In contrast to some alternative belief calculi or logical formalisms 
based on discrete algebraic structures, the FBST is based on continuous 
statistical models. This makes it easy to support concepts like  sharp
hypotheses, asymptotic convergence and stability,  and these  are
essential concepts in the representation of eigen-solutions.  
 The same chapter presents cognitive constructivism as a coherent 
epistemological framework that is compatible with the FBST formalism, 
and vice-versa. 
 I will refer to this setting as the Cognitive Constructivism plus
FBST formalism, or CogCon+FBST framework for short.

 The discussion in Chapter 1 raised some interesting questions, some
of which we will try to answer in the present chapter.  
 The first question relates to the role and the importance of language
in the emergence of eigen-solutions and is discussed in section 2. 
 In answering it, we make extensive use of the William Rasch ``two-front
war'' metaphor of cognitive constructivism, as exposed in Rasch (2000).
 As explained in section 4, this is the war against dogmatic realism at
one front, and against skepticism or solipsism, at the second. 
 The results of the first part of the paper are summarized in section 5. 
 To illustrate his arguments, Rasch uses some ideas of Niels Bohr
concerning quantum mechanics. In section 3, we use some of the same
ideas to give concrete examples of the topics under discussion. 
  The importance (and also the mystery) related to the role of 
 language in the practice of science was one of the major concerns  
 of Bohr's philosophical writings, see Bohr (1987, I-IV), 
 as exemplified by his  famous ``dirty dishes'' metaphor:     
 
 \begin{quotation}
 \noindent  
 {\it ``Washing dishes and language can in some respects be compared. 
  We have dirty dishwater and dirty towels and nevertheless finally
 succeed in getting the plates and glasses clean. 
  Likewise, we have unclear terms and a logic limited in an unknown way
 in its field of application -- but nevertheless we succeed in using it 
 to bring clearness to our understanding of nature.''} Bohr (2007).  
 \end{quotation}

 The second question, posed by S{\o}ren Brier, which asks whether the 
CogCon+FBST framework is compatible with and can benefit from the
concepts of Semiotics and Peircean philosophy, is addressed in section
6. In section 7 I present my final remarks.

 Before ending this section a few key definitions related to the 
concept of eigen-solution are reviewed. 
 As stated in Maturana and Varela (1980, p.10), the concept of recurrent
state is the key to understand the concept of cognitive domain in an
autopoietic system. 

 \begin{quotation}
 {\it ``Living systems as units of interaction specified by their
  conditions of being living systems cannot enter into interactions that
  are not specified by their organization. 
 The circularity of their organization continuously brings them back
to the same internal state (same with respect to the cyclic process).
Each internal state requires that certain conditions (interactions with
the environment) be satisfied in order to proceed to the next state.
Thus the circular organization implies the prediction that an
interaction that took place once will take place again.  
   If this does not happen the system maintains its integrity (identity
  with respect to the observer) and enters into a new prediction. In a
  continuously changing environment these predictions can only be
  successful if the environment does no change in that which is predicted.
  Accordingly,   the predictions implied in the organization of the living
  system are not predictions of particular events, but of classes of
  inter-actions. 
 Every interaction is a particular interaction, but every prediction is
a prediction of a class of interactions that is defined by those
features of its elements that will allow the living system to retain its
circular organization after the interaction, and thus, to interact
again. This makes living systems inferential systems, and their domain
of interactions a cognitive domain.''} 
 \end{quotation}

 The epistemological importance of this circular (cyclic or recursive)
regenerative processes and their eigen (auto, equilibrium, fixed,
homeostatic, invariant, recurrent, recursive) -states, both in concrete
and abstract autopoietic systems, are further investigated in Foerster
and Segal (2001, p.145, 127-128):

 \begin{quotation}
 {\it ``The meaning of recursion is to run through one's own path again. 
 One of its results is that under certain conditions there exist indeed
solutions which, when reentered into the formalism, produce again the
same solution. These are called ``eigen-values'', ``eigen-functions'',
``eigen-behaviors'', etc., depending on which domain this formation is
applied - in the domain of numbers, in functions, in behaviors, etc.''} 
 \end{quotation}

 \begin{quotation}
 {\it ``Objects are tokens for eigen-behaviors. Tokens stand for
something else. In exchange for money (a token itself for gold held by
one's government, but unfortunately no longer redeemable), tokens are
used to gain admittance to the subway or to play pinball machines. In
the cognitive realm, objects are the token names we give to our
eigen-behavior. 
  When you speak about a ball, you are talking about the
  experience arising from your recursive sensorimotor behavior when
  interacting with that something you call a ball. The ``ball'' as object
  becomes a token in our experience and language for that behavior which
  you know how to do when you handle a ball. 
 This is the constructivist's insight into what takes place when we 
talk about our experience with objects.''} 
 \end{quotation}

 Von Foerster also establishes several essential attributes of
these eigen-solutions, as quoted in the following paragraph from
Foerster (2003c, p.266). 
 These essential attributes can be translated into very specific 
mathematical properties, that are of prime importance when  
investigating several aspects of the CogCon+FBST framework.

 \begin{quotation}
 \noindent {\it ``Eigenvalues have been found ontologically to be
discrete, stable, separable and composable, while ontogenetically to
arise as equilibria that determine themselves through circular
processes. Ontologically, Eigenvalues and objects, and likewise,
ontogenetically, stable behavior and the manifestation of a subject's
``grasp'' of an object cannot be distinguished.''} 
 \end{quotation}

 \section{Eigen-solutions and Language}
 \markboth{CHAPTER 2: LANGUAGE AND SELF-REFERENCE}
  {2.2 \ EIGEN-SOLUTIONS AND LANGUAGE}

 Goudsmit (1998, sec.2.3.3, Objects as warrants for eigenvalues), finds
an apparent disagreement between the form in which eigen-solutions
emerge, according to von Foster and Maturana:  

 \begin{quotation}
 \noindent {\it ``Generally, von Foersters concept of eigenvalue
concerns the value of a function after a repeated (iterative)
application of a particular operation. ... 

 \noindent This may eventually result in a stable performance, which is
an eigenvalue of the observers behavior. The emerging objects are
warrants of the existence of these eigenvalues.

 \noindent ... contrary to von Foerster, Maturana considers the
consensuality of distinctions as necessary for the bringing forth of
objects. It is through the attainment of consensual distinctions that
individuals are able to create objects in language. 
 ''} 
 \end{quotation}

 Confirmation for the position attributed by Goudsmit to von Foerster
can be found in several of his articles. In Foerster (2003a, p.3), for
example, one finds:  

 \begin{quotation}
 \noindent {\it ``... I propose to continue the use of the term
`self-organizing system,' whilst being aware of the fact that this term
becomes meaningless, unless the system is in close contact with an
environment, which possesses available energy and order, and with which
our system is in a state of perpetual interaction, such that it somehow
manages to `live' on the expenses of this environment. ... 

 \noindent ... So both the self-organizing system plus the energy and
order of the environment have to be given some kind of pre-given
objective reality for this view points to function."}   \end{quotation}

 Confirmation for the position attributed by Goudsmit to Maturana can
also be found in several of his articles. In Maturana (1988,sec.9.iv), 
for example, one finds:

 \begin{quotation}
 \noindent {\it ``Objectivity. Objects arise in language as consensual
coordinations of actions that in a domain of consensual distinctions are
tokens for more basic coordinations of actions, which they obscure.
Without language and outside language there are no objects because
objects only arise as consensual coordinations of actions in the
recursion of consensual coordinations of actions that languaging is. For
living systems that do not operate in language there are no objects; or
in other words, objects are not part of their cognitive domains. ...
Objects are operational relations in languaging.''}  
 \end{quotation}

 The standpoint of Maturana is further characterized in the following 
paragraphs from Brier (2005, p.374):

 \begin{quotation}  
 \noindent  
 {\it ``The process of human knowing, is the process in which we,
through languaging, create the difference between the world and
ourselves; between the self and the non-self, and thereby, to some
extent, create the world by creating ourselves. But we do it by relating
to a common reality which is in some way before we made the difference
between `the world' and `ourselves' make a difference, and we do it on
some kind of implicit belief in a basic kind of order `beneath it all'.
 I do agree that it does not make sense to claim that the world exists
completely independently of us. But on the other hand it does not make
sense to claim that it is a pure product of our explanations or
conscious imagination.'' 

 \noindent 
 ``...it is clear that we do not create the trees and the mountains
through our experiencing or conversation alone. But Maturana is close to
claim that this is what we do.''}  
 \end{quotation}

 In order to understand the above comments, one must realize that
Maturana's viewpoints, or at least his rhetoric, changed greatly over
time, ranging from the ponderate and precise statements in Maturana
and Varela (1980), to some extreme positions assumed in 
Maturana (1991, p.36-44)), see next paragraph. 
 Maturana must have had in mind the celebrated quote by 
Albert Einstein at the beginning of this chapter.

 \begin{quotation} 
 {\it ``Einstein said, and many other scientists have agreed with him,
that scientific theories are free creations of the human mind, and he
marveled that through them one could understand the universe. 
 The criterion of validation of scientific explanation as operations in
the praxis of living of the observer, however, permit us to see how it
is that the first reflection of Einstein is valid, and how it is that
there is nothing marvelous in that it is so.''  

 \noindent 
 ``Scientific explanations arise operationally as generative mechanisms
accepted by us as scientists through operations that do not entail or 
imply any supposition about an independent reality, so that in fact 
there is no confrontation with one, nor is it necessary to have one 
even if we believe that we can have one.''

 \noindent 
 ``Quantification (or measurements) and predictions can be used in the 
generation of a scientific explanation but do not constitute the source
of its validity. 
 The notions of falsifiability (Popper), verificability, or confirmation
would apply to the validation of scientific knowledge only if this were
a  cognitive domain that revealed, directly or indirectly, by denotation
or connotation, a transcendental reality independent of what the
observer does...''

 \noindent 
 ``Nature is an explanatory proposition of our experience with elements 
of our experience. Indeed, we human beings constitute nature with our
explaining, and with our scientific explaining we constitute nature as 
the domain in which we exist as human beings (or languaging living
systems).''} 
 \end{quotation}

 Brier (2005, p.375) further contrasts the standpoint of Maturana 
with that of von Foerster:

 \begin{quotation} 
 \noindent 
 {\it ``Von Foerster is more aware of the philosophical demand that to
put up a new epistemological position one has to deal with the problem
of solipsism and of pure social constructivism.'' 

 \noindent 
 ``The Eigenfunctions do not just come out of the blue. In some,
yet only dimly viewed, way the existence of nature and its `things'
and our existence are intertwined in such a way that makes it very
difficult to talk about. Von Foerster realizes that to accept the
reality of the biological systems of the observer leads into further
acceptance about the structure of the environment.''} 	
 \end{quotation}

 While the position adopted by von Foerster appears to be more realistic
or objective, the one adopted by Maturana  seems more Idealistic or
(inter) subjective. Can these two different positions, which may seem so
discrepant, be reconciled? Do we have to chose between an idealistic or
a realistic position, or can we rather have both? 
 This is one of the questions we address in the next sections. 
  
 In Chapter 1 we used an example of physical eigen-solution (physical
invariant) to illustrate the ideas in discussion, namely, the speed of
light constant, $c$. Historically, this example is tied to the birth of
Special Relativity theory, and the debacle of classical physics. 
 In this chapter we will illustrate them with another important
historical example, namely, the Einstein-Podolsky-Rosen paradox.
Historically, this example is tied to questions concerning the
interpretation of quantum mechanics. This is one of the main topics of
the next section.

 \section{The Languages of Science}
 \markboth{CHAPTER 2: LANGUAGE AND SELF-REFERENCE}
  {2.3 \ LANGUAGES OF SCIENCE}

 At the end of the 19th century, classical physics was the serene
sovereign of science. Its glory was consensual and uncontroversial.
However, at the beginning of the 20th century, a few experimental
results challenged the explanatory power of classical physics. 
 The problems appeared in two major fronts that, from a historical
perspective, can  be linked to the theories (at that time still non
existent) of Special Relativity and quantum mechanics. 

 At that time, the general perception of the scientific community was
that these few open problems could, should and would be accommodated in
the framework of classical physics. Crafting sophisticated structural
models such as those for the structure of ether  (the medium in which
light was supposed to propagate), and those for atomic structure,
was typical of the effort to circumvent these open problems by artfully
maneuvering classical physics. But physics and engineering laboratories
insisted, building up a barrage of new and challenging experimental
results.
 
 The difficulties with the explanations offered by classical physics not
only persisted, but also grew in number and strength.  In 1940 the
consensus was that classical physics had been brutally defeated, and
Relativity and quantum mechanics were acclaimed as the new sovereigns.
Let us closely examine some facts concerning the development of quantum
mechanics (QM).      
 
 One of the first steps in the direction of a comprehensive QM theory
was given in 1924 by Louis de Broglie, who postulated the particle-wave
duality principle, which states that every moving particle has an
associated pilot wave of wavelength $\lambda=h/mv$, where $h$ is
Planck's constant and $mv$ is the particle's momentum, i.e., the product
of its mass and velocity.
 In 1926 Erwin Schr\"{o}dinger stated his wave equation, capable of
explaining all known quantic phenomena, and predicting  several new ones
that where latter confirmed by new experiments. Schr\"{o}dinger  theory
is known as Orthodox QM, see Tomonaga (1962) and Pais (1988) for 
detailed historical accounts. Orthodox QM uses a mathematical formalism
based on a complex wave equation, and shares much of the descriptive
language of de Broglie's particle-wave duality principle. 
 
 There is, however, something odd in the wave-particle descriptions of
orthodox QM. When describing a model we speak of each side of a double
faced wave-particle entity, as if each side existed by itself, and then
inextricably fuse them together in  the mathematical formalism. Quoting
Cohen (1989, p.87),  

 \begin{quotation}
 \noindent {\it ``Notice how our language shapes our imagination. To say
that a particle is moving in a straight line really means that we can
set up particle detectors along the straight line and observe the
signals they send. These  signals would be consistent with a model of
the particle as a single chunk of mass moving (back and forth) in
accordance with Newtonian particle physics. It is important to emphasize
that we are not claiming that we know what the particle is, but only
what we would observe  if we set up those particle detectors.''}
 \end{quotation}

 From Schroedinger's equation we can derive Heisenberg's uncertainty
principle, which states that we can not go around measuring everything
we want until we pin down every single detail about (the classical
entities in our wave-particle model of) reality. One instance of the
Heisenberg uncertainty principle states that we can not simultaneously
measure a particle position and momentum beyond a certain accuracy. 
 One way of interpreting this instance of the Heisenberg uncertainty
principle goes as follows: In classical Newtonian physics our particles
are ``big enough'' so that our  measurement devices can obtain the
information we need about the particle without disturbing it. In QM, on
the other hand, the particles are so small that the measurement
operation will always disturb the particle. For example, the light we
have to use in order to illuminate the scene, so we can see where 
the particle is, has to be  so strong, relative to the particle size,
that it ``blows'' the particle away changing its velocity. 
 The consequence is that we cannot (neither in practice, nor even in
principle) simultaneously measure with arbitrary precision, both the
particle's position and momentum. Hence, we have to learn how to tame
our imagination and constrain our language.

 The need to exercise a strict discipline over what kinds of statements
to use was a lesson learned by 20th century physics - a lesson that
mathematics had to learn a bit earlier.  A classical example from set
theory of a statement that cannot be allowed is the Russell's catalog
(class, set), defined in Robert (1988, p.x) as: 

 \begin{quotation}
 \noindent {\it ``The `catalogue of all catalogues not mentioning
themselves.' Should one include this catalogue in itself? ... Both
decisions lead to a contradiction!''} 
 \end{quotation}

 Robert (1988) indicates several ways to avoiding this paradox (or
antinomy). All of them imply imposing a (very reasonable) set of rules
on how to form valid statements. Under any of these rules, Russell's
definition becomes an invalid or ill posed statement and, as such,
should be disregarded,   
 see Halmos (1998, ch.1 and 2) and Dugundji (1966, ch.1) for 
 introductory texts and Aczel (1988) for an alternative view.  
 Measure theory (of Borel, Lebesgue, Haar, etc.) was a fundamental
achievement of 20th century  mathematics. It defines measures (notions
such as mass, volume and probability) for parts of $R^n$. However not
all parts of $R^n$ are included, and we must refrain of speaking about
the measure of
 inadmissible (non-measurable) sets,  see Ulam (1943) for a short
article, Kolmogorov and Fomin (1982) for a standard text, and Nachbin
(1965) and Bernardo (1993) for extensions pertinent to the FBST 
formalism. 
 The main subject in Robert (1988) is Non Standard  Analysis, a form of
extending the languages of both Set Theory and Real Analysis, see the
observations in section 6.6 and also Davis (1977, sec.3.4), 
Goldblatt (1998) and Nelson (1987).

 All the preceding examples of mathematical languages have one thing in
common: When crafting a specific language, one has to carefully define
what kinds of statements are accepted as valid ones. Proper use of the
language must be constrained to valid statements. Such constraints are
necessary in order to preserve language coherence.   
    
 The issue of what kinds of statements should be accepted as valid in QM
is an interesting and still subsisting issue, epitomized by the famous
debate at the Brussels Solvay conference of 1930 between Niels Bohr and
his friend and opponent Albert Einstein.  Ruhla (1992, ch.7 and 8) and
Baggott (1992, under the topic hidden variables) give very intuitive
reviews of the subject, requiring minimal mathematical expertise.
 Without the details concerning the physics involved, one can describe
the debate as: While  Bohr suggested very strict rules for admissible
statements in QM, Einstein advocated for more amiable ones. In 1935
Einstein,  Podolsky and Rosen suggested a ``gedankenexperiment'', known
as the EPR paradox, as a compelling argument supporting Einstein's point
of view. D.Bohm, in 1952 and J.Bell, in 1964, contributed to the debate
by showing that the EPR paradox could lead to concrete experiments
providing a way to settle the debate on empirical grounds. It was only
in 1972 that the first EPR experiment could be performed in practice.
The observational evidence from these experiments seems to favor Bohr's
point of view!   
 
  One of today's standard formalisms for QM is Abstract QM, see Hughes
 (1992) or Chester (1987) for a readable text and Cohen (1989) 
 for a concise and formal treatment. 
  For an alternative formalism based on Niels Bohr's concept of 
 complementarity, see Bohr (1987, I-IV) and Costa and Krause (2004).     
  Other formalisms may also become usefull, see for example  
 Kolmanovskii and Nosov (1986, sec.2.3) and Zubov (1983). 
 Abstract QM, which is very clean and efficient, can be stratified in
two layers. In the first layer, all basic calculations are carried out
using an algebra of operators in (Rigged) Hilbert spaces. In a second
layer, the results of these calculations are interpreted as
probabilities of obtaining specific results in physical measurements,
see also Rijsbergen (2004). One advantage of using the stratified
structure of abstract QM is that it naturally avoids (most of) the
danger of forming invalid statements in QM language. 
 Cohen (1989, p.vii) provides the following historical summary:  

 \begin{quotation}
 \noindent {\it ``Historically, ... quantum mechanics developed in three
stages. First came a collection of ad hoc assumptions and then a
cookbook of equations known as (orthodox) quantum mechanics. The
equations and their philosophical underpinning were then collected into
a model based on mathematics of Hilbert space. From the Hilbert space
model came the abstraction of quantum logics.''}  
 \end{quotation}

 From the above historical comments we draw the following conclusions: 

 \pagebreak 

 \begin{quotation}
 \noindent {\bf 3.1.} Each of the QM formalisms discussed in this
section, namely, de Broglie wave-particle duality principle,
Schr\"{o}dinger orthodox QM and Hilbert space abstract QM, operates 
like a language. Maturana stated that objects arise in language. 
 He seems to be right.

 \noindent {\bf 3.2.} It seems also that new languages must be created
(or discovered) to provide us the objects corresponding to the structure
of the environment, as stated by von Foerster.  

 \noindent {\bf 3.3.} Exercising a strict discipline concerning what
kinds of statements can be used in a given language and context, seems
to be vital in many areas.  

 \noindent {\bf 3.4.} It is far from trivial to create, craft, discover,
find and/or use a language  so that ``it works'', providing us the
``right'' objects (eigen-solutions). 

 \noindent {\bf 3.5.} Even when everything looks (for the entire
community) fine and well, new empirical evidence can bring our theories
down as a castle of cards.
 \end{quotation}

 As indicated by an anonymous referee, abstract formalisms or
languages do not exist in a vacuum, but sit on top of (or are embedded
in) natural (or less abstract) languages.
 This bring us to the interesting and highly relevant issues of
hierarchical language structures and constructive ladders of
objects, including interdependence analyses  between objects at 
different levels of such complex structures,  see Piaget (1975) for
an early reference. 
 For a recent concrete example of the scientific relevance of such
interdependences in the field of Psychology, using a Factor Analysis
statistical model, see Shedler and Westen (2004, 2005);   
 These issues are among of the main topics addressed in 
 chapter 3 and forthcoming articles.

 \section{The Self-Reference Paradox}
 \markboth{CHAPTER 2: LANGUAGE AND SELF-REFERENCE}
  {2.1 \ SELF-REFERENCE PARADOX}

 The conclusions established in the previous section may look
reasonable. In 3.4, however, what exactly are the ``right'' objects?
 Clearly, the ``right'' objects are ``those'' objects we more or less 
clearly see and can point at, using as reference language the language
we currently use. 

 There! I have just fallen, head-on, into the quicksands of the
self-reference paradox. Don't worry (or do worry), but note this: 
 The self-reference paradox is unavoidable, 
 especially as long as we use English or any other natural human language.  

 Rasch (2000, p.73,85) has produced a very good description of the
self-reference paradox and some of its consequences: 

 \begin{quotation}
 \noindent {\it ``having it both ways seems a necessary consequence...
One cannot just have it dogmatically one way, nor skeptically the
other... One oscillates, therefore, between the two positions, neither
denying reality nor denying reality's essentially constructed nature.
One calls this not idealism or realism, but (cognitive)
constructivism.''}  

 \noindent {\it ``What do we call this oscillation? We call it paradox.
Self - reference and paradox - sort of like love and marriage, horse and
carriage.''}  
 \end{quotation}

 Cognitive Constructivism implies a double rejection: 
 That of a solipsist denial of reality, and that of any dogmatic
knowledge of the same reality. Rasch uses the ``two front war'' metaphor
to describe this double rejection. Carrying the metaphor a bit further,
the enemies of cognitive constructivism could be portrayed, 
or caricatured, as:

 \begin{quotation}
 \noindent - Dogmatism despotically requires us to believe in its
(latest) theory. Its statements and reasons should be passively accepted
with fanatic resignation as infallible truth; 

 \noindent - Solipsism's anarchic distrust wishes to preclude any
established order in the world. Solipsism wishes to transform us into
autistic skeptics, incapable of establishing any stable knowledge about
the environment in which we live.   

 \noindent  We refer to Caygill (1995, dogmatism) for a historical
perspective on the Kantian use of some of the above terms.   

 \end{quotation}

 Any military strategist will be aware of the danger in the oscillation
described by Rasch, which alternately exposes a weak front. The enemy at
our strong front will be subjugated, but the enemy at our weak front
will hit us hard. Rasch sees a solution to this conundrum, even
recognizing that this solution may be difficult to achieve, 
 Rasch (2000, p.85): 

 \begin{quotation}
 \noindent {\it ``There is a third choice: to locate oneself directly on
the invisible line that must be drawn for there to be a distinction 
 mind / body (system / environment) in the first place. 
  Yet when one attempts to land on that perfect center, one finds
oneself oscillating wildly from side to side, perhaps preferring the
mind (system) side, but over compensating to the body (environment) side
- or vice versa. 

 \noindent 
  The history of post-Kantian German idealism is a history of the failed 
 search for this perfect middle, this origin or neutral ground outside 
 both mind and body that would nevertheless actualize itself as a perfect
 transparent mind/body within history. Thus, much of contemporary
 philosophy  that both follows and rejects that tradition has become 
 fascinated by, even if trapped in, the mind/body oscillation.''}    
 \end{quotation}

  So, the question is: How do we land on Rasch' fine (invisible) line,
 finding the perfect center and avoiding dangerous oscillations? 
  This is the topic of the next section.

 \section{Objective Idealism and Pragmatism}
 \markboth{CHAPTER 2: LANGUAGE AND SELF-REFERENCE}
  {2.5 \ OBJECTIVE IDEALISM AND PRAGMATISM}

 We are now ready for a few definitions of basic epistemological terms.
These definitions should help us build epistemic statements in a clear
and coherent form according to the CogCon+FBST perspective.  

 \begin{quotation} 
 \noindent {\bf 5.1. Known (knowable) Object:} An actual (potential)
eigen-solution of a given system's interaction with its environment. 
 In the sequel, we may use a somewhat more friendly terminology by 
simply using the term Object.
 
 \noindent {\bf 5.2. Objective (how, less, more):} Degree of conformance
of an object to the essential attributes of an eigen-solution. 

 \noindent {\bf 5.3.  Reality:} A (maximal) set of objects, as
recognized by a given system, when interacting with single objects or
with compositions of objects in that set. 

 \noindent {\bf 5.4. Idealism:} Belief that a system's knowledge of an 
object is always dependent on the systems' autopoietic relations. 

 \noindent {\bf 5.5. Realism:} Belief that a system's knowledge of an 
object is always dependent on the environment's constraints. 
 
 \noindent {\bf 5.6. Solipsism, Skepticism:} Idealism without Realism. 

 \noindent {\bf 5.7. Dogmatic Realism:} Realism without Idealism.  

 \noindent {\bf 5.8. Realistic or Objective Idealism:} Idealism
and Realism. 
    
 \noindent {\bf 5.9. ``Something in itself'':} 
 This expression, used in reference to a specific object, 
 is a marker or label for ill posed statements. 
 \end{quotation}

 Cog-Con+FBST assumes an objective and idealistic epistemology.
 Definition 5.9 labels some ill posed dogmatic statements. 
 Often, the description of the method used to access   
 something in itself  looks like: 
 
 - Something that an observer would observe if the (same)
observer did not exist, or   

 - Something that an observer could observe if he made no
observations, or     

 - Something that an observer should observe in the environment
without interacting with it (or disturbing it in any way),      
 and many other equally nonsensical variations.

 Some of the readers may not like this form of labeling this kind of
invalid statement, preferring to use, instead, a more elaborate
terminology, such as ``object in parenthesis'' (approximately) as
object, ``object without parenthesis'' (approximately) as  
something in itself, etc.
 There may be good reasons for doing so, for example, this elaborate
language has the advantage of automatically stressing the differences
between constructivist and dogmatic epistemologies, see Maturana (1988),
Maturana and Poerksen (2004) and Steier (1991). 
 Nevertheless, we have chosen our definitions in agreement with some
very pragmatic advice given in Bopry (2002):   
  
 \begin{quotation}
 \noindent {\it ``Objectivity as defined by a (dogmatic) realist
epistemology may not exist within a constructivist epistemology; but,
part of making that alternative epistemology acceptable is gaining
general acceptance of its terminology. As long as the common use of the
terms is at odds with the concepts of an epistemological position, that
position is at a disadvantage.
 Alternative forms of inquiry need to coopt terminology in a way that is
consistent with its own epistemology. I suggest that this is not so
difficult. The term objective can be taken back...''} 
 \end{quotation}
 
 Among the definitions 5.1 to 5.9, definition 5.2 plays a key role. 
 It allows us to say how well an eigen-solution manifests von Foerster's
essential attributes, and consequently, how good (objective) is our
knowledge of it. However, the degree of objectivity can not be assessed
in the abstract, it must be assessed by the means and methods of a given
empirical science, namely the one within which the eigen solution is
presented. Hence, definition 5.2 relies on an ``operational approach'',
and not on metaphysical arguments. Such an operational approach may be
viewed with disdain by some philosophical schools. Nevertheless, for
C.S.Peirce it is


 \begin{center}
 {\it ``The Kernel of Pragmatism''}, CP 5.464-465:  
 \end{center}

 \begin{quotation}
 \noindent {\it ``Suffice it to say once more that pragmatism is, in
itself, no doctrine of metaphysics, no attempt to determine any truth of
things. It is merely a method of ascertaining the meanings of hard words
and of abstract concepts. ...  
 All pragmatists will further agree that their method of ascertaining
the meanings of words and concepts is no other than that experimental
method by which all the successful sciences (in which number nobody in
his senses would include metaphysics) have reached the degrees of
certainty that are severally proper to them today; this experimental
method being itself nothing but a particular application of an older
logical rule, `By their fruits ye shall know them'. ''}  
 \end{quotation}

 Definition 5.2 also requires a belief calculus specifically designed to
measure the statistical significance, that is, the degree of
support of empirical data to the existence of an eigen-solution.
 In Chapter 1 we showed why confirming the existence of an
eigen-solution naturally corresponds to testing a sharp statistical
hypotheses,  and why the mathematical properties of FBST e-values
correspond to the essential attributes of an eigen-solution as stated by
von Foerster.  
 In this sense, the FBST calculus is perfectly
adequate to support the use of the term Objective and correlated terms
in scientific language. Among the most important properties of the
e-value mentioned in Chapter 1 and Appendix A, we find: 

 \noindent {\bf Continuity:} Give a measure of significance that is
smooth, i.e. {\it continuous and differentiable}, on the hypothesis
parameters and the sample statistics, under appropriate regularity
conditions of the statistical model.    

 \noindent {\bf Consistency:} Provide a {\it consistent}, that is, 
 asymptotically convergent significance measure for a given sharp 
 hypothesis.

 Therefore, the FBST calculus is a formalism that allow us to assess,
continuously and consistently, the objectivity of an eigen-solution, 
by means of a convergent significance measure, see Chapter 1. 
 We should stress, once more, that achieving comparable goals using 
alternative formalisms based on discrete algebraic structures may be, 
in general, rather difficult.   
 Hence, our answer to the question of how to land on Rasch's perfect
center is: Replace unstable oscillation for stable convergence!

 Any dispute about objectivity (epistemic quality or value of an object
of knowledge), should be critically examined and evaluated within this
pragmatic program. This program (in the Luhmann's sense) includes the
means and methods of the empirical science in which the object of
knowledge is presented, and the FBST belief calculus, used to evaluate
the empirical support of an object, given the available experimental
data.

 Even if over optimistic (actually hopelessly utopic), it is worth
restating Leibniz' flag of {\it Calculemus}, as found in 
 Gerhardt (1890, v.7, p.64-65):   

 \begin{quotation}
 \noindent {\it ``Quo facto, quando orientur controversiae, non magis
disputatione opus erit inter duos philosophos, quam inter duos
Computistas. Sufficiet enim calamos in manus sumere sedereque ad abacos,
et sibi mutuo (accito si placet amico) dicere: Calculemus.''} 
 \end{quotation}

 A contemporary translation could read: {\it Actually, if controversies
were to arise, there would be no more need for dispute between two
philosophers, rather than between two statisticians. For them it would
suffice to reach their computers and, in friendly understanding, say to
each other: Let us calculate!}


 \section{The Philosophy of C.S.Peirce}
 \markboth{CHAPTER 2: LANGUAGE AND SELF-REFERENCE}
  {2.6 \ PHILOSOPHY OF C.S.PEIRCE}

 In the previous sections we presented an epistemological perspective
based on  a pragmatic objective idealism. Objective idealism and
pragmatism are also distinctive characteristics of the philosophy of
C.S.Peirce. 
 Hence the following question, posed by S{\o}ren Brier,  
that we examine in this section:   
 Is the CogCon+FBST framework compatible with and can it benefit 
from the concepts of Semiotics and Peircean philosophy?  

 In Chapter 1 we had already explored the idea that  eigen-solutions,
as discrete entities, can be named, i.e., become  signs in a language
system, as pointed by von Foerster in Segal (2001, p.128):

 \begin{quotation}
 \noindent {\it 
 ``There is an additional point I want to make, an important point.
 Out of an infinite continuum of possibilities, recursive operations
carve out a precise set of discrete solutions. Eigen-behavior generates
discrete, identifiable entities. Producing discreteness out of infinite
variety has incredibly important consequences. It permits us to begin
naming things. Language is the possibility of carving out of an infinite
number of possible experiences those experiences which allow stable
interactions of your-self with yourself.''} 
 \end{quotation}

 We believe that the process of recursively ``discovering'' objects of
knowledge, identifying them by signs in language systems, and using
these languages to ``think'' and structure our lives as self-concious
beings, is the key for understanding concepts such as signification and
meaning. These ideas are explored, in a great variety of contexts, in
Bakken and Hernes (2002), Brier (1995), Ceruti (1989), Efran et al.
(1990), Eibel-Eibesfeldt (1970), Ibri (1992), Piaget (1975), Wenger et
al. (1999), Winograd and Flores (1987) and many others. 
 Conceivably, the key underlying  common principle is stated in 
Brier (2005, p.395):

 \begin{quotation}
 \noindent {\it ``The key to the understanding of understanding,
consciousness, and communication is that both the animals and we humans
live in a self-organized signification sphere which we not only project
around us but also project deep inside our systems. 
 Von Uexk\"{u}ll calls it ``Innenwelt'' (Brier 2001). 
 The organization of signs and the meaning they get through the habits
of mind and body follow very much the principles of second order
cybernetics in that they produce their own Eigenvalues of sign and
meaning and thereby create their own internal mental organization. 
 I call this realm of possible sign processes for the signification
sphere. In humans these signs are organized into language through social
self-conscious communication, and accordingly our universe is organized
also as and through texts. 
 But of course that is not an explanation of meaning.''}  
 \end{quotation}

 When studying the organization of self-conscious beings and trying to
understand semantic concepts such as signification and meaning, or
teleological concepts such as finality, intent and purpose, we move
towards domains concerning systems of increasing complexity that are
organized as higher hierarchical structures, like the domains of
phenomenological, psychological or  sociological sciences.
 In so doing, we leave the domains of natural and technical sciences
behind, at least for a moment, see  Brent and Bruck (2006) and Muggleton
(2006), in last month's issue of {\it Nature} (March 2006, when this
article was written), for two perspectives on future developments.

 As observed in Brier (2001), the perception of the  objects of
knowledge, changes from more objective or realistic to more idealistic
or (inter) subjective as we progress to higher hierarchical levels.
 Nevertheless, we believe that the fundamental nature of objects of
knowledge as eigen-solutions, with all the essential attributes 
pointed out by von Foerster, remains just the same. 
 Therefore, a sign, as understood in the CogCon+FBST framework, always
stands for the following triad:

 \begin{quotation}
 \noindent {\bf S-1.} 
 Some perceived aspects, characteristics, etc.,   
 concerning the  organization of the autopoietic system. 

 \noindent {\bf S-2.}  
 Some perceived aspects, characteristics, etc.,  
 concerning the  structure of the system's environment. 

 \noindent {\bf S-3.} 
 Some object (discrete, separable, stable and composable
eigen-solution based on the particular aspects stated in S-1 and S-2)
concerning the interaction of the autopoietic system with its
environment.   
 \end{quotation}


 This triadic character of signs bring us, once again, close to the
semiotic theory of C.S.Peirce, offering many opportunities for further
theoretical and applied research. 
 For example, we are currently using statistical psychometric
analyses in an applied semiotic project for the development of software
user interfaces, for related examples see Ferreira (2006). 
 We defer, however, the exploration of these opportunities to
forthcoming articles.

 In the remainder of this section we focus on a more basic investigation
that, we believe, is a necessary preliminary step that must be
undertaken in order to acquire a clear conceptual horizon that
will assist a sound and steady progress in our future research.
 The purpose of this investigation is to find out whether the CogCon+FBST
framework can find a truly compatible ground in the basic concepts of
Peircean philosophy.
 We proceed establishing a conceptual mapping of the fundamental
concepts used to define the CogCon+FBST epistemological framework into
analogous concepts in Peircean philosophy.
 Before we start, however, a word of caution: The work of C.S.Peirce is
extremely rich, and open to many alternative interpretations. 
 Our goal is to establish the compatibility of CogCon+FBST with one
possible interpretation, and not to ascertain reductionist deductions,   
in any direction.

 The FBST is a Continuous Statistical formalism. Our first step in
constructing this conceptual mapping addresses the following questions:
 Is such a formalism amenable to a Perircean perspective? If so, which
concepts in Peircean philosophy can support the use of such a formalism?

 \noindent {\bf 6.1 Probability and Statistics:} The FBST is a
probability theory based statistical formalism. Can the probabilistic
concepts of the FBST find the necessary support in concepts of Peircean
philosophy? We believe that Tychism is such a concept in Peircean
philosophy, providing the first element in our conceptual mapping. In CP
6.201 Tychism is defined as: 

 \begin{center}
 {\it ``... the doctrine that absolute chance is a factor of the
universe.''}  
 \end{center}

 \noindent {\bf 6.2 Continuity:} As stated in the previous section, the
CogCon+FBST program pursues the stable convergence of the epistemic
e-values given by the FBST formalism. The fact that FBST is a belief
calculus based on continuous mathematics is essential for its
consistency and convergence properties. Again we have to ask: Does the
continuity concept used in the FBST formalism have an analogous concept
in Peircean philosophy? We believe that the analogy can be established
with the concept of Synechism, thus providing the second element in our
conceptual mapping.

 In CP 6.169 synechism is defined as:

 \begin{quotation}
 \noindent 
 {\it ``that tendency of philosophical thought which insists upon the
idea of continuity as of prime importance in philosophy and, in
particular, upon the necessity of hypotheses involving true
continuity.''}    
 \end{quotation}

 \noindent {\bf 6.3 Eigen-Solutions:} A key epistemological concept in
the CogCon +FBST perspective is the notion of eigen-solution. Although
the system theoretic concept of Eigen-solution cannot possibly have an
exact correspondent in Peirce philosophy, we believe that Peirce's 
fundamental concept of ``Habit'' or ``Insistency'' offers an  adequate
analog. Habit, and reality, are defined as: 

 \begin{center}
 {\it ``The existence of things consists in their regular behavior.''},
CP 1.411. 
 \end{center}

 \begin{quotation}
 \noindent {\it ``Reality is insistency. That is what we mean by
`reality'.  It is the brute irrational insistency that forces us to
acknowledge the reality of what we experience, that gives us our
conviction of any singular.''}, CP 6.340.  
 \end{quotation}

 However, the CogCon+FBST concept of eigen-solution is characterized by
von Foerster by several essential attributes. Consequently, 
 in order that the conceptual mapping under construction can be coherent, 
 these characteristics have to be mapped accordingly. In the following
paragraphs we show that the essential attributes of sharpness
(discreteness), stability and compositionality can indeed be adequately
represented.

 \noindent {\bf 6.3a Sharpness:} The first essential attribute of
eigen-solutions stated by von Foerster is discreteness or sharpness. As
stated in Chapter 1, it is important to realize that, in the sequel,
the term `discrete', used by von Foerster to qualify eigen-solutions in
general, should be replaced, depending on the specific context, by terms
such as lower-dimensional, precise, sharp, singular, etc. 
 As physical laws or physical invariants, sharp hypotheses are formulated 
as mathematical equations.

 Can Peircean philosophy offer a good support for sharp hypotheses?
Again we believe that the answer is in the affirmative. The following
quotations should make that clear. The first three passages are taken
from Ibri (1992, p.84-85) and the next two from CP, 1.487 and CP 1.415,
see also NEM 4, p.136-137 and CP 6.203. 

 \begin{quotation}
 \noindent {\it ``an object (a thing) IS only in comparison with a
continuum of possibilities from which it was selected.''}  

 \noindent {\it ``Existence involves choice; the dice of infinite faces,
from potential to actual, will have the concreteness of one of them.''} 

 \noindent {\it ``...as a plane is a bi--dimensional singularity,
relative to a tri-dimensional space, a line in a plane is a topic
discontinuity, but each of this elements is continuous in its proper
dimension.''} 

 \noindent {\it`` Whatever is real is the law of something less real.
Stuart Mill defined matter as a permanent possibility of sensation. What
is a permanent possibility but a law?''}   

 \noindent {\it ``In fact, habits, from the mode of their formation,
necessarily consist in the permanence of some relation, and therefore,
on this theory, each law of nature would consist in some permanence,
such as the permanence of mass, momentum, and energy. In this respect,
the theory suits the facts admirably.''}  
 \end{quotation}
       
 \noindent {\bf 6.3b Stability:} The second essential attribute of
eigen-solutions stated by von Foerster is stability. As stated in Stern
(2005), a stable eigen-solution of an operator, defined by a fixed-point
or invariance equation, can be found (built or computed) as the limit of
a sequence of recursive applications of the operator. Under appropriate
conditions (such as within a domain of attraction, for instance) the
process convergence and its limiting eigen-solution will not depend on
the starting point.

A similar notion of stability for an object-sign complex is given by Peirce. As stated in CP 1.339: 

 \begin{quotation}
 \noindent {\it ``That for which it (a sign) stands is called its
object; that which it conveys, its meaning; and the idea to which it
gives rise, its interpretant. The object of representation can be
nothing but a representation of which the first representation is the
interpretant. But an endless series of representations, each
representing the one behind it, may be conceived to have an absolute
object at its limit.''} 
 \end{quotation}

 \noindent {\bf 6.3c Compositionality:} The third essential attribute of
eigen-solutions stated by von Foerster is compositionality. As stated in
Chapter 1 and Appendix A, compositionality properties
concern the relationship between the credibility, or truth value, of a 
complex hypothesis, $H$, and those of  its elementary constituents,
$H^j$, $j=1\ldots k$.
 Compositionality is at the very heart of any theory of language, see
Noeth (1995). 
 As an example of compositionality, see CP 1.366 and CP 6.23. 
 Peirce discusses the composition of forces, that is, how the components 
are combined using the parallelogram law.  


 \begin{quotation}
 \noindent {\it ``If two forces are combined according to the
parallelogram of forces, their resultant is a real third... 
 Thus, intelligibility, or reason objectified, is what makes Thirdness
 genuine.''}. 

 \noindent {\it ``A physical law is absolute. What it requires is an
exact relation. Thus, a physical force introduces into a motion a
component motion to be combined with the rest by the parallelogram of
forces;''}.  
 \end{quotation}

 In order to establish a minimal mapping, there are two more concepts in
CogCon+FBST to which we must assign adequate analogs in Peircean
philosophy.

 \noindent {\bf 6.4 Extra variability:} In Chapter 1 the importance
of incorporating all sources of noise and fluctuation, i.e., all the
extra variability statistically significant to the problem under study,
into the statistical model is analyzed. The following excerpt from CP
1.175 indicates that Peirce's notion of falibillism may be used to
express the need for allowing and embracing all relevant (and in
practice inevitable) sources of extra variability. According to Peirce,
falibilism is {\it ``the doctrine that there is no absolute certainty in
knowledge''}.

 \begin{quotation}
 \noindent {\it ``There is no difficulty in conceiving existence as a
matter of degree. The reality of things consists in their persistent
forcing themselves upon our recognition. If a thing has no such
persistence, it is a mere dream. Reality, then, is persistence, is
regularity.  ... as things (are) more regular, more persistent, they
(are) less dreamy and more real. Fallibilism will at least provide a big
pigeon-hole for facts bearing on that theory.''} 
 \end{quotation}

 \noindent {\bf 6.5 - Bayesian statistics:} FBST is an Unorthodox
Bayesian statistical formalism. Peirce has a strong and unfavorable
opinion about Laplace's theory of inverse probabilities.  

 \begin{quotation}
 \noindent {\it ``...the majority of mathematical treatises on
probability follow Laplace in results to which a very unclear conception
of probability led him. ... This is an error often appearing in the
books under the head of `inverse probabilities'.''} CP 2.785. 
 \end{quotation}

 Due to his theory of inverse probabilities, Laplace is considered one
of the earliest precursors of modern Bayesian statistics. Is there a
conflict between CogCon+FBST and Peirce's philosophy? We believe that a
careful analysis of Peirce arguments not only dissipates potential
conflicts, but also reinforces some of the arguments used in Chapter 1. 
  
 Two main arguments are presented by Peirce against Laplace's inverse
probabilities. In the following paragraphs we will identify these
 arguments and present an up-to-date analysis based on the FBST
 (unorthodox) Bayesian view:   

 \noindent {\bf 6.5a - Dogmatic priors vs. Symmetry and 
            Maximum Entropy arguments:} 

 \begin{quotation}
 \noindent {\it ``Laplace maintains that it is possible to draw a
necessary conclusion regarding the probability of a particular
determination of an event based on not knowing anything at all 
[about it]; that is, based on nothing. ... 
 Laplace holds that for every man there is one law (and necessarily but
one) of dissection of each continuum of alternatives so that all the
parts shall seem to that man to be `\'{e}galement possibles' in a
quantitative sense, antecedently to all information.''},  CP 2.764. 
 \end{quotation}

 The dogmatic rhetoric used at the time of Laplace to justify ad hoc
prior distributions can easily backfire, as it apparently did for
Peirce. Contemporary arguments for the choice of prior distributions are
based on MaxEnt formalism or symmetry relations, see Dugdale (1996),
Eaton (1989), Kapur (1989) and Nachbin (1965). Contemporary arguments
also examine the initial choice of priors by sensitivity analysis, for
finite samples, and give asymptotic dissipation theorems for large
samples, see DeGroot (1970), Gelman et al. (2003) and Stern (2004).  
 We can only hope that Peirce would be pleased with the contemporary
state of the art. These powerful theories have rendered ad hoc priors
unnecessary, and shed early dogmatic arguments into oblivion. 

 \noindent {\bf 6.5b- Assignment of probabilities to (sharp) hypotheses
vs. FBST possibilistic support structures:}

 \begin{quotation}
 \noindent {\it ``Laplace was of the opinion that the affirmative
experiments impart a definite probability to the theory; and that
doctrine is taught in most books on probability to this day, although it
leads to the most ridiculous results, and is inherently
self-contradictory. It rests on a very confused notion of what
probability is. Probability applies to the question whether a specified
kind of event will occur when certain predetermined conditions are
fulfilled; and it is the ratio of the number of times in the long run in
which that specified result would follow upon the fulfillment of those
conditions to the total number of times in which those conditions were
fulfilled in the course of experience.''}, CP 5.169. 
 \end{quotation}

 In the second part of the above excerpt Peirce expresses a classical
(frequentist) understanding of having probability in the sample space,
and not in the parameter space, that is, he admits predictive probability
statements but does not admit epistemic  probability statements. The
FBST is a Bayesian formalism that uses both predictive and epistemic
probability statements, as explained in Chapter 1. 
 However, when we examine the reason presented by Peirce for adopting
this position, in the first part of the excerpt, we find a remarkable
coincidence with the arguments presented in Stern (2003, 2004, 2006,
2007) against the orthodox Bayesian methodology for testing sharp
hypotheses: 
 The FBST {\it does not}  attribute a {\it probability} to the theory
(sharp hypothesis) being tested, as do orthodox Bayesian tests, but
rather a degree of {\it possibility}. In Stern (2003, 2004, 2006, 2007)
we analyze  procedures that attribute a probability to a given theory,
and came to the exact same conclusion as Pierce did, namely, those
procedures are absurd.

 \noindent {\bf 6.6 Measure Theory:} Let us now return to the Peircean
concept of Synechism, to discuss a technical point of contention
between orthodox Bayesian statistics and the FBST unorthodox Bayesian
approach. The FBST formalism relies on some form of Measure theory, see
comments in section 3. 
 De Finetti, the founding father of the orthodox school of Bayesian
statistics, feels very uncomfortable having to admit the existence of
non-measurable sets when using measure theory in dealing with
probabilities, in which valid statements are called events, 
 see Finetti (1975, 3.11, 4.18, 6.3 and appendix). 
 Dubins and Savage (1976, p.8) present
similar objections, using the colorful gambling metaphors that are so
characteristic of orthodox (decision theoretic) Bayesian statistics. In
order to escape the constraint of having non-measurable sets, 
 de Finetti (1975, v.2, p.259) readily proposes a deal: to trade off other
 standard properties of a measure, like countable ($\sigma$) additivity: 

 \begin{quotation}
 \noindent {\it ``Events are restricted to be merely a subclass
(technically a $\sigma$-ring with some further conditions) of the class
of all subsets of the base space. In order to make $\sigma$-additivity
possible, but without any real reason that could justify saying to one
set `you are an event', and to another `you are not'.''} 
 \end{quotation}

 In order to proceed with our analysis, we have to search for the roots
of de Finetti's argument, roots that, we believe, lay outside de
Finetti's own theory, for they hinge on the perceived structure of the
continuum. Bell (1998, p.2), states: 
 
 \begin{quotation}
 \noindent {\it ``the generally accepted set-theoretical formulation of
mathematics (is one) in which all mathematical entities, being
synthesized from collections of individuals, are ultimately of a
discrete or punctate nature. This punctate character is possessed in
particular by the set supporting the `continuum' of real numbers - the
`arithmetical continuum'.''} 
 \end{quotation}

 Among the alternatives to arithmetical punctiform perspectives of the
continuum, there are more geometrical perspectives. 
 Such geometrical perspectives allow us to use an arithmetical set as a
coordinate (localization) system in the continuum, but the `ultimate
parts' of the continuum, called infinitesimals, are essentially
nonpunctiform, i.e. non point like. 
 Among the proponents of infinitesimal perspectives for the continuum
one should mention G.W.Leibniz, I.Kant, C.S.Peirce, H.Poincar\'{e}, 
L.E.J.Brouwer, H.Weyl, R.Thom, F.W.Lawvere, A.Robinson, E.Nelson, 
and many others. 
 Excellent historical reviews are presented in 
 Bell (1998 and 2005), a general view, and  
 Robertson (2001), for the ideas of C.S.Peirce. 
 In the infinitesimal perspective, see Bell (1998, p.3), 

 \begin{quotation}
 \noindent {\it ``any of its (the continuum) connected parts is also a
continuum and, accordingly, divisible. A point, on the other hand,  is
by its nature not divisible, and so (as stated by Leibniz) cannot be
part of the continuum.''}   
 \end{quotation}

 In Peirce doctrine of synechism, the infinitesimal geometrical
structure of the continuum acts like {\it `` the `glue' causing points
on a continuous line to lose their individual identity.''}, see Bell
(1998, p.208, 211). According to Peirce, {\it `` The very word continuity
implies that the instants of time or the points of a line are everywhere
welded together.''} 

 De Finetti's argument on non-measurable sets implicitly assumes that
all point subsets of $R^n$ have equal standing, i.e., that the continuum
has no structure. Under the arithmetical punctiform perspective of the
continuum, de Finetti's objection makes perfect sense, and we should
abstain from measure theory or alternative formalisms, as does orthodox
Bayesian statistics. This is how Peirce's concept of synechism helps us
to overcome a major obstacle (for the FBST) presented by orthodox
Bayesian philosophy, namely, the objections against the use of measure
theory.

 At this point it should be clear that my answer to Brier's question is
emphatically affirmative. From Brier's comments and suggestions it is
also clear 
 how well he knew the answer when he asked me the
question. As a maieutic teacher however, he let me look for the answers
my own way. I can only thank him for the invitation that brought me for
the first time into contact with the beautiful world of semiotics and
Peircean philosophy.

 \section{Final Remarks} 
 \markboth{CHAPTER 2: LANGUAGE AND SELF-REFERENCE}
  {2.7 \ FINAL REMARKS}

 The physician  Rambam, Moshe ben Maimon (1135--1204) of 
 (the then caliphate of) Cordoba, wrote Shmona Perakim, a book on 
 psychology (medical procedures for healing the human soul) based on 
 fundamental principles exposed by Aristotle in Nicomachean Ethics, 
 see Olitzky (2000) and Rackham (1926). 
 Rambam explains how the health of the human soul depends on always
finding the straight path (derech y'shara) or golden way  (shvil
ha-zahav), at the perfect center between the two opposite extremes of
excess (odef) and scarcity (choser), see 
 Maimonides (2001, v.1: Knowledge, ch.2: Temperaments, sec.1,2):     

 \begin{quotation}
 \noindent {\it ``The straight path is the middle one, that is
equidistant from both extremes.... Neither should a man be a clown or
jokester, nor sad or mourning, but he should be happy all his days in
serenity and pleasantness. And so with all the other qualities a man
possesses. This is the way of the scholars. Every man whose virtues
reflect the middle, is called a chacham... a wise man.''}  
 \end{quotation}

 Rambam explains that a (always imperfect) human soul, at a given time
and situation, may be more prone to fall victim of one extreme than to
its opposite, and should try to protect itself accordingly. One way of
achieving this protection is to offset its position in order to
(slightly over-) compensate for an existing or anticipated bias. 

 At the dawn of the 20th century, humanity had in classical physics a
paradigm of science handing out unquestionable truth, and faced the
brutality of many totalitarian states. 
 Dogmatism had the upper hand, and we had to protect
ourselves accordingly.

 At the beginning of the 21st century we are enjoying the comforts of an
hyperactive economy that seems to be blind to the constraints imposed by
our ecological environment, and our children are being threatened by
autistic alienation through the virtual reality of their video games. 
 It may be the turn of (an apathetic form of) solipsism.      

 Finally, Rambam warns us about a common mistake: Protective offsets may
be a useful precautionary tactic, or even a good therapeutic strategy,
but should never be considered as a virtue per se. 
 The virtuous path is the straight path, neither left of it nor right of
it, but at the perfect center.  


%% file: CAPE3.TEX
  
  \chapter{Decoupling, Randomization, Sparsity,  
           and Objective Inference} 



 {\flushright 
 {\it  
 ``The light dove, that at her free flight cleaves the air,\ \mbox{} \\
 therefore feeling its resistance, could perhaps imagine \ \mbox{} \\ 
 that she would succeed even better in the empty space.'' 
 } 




 Immanuel Kant (1724-1804), \\ 
 Critique of Pure Reason (1787, B-8). 

 \mbox{} 

 {\it  
 Step by step the ladder is ascended. 
 } 


 George Herbert (1593 - 1633), \\ 
 Jacula Prudentium (1651).  

 \mbox{} 

 }

 \section{Introduction}
 \markboth{CHAPTER 3: DECOUPLING AND OBJECTIVE INFERENCE}
 {3.1 \ INTRODUCTION}

  H.von Foerster characterizes ``known'' objects as eigen-solutions
for an autopoietic system, that is, as discrete (sharp), separable
(decoupled), stable and composable states of the interaction of the
system with its environment. 
 Previous chapters have presented the Full Bayesian  Significance Test
(FBST) as a mathematical formalism specifically designed  to access the
support for sharp statistical hypotheses, and have shown that these
hypotheses correspond, from a constructivist perspective, to systemic
eigen-solutions in the  practice of science, as seen in chapter 1. 
 In this chapter, the role and importance of one of these four essential 
attributes indicated by von Foerster, namely, separation or decoupling, 
is studied.

 Decoupling is the general principle that allows us to understand the 
world step by step, `looking' at it a piece at a time, localizing 
single features, isolating basic components or identifying simple 
objects, out of the immense complexity of the whole universe.   
 In statistical models, decoupling is often introduced by means of 
no association assumptions, such as independence, zero covariance, etc. 
 In this context, decoupling relations are sharp statistical hypotheses 
that can be tested, see for example Stern and Zacks (2002). 
 Decoupling relations in statistical models can also be introduced a
priori by means of special Design of Statistical Experiments (DSEs) 
techniques, the best known of which being randomization.

 In chapter 2 the general meaning of the term ``Objective'' 
 (how, less, more) is defined  as the ``degree of conformance of an
object to the essential attributes  of an eigen-solution''. 
 One of the common uses of the word objective, as opposed to 
``subjective'', stresses the decoupling or separation of a given systemic 
eigen-solution, such as an object of a scientific program, from the 
peculiarities of a second system, such as a specific human observer. 
 It is this restricted meaning, focusing on the decoupling property of
systemic eigen-solutions, that justifies the use of the term objective
in this chapter's title.

 The decoupling principle, and one of its most celebrated examples in 
Physics, the vibrating chord, are presented in section 2.  
  In the vibrating chord model, a basic linear algebra operation, the
eigen-value factorization, is the key to obtain the decoupling operator. 
 In addition, the importance of eigen-solutions and  decoupling
operations are discussed from a constructivist epistemological
perspective. 
 Herein, we shall focus on decoupling operators related to an other  
basic linear algebra operation, namely, the Cholesky factorization.
 In section 3 we show how Cholesky factorization can be used to 
decouple covariance structure models. 
 In section 4, Simpson's paradox and some strategies for DSEs, 
such as control and randomization, are discussed. 
 These strategies can be used to induce independence relations, that 
are expressed into the sparsity structure of the model, which can, 
in turn, be used for efficient decoupling.  
 In section 5, the role  of C.S.Peirce in the introduction of  control
and randomization in DSEs is reviewed from an historical perspective. 
 This revision will help us set the stage for the discussion, in section 6, 
of a controversial issue: randomization in  Bayesian Statistics.   
 In section 7 some epistemological consequences of randomization,
are discussed and the underlying themata of constructivism and objective
knowledge are revisited. 

 The Cholesky factorization operator is presented in section 3, 
 in conjunction with the computational concepts of sparse and structured 
 matrices. 
 Covariance structure and Bayesian networks are some of the most basic
and widely used statistical models. Therefore, understanding their 
decoupling properties is important, not only from a computational point
of view, but also from the theoretical and a epistemological perspective.  
 Furthermore, one could argue that the usefulness of these statistical
models are due exactly to their decoupling properties.  
 Final remarks are presented in section 8.

 \section{The Decoupling Principle}
 \markboth{CHAPTER 3: DECOUPLING AND OBJECTIVE INFERENCE}
  {3.2 \  DECOUPLING PRINCIPLE}

  Understanding the entire universe, with all its intricate 
constituents, relations and interconnections, can be a daunting task, 
as stated by Schlick (1979, v.1, p.292): 

  \begin{quote} 
  {\it `` The most important (of these) difficulties arises from the 
recognition of the unending linkage of all natural processes one with 
another. 
 Its effect is that, on an exact view, every occurrence in the  world is
dependent on every other; the fall of a leaf is ultimately influenced 
by the motions of the stars, and it would be a task utterly beyond
fulfillment to assign its `cause' with absolute completeness to  any given
process that we suppose determined down to the last detail. 
 For this purpose we should have to adduce nothing less than all of 
the circumstances of the universe that have so far occurred. 

 Now fortunately this boundlessness is at once considerably restricted 
by experience, which teaches us that the reciprocal interdependence of 
all events in nature is subject to certain easy formulable
conditions.''}  
 \end{quote}

 L.Sadun has written an exceptionally clear book on linear algebra,
emphasizing the idea of decoupling, i.e. the strategy of  breaking
down complicated multivariate systems into simple `modes',  by a suitable 
change of coordinates, see also Rijsbergen (2004).  
 Sadun (2001, p.1) states the goal of his book as follows: 

 \begin{quote}
 {\it 
  ``In this book we cover a variety of linear  evolution equations, 
 beginning with the simplest  equations in one variable, moving to 
 coupled equations in several variables, and culminating in problems 
 such as wave propagation that involve an infinite number of degrees  
 of freedom. 
  Along the way we develop techniques, such as Fourier analysis, that 
 allow us to decouple the equations into a set of scalar equations that 
 we already know how to solve. 

  The general strategy is always the same. When faced with coupled 
 equations involving variables $x_1,\ldots,x_n$, we define new variables 
 $y_1,\ldots,y_n$. These variables can always be chosen so that the 
 evolution of $y_1$ depends only of $y_1$ (and not on $y_2,\ldots,y_n$), 
 the evolution of $y_2$ depends only of $y_2$, and so on. 
 To find $x_1(t),\ldots,x_n(t)$ in terms of the initial conditions 
 $x_1(0),\ldots,x_n(0)$, we convert $x(0)$ to $y(0)$, then solve for 
 $y(t)$, then convert to $x(t)$.}   
 \end{quote}

 As an example of paramount theoretical and historical importance in 
 Physics, we consider the discrete chord. 
 The chord is kept at tension $h$, with $n$ particles of mass $m$ 
 at equally spaced positions $js$, $j=1\ldots n$. 
 The extremes of the chord, at positions $0$ and $(n+1)s$, are kept fixed, 
 and $x=[x_1,x_2,\ldots,x_n]'$ denote the particles' vertical displacements,  
 see French (1974, ch.5 Coupled oscillators and normal modes, p.119-160), 
 Marion (1999, ch.9) and 
 Franklin (1968, ch.7), 
 Figure 1 shows the discrete chord for $n=2$.

 \begin{figure}[hbt] 
 \centerline{\includegraphics*[height=4.0in, width=5.0in, angle=0]{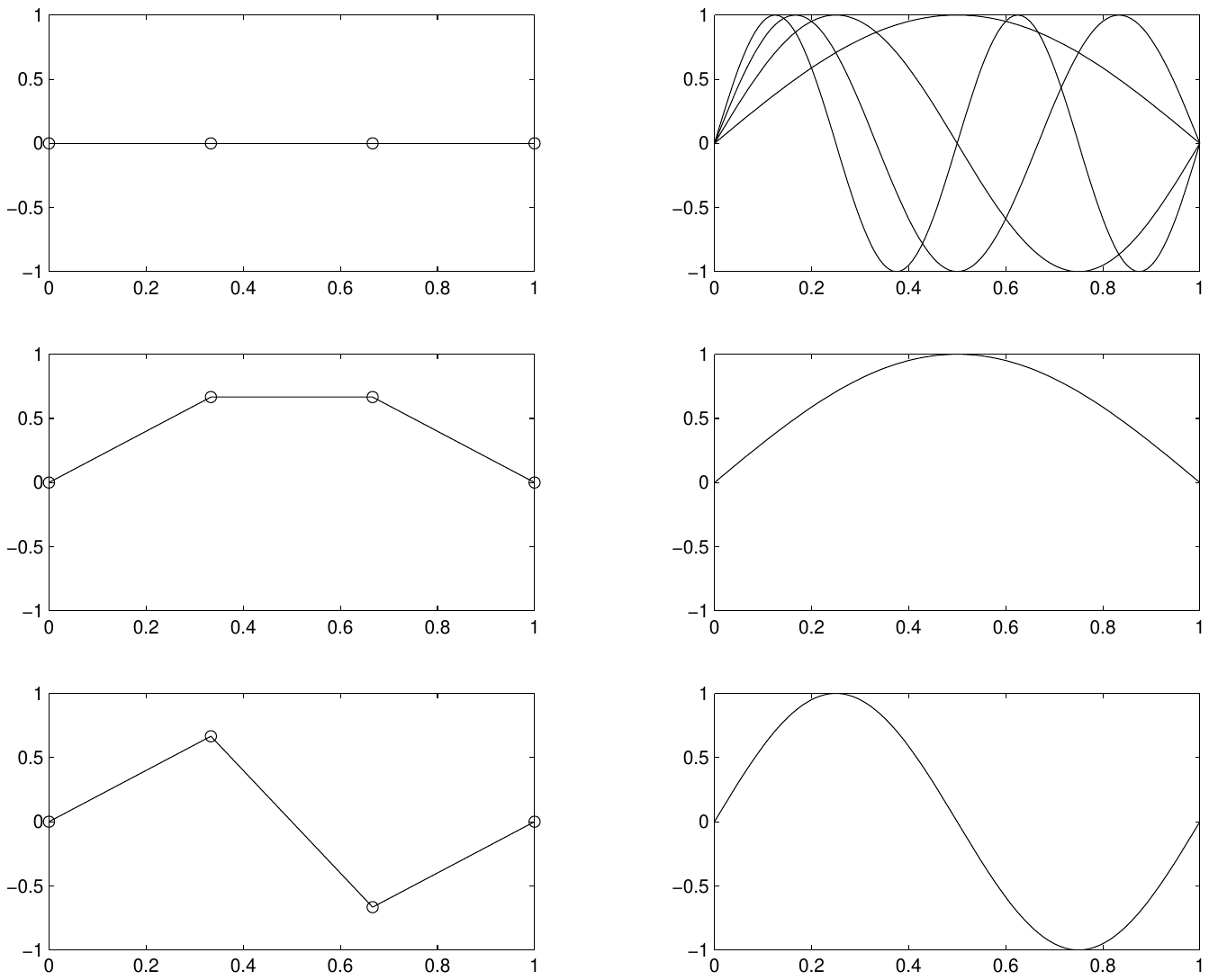}} 
 \centerline{Figure 1: Eigen-Solutions of Continuous and Discrete Chords.} 
 \end{figure}

 The second order differential equation of classical mechanics, below, 
privides a linear approximation for the discrete chord system's dynamics:   
  \[ 
    \ddot{x} +Kx =0 \ \ , \ \ 
    K= w_0^2 \left[ \begin{array}{cccccc}
          2 &     -1 &      0 &      0 & \cdots &      0 \\ 
         -1 &      2 &     -1 &      0 & \cdots &      0 \\ 
          0 &     -1 &      2 &     -1 & \ddots & \vdots \\ 
          0 &      0 &     -1 & \ddots & \ddots &      0  \\ 
     \vdots & \vdots & \ddots & \ddots &      2 &     -1 \\       
          0 &      0 & \cdots &      0 &     -1 &      2 
    \end{array} \right] 
    \ \ , \ \ w_0^2 = \frac{h}{ms} \ \ . 
  \]   

  As it is, the discrete chord differential equation is difficult to
 solve, since the $n$ coordinates of vector $x$ are coupled by matrix 
 $K$. In the following paragraphs we show how to decouple this
 differential equation.  

 Suppose that an orthogonal matrix $Q$ is known to diagonalize matrix $K$, 
 that is, $Q^{-1}=Q'$, and    
 $Q'KQ=D=\diag(d)$, $d=[d_1,d_2,\ldots,d_n]'$. 
 After pre-multiplying the above differential equation by $Q'$, 
 we obtain the matrix equation 
 \[ 
    Q'(Q\ddot{y}) +Q'K(Qy) = I\ddot{y} +Dy =0 
 \] 
 which is equivalent to the $n$ decoupled scalar equations 
 for harmonic oscillators, $\ddot{y}_k +d_k y_k =0$, 
 in the new `normal' coordinates, $y=Q'x$. 
  The solution of each harmonic oscillator, as a function of 
 time, $t$, has the form 
 $y_k(t) = \sin(\varphi_k +w_k t)$, 
 with phase $0\leq \varphi_k \leq 2\pi$ and 
 angular frequency $w_k=\sqrt{d_k}$.

  The columns of matrix $Q$, the decoupling operator, are the
eigenvectors of  matrix $K$, which are, as one can easily check,
multiples of the un-normalized vectors $z^k$. 
 Their corresponding eigenvalues, $d_k=w_k^2$, for $j,k=1\ldots n$, 
are given by   
 \[ 
   z_j^k = \sin \left( \frac{jk\pi}{n+1} \right)    
   \ \ , \ \ 
   w_k = 2 w_0 \sin \left( \frac{k\pi}{2(n+1)} \right) 
   \ \ . 
 \]

 The decoupled modes of oscillation, for $n=2$, are depicted in Figure 1.    
 They are called `normal' modes in physics, `standing' modes in 
engineering, and eigen-solutions in mathematics. 
 The discrete chord with $n$ particles will have $n$ normal modes, and  
 the limit case, $n\rightarrow \infty$, is called the continuous chord. 
 The normal modes of the continuous chord are given by trigonometric 
functions, the first few of which are depicted in Figure 1. 
 They are also called `standing' waves or eigen-functions of the chord,
and constitute the basis of Fourier analysis. 

 In either the discrete or the continuous chord, we can `excite',  i.e.
give energy or `put in motion', one of the normal modes,  without
affecting any other normal mode. 
 This is the physical meaning of decoupling, i.e. to have `separate' 
eigen-solutions. 
 Since the differential equation describing the system is linear, 
distinct normal modes can also be superposed.  
 This is called the `superposition' principle, which renders the 
compositionality rule for the eigen-solutions of the chord. 

 In the original coordinate system, $x$, coupling made it hard 
to follow the system's evolution. 
 In the normal coordinate system, $y$, based on the system's
eigen-solutions, decoupling and superposition made it easier to
understand the system behavior.   
 But are these eigen-solutions ``just'' a formal basis for an
alternative coordinate system,  or do they represent ``real objects'' 
within the system under study? 
       
 Obviously, this is not a mathematical or physical question, but rather
an  epistemological one. From a constructivist perspective,  we can
consider these eigen-solutions ``objectively known''  entities in the
system. Nevertheless, the meaning of the term objective in a 
constructivist epistemology is distinct from its meaning in a  dogmatic
realist epistemology, as explained in Stern (2006b, 2007a,b). 
   
 From a constructivist perspective, systemic eigen-solutions can be 
identified and ``named'' by an observer. 
 Indeed, the eigen-solutions of the  vibrating chord have been
identified and named thousands of years before mankind knew anything 
about differential equations. 
 The eigen-values of the chord are known in music as the `fundamental
tone'  and its `higher harmonics', and constitute the basis for all
known musical systems, see Benade (1992).

 The linear model for the vibrating chord is a paradigmatic example of 
the fact that, despite the simplicity to understand and manipulate, 
linear models often give excellent approximations for complex systems. 
 Also, since linear operators are represented by matrices in standard
matrix algebra, the importance of certain matrix operations in the 
decoupling of such models should not be surprising at all. 
 In the vibrating chord model, the eigen-value factorization, $K=QDQ'$, 
was the key to obtain the decoupling operator, $Q$. 
 The eigen-value factorization plays the same role in many important 
statistical procedures, such as spectral analysis of time series, wavelet 
signal analysis, and kernel methods.  

 Related operations of linear algebra, like Singular Value
 Decomposition, SVD, and Nonnegative Matrix  Factorizations, NNMF, are
important in principal components analysis and latent structure models, 
 see for example 
 Bertsekas and Tsitsiklis (1989), Censor and S.A.Zenios (1998),
 Cichocki et al. (2006), Dhillon and Sra (2005) and Hoyer (2004). 
  Distinct decoupling operators have distinct characteristics, 
 relying upon stronger or weaker structural properties of the model, 
 requiring more or less computational work, and having different 
 capabilities for handling sparse data.   
   
 In this chapter, we will be mainly interested in the decoupling of
statistical models. More precisely, we shall focus on decoupling methods 
related to an important basic linear algebra operation, namely, 
the Cholesky factorization.
 In the next section we show how Cholesky factorization can be used to 
decouple covariance structure statistical models.

 The decoupling principle emerges, sometimes with different
denominations, in virtually every area of the hard sciences. In Systems
Theory and Mathematical Programming, for example, it arises under the
name of Decomposition Methods. In the optimization of large systems,
for example, there are two basic  approaches to decomposition:  

 - High level methods focus on the  underlying structure of the
optimization problems. High level decomposition strategies replace the
original large or complex  problem by several hierarchically
interconnected small or simple optimization problems, see for example 
 Geoffrion (1972), Lasdon (1970) and Wismer (1971).     
 
 - Low level methods look at the matrix representation of the
optimization problems. Low level decomposition strategies benefit from 
tailor made computational linear algebra subroutines to take advantage 
of the underlying sparse matrix structure. 
 Some of these techniques are discussed in the next section.

 \section{Covariance Structure Models}
 \markboth{CHAPTER 3: DECOUPLING AND OBJECTIVE INFERENCE}
  {3.3 \ COVARINCE STRUCTURE MODELS}

 Covariance structure, multivariate regression, Kalman filter and
several other related linear statistical models are widely used in the
practice of science. 
 They provide a powerful analytical tool in which the association, 
coupling or dependence between multiple variables is represented by 
covariance matrices, as briefly noted in the next paragraphs. 
 These models are simple to manipulate and interpret, and can  be
implemented using efficient computational algorithms capable of 
handling millions of (sparsely coupled) variables. 
 In this and the next sections, it is shown how such desirable 
characteristics of covariance models ultimately rely upon some basic 
properties of its decoupling operators.        
 
 Given a (vector) random variable, $x$, its covariance matrix, $V$,   
is defined as the expected square distance to its expected (mean) value, 
$\beta$, that is,   
 \[ 
    \beta= E(x) \ , \ \ V=\Cov(x)=E((x-\beta)\otimes(x-\beta)') \ . 
 \] 
 The diagonal elements, or variances,  
 $\Var(x_i)=V_{i,i}$,  
 give the most usual scalar measure of error, dispersion or uncertainty 
 used in statistics, while the off diagonal elements, 
 $\Cov(x_i,x_j)=V_{i,j}$, 
 give a measure of association between two scalar random variables, 
 $x_i$ and $x_j$, see Hocking (1985) for a general reference. 
 
  Also recall that since the expectation operator, $E$, is linear,
 that is,  $E(Ax+b)=AE(x)+b$ for any random vector $x$, matrix $A$ and 
 vector $b$, we have    
 \[ 
 \Cov(Ax+b)=A\Cov(x)A' \ . 
 \]  

 The standard deviation, $\sigma_i=\sqrt{V_{i,i}}$, is a dispersion 
 measure given in the same unit as $x$, and the correlation, 
 $C_{i,j}=V_{i,j}/\sigma_i\sigma_j$, is a measure of association  
 normalized in the $[-1,1]$ interval. 

  As it is usual in the covariance structure literature,  
 we can write a covariance matrix as 
 $V(\gamma)=\sum \gamma_t G^t$, in which the matrices 
 $G^t$ constitute a basis for the space of symmetric matrices 
 of dimension $n\times n$, see Lauretto et al. (2002). 
  For example, for dimension $n=4$,      
 \[ 
 V(\gamma) = \sum_{t=1}^{10} \gamma_t\, G^t \, = \,  
 \left[ \begin{array}{cccc} 
 \gamma_1 & \gamma_5    & \gamma_7 & \gamma_8 \\  
 \gamma_5 & \gamma_2    & \gamma_9 & \gamma_{10} \\  
 \gamma_7 & \gamma_9    & \gamma_3 & \gamma_6 \\  
 \gamma_8 & \gamma_{10} & \gamma_6 & \gamma_4 \\    
 \end{array} \right]  \ .    
 \]            
 Using the above notation, we can easily express  hypotheses concerning 
structural properties, including  sparsity patterns, in the standard 
form of vector functional equations, $h(\beta,\gamma)=0$. 
 Details on how to use the FBST to test such general hypotheses 
 in some particular settings can be found in Lauretto et al. (2002). 

 Once we have established the structural properties of the model, 
we can estimate the parameters $\beta$ and $\gamma$ accordingly. 
 Following the general line of investigation adopted herein, a 
question that arises naturally is: 
 How can we decouple the estimated model? 

  One possible answer to this question can be given in terms of the 
 Cholesky factorization, $LL'=V$ where $L$ is lower triangular.  
 Such a factorization is available for any full rank symmetric matrix 
 $V$, as shown in Golub and van Loan (1989). 
  Let $V=LL'$ be the Cholesky factorization of the covariance 
 matrix, $V$, and let us consider the transformation of variables 
 $y =L^{-1}x$, or $x =Ly$. 
  The covariance matrix of the new variables can be computed as 
 $\Cov(y)=L^{-1}VL^{-t}=L^{-1}LL'L^{-t}=I$. 
 Hence, the transformed model has been decoupled, 
 i.e., has uncorrelated random components.

  Let us consider a simple 
 numerical example of Cholesky factorization: 
 \[ 
 V = \left[ \begin{array}{cccc} 
 1 & 1 & 0 & 0 \\ 
 1 & 2 & 0 & 0 \\ 
 0 & 0 & 4 & 4 \\ 
 0 & 0 & 4 & 8 
 \end{array} \right]  \ , \ \ 
 L = \left[ \begin{array}{cccc} 
 1 & 0 & 0 & 0 \\ 
 1 & 1 & 0 & 0 \\ 
 0 & 0 & 2 & 0 \\ 
 0 & 0 & 2 & 2 
 \end{array} \right]  \ , \ \ 
 V = LL' \ .    
 \]            
  This example of Cholesky factorization has some peculiarities: 
 The matrix $V$ is {\it sparse}, i.e., it has several zero elements. 
 In contrast, a matrix with few or no zero elements is said to be 
 {\it dense}. 
 Matrix $V$ in the example is also {\it structured}, i.e., the zeros are
arraged in a nice pattern, in this example, a $2\times 2$ off diagonal
block. 
  In this example, the Cholesky factor, $L \g LL'=V$, 
 preserves the sparsity and structure of $V$, that is, no position with 
 a zero in $V$ is {\it filled} with a non-zero in $L$. 
  A factorization (or ellimination) resulting in no fill in is called  
  {\it perfect}.   
  Perfect eliminations are not always possible, however, there are 
 several techniques that can be used to obtain sparse (and structured) 
 Cholesky factorizations in which the fill in is minimized, 
 that is, the sparsity of the Cholesky factor is maximized. 
 Pertinent references on sparse factorizations include 
 Blair and B.Peyton (1993), Bunch and D.J.Rose (1976) 
 George et al. (1978, 1981, 1989, 1993), Golumbic (1980), 
 Pissanetzky (1984), Rose (1972), Rose and Willoughby (1972), 
 Stern (1992,1994), Stern and Vavasis (1993,1994) and 
 van der Vorst and van Dooren (1990).

 Large models may have millions of sparsely coupled variables. 
 A sparse and structured factorization of such a model gives a 
`simple' decoupling operator, $L$.  
 This is a matter of vital importance when designing efficient 
computational procedures. 
 In practice, large models can only be computed with the help 
of these techniques. 
 An other important class of statistical models, Bayesian Networks,     
relies on sparse factorization techniques that, from an abstract 
graph theoretical perspective, are almost identical to sparse 
Cholesky factorization, see for example 
 Lauritzen (2006) and Stern (2006a, sec.9-11).

 In the next section we continue to examine the role of  covariance, 
or more general forms of association, in statistical modeling. 
 On particular, we examine some situations leading to spurious 
associations, destroying a  model's presumed sparsity and structure.  
 In the following sections we review, from an historical and
epistemological  perspective, some techniques of Design of Statistical
Experiments (DSE), used to induce (no) association relations in
statistical models. These relations translate into sparsity and
structural patterns that, in turn, can be used by efficient
factorization algorithms.

 \section{Simpson's Paradox and the Control   
          of Confounding Variables}
 \markboth{CHAPTER 3: DECOUPLING AND OBJECTIVE INFERENCE}
  {3.4 \ CONTROL OF CONFOUNDING VARIABLES}

 Lindley (1991, p.47-48) illustrates Simpson's paradox with a medical
trial  example. From 80 patients in the study, 40 received treatment, T,  
and 40  received a placebo with no effect, NT. 
 Some patients recovered from their illness, R, and some did not, NR. 
 The recovery rates, R\%, are given in Table 1, where the experimental 
data is shown, both in aggregate form for All patients, and 
separated or disaggregated according to Sex. 
 Looking at the table one concludes that the treatment is
bad for either male or female patients, but good for all of them together! 
 This is the Simpson's Paradox: The association between two variables, 
 T and R in Lindley's example, is reversed if the data is 
 aggregated / disaggregated over a {\it confounding} variable, 
 Sex in Lindley's example.

 \begin{center} 
 {
 {Table 1: Simpson's Paradox.} 

 \begin{tabular}{c c r r r r } 
 \hline 
  Sex  & T  &  R & NR & Tot & R\% \\ 
 \hline
  All  & T  & 20 & 20 & 40 & 50\% \\ 
  All  & NT & 16 & 24 & 40 & 40\% \\ 
 \hline 
  Male & T  & 18 & 12 & 30 & 60\% \\ 
  Male & NT &  7 &  3 & 10 & 70\% \\ 

 \hline 
  Fem & T  &  2 &  8 & 10 & 20\% \\ 
  Fem & NT &  9 & 21 & 30 & 30\% \\ 
 \hline 
 \end{tabular} 
 } 
 \end{center}

 Lindley provides the following scenario for the situation illustrated by 
this example:  
 The physician responsible for the experiment did not trust the 
treatment and also was aware that the illness under study affects females 
most severely. Hence, he decided to try it mainly on males, who would
probably recover anyway. 
 This illustrates the general Simpson's paradox situation, generated by
the association of the confounding variable with both the explained
and one (or more) of the explaining variables. 
   Additional references on several aspects related to the Simpson 
  paradox include Blyth (1972), Cobb (1998), Good and Mittal (1987), 
  Gotzsche (2002), Greenland et al. (1999, 2001), Heydtmann (2002), 
  Hinkelmann (1984), Pearl (2004) and Reintjes et al. (2000).

 The obvious question then is: How can we design a statistical experiment 
in order to avoid spurious associations?  

 Two strategies are self-evident: 
 \begin{enumerate} 
 \item 
  Control possible confounding variables in order to impose some form of    
  invariance (constancy, equality) in the experiment, or 
 
 \item 
  Measure possible confounding variables so that the relevant ones 
  can be included in the statistical model. 
 \end{enumerate}

 The simplest form of the first strategy would be to test the treatment 
in a set of `clones', individuals that are, using the words of  
 Fisher (1966, sec.9, Randomization; the Physical Basis of 
 Validity of the Test, p.17-19),  
 \begin{quote}
 {\it ``exactly alike, in every respect except that to be tested''}, 
 \end{quote}
 This strategy, however, is too strict.  
 Even if feasible, the conclusions of the study would only apply to
 the `clone population', not to individuals from a population with 
 natural variability. 

 A more general form of the first strategy in known as blocking, 
 defined in 
 Box et al. (1978, p.102-103, Sec.4.3, Blocking and Randomization) as:  

 \begin{quote}
 {\it ``The device of pairing observations is a special case of
`blocking' that has important applications in many kinds of experiments.
 A block is a portion of the experimental material (the two shoes of one
boy in this example) that is expected to be more homogeneous than the
aggregate (all shoes of all the boys). By confining treatment comparisons
within such blocks, greater precision can often be obtained.''}
 \end{quote}

 Blocking is a very important strategy in the  
 design of statistical experiments (DSEs), used to increase, 
 whenever possible, the precision of the study's conclusions.

 As for the second strategy, it looks a sure thing! 
 No statistician would ever refuse more information, 
 in a larger and richer data bank.   

 Nevertheless, we have to ask whether we want to control and/or measure 
 SOME of the possibly confounding variables, i.e. those perceived as the 
 most important or even those we are aware of, or ALL of them?

 Keeping everything under control in a statistical experiment 
 (or in life in general) constitutes, in the words of Fisher,   

 \begin{quote}
 {\it ``a totally impossible requirement in our example, and equally 
  in all other forms of experimentation''}. 
 \end{quote}

 Not only the cost and complexity of trying to do so for a very large
set of variables would be prohibitive in any practical circumstance, 
but also 

 \begin{quote}
 {\it ``it would be impossible to present an exhaustive list of such
possible differences (variables) appropriate for any one kind of
experiment, because the uncontrolled causes which may influence the
result are always strictly innumerable''}.  
 \end{quote}

 Modern theory of DSEs offers a way out of this conundrum that, 
in its most concise form, see  Box et al. (1978, p.102-103), 
can be stated as: 

 - Control what you can, and randomize what you can not. 

 Randomization, as defined by  Hacking (1988, p.428), is  
    
 \begin{quote}
 {\it ``(the) notion of random assignment of treatment to a subset 
of the plots or persons, leaving the rest as controls. ... 
 I shall speak of an experiment using randomization in this way as 
involving a randomized design. ... } \\ 
 {\it ... There is a related but distinguishable idea of (random)
representative sampling.''}    
 \end{quote}

 As it is usual in the statistical literature, Hacking distinguishes 
between two intended uses of randomization, namely random design and 
random sampling. 
 Random design aims to eliminate bias  coming from systematic design 
problems, including  several forms of uncontrolled influence, 
either conscious or unconscious, received from and exerted by agents
participating in the experiment. 
 Random sampling, on the other hand, is intended to justify, somehow,  
assumptions concerning the functional form of a distribution in  the
statistical model of the experiment.  
 The distinction between random design and random  sampling 
will be kept here, even though, as briefly  mentioned
in section 6, a deeper probabilistic analysis of  randomization
shows that, from a theoretical point of view, the two concepts can
greatly overlap.

 Our immediate interest in randomization (and control) is on  whether it
can assist the design  of experiments by inducing independence relations. 
 This strategy is pinpointed in the following quote from 
 Pearl (2000, p. 340,348.
  Epilogue: The Art and Science of Cause and Effect):

 \begin{quote}
 {\it ``...Fisher's `randomized experiment'...
  consists of two parts, `randomization' and intervention'.''}

 {\it ``Intervention means that we change the natural behavior of the
  individual: we separate subjects into two groups, called treatment and
  control, and we convince the subjects to obey the experimental policy.
   We assign treatment to some patients who, under normal circumstances,
  will not seek treatment, and give placebo to patients who otherwise
  would receive treatment. That, in our new vocabulary, means `surgery'
  - we are severing one functional link and replacing it with another.
   Fisher's great insight was that connecting the new link to a random
  coin flip `guarantees'   that the link we wish to break is actually broken.
   The reason is that a random coin is assumed to be unaffected by
  anything we can measure on macroscopic level...''}
  \end{quote}

 \section{C.S.Peirce and Randomization}
 \markboth{CHAPTER 3: DECOUPLING AND OBJECTIVE INFERENCE}
  {3.5 \ C.S.PEIRCE AND RANDOMIZATION}

 We believe that many fine points about the role of randomization in the
DSEs can be better understood by following its development from an 
historical perspective. This is the topic of this section.

 In the period of 1850 to 1880 the quantitative analysis of human 
sensation in response to physical (tactile, acoustic or visual) 
stimuli,   was the main goal of `psychophysics'. 
 A typical hypothesis in this research program was Fechner's law, 
see Hernstein and Boring (1966, p.72), which stated that,   

 \begin{quote}
 {\it ``The magnitude of sensation ($\gamma$) is not proportional to the
absolute value of the stimulus ($\beta$), but rather to the logarithm of
the  magnitude of the stimulus when this is expresses in terms of its
threshold  value ($b$), i.e. that magnitude considered as unit at which
the sensation  begins and disappears.''}  
 \end{quote}

 In modern mathematical notation, \ 
 $\gamma= k \log(\beta /b) \; I(\beta>b)$.

 In his psychophysical experiments Fechner tested his own ability to
distinguish the strongest in a pair of stimuli. For example, he  
would prepare two objects of masses $\mu$ and $\mu+\delta$,   
 and later on he would lift them, and `answer' which one appeared to him
to be the heaviest. A quantitative analysis would latter relate the
proportion of right and wrong answers with the values of $\mu$ and
$\delta$,
 see Stigler (1986, ch.7, Psychophysics as a Counterpoint, p.239-261).    
 Fechner was well aware of the potential difficulties resulting from 
the fact that the experiments where not performed blindly, that is, 
since he prepared the experiment himself, he could know in advance the
right answer. Nevertheless, he claimed to be able to control himself,  
be objective, and overcome this difficulty.   

 According to Dehue (1997), in the decade of 1870, G.E.M\"{u}ller and 
several researchers at T\"{u}bingen and G\"{o}ttingen Universities,
began to improve  the design of psychophysical experiments. 
 The first major improvement was blinding: the stimuli were prepared 
or administered by an `Experimenter' or `Operator' and applied to a
distinct person, the `Observer', `Patient' or `Subject', who was kept
unaware of the actual intensity of the  stimuli. 

 The second major improvement was the precaution of presenting the 
stimuli in `irregular order' (buntem Wechsel). This irregularity was 
introduced  to prevent the patient from becoming habituated to  patterns
in the sequence of stimuli presented to him or,  in other words,  
to keep him to form building expectations and guessing  
the right answers. 
 Nevertheless, there was, at that time, neither a general theory
defining `irregularity', nor a systematic method for providing an
`irregular order'.

 In 1885, Charles Saunders Peirce and his student Joseph Jastrow
presented randomization as a practical solution, in this context, 
to the question of irregularity, that is, systematic randomization  
should prevent any effective guessing by the patient, 
 see Hacking (1988, III. Psychophysics: Peirce's at Work, p.431-434). 
 Peirce was in fact insisting on `exchangeability',  a key
notion in the analysis of randomization in modern statistics and, most
specially, in Bayesian statistics, that will be discussed in
the next section. 

 Peirce also struggled with the dilemma of allowing or not, in the course 
of the experiment, sequences that do not `appear' to be random. 
 His conclusions, see Peirce and Jastrow (1884, p.122), are, once more,   
 precursors to De Finetti's  concept of exchangeability:  

 \begin{quote}
 {\it ``The pack (of playing-cards) was well shuffled, and, the operator 
and subject having taken their places, the operator was governed by the 
color of the successive cards ... 

 A slight disadvantage in this mode of  proceeding arises from the long
runs of  one particular kind of change, which would occasionally be
produced  by chance and would tend to confuse the mind of the subject. 
 But it seems clear that this disadvantage was less than that which 
would have been occasioned by his knowing that there would be no 
such long runs if any means had been taken to prevent them.''} 
 \end{quote}

 Regardless of its importance, Peirce's solution of randomization 
was not accepted by his contemporaries, fell into
oblivion, and was almost forgotten,  until it reappeared much latter in
the work of R.A.Fisher. 
 We believe that there are several entangled reasons to explain such a   
twisted historical process.  
 The psychopysics community raised objections against some of the
hypotheses, and also against some methodological
aspects presented in Peirce's paper. 
 Besides, there is also a confounding factor generated by a second role
played by randomization in Peirce's paper, namely, `randomization to
measure faint effects'.   
 We shall briefly discuss these aspects in the next paragraphs.  

 Fechner assumed the existence of a threshold (Schwelle), $b$, bellow
which small differences could no longer be discerned. 
 Peirce wanted to refute the existence of this threshold assuming, 
instead, a continuously decreasing sensitivity to smaller and smaller
differences. 
 We should remark that for Peirce this should not have been a fortuitous
hypothesis, since it can be related to his general philosophical ideas,
most specially with the concept of synechism, 
 see chapter 2, Hartshorne et al. (1992) and Eisele (1976).

 Peirce postulated that the patients' sensitivity could be adequately
measured by the probability of correct answers, even when  the
difference was too faint to be consciously discerned by the same
patients.  
 Hence, in experiments similar to Fechner's, Peirce asked the patient 
always to guess the correct answer. 
 Peirce also asked the patient to give the answer a confidence score 
from 0 to 3. 
 Peirce analyzed his experimental data and derived empirical formulae 
relating the (rounded) `subjective' confidence scores, $m$, 
and the `objective' probability of correct answers, $p$, 
 as in Peirce and Jastrow (1884, p.122): 

 \begin{quote}
 {\it `` The average marks seem to conform to the formula 
 $m= c \log(p/(1-p))$, where $m$ denotes the degree of confidence on the
 scale, $p$ denotes the probability of the answer being right, and $c$ 
 is a constant which may be called the index of confidence.''} 
 \end{quote}
 
 At the time of Peirce's experiments, the psychophysical community gave 
great importance to the analysis of the patient's subjective 
`introspections'. 
 According to this view, Peirce's experiments were  criticized by asking
the patient to guess the correct answer even when  he expressed low
confidence. 
 Of course, if one understands Peirce's research program, it is 
clear that that the experimental design he used is perfectly coherent. 
 Unfortunately, this was not the judgment of his contemporaries. 

 The same techniques and experimental designs used by Peirce were 
subsequently used by several researchers in attempts to measure faint
effects, including effects produced by `below the consciousness threshold', 
sub-conscious, or sub-liminal stimuli. Some of these studies were 
really misconceived, and that may have been yet another contributing 
factor for the reactions against the use of randomization. 
 Whatever the explanation might be, Peirce's paper fell into oblivion, 
and the progress of DSEs was delayed by half a century.

 \section{Bayesian Analysis of Randomization}
 \markboth{CHAPTER 3: DECOUPLING AND OBJECTIVE INFERENCE}
  {3.6 \ BAYESIAN ANALYSIS OF RANDOMIZATION}

 The work of Ronald Aylmer Fisher can undoubtedly be held responsible for
disseminating the modern approach to DSEs, including randomization, to
almost any area of empirical research, see for example 
 Fisher (1926, 1935). 
 The idea of randomization, however, was later contested by some
members of the Bayesian school.   
  Commenting on the use of randomization after Fisher, 
Hacking (1988, p.429-430), states: 

 \begin{quote}
 {\it ``Undoubtedly Fisher won the day, at least for the following 
generation, but then a new, although not completely unrelated, challenge
to randomized design arose. This came from the revival of the `Bayesian'
school, typically associated with L.J.Savage's theory of what he called
personal probability. Here the object is to form an initial assessment
of one's personal beliefs about a subject and to modify them in the 
light of experience and a theoretical analysis formally modeled by the 
calculus of probability and a theory of personal utility. 
 It is widely held to be an almost immediate consequence of this approach 
that randomization is of no value at all (except perhaps to eliminate 
some kind of fraud).''} 
 \end{quote}

 This erroneous notion of incompatibility between the use of randomization 
and Bayesian statistics in now completely outdated. 
 One of the most prestigious textbooks in contemporary Bayesian statistics,
 see Gelman et al. (2003, ch.7, p.198),  states:

 \begin{quote}
 {\it ``A naive student of Bayesian inference might claim that because 
all inference is conditional on the observed data, it makes no
difference how those data were collected. This misplaced appeal to the
likelihood principle would assert that given 
 (1) a fixed model (including the  prior distribution) for the
 underlying data and 
 (2) fixed observed values  of the data, 
 Bayesian inference is determined regardless of the design for the 
collection of the data. 
 Under this view there would be no formal role for randomization in 
either sample surveys or experiments. The essential flaw in the argument 
is that a complete definition of `the observed data' should include 
information on how the observed values arose, and in many situations 
such information has a direct bearing on how these values should be 
interpreted. Formally then, the data analyst needs to incorporate 
the information describing the data collection process in the 
probability model used for analysis.''} 
 \end{quote}

 Indeed, the classical argument using the likelihood principle against 
randomization in the DSEs, assumes a fixed, given statistical model and, 
as concisely stated by Kempthorne (1977, p.16):

 \begin{quote}
 {\it ``The assertion that one does not need randomization in the
 context of the assumed (linear) model (above) is an empty one because an 
 intrinsic role of randomization is to `insure` against model 
 inadequacies.''} 
 \end{quote}

 Gelman et al. (2003, ch.7, p.223-225) proceeds offering a much deeper
analysis of the role of randomization  from a Bayesian perspective, 
 see also Rubin (1978).  
 The key concept of ``ignorable design'' specifies decoupling conditions
between the sampling (or censoring)  process, described by an indicator 
variable, $I$, and the distribution  of the observed variables,
$y_{obs}$. If the experiment has an  ignorable design, we can build a
statistical model that explicitly considers $y_{obs}$ alone. 
 Finally, it is ironic that perhaps one of the best arguments for 
incorporating randomization in Bayesian experimental design is a 
consequence of de Finetti theorem for exchangeability. 
 As mentioned in section 4, this argument also blurres the distinction 
between the concepts of randomized design and randomized sampling. 
 We quote, once again, from Gelman et al. (2003, ch.7, p.223-225): 

 \begin{quote}
 {\it ``How does randomization fit into this picture? 
 First, consider the situation with no fully observed covariates $x$, 
in which case the `only' way to have an invariant to permutation design 
 - is to randomize.''}    

 {\it``In this sense, there is a benefit to using different patterns of
treatment assignment  for different experiments; if nothing else about
the experiments is specified, they are exchangeable, and the global
treatment assignment is  necessarily randomized over the set of 
experiments.''}  
 \end{quote}

 \section{Randomization, Epistemic Considerations}
 \markboth{CHAPTER 3: DECOUPLING AND OBJECTIVE INFERENCE}
  {3.7 \ RANDOMIZATION, EPISTEMIC CONSIDERATIONS}

 Several researchers currently concerned with epistemological questions
in Bayesian statistics are engaged in a reductionist program  dedicated
to translate every statistical test or inference problem into a 
decision theoretic procedure. 
  One of the main proponents and early contributors to this program, 
 but one who also had a much broader perspective, clearly articulating 
 his epistemological insights and motivations, was Bruno de Finetti.

 In statistical models our knowledge of the world is encoded in 
probability distributions. Hence, it is vital to clarify the 
epistemological or ontological status of probability. 
  Let us examine de Finetti's position, based on his own words, beginning
with Finetti (1972, p.189) and  Finetti (1980, p.212): 
 
 \begin{quote} 
 {\it ``Any assertion concerning probabilities of events is merely the 
 expression of somebody's opinion and not itself an event. There is no 
 meaning, therefore, in asking whether such an assertion is true or 
 false, or more or less probable.''} 

 {\it ``Each individual making a `coherent' evaluation of probability 
(in the sense I shall define later) and desiring it to be `objectively 
exact', does not hurt anyone: everyone will agree that this is his 
subjective evaluation and his `objectivist' statement will be a harmless 
boast in the eyes of the subjectivist, while it will be judged as true 
or false by the objectivist who agree with it or who, on the other hand, 
had a different one. This is a general fact, which is obvious but 
insignificant: `Each in his own way.' \ ''}  
 \end{quote}

 Solipsism, from the Latin solus (alone) +ipse (self), can be defined  as
the epistemological thesis that the individual's subjective states of mind
are the only proper or possible basis of knowledge. 
 Metaphysical solipsism goes even further, stating that nothing 
really `exists' outside of one's own mind. 
 From the two above quotations, it is clear that de Finetti stands, if
not from a metaphysical, at least from a  epistemological perspective,
as a true solipsist.  
 This goes farther than many theorists of the Bayesian subjectivist
school would venture, but de Finetti charges ahead, with a program that
is not only anti-realist, but also anti-idealist. 
 In (1974, V1, Sec.1.11, p.21,22, The Tyranny of Language),  de Finetti 
launches a full-fledged attack against the vain and futile desire for 
any objective knowledge: 

 \begin{quote} 
 {\it ``Much more serious is the reluctance to abandon the inveterate
tendency of the savages to objectivize and mythologize everything (1);
a tendency that, unfortunately, has been, and is, favored by many more
philosophers than have struggled to free us from it (2).

 (1) The main responsibility for the objectivizationistic fetters inflicted
on thought by everyday language rests with the verb `to be' or `to exist',
and this is why we drew attention to it in the exemplifying sentences.
From it derives the swarm of pseudoproblems from `to be or not to be', to
`cogito ergo sum', from the existence of `cosmic ether' to that of
`philosophical dogmas'.

 (2) This is what distinguishes acute minds, who enlivened thought and
stimulated its progress, from narrow-minded spirits who mortified and
tried to mummify it ...
`great thinkers' (like Socrates and Hume) and `school philosophers'
(like Plato and Kant).}
 \end{quote} 

 De Finetti was also aware of the dangers of `objective contamination', 
that is, any `objective' (probabilistic) statement can potentially
`infect'  and spread its objectivity to other statements, see De Finetti 
 (1974, V2, Sec.7.5.7, p.41-42, Explanations based on `homogeneity'): 

 \begin{quote} 
 {\it ``There is no way, however, in which the individual can avoid the
burden of his own evaluations. The key can not be found that will unlock
the enchanted garden wherein, among the fairy-rings and the shrubs of
magic wands, beneath the trees laden with monads and noumena, blossom
forth the flowers of `Probabilitas realis'. With the fabulous blooms
safely in our button-holes we would be spared the necessity of forming
opinions, and the heavy loads we bear upon our necks would be rendered
superfluous once and for all.''}
 \end{quote}

 As we have seen in the last sections, a randomization device is built
so to provide legitimate  `objective' probabilistic statements
about some events, and  randomization  procedures in DSEs  are conceived 
exactly in order to spread this objectivity around.

 I.J.Good was an other leading figure of the early days of the Bayesian
revival  movement. Contrary to de Finetti, Good has always been aware of
the  dangers of an extreme subjectivist position, see for example 
 Good (1983, Ch.8 Random Thoughts about Randomness, p.93): 

 \begin{quote} 
 {\it `` Some of you might have expected me, as a confirmed Bayesian, 
to restrict the meaning of the word `probability' to subjective (personal)
probability. That I have not done so is because I tend to believe that 
physical probability exists and is in any case a useful concept. 
 I think  physical probability can be measured  only with the help of 
subjective  probability, whereas de Finetti believes that it can be
`defined' in  terms of subjective probability. 
 De Finetti showed that if a person has a consistent set of subjective 
or logical probabilities, then he will behave `as if' there were physical 
probabilities, where the physical probability has an initial subjective 
probability distribution. 
 It seems to me that, if we are going to act if  the physical
probability exists, then we don't lose anything practical  if we assume
it really `does' exist. 
 In fact I am not sure that existence means more than there are no
conceivable circumstances in which the assumption  of existence would be
misleading. But this is perhaps too glib a  definition. 
 The philosophical impact of de Finetti's theorem is that  it supports
the view that solipsism cannot be logically disproved.  Perhaps it is
the mathematical theorem with most potential philosophical  impact.''} 
 \end{quote}

 In our terminology we would have used the expression 
 `objective probability' instead of Good's expression, 
 `physical probability'. 
 In 1962 Good edited a collection of speculative essays, including some
 on the foundations  of statistics. The following short essay by
 Christopher S.O'D.Scott  offers an almost direct answer to de Finetti, 
 see Good (1962, sec.114, p.364-365): 

 \begin{quote} 
 {\it ``Scientific Inference: 
 You are given a large number of identical inscrutable boxes. 
 You are to select one, the `target box', by any means you wish which 
does not involve opening any boxes, and you then have to say something 
about is in it. You may do this by any means you wish which does not
involve opening the target box. 

 This apparent miracle can easily be performed. You only have to select
the target box at random, and then open a random sample of other boxes. 
 The contents of the sample boxes enable you to make an estimate of the 
contents of the target box which will be better than a chance guess. 
 To  take an extreme case, if none of the sample boxes contains a rabbit 
and your sample is large, you can state with considerable confidence: 
 `The target box does not contain a rabbit.' 
 In saying this, you make no assumption whatever about the principles 
which may have been used in filling the boxes.

 This process epitomizes scientific induction at its simplest, which is 
the basis of all scientific inference. It depends only on the existence 
of a method of randomization, that is, on the assumption that events can 
be found which are unrelated (or almost) to other given events. 

 It is usually thought that scientific inference depends upon nature 
being orderly. The above shows that a seemingly weaker condition will 
suffice: Scientific inference depends upon our knowing ways in which 
nature is disorderly.''} 
 \end{quote}

 In the preceding chapters we discussed general conditions validating 
objective knowledge, from a constructivist epistemological perspective. 
 In this chapter we discuss the use of randomization devices,  that can
generate observable events with distribution that are independent of the
distribution of any event relevant to a given statistical study. 
 For example, the statistical study could be concerned with the reaction
of human patients affected by a given disease to alternative medical 
treatments,  whereas  a ``good'' randomization device could be a generic
`coin flipping machine', like a regular dice or a mechanical roulette
borrowed from a casino. The randomization device could also be a 
sophisticated apparatus detecting flips (state transitions) in some 
quantum system, with transitions probabilities known with a relative 
precision of one over a trillion.    

 So far in this chapter we have seen how well can decoupling strategies
used in the DSEs, including randomization procedures, help us to perform
robust statistical inference and,  in doing so, escape, from a pragmatic
perspective, the solipsist burdens of an extreme subjectivist position. 
 The same techniques can induce no association relations, generating 
sparse or structured statistical models.   
 No association hypotheses can then be tested, confirming (or not) 
such sparse or structured patterns in the statistical model.

 \section{Final Remarks}
 \markboth{CHAPTER 3: DECOUPLING AND OBJECTIVE INFERENCE}
  {3.8 \ FINAL REMARKS}

 As analyzed in this chapter, the randomization method, introduced by
 C.S.Peirce and J.Jastrow (1884), is the fundamental decoupling
technique used in the design of statistical  experiments (DSEs).
 Nevertheless, only after the work of R.A.Fisher (1935), were     
randomized designs used regularly in practice. Today, randomization 
is one of the   basic backbones of statistical theory and methods. 
 Meanwhile, the pioneering work of Peirce had been virtually forgotten
by the Statistics community, until rediscovered by the historical
research of  Stigler (1978) and Hacking (1988).
 Nevertheless, even today, the work of Peirce is presented as an
isolated and ad hoc contribution.
 As briefly indicated in section 5, it is plausible that Peirce and
Jastrow's experimental and methodological work could have had
motivations related to more general ideas of Peircean philosophy. 
 In particular, we believe that the faint effects psychophysical 
hypothesis can be liked to the concept of synechism, while the 
randomized design solution can be embedded in the epistemological
framework of Peirce's objective idealism. 
 We believe that these topics deserve the attention of further research.

 In this chapter we have examined some aspects of DSEs, such as blocking,
control and randomization, from an epistemological  perspective. 
 However, in many applications, most noticeably in medical studies, 
several other aspects have to be taken into account, including the well 
being of the patients taking part in the study. 
 In our view, such complex situations require a thorough, open and honest 
discussion of all the moral and ethical aspects involved. 
 Typically they also demand sound protocols and complex statistical 
models, suited to the fine quantitative analyses needed to   
balance multiple objectives and competing  goals. 
  %
  For the Placebo, Nocebo, Kluge Hans, and similar effects, and the 
 importance of blinding and randomization in  clinical trials, 
 see Kotz et al. (2005), under the entries 
 Clinical Trials I, by N.E.Breslow, v.2, p.981-989, and 
 Clinical Trials II, by  R.Simon,   v.2, p.989-998.   
 For additional references on statistical randomization procedures,  
 see Folks (1984), Kadane and Seidenfeld (1990), 
 Kaptchuk and Kerr (2004), Karlowski et al. (1975), 
 Kempthorne (1977, 1980), Noseworthy et al. (1994),   
 Pfeffermann et al. (1998) and Skinner and Chambers (2003). 

%% file: CAPE4.TEX
  \chapter{Metaphor and Metaphysics:  
         The Subjective Side of Science}


 {\flushright 
  {\it  
  ``Why? -  That is what my name asks! \mbox{} \\
    And there He blessed him.''  \mbox{}  \\ 
  {\em Genesis, XXXII, 30.} \ \  \mbox{} \\ 
  }
 }

 {\flushright 
  {\it  
  ``Metaphor is perhaps one of man's most fruitful potecialities. \mbox{} \\
  Its efficacy verges on magic, and it seems a tool for creation \ \mbox{} \\ 
  which God forgot inside His creatures when He made them.'' \mbox{}  \\ 
  {\em Jos\'{e} Ortega y Gasset, The Dehumanization of Art, 1925.}  
  \ \  \mbox{} \\ 
  }
 }

 {\flushright 
  {\it  
 ``There is nothing as practical as a good theory.'' \mbox{} \\ 
  {\em Attrituted to Ludwig Boltzmann (1844-1906).} \ \  \mbox{} \\ 
  }
 }



 \mbox{}

 \section{Introduction}

 In this chapter we proceed with the exploration of the Cognitive
Constructivism  epistemological framework (Cog-Con), continuing the
previous work developed in previous chapters,  
 and briefly reviewed in section 5.    
 In the previous chapters, we analyzed questions concerning {\it How} 
objects (eigen-solutions) emerge, that is, How they (eigen-solutions)
become known in the interaction processes of a system with its environment. 
 These questions had to do with laws, patterns, etc., expressed as sharp
or precise hypotheses, and we argued that statistical hypothesis testing
plays an important role in their validation.

 It is then natural to ask - {\it Why?}  
 Why do these objects are (the way they are) and interact the way they do? 
 Why-questions claim for a causal nexus in a chain of events. 
 Therefore, their answers must be theoretical constructs based on
interpretations of the laws used to describe these events. 
 This chapter 
is devoted to the investigation of these issues. Likewise, the interplay
between the How and Why levels of inquiry which, in the constructivist
perspective, are not neatly stratified in separate hierarchical layers,
but interact in complex (often circular) patterns, will also be
analyzed. As in the previous chapters, the discussion is illustrated by
concrete mathematical models. In the process, we raise some interesting
questions related to the practice of statistical modeling.

 Sections 2 examines the dictum ``Statistics is Prediction''. 
 The importance of accurate prediction is obvious for any 
statistics practitioner, but is that all there is?     
 The investigation on the importance of model interpretability begins in
section 3, the rhetorical power of mathematical models, self-fulfilling
prophecies and some related issues are discussed and a practical
consulting case in Finance, concerning the detection  of trading
opportunities for intraday operations in both the {\it BOVESPA} and {\it
BM\&F} financial markets is presented. In this example, the REAL
classification tree algorithm, a statistical technique presented in
Lauretto et al. (1998), is used.

 Section 4 is devoted to the issue of language dependence. Therein, the
investigation on model interpretability continues with an analysis of
the eternal counterpointing issues of models for prediction and models
for insight. An example from Psychology, concerning dimensional
personality models is also presented. These models are based on a
dimension reduction technique known as Factor Analysis.

In section 6, the necessary or ``only world'' vs. optimal or ``best
world'' formulations of optics and mechanics are discussed. Simple
examples related to the calculus of variations, are presented, which
abridge the epistemological discussion in the following sections.
Section 7 discusses efficient and final causal relations, teleological
explanations, necessary and best world  arguments, and the possibility
or desirability of having  multiple interpretations for the same model
or multiple models for the same phenomenon. In section 8, the form of
modern metaphysical arguments in  the construction of physical theories
is addressed.

In section 9, some simple but widely applicable models based on
averages computed over all ``possible worlds'', or more
specifically, path integrals over all possible trajectories of a
system, are presented. The first example in this section relates to the
linear system Monte Carlo solution to the Dirichlet problem, a technique
driven by a stochastic process known as Gaussian Random Walk or Brownian
Motion. Section 9 also points out to a generalization of this process
known as Fractional Brownian Motion.
  In sections 7 to 9 we also try to examine the interrelations between 
 ``only world'',  ``best world'' and ``possible worlds''  forms of 
explanation, as well as their role and purpose in the light of
cognitive constructivism,  since they are at the core of modern
metaphysics.

Section 10 discusses how hypothetical models, mathematical equations,
etc., relate to the ``true nature'' of ``real objects''. The importance
of this relationship in the history of science is illustrated therein
with two cases: The Galileo affair, and the atomic or molecular
hypothesis, as presented by L.Boltzmann, A.Einstein and J.Perrin. In
section 11 our final remarks are presented.

 All discussions in the paper are motivated with illustrative 
examples, and these examples follow an approximate order from soft to 
hard science. The example of Psychology presented at section 4, together
with the corresponding Factor Analysis modeling technique, is at an
intermediate point  of this soft-hard scale, making it a natural place 
for making a pause, taking a deep breath, and trying to get a bird's 
eye view of the panorama. 
 Section 5 reviews some concepts of Cog-Con ontology defined in previous
chapters, and discusses some insights on Cog-Con metaphysics.

 \section{Statistics is Prediction. Is that all there is?}

As a first example for discussion, we present a consulting case in
finance. The goal of this project was to implement a model for the
detection of trading opportunities for intraday operations in both the
{\it BOVESPA} and the {\it BM\&F} financial markets. For details we
refer to Lauretto et al. (1998). The first algorithms implemented were
based on Polynomial Networks, as presented in Farlow (1984) and Madala
and Ivakhnenko (1994), combined with standard time series pre-processing
analysis techniques such as de-trending, de-seasonalization,
differencing, stabilization and linear transformation, as exposed in Box
and Jenkins (1976) and Brockwell and Davis (1991). 
 A similar model is presented in Lauretto et al. (2009). 
 The predictive power of the Polynomial Network model was considered
good enough to render a profitable return / risk performance.
 
 According to the decision theoretic theory, 
 and its gambling metaphor  
 as presented in section 1.5, 
 the fundamental purpose of a statistical model 
is to help the user in a specific gambling operation, or decision problem.  
 Hence, at least according to the orthodox Bayesian view, predictive 
power is the basic criterion to judge the quality of a statistical model.  
 This conclusion is accepted with no reservations by most experts in 
decision theory, orthodox Bayesian epistemologists, and even by many  
general practitioners. 
 As typical examples, consider the following statements: 
 \begin{quotation} 
 {\it ``We assume that the primary aim of [statistical] analysis is 
  prediction.''} Robert (1995, p.456). 

 {\it ``Although association with theory is reassuring, it does not 
 mean that a statistical fitted model is more true or more useful. 
 All models should stand or fall based on their predictive power.''} 
 Newman and Strojan (1998, p.168).

 {\it ``The only useful function of a statistician is to make predictions, 
 and thus to provide a basis for action.''} 
 W.E.Demming, as quoted in W.A.Wallis (1980).

 {\it ``It is my contention that the ultimate aim of any statistical
   analysis is to forecast, and that this determines which techniques 
   apply in particular circumstances... 
    The idea that statistics is all about making forecasts based on  
   probabilistic models of `reality' provides a unified approach to the
   subject. In the literary sense, it provides a consistent authorial
   `voice'... 
   the underlying purpose, often implicit rather than explicit, of every 
   statistical analysis is to forecast future values of a variable.''} 
   A.L.McLean (1998). 
 \end{quotation}

 Few theaters of operation so closely resemble a real casino as the 
stock market, hence, we were convinced that our model would be a success.   
 Unfortunately, our Polynomial Network model was not well accepted by
the client, that is, it was seldomly used for actual trading. 
 The main complaint was the model's lack of interpretability. 
 The model was perceived as cryptic, a ``black box'' capable of
selecting strategic operations and computing predicted margins and
success rates, but incapable of providing an explanation of {\it Why} 
the selection was recommended in the particular juncture. 
 This state of affairs was quite frustrating indeed: 
 First, the client had never explicitly required such functionality
during the specification stage of the project, hence the model was 
not conceived to provide explanatory statements.  
 Second, as a fresh Ph.D. in Operations Research, I was well trained in
the minutiae of Measure Theory and Hilbert Spaces, but had very little 
experience on how to make a model that could be easily interpreted by
somebody else.  
 Nevertheless, since (good) costumers are always right, a second model
was specified, developed and implemented, as explained in the next 
section.

 \section{Rhetoric and Self-Fulfilling Prophecies}

 The first step to develop a new model for the problem presented in the
last section, was to find out what the client meant by an interpretable
model. After a few brainstorm sessions with the client, we narrowed
it down to two main conditions: understandable I/O  and  understandable
rules. The first condition (understandable I/O) called for the model's
input and output data to be already known, familiar or directly
interpretable. The second condition (understandable rules) called for
the model's transformation functions, re-presentation maps or derivation
rules to be also based in already known, familiar or directly
interpretable principles.

 Technical Indicators, derived from pre-processed price and volume trading
data, constituted the input to the second model. Further details on their
nature will be given later in this section. For now, it is enough to
know that they are widely used in financial markets, and that the client
possessed ample expertise in technical analysis. 
 The model's statistical data processing, on the other hand, was based
on a classification tree algorithm specially developed for the
application - the Real Attribute Learning Algorithm, or REAL, as
presented in Lauretto et al. (1998). 
 For general classification tree algorithms, we refer to Breiman (1993),
 Denison et al. (2002),  Michie et al. (1994), Mueller and Wysotzki
(1994), and Unger and Wysotzki (1981).

 The REAL based model turned out to be very successful. In fact,
statistically, it performed almost as well as the Polynomial Network
model, under the performance metric specified in Lauretto et al. (1998).
 Moreover, when combined with a final interpretive analysis and go-ahead
decision from the traders, the REAL based model performed better than
the Polynomial Network model. The model was finally put into actual use,
once it was perceived as interpretable and understandable. Since a large
part of our consulting fees depended on the results in actual trading,
this was an important condition for getting fair economical compensation
for all this intellectual endeavor.

 As already mentioned, we were intrigued at the time (and still are) by
many aspects related to model interpretation and understanding. In this
section we begin to analyse this and other similar issues. Concerning
first the very need for explanations: Humans seem to be always avid for
explanations. They need them in order to carry out their deeds, and they
want them to be based on already known schemata, as acknowledged in
Damodaran (2003,ch.7,p.17):

  \begin{quotation} 
 {\it ``The Need for Anchors:
 When confronted with decisions, it is human nature to begin with the
familiar and use it to make judgments. ...

 The Power of the Story: For better or worse, human actions tend to be
based not on quantitative factors but on story telling. People tend to
look for simple reasons for their decisions, and will often base their
decision on whether these reasons exist.''}
 \end{quotation} 

 The rhetorical purpose and power of statistical models have been able
to conquer, within the statistical literature, only a small fraction of
its relative importance in the consulting practice. There are,
nevertheless, some remarkable exceptions, as see for example, in 
 Abelson (1995, p.xiii):
  \begin{quotation} 
 {\it 
 ``The purpose  of statistics is to organize a useful argument from
quantitative evidence,  using a form of principled rhetoric. 
 The word {\em principled} is crucial. Just because rhetoric is
unavoidable, indeed acceptable, in statistical presentations does not
mean that you should say anything you please.'' 
 }

 {\it ``Beyond the rhetorical function, statistical analysis also has a
narrative role. Meaningful research tells a story with some point to it,
and statistics can sharpen the story.''}
 \end{quotation} 
 
 Let us now turn our attention to the inputs to the REAL based model,
the Technical Indicators, also known as Charting Patterns. For a general
description, see Damodaran (2003,ch.7). For some of the indicators used
in the REAL project, see Colby (1988) and Murphy (1986). Technical
indicators are primarily interpreted as behavioral patterns in the
markets or, more appropriately, as behavioral patterns of the market
players. Damodaran defines five groups that categorize the indicators
according to the dominant aspects of the behavioral pattern. A concise
description of these five groups of indicators is given in Damodaran
(2003,ch.7,p.46-47):
 
 \begin{quotation} 
 1 -  External Forces / Large Scale Indicators: 
 {\it ``If you believe that there are {\em long-term} cycles in stock
prices, your investment strategy may be driven by the cycle you
subscribe to and where you believe you are in the cycle.''}

 2 - Lead / Follow Indicators: 
 {\it ``If you believe that there are some traders who trade ahead of
the market, either because they have better analysis tools or
information, your indicators will follow these traders
 - specialist short sales and insider buying/selling, for instance - 
 with the objective of piggy-backing on their trades.''}

 3 - Persistence / Momentum Indicators: 
 {\it ``With momentum indicators, such as relative strength and trend
lines, you are assuming that markets often learn slowly and that it
takes time for prices to adjust to true values.''}

 4 - Contrarian / Over Reaction Indicators: 
 {\it``Contrarian indicators such as mutual fund holdings or odd lot
ratios, where you track what investors are buying and selling with the
intention of doing the opposite, are grounded in the belief that markets
over react.''}

 5- Change of Mind / Price-Value Volatility Indicators: 
 {\it ``A number of technical indicators are built on the presumption
that investors often change their views collectively, causing shifts in
demand and prices, and that patterns in charts - support and resistance
lines, price relative to a moving average- can predict these changes.''}
  \end{quotation} 

 At this point, it is important to emphasize the dual nature of
technical indicators: They disclose some things that may be happening
with the trading market and also some things that may be happening with
the traders themselves. In other words, they portray dynamical patterns
of the market that reflect behavioral patterns of the traders.




 Two characteristics of the REAL based model, of vital importance to the
success in the consulting case presented, relate to rhetorical and
psychological aspects that have been commented so far:

 - Its good predictive and rhetorical  power, which motivated the client
to trade on the basis of the analyses provide by the model;

 - The possibility of combining and integrating the analyses provided by
the model with expert opinion.

 Technical indicators often carry the blame of being based in
self-fulfilling prophecies, over-simplified formulas, superficial and
naive behavioral patterns, unsound economic grounds, etc. From a
pragmatic perspective, market analysts do not usually care about
technical analysis compatibility with sound economic theories,
mathematical sophistication, etc. Its ability to detect trading
opportunities is what counts. From a conceptual perspective, each of
these analyses does tell a story about a cyclic reinforcement or
correction (positive or negative feed-back) mechanism in the financial
system. What is peculiar about self-fulfilling prophecies is that the
collective story telling activity is a vital link in the feed-back
mechanism. It is not surprising then that the market players' perception
of how good the story itself looks should play an important role in
fortelling whether the prophecy will come true. 
 From this perspective one can understand the statement in 
 Murphy (1986, p.19):

  \begin{quotation} 
 {\it ``The self-fulfilling prophecy (argument) is generally listed as a
criticism of charting. It might be more appropriate to label it as a
compliment.''}
 \end{quotation} 

 The importance of the psychological aspects of the models studied in
this section motivate us to take a look, in the sequel, at some
psychological models of personality.

 \section{Language, Metaphor and Insight}

 In chapter 1, the dual role played by Statistics in scientific
research, namely, predicting experimental events and testing hypotheses,
was pointed out. It was also emphasized that, under a constructivist
perspective, these hypotheses are often expressed as equations of a
mathematical model. In the last section we began to investigate the
importance of the interpretability of these models. 
 The main goal of this section is to further investigate subjective
aspects of a statistical or mathematical model, specifically, the
understanding or {\it insight} it provides.

 We start with three diffent versions of the well-known motto of Richard
Hamming: 

 - {\it ``The purpose of models is insight, not numbers.''}

 - {\it ``The purpose of computing is insight, not numbers.''}

 - {\it ``The purpose of numbers is insight, not numbers.''}

 Dictionary definitions of Insight include:

 - A penetrating, deep or clear perception of a complex situation;

 - Grasping the inner or hidden nature of things;

 - An intuitive or sudden understanding.

 The illustrative case presented in this section is based on
psychological models of personality. Many of these models rely on
symmetric configurations known as ``mandala'' schemata, see for example
Jung (1968), and a good example is provided by the five elements model
of traditional Chinese alchemy and their associated personality traits:

 1- Fire: Extroverted, emotional, emphatic, self-aware, sociable,
eloquent.

 2- Earth: Caring, supporting, stable, protective, worried, attached.

 3- Metal: Analytical, controlling, logical, meticulous, precise, zealous.

 4- Water: Anxious, deep, insecure, introspective, honest, nervous.

 5- Wood: Angry, assertive, creative, decisive, frustrated, leading.

 Interactions between elements are conceived as a double feed-back cycle,
represented by a pentagram inscribed in a pentagon. The pentagon  
or external cycle represent the creation, stimulus or positive feed-back
in the system, while the pentagram or internal cycle represent
the destruction, control or negative feed-back in the system. The
traditional representation of these systemic generative mechanisms or
causal relations are:

 Pentagon:
 fire $\left[ calcinates\ to\ \right>$ earth
 $\left[harbors\right>$ metal $\left[condenses\right>$ water
 $\left[nourishes\right>$ wood $\left[fuels\right>$ fire.

 Pentagram:
 fire $\left[melts\right>$ metal $\left[cuts\right>$ wood
 $\left[incorporates\right>$ earth $\left[absorbs\right>$
 water \\ $\left[extinguishes\right>$ fire.

 This double feed-back structure allows the representation of system
with complex interconnections and nontrivial dynamical properties. In
fact, the systemic interconnections are considered the key for
understanding a general five-element model, rather than any superficial
analogy with the five elements' traditional labels. 

 It is an entertaining exercise to compare and relate the five
alchemical elements listed above with the five groups of technical
indicators presented in the last section, or with the big-five
personality factors presented next,  
 even if some of these models are considered pre-scientific. 
 Why, for example, do these models employ exactly five factors? 
 That is, why is it that ``four are few and six are many''? 
 Is there an implicit mechanism in the model, 
 see Hargittai (1992), Hotchkiss (1998) or Philips (1995, ch.2), 
 or is this an empirical statement supported by research data?   

 Scientific psychometric models must be based on solid statistical
analysis of testable hypotheses. Factor Analysis has been one of the
preferred techniques used in the construction of modern psychometric
models and it is the one used in the examples we discuss next. 
 In section C.5, the basic structure of factor analysis
statistical  models is reviewed.

 In Allport and Odbert (1936) the authors presented their Lexical
Hypothesis. According to them, important aspects of human life
correspond to words in the spoken language. Also the number of
corresponding terms in the lexicon is supposed to reflect the importance
of each aspect:

  \begin{quotation} 
 {\it ``Those individual differences that are most salient and socially
relevant in peoples lives will eventually become encoded into their
language; the more important such a difference, the more likely is it to
become expressed as a single word.''}
 \end{quotation} 

 One of the most widely used factor model takes into account five
factors or personality traits. These are the five dimensions of the
``OCEAN'' personality model or ``big-five''. Further details on the
meaning of these factors can be found in Shelder and Westen (2004), from
the list of the most relevant factor loadings. The ``OCEAN'' labels,
ordered according to their statistical relevance, are:

 1- Extraversion, Energy, Enthusiasm;

 2- Agreeableness, Altruism, Affection;

 3- Conscientiousness, Control, Constraint;

 4- Neuroticism, Negative Affectivity, Nervousness;

 5- Openness, Originality, Open-mindedness.

 Subsequent studies pointed to the ``existence'' of more factors, for a
review of several of such models, see Widiger and Simonsen (2005).
Herein, we focus our attention in the 12-factor model of Shelder and
Westen. We remark, however, that the publication of the 12-factor
model, fired an inflamed literary debate concerning the necessity 
(or not) of more than 5 factors. 
 In the quotation below, Shelder and Westen (2004, p.1752-1753) pinpoint
the issue of language dependence  in the description of reality, an
issue of paramount importance in cognitive constructivism and one of the
main topics analyzed in this section.

  \begin{quotation} 
 {\it ``Applying the Lexical Hypothesis to Personality Disorders:

 Ultimately, the five-factor model is a model of personality derived
from the constructs and observations of lay-people, and it provides an
excellent map of the domains of personality to which the average
layperson attends. However, the present findings suggest that the
five-factor model is not sufficiently comprehensive for describing
personality disorders or sophisticated enough for clinical purposes.

 In contrast to laypeople, practicing clinicians devote their
professional lives to understanding the intricacies of personality. They
develop intimate knowledge of others lives and inner experience in ways
that may not be possible in everyday social interaction. Moreover, they
treat patients with variants of personality pathology that laypeople
encounter only infrequently (and are likely to avoid when they do
encounter it). One would therefore expect expert clinicians to develop
constructs more differentiated than those of lay observers.

 Indeed, if this were not true, it would violate the lexical hypothesis
on which the five-factor model rests: that language evolves over time to
reflect what is important. To the extent that mental health
professionals observe personality with particular goals and expertise,
and observe the more pathological end of the personality spectrum, the
constructs they consider important should differ from those of the
average layperson.''}
  \end{quotation} 

 The issue of language dependence is very important in cognitive
constructivism. For further discussion, see Maturana (1988, 1991).
 Thus far we have stressed the lexical aspect of language, that is, the
importance of the available vocabulary in our description of reality. 
 In the remaining part of this section we shall focus on the symbolic
or figurative use of the language constructs in these descriptions. We
proceed by examining in more detail the factor analysis model.

 Factor analysis is a dimension reduction technique. 
 Its application renders a `simple' object, the factor model,
capable of efficiently ``coding'', into a space of reduced dimension,
a complex `real' object from a full or high dimensional space. 
 In other words, a dimension reduction technique presumes some form of 
valid knowledge transference, back and forth the complex  (high
dimensional) object and its simple (low dimensional) model. 
 Hence, the process of using and interpreting factor analysis
models can be conceived as metaphorical. 
 Recall that the Greek word
metaphor stands for transport or transfer, so that a linguistic metaphor
transfers some of the characteristics of one object, called the source
or vehicle, into a second distinct object, called the target, tenant or
topic; for a comprehensive reference see Lakoff and Johnson (2003).

 For reasons which are similar to those studied in the last section,
most users of a personality model require it to be statistically sound.
Many of them further demand it to be interpretable, in order to provide
good insights to their patient's personality and problems. A good model
should not only be useful in predicting recovery rates or drug
effectiveness, but also help in supplying good counseling or
therapeutics.

 Paraphrasing Vega-Rodr\'{\i}guez (1998):

 {\it The metaphorical mechanism should provide an articulation point
between the empirical and the hypothetical, the rational and the
intuitive, between calculation and insight.}

 The main reason for choosing factor analysis to illustrate this section
is its capability of efficiently and transparently building sound
statistical models that, at the same time, provide intuitive 
interpretations. 
 While soundness is the result of ``estimation and identification
tools'', such as ML (maximum likelihood) or MAP (maximum a posteriori)
optimization, hypothesis testing and model selection, interpretableness
results from ``representation tools'', such as orthogonal and oblique
factor rotation techniques.

 Factor rotation tools are meant to reconfigure the structure of a given
factor analysis model, so as to maintain its probabilistic explanatory
power while maximizing its heuristic explanatory power.  
 Factor rotations are performed to implement an objective optimization
criteria, such as sparsity or entropy maximization. The optimal solution
(for each criterion) is unique and hoped to enhance model
interpretability a great deal.

 \section{Constructive Ontology and Metaphysics}

 How important heuristic arguments are in other areas of science? 
 Should statistical or mathematical models play a similar rhetorical
role in other fields of application? 
 We will try to answer these questions by discussing the role played by
similar heuristic arguments in physics. 
 In sections 2 and 3 we dealt with application areas in which  text(ure)
 manufacture comprised, to a great extent, the very spinning of the
threads. 
 Nevertheless, one can have the false impression that the constructivist
approach suits better high level, soft science areas, rather than low level,
rock bottom Physics. This widely spread misconception is certainly not
the case. In sections 7 through 10 we analyze the role played in science by
metaphysics, a very special form of heuristic argumentation.

 The example  presented in section 4, together with the corresponding
Factor Analysis modeling technique, is at an intermediate point  of the
soft-hard science scale used herein to (approximately) order the examples. 
 Therefore, as previously stated in the introduction, we shall use 
 section the current section to make a pause in the exposition, take a
deep breath, and try to get a bird's eye view of the scenario. 
 This section also  reviews some concepts of Cog-Con ontology defined in
previous chapters and discusses some insights on Cog-Con metaphysics.

 The Cog-Con framework rests upon two basic metaphors: 
  the Heinz von Forster's metaphor of 
 {\it Object as token for an eigensolution}, which is
 the key to Cog-Con ontology, and    
 the Humberto Maturana and Francisco Varela's metaphor
 of {\it Autopoiesis and cognition}, 
 the key to Cog-Con metaphysics. 
 Below we review these two metaphors, as they where used in chapter 1.

 \subsubsection*{Autopoiesis and Cognition}

 Autopietic systems are non-equilibrium (dissipative) dynamical systems
exhibiting (meta) stable structures, whose organization remains 
invariant over (long periods of) time, despite the frequent substitution
of their components. 
 Moreover, these components are produced by the same structures they 
regenerate. 
 As an example, take the macromolecular population of a single cell,
which can be renewed thousands of times during its lifetime, see
Bertalanffy (1969). 
 However, in spite of the fact that autopoiesis was a metaphor developed
to suit the essential  characteristics of organic life, the concept of
autopoietic  system has been applied in the analysis of many other
concrete or abstract  autonomous systems such as social systems and
corporate organizations, 
 see for example Luhmann (1989) and Zelleny (1980).

 The regeneration processes in the autopoietic system production network 
require the acquisition of resources such as new materials, 
energy and neg-entopy (order), from the system's environment. 
 Efficient acquisition of the needed resources demands selective 
(inter)actions which, in turn, must be based on suitable
inferential processes (predictions). 
 Moreover, these inferential processes characterize the agent's 
domain of interaction as a cognitive domain. 
 For more details see the comments in chapter 1 and, more
importantly,  the original statements in  
 Maturana and Varela (1980, p.10):

  
  \begin{quote} 
 {``The circularity of their organization continuously brings them back
to the same internal state (same with respect to the cyclic process). ... 
 Thus the circular organization implies the prediction that an 
interaction that took place once will take place again. ...    
 Accordingly,   the predictions implied in the organization of the living
system are not predictions of particular events, but of classes of
inter-actions. ... 
 This makes living systems, inferential systems, and their domain
of interactions a cognitive domain.''}  
  \end{quote}

 \subsubsection*{Object as Tokens for Eigen-Solutions}

 The circular (cyclic or recursive) characteristic of autopoietic 
regenerative processes and their eigen (auto, equilibrium, fixed,
homeostatic, invariant, recurrent, recursive) -states, both in concrete
and abstract autopoietic systems, are investigated in 
 Foerster (2003) and Segal (2001).

  \begin{quote} 
 { ``The meaning of recursion is to run through one's own path again. 
 One of its results is that under certain conditions there exist indeed
solutions which, when reentered into the formalism, produce again the
same solution. These are called ``eigen-values'', ``eigen-functions'',
``eigen-behaviors'', etc., depending on which domain this formation is
applied - in the domain of numbers, in functions, in behaviors, etc.''} 
 Segal (2001, p.145).  
 \end{quote} 

 The concept of eigen-solution for an autopoietic system is the key to
distinguish specific objects in a cognitive domain.

  \begin{quote} 
 { ``Objects are tokens for eigen-behaviors. Tokens stand for
something else. 
 In exchange for money (a token itself for gold held by
 one's government, but unfortunately no longer redeemable), tokens are
 used to gain admittance to the subway or to play pinball machines. 
 In the cognitive realm, objects are the token names we give to our
 eigen-behavior. ...     
  When you speak about a ball, you are talking about the
  experience arising from your recursive sensorimotor behavior when
  interacting with that something you call a ball. The ``ball'' as object
  becomes a token in our experience and language for that behavior which
  you know how to do when you handle a ball. 
 This is the constructivist's insight into what takes place when we 
talk about our experience with objects.''} 
 Segal (2001, p.127). 
 \end{quote} 

 Furthermore, von Foerster establishes four essential attributes of
eigen-solutions: 

 \begin{quote} 
 { ``Eigenvalues have been found ontologically to be discrete, stable,
separable and composable, while ontogenetically to arise as equilibria
that determine themselves through circular processes. Ontologically,
Eigenvalues and objects, and likewise, ontogenetically, stable behavior
and the manifestation of a subject's ``grasp'' of an object cannot be
distinguished.''} 
 Foerster (2003, p.266). 
   \end{quote}

 \subsection*{Constructive Ontology}

 The Cog-Con framework also includes the following conception of
reality and some related terms, as defined in chapter 2: 

 \begin{quotation} 
 \noindent {\it 1. Known (knowable) Object:} An actual (potential)
eigen-solution of a given system's interaction with its environment. 
 In the sequel, we may use a somewhat more friendly terminology by 
simply using the term Object.
 
 \noindent {\it 2. Objective (how, less, more):} Degree of conformance
of an object to the essential attributes of an eigen-solution 
 (to be precise, stable, separable and composable). 

 \noindent {\it 3.  Reality:} A (maximal) set of objects, as
recognized by a given system, when interacting with single objects or
with compositions of objects in that set. 
 \end{quotation}

 The Cog-Con framework assumes that an object is 
always observed by an observer, just like a living organism or a more 
abstract system, interacting with its environment. 
 Therefore, this framework asserts that the manifestation of the
corresponding eigen-solution and the properties of
the object are respectively driven and specified by both the system and its environment. 
 More concisely, Cog-Con sustains: 

 \begin{quotation}
 \noindent {\it 4. Idealism:} The belief that a system's knowledge of an 
object is always dependent on the systems' autopoietic relations. 

 \noindent {\it 5. Realism:} The belief that a system's knowledge of an 
object is always dependent on the environment's constraints. 
 \end{quotation}

  Consequently, the Cog-Con perspective requires a fine equilibrium, 
called  {\it Realistic or Objective Idealism}. 
 {\it Solipsism or Skepticism} are symptoms of an epistemological
analyses that loose the proper balance by putting too much weight on 
the idealistic side. Conversely, 
 {\it Dogmatic Realism} is a symptom of an epistemological
analyses that loose the proper balance by putting too much weight on 
the realistic side.  
 Dogmatic realism has been, from the Cog-Con perspective, 
a very common (but mistaken) position in modern epistemology. 
 Therefore, it is useful to have a specific expression, namely, 
 {\it something in itself} to be used as a marker or label for such ill posed dogmatic statements. 
 The method used to access   
 something in itself is often described as: 
 - Something that an observer would observe if the (same)
observer did not exist, or   
 - Something that an observer could observe if he made no
observations, or  
 - Something that an observer should observe in the environment
without interacting with it (or disturbing it in any way),      
 and many other equally senseless variations.

 \begin{table}[tb] 
 \begin{center}
 \begin{tabular}{c c c c c} 

 Experiment & &     & & Theory \\  \\   

 Operation-  & $\Leftarrow$ & Experiment  

                       &  $\Leftarrow$ &  Hypotheses   \\ 
 alization & & design   & &  formulation                \\ 
   $\Downarrow$   & &               & & $\Uparrow$        \\  
 Effects       &  \multicolumn{3}{c}{True/False} & Creative  \\ 
 observation    &  \multicolumn{3}{c}{eigen-solution} & interpretation  \\     

   $\Downarrow$   & &               & & $\Uparrow$         \\ 
 Data    & & Mnemetic & &  Statistical   \\ 
 acquisition  & $\Rightarrow$ & explanation  
 & $\Rightarrow$ & analysis \\  \\ 
 
 \multicolumn{2}{l}{Sample space} & & 
 \multicolumn{2}{r}{Parameter space} 
 \end{tabular} 
 \mbox{} \\ \mbox{} \\ 
 \centerline{Figure 1: Scientific production diagram.}
 \end{center}    
 \end{table}

 Although the application of the Cog-Con framework is as general 
as that of autopoiesis, this paper is focused on scientific activities.  
 The interpretation of scientific knowledge as an eigensolution of a
research process is part of a Cog-Con approach to epistemology.
 Figure 1 presents an idealized structure and dynamics of knowledge
production, see Krohn and K\"{u}ppers (1990) and chapters 1 and 6.   
 The diagram represents, on the Experiment side (left column) the  
laboratory or field operations of an empirical science, where experiments
are designed and built, observable effects are generated and
measured, and an experimental data bank is assembled. 
 On the Theory side (right column), the diagram represents the
theoretical work of statistical analysis, interpretation and (hopefully)
understanding according to accepted patterns. If necessary, new
hypotheses (including whole new theories) are formulated, motivating the
design of new experiments. 
 Theory and experimentation constitute a double feed-back cycle making it clear
that the design of experiments is guided by the existing theory and its
interpretation, which, in turn, must be constantly checked, adapted or
modified in order to cope with the observed experiments. The whole
system constituting an autopoietic unit.


 \subsubsection*{Fact or Fiction?}

 At this point it is useful to (re)turn our attention to a specific model, 
 namely, factor analysis, as discussed in section 4, and consider the 
 following questions raised by Brian Everitt (1984, p.92, emphases are ours) 
 concerning  the appropriate interpretation of factors:

 \begin{quote}   
 {`` {\bf Latent variables - fact or fiction?}  
 One of the major criticisms of factor analysis has been the 
tendency for investigators to give names to factors, and 
subsequently, to  imply that these  factors 
 {\bf have a reality of their own over and above the  manifest variables.} 
 This tendency continues with the use of  
 {the term latent variables} since it 
 {suggests that they are existing variables} and that there 
 is simply a problem of how they should be measured. 
 In truth, of course, latent variables will never be anything 
more than is contained in the observed variables and will never 
be anything beyond what has been specified in the model. 
 For example, in the statement that verbal ability is whatever 
certain test have in common, the empirical meaning is nothing 
more than a shorthand for the observations of the correlations. 
 It does not mean that verbal ability is a variable that is measurable 
in any manifest sense. 
 However, the concept of latent variable may still be extremely helpful.
 A scientist may have a number of hypothetical constructs in terms of 
which some theory is formulated, and he is willing to assume that the 
latent variables used in specifying the structural models of interest 
are the operational equivalents to theoretical constructs. 
 As long as it is remembered that in most cases there is no empirical 
way to prove this correspondence, then such an approach can lead to 
interesting and informative {\bf theoretical insights}.''} 
 \end{quote}

 Ontology is a term used in philosophy in reference to a systematic
account of {\it existence} or {\it reality}. 
 We have already established the Cog-Con approach to objects as 
tokens for eigen-solutions, and explained their four essential 
attributes, namely, discreteness (preciseness, sharpness or exactness), 
stablity, separability and composability.
 Therefore, in the Cog-Con framefwork, accessing the ontological 
status of an object, or to say how objective it is, is to 
ascertain how well it manifests the four essential attributes 
of an eigen-solution.

  The Full Bayesian Significance Test, or FBST, is a  possibilistic
 belief calculus, based on (posterior) probabilistic measures, that was 
 conceived as a statistical significance test to access the 
 objectivity of an eigen-solution, that is, to measure how well a 
 given object manifests or conforms to von Foerster's four essential 
 attributes. 
  The FBST belief or credal value, $\mbox{ev}(H \g X)$, 
 the  {\it e-value} of hypothesis $H$ given the observed data $X$, 
 is interpreted as 
 the {\it epistemic value} of hypothesis $H$ (given $X$), or  
 the {\it evidence value} of data $X$ (supporting $H$). 
  The formal definition of the FBST and several of its implementation in 
 specific problems can be found in the author's previous publications, 
 and are reviwed in appendix A.

 \subsubsection*{Greek or Latin? Latent or Manifest?} 

 We have already discussed the ontological status of an object.  
 This discussion assumes testing hypotheses in a statistical model which, 
 in order to built, one must know how to distinguish 
 concrete measurable entities from abstract concepts, 
 observed values from model parameters, 
 latent from manifest variables, etc. 
 When designing and conducting an experiment, 
 a scientist must have a well defined a statistical model,  
 and keep these distinctions crisp and clear.    
 This is so important in the experimental sciences that statisticians  
 have the habit of using Latin letters for observables, 
 and Greek letters for parameters.  
  When a statistician questions whether a letter is
 Latin or Greek, 
 he or she is not asking for help with foreign alphabets, 
 but rather seeking information about the aforementioned distinctions.

 According to the positivist philosophical school,   
 measurable entities, observed values, manifest variables, etc. 
 are the true, first class entities of a hard science, while   
 abstract concepts, model parameters, latent variables, etc. 
 should be considered second class entities. 
 One reason for downgrading the later class is that the positivist 
 school assumes a nominalist perspective. 
  Nominalism (at least in its strictest form) considers abstract concepts 
 as mere names {\it (nomina)},  that may stand as proxy for a ``really
 existing item'', denoting ``that singular thing''  
 {\it (supponere pro illa re singulari)}.  
  The Cog-Con perspective plays no role in the positivist dream. 
  This issue will be further investigated in the next sub-section, 
 as well as in sections 8 and 11. 
 For now we offer the following argument:

 Although for a given model, the aforementioned distinctions  between
what to write using Latin or Greek letters should be  always crisp and
clear, we may have to simultaneously work with several models. 
 For example, we may need to use several models hierarchically organized 
to cope with phenomena at different scales or 
levels of granularity, like models in physics, chemistry, 
biology, and psychology, see chapters 5 and 6. 
 We may also need different models for competing theories trying 
to explain a given phenomenon. 
 Finally, we may need different models providing equivalent or 
compatible laws to given phenomena that, nevertheless, use 
distinct theoretical approaches, see section 8, 9 and 10.     
 The positivist dream quickly turns into a nightmare when one 
realizes that an entity corresponding to a Greek letter variable 
in one model corresponds to a Latin letter variable in another, 
and vice-versa. 

 It is also important to realize that in the Cog-Con approach 
the ontological status of an object is a reference to the properties 
of the corresponding eigen-solution emerging in a cyclic process. 
 This leads to an intrinsically dynamic approach to ontology, in sharp 
contrast with other analyses based on static categories.  
 A consequence of this dynamical setting is that in the Cog-Con approach
a statement about the ontological status of a single element or isolated
component in a process is an indirect reference to its role in the
emergence of the corresponding eigen-solution. 
 Equivalent or similar elements may play very different roles in
distinct processes. Such distinct or multiple roles will not pose
conceptual difficulties to the Cog-Con framework as long as the 
corresponding (statistical) models are clearly stated and well defined. 
 For interesting examples of this situation, typical of modular and 
hierarchical architectures, hypercyclical organization, and emergent 
properties, see chapters 5 and 6.

 \subsection*{Constructive Metaphysics}

 Metaphysics, in its gnosiological sense, is a philosophical term we
use to refer to a systematic account of possible forms of understanding,
valid forms of explanation or rational principles of intelligibility. 
 In science, such explanations are often well represented in a schematic 
diagram describing the organization of a conceptual network. 
 A link in such a diagram expresses a theoretical relation like, for
example, a causal nexus, that is, a cause and effect relation. 
 In modern science, such explanations must also include the symbolic
derivation of scientific hypotheses from general scientific laws, the 
formulation of new laws in an existing theory, and even the conception
of new theories, as well as their general understanding based on
general metaphysical principles.  

 In this context, it is natural to ask questions like: 
 What do we mean by the intuitive quality or theoretical importance
 of a concept or, more generally, of a sub-network?  
 How interesting are the insights we gain from it? 
 How can we access its explanatory power or heuristic value? 
 We will try to answer these questions in the following sections, 
 most specially in section 8, on modern metaphysics. 
  In this section we provide only a preliminary discussion of the importance 
of metaphysical entities in the constructivist perspective.

 We now return to Humberto Maturana and Francisco Varela's metaphor of
autopoiesis and cognition. As stated at the beginning of this section
this metaphor is the key for Cog-Con metaphysics. 
 From details of this metaphor we conclude that the autopoietic
relations of a system not only define  who or what it ``is'', but also
limit the class of interactions in which  it can possibly engage or the
class of events it can possibly perceive. 
 An adaptive system can learn, that is, it can  reconfigure its
internal organization, reshape its architecture,  in order to enlarge
its scope of inference or make better predictions.    
 Nevertheless, learning is an evolutive process, and any evolutionary 
path to the future has to progress from the system's  present (or initial)
configuration. 
 From the above considerations it is clear that, from a constructivist 
perspective, the specification of autopoietic relations are of vital
importance since they literally define the scope and possibilities 
of the system's life.

 \subsubsection*{Theoretical Insights}

 Cog-Con approaches science as an autopoietic system whose organization  
is coded by symbolic laws, causal relations, and metaphysical principles. 
 Consequently, we must give them the greatest importance.   
 Nevertheless, such metaphysical entities are even more abstract
than the latent variables discussed in the last
subsection. In contrast with the constructivist approach, the
positivist school is thus quite hostile to metaphysical concepts.

 In the Cog-Con perspective, metaphysics provides meaning to objects in
a give reality,  explaining  {\it why} the corresponding eigen-solutions
manifest themselves the way they do. 
 Accordingly, theoretical concepts become building blocks in the coding
of systemic knowledge and reference marks in the mapping of the systems
environment. Conceptual relations are translated into inference tools, 
thus becoming, by definition, the basis of autopoietic cognition. 
 In the Cog-Con perspective, better understanding will strengthen a 
given theoretical architecture or entail its evolution.  
 In so doing, the importance of the pertinent concepts is enhanced, 
their scope is enlarged and their utility increased. 
 The whole process enables richer and wider connections in the web of
knowledge, embedding theory even deeper in the system's
life, revealing more links in the great chain of being!

 \section{Necessary and Best Worlds}

 In sections 7 through 10 we analyze the role played in modern science
by metaphysics, a very special form of heuristic argumentation. 
 Such arguments often explain why a system follows a given trajectory 
or evolves along a given path. 
 These arguments may explain why a system must follow a necessary path  
or is effectively forced along a single trajectory; 
 these are ``only world'' explanations. 
 Teleological arguments explain why a system chooses the best trajectory
according to some optimality criterion;  
 these are ``best world'' explanations. 
 Stochastic or integral arguments explain why the system evolution takes 
into account, including, averaging, summing or integrating over, all 
possible or admissible trajectories; 
 these are ``possible worlds'' explanations. 

  In sections 7 to 9 we also try to examine the interrelations between 
 ``only world'',  ``best world'' and ``possible worlds''  forms of 
explanation, as well as their role and purpose in the light of
cognitive constructivism,  since they are at the core of modern
metaphysics.  
 We begin this journey by studying in this section a simple and
seemingly innocent mathematical puzzle.  
 The puzzle, which will be solved directly by elementary
calculus, is in fact used by Richard Feynman as an allegory to present  
an important variational problem. 

 Consider a beach with shore line represented by $x=a$, in the standard 
Cartesian plane. 
 A lifeguard, at position $(x,y)=(0,0)$, spots a person drowning at
position $(x,y)=(a+b,d)$. 
 While on the athletic track the lifeguard car can run at top speed 
 $c$, on the sand it can run at speed $c/\nu_1$. 
 Once in the water, the lifeguard can only swim at speed 
 $c/\nu_2$, $1< \nu_1< \nu_2$. 
 Letting $(x,y)=(a,y(a))$ be the point where he enters the water, what
is the optimal value $y(a)=z$ if he wants to reach position $(a+b,d)$ as
fast as possible?

 Since the shortest path in an homogenous medium is a straight line, the
optimal trajectory is a broken line, from $(0,0)$ to $(a,z)$, and then
from $(a,z)$ to $(a+b,d)$. The total travel time is $J(z)/c$, where
 \[
  J(z)= \nu_1 \sqrt{a^2+z^2} +\nu_2 \sqrt{b^2+(d-z)^2} \ .
 \]
 Since we want $J(z)$ at a minimum, we set
 \[
  \frac{dJ}{dz}=
   \nu_1 \frac{-2z}{2\sqrt{a^2+z^2}}
  +\nu_2 \frac{-2(d-z)}{2\sqrt{b^2+(d-z)^2}} =0 \ ,
 \]
 so that, we should have
 \[
   \nu_1 \sin(\theta_1) = \nu_2 \sin(\theta_2) \ .
 \]

 Professional lifeguards claim that this simple model can be improved by
dividing the sand in a dry band, $V_1$, and a wet band, $V_2$, and the
water in a shallow band, $V_3$, and a deep band, $V_4$, with respective
different media `resistance' indices, 
 $\nu_1 , \nu_2 , \nu_3 , \nu_4$, satisfying 
 $\nu_4> \nu_3> \nu_1 >\nu_2 >1$. 
 Although the solution for the improved model can be similarly obtained,
a general formalism to solve `variational' problems of this kind exists
which is known as the Euler-Lagrange equation. For an instructive
introduction see Krasnov et al. (1973), Leech (1963) and Marion (1970).

 The trigonometric relation, $\nu(x) \sin(\theta)=K$, obtained in the
last equation, is known in optics as Snell-Descartes' law. 
 It explains the refraction (bending) of a light ray incident to a
surface separating two distinct optic media. In this relation, $\nu$ is
the medium refraction index. 
 The variational problem solved above was proposed by Pierre de Fermat
in 1662 to `explain' Snell-Descartes' law. Fermat's {\it principle of
least time} states that a ray of light, going from one point to another,
follows the path which is traversed in the smallest time.

 Notice that Fermat enounced this principle {\it before} any measurement
of the speed of light. The first quantitative estimate of the speed of
light, in sidereal space, was obtained by O. Roemer in 1676. He measured
the Doppler effect on the period of Io, a satellite of Jupiter
discovered by Galileo in 1610. More precisely, he measured the violet
and red shifts, i.e., the variation for shorter and longer in the
observed periods of Io, as the Earth traveled in its orbit towards and
away from Jupiter. 
 Roemer's final estimate was $c=1au/11'$, that is, one astronomial unit
 (the length of the semi-major axis of the earth's elliptical orbit
 around the sun, approximately 150 million kilometres) per $11$ minutes. 
 Today's value is around $1au/8'20''$. 
 The first direct measurements of the comparative speed of
light in distinct material media (air and water) were obtained by 
L\'{e}on J.B.Foucault, almost two centuries latter, in 1850, using a
rotating mirror device. For details, see Tobin (1993) and Jaffe (1960).
For a historical perspective of several competing theories of light we
refer to Ronchi (1970) and Sabra (1981).

 Snell-Descartes' ``law'' is an example of mathematical model that
dictates a ``necessary  world'', stating, plain and simple, how things
``have to be''. In contrast, Fermat's ``principle'' is a theoretical
construct that elects a ``best world'' according to some criterion used
to compare ``possible worlds''.

 Fermat's principle is formulated minimizing the integral of $ds=1/dt$.
In a similar way, Leibniz, Euler, Mauperius, Lagrange, Jacobi,
Hamilton, and many others were able to reformulate Newtonian mechanics,
minimizing the integral of a quantity called {\it action},  
 $ds= L\; dt$, where the Lagrangian, $L$, is the difference between 
 the kinetic energy (Leibniz' vis viva), $(1/2)mv^2$, and 
 the potential energy of the system (Leibniz' vis morta). 
 Hence, these formulations are called in physics principles of minimum
action or principles of least action.

 \section{Efficient and Final Causes}

 At the XVII century, several models of light and its propagation 
were developed to explain Snell-Descartes' law, see Sabra (1981).  
 The discussion of these models, and the necessary versus best world
formulations of optics and mechanics discussed in the last section are
historically connected to the discussion of the metaphysical concepts of
efficient and final causes.

 This terminology dates back to Aristotle, who distinguishes, in
Metaphysics, four forms of causation, that is, four types of answers
that can be given to a Why-question. Namely:

 - Material cause: Because it is made of, or its constituent parts are ...

 - Formal cause: Because it has the form of, or is shaped like ...

 - Efficient cause: Because it is produced, or accomplished by ...

 - Final cause: Because it is intended to, or has the purpose of ...

 Efficient and final causes are the subject of this section. For a
general overview of the theme in the history of 17th and 18th century
Physics, see Brunet (1938), Dugas (1988), Pulte (1989), Goldstine
(1980), Wiegel (1986) and Yourgrau and Mandelstam (1979).

 Newtonian mechanics is formulated only in terms of efficient causes -
an existing force acts on a particle (or body) producing a movement
described by the Newtonian differential equations. Least action
principles, on the other hand, are formulated through the use of a final
cause: the trajectory followed by the particle (or light ray) is that
which optimizes a certain characteristic, given its original and final
positions. 
 This is why these formulations are also called teleological,
 from the Greek $\tau \epsilon \lambda o \varsigma$, 
 aim, goal or purpose. 
  A general discussion of teleological principles in this context was
presented by Leibniz in his Specimen Dynamicum of 1695, a translation of
which appears in Loemker (1969, p.442).
 \begin{quotation} 
 {\it ``In fact, as I have shown by the remarkable example of the
 principles of optics,
 ....(that) final causes may be introduced with great fruitfulness even
into the special problems of physics, not merely to increase our
admiration for the most beautiful works of the supreme Author, but also
to help us make predictions by means of them which could not be as
apparent, except perhaps hypothetically, through the use of efficient
cause... It must be maintained in general that all existent facts can be
explained in two ways - through a kingdom of power or efficient causes
and through a kingdom of wisdom or final causes... Thus these two
kingdoms everywhere permeate each other, yet their laws are never
confused and never disturbed, so the maximum in the kingdom of power,
and the best in the kingdom of wisdom, take place together.''}
 \end{quotation}

 Euler and Maupertuis generalized the arguments of Fermat and Leibniz,
deriving Newtonian mechanics from the least action principle.The
Principle of Least Action, was stated in Maupertuis (1756, IV, p.36), as
his {\it Lois du Mouvement, Principe G\'{e}n\'{e}ral},

 \begin{quotation} 
 {\it ``Laws of Movement, General Principle:

  When a change occurs in Nature, the quantity of action necessary for
that change is as small as possible.

  The quantity of action is the product of the mass of the bodies times
their speed and the distance they travel. When a body is transported
from one place to another, the action is proportional to the mass of the
body, to its speed and to the distance over which it is transported.''}
 \end{quotation} 

 Maupertuis also used the same theological arguments of Leibniz
regarding the harmony between efficient and final causes. In Maupertuis
(1756, IV, p.20-23 of {\it Accord de Diff\'{e}rents Lois de la Nature,
qui avoient jusqu'ici paru incompatibles}), for example, we find:
  \begin{quotation} 
 {\it ``Accord Between Different Laws of Nature, that seemed
incompatible. ...

 I know the distaste that many mathematicians have for final causes
applied to physics, a distaste that I share up to some point. I admit,
it is risky to introduce such elements; their use is dangerous, as shown
by the errors made by Fermat (and Leibniz(?)) in following them.
Nevertheless, it is perhaps not the principle that is dangerous, but
rather the hastiness in taking as a basic principle that which is merely
a consequence of a basic principle.

  One cannot doubt that everything is governed by a supreme Being who
has imposed forces on material objects, forces that show his power, just
as he has fated those objects to execute actions that demonstrate his
wisdom. The harmony between these two attributes is so perfect, that
undoubtedly all the effects of Nature could be derived from each one
taken separately. A blind and deterministic mechanics follows the plans
of a perfectly clear and free Intellect. If our spirits were
sufficiently vast, we would also see the causes of all physical effects,
either by studying the properties of material bodies or by studying what
would most suitable for them to do.

  The first type of studies is more within our power, but does not take
us far. The second type may lead us stray, since we do not know enough
of the goals of Nature and we can be mistaken about the quantity that is
truly the expense of Nature in producing its effects.

  To unify the certainty of our research with its breadth, it is
necessary to use both types of study. Let us calculate the motion of
bodies, but also consult the plans of the Intelligence that makes them
move.

  It seems that the ancient philosophers made the first attempts at this
sort of science, in looking for metaphysical relationships between
numbers and material bodies. When they said that God occupies himself
with geometry, they surely meant that He unites in that science the
works of His power with the perspectives of His wisdom.''}
 \end{quotation} 

 Some of the metaphysical explanation given by Leibniz and Maupertuis
are based on theological arguments which can be regarded as late
inheritances of medieval philosophy. 
 This form of metaphysical argument, however, faded away from the
mainstream of science after the 18th century. Nevertheless, in the
following century, the (many variations of the) least action principle
disclosed more powerful formalisms and found several new applications in
physics. For details, see Goldstine (1980) and Wiegel (1986). As stated
in Yourgrau and Mandelstam (1979, ch.14 of
 {\it The Significance of Variational Principles in Natural Philosophy}),

 \begin{quotation} 
 {\it ``Towards the end of the (XIX) century, Helmholtz invoked, on
purely  scientific grounds, the principle of least action as a unifying
scientific  natural law, a `leit-motif' dominating the whole of physics,
Helmholtz (1887).

 `From these facts we may even now draw the conclusion that the domain 
 of validity of the principle of least action has reached far beyond the
boundaries of the mechanics of ponderable bodies. Maupertuis' high hopes
for the absolute general validity of his principle  appear to be
approaching their fulfillment, however slender the mechanical  proofs
and however contradictory the metaphysical speculation which the author
himself could at the time adduce in support of his new principle. Even
at this stage, it can be considered as highly probable that it is the
universal law pertaining to all processes in nature. ... In any case,
the general validity of the principle of least action seems to me
assured, since it may claim a higher place as a heuristic and guiding
principle in our endeavor to formulate the laws governing new classes of
phenomena. Helmholtz (1887).' \  ''}
  \end{quotation}

 \section{Modern Metaphysics}

 In this section we continue the investigation on the use and nature of
metaphysical principles in theoretical Physics. Like many others
adjectives, the word metaphysical has acquired both a positive
(meliorative, eulogistic, appreciative) and a negative (pejorative,
derogatory, unappreciative) connotation.

 Logical positivism or logical empiricism was a mainstream school in the
philosophy of science of the early 20th century. One of the objectives
of the positivist school was to build science from empirical
(observable) concepts only. 
 According to this point of view every metaphysical, that is,
non-empirical or non-directly observable, entity  is cognitively
meaningless and all teleological principles were perceived to fall in
this category.

 Teleological arguments were also perceived as problematic in Biology
and related fields due to the frequent abuse of phony teleological
arguments, usually in the form of crude fallacies or obvious
tautologies, given to provide support to whatever statement in need.
Maupertuis, the proponent of the first general least action principle,
himself, was aware of such problems, as clearly stated in the text of
his quoted in the previous section. Why then did important theoretical
physicists insist in keeping teleological arguments and other kinds of
principles perceived as metaphysical among the regular 
 tools of the trade?

 Yourgrau and Mandelstam (1979, p.10) emphasize the heuristic importance
of metaphysical principles in the early development of prominent
physical theories:

 \begin{quotation} 
 {\it ``In conformity with the scope of our subject, the speculative
facets of  the thinkers under review have been emphasized. Historically
by far more  consequential were the positive contributions to natural
science, contributions which transferred the emphasis from `a priori'
reasoning  to theories based upon observation and experiment. Hence,
while the future exponents of least principles may have been guided in
their metaphysical outlook (1) by the idealistic background we have
described, they had, nevertheless, to present their formulations in such
fashion that the data of experience would thus be explained. A
systematic scrutiny of the individual chronological stages in the
evolution of minimum principles can furnish us with profound insight
into continuous transformation of a metaphysical canon to an exact
natural law.

 (1) By `metaphysical outlook' we comprehend nothing but those general
assumptions which are accepted by the scientist.''}
 \end{quotation}

 The definition of Metaphysics used by Yougrau is perhaps a bit too
vague, or too humble. We believe that a deeper understanding of the role
played by metaphysics in modern theoretical physics can be found
(emphases are ours) in Einstein (1950):
  \begin{quotation} 
 {\it ``We have become acquainted with concepts and general relations
that enable us to comprehend an immense range of experiences and make
them {\bf accessible to mathematical treatment}. ...

 (but) Why do we devise theories at all? The answer to the latter
question is simply: Because we enjoy {\bf comprehending}, i.e., reducing
phenomena by the process of logic to something already known or
(apparently) evident. ...

  This is the striving toward {\bf unification and simplification 
of the premi-ses of the theory} as a whole (Mach's principle of economy,
interpreted as a logical principle). ...

  There exists a passion for comprehension, just as there exists a
passion for music. That passion is rather common in children, but gets
lost in most people later on. Without this passion, there would be
neither mathematics nor natural science. Time and again the passion for
understanding has led to the illusion that man is able to comprehend the
objective world rationally, by pure thought, without any empirical
foundations-in short, by metaphysics. I believe that every true theorist
is a kind of tamed metaphysicist, no matter how pure a `positivist' he
may fancy himself. The metaphysicist believes that the logically simple
is also the real. The tamed metaphysicist believes that not all that is
logically simple is embodied in experienced reality, but that the
totality of all sensory experience can be `comprehended' on the basis of
a conceptual system built on premises of great simplicity. The skeptic
will say that this is a `miracle creed.' Admittedly so, but it is a
miracle creed which has been borne out to an amazing extent by the
development of science.''}
 \end{quotation}

 Even more resolute statements are made by Max Planck (emphases are
ours) in the encyclopedia Die Kultur der Gegenwart (1915, p.68), and in
Planck (1915, p.71-72):
  \begin{quotation} 
 {\it ``As long as there exists physical science, its highest desirable
goal had been the solution of the problem to {\bf integrate all natural
phenomena} observed and still to be observed into a {\bf single simple 
principle} which {\bf permits to calculate} all past and, in particular, 
all future processes from the present ones. It is natural that this goal 
has not been reached to date, nor ever will it be reached entirely. 
 It is well possible, however, to approach it more and more, and the
history of theoretical physics demonstrates that on this way a rich
number of important successes could already be gained; which clearly
indicates that this ideal problem is not merely utopical, but eminently
fertile. Among the more or less general laws which manifest the
achievements of physical science in the course of the last centuries,
the Principle of Least Action is probably the one which, as regards form
and content, may claim to come nearest to that final ideal goal of
theoretical research.''}

 {\it ``Who instead seeks for higher connections within the system of
natural laws which are most easy to survey, in the interest of the
aspired harmony will, from the outset, also admit those means, such as
reference to the events at later instances of time, which are not
utterly necessary for the complete description of natural processes, but
which are {\bf easy to handle} and can be {\bf interpreted 
intuitively}.''}
 \end{quotation}

 From the last quoted statements of Einstein and Planck we can draw the
following four points list of motivations for the use of (or for
defining the characteristics of) good metaphysical principles: 
 
 1- Simplicity; \ \ 
 2- Generality; \ \ 
 3- Interpretability; and \ \ 
 
 4- Derivation of powerful and easy to handle (calculate, compute) 
    symbolic (mathematical) formalisms.

 The first three these points are very similar to the characteristics of
good metaphorical arguments, as analyzed in section 3. 
 In this particular context, generality means the ability of crossing
over different areas or transferring knowledge between multiple fields to
integrate the understanding of different natural phenomena.
 Since the least action principle clearly conforms with all four
criteria in the above list, it is easy to understand why it is so
endeared by physicists, despite the objections to its teleological
nature.

  Up to this point we have been arguing that the laws of mechanics
in integral form, stated in terms of the least action principle, and its
associated teleological metaphysical concepts, should be accepted along
side with the ``standard'' formulation of mechanics in differential
form, that is, the differential equations of Newtonian mechanics.
 However, Schlick (1979, V.1, p.297) proposes a complete inversion of the
empirical / metaphysical status of the two formulations, 
 see also Muntean (2006) and St\"{o}ltzner (2003). 
 According to Schlick's view, while the integral or macro-law
formulation has its grounds in observable quantities, the differential
or micro-law formulation is based on non-empirical concepts:
  \begin{quotation} 
 {\it ``That the event at a point depends only on those processes
occurring in its immediate temporal and special neighborhood, is
expressed in the fact that space and time appear in the formulae of
natural laws as infinitely small quantities; these formulae, that is,
are differential equations. We can also describe them in a readily
intelligible terminology as {\em micro-laws}. Through the mathematical
process of integration, there emerge from them the {\em macro-laws} (or
integral laws), which now state natural dependencies in their extension
over spatial and temporal distances. Only the latter fall within
experience, for the infinitely small is not observable. 
 The differential laws prevailing in nature can therefore be conjectured
and inferred only from the integral laws, and these inferences are
never, strictly speaking, univocal, since one can always account for the
observed macro-laws by various hypotheses about the underlying
micro-laws. Among the various possibilities we naturally choose that
marked by the greatest simplicity. It is the final aim of exact science
to reduce all events to the fewest and simplest possible differential
laws.''}
  \end{quotation} 

 From this and other examples presented in sections 6 to 9, we come to
the conclusion that metaphysical concepts are unavoidable, regardless of
the formulation in use. Positivists, on the other hand, envision the
exclusive use of metaphysical free scientific concepts, with grounds on
pure empirical experience. At the end, it seems that the later devote
themselves to the worthless pursuit of chasing chimeras. Moreover,
metaphysical arguments are essential to build our intuition. Without
intuition, physical reasoning would be  downgraded to merely cranking
the formalism, either by algebraic manipulation of the symbolic
machinery or by sheer number crunching. Planck (1950, p.171-172), states
that: 
 \begin{quotation} 
 {\it ``To be sure, it must be agreed that the positivistic outlook
possesses a distinctive value; for it is instrumental to a conceptual
clarification of the significance of physical laws, to a separation of
that which is empirically proven from that which is not, to an
elimination of emotional prejudices nurtured solely by customary views,
and it thus helps to clear the road for the onward drive of research.
But Positivism lacks the driving force for serving as a leader on this
road. True, it is able to eliminate obstacles, but it cannot turn them
into a productive factors. For its activity is essentially critical, its
glace is directed backward. But progress, advancement requires new
associations of ideas and new queries, not based on the results of
measurements alone, but going beyond them, and toward such things the
fundamental attitude of Positivism is one of aloofness.

 Therefore, up to quite recently, positivists of all hues have also put
up the strongest resistance to the introduction of atomic hypotheses
.... ''}
  \end{quotation}

 At this point it is opportune to remember Kant's 
 allegory of breathing,  
 that offers a couterpoint in contrast and complement to his 
 allegory of the dove 
 (Prolegomena to Any Future Metaphysics;  
 How Is Metaphysics Possible As a Science?):  

 \begin{quotation} 
 {\it ``That the human mind will ever give up metaphysical
 researches is as little to be expected as that we should prefer
 to give up breathing altogether, to avoid inhaling impure air.''}  
 \end{quotation}

 \section{Averaging over All Possible Worlds}

 The last example quoted by Planck provides yet another excellent
illustration to enlighten not only the issue currently under discussion,
but also other topics we want to address. In the next section we shortly
introduce one of the most important models related to the debate
concerning the atomic hypothesis, namely, Brownian motion.

 We are interested in the Dirichlet problem of describing the steady
state temperature at a two dimensional plate, given the temperature at
its border. The partial differential equation that the temperatures,
$u(x,y)$, must obey in the Dirichlet problem is known as the
2-dimensional Laplace equation,
 \[
    \div \ \grad \ u  =
    \frac{\del^2 u}{\del x^2} +\frac{\del^2 u}{\del y^2} = 0
     \ ,
 \]
 as in Butkov (1968, Ch.8).

 From elementary calculus, see Demidovich and Maron (1976), we have the
forward and backward finite difference approximations for a partial
derivative,
 \[
    \frac{\del u}{\del x} \approx
     \frac{u(x+h,y) -u(x,y)}{h} \approx
     \frac{u(x,y) -u(x-h,y)}{h}  \ .
 \]
 Using these approximations twice, we obtain the symmetric or central
finite difference approximation for the second derivatives,
 \[
    \frac{\del^2 u}{\del x^2} \approx
     \frac{u(x+h,y) -2u(x,y) +u(x-h,y)}{h^2}  \ ,
 \]
 \[
    \frac{\del^2 u}{\del y^2} \approx
     \frac{u(x,y+h) -2u(x,y) +u(x,y-h)}{h^2}  \ .
 \]
 Substitution in the Laplace equation gives the ``next neighbors' mean
value'' equation,
 \[
    u(x,y)= \frac{1}{4} \left(
     u(x+h,y) +u(x-h,y) +u(x,y+h) +u(x,y-h) \right) \ .
 \]

 From the last equation we can set a linear system for the temperatures
in a rectangular grid. The unknown variables, in the left hand side, are
the temperatures at the interior points of the grid, in the right hand
side we have the known temperatures at the boundary points.

 From the temperatures at the four neighboring points of a given grid
point, $[x,y]$, an estimate of the temperature, $u(x,y)$, at this point
is the expected value of the random variable $Z(x,y)$ whose value is
uniformly sampled from 
 \[
   \left\{ u(x+h,y), u(x-h,y), u(x,y+h), u(x,y-h) \right\} \ ,
 \]
 the north, south, east and west neighbors. Also, if we did not know the
temperature at the neighboring point sampled, we could estimate the
neighbor's temperature by sampling one of the neighbor's neighbors.
Using this argument recursively, we could estimate the temperature
$u(x,y)$ through the following Monte Carlo algorithm:

 Consider a ``particle'' undergoing a symmetric random 
 (or drunken sailor) walk, that is, a stochastic trajectory, 
 $T=[T(1),\ldots T(m)]$, such that starting at position 
 $T(1)=[x(1),y(1)]$, it jumps to positions 
 $T(1),T(2),\ldots T(m)$ by uniformly sampling among the neighboring
points of its current position, until it eventually hits the boundary.
 More precisely, from a given position, $T(k)=[x(k),y(k)]$, at step $k$,
the particle will equally likely jump to one of its neighboring
positions 
 at step $k+1$, that is, 
 \[ 
    T(k+1)= \left[ x(k+1),y(k+1) \right] 
 \]  
 is randomly selected from the set 
 \[
 \left\{ \ \left[ x(k)+h,y(k) \right], \left[ x(k)-h,y(k) \right], 
 \left[ x(k),y(k)+h \right], \left[ x(k),y(k)-h \right] \ \right\} \ .
 \] 
 The journey ends when a boundary point, $T(m)=[x(m),y(m)]$, is hit by 
 ``particle'' at (random) step $m$. Defining the random variable
 $Z(T)=u(x(m),y(m))$, it can be shown that the expected value of $Z(T)$,
for $T$ starting at $T(1)=[x(1),y(1)]$, equals $u(x(1),y(1))$, the
solution to the Dirichlet problem at $[x(1),y(1)]$. 

 The above algorithm is only a particular case of more general Monte
Carlo algorithms for solving linear systems. 
 For details see 
 Demidovich and Maron (1976), Hammersley and Handscomb (1964), 
 Halton (1970) and Ripley (1987).
 Hence, these Monte Carlo algorithms allow us to obtain the
solution of many continuous problems in terms of an expected (average)
value of a discrete stochastic flow of particles. 
 More precisely, efficient Monte Carlo algorithms are available for
solving linear systems, and many of the mathematical models in Physics,
or science in general, are (or can be approximated by) linear equations.
Consequently, one should not be surprised to find physical models
interpretations in terms of particle flows.

 In 1827, Robert Brown observed the movement of plant spores (pollen)
immersed in water. He noted that the spores were in perpetual movement,
following an erratic or chaotic path. Since the motion persisted over
long periods of time on different liquid media and powder particles of
inorganic minerals also exhibited the same motion pattern, he
 discarded the hypothesis of live or self propelled motion.
This ``Brownian motion'' was the object of several subsequent studies,
linking the intensity of the motion to the temperature of the liquid
medium. For further readings, see Brush (1968) and Haw (2002).

 In 1905 Einstein published a paper in which he explains Brownian motion
as a fluctuation phenomenon caused by the collision of individual water
molecules with the particle in suspension. Using a simplified argument,
we can model the particle's motion by a random path in a rectangular
grid, like the one used to solve the Dirichlet problem. In this model,
each step is interpreted as a molecule collision with the particle,
causing it to move, equally likely, to the north, south, east or west.
 The stating the formal mathematical properties of this stochastic
process, known as a random walk, was one of the many scientific 
contribution of Norbert Wiener, one of the forefathers of Cybernetics, 
see Wiener (1989). 
 For good reviews, see Beran (1994) and Embrechts (2002). 
 For an elementary introduction, see Berg (1993), Lemons (2002)
MacDonald (1962) and Mikosch (1998).

 A basic assumption of the random walk model is that distinct collisions
or moves made by the particle are uncorrelated. Let us consider the one
dimensional random walk process, where a particle, initially positioned
at the origin, $y_0=0$, undergoes incremental unitary steps, that is,
$y_{t+1} = y_{t}+x_{t}$, and $x_t=\pm 1$. The steps are assumed
unbiased, and uncorrelated, that is, $E(x_t)=0$ and $\Cov(x_s,x_t)=0$.
Also, $\Var(x_t)=1$. From the linearity of the expectation operator, we
conclude that $E(y_t)=0$. 
 Also
 \[
   E(y_t^2) = E\left( \ssum_{j=1}^{t} x_j \right)^2
   = E\ssum_{j=1}^t x_j^2 \; +E\ssum_{j \neq k} x_j x_k
   = t +0 = t \ ,
 \]
 so that at time $t$, the standard deviation of the particle's position is
 \[
    \sqrt{E(y_t^2)} = t^H, \ \mbox{for} \ H=\frac{1}{2} \ .
 \]

 From this simple model an important characteristic, expressed as a
sharp statistical hypothesis to be experimentally verified, can be
derived: Brownian motion is a self-similar process, with scaling factor,
or Hurst exponent, $H=1/2$. One possible interpretation of the last
statement is that, in other to make coherent observations of a Brownian
motion, if time is rescaled by a  factor $\phi$, then space should also
be rescaled by a factor $\phi^H$. The generalization of this stochastic
process for $0<H<1$, is known as fractional Brownian motion. 

 The sharp hypothesis $H=1/2$ takes us back to the eternal underlying
theme of system coupling / decoupling. While regular Brownian motion was
built under the essential axiom of decoupled (uncorrelated) increments
over non-overlapping time intervals, the relaxation of this condition,
without sacrificing self-similarity, leads to long range correlations. 
 For fresh insight, see the original work of 
 Paul Levy (1925, 1948, 1954, 1970) and Beno\^{\i}t Mandelbrot (1983);   
 for a textbook, see Beran (1994) and Embrechts (2002).

 As we have seen in this section, regular Brownian motion can be very
useful in modeling the low level processes often found in disorganized
physical systems.  However, in several phenomena related to living
organisms or systems, long range correlations are exhibited. This is the
case, for example, in the study of many complex or (self) organized
systems, such as colloids or liquid crystals, found in soft matter
science, in the development of embryos or social and urban systems, in
electrocardiography, electroencephalography or other monitoring of
biological signals procedures. Modeling in many of these areas can,
nevertheless, benefit from the techniques of fractional Brownian motion,
as seen in Addison (1997), Beran (1994), Bunde and Havlin (1994),
Embrechts (2002) and Feder (1988). Some of the epistemological
consequences of the mathematical and computational models introduced in
this section are commented in the following section.

 \section{Hypothetical versus Factual Models}

 The Monte Carlo algorithms introduced in the last section are based on
the stochastic flow of particles. Yet, these particles can be regarded
as mere imaginary entities in a computational procedure. On the other
hand, some models based on similar ideas, such as the kinetic theories
of gases, or the random walk model for the Brownian motion, seem to give
these particles a higher ontological status. It is thus worthwhile to
discuss the epistemological or ontological status of an entity in a
computational procedure, like the particles in the above example.

 This discussion is not as trivial, innocent and harmless, at it may
seem at first sight. In 1632 Galileo Galilei published in Florence his
Dialogue Concerning the Two Main World Systems. At that time it was
necessary to have a license to publish a book, the {\it imprimatur}.
Galileo had obtained the imprimatur from the ecclesiastical authorities
two years earlier, under the explicit condition that some of the theses
presented in the book, dangerously close to the heliocentric heretical
ideas of Nicolas Copernicus, should be presented as a ``hypothetical
model'' or as a ``calculation expedient'' as opposed to  the
``truthful'' or ``factual'' description of ``reality''.

 Galileo not only failed to fulfill the imposed condition, but also
ridiculed the official doctrine. He presented his theories in a dialogue
form. In these dialogues, Simplicio, the character defending the
orthodox geocentric ideas of Aristotle and Ptolemy, was constantly
mocked by his opponent, Salviati, a zealot of the views of Galileo. In
1633 Galileo was prosecuted by the Roman Inquisition, under the
accusation of making heretical statements, as quoted from Santillana
(1955, p.306-310):
  \begin{quotation} 
 {\it ``The proposition that the Sun is the center of the world and does
not move from its place is absurd and false philosophically and formally
heretical, because it is expressly contrary to Holy Scripture. The
proposition that the Earth is not the center of the world and immovable
but that it moves, and also with a diurnal motion, is equally absurd and
false philosophically and theologically considered at least erroneous in
faith.''}
  \end{quotation} 

 In the Italian renaissance, one of the most open and enlighten
societies of its time, but still within a pre-modern era, where
subsystems were only incipient and not clearly differentiated, the
consequences of mixing scientific and religious arguments could be
daring. 
 Galileo even uses some arguments that resemble the concept of systemic 
differentiation, for example: 
  \begin{quotation} 
 {\it ``Therefore, it would perhaps be wise and useful advice not to 
add without necessity to the articles pertaining to salvation and to 
the definition of faith, against the firmness of which there is no danger 
that any valid and effective doctrine could ever emerge. 
 If this is so, it would really cause confusion to add them upon request 
from persons about whom not only do we not know whether they speak with 
heavenly inspiration, but we clearly see they are deficient in the 
intelligence necessary first to understand and then to criticize the 
demonstrations by which the most acute sciences proceed in confirming 
similar conclusions.''} Finocchiaro (1991, p.97).
  \end{quotation} 

 The paragraph above is from a letter of 1615 from Galileo to 
 Her Serene Highness Grand Duchess Cristina but, as usual, 
 Galileo's rhetoric is anything but serene. 
 In 1633 Galileo is sentenced to prison for an indefinite term. 
 After he abjures his allegedly heretical statements, the sentence is
commuted to house-arrest at his villa. Legend has it that, after his
formal abjuration, Galileo muttered the now celebrated phrase, 
  \begin{quotation} 
 {\it Eppur si mouve}, 
 ``But indeed it (the earth) moves (around the sun)''.
  \end{quotation}

 Around 1610 Galileo built a telescope (an invention coming from 
Netherland) that he used for astronomical observations. Among his
findings were four satellites to planet Jupiter, namely, Io, Europa,
Ganymedes and Callisto. He also observed phases (such as the lunar
phases) exhibited by planet Venus. Both facts are either compatible or
explained by the Copernican heliocentric  theory, but problematic or
incompatible with the orthodox Ptolemaic geocentric theory. During his
trial, Galileo tried to use these observations to corroborate his
theories, but the judges would not, literally, even `look' at them. The
church's chief astronomer, Christopher Clavius, refused to look through
Galileo's  telescope, stating that there was no point in `seeing' some
objects through an instrument that had been made just in order to
`create' them. 
 Nevertheless, only a few years after the trial, the same Clavius was
building fine telescopes,   
 used to make 
 new astronomical observations. 
 He took care, of course, not to upset his boss with ``theologically
incorrect'' explanations for  what he was observing.

 From the late 19th century to 1905 the world witnessed yet another
trial, perhaps not so famous, but even more dramatic. Namely, that of
the atomistic ideas of Ludwig Boltzmann. For a excellent biography of
Boltzmann, intertwined (as it ought to be) with the history of his
scientific ideas, see Cercignani (1998). The final verdict on this
controversy was given by Albert Einstein  in his {\it annus mirabilis}
paper about Brownian Motion, together with the subsequent experimental
work of Jean Perrin. For details see Einstein (1956) and Perrin (1950).
A simplified version of these models was presented in the previous
section, including a ``testable'' sharp statistical hypothesis, $H=1/2$,
to empirically check the theory. As quoted in Brush (1968), in his
Autobiographical Notes, Einstein states that:
 \begin{quotation} 
 {\it ``The agreement of these considerations with experience together
with Planck's determination of the true molecular size from the law of
radiation (for high temperatures) convinced the skeptics, who were quite
numerous at that time (Ostwald, Mach) of the reality of atoms. The
antipathy of these scholars towards atomic theory can indubitably be
traced back to their positivistic philosophical attitude. This is an
interesting example of the fact that even scholars of audacious spirit
and fine instinct  can be obscured in the interpretation of facts by
philosophical prejudices. The prejudice  - which has by no means died
out in the meantime - consists in the faith that facts themselves can
and should yield scientific knowledge without free conceptual
construction.

 Such misconception is possible only because one does not easily become
aware of the free choice of such concepts, which through verification
and long usage, appear to be immediately connected with the empirical
material''}
 \end{quotation}

 Let us follow Perrin's perception of the ``empirical connection''
between the concepts used in the molecular theory, which contrasted to
that of the rival energetic theory, during the first decade of the 20th
century.
 In 1903 Perrin was already an advocate of the molecular hypothesis, as
can be seen in Perrin (1903). According to Brush (1968, p.30-31), Perrin
refused the positivist demand for using only directly observable
entities. Perrin referred to an analogous situation in biology where,
 \begin{quotation} 
 {\it ``the germ theory of disease might have been developed and
successfully tested before the invention of the microscope; the microbes
would have been hypothetical entities, yet, as we know now, they could
eventually be observed.''}
  \end{quotation}

 But only three years latter, was Perrin (1906) confident enough to
reverse the attack, accusing the energetic view rivaling the atomic
theory, of having ``degenerated into a pseudo-religious cult''. It was
the energetic theory, claimed Perrin, that was making use of
non-observable entities! To begin with, Classical thermodynamics had a
differential formulation, with the functions describing the evolution of
a system assumed to be continuous and differentiable (notice the
similarity between the argument of Perrin and that of Schlick, presented
in section 8). Perrin based his argument of the contemporary evolution
of mathematical analysis when, until late in the 20th century,
continuous functions were naturally assumed to be differentiable.
Nevertheless, the development of mathematical analysis, on the turn to
the 20th century, proved this to be a rather naive assumption. Referring
to this background material, Perrin argues:
  \begin{quotation} 
 {\it ``But they still thought the only interesting functions were the
ones that can be differentiated. Now, however, an important school,
developing with rigor the notion of continuity, has created a new
mathematics, within which the old theory of functions is only the study
(profound, to be sure) of a group of singular cases. It is curves with
derivatives that are now the exception; or, if one prefers the
geometrical language, curves with no tangents at any point become the
rule, while familiar regular curves become some kind of curiosities, 
doubtless interesting, but still very special.''}
 \end{quotation} 

 In three more years, even former opponents were joining the ranks of the
atomic theory. As W.Nernst (1909, 6th.ed., p.212) puts it:
 \begin{quotation} 
 {\it ``In view of the {\em ocular} confirmation of the picture which
the kinetic theory provides us of the world of molecules, one must admit
that this theory begins to lose its hypothetical character.''}
 \end{quotation} 

 \section{Magic, Miracles and Final Remarks}

 In several incidents analyzed in the last sections, one can repeatedly
find the occurrence of  theoretical ``phase transitions'' in the history
of science. In these transitions, we observe a dominant and strongly
supported theory being challenged by an alternative point of view. 
 In a first moment, the cheerleaders of the dominant group come up with
a variety of ``disqualifying arguments'', to show why the underdog
theory, plagued by phony concepts and faulty constructions, should not
even be considered as a serious contestant. 
 In an second moment, the alternative theory is kept alive by a small
minority, that is able to foster its progress. 
 In a third and final moment, the alternative theory becomes, quite
abruptly, the dominant view, and many wonder how is it that the old, now
abandoned theory, could ever had so much support. This process is
captured in the following quotation, from the preface to the first
edition of Schopenhauer (1818):
 %
 %
 %
  \begin{quotation} 
 {\it ``To truth only a brief celebration of victory is allowed between
the two long periods during which it is condemned as paradoxical, or
disparaged as trivial.''}
  \end{quotation}  

 Perhaps this is the basis for the gloomier statement found in Planck
(1950, p.33-34):
 \begin{quotation} 
 {\it ``A new scientific truth does not triumph by convincing its
opponents and by making them see the light, but rather because its
opponents eventually die, and a new generation grows up that is familiar
with it.''}
 \end{quotation} 

As for the abruptness of the transition between the two phases,
representing the two theoretical paradigms, this is a phenomenon that
has been extensively studied, from sociological, systemic and historical
perspectives, by Thomas Kuhn (1996, 1977). See also Hoyningen-Huene
(1993) and Lakatos (1978a,b). For similar ideas presented within an
approach closer to the orthodox Bayesian theory, see Zupan (1991).

 We finish this section with a quick and simple alternative explanation,
possibly just as a hint, that I believe can shed some light on the
nature of this phenomenon. Elucidations of this kind were used many
times by von Foerster (2003,b,e) who was, among many other things, a
skilful magician and illusionist.

 An Ambigram, or ambiguous picture, is a picture that can be looked at
in two (or more) different ways. Looking at an ambigram, the observer's
interpretation or re-solution of the image can be attracted to one of
two or more distinct eigen-solutions. A memorable instance of an
ambigram is the Duck-Rabbit, born in 1892, in the humble pages of the
German tabloid Fliegende Bl\"{a}tter. It was studied in 1899 by the
psychologist Joseph Jastrow in an article antecipating several aspects
of cognitive constructivism, and finally made famous by the philosopher
Ludwig Wittgenstein in 1953. For a historical account of this ambigram,
see Kihlstrom (2006), as  well as several nice figures. In case anyone
wonders, Jastrow was Peirce's Ph.D. student and coauthor of the 1885
paper introducing randomization, and Wittgenstein is no other than von
Foster's uncle Ludwig.

 According to Jastrow (1899), an ambigram demonstrates how
  \begin{quotation} 
 {\it ``True seeing, observing, is a double process, partly objective or
 outward - the thing seen and the retina  - and partly subjective or
inward - the picture mysteriously transferred to the mind's
representative, the brain, and there received and affiliated with other
images.''}
  \end{quotation} 
 Still according to Jastrow, in an ambigram,
  \begin{quotation} 
 {\it ``...a single outward  impression changes its character according
as it is viewed as representing  one thing or another. In general we see
the same thing all the time,  and the image on the retina does not
change. But as we shift the attention  from one portion of the view to
another, or as we view it with a different  mental conception of what
the figure represents, it assumes a different aspect,  and to our mental
eye becomes becomes quite a different thing.''}
 \end{quotation} 

 Jastrow also describes some characteristics of the mental process of
shifting between the eigen-solutions of an ambigram, that is, how in
{\it ``The Mind's Eye''} one changes from one interpretation to the
other. Two of these characteristics are specially interesting in our
context:

 First, in the beginning, \ 
 {\it ``It may require a little effort to bring about this change, but
it is very marked when once realized.''}

 Second, after both interpretations are known, \  
 {\it ``Most observers find it difficult to hold either interpretation
steadily, the fluctuation being frequent, and coming as a surprise.''}

 The first characteristic can help us understand either Nernst's
``ocular readiness''  or, in contrast,  Clavius' ``ocular blindness''.
After all, the satellites of Jupiter were quite tangible objects, ready
to be watched through Galileo's telescope, whereas the grains of
colloidal suspension that could be observed with the lunette of Perrin's
apparatus provided a much more indirect evidence for the existence of
molecules. Or maybe not, after all, it all depends on what one is
capable, ready, or willing to see...

 The second characteristic can help us understand  Leibniz' and
Maupertuis' willingness to accommodate and harmonize two alternative
explanations for a single phenomenon, that is, to have effective and
final causes, or micro and macro versions of physical laws.

 Yet, the existence of sharp, stable, separable and composable
eigen-solutions for the scientific system in its interaction with its
environment, goes far beyond our individual or collective desire to have
them there.

 These eigen solutions are the basis upon which technology builds much
of the world we live in. How well do the eigen-solutions used in these
technological gadgets conform with von Foerster criteria? Well, the
machine I am using to write this chapter has a 2003 Intel Pentium CPU
carved on a silicon waffle with a ``precision'' of 0.000,000,1m,
and is ``composed'' by about 50 million transistors. This CPU has a
clock of 1GHz, so that each and every one of the transistors in this
composition must operate synchronously to a fraction of a thousandth of
a thousandth of a thousandth of a second!

 And how well do the eigen-solutions expressed as fundamental physical
constants, upon which technological projects rely, conform with von
Foerster criteria? Again, some of these constants are known up to a
precision (relative standard uncertainty) of 0.000,000,001, that
is, a thousandth of a thousandth of a thousandth! The world wide web
site of the United States' National Institute of Standards and
Technology, at 
 {\tt www.physics.nist.gov,} gives an encyclopaedic view of  these
constants and their inter-relations. Planck (1950, Ch.6) comments on
their epistemological significance.

 But far beyond their practical utility or even their scientific
interest, the existence of these eigen-solutions are not magical
illusions, but true miracles. Why ``true'' miracles? Because the more
they are explained and the better they are understood, the more
wonderful they become!

%% file: CAPE5.TEX
 \chapter{Complex Structures, Modularity, 
          and Stochastic Evolution}    


 \mbox{} 

 {\flushright
 
  {\it 
    ``Hierarchy, I shall argue, is one of the central struc- \\ 
   tural schemes that the architect of complexity uses.'' } 
  
 {\it ``The time required for the evolution of a complex form \\
 from simple elements depends critically on the number and \\ 
 distribution of potential intermediate stable subassemblies.'' 
  }
 
  Herbert Simon (1916-2001), \\   
  The Sciences of the Artificial. 

 \mbox{}

 {\it  
 ``In order to make some sense here, we must keep an \\ 
 open mind about the possibility that for sufficiently \mbox{} \\  
 complex systems, amplitudes become probabilities.
 } 


 Richard Feynman (1918-1988), \\ 
 Lecture notes on Gravitation.

 }

 \section{Introduction}
 \markboth{CHAPTER 5: MODULARITY AND STOCHASTIC EVOLUTION}
  {5.1 \ INTRODUCTION}

  The expression {\it stochastic evolution} may seem an oxymoron. After
all, evolution indicates progress towards complexity and order, while a
stochastic (probabilistic, random) process seems to be only capable of
generating confusion or disorder. The etymology of the word stochastic,
from $\sigma \tau o \chi o \varsigma$, meaning {\it aim}, {\it goal} or
{\it target}, and its current use, meaning {\it chancy} or {\it noisy},
seems to incorporate this apparent contradiction. An alternative use of
the same root, $\sigma \tau o \chi \alpha \sigma \tau \iota \kappa o
\varsigma$ meaning {\it skillful at guessing, conjecturing, or divining
the truth}, may offer a bridge between the two meanings. 

  The main goal of this chapter is to study how the concepts of
stochastic process and evolution of complex systems can be reconciled.
Sections 2 and 3 examine two prototypical algorithms: Simulated
Annealing and Genetic Programming. The ideas behind these two algorithms
will be used as a basis for most of the arguments used in this chapter.
The mathematical details of some of these algorithms are presented in
appendix H. Section 4 presents the concept of modularity, and explains
its importance in the evolution of complex systems. 

 While sections 2, 3 and 4 are devoted to the study of general systems,
including applications to biological organisms and technological
devices, section 5 pays closer attention to the evolution of complex
hypotheses and scientific theories. Section 5 also examines the idea of
complementarity, developed by the physicist and philosopher Niels Bohr
as a general framework for the reconciliation of two concepts that
appear to be incompatible but are, at the same time, indispensable to
the understanding of a given system. Section 6 explores the connection
between complementarity and probability, presenting Heisenberg's
uncertainty principle. Section 7 extends the discussion to general
theories of evolution and returns to the pervasive theme of
probabilistic causation. Section 8 presents our final remarks.

 \section{The Ergodic Path: One for All}
 \markboth{CHAPTER 5: MODULARITY AND STOCHASTIC EVOLUTION}
  {5.2 \ THE ERGODIC PATH}

 Most human societies are organized as hierarchical structures.
Universities are organized in research groups, departments, institutes
and schools; Armies in platoons, battalions, regiments and brigades; and
so on. This has been the way of doing business as described in the
earliest historical records. Deuteronomy (1:15) describes the ancient
hierarchical structure of Israel:   
 \begin{quotation} 
 ``So I took the heads (ROSh) of your tribes, men wise and known, and
made them heads over you, leaders (ShR) of thousands , hundreds, fifties
and tens, and officers (ShTR) for your tribes.'' 
 \end{quotation}   

 This verse gives us some idea of the criteria used to appoint leaders
(knowledge and wisdom), but give us no hint on the criteria and methods
used to form the groups (of 10, 50, 100 and 1000). Perhaps that was
obvious from the family and tribal structure already in place. There are
many situations, however, where organizing groups to obtain an optimal
structure is far from trivial. In this section we study such a case: the
block partition problem.

 \subsection{Block Partitions} 

 The matrix block partition problem arises in many practical situations
in engineering design, operations research and management science. In
some applications, the elements of a rectangular matrix, $A$, may 
represent the interaction between people, corresponding to columns, and
activities, corresponding to rows, that is, $A_i^j$, the element in row
$i$ and column $j$, represents the intensity of the interaction between
person $j$ and activity $i$. The block partition problem asks for an
optimal ordering or permutation of rows and columns taking the permuted
matrix to Block Angular Form (BAF), so that each one of $b$ diagonal
blocks  bundles a group of strongly coupled people and activities. 
 Only a small number of activities are leaft outside the diagonal
blocks, in a special $(b+1)$-th block of residual rows. 
 Also, only a small number of people interact with more than one of the
$b$ diagonal activities, these corespond to residual columns, 
 see Figure 1.  
 %
 {
 \[ 
 \left[ \begin{array}{cccccccccccccc} 
 1 &   &   &   &   & 1 & 1 &   &   &   &   & 1 &   & 1 \\  
 1 & 1 &   &   &   &   & 1 &   &   &   &   & 1 & 1 &   \\ 
   &   & 1 &   & 1 &   & 1 &   &   &   &   & 1 & 1 & 1 \\ 
   &   &   & 1 &   & 1 & 1 &   &   &   &   &   &   & 1 \\   
 1 & 1 & 1 & 1 & 1 & 1 & 1 &   &   &   &   & 1 & 1 & 1 \\   
   &   &   &   &   &   &   & 2 &   & 2 & 2 &   & 2 & 2 \\ 
   &   &   &   &   &   &   & 2 & 2 &   & 2 & 2 &   & 2 \\  
   &   &   &   &   &   &   & 2 & 2 & 2 &   & 2 & 2 &   \\  
 3 &   & 3 &   & 3 &   & 3 &   & 3 &   & 3 & 3 & 3 & 3 \\ 
   & 3 &   & 3 &   & 3 & 3 & 3 &   & 3 & 3 & 3 & 3 & 3 \\ 
 \end{array} \right]  
 \ \ \ \ , \ \ \ \ \ \ 
 \left[ \begin{array}{ccccccc} 
 1 &   &   &   &   & 1 & 1  \\  
 1 & 1 &   &   &   &   & 1  \\ 
   &   & 2 &   & 2 &   & 2  \\ 
   &   &   & 2 &   & 2 & 2  \\   
 3 & 3 & 3 & 3 & 3 & 3 & 3  \\ 
 \end{array} \right]  
 \] 
 }
 \centerline{Figure 1a,b: Two Matrices in Block Angular Form.}

 A matrix in BAF is in Row Block Angular Form (RBAF) if it has only
residual rows, and is in Column Block Angular Form (CBAF) if it has only
residual columns. Each angular block can, in turn, exhibit again a BAF,
thus creating a recursive or Nested Block Angular Form (NBAF). Figure 1a
exhibits a matrix in NBAF. In this figure, zero elements of the matrix
are represented by blanck spaces. The number at the position of a
non-zero element (NZE) is not the corresponding matrix element's value,
but rather a class tag or ``color'' indicating the block to which the
row belongs. Residual rows receive the special color $b+1$. The
first block has a nested CBAF structure, shown in Figure 1b. For the
sake of simplicity, this chapter will focus on the BAF partition
problem, although all our conclusions can be generalized to the NBAF
case.

 We motivate the block partition problem further with an application
related to numerical linear algebra. Gaussian elimination is the name of
a simple method for solving linear systems of order $n$, by reducing the
matrix of the original system to (upper) triangular form. This is
accomplished by successively subtracting multiples of the row $1$
through $n$ from the rows bellow them, so as to eliminate (zero) the 
elements below each diagonal element (or pivot element). The example in
Figure 2 illustrates the Gaussian elimination algorithm, where the
original system, $Ax=b$, is transformed into an upper triangular system,
$Ux=c$. The matrix $L$ stores the multipliers used in the process. Each
multiplier is stored at the position of the element it was used to
eliminate, that is, at the position of the zero it was used to  create.
It is easy to check that $A=LU$, hence the alternative name of the
algorithm: LU Factorization. 

 The example in Figure 2 also displays some structural peculiarities.
Matrix $A$ is in BAF, with two diagonal blocks,  one residual row (at
the bottom or south side of the matrix) and one residual column (at the
right or east side of the matrix). This structure is  preserved in the
$L$ and $U$ factors. This structure and its preservation is of paramount
importance in the design of efficient factorization algorithms. Notice
that the elimination process in Figure 2 can be done in parallel. That
is, the factorization of each diagonal block can be done independently
of and simultaneously with the factorization of the other blocks, for
more details see Stern and Vavasis (1994). 
 %
 { 
 \[  
     \left[ \begin{array}{ccc|ccc|c}   
     1  &  2  &  3  &  0  &  0  &  0  &  1  \\  
     1  &  6  &  8  &  0  &  0  &  0  &  3  \\  
     2  &  8  & 17  &  0  &  0  &  0  &  7  \\  
     \hline  
     0  &  0  &  0  &  2  &  3  &  4  &  4  \\  
     0  &  0  &  0  &  4  & 11  & 14  & 13  \\  
     0  &  0  &  0  &  4  & 16  & 27  & 24  \\  
     \hline 
     1  & 10  & 31  &  8  & 37  & 88  & 98   \\  
  \end{array}   \right]  
   \ = \   
     \left[ \begin{array}{ccc|ccc|c}   
     1  &  0  &  0  &  0  &  0  &  0  &  0  \\  
     1  &  1  &  0  &  0  &  0  &  0  &  0  \\  
     2  &  2  &  1  &  0  &  0  &  0  &  0  \\  
     \hline  
     0  &  0  &  0  &  1  &  0  &  0  &  0  \\  
     0  &  0  &  0  &  2  &  1  &  0  &  0  \\  
     0  &  0  &  0  &  2  &  2  &  1  &  0  \\  
     \hline 
     1  &  2  &  3  &  4  &  5  &  6  &  1   \\  
  \end{array}  \right] 
     \    
     \left[ \begin{array}{ccc|ccc|c}   
     1  &  2  &  3  &  0  &  0  &  0  &  1  \\  
     0  &  4  &  5  &  0  &  0  &  0  &  2  \\  
     0  &  0  &  6  &  0  &  0  &  0  &  3  \\  
     \hline  
     0  &  0  &  0  &  2  &  3  &  4  &  4  \\  
     0  &  0  &  0  &  0  &  5  &  6  &  5  \\  
     0  &  0  &  0  &  0  &  0  &  7  &  6  \\  
     \hline 
     0  &  0  &  0  &  0  &  0  &  0  &  7  \\  
  \end{array} \right]  
 \] 
 }
 \centerline{Figure 2: A=LU Factorization of CBAF Matrix} \\ 

 A classic combinatorial formulation for the CBAF partition problem, for
a rectangular matrix $A$, $m$ by $n$, is the  Hypergraph Partition
Problem (HPP). In the HPP formulation, we paint all nonzero elements
(NZE's) in a vertex $i \in \{1, \ldots ,m\}$, (corresponding to row
$A_i$) with a color $x_i \in \{1, \ldots ,b\}$. The color $q^j(x)$ of an
edge  $j \in \{1, \ldots ,n\}$, (corresponding to column $A^j$) is then
the set of all its NZE's colors. Multicolored edges of the  hypergraph
(corresponding to columns of the matrix containing NZE's of several
colors) are the residual columns in the CBAF. The formulation for the
general BAF problem also allows some residual rows to 
receive the special color $b+1$. 

 The BAF applications typically require: \\ 
 1. Roughly the same number of rows in each block. \\ 
 2.  Only a few residual rows or columns.  \\ 
 From 1 and 2 it is natural to consider the minimization of
the objective or cost function
 \[ 
 f(x) = \alpha \ssum_{k=1}^{b}{h_k(x)}^2 + \beta c(x) + \gamma r(x) 
  \ , \ \  h_k(x) =  s_k(x) - m/b \ , 
 \] 
 \[ 
 q^j(x) = \left\{ k \in \left\{1, \ldots ,b \right\} \; : \;   
     \exists i ,\; A_i^j\neq 0 \wedge x_i=k \right\} \ , \ \ 
 s_k(x) = \left| \left\{ i \in \left\{ 1, \ldots ,m \right\} \; : \    
               x_i=k \right\} \right| \ , 
 \] 
 \[  
 c(x) = \left| \left\{ j \in \left\{ 1, \ldots ,n \right\} \; : \   
               |q^j(x)| \geq 2 \right\} \right| \ , \ \ 
 r(x) = \left| \left\{ i \in \left\{ 1, \ldots ,m \right\} \; : \  
               x_i=b+1 \right\} \right| \ .  
 \] 

 \noindent The term $c(x)$ is the number of residual columns, and the
term $r(x)$ is the number of residual rows. The constraint functions
$h_k(x)$ measure the deviation of each block from the ideal size $m/b$.
Since we want to enforce these constraints only approximately, we use
quadratic penalty functions, $h_k(x)^2$, that (only) penalize large
deviations. If we wanted to enforce the constraints more strictly, we
could use exact penalty functions, like $|h_k(x)|$, that penalize even
small  deviations, see Bertzekas and Tsitsiklis (1989) and Luenberger
(1984).

 \subsection{Simulated Annealing}

 The HPP stated in the last section is very difficult to solve exactly.
Technically it is an NP-hard problem, see Cook (1997).  Consequently, we
try to develop heuristic procedures to find approximate or almost
optimal solutions. Simulated Annealing (SA) is a powerful
meta-heuristic, well suited to solve many combinatorial problems. The
theory behind SA also has profound epistemological implications, that we
explore latter on in this chapter.     

 The first step to define an SA procedure is to define a neighborhood
structure in the problem's state or configuration space.  The
neighborhood, $N(x)$, of a given initial state, $x$, is the set of
states, $y$, that can be reached from $x$, by a single move. In the HPP,
a single move is defined as changing the color of a single row, $x_i
\longmapsto y_i$.

 In this problem, the neighborhood size is therefore the same, for any
state $x$, namely, the product of the number of rows and colors, that
is, $|N(x)|=mb$ for CBAF, and $|N(x)|=m(b+1)$ for BAF. This neighborhood
structure provides good mobility in the state space, in the sense that
it is easy to find a path (made by a succession of single moves) from
any chosen initial state, $x$, to any other final state, $y$. This
property is called irreducibility or strong connectivity. There is also
a second technical requirements for good mobility, namely, this set of
paths should be aperiodic. If the length (the number of single moves) of
any path from $x$ to $y$ is a multiple of an integer $k>1$, $k$ is
called the period of this set. Further details are given in appendix
H.1. 

 In an SA, it is convenient to have an easy way to update the cost
function, computed at a given state, $x$, to the cost of a  neighboring
state, $y$. The column color weight matrix, $W$, is defined so that the
element $W_k^j$ counts the number of NZE's in column $j$ (in rows) of
color $k$, that is, 
 \[ 
    W_k^j \equiv \left| \left\{ A_i^j \g A_i^j \neq 0 \wedge 
    x_i=k  \right\} \right| \ .  
 \]
 The weight matrix can be easily updated at any single move and, from $W$, 
it is easy to compute the cost function or a cost differential,    
 \[ 
    \delta \equiv  f(y) - f(x) \ .  
 \]
 
 The internal loop of the SA is a Metropolis sampler, where single moves
are chosen at random (uniformly among any possible move) and then
accepted with the Metropolis probability,     
 \[ 
    M(\delta,\theta ) \equiv \left\{ \begin{array}{ll}
    1 \ , & \mbox{if\ } \delta \leq 0 \ ;  \\
    \exp( -\theta \; \delta ) \ , \ \ & \mbox{if\ } \delta \geq 0 \ \ . 
    \end{array} \right.   
 \] 
 The parameter $\theta$ is known as the inverse temperature, which has a
natural interpretation in statistical physics, see MacDonald (2006),
Nash(1974) and Rosenfeld (2005), for intuitive introductions, and
Thompson (1972) for a rigorous text. 

 The Gibbs distribution, $g(\theta)^x$, is the invariant distribution
for the  Metropolis sampling process, given by
 \[ 
    g(\theta)^x = \frac{1}{Z(\theta)} \exp( -\theta f(x) )   
      \ \ \mbox{with} \ \ \ 
    Z(\theta) = \ssum_{x} \exp( -\theta f(x) ) \ .   
 \] 
 The symbol $g(\theta)$ represents a row vector, where the column index,
$x$, spans the possible states of the system.

 Consider a system prepared (shuffled) in such a way that the
probability of starting the system in initial state $x$ is
$g(\theta)^x$. If we move the system to a neighboring state, $y$,
according to the Metropolis sampling procedure, the invariance property
of the Gibbs distribution assures that the probability that the system
will land (after the move) in any given state, $y$, is $g(\theta)^y$,
that is, the probability distribution of the final (after the move)
state remains unchanged.

 Under appropriate regularity conditions, see appendix H.1, the process
is also ergodic. Ergodicity means that even if the system is prepared
(shuffled) with an arbitrary probability distribution, $v(0)$, for the
initial state, for example, the uniform distribution, the probability
distribution, $v(t)$, of the final system state after $t$ moves chosen
according to the Metropolis sampling procedure will be sufficiently
close to $g(\theta)$ for sufficiently large $t$. In other words, the
probability distribution of the final system state converges to the
process' invariant distribution. Consequently, we can find out the
process' invariant distribution by following, for a long time, the
trajectory of a single system evolving according to to the Metropolis
sampling procedure. Hence the expression, The Ergodic Path: One for All.
From the history of an individual system we can recover important
information about the whole process guiding its evolution.
 
 Let us now study how the Metropolis process can help us finding the
optimal (minimum cost) configuration for such a system. The behavior of
the Gibbs distribution, $g(\theta)$, changes according to the inverse
temperature parameter, $\theta$:  \\ 
 - In the high temperature extreme, $1/\theta \rightarrow \infty$, the
  Gibbs distribution approaches the uniform distribution. \\ 
 - In the low temperature extreme, $1/\theta \rightarrow 0$, the Gibbs
   distribution is concentrated in the states with minimum cost only.  

 Correspondingly the Metropolis process behaves as follows: \\  
 - At the high temperature extreme, the Metropolis process becomes
insensitive to the value of the cost function,  wandering (uniformly) at
random in the state space. \\
 - At the low temperature extreme, the Metropolis process becomes very
sensitive to the value of the cost function,  accepting only downhill
moves, until it reaches a local optimum. 

 The central idea of SA involves the use intermediate temperatures: \\ 
 - At the beginning use high temperatures, in order to escape the local
optima, see Figure 3a (L), placing the process at the deepest valley,
and \\ 
 - At the end use low temperatures, in order to converge to the global
optimum (the local optimum at the deepest valley), see Figure 3a (G). 

 \begin{figure}[!h]
 \centerline{\includegraphics*[angle=0, width=6.5in, height=3.0in]{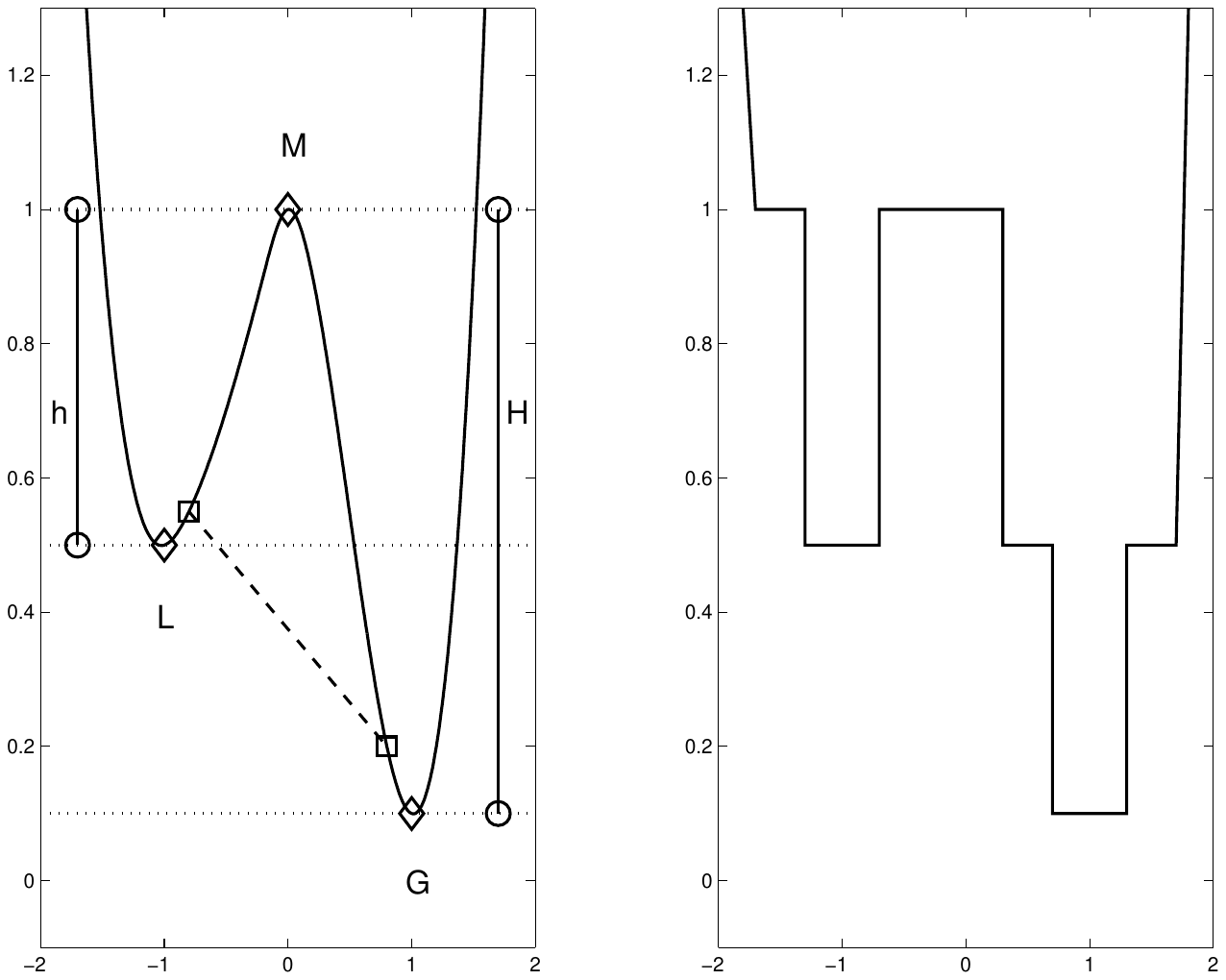}} 
 \vspace{-0.7cm}  
 \centerline{Figure 3a: L,G- Local and global minimum; M- Maximum;} 
 \centerline{S- Short-cut; h,H- Local and global escape energy.} 
 \centerline{Figure 3b: A difficult problem, 
             with steep cliffs and flat plateaus.} 
 \end{figure}

 The secret to play this trick is in the external loop of the SA
algorithm, the Cooling Schedule. The cooling schedule initiates the
temperature high enough so that most of the proposed moves are accepted,
and then slowly cools down the process, until it freezes at an optimum
state. The theory of SA is presented in appendix H.1. 

 The most important result concerning the theory of SA, states that,
under appropriate regularity conditions, the process converges to the
system's optimal solution as long as we use the Logarithmic Cooling
Schedule. This schedule draws the $t$-th move according to Metropolis
process using temperature 

 \[ 
   \theta(t)= \frac{1}{n\Delta}\ln(t) \ ,    
 \]  
 where $\Delta$ is the maximum objective function differential in a
single move and $n$ is the minimum number of steps needed to connect any
two states. Hence, the {\it cooling constant}, $n\Delta$ can be
interpreted as an estimate of how high a mountain we may need to climb
in order to reach the optimal position, see Figure 3a(h). 
  
 Practical implementations of SA usually cool the temperature
geometrically, $\theta \leftarrow (1+\epsilon) \theta$,   after each
batch of Metropolis sampling. The SA is terminated when it freezes, that
is, when the acceptance rate in the Metropolis sampling drops below a
pre-established threshold. 
 Further details on such an implementation are given in the next
section.

 \subsection{Heuristic Acceleration}

 The Standard Simulated Annealing (SSA), described in the last section,
behaves poorly in the BAF problem mainly because it is very difficult to
sense the proximity of low cost states, see Figure 3b, that is,
 \begin{enumerate}
 \item Most of the neighbors of a low cost state, $x$, may have much
higher costs; and
 \item The problem is highly degenerate in the sense that there are
states, $x$, with a large (sub) neighborhood of equal cost states,
$S(x)=\{y\in N(x) \g f(y)=f(x)\}$. In this case, even rejecting all the
proposals that would take us out of $S$, would still give us a
significant acceptance rate.
 \end{enumerate}

 Difficulty 2, in particular, implies the failure of the SSA termination
criterion: A degenerate local minimum (or meta-stable minimum) could
trap the SSA into forever, sustaining an acceptance rate above the
established threshold. 

 The best way we found to overcome these difficulties is to use a
heuristic temperature-dependent cost function, designed to accelerate
the SA convergence to the global optimum and to avoid premature
convergence to locally optimal solutions: 
 \[
  f(x,\mu(\theta)) \equiv \; f(x)\; 
   +\frac{1}{\mu(\theta)} u(x) \ , \ \ 
  u(x) \equiv \; \sum_{j,|q^j(x)| > 1} |q^j(x)|  \ . 
 \]

 The state dependent factor in the additional term of the cost function,
$u(x)$, can be interpreted as an heuristic merit or penalty function that 
rewards multicolored columns for using fewer colors. 
 This penalty function, and some possible variants, have the effect of
softening the landscape, eroding sharp edges, such as in Figure 3b, into
rounded hills and valleys, such as in Figure 3a. The actual functional
form of this penalty function is inspired by the {\em tally function}
used in the $P3$ heuristic of Hellerman and Rarick (1971) for sparse
$LU$ factorization. The temperature dependent parameter, $\mu(\theta)$,
gives the inverse weight of the heuristic penalty function in the cost
function  $f(x,\mu)$ .

 Function $f(x,\mu )$ also has the following properties: \ (1) $f(x,0) =
f(x)$; \ (2) $f(x,\mu )$ is linear in $1/\mu$. Properties 1 and 2
suggest that we can cool the weight $1/\mu$ as we cool the temperature,
much in the same way we control a parameter of the barrier functions in
some constrained optimization algorithms, see McCormick (1983).

 A possible implementation of this Heuristic Simulated Annealing, HSA,
is as follows:
 \begin{itemize}
 \item Initialize parameters $\mu$ and $\theta$, set a random partition,
$x$, and initialize the auxiliary variables $W$, $q$, $c$, $r$, $s$, and
the cost and penalty functions, $f$ and $h$;
 \item For each proposed move, $x\rightarrow y$, compute the cost
differentials 
 \[ 
 \delta_{0} = f(y) - f(x) \ \mbox{and} \ \ \delta_{\mu} = 
     f(y,\mu ) - f(x,\mu ) \ . 
 \] 
 \item Accept the move with the Metropolis probability,
$M(\delta_{\mu},\theta)$. If the move is accepted, update $x$, $W$, $q$,
$c$, $r$, $s$, $f$ and $h$;   
 \item After each batch of Metropolis sampling steps, perform a cooling
step update   
 \[ 
   \theta \leftarrow (1+\epsilon_1)\theta \ , \ \ 
   \mu \leftarrow (1+\epsilon_2)\mu \ , \ \ 
   0 < \epsilon_1 < \epsilon_2 << 1 \ . 
 \]  
 \end{itemize} 

  Computational experiments show that the HSA successfully overcomes the
difficulties undergone by the SSA, as shown in Stern (1991). As far as
we know, this was the first time this kind of perturbative heuristic has
been considered for SA. Pflug (1996) gives a detailed analysis for the
convergence of such perturbed processes. These results are shortly
reviewed is section H.1.

 In the next section we are going to extend the idea of stochastic
optimization to that of evolution of populations, following insights
from biology. In zoology, there are many examples of heuristic merit or 
penalty functions, often called fitness or viability indicators, that
are used as auxiliary objective functions in mate selection, see Miller
(2000, 2001) and Zahavi (1975). 
 The most famous example of such an indicator, the peacock's tail, was
given by Charles Darwin himself, who stated: 
 {\it ``The sight of a feather in a peacock's tail, 
 whenever I gaze at it, makes me feel sick!''}      
 For Darwin, this case was an apparent counterexample to natural
selection, since the large and beautiful feathers have no  adaptive
value for survival but are, quite on the contrary, a handicap to the
peacock's camouflage and flying abilities.  
  However, the theory presented in this section give us a key to 
unlock this mystery and understand the tale of the peacock's tail.

 \section{The Way of Sex: All for One}
 \markboth{CHAPTER 5: MODULARITY AND STOCHASTIC EVOLUTION}
  {5.3 \ THE WAY OF SEX}

 From the interpretation of the cooling constant given in the last
section, it is clear that we would have a lower constant, resulting in a
faster cooling schedule, if we used a richer set of single moves.
Specially, if the additional moves could provide short-cuts in the
configuration space, as the moves indicated by the dashed line in Figure 3a. 
 This is one of the arguments that can be used to motivate another
important class of stochastic evolution algorithms. Namely, Genetic
Programming, the subject of the following sections. We will focus on a
special class of problems known as functional trees. The general
conclusions, however, remain valid in many other applications.

\subsection{Functional Trees}

 In this section, we deal with methods of finding the correct
specification of a complex function. This complex function must be
composed recursively from a finite set, $OP=\{ op_1, op_2, \ldots op_p
\}$, of primitive functions or operators, and from a set, $A=\{ a_1,
a_2, \ldots \}$, of atoms. The $k$-th operator, $op_k$, takes a specific
number, $r(k)$, of arguments, also known as the arity of $op_k$. We use
three representations for (the value returned by) the operator $op_k$
computed on the arguments $x_1,x_2,\ldots x_{r(k)}$ :  
 \[ 
    op_k(x_1, \ldots x_{r(k)} )  
    \ \ , \ \ \ \ \  
    \begin{array}{ccc}
     & op_k & \\ 
     \slash &  & \backslash \\ 
     x_1 & \ldots & x_{r(k)} \\ 
    \end{array}  
    \ \ , \ \ \ \ \  
    \left( op_k \;  x_1 \;  \ldots \; x_{r(k)} \right) \ .
 \] 
 The first is the usual form of representing a function in mathematics;
the second is the tree representation, which displays the operator and
their arguments as a tree; and the third is the prefix, preorder or LISP
style representation, which is a compact form of the tree
representation.

 As a first problem, let us consider the specification of a Boolean
function of $q$ variables, $f(x_1,\ldots x_q)$, to mach a  target table,
$g(x_1,\ldots x_q)$, see Angeline (1996) and Banzhaf el al. (1998). The
primitive set of operators and atoms for this problem are:  

 \[ 
   OP=\left\{\sim, \wedge, \vee, \rightarrow, \odot, \otimes \right\} 
    \ \ \mbox{and} \ \ 
   A=\left\{x_1, \ldots x_q, 0, 1 \right\} \ .    
 \] 
 Notice that while the first operator (not) is unary, the last five
(and, or, imply, nand, xor) are binary. 

 \[ 
 \begin{array}{|c|c|c|c|c|c|c|c|} 
  \hline  
  x & y & \sim x & x \wedge y & x \vee y & x \rightarrow y &     
  x \odot y & x \otimes y \\ 
  \hline 
  0 & 0 & 1 & 0 & 0 & 1 & 1 & 0 \\   
  0 & 1 & 1 & 0 & 1 & 1 & 0 & 1 \\   
  1 & 0 & 0 & 0 & 1 & 0 & 0 & 1 \\   
  1 & 1 & 0 & 1 & 1 & 1 & 0 & 0 \\   
  \hline 
 \end{array}
 \]

 The set, $OP$, of Boolean operators defined above is clearly redundant.
Notice, for example, that 
 \[ 
   x_1 \rightarrow x_2 = \sim ( x_1 \wedge \sim x_2 ) 
   \ , \ \ 
   \sim x_1 = x_1 \odot x_1 
   \ \ \mbox{and} \ \  
   x_1 \wedge x_1 = \sim (x_1 \odot x_2) \ . 
 \] 
 This redundancy may, nevertheless, facilitate the search for the best
configuration in the problem's functional space. 

 Example 1a shows a target table, $g(a,b,c)$. As it is usual when the
target function is an experimentally observed variable, the target
function is {\it not} completely specified. Unspecified values in the
target table are indicated by the don't-care symbol $*$. The two
solutions, $f_1$ and $f_2$, match the table in all specified cases.
Solution $f_1$, however, is simpler and for that may be preferred, see
section 4 for further comments. 

 \[ 
  \begin{array}{|c|c|c|c|c|c|} 
  \hline 
  a & b & c & g & f_1 & f_2 \\ 
  \hline 
  0 & 0 & 0 & 1 & 1   & 1   \\   
  0 & 0 & 1 & 1 & 1   & 1   \\   
  0 & 1 & 0 & * & 1   & 0   \\   
  0 & 1 & 1 & * & 1   & 0   \\   
  1 & 0 & 0 & 0 & 0   & 0   \\   
  1 & 0 & 1 & 1 & 1   & 1   \\   
  1 & 1 & 0 & 0 & 0   & 0   \\   
  1 & 1 & 1 & 1 & 1   & 1   \\    
  \hline 
  \end{array} 
  \ \ \mbox{} \ \ \ \mbox{} \ \   
  \begin{array}{ccc}   
    & f_1 & \\ 
    & | & \\ 
    & \vee & \\ 
    \slash & & \backslash \\ 
    \sim & & |  \\ 
     | & & | \\ 
     a & & c \\ 
  \end{array} 
  \ \ \mbox{} \ \ \ \mbox{} \ \   
  \begin{array}{ccccccc}   
    & & & f_2 & & & \\ 
    & & & | & & & \\ 
    & & & \vee & & & \\ 
    & & \slash & & \backslash & & \\ 
    & \wedge & & & & \wedge & \\ 
    \slash & & \backslash & & \slash & & \backslash \\   
    \sim & &  \sim & & | & & | \\ 
    | & & | & & | & & | \\   
    a & & b & & a & & c \\   
  \end{array} 
 \]   

 \[  
  f_1= (\sim A) \vee C \ , \ \  
  f_2= (\sim A \wedge \sim B) \vee (A \wedge C) \ . 
  \] 
  \[  
  f_1= (\vee \; (\sim A) \; C) \ , \ \  
  f_2= ( \vee \; (\wedge \; (\sim A) \; (\sim B)) \; (A \wedge C)) \ . 
 \] 
 \centerline{Example 1a: Two Boolean functional trees for the target 
  $g(a,b,c)$.} 

 As a second problem, let us consider the specification of a function
for an integer numerical sequence, such as the Fibonacci sequence,
presented in Koza (1983).
 \[ 
    g(j) \equiv \left\{ \begin{array}{ll}
    j \ , & \mbox{if\ } j=0 \vee j=1\ ;  \\
    g(j-1)+g(j-2) \ , \ \ & \mbox{if\ } j \geq 2 \ \ . 
    \end{array} \right.   
 \] 
 The following array, $g^j$, $0\leq j \leq 20$, lists the first $21$
elements of the Fibonacci sequence.     
 \[ 
   g=\left[ 
   0, 1, 1, 2, 3, 5, 8, 13, 21, 34, 55, 89, 144, 233, 377, 610, 
   987, 1597, 2584, 4181, 6765   
   \right] \ . 
 \] 

 In this problem, the primitive set of operators and atoms are:  
 \[ 
   OP=\left\{ +, -, \times, \sigma \right\} \ , \ \ 
   A=\left\{j, 0, 1 \right\} \ ,   
 \] 
 where $j$ in an integer number, and the first three operators are the
usual arithmetic operators. The specified function is used to compute
the first $n+1$ elements of the array $f^j$, seeking to mach the target
array $g^j$, $0\leq j\leq n$. The last primitive function is the
recursive operator, $\sigma(i,d)$, that behaves as follows: When
computing the $j$-th element, $f(j)$, $\sigma(i,d)$ returns the already
computed element $f^i$, if $i$ is in the range, $0\leq i < j$, or a
default value, $d$, if $i$ is out of the range. 

 In the functional space of this problem, possible specifications for
the Fibonacci function in prefix representation, are  
 \[ 
    (+ \; (\sigma \; (- \; j \; 1) \; 1) \; 
    (\sigma \; (- \; j \; (+ \; 1 \; 1) \; 0))) 
    \ , \ \ \ 
    (+ \; (\sigma \; (- \; j \; 1) \; 1) \; 
    (+ \; 0 \; (\sigma \; (- \; j \; (+ \; 1 \; 1) \; 0)))) \ . 
 \]    
 \centerline{Example 2a: Two functional trees for the Fibonacci sequence.} 

 Since the two expressions above are functionally equivalent, the first
one may be preferable for being simpler, see section 4 for further
comments.

 As a third problem, we mention Polynomial Network models. These
functional trees use as primitive operators linear, quadratic or cubic
polynomials in one, two or three variables. For several examples and
algorithmic details, see Farlow (1984), Madala and Ivakhnenko (1994) and
Nikolaev and Iba (2006). Figure 4 shows a simple network used for sales
forcast, a detailed report is given in Lauretto et al. (1995). Variable
$x_5$ is a magazine's sales forecast obtained by a VARMA time series
model using historic sales, econometric and calendric data. Variables
$x_1$ to $x_4$ are qualitative variables (in the scale: Bad, Weak,
Average, God, Excellent) to assess the appeal or attractiveness of an
individual issues of the magazine, namely: (1) cover impact; (2)
editorial content; (3) promotional items; and (4) point of sale
marketing.
 \[
 \dgARROWPARTS=3 
 \begin{diagram}
 \node[4]{}  \\ 
 \node[4]{\Ovalbox{\Ovalbox{\Ovalbox{3}}}} \arrow{n} \\   
 \node[2]{\Ovalbox{1}} \arrow{nee}   
 \node[4]{\Ovalbox{2}} \arrow{nww} \\ 
 \node{x_1} \arrow{ne} \node{x_2} \arrow{n} \node{x_3} \arrow{nw} 
 \node{x_5} \arrow[2]{n}
 \node{x_3} \arrow{ne} \node{} \node{x_4} \arrow{nw} \\    
 \end{diagram} 
 \] 
 \centerline{Figure 4: Polynomial Network.}  
 \centerline{Rings on a node: 1- Linear; 2- (incomplete) 
    Quadratic; 3- (incomplete) Cubic.} 

 Of course, the optimization of a Polynomial Network is far more complex 
than the optimization of Boolean or algebraic etworks, since 
not only topology has to be optimized (identification problem), 
but also, given a topology, the parameters of the polynomial function 
have to be optimaized (estimation problem). 
 Parameter optimization of sub-trees can be based on Tikhonov regularization, 
ridge regression, steepest descent or Partan gradient rules. 
 For several examples and algorithmic details, see Farlow (1984), Madala
and Ivakhnenko (1994), Nikolaev and Iba (2001, 2003, 2006), and 
Stern (2008).

 \subsection{Genetic Programming} 

 Starting from a given random tree, one can start an SA type search in
the problem's (topological) space. In GP terminology, the individual's
functional specification is called its {\it genotype}. the individual's
expressed behavior, or computed solutions, is called its {\it
phenotype}. Changing a genotype to a neighboring one is called a {\it
mutation}. The quality of a phenotype, its performance, merit or
adaptation, is measured by a {\it fitness} function.    
 
 While SA looks at the evolution of a single individual, GP looks at the
evolution of a population. A time parameter, $t$, indexes the successive
generations of the evolving population. In GP, individuals typically
have short lives, surviving only a few generations before dying.
Meanwhile, populations may evolve for a very long time.

 In GP an individual may, during its ephemeral life, share information,
that is, swap (copies) of its (partial) genome, with other individuals.
This genomic sharing process is called {\it sex}. In GP an individual,
called a {\it parent}, may also participate in the creation of a new
individual, called its {\it child}, in a process called {\it
reproduction}. In the reproduction process, an individual gives
(partial) copies of its genotype to its offspring. Reproduction
involving only one parent is called asexual, otherwise it is called a
sexual reproduction.  

 In the following list, a set of possible mutation and sex operators are
given:  

 1- Point leaf mutation: Replace a leaf atom by an other atom. 

 2- Point operator mutation: Replace a node operator by a compatible operator. 

 3- Shrink mutation: Replace a sub-tree by a leaf with a single atom. 

 4- Grow mutation: Replace the atom at a leaf by a random tree.  

 5- Permutation: Change the order of the children of a given node. 
 
 6- Gene duplication: Replace a leaf by a copy of a sub-tree. 

 7- Gene inversion: Switch two sub-trees.

 8- Crossover: Share or exchange sub-trees between individuals.

  The first five operators, involving only one sub-tree, are sometimes
called (proper) mutations, while the last three operators, involving two
or more separate sub-trees, are called recombinations. Also notice that
the first seven operators involve only one individual, while crossover
involves two or more. This list of mutation and recombination operators
is redundant but, again, this redundancy may also facilitate the search
for the best configuration in the problem's functional space.

 We should mention that the terms used to name these operators are not
standard in the field of GP, and even less so in biology, genetics,
zoology and botany. We should also mention that the forms of GP
presented in this section, do not explore the possibility of allowing
individuals to carry a (redundant) set of two or more homologous
(similar but not identical) specifications (genes), a phenomenon known
as diploidy or multiploidy. Diploidy is common in eukaryotic
(biological) life, and can provide a much richer structure and better
performance to GP.      
 
  Sexual reproduction can be performed by crossover, with parents giving
(partial) copies of their genome to the children.   The following
examples show a pair of parents and children generated by a
single crossover, for some of the problems considered in the last
section. A square parenthesis in the prefix representation indicates a
crossover point. The tree representation would indicate the same
crossover points by broken edges ($=$). Notice that in these examples  
the is a child corresponding to a solution presented in the last section. 
 \[ 
  \begin{array}{ccccc}   
    & & & f_1 & \\ 
    & & & | & \\ 
    & & & \vee & \\ 
    & & \slash & & \backslash \\ 
    & \wedge & & & = \\ 
    \slash & & \backslash & & | \\   
    \sim & &  \sim & & a \\ 
    | & & | & &  \\   
    a & & b & &  \\   
  \end{array} 
  \ \ \mbox{} \ \ \ \mbox{} \ \   
  \begin{array}{ccccc}   
    & f_2 & & & \\ 
    & | & & & \\ 
    & \vee & & & \\ 
    \slash & & = & & \\ 
    | & & & \wedge & \\ 
    b & & \slash & & \backslash \\   
      & & | & & | \\ 
      & & | & & | \\   
      & & a & & c \\   
  \end{array} 
 \ \ \Rightarrow \ \ 
  \begin{array}{ccccccc}   
    & & & f_3 & & & \\ 
    & & & | & & & \\ 
    & & & \vee & & & \\ 
    & & \slash & & \backslash & & \\ 
    & \wedge & & & & \wedge & \\ 
    \slash & & \backslash & & \slash & & \backslash \\   
    \sim & &  \sim & & | & & | \\ 
    | & & | & & | & & | \\   
    a & & b & & a & & c \\   
  \end{array} 
 \]   
    
 \centerline{Example 1b: Crossover between Boolean functional trees.} 
 
 \[ 
    \mbox{Parents:} \ \ \  
    (* \; [\sigma \; (- \; j \; 1) \; 1] \; (* \; j \; j))  
    \ , \ \ 
    (+ \; (\sigma \; (- \; j \; (+ \; 1 \; 1) \; 0)) \; 
    [- \; j \; 1] \; ) \ ; 
 \] 
 \[ 
    \mbox{Children:} \ \ \  
    (* \; [- \; j \; 1] \; (* \; j \; j))  
    \ , \ \ 
    (+ \; (\sigma \; (- \; j \; (+ \; 1 \; 1) \; 0)) \;  
    [\sigma \; (- \; j \; 1) \; 1] \; ) \ . 
 \]  
 \centerline{Example 2b: Crossover between arithmetic functional trees.} 
 
 Finally, the reproduction and survival selection processes in GP assume
that individuals are chosen from the general population according to
sampling probabilities called the {\it mating} (or representation)
distribution and the {\it survival} distribution, respectively. Some
general policies used to specify these probability distributions, based
on the individual's fitness, are given below:

 1- Top Rank Selection: The highest ranking (best fit) individual is
selected. 

 2- High Pressure Selection: An individual is selected from the
population with a probability that increases sharply (super-linearly)
with its fitness or fitness' rank.   

 3- Fitness Proportional Selection: An individual is selected from the
population with a probability that is proportional to its fitness.   

 4- Rank Proportional Selection: An individual is selected from the
population with a probability that is proportional to its fitness' rank.

 5- Low Pressure Selection: An individual is selected from the
population with a probability that increases modestly (sub-linearly)
with its fitness or fitness' rank.

 6- Tournament Selection: A small subset is sampled at random
(uniformly) from the population, from which the best (one or two)
individuals are selected.  

 7- Uniform Selection: An individual is selected from the population
with uniform probability.

 These processes are supposed to mimic biological selection mechanisms,
including sexual differentiation, like male and female (alleged)
behavior, familiar and other sub-population structures, etc.

 \subsection{Schemata and Parallelism}

 A possible motivation for developing populational evolutionary
algorithms like GP, instead of single individual evolutionary
algorithms, like straight SA, is to consider a richer and better
neighborhood structure. The additional moves made available should
provide short-cuts in the problem's configuration space, lowering the
cooling constant and allowing a faster convergence of the algorithm.

 The {\it intrinsic parallelism} argument, first presented in Holland
(1975), proves that, under appropriate conditions, GP is likely to
succeed in providing such a rich neighborhood structure. The
mathematical analysis of this argument is presented in section H.2, see
also Reeves (1993, Ch.4 Genetic Algorithms). According to Reeves, 
 \begin{quotation}   
 {\it ``The underlying concept Holland used to develop a theoretical
analysis of his GA [GP] was that of {\em schema}. The word comes from the
past tense of the Greek verb $\epsilon \chi \omega$, {\em echo}, to
have, whence it came to mean shape or form; its plural is {\em
schemata}.''} (p.154)   
 \end{quotation} 

 Schemata are partially specified patterns in a program, like partially
specified segments of prefix expressions, or partial code for functional
sub-trees. The {\em length} and {\em order} of a schema are the {\em
distance} between the first and last defined position on the schema, and
the number of defined positions, respectively, see section H.2. The
Intrinsic Parallelism theorem states that the number of schemata (of
order $l$ and length $2l$, in binary coded programs, in individuals of
size $n$) present in a population of size $m$, is proportional $m^3$.
The crossover operator enriches the neighborhood of an individual with
the schemata present in other individuals of the population. If, as
suggested by the implicit parallelism theorem, the number of such
schemata is large, GP is likely to be an effective strategy. 

 Schaffer (1987, p.89), celebrates this theorem stating that: 
 \begin{quotation} 
 {\it ``this [intrinsic parallelism] constitutes the only known example
of combinatorial explosion working to advantage instead of disadvantage.''}
 \end{quotation}

 Indeed, Schaffer has ample reason to praise Holland's result.
Nevertheless, we must analyze this important theorem carefully, in order
to understand its consequences correctly. In particular, we should pay
close attention to the unit, $u$, used to measure the population size,
$m$. As shown in detail in section H.2, this unit, $u=2^l$, is itself
exponential in the schemata order. Therefore, the combinatorial
explosion works to our advantage as long as we use short schemata,
relative to the log-size of the population. This situation is described
by Reeves as:   
 \begin{quotation} 
  {\it ``Thus the ideal situation for a GA [GP] are those where short,
low-order schemata combine with each other to form better and better
solutions. The assumption that this will work is called by Goldberg
(1989) the {\em building-block hypothesis}. Empirical evidence is strong
that this is a reasonable assumption in many problems.''} (p.158) 
 \end{quotation} 

 One key question we must face in order to design a successful GP
application is, therefore: How then can we organize our working space so
that our programming effort can rely on short schemata? 

 The solution to this question is well known to computer scientists and
software engineers: Organize the programs hierarchically (recursively)
as self-contained (encapsulated) building-blocks (modules, functions,
objects, sub-routines, etc.). The next section is dedicated to the study
of modular organization, and its spontaneous emergence in complex
systems.

 \section{Simple Life: Small is Beautiful}
 \markboth{CHAPTER 5: MODULARITY AND STOCHASTIC EVOLUTION}
  {5.4 \ SIMPLE LIFE}

 The biological world is an endless source of inspiration for
improvements and variations in GP (of course, one should also be 
careful not to be carried away by superficial analogies). A nice
anthology of introductory articles can be found in the book by Michod
and Levin (1988), {\it The Evolution of Sex: An Examination of Current
Ideas}. Let us begin this section with an interesting biological
example.

 It is a well known phenomenon that bacteria can develop antibiotic
resistance. Among the most common mechanisms conferring resistance to
new antibiotics, one can list: Agents that modify or destroy the
antibiotic molecular structure; Agents that modify or protect the
antibiotic targets; New pathways offering alternatives to those blocked
by the antibiotic action; etc. However, all these mechanisms entail a
fitness cost to the modified individuals. At the very least, there is
the cost of complexity, that is, the cost of building and maintaining
these new mechanisms. Hence, if the selective pressure of the antibiotic
presence is interrupted, resistant bacterial populations will often
revert to non-resistant, see for example Bj\"{o}rkholm et al. (2001). 

 This biological example can be interpreted as the embodiment of {\it
Okcam's razor} or {\it lex parsimoniae}, an epistemological principle
stated by the 14th-century English logician friar William of Ockham, in
the following forms: 

 - {\it Entia non sunt multiplicanda praeter necessitate,} \ or  

 - {\it Pluralitas non est ponenda sine neccesitate.}   

 \noindent that is, entities should not be created or multiplied without
necessity. 

 In section 4.1 we will see how well this principle applies to
statistical models, and how it can be enforced. In section 4.2 we will
examine {\it introns}, a phenomenon that at first glance appears to
contradict Okcam's razor. Nevertheless, we will also see how introns
allow building blocks to appear spontaneously as an emergent feature in
GP.

 \subsection{Overfitting and Regularization}

 This section discusses the use of Okcam's razor in statistical
modeling. As an illustrative example, we use a standard normal multiple
linear regression model. This model states that $y = X\beta +u$, $X\;
n\times k$, where $n$ is the number of observations, $k$ is the number
of independent variables, $\beta\in \: ]-\infty, \infty[^k$ is the
vector of regression coefficients, and $u$ is a Gaussian white noise
such that $E(u)=0$ and $\Cov(u)=\sigma^2 I$, $\sigma \in [0, \infty[$,
see DeGroot (1970), Hocking (1985) and Zellner (1971). Using the
standard diffuse prior $p(\beta,\sigma)= 1/\sigma$, the joint posterior
probability density, $f(\beta,\sigma\g y,X)$, and the MAP (maximum a
posteriori) estimators for the parameters are given by:
 \begin{eqnarray*} 
 f(\beta,\sigma\g y,X) &=& 
 \frac{1}{\sigma^{n+1}} \exp( -\frac{1}{2\sigma^2} (\: (n-k)s^2 
            +(\beta -\hat{\beta})'X'X(\beta -\hat{\beta}) \: ) \: )\; ,\\ 
 \hat{\beta} &=& (X'X)^{-1}X'y\; ,\\ 
 \hat{y} &=& X\hat{\beta}\; ,\\   
 s^2 &=& (y -\hat{y})'(y -\hat{y})/(n-k)\; .\\ 
 \end{eqnarray*}     

 In the polynomial multiple linear regression model of order $k$, the
dependent variable $y$ is explained by the powers $0$ through $k$ of the
independent variable $x$, i.e., the regression matrix element at 
 row $i$ and column $j$ is $X_i^j = (x_i)^{j-1}$, 
 $i=1\ldots n$, $j=1\ldots k+1$. 
 Note that the model of order $k$ has dimension
 $d=k+2$, with parameters $\beta_0, \beta_1,\ldots \beta_k,\:
 \mbox{and}\: \sigma$. 

 In the classical example presented in Sakamoto et al. (1986, ch.8), we
want to fit a linear regression polynomial model of order $k$, $$y=
\beta_0 \uno+\beta_1x +\beta_2x^2 \ldots +\beta_kx^k +N(0,\sigma I)$$
through the $n=21$ points, $(x_i,y_i)$, in Table 1. 
 \begin{table}[h]  
 \begin{center} 
 \begin{tabular}{|r r r|r r r|r r r|} 
 \hline 
 $i$ & $x_i$ & $y_i$ & $i$ & $x_i$ & $y_i$ & $i$ & $x_i$ & $y_i$ \\      
 \hline 
  1 & 0.00 &  0.125 &  8 & 0.35 & -0.135 & 15 & 0.70 &  0.035 \\  
  2 & 0.05 &  0.156 &  9 & 0.40 &  0.105 & 16 & 0.75 &  0.327 \\ 
  3 & 0.10 &  0.193 & 10 & 0.45 &  0.131 & 17 & 0.80 &  0.061 \\   
  4 & 0.15 & -0.032 & 11 & 0.50 &  0.154 & 18 & 0.85 &  0.383 \\    
  5 & 0.20 & -0.075 & 12 & 0.55 &  0.114 & 19 & 0.90 &  0.357 \\  
  6 & 0.25 & -0.064 & 13 & 0.60 & -0.094 & 20 & 0.95 &  0.605 \\  
  7 & 0.30 &  0.006 & 14 & 0.65 &  0.215 & 21 & 1.00 &  0.499 \\ 
 \hline  
 \end{tabular} 
 \end{center}   
 \caption{Sakamoto's data set for polynomial model} 
 \end{table} 
 
 This example was produced by Sakamoto simulating the i.i.d. stochastic
process
 \[ 
   y_i  = g(x_i) +0.1*N(0,1) \ , \ \ 
   g(x) = \exp((x -0.3)^2) -1 \ ,  
 \]  
 where the target function, $g(x)$, cannot be expressed exactly as a
finite order linear regression polynomial model. 

 Figure 5 presents the target function in the example's range, the data
set (Sakamoto's set in 5a and a second set generated by the same
stochastic process in 5b), and the regression polynomials of orders 0
through 5. In this example, all the available data points are used to
fit the model. An alternative procedure would be to divide the available
data in two sets, the {\it training set}, used to adjust the model, and
the {\it test set}, used to test the model's predictive or extrapolation
power.   
 \begin{figure}[h]
  \centerline{\includegraphics*[angle=0, width=7.6in, 
              height=5.5in]{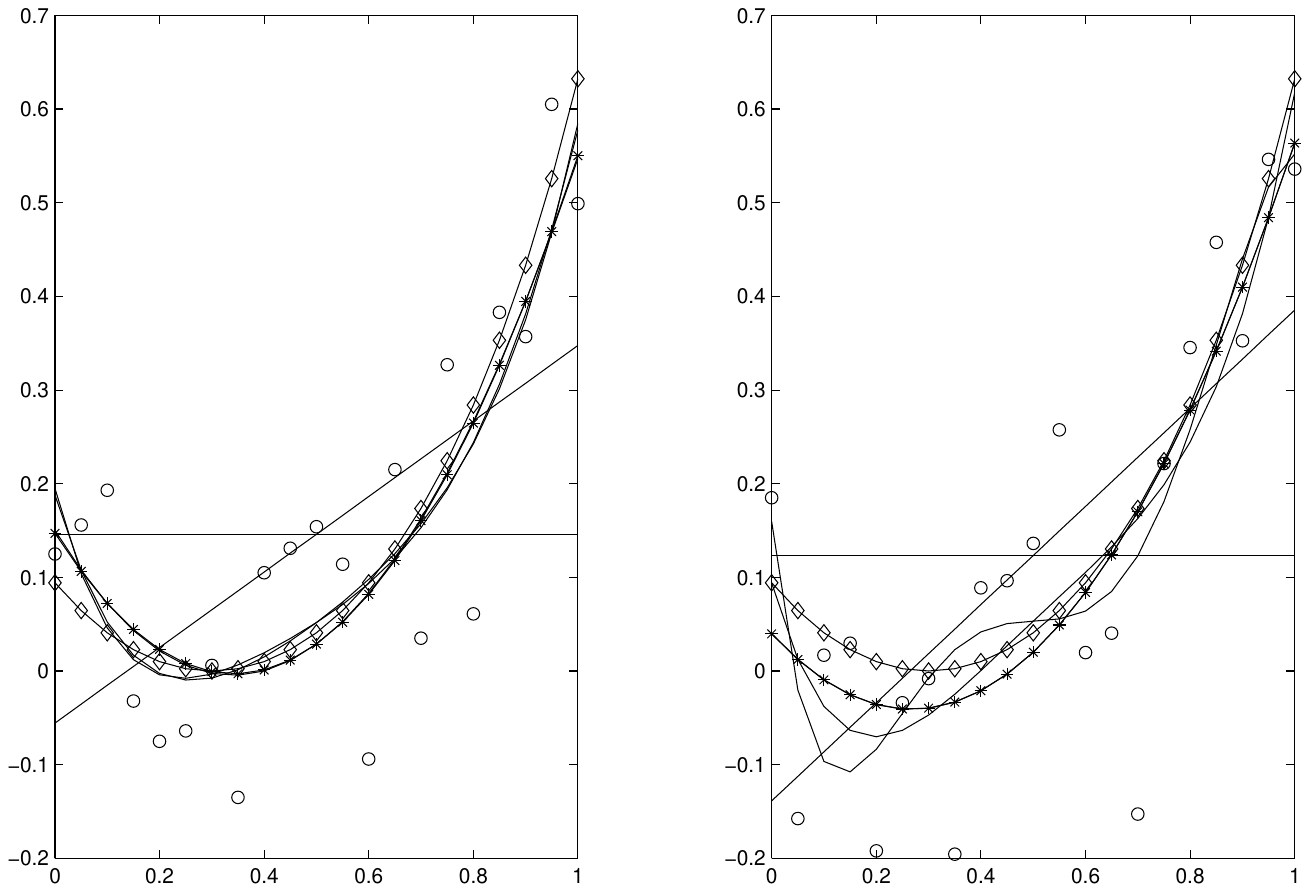}} 
  \mbox{} \vspace{-0.9cm} 

  \centerline{Figure 5a,b: Target function, data points, and  
             polynomial regressions of order 0 to 5;} 
  \centerline{$\circ$: Data points; $\diamond$: Target function;  
             $*$: Best (quadratic) polynomial regression.} 
 \end{figure}

 Just by visual inspection, one can come to the following conclusions:  

 - If the model is too simple, it fails to capture important information
available in the data, making poor predictions.   

 - If the model is too complex, it {\it overfits} the training data,
that is, the curve $f(t)$ tends to become an interpolation curve, but
the curve becomes unstable and predicted values become meaningless.  

 The polynomial regression model family presented in the example is
typical, in the sense that it offers a class o models of increasing
dimension, or complexity. This poses a {\it model selection} problem,
that is, deciding, among all models in the family, the ``best'' adapted
to the data. It is natural to look for a model that accomplishes a small
empirical error, the estimated model error in the training data,
$R_{emp}$. A regression model is estimated by minimizing the 2-norm
empirical error. However, we cannot select the ``best'' model based only
on the empirical error, because we would usually select a model of very
high complexity. In general, when the dimensionality of the model is
high enough, the empirical error can be made equal to zero by simple
interpolation. It is a well known fact in statistics (or learning
theory), that the prediction (or generalization) power of such high
dimension models is poor. Therefore the selection criterion has to
penalize also the model   dimension. This is known as a {\it
regularization} mechanism.

 Some model selection criteria define $R_{pen} = r(d,n) R_{emp}$ as a
penalized (or regularized) error, using a regularization factor, $r(d,n)$, 
where $d$ is the model dimension and $n$ the number of training data points.  
 Common regularization factors, using $p=(d/n)$, are:
 \begin{itemize} 
 \item Akaike's final prediction error: $\mbox{FPE}=(1+p)/(1-p)$,  
 \item Schartz' Bayesian criterion: $\mbox{SBC}= 1 +\ln(n)p/(2-2p)$,  
 \item Generalized cross validation: $\mbox{GCV}=(1-p)^{-2}$,  
 \item Shibata model selector: $\mbox{SMS}=1+2p$,
 \end{itemize} 
 All these regularization factors are supported by
theoretical arguments as well as by empirical performance; other common
regularization methods are Akaike information criterion (AIC), and
Vapnik-Chervonenkis (VC) prediction error. For more details, see Akaike
(1970 and 1974), Barron (1984), Breiman (1984), Cherkassky (1998),
Craven (1979),  Michie (1994), Mueller (1994), Shibata (1981), Swartz
(1978), Unger (1981) and Vapnik (1995, 1998).  

 We can also use the FBST as a model selection criterion by testing the
hypothesis of some of its parameters being null, as detailed in Pereira
and Stern (2001). The FBST version of Okcam's razor states: 

  \noindent 
 - {\it Do not include in the model a new parameter unless there is
strong evidence that it is not null.}

 Table 2 presents the empirical error, $\mbox{EMP}=||y-\hat{y}||_2^2/n$,
for models of order $k$ ranging from $0$ to $5$, several regularization
criteria previously mentioned as well as the Akaike information
criterion (AIC), as computed by Sakamoto. Table 2 also presents the
$e$-value supporting the hypothesis $H:\beta_k=0$, that is, the
hypothesis stating that the model is in fact of order $k-1$. 
  
 \begin{table}[h]  
 \begin{center} 
 \begin{tabular}{|c|c|c|c|c|c|c|c|} 
 \hline 
 Order & EMP & FPE & SBC & GCV & SMS & AIC & FBST \\  
 \hline 
 0 & 0.03712 & 0.04494 & 0.04307 & 0.04535 & 0.04419 & -07.25 & 0.00 \\ 
 1 & 0.02223 & 0.02964 & 0.02787 & 0.03025 & 0.02858 & -20.35 & 0.00 \\ 
 2 & 0.01130 & 0.01661 & 0.01534 & 0.01724 & 0.01560 & -32.13 & 0.00 \\ 
 3 & 0.01129 & 0.01835 & 0.01667 & 0.01946 & 0.01667 & -30.80 & 1.00 \\ 
 4 & 0.01088 & 0.01959 & 0.01751 & 0.02133 & 0.01710 & -29.79 & 0.99 \\ 
 5 & 0.01087 & 0.02173 & 0.01913 & 0.02445 & 0.01811 & -27.86 & 1.00 \\ 
 \hline 
 \end{tabular} 
 \caption{Selection Criteria for the Polynomial Model} 
 \end{center} 
 \end{table} 

 Alternative approaches to regularization are given by Jorma Rissanen's
MDL (minimum description length) and Chris Wallace's MML (minimum
message length). Following an old idea of Andrey Kolmogorov, these
criteria make direct use of a program's code-length as a measure of
complexity, see  Rissanen (1978, 1989), Wallace and Boulton (1968) and
Wallace and Dowe (1999).

 \subsection{Building Blocks and Modularity}

 As seen in section 3, GP can produce polynomial networks that are
very similar to the polynomial regression models presented in the last
section. The main difference between the polynomial networks and the
regression models lies in their generation process: While the regression
models are computed by a deterministic algorithm, the GP networks are
generated by a random evolutionary search. However, if one uses
compatible measures of performance for the GP fitness function and the
regression (penalized or regularized) error, one could expect GP to
produce networks that somehow fulfill Okcam's parsimony principle. 

 Surprisingly, this is not so. P. Angeline (1994, 1996) noted that GP
generated networks typically contain large segments of {\it extraneous}
code, that is, code segments that, if removed, do not (significantly)
alter the solution computed by the network. Trivial examples of
extraneous code segments are $(+\; s\; 0)$ and $(*\; s\; 1)$, where $s$
is a sub-expression. By their very definition, extraneous code segments
cannot (significantly) contribute to an individual's fitness, and hence
to its survival or mating probabilities. However, Angeline noticed that
the presence of extraneous code could significantly contribute to the
expected fitness of the individual's descendents! Apparently, the role
of these (sometimes very large) patches of inert code is to isolate
important blocks of working code, and to protect these blocks from being
broken at recombination (destructive crossover).     

 In biological organisms, the genetic code of eukaryots exhibits similar
regions of code (DNA) that are or are not expressed in protein
synthesis; these regions are called {\it exons} and {\it introns},
respectively. Introns do not directly code amino-acid sequences in
proteins, nevertheless, they seem to have an important role in the
meta-control of the genetic material expression and reproduction. 

 Subsequent work of several authors tried to incorporate meta-control
parameters to GP. Iba and Sato (1993, p.548), for example, propose a
meta-level strategy for GP based on a self-referential representation,
where  
 \begin{quotation} 
 {\it ``[a] self-referential representation maintains a
meta-description, or meta-prescription, for crossover. This meta-genetic
descriptions are allowed to co-evolve with the gene pool. Hence, genetic
and meta-genetic code variations are jointly selected. How well the
genetic code is adapted to the environment is translated by the merit or
objective function which, in turn, is used for the immediate, short-term
or individual selection process. How well the genetic and meta-genetic
code are adapted to each other impacts on the system's evolvability, a
characteristic of paramount importance in long-run survival of the
species.''}  
 \end{quotation} 

 Functional trees, for example, can incorporate edge annotations,
like probability weights, linkage compatibility or affinity,
etc. Such annotations are meta-parameters used to control the
recombination of the sub-tree directly bellow a given edge. 
 For example, weights may be used to specify the probability that a
recombination takes place at that edge, while linkage compatibility or
affinity tags may be used to identify homologous or compatible
genes, specifying the possibility or probability of swapping two
sub-trees. 
 Other annotations, like context labels, variable type, etc., may
provide additional information about the possibility or probability of
recombination or crossover, the need of type-cast operations, etc. 
 When such metacontrol annotations coevolve in the stochastic optimization
process, they may be interpreted as a spontaneusly emergent semantics. 
 Any semantic information may, in turn, be used in the design of 
acceleration procedures based on heuristic merit functions, 
like the example studied in section 5.2.3.

 Banzahf (1998, ch.6, p.164), gives a simple example of functional tree
annotation:
 \begin{quotation} 
  {\it ``Recently, we introduced the explicitly defined introns (EDI)
into GP. An integer value is stored between every two nodes in the GP
individual. This integer value is referred as the EDI value (EDIV). The
crossover operator is changed so that the probability that crossover
occurs between any two nodes in the GP program is proportional to the
integer value between the nodes. That is, the EDIV integer value
strongly influences the crossover sites chosen by the modified GP
algorithm, Nordin et al. (1996). 

 The idea behind EDIVs was to allow the EDIV vector to evolve during the
GP run to identify the building blocks in the individual as an {\em
emergent} phenomenon. Nature may have managed to identify genes and to
protect them against crossover in a similar manner. Perhaps if we gave
the GP algorithm the tools to do the same thing, GP, too, would learn
how to identify and protect the building blocks. If so, we would predict
that the EDIV values within a good building block should become low and,
outside the good block, high.''}  
 \end{quotation} 

 Let us finish this section presenting two interpretations for the role
of modularity in genetic evolutionary processes. This interpretations
are common in biology, computer science and engineering, an indication
that they provide powerful insights. These two metaphors are commonly
referred to as: 

 - New {\it technology dissemination} or {\it component design
substitution}, and 

 - {\it Damage control} or {\it repair mechanism}.

 The first interpretation is perhaps the more evident. In a modular
system, a new design for an old component can be easily incorporated
and, if successful, be rapidly disseminated. A classical example is the
replacement of mechanical carburetors by electronic injection as the
standard technology for this component of gasoline engines in the
automotive industry. The large assortment of {\it upgrade kits}
available in any automotive or computer store gives a strong evidence of
how much these industries rely on modular design.
 The second interpretation explains the possibility for the ``continued
evolution of germlines otherwise destined to extinction'', see Michod
and Levin (1988). A classic illustration related to the damage control
and repair mechanisms offered by modular organization is given by the
Hora and Tempus parable of Simon (1996), presented in section 6.4. 

 The lessons learned in this section may
be captured by the following dicta of Herbert Simon:  
 \begin{quotation} 
 {\it ``The time required for the evolution of a complex form from
simple elements depends critically on the number and distribution of
potential intermediate stable subassemblies.''} \ Simon (1996, p.190).  

 \mbox{} 

  {\it ``Hierarchy, I shall argue, is one of the central structural
schemes that the architect of complexity uses.''} \ Simon (1996, p.184).

 \end{quotation}

 \section{Evolution of Theories}
 \markboth{CHAPTER 5: MODULARITY AND STOCHASTIC EVOLUTION}
  {5.5 \ EVOLUTION OF THEORIES}

 The last sections presented a general framework for the stochastic
evolution of complex systems. Figure 6 presents a systemic diagram of
biological production, according to this framework. This diagram, is
also compatible with the current biological theories of life evolution,
provided it is considered as a schematic simplification focusing on our
particular interests. 

 The comparison of this biological production diagram with the
scientific production diagram presented in section 1.5. motivates
several analogies which may receive further encouragement from a comment
by Davis and Steenstrup (1987, p.2):  
 \begin{quotation} 
 {\it ``The metaphor underlying genetic algorithms is that of natural
evolution. In evolution, the problem each species faces is one of
searching for beneficial adaptations to a complicated and changing
environment. The `knowledge' that each species has gained is embodied in
the makeup of the chromosomes of its members.''}  
 \end{quotation}

 According to this view, computational (or biological genetic) programs
are perceived as coded knowledge acquired by a population. An immediate
generalization of this idea is to consider the evolution of other
corpora of knowledge, embodied in a variety of media. Our main interest,
given the scope of this book, is in the evolution of scientific theories
and their supporting statistical models. This is the topic discussed in
this and the next sections. For some very interesting qualitative
analyses related to this subject see Richards (1989, appendix II) and
Lakatos (1978a,b). 

 Section 5.1 considers several ways in which statistical models can be
nested, mixed and separated. It also analyzes the series-parallel
composition of several simpler and (nearly) independent models.  
 Section 5.2 is devoted to complementary models. Complementarity is a
basic form of model composition in quantum mechanics that has received,
so far, little attention in other application areas. All these forms of
model transformation and combination should provide a basic set of
mutations and recombination operators in an abstract modeling space. In
this section we focus on the statistical operations themselves, leaving
some of the required epistemological analyses and historical comments to
sections 6 and 7.

 \begin{table}[ht] 
 \begin{center} 
 \begin{tabular}{c c c c c} 

 \multicolumn{2}{l}{Somatic behavior} & & 
 \multicolumn{2}{r}{Genetic code \mbox{}} \\  \\ 

 Individual  & $\Leftarrow$ & Epigenetic     
                       &  $\Leftarrow$ &  Individual   \\ 
 ontogenesis  & &  development   & &   genotypes       \\ 
   $\Downarrow$   & &               & & $\Uparrow$        \\  
 Fittest      &  \multicolumn{3}{c}{Live/Dead} &  Genetic    \\ 
 survival    &  \multicolumn{3}{c}{organism}   & Mutation  \\     

   $\Downarrow$   & &               & & $\Uparrow$         \\ 
 Mating    & & Reproductive & &  Sexual   \\ 
 competition  & $\Rightarrow$ & representation  
 & $\Rightarrow$ & recombination \\  \\ 
 
 \multicolumn{2}{l}{Phenotypic space} & & 
 \multicolumn{2}{r}{Genotypic space} 
 \end{tabular} 
 \mbox{} \\ \mbox{} \\ 
 \centerline{Figure 6: Biological production diagram.}
 \end{center}    
 \end{table}

 \subsection{Nested, Mixture, Separate and Series-Parallel Models}

 In this subsection we use some examples involving the (two-parameter)
Weibull (W2) and Gompertz (G2) probability models. The hazard (or
failure rate) functions, $h_{W2}$ and $h_{G2}$, the reliability (or
survival) function, $r_{W2}$ and $r_{G2}$, and the density function,
$f_{W2}$ and $f_{G2}$, of these models are given by:
 \[ 
   h_{W2}(x \g \beta, \gamma) = \frac{\beta}{\gamma^\beta} x^{\beta-1}     
   \ ; \ \   
   r_{W2} = \exp\left( -\left(\frac{x}{\gamma}\right)^\beta \right) 
   \ ; \ \   
   f_{W2} = \frac{\beta}{\gamma^\beta} x^{\beta-1}  
          \exp \left(-\left( \frac{x}{\gamma} \right)^\beta \right) \ ;  
 \] 
 \[ 
   h_{G2}(x \g \alpha, \lambda) = \lambda \alpha^x 
   \ ; \ \   
   r_{G2} = \exp\left( \frac{-\lambda}{\log \alpha}
   \left( \alpha^x -1 \right)  \right)   
   \ ; \ \ \mbox{and} \ \ 
   f_{G2} = \lambda \alpha^x \exp \left( 
    \frac{-\lambda}{\log \alpha} \left(\alpha^x - 1\right) \right) \ . 
 \] 
 The parameters: $\beta$ and $\gamma$ for the Weibull model; and
$\lambda$ and $\alpha$ for the Gompertz model, are known, respectively,
as the scale and shape parameters. Notice that $h=f/r$, and $r=1-F$,
that is, the reliability function is the complement of the cumulative
distribution function $F$. 

 These probability models are used in reliability theory to study the
characteristics of the survival (or life) time of a system, until it
first fails (or dies). It can be shown, see Barlow and Prochan (1981),
Gavrilov (1991, 2001) and appendix H.3, that the Weibull distribution is
adequate to describe the survival time of many allopoietic, manufactured
or industrial systems, while the Gompertz distribution is adequate to
describe the life time of many autopoietic, biological or organic
systems. In this setting, the key difference between autopoietic and
allopoietic systems is the nature of their ontogenesis or assembling
process, as described in the next paragraphs. Reasonable assumptions
concerning the systems' ontogenesis will render either the Weibull or
the Gompertz distributions as asymptotic eigen-solutions.

 The assemblage of allopoietic systems is assumed to be subject to rigid
protocols of quality control, to assure that parts and components as
well as the final product, work properly. The good quality of the parts
and components allows the use of efficient projects and streamlined
designs, with little redundancy and practically no waste. Project
optimization provides the designer of such products the means to
minimize the use of space, material resources, and even assembling time.
The lack of redundancy, however, implies that the failure of just a few
or even only one small component can disable the system.

 In contrast, an autopoietic system is assumed to be self-assembled. The
very nature of organic ontogenesis does not allow for strict quality
control. For example, in embryonic or fetal development there is not an
opportunity to check individual cells, nor is there a mechanism for
replacing defective ones. The viability of autopoietic systems relies
not on quality control of individual components, but on massive
redundancy and parallelism.

 Let us now examine more closely some of the details of these
statistical models. In so doing we will also be able to explain several
modes of model composition. 

 In the Weibull model, the scale parameter, $\gamma$, is approximately
the 63rd lifetime percentile, regardless of the value of the shape
parameter. By altering its shape parameter, $\beta$, the (two-parameter)
Weibull distribution can take a variety of forms, see Figure 7 and
Dodson(1994). Some particular values of the shape parameter are
important special cases:
 for $\beta=1$,   it is the exponential distribution; 
 for $\beta=2$,   it is the Rayleigh distribution;   
 for $\beta=2.5$, it approximates the lognormal distribution;   
 for $\beta=3.6$, it approximates the normal distribution; and    
 for $\beta=5.0$, it approximates the peaked normal distribution.   
 The flexibility of the Weibull distribution makes it very useful for
empirical modeling, specially in quality control and reliability. The
regions $\beta<1$, $\beta=1$, and $\beta>1$ correspond to decreasing,
constant and increasing hazard rates. These three regions are also known
as infant mortality, memoryless, and wearout. In the limit case
$\beta=1$, the Weibull degenerates into the Exponential distribution.
This (no) aging regime represents a simple element with no structure
exhibiting, therefore, the memoryless property of constant failure rate,
$h_E(x\g \gamma) = 1/\gamma$.

 \begin{figure}[!h]
  \centerline{\includegraphics*[angle=0, width=8.0in, height=3.5in]{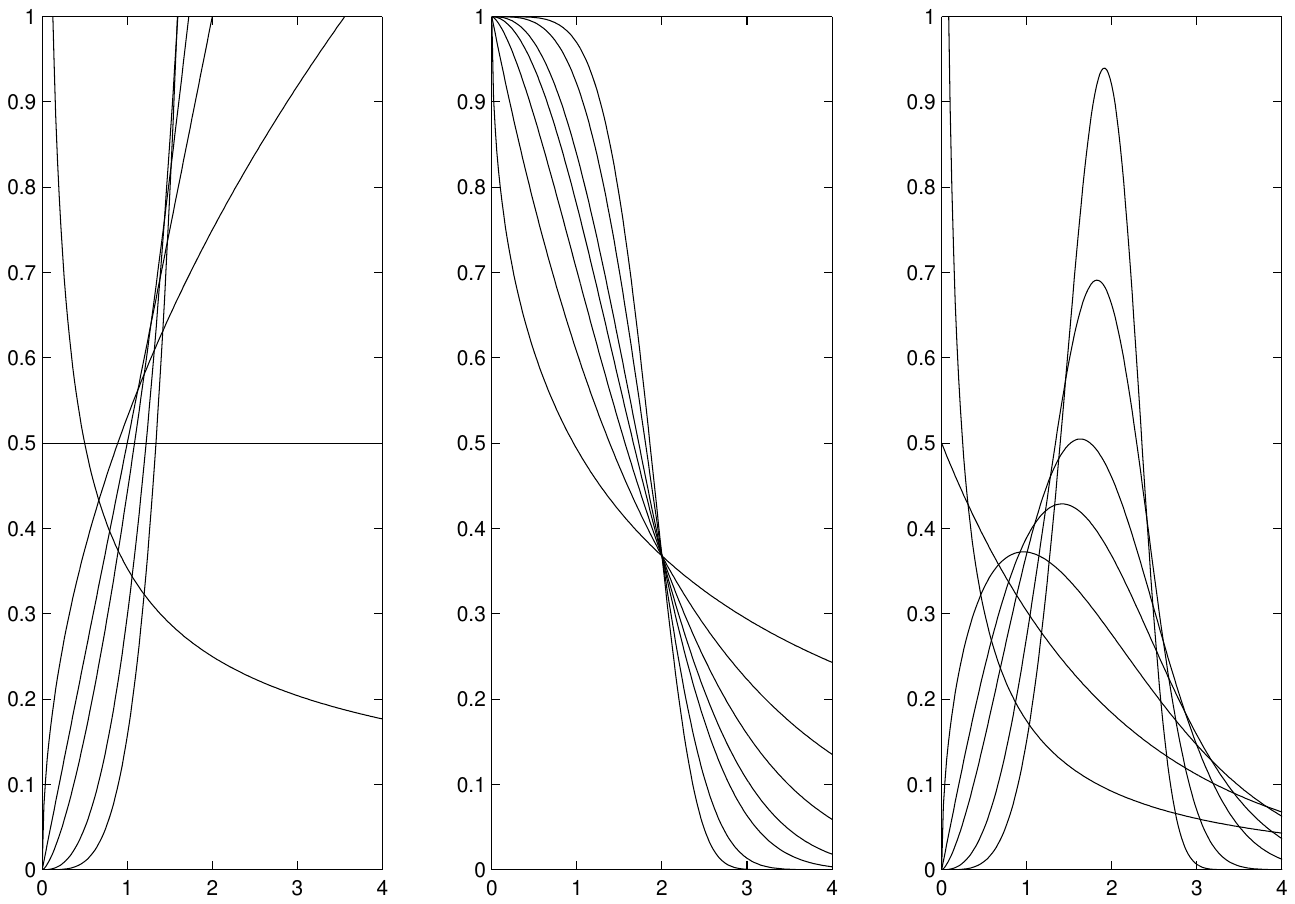}
   \mbox{}} 
  \vspace{-1.0cm} 
 \centerline{Figure 7: Shapes of the Weibull Distribution, $h$, $r$ and $f$.} 
 \centerline{Parameters: $\gamma=1$,\; $\beta=0.5, 1.0, 1.5, 2.0, 2.5, 3.6, 5.0$.} 
  \label{fig:bowtie}
 \end{figure}

 The affine transformation $x=x'+\alpha$ leads to the (three parameter)
Truncated Weibull distribution. A location (or threshold) parameter,
$\alpha>0$ represents beginning observation of a Truncated Weibull
variate at $t=0$, after it has already survived the period
$[-\alpha,0[$. For the sake of comparison, the reliability functions of
the (one-parameter) Exponential, and the two and three-parameter Weibull
distributions are  given next;  
 \[ 
   r_{E}(x \g \gamma) = \exp\left( -\left(\frac{x}{\gamma}\right) \right) 
   \ ; \ \   
   r_{W2}(x \g \beta, \gamma) = 
         \exp\left( -\left(\frac{x}{\gamma}\right)^\beta \right) 
   \ ; 
 \] 
 \[    
   r_{W3}(x \g \alpha, \beta, \gamma) = 
     \frac{1}{r_{W2}(\alpha \g \beta, \gamma)} 
   \exp\left( -\left(\frac{x+\alpha}{\gamma}\right)^\beta \right) 
   \ . 
 \] 

 In the example at hand we have three {\it nested models}, in which a
distribution with less parameters (or degrees of freedom) is a special
case (a sub-manifold in the parameter space) of a distribution with more
parameters (or degrees of freedom): The (one-parameter) Exponential
distribution is a special case of the (two-parameter) Weibull
distribution which, in turn, is a special case of the (three-parameter)
Truncated Weibull distribution. Nesting is one of the basic modes of
relating different statistical models. For examples of the FBST used for
model selection in nested models see Ironi at al. (2002), Lauretto et
al. (2003), Stern and Zacks (2002).  

 The (two-parameter) Weibull distribution has also an important
theoretical property: 
 Its functional form is invariant by serial composition. 
 If $n$ i.i.d. random variables have Weibull distribution, 
 $X_i \sim f(x\g \beta,\gamma)$,
 then the first failure is a Weibull variate with characteristic life
 $\gamma/n^{1/\beta}$, i.e. $X_{[1,n]} \sim f(x\g \beta,\gamma/n^{1/\beta})$. 
 This is a key property for its characterization as a stable
distribution, that is, for the characterization of the Weibull
distribution as an (asymptotic) eigensolution. For applications  
in the context of extreme value theory, see Barlow and Prochan (1981).

 While a series system fails when its first element fails, a parallel
system fails when its last element fails. Figure 8 gives the standard
graphical representation of series and parallel systems. This
representation is inspired in circuit theory: While in a serial system
the current flow is cut if a single element is cut, in a parallel system
the current flow is cut only if all elements are cut. Series / parallel
composition are the two basic modes used in Reliability Engineering for
structuring and analyzing complex systems. Some of the statistical
properties of these structures are captured in the form of algebraic
lattices, see Barlow and Prochan (1981) and Kaufmann et al. (1977). Some
of these properties are similar or analog to the compositional rules
analyzed in section A.4 and Borges and Stern (2007). The
characterization of the Gompertz as a limit distribution for parallel
systems is given in appendix H.3, following Gavrilov (1991).  

 \[
 \dgARROWPARTS=3 
 \dgHORIZPAD=0.0em 
 \dgVERTPAD=0.0ex  
 \begin{diagram}
 \node{} \arrow{e,-}  \node{\Ovalbox{1}} \arrow{e,-} \node{\Ovalbox{2}}    
   \arrow{e,-} \node{\Ovalbox{3}} \arrow{e,-} \\ 
 \node{}                 \node{}  \arrow{s,-} \arrow{e,-} 
     \node{\Ovalbox{1}} \arrow{e,-} \node{} \arrow{s,-} \\ 
 \node{}    \arrow{e,-}  \node{}  \arrow{s,-} \arrow{e,-}   
     \node{\Ovalbox{2}} \arrow{e,-} \node{} \arrow{s,-} \arrow{e,-} \\ 
 \node{}                 \node{}              \arrow{e,-} 
     \node{\Ovalbox{3}} \arrow{e,-} \node{}                 
  \end{diagram} 
 \] 
 \centerline{Figure 8: Series and Parallel Systems.}

 Imagine a situation in which a scientist receives a data bank of
observed individual lifetimes in a population. The scientist also knows
that all individuals in the population are of the same nature, that is,
the population is either entirely allopoietic, $H_1$, or autopoietic,
$H_2$. Since hypotheses $H_1$ and $H_2$ imply life distributions with
distinct functional forms, the scientist could use his/her observed life
data to decide which hypothesis is correct (or more adequate). This
situation is known in statistics as the problem of {\it separate
hypotheses}. The scientist could also be faced with a mixed population,
a situation in with a fraction $w_1$ of the individuals are allopoietic,
and a fraction $w_2$ of the individuals are autopoietic. In this
situation the scientist could use his/her observed data to infer the
fractions or {\it weights}, $w_1$ and $w_2$, in the {\it mixture model.}  

 For mixture models in the general, the p.d.f. of the data is a convex
linear combination of fixed candidate densities. Writting the model's
vector parameter as $\theta = [w, \psi_1,\ldots \psi_m],$
 \[ 
    f(x \g \theta) = w_1 f_1(x \g \psi_1) +\ldots + w_m f_m(x \g \psi_m) 
     \ , \ \  w \geq 0 \g w \uno = 1 \ , 
 \]
 and the model's likelihood function is
 \[ 
     f(X \g \theta) = 
      \pprod_{j=1}^{n} \ssum_{k=1}^{m} w_k f_k(x_j \g \psi_k) \ . 
 \]

 \begin{center} 
 \centerline
 {\includegraphics*[height=3.5in, width=7.0in, angle=0]{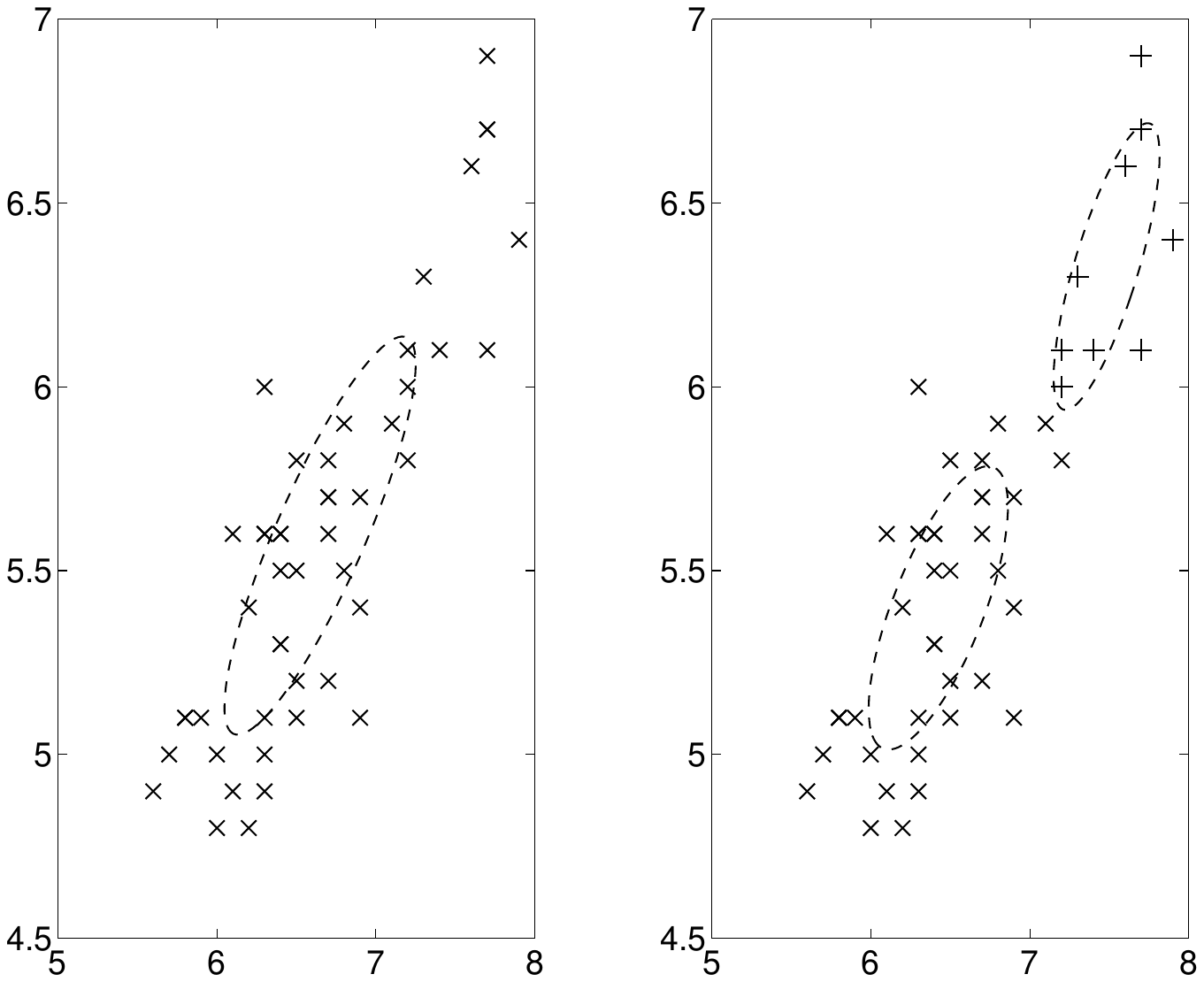}} 
 \centerline{Figure 9a,b: Mixture models with 1 and 2 
             bivariate-Normal components.} 
 \end{center} 

 In mixture analysis for unsupervised classification, we assume that the
data comes from two or more subpopulations (classes), distributed under
distinct densities. Statistical mixture models may also be able to infer
the classification probabilities for each data point, see Figure 9. In a
{\it heterogeneous mixture} model, the components in the mixture have
distinct functional forms. In a {\it homogeneous mixture} model, all
components in the mixture have the same functional form. For several
applications of these models, see Fraley (1999), Lauretto et al. (2006,
2007), Robert (1996) and Stephens (1997).

 \subsection{Complementary Models}

 According to Bohr, the word {\it Complementarity} is used    

 \begin{quotation} 
 {\it ``...to characterize the relationship between experiences obtained
by different experimental arrangements and visualizable only by mutually
exclusive ideas...''}. (N.Bohr II, Natural Philosophy and Human
Cultures, p.30)

 {\it ``Information regarding the behavior of an object obtained under
definite experimental conditions may, however, 
 ...be adequately characterized as complementary to any information
about the same object obtained by some other experimental arrangement
excluding the fulfillment of the first conditions. Although such kinds
of information cannot be combined into a single picture by means of
ordinary concepts, they represent indeed equally essential aspects of
any knowledge of the object in question which can be obtained in this
domain.''} (Bohr 1938, p.26).  
 \end{quotation} 

 In quantum Mechanics, at least from a historical perspective, the most
important complementarity relations are those implied by the
wave-particle complementarity or duality principle. We have mentioned
these complementarity relations in section 3.3, and we will examine them
again in sections 6 and 7. This principle states that microparticles
exhibit the properties of both particle and waves, even considering
that, in classical physics, these categories are mutually exclusive. At
the dawn of the XX century, physics had an assortment of phenomena that
could not be appropriately explained by classical physics. In order to
explain one of these phenomena, known as the photoelectric effect,
Albert Einstein postulated in 1905, annus mirabilis, a model in which
light, conceived in classical physics as electro-magnetic waves, should
also be seen as a rain of tiny particles, now called photons. Einstein
basic hypothesis was that a photon's energy is proportional to the
light's frequency, $E=h\nu$, where the proportionality constant, $h$, is
Planck's constant.

 In 1924, Louis de Broglie generalized Einstein's hypotheses. Using
Einstein's relativistic relation, $E=mc^2$, the photon's wavelength,
$\lambda=c/\nu$, can be written as $\lambda=h/(mc)$, where $m=E/c^2$ is
the effective mass attributed to the photon. A moving particle's moment
is defined as the product of its mass and velocity, $p=mv$. Hence, de
Broglie conjectured that any moving particle has associated to itself a
``pilot wave'' of wavelength $\lambda=h/p=h/(mv)$, see Broglie (1946,
ch.IV, Wave Mechanics) for the original argument. Just two years later,
in 1926, Erwin Sch\"{o}rdinger published the paper ``Quantization as an
Eigenvalue Problem'', further generalizing these ideas into his
(Sch\"{o}rdinger's) wave equation, the basis for a general theory of
Quantum Mechanics, see next section. The details of the early
developments of Quantum Mechanics can be found in Tomonaga (1962) and
Pais (1988, ch.12), but from this brief history it is clear that the 
general idea of complementarity was a cornerstone in the birth of modern
physics.

 Nevertheless Bohr believed that complementarity could be a useful
concept in many other areas. Folse (1985) gives an interesting essay
about Bohr's ideas on complementarity, including its application to
fields outside quantum mechanics. Possible examples of such applications
are given next:

 \begin{quotation} 
 {\it ``...the lesson with respect to the role which the tools of
observation play in defining the elementary physical concepts gives a
clue to the logical applications of notions like purposiveness foreign
to physics, but lending themselves so readily to the description of
organic phenomena. Indeed, on this background it is evident that the
attitudes termed mechanistic and finalistic do not present contradictory
views on biological problems, but rather stress the mutually exhaustive
observational conditions equally indispensable in our search for an ever
richer description of life.''} (Bohr II, Physical Science and Problems
of Life, p.100).  
 \end{quotation} 

 \begin{quotation} 
 {\it ``For describing our mental activity, we require, on one hand, an
objectively given content to be placed in opposition to a perceiving
subject, while, on the other hand, as is already implied in such an
assertion, no sharp separation between object and subject can be
maintained, since the perceiving subject also belongs to our mental
content. From these circumstances follows not only the relative meaning
of every concept, or rather of every word, the meaning depending upon
our arbitrary choice of view point, but also we must, in general, be
prepared to accept the fact that a complete elucidation of one and the
same object may require diverse points of view which defy a unique
description. Indeed, strictly speaking, the conscious analysis of any
concept stands in a relation of exclusion to its immediate application.
The necessity of taking recourse to a complementarity, or reciprocal,
mode of description is perhaps most familiar to us from psychological
problems. In opposition to this, the feature which characterizes the
so-called exact sciences is, in general, the attempt to attain to
uniqueness by avoiding all reference to the perceiving subject. This
endeavor is found most consciously, perhaps, in the mathematical
symbolism which sets up for our contemplation an ideal of objectivity to
the attainment of which scarcely any limits are set, so long as we
remain within a self-contained field of applied logic. In the natural
sciences proper, however, there can be no question of a strictly
self-contained field of application of the logical principles, since we
must continually count on the appearance of new facts, the inclusion of
which within the compass of our earlier experience may require a
revision of our fundamental concepts.} (Bohr I, The Quantum of Action,
p.96-97).  
 \end{quotation} 

 Examining some basic concepts of quantum mechanics, L.V.Tarasov (1980,
p.153) poses a question concerning the concept of complementarity that
is very pertinent in our context:
 \begin{quotation} 
 {\it ``A microparticle is neither a corpuscle, nor a wave, but still we
employ both these images, which mutually exclude each other, for
describing a microparticle. ... Naturally, this could give rise to a
ticklish question: Doesn't this mean an alienation of the image from the
object, which is fraught with a transition to the position of
subjectivism? A negative answer to this question is given by the
principle of complementarity itself. From the position of this
principle, pictures mutually excluding one another are used as mutually
complementary pictures, adequately representing various sides of the
objective reality called the  microparticle.''}      
 \end{quotation} 
 Even considering that Tarasov makes his point from a very different
epistemological perspective, his statement fits admirably well into our
constructivist framework. Within it the objectivity of a complementarity
model can be interpreted as follows: Although complementary, the several
views employed to describe an object should still render objective
(epistemic) eigensolutions. As always, the whole model will be
considered as objective as these well characterized eigensolutions, that
is, sharp, stable, separable and composable, as examined in detail in
section 3.5. Of course, the compositionality rules for a given theory or
model must be given by an appropriate formalism. Such a formalism must
include a full specification of compatibility / incompatibility rules
for axioms or statements in the theory or model. For an example of this
kind of formalism, see Costa and Krause (2004).

\section{Varieties of Probability}
\markboth{CHAPTER 5: MODULARITY AND STOCHASTIC EVOLUTION}
  {5.6 \ VARIETIES OF PROBABILITIES}

 This section presents some basic ideas of Quantum Mechanics, providing
simple heuristic derivations for a few of its basic principles. Its main
objective is to discuss the impact of Quantum Mechanics on the concept
and interpretation of probability models.

 \subsection{Heisenberg's Uncertainty Principle}

 In this section we present Werner Heisenberg's uncertainty principle,
derived directly from de Broglie's wave-particle complementarity principle.

 A particle with a precise moment, $p$, has associated to it a pilot
wave that is monochromatic, that is, has a single wavelength, $\lambda$.
Hence, this wave is homogeneously distributed in space. Let us think of
a particle with an uncertain moment, specified by a probability
distribution, $\phi(p)$. What would the distribution, $\psi(x)$, of the
location of its associated pilot wave, be? Assuming that the composition
rule for pilot waves is the standard linear superposition principle, see
Section 4.2, the answer to this question is given by the mathematics of
Fourier series and transforms, see Butkov (1968, ch.4 and 7), 
Byron and Fuller (1969, ch.4 and 5) or Sadun (2001, ch.8 and 10).  

 The Fourier synthesis of a function, $f(x)$, in the interval $[0,L]$ is
given by the Fourier series
 \[ 
    f(x)= \frac{a_0}{2} +\ssum_{n=1}^\infty \left( 
      a_n \cos\left(\frac{n2\pi x}{L} \right) 
     +b_n \sin\left(\frac{n2\pi x}{L} \right) \right) \ \ , 
  \]
 \[
   a_n= \frac{2}{L} \int_{0}^{L} f(x) \cos\left(\frac{n2\pi x}{L} \right) dx  
   \ \ , \ \ \  
   b_n= \frac{2}{L} \int_{0}^{L} f(x) \sin\left(\frac{n2\pi x}{L} \right) dx  
   \ \ .  
 \]

 The following examples give the Fourier series for the rectangular 
 and triangular spike functions, $R_{2h}(x)$ and $T_{2h}(x)$. 
 In order to obtain simpler expressions, the spikes are 
 presented at the center of interval $[-\pi,+\pi]$, the standard 
 interval of length $L=2\pi$ shifted to be centered at the origin.
 Figure 10a displays the first 5 even harmonics, $\cos(nx)$, for wave
number $n=1\ldots 5$,  Figure 10b displays
the Fourier coefficients, $a_n$, in the synthesis of the triangular
spike $T_{2h}(x)$, for $h=1.0$. 
  Figures 10c and 10d display the triangular spike and its Fourier
syntheses with the first 2 and the first 5 harmonics.

 \begin{figure}[h] 
  \centerline{\includegraphics*[angle=0, width=6.5in]{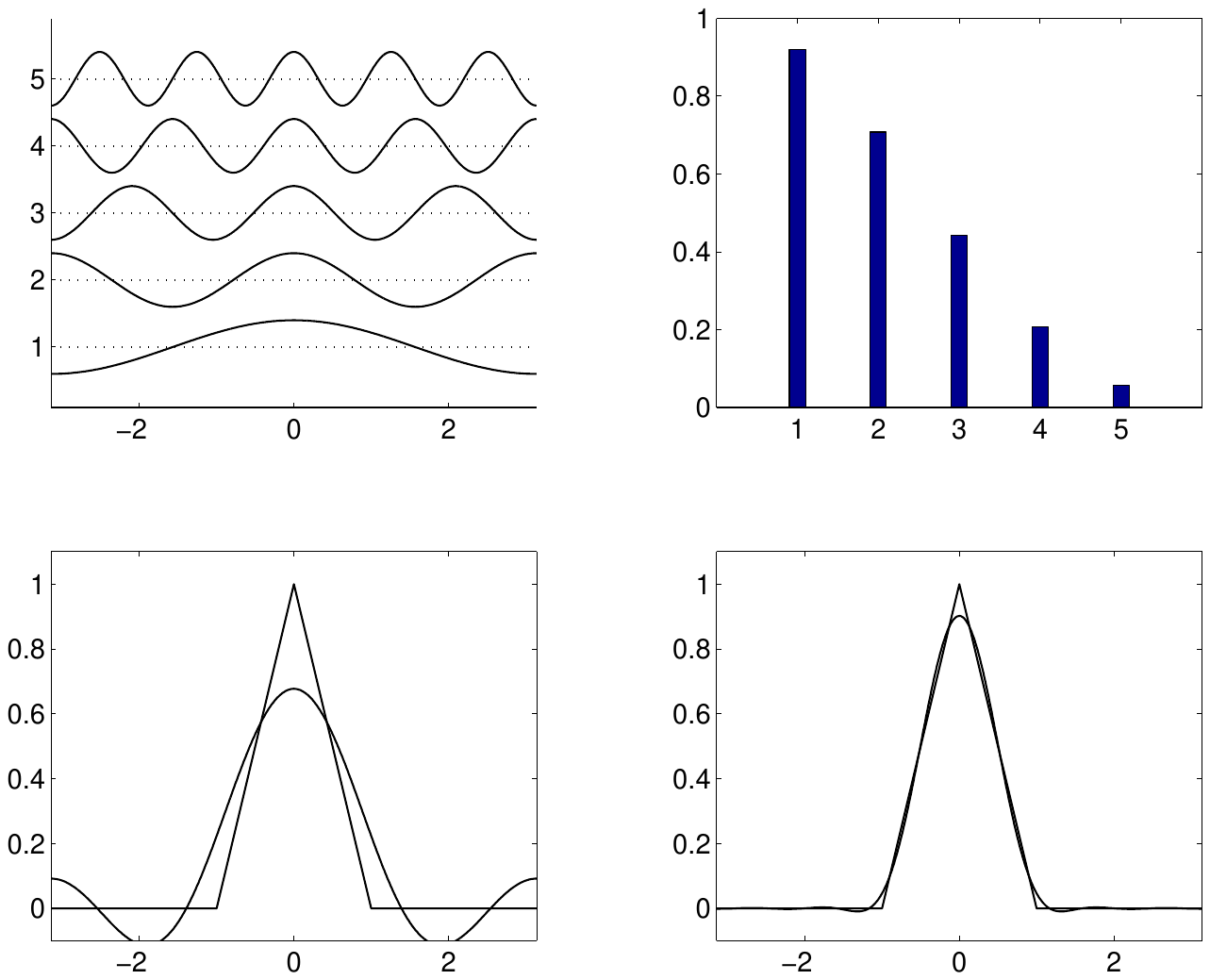}} 
  \vspace{-0.5cm} 
 \centerline{Figure 10: Monochromatic Waves and Superposition Packets.} 
 \end{figure}

 \[
     R_{2h}(x)= \left\{ \begin{array}{l} 
       1 \ , \ \ \mbox{if} \   -h < x <  +h  , \\ 
       0 \ , \ \ \mbox{otherwise in} \ \  [-\pi,\pi] .          
       \end{array} \right. 
       \ \ = \ \    
      \frac{2h}{\pi} +\frac{2}{\pi}\ssum_{n=1}^{\infty} 
        \frac{\sin(nh)}{n} \cos(nx) . 
 \]
 \[
     T_{2h}(x)= \left\{ \begin{array}{l} 
       1-|x|/h \ , \ \ \mbox{if} \   |x|<h , \\ 
       0 \ , \ \ \mbox{otherwise in} \ \  [-\pi,\pi] .          
       \end{array} \right. 
       \ \  = \ \   
       \frac{h}{2\pi}  
        +\frac{4h}{\pi}\ssum_{n=1}^{\infty} \left( 
        \frac{\sin(nh)}{nh} \right)^2 \cos(nx) . 
 \]

  It is also possible to express the Fourier series in complex form.
Using the complex exponential notation, $\exp(ix)= \cos(x) +i\sin(x)$,
we write
 \[ 
    f(x)= \ssum_{-\infty}^{+\infty} c_n e^{in2\pi x/L} 
    \ \ , \ \ \ 
   c_n = \frac{1}{L} \int_{0}^{L} f(x) e^{-in2\pi x/L} dx 
    \ \ .   
 \] 
 The trigonometric and complex exponential Fourier coefficients are
related as follows
 \[ 
   c_0= \frac{1}{2}a_0 \ \ , \ \ \ 
   c_n= \frac{1}{2} ( a_n -ib_n ) \ \ , \ \ \ 
   c_{-n}= \frac{1}{2}( a_{n} +ib_{n} ) \ \ , \ \ \ 
   n= 1,\ldots \infty \ \ . 
 \]  
 The complex form is more symmetric and elegant. In particular, the
orthogonality relations,
 \[ 
   \int_{0}^{L} e^{in2\pi x/L} e^{-im2\pi x/L} dx = 
   \int_{0}^{L} e^{i(n-m)2\pi x/L} dx = 
   \left\{ \begin{array}{ll} 
   L \ , & \mbox{if} \  n=m \ , \\ 
   0  \ , & \mbox{if} \ n\neq m \ . 
   \end{array} \right.  \ \ ,  
 \] 
 are the key for interpreting the set of complex exponentials, $\{
e^{in2\pi x/L} \}$, for wave numbers $n= -\infty \ldots +\infty$, as an
orthogonal basis for the appropriate functional vector space in the
interval $[0,L]$. 

  If we want to synthetize functions in the entire real line, not just
in a finite interval, we must replace Fourier series by Fourier
transforms. The Fourier transform, $\overha{f}(k)$, of a function,
$f(x)$, and its inverse transform are defined, respectively, by
 \[ 
   \overha{f}(k)= \frac{1}{\sqrt{2\pi}} \int_{-\infty}^{\infty} 
         f(x) \exp(-ikx) dx \ \ \mbox{and} \ \   
   f(x)= \frac{1}{\sqrt{2\pi}} \int_{-\infty}^{\infty} 
         \overha{f}(k) \exp(ikx) dk \ .    
 \] 

 In the Fourier transform the propagation number (or angular frequency),
$k=n2\pi /L$, replaces the wave number, $n$, used in the Fourier series.
The new normalization constants are defined to stress the duality
between the complementary representations of the function in state and
frequency spaces, $x$ and $k$.  
 
 As an important example, let us compute the Fourier transform, of a
Gaussian distribution with mean $\mu=0$ and standard deviation (uncertainty) 
$\sigma_x=\sigma$:
 \[ 
    f(x)= \frac{1}{\sigma \sqrt{2\pi}}   
    \exp\left( \frac{-x^2}{2\sigma^2} \right) \ , \ \   
    \overha{f}(k)= \frac{\sigma}{\sqrt{2\pi}}  
    \exp\left( \frac{-\sigma^2 k^2}{2} \right) \ .    
 \] 
 This computation can be checked using the analytic formula of the
Gaussian integral,
 \[ 
      \int_{-\infty}^{\infty} \exp\left( -ax^2 +bx +c \right) dx  
    = \sqrt{\frac{\pi}{a}} \exp\left( \frac{b^2-4ac}{4a} \right)
    \ .   
 \] 

 \begin{figure}[h] 
  \centerline{\includegraphics*[angle=0, width=5.0in, 
               height=2.5in]{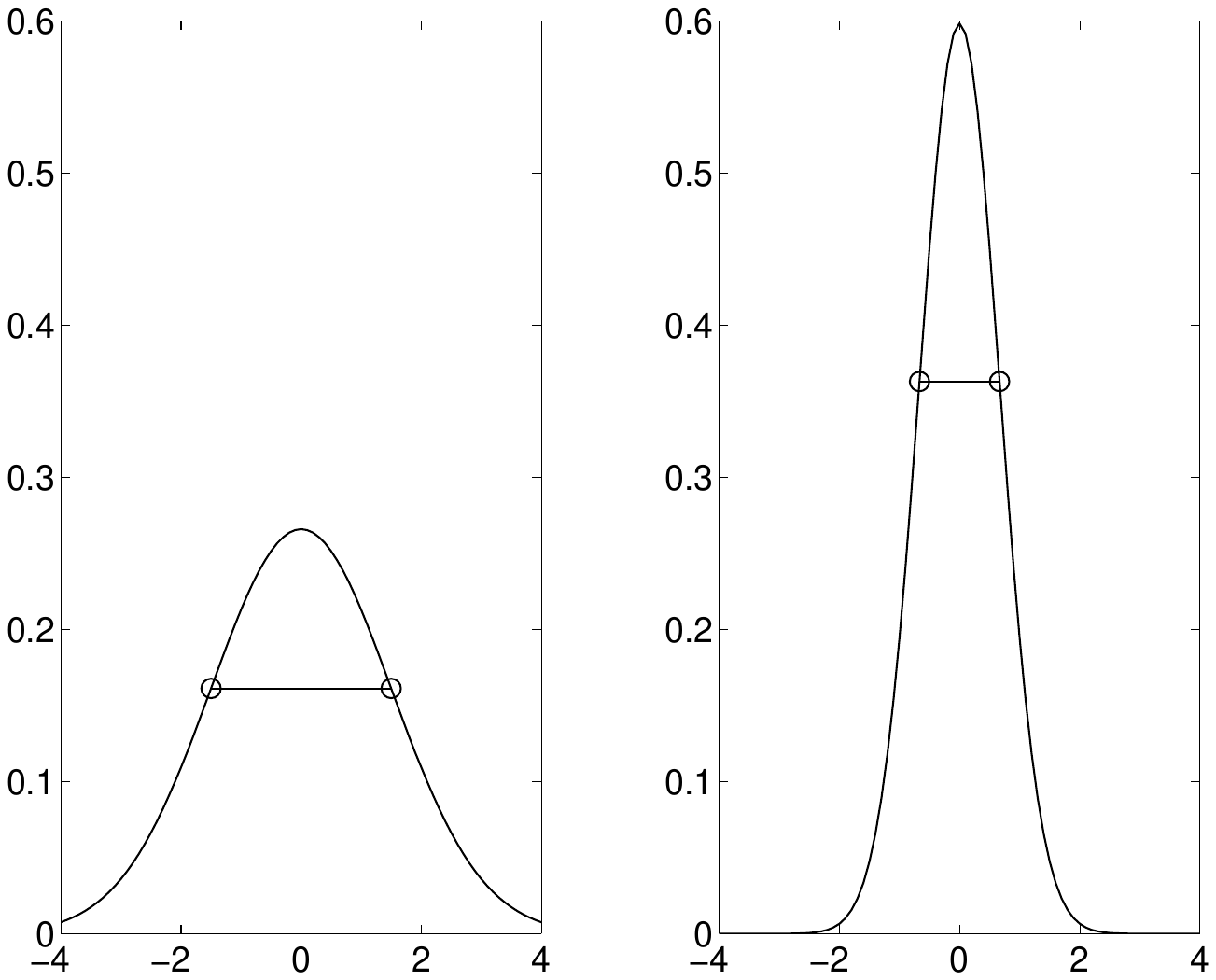}} 
  \vspace{-0.5cm} 
 \centerline{Figure 11: Uncertainty Relation for Fourier Conjugates.} 
 \end{figure}

 Hence, the Fourier transform of a Gaussian distribution with standard
deviation $\sigma_x=\sigma$, is again a Gaussian distribution, but
with standard deviation $\sigma_k=1/\sigma$, that is, $\sigma_x\;
\sigma_k =1$. Figure 11 displays the case $\sigma=1.5$. It is also
possible to show that this example is a best case, in the sense that,
for any other function, $f(x)$, the standard deviations of the
conjugate functions, $f(x)$ and $\overha{f}(k)$, obey the inequality of
the uncertainty principle, $\sigma_x\; \sigma_k \geq 1$, see Sadun 
(2001, sec.10.5).

 In the context of Quantum Mechanics, the best known instance of the
uncertainty principle gives a lower bound on the product of the standard
deviations of the position and momentum of a particle,
 \[ 
    \sigma_x \; \sigma_p \geq \frac{\hbar}{2} \ , \ \ 
    \hbar=\frac{h}{2\pi} \ , \ \ 
    h= 6.62606896(33)E-34 J s 
    \ \ .  
 \] 
 Heisenberg's bound is written as a function of the moment, $p$, instead
of the frequency, $k$; this is why in the right hand side of the
inequality we have half the reduced Planck's constant, $\hbar/2$,
instead of $1$, as in Fourier transform conjugate functions.

 Planck's constant dimension is that of action, an energy-time product,
like joule-second or electron-volt-second. The values above present the
best current (2006) estimates for this fundamental physical constants,
in the format recommended by the Committee on Data for Science and
Technology, CODATA. The two digits in parentheses denote the standard
deviation of the last two significant digits of the constant's value.
The importance of this constant and its representation are further
analyzed in the next sections.

 \subsection{Schr\"{o}dinger's Wave Equation} 

 In the last sections we have analyzed de Broglie's complementarity
principle, which states that any moving particle has associated to
itself a ``pilot wave'' of wavelength  $\lambda=h/mv$. In section 4.2 we
analyzed some of the basic properties of the classical wave equation,
displayed below on the left hand side:   
 \[
    \frac{d^2 \psi}{dx^2} +\omega^2 \; \psi =0 \ \ , \ \ \    
    \omega^2 = \frac{2m}{\hbar} \left(E-V(x)\right) \ .    
 \]
   
 In the classical equation, $\omega = 2\pi/\lambda$ is the wave's
angular frequency. What should a quantum wave equation equation look
like? Schr\"{o}dinger's idea was to replace the classical wavelength by
de Broglie's, that is, to use  $\omega=2\pi mv/h$. Using the definition
of the kinetic energy of a particle, $T=(1/2)mv^2$, and its relation to
$V(x)$ and $E$, the particle's  potential and total energy, $T=E-V(x)$,
we find the expression for $\omega^2$ displayed above on the right.

 This is Schr\"{o}dinger's (time independent) wave equation, which
established a firm basis for the development of Quantum Mechanics, also
known in its early days as ``wave mechanics''. One of the immediate
successes of Quantum Mechanics was to provide elegant explanations,
based of physical first principles, to many known empirical facts of
chemistry, like the properties of the periodic table, molecular
geometry, etc. 
 Among the books providing accessible introductions to QM we mention:
 the very nice elementary text by Enge (1972),  the concise introduction
by Landshoff (1998), McGervey (1995) which  focus on wave mechanics,
and Heitler (1956) which focus on quantum chemistry. 

 Quantum Mechanics was also the basis for the development of completely
new technologies. Among the most distinguished examples are solid-state
or condensed matter electronic devices such as transistors, integrated
circuits, lasers, liquid crystals, etc.. These devices constitute, in
turn, the basic components of modern digital computers. Finally, one can
argue that computer based information processing tools are among the
most revolutionary technologies introduced in human society, having had
an impact in its organization comparable only to a handful of other
technologies (perhaps the steam and internal combustion engines, or
electric power), see XX (20xx).
   
 Nevertheless, all this success was not for free. Quantum Mechanics
required the re-thinking and re-interpretation of some of the most
fundamental concepts of science. In this and the next sections we
analyze the impact of Quantum Mechanics on the most important concept of
statistical science, namely, probability.    

 Although Scr\"{o}dinger arrived at the appropriate functional form of a
wave equation for Quantum Mechanics, the adequate interpretation for the
wave function, $\psi$, was given only a few months later by Max Born.
According to Born's interpretation: The probability density of
``finding'' the particle at position $x$, is proportional to the square
of the wave function absolute amplitude, $|\psi(x)|^2$. Since, in the
general case, $\psi$ is a complex function, the last quantity can also
be written as the product of the wave function by its complex conjugate,
that is, $|\psi(x)|^2=\psi^* \psi$.  
        
 From this interpretation of the wave function, we can understand Max
Born's formulation of `the core metaphor of wave mechanics', as quoted
in Pais (1988, ch.12, sec.d, p.258), 
 \begin{quotation} 
 {\it ``The essence of wave mechanics: `The motion of particles follows
probability laws but the probability itself propagates according to the
law of causality.''} 
 \end{quotation} 

  This is a revolutionary interpretation, that attributes to the concept
of probability a new and distinct `objective' character. Hence, it is
interesting to have some insight on the genesis of Born's
interpretation. Born's own recollections are presented at Pais (1988,
ch.12, sec.d, p.258-259):
 \begin{quotation}
  {\it ``What made Born take his step?  

  In 1954 Born was awarded the Nobel Prize `for his fundamental
research, specially for his statistical interpretation of the wave
function'. In his acceptance speech Born, then in his seventies,
ascribed his inspiration for the statistical interpretation to `an idea
of Einstein's [who] had tried to make the duality of particles -
light-quanta or photons - and waves comprehensible by interpreting the
square of the optical wave amplitudes as probability density for the
occurrence of photons. This concept could at once be carried over to the
$\psi$-function: $\mbox{}\; |\psi|^2$ ought to represent the probability
density of electrons.'\ ''}      
 \end{quotation}

 \subsection{Classic and Quantum Probability}

 One of the favorite metaphors used by the orthodox Bayesian school
describes the scientist's work as a game against nature, with the
objective of scoring a good guess on ``nature's true state''. 
 Implicit in this metaphor is the assumption that such a ``true state of
nature'' exists and is, at least in principle, accessible. 
 In this paradigm, omniscience is usually a matter of money, that is,
with enough economic resources all pertinent information can, at least
in principle, be acquired, see Blackwell and Girshick (1954), for
example. 

 \begin{quotation}  
 {\it ``Statistics {\em can} be viewed as a game against nature.'' 
 (p.75).  

 ``...games where one of the players is not faced with an intelligent
opponent but rather with an unknown state of nature.'' (p.121).   

 ``The same theory that served to delineate optimal strategies in games
played against an intelligent opponent will serve to delineate classes
of optimal strategies in games played against nature.'' (p.123). 

 ``What prevents the statistician from getting full knowledge of
$\omega$ [the state of nature] by unlimited experimentation is the cost
of experiments.'' (p.78). }
 \end{quotation} 

 This paradigm seems incompatible with, or at least very unfriendly to,
Born's probabilistic interpretation of Quantum Mechanics and
Heisenberg's uncertainty principle. We believe that, in the context of
quantum mechanics, the strictly subjective interpretation of probability
is, please forgive the pun, a very risky metaphor, and that pushing this
metaphor where it does not belong will lead to endless paradoxes. In
Chapter 7 of his book, {\it The Physics of Chance}, for example, Charles
Ruhla presents the  adventures of the simple-minded hero Monsieur de La
Palice, struggling to understand some basic quantum experiments.   

 For a strict subjectivist the situation is even worse, and the use of 
Quantum Mechanics is at risk of being considered illegal.  
 A statement giving the current best estimate of $h$ (Planck's constant) 
toghether with its standard deviation was presented in section 5.6.1. 
 Since $h$ appears at the right hand side of Heisenberg's uncertainty 
principle, an uncertainty about the value of $h$ implies a second order 
uncertainty. 
 The propagation of the uncertainty about the value of fundamental physical 
constants generates similar second order probabilistic statements about 
the detection, mesurement or observation of quantum phenomena.    
 For example, section 5.7.2 arrives at statements giving the 
(probabilistic) uncertainty of (probabilistic) transition rates. 
 All these are  prototypical examples of statements that are 
categorically forbidden in orthodox Bayesian statistics, as
bombastically proclaimed in the following quotations from Finetti (1977,
p.1,5 and 1972, p.190), see also Mosleh and Bier (1996) and Wechsler et
al. (2005).
  \begin{quotation}
   {\it ``Does it make sense to ask what is the probability that the
probability of a given event has a given value, $p_i$ ? ... It makes
{\em no} sense to state that the probability of an event $E$ is to be
regarded as {\em unknown} in that its {\em true} value is one of the
$p_i$'s, but we do not know which one.''   
    
   ``Speaking of unknown probabilities [or of probability of a
probability] must be forbidden as meaningless.''}  
 \end{quotation}
 
 A similar statement of de Finetti was analyzed in section 4.7. Such an
awkward position, at least for a modern physicist, was seen by the
founding fathers of orthodox Bayesian statistics as an unavoidable
consequence of the subjectivist doctrine, according to which,
 \begin{quotation} 
  {\it ``Probabilities are states of mind, not of nature.''} Savage
(1981, p.674). 
 \end{quotation} 

 From a constructivist perspective, fundamental physical constants, 
including of course Planck's constant, correspond to very objective 
(very sharp, stable, separable and composable) eigenvalues 
of Physics' research program, and it is perfectly
admissible to speak about the uncertainty of their estimated values. 
 Of course that is what physicists need to do, and have done for almost
a century, regardless of being disapproved by the Bayesian orthodoxy
(theoretically coherent, but understandably very shy and timid). There
have also been some attempts to reconcile a strict subjectivist position
with modern physics, through long and sophisticated translations of
simple ``crude'' statements like the ones quoted above. Some of these
translations are as bizarre and / or intricately involved as similar
attempts to translate epistemic probabilistic statements that are
categorically forbidden in frequentist statistics into ``acceptable'' 
frequentist probabilistic statements, see section 2.5 and  Rouanet et
al. (1998, Preamble). Richard Feynman (2002, p.14), makes the following
comments on some ideas behind some of such interpretations: 
 \begin{quotation}  
 {\it ``Now, the philosophical question before us is, when we make an
observation of our track in the past, does the result of our observation
become real in the same sense that the final state would be defined if
an outside observer were to make the observation? This is al very
confusing, especially when we consider that even though we may
consistently consider ourselves always to be the outside observer when
we look at the rest of the world, the rest of the world is at the same
time observing us, and that often we agree on what we see in each other.
Does this mean that my observations become real only when I observe an
observer observing something as it happens? This is an horrible
viewpoint. Do you seriously entertain the thought that without observer
there is no reality? Which observer? Any observer? Is a fly an observer?
Is a star an observer? Was there no reality before 109 B.C. before life
began? Or are you the observer? Then there is no reality to the world
after you are dead? I know a number of otherwise respectable physicists
who have bought life insurance. By what philosophy will the universe
without man be understood?

 In order to make some sense here, we must keep an open mind about the
possibility that for sufficiently complex systems, amplitudes become
probabilities....''}   
 \end{quotation} 

 In order to provide deeper insight on the meaning of Heisenberg's
uncertainty principle, let us link it to Noether's theorems, already
discussed in section 2.8.1. The central point of Noether's theorems lies
in the existence of an invariant physical quantity for each continuous
symmetry group in a physical theory. Heisenberg's uncertainty relation,
presented in section 6.1, sets a bound on the accuracy with which we
can access, by means of physical measurements, such symmetry / invariant
dual or conjugate pairs. This point is further analyzed by Bohr:  
 \begin{quotation} 
 {\it``...we admire Planck's happy intuition in coining the term
`quantum of action' which directly indicates a renunciation of the
action principle, the central position of which in the classical
description of nature he himself has emphasized on more than one
occasion. This principle symbolizes, as it were, the peculiar reciprocal
symmetry relation between the space-time description and the laws of
conservation of energy and momentum, the great fruitfulness of which,
already in classical physics, depends upon the fact that one may
extensively apply them without following the course of the phenomena in
space and time.''} (p.94 or 210).

 {\it ``Indeed, the inevitability of using, for atomic phenomena, a mode
of description which is fundamentally statistical arises from a closer
investigation of the information which we are able to obtain by direct
measurement of these phenomena and the meaning we may ascribe, in this
connection, to the application of the fundamental physical concepts...
 
 Such considerations lead immediately to the reciprocal uncertainty
relations set up by Heisenberg and applied by him as the basis of a
thorough investigation of the logical consistency of quantum
mechanics.''} (p.113-114 or 247-248).
 \end{quotation} 

 In the article {\it Space-Time Continuity and Atomic Physics}, Bohr
(1935, p.370) further explores the relation between quantization and our
use of probabilistic language:   
 \begin{quotation} 
 {\it ``With the forgoing analysis we have described the new point of
view brought forward by the quantum theory. Sometimes one has described
it as leaving aside the idea of causality. I think we should rather say
that in the quantum theory we try to express some laws of nature that
lie so deep that they can not be visualized, or, which cannot be
accounted for by the usual description in terms of motion. This state of
affairs brings about the fact that we must use to a great extent
statistical methods and speak of nature making choices between
possibilities.''} 
 \end{quotation} 

 The correct interpretation of probability has been one of the key
conceptual problems of modern physics. The importance of this problem
can be further appreciated in the following statement of Paul Dirac,
found in (Pais 1986, p.255), regarding the early development of quantum
mechanics:  
 \begin{quotation}
 {\it ``This problem of getting the interpretation proved to be rather
more difficult than just working out the equations.'' } 
 \end{quotation}

 The ``correct'' interpretation or ``best'' metaphysics for quantum  
mechanics, including the ontological and epistemological status of
probability and the understanding of its role  in the theory, is an area
of strong academic interest and current research, see for example Albert
(1993, ch.7) for an exposition of David Bohm's interpretation of QM.  
 Richard Feynman's path integral formalism, see for example 
Feynman and Hibbs (1965), Honerkamp (1993) and Wiegel (1986), 
makes it possible to support other alternative interpretations.   

 Perhaps the most important lesson to be learned from this section is
that one must be aware of the several possible meanings and
interpretations of the concept of probability, and that distinct
situations may require or benefit from distinct approaches. In the best
spirit of complementarity, we should even consider the possibility of
studying the same situation under different perspectives, each one of
them providing a positive and irreplaceable contribution to our
understanding of a whole that is beyond the grasp of a single picture   
  \footnote{   
    The following quote was brought to my attention by Jean-Yves Beziau: 
   ``The ordinary man has always been sane because 
    the ordinary man has always been a mystic... 
    He has always cared for truth more than for consistency. 
    If he saw two truths that seemed to contradict each other, 
    he would take the two truths and the contradiction along with them.''
    Gilbert Keith Chesterton (1874 - 1936).  
   }. 

 \section{Theories of Evolution}
 \markboth{CHAPTER 5: MODULARITY AND STOCHASTIC EVOLUTION}
  {5.7 \ THEORIES OF EVOLUTION}
 
 The objective of this section is to highlight the importance of three
key concepts that are essential to modern theories explaining the
evolution of complex systems, and to follow some points in their
development and interconnection, namely: (1) the systemic view; (2)
modularity; and (3) stochastic evolution and/or probabilistic causation.
Probabilistic causation is by far the most troublesome of these
concepts. It is absolutely essential, at least in the framework
presented in this chapter, to the evolution of complex systems, on one
hand, but it was not easy for stochastic evolution to make its way as a
``legitimate'' concept in modern science, on the other. We believe that
the historical progress and acceptance of the ontological status of
these probabilistic concepts is closely related to the evolution of
epistemological frameworks that can, in turn, strongly influence and be
influenced by the corresponding statistical theories giving them
operational support.

 \subsection{Systemic View and Probabilistic Causation} 

 The systemic view has always been part of the biological thinking. The
teleomechanics school gave particular importance to a systemic view of
living organisms, see Lenoir (1989) for an excellent historical account.
As quoted in Lenoir (1989, p.220,221), for example, the XVIII century
biologist C. Reichert states:
 \begin{quotation} 
 {\it ``...`we have a systemic product before us,... in which the
intimate interconnections of the constituent parts have reached their
highest degree. When we think about a system, we normally picture
ourselves precisely this form of systematic product. Concerning  such
systems Kant said that the parts only exist with reference to the whole
and the whole, on the other hand, only appears to exist for the sake of
the parts.'  
 
  In order to investigate the systematic character of biological
organisms Reichert reminded the readers that it was necessary to have a
method appropriate to the subject...  Reichert could envision only one
method to the investigation of the living organism which avoids
disrupting the intimate interconnections of its parts: 

 `The systematist is aware both that he proceeds genetically and that he
must proceed genetically. He is aware that the structure on an organism
consists in the systematic division or dissection of the germ, which
receives a particular systematic unity through inheritance, makes it
explicit through development and transmits it further through
procreation.' \ ''}  
 \end{quotation}

 These statements express one of the core methodological doctrines of
the teleomechanics school, namely, that to understand the systemic
character of the organism, one must examine its development. The
systemic approach of the teleomechanics school greatly contributed to
the study of many fields in ``Biology'' (a word coined within this
school), facilitating complex analyses and multiscale interconnections.
C.F.Kielmeyer, another great representative of the teleomechanics
school, for example, linked individual and populational developments in
his celebrated biogenic, parallelism, or recapitulation 
principle of Embryology:
 \begin{quotation} 
 {\it ``Ontogeny recapitulates phylogeny.''} 
 \end{quotation}
 The teleomechanics research program, however, could never overcome
(perceived) incompatibility conflicts among some of its basic
principles, such as, for example, the conflict between the teleological
organization of organic systems, on one hand, and the need to use only
scientifically accepted forms of causal explanation, on the other. 
 Consequently the scientists in this program found themselves struggling
between deterministic reductionist mechanisms and vitalistic explanations, 
both unable to offer significant scientific knowledge or acceptable 
understanding for the phenomena in study.

 According to the framework for evolution presented in this chapter, the
diagnostic for this failure is quite obvious, namely, the lack of key
conceptual probabilistic ingredients. This situation is analyzed in
Lenoir (1989, p.239-241):
 \begin{quotation} 
 {\it ``Only in a universe operating according to probabilistic laws, a
universe grounded in non-deterministic causal processes, is it possible
to harmonize the evolution of sequences of more highly organized beings
with the principles of mechanics. 

 Two paths lay open for providing a consistent and rigorous solution to
this dilemma. One alternative is that of twentieth century science. It
is simply to abandon the classical notion of cause in favor of a
non-deterministic conception of causality. In the late nineteenth
century this was not an acceptable strategy. To be sure statistical
methods were being introduced into physics with great success, but prior
to the quantum  revolution in mechanics no one was prepared to assert
the probabilistic nature of physical causes.... 
 
 A second solution to this dilemma is that proposed by teleomechanists.
According to this interpretation rigidly determined causality can be
retained, but then limits must be placed on the analysis of the ultimate
origins of biological organization, and certain ground states of
purposive or zweckm\"{a}ssig organization must be introduced.   
 
 In the final analysis the only resolution of their impasse was the
construction of an entirely new set of conceptual foundations for both
the biological and the physical sciences which could cut the Gordian
knot of chance and necessity.''}  
 \end{quotation}

 The breakthrough of introducing stochastic dynamics in modern theories
of evolution is perhaps the greatest merit of Charles Darwin. According
to Peirce (1893, 183-184):
 \begin{quotation} 
 {\it ``(In) {\em The origin of Species} published toward the end of
1859... the idea that chance begets order, which is one of the
cornerstones of modern physics... was at that time put into its clearest
light.''} 
 \end{quotation} 

 The role of probability in Darwin's theories can be best appreciated in
his own words:
 \begin{quotation} 
 {\it ``Throughout this chapter and elsewhere I have spoken of selection
as the paramount power, yet its action absolutely depends on what we in
our ignorance call spontaneous or accidental variability. Let an
architect be compelled to build an edifice with uncut stones, fallen
from a precipice. The shape of each fragment may be called accidental;
yet the shape of each has been determined by the force of gravity, the
nature of the rock, and the slope of the precipice, -events and
circumstances, all of which depend on natural laws ; but there is no
relation between these laws and the purpose for which each fragment is
used by the builder. In the same manner the variations of each creature
are determined by fixed and immutable laws; but these bear no relation
to the living structure which is slowly built up through the power of
selection, whether this be natural or artificial selection.

  If our architect succeeded in rearing a noble edifice, using the rough
wedge-shaped fragments for the arches, the longer stones for the
lintels, and so forth, we should admire his skill even in a higher
degree than if he had used stones shaped for the purpose. So it is with
selection, whether applied by man or by nature; for although variability
is indispensably necessary, yet, when we look at some highly complex and
excellently adapted organism, variability sinks to a quite subordinate
position in importance in comparison with selection, in the same manner
as the shape of each fragment used by our supposed architect is
unimportant in comparison with his skill.''} Darwin (1887, ch.XXI,
p.236)
 \end{quotation}

  In the above passage, the importance given to the systemic view, that
is, to the  {\it living structure of the organism} is evident. At the
same time, randomness is added as an essential provider of raw materials
in the evolutionary process. However, there are some important points of
divergence between the way randomness plays a role in Darwinian
evolution, and in contemporary theories. We highlight three of them: (1)
Darwin uses only pseudo-randomness; (2) Genetic and somatic components
of variation are not clearly distinguished; (3) Darwinian variations are
continuous. Let us examine these three points more carefully: 

 1- Darwin used pseudo-randomness, not essential uncertainty. S.J.Gould
(p.684) assesses this point is as follows: 
 \begin{quotation}
 {\it ``The Victorian age, basking in triumph of an industrial and
military might rooted in technology and mechanical engineering, granted
little conceptual space to random events... Darwin got into enough
trouble by invoking randomness for sources of raw material; he wasn't
about to propose stochastic causes for {\em change} as well!''}
 \end{quotation} 

 As far as biological evolution is concerned, pseudo-randomness, as
introduced by Darwin, is perfectly acceptable. The real need for the
notion of ``true'' or objective probability, as in Quantum Mechanics,
was still a few decades in the future. 
 
 2- Darwin didn't have a clear distinction between somatic versus
genetic, or external versus internal, causes of variations. Winther
(2000, p.425), makes the following comments: 
 \begin{quotation} 
 {\it ``Darwin's ideas on variation, hereditarity, and development
differ significantly from twentieth-century views. First, Darwin held
that environmental changes, acting on the reproductive organs or the
body, were necessary to generate variation. Second, hereditarity was a
developmental, not a transmitional process...''}   
 \end{quotation} 

 At the time of Darwin, the available technology could not, of course,
reveal the bio-chemical mechanisms of heredity. Nevertheless, scientists
like Hugo de Vries and Erwing Schr\"{o}dinger have had powerful insight
on this mechanisms, even before the necessary technology became
available. de Vries (1900), for example, advanced the following
hypotheses: 
 \begin{quotation} 
 {\it ``1. Protoplasm is made up of numerous small units, which are
bearers of the hereditarity characters. 2. These units are to be
regarded as identical with molecules.''}  
 \end{quotation} 

 In his book {\it What is Life}, Schr\"{o}dinger (1945) advanced more
detailed hypotheses about the genetic coding mechanisms, based on far
reaching theoretical insights provided by quantum mechanics. This small
book was a declared source of inspiration for both James Watson and
Francis Crick, who, in 1953, discovered the double-helix molecular
structure of DNA, opening the possibility of deciphering the genetic
code and its expression mechanisms. 
    
 3- Continuous variations. From several passages of Darwin's works, it
is clear that he saw actual variations as coming from a continuum of
potential possibilities:  
 \begin{quotation} 
 {\it ``[as] I have attempted to show in my work on variation... they
[are] extremely slight and gradual.''} 
 Darwin (1959, p.86). 

 {\it ``On the slow and successive appearence of new species: 
 ...organic beings accord best with the common view of the
 immutability of species, or with that of their slow and gradual
 modification, through variation and natural selection.''} 
 Darwin (1959, p.167). 

 {\it ``It is indeed manifest that multitudes of species are related in
the closest manner to other species that still exist, or have lately
existed; and it will hardly be maintained that such species have been
developed in an abrupt or sudden manner. Nor should it be forgotten,
when we look to the special parts of allied species, instead of to
distinct species, that numerous and wonderfully fine graduations can be
traced, connecting together widely different structures.''} 
 Darwin (1959, p.117). 
 \end{quotation} 

 The first modern reference for discrete or modular genetic variations
can be found in the work of Gregor Mendel (1865), see next paragraph. It
was unfortunate that the ideas of Mendel, working at a secluded
monastery in Br\"{u}nn (Brno), were not immediately appreciated. 
 For a contemporary view of evolution and modularity, 
 see Margulis (1999) and Margulis and Sagan (2003). 
 \begin{quotation} 
 {\it ``The Forms of the Hybrids: 

 With some characters...    
 one of the two parental characters is so preponderant that it is
difficult, or quite impossible, to detect the other in the hybrid.

  This is precisely the case with the Pea hybrids. In the case of each
of the 7 crosses the hybrid-character resembles that of one of the
parental forms so closely that the other either escapes observation
completely or cannot be detected with certainty. This circumstance is of
great importance in the determination and classification of the forms
under which the offspring of the hybrids appear. Henceforth in this
paper those characters which are transmitted entire, or almost unchanged
in the hybridization, and therefore in themselves constitute the
characters of the hybrid, are termed the dominant, and those which
become latent in the process recessive. The expression "recessive" has
been chosen because the characters thereby designated withdraw or
entirely disappear in the hybrids, but nevertheless reappear unchanged
in their progeny, as will be demonstrated later on.''} 
 \end{quotation} 

 The third point of divergence, variations discreteness, is, of course,
closely linked with the second, the nature of genetic coding. However,
its implications are much deeper, as examined in the next section.

 \subsection{Modularity Requires Quantization} 

 The ideas of Herbert Simon about modularity, examined in section 3.2,
seem to receive empirical support from anywhere we look in the
biological world. Ksenzhek and Volkov (1998, p.80), also quoted in Souza
and Manzatto (2000), for example, gives the following example from
Botany: 
 \begin{quotation} 
 {\it ``A plant is a complicated, multilevel, hierarchical system, which
provides  a very high degree of integration, beginning from the
elementary process  of catching light quanta and ultimately resulting in
the functioning of a macroscopic plant as an entire organism. The
hierarchical structure of plants may be examined in a variety of
aspects. (The following) table shows seven hierarchical levels of mass
and energy.''}
 \end{quotation} 

 \begin{table}[bt] 
 \begin{center}
 {\small 
 \begin{tabular}{ccccc} 

 \hline 

 Level & Size (m)  & Structure  & Transfer mechanism & Integration \\ 

 \hline 

 1 & $2E-8$ & Chlorophyll antenna &  Resonant exitons & 300  \\ 

 2 & $5E-7$ & Tylakoid membrane & Electrochemical & 500  \\ 

 3 & $5E-6$ & Cloroplast & Diffusion & 50  \\ 

 4 & $5E-5$ & Cell & Diffusion, cyclosis & 50  \\ 

 5 & $2E-4$ & Areolae & Diffusion & 1E3  \\ 

 6 & $1E-1$ & Leaf & Hydraulics & 1E5  \\ 

 7 & $1-10$ & Tree & Hydraulics & 1E4  \\ 

 \hline 

 \end{tabular} 
 }

 \mbox{} \\ 

 \centerline{Table 3. Plant Energetics Hierachical and Modular Structure.} 
 \end{center}    
\end{table} 

 As an example of how to interpret this table, we give further details
concerning its first line: in a thylakoid membrane, about 300
chlorophyll molecules act like an antenna in a reaction center or
photosynthetic unit, capable of absorbing light quanta at a rate of
about 1K cycles / second. This energy conversion cycle absorbs photons
of about $1.8eV$ (430 Hz or 700nm), synthesizing compounds,
carbohydrates and oxygen, at an energy level of about $1.2eV$ higher
than its input compounds, carbon dioxide and water.   

 Ksenzhek and Volkov (1998, p.80), see next quotation, also makes an
important remark concerning the need for a specific and non-reductionist
interpretation of each line in the above table, or structural level in
the organism. For related aspects in Biology, see Buss (2007). Niels
Bohr (1987b, Light and Life, p.3-12; Biology and Atomic Physics, p.13-22) 
presents a similar argument based on the general
concept of complementarity.
 \begin{quotation} 
 {\it ``It should be noted that any hierarchical level that is above
another  level cannot be considered as the simple sum of the elements
belonging  to that lower level. In all cases, each step from a given
level of the  hierarchical staircase to the next one is followed by the
development of  new features not inherent in the elements of the lower
level.''}
 \end{quotation}

 Table 2 stops at somewhat arbitrary levels and could be extended
further up or down. Higher levels in the table would enter the domains
of Ecology. Lower levels would penetrate the domains of Chemistry, and
then Physics. At this point, we make an astonishing observation:
Classical Physics {\em cannot} accommodate stable atomic models.
Classical Physics gives {\em no} support for discreteness or modularity
of any kind. Hence, our modular view of the world would be, within
classical Physics, a giant with feet of clay! Werner Heisenberg (1958,
p.5,6) describes the situation as follows: 
 \begin{quotation} 
  {\it ``In 1911 Rutherford's observations... resulted in his famous
atomic model. The atom is pictured as consisting of a nucleus, which is
positively charged and contains nearly the total mass of the atom, and
electrons, which circle around the nucleus like planets circle
around the sun. The chemical bond between atoms of different elements is
explained as an interaction between the outer electrons of the
neighboring atoms; it has not directly to do with the atomic nucleus.
The nucleus determines the chemical behavior of the atom through its
charge which in turn fixes the number of electrons in the neutral atom.
Initially this model of the atom could not explain the most
characteristic feature of the atom, its enormous stability. No planetary
system following the laws of Newton's mechanics would ever go back to
its original configuration after a collision with another such system.
But an atom of the element carbon, for instance, will still remain a
carbon atom after any collision or interaction in chemical binding. 

   The explanation of this unusual stability was given by Bohr in 1913,
through the application of Planck's quantum hypothesis. An atom can
change its energy only by discrete energy quanta, this must mean that
the atom can exist only in discrete stationary states, the lowest of
which is the normal state of the atom. Therefore, after any kind of
interaction, the atom will finally always fall back into its normal
state.''}
 \end{quotation}

  \mbox{} 
  \vspace{-0.0cm} 
  \centerline{\includegraphics*[angle=0, height=3.0in,  width=6.5in,  
           viewport=120 150 840 480 , clip]{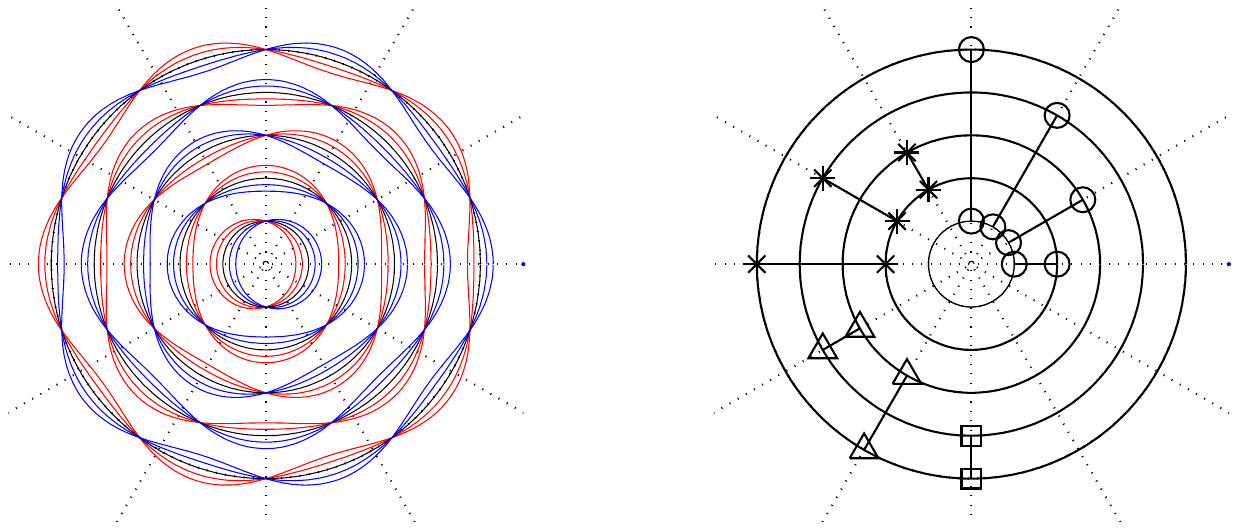}} 
  \vspace{-0.0cm} 
 \centerline{Figure 12: Orbital Eigensolutions for the Hydrogen Atom.} 
 \centerline{Figure 13: Orbital Transitions for Hydrogen Spectral Lines;} 
 \centerline{Series: Lyman, $n=1$; Balmer, $n=2$; Paschen, $n=3$; 
     $m=n+1,\ldots \infty$.}  
 
 \mbox{} 


 Bohr's model is based on the quantization of the angular momentum of
the electron in the planetary atomic model. The wave-particle duality
metaphor can give us a simple visualization of Bohr's model. As already
mentioned in section 4.2, a string of length $L$ with two fixed ends can
only accommodate (stationary) waves whose (half) wavelength are a
multiple of the string length, i.e. $L=n\lambda$, $n=1,2,3,\ldots$. The
first one ($n=1$, longer wavelength, lower frequency) is called the
fundamental frequency of the string, and the others ($n=2,3\ldots$,
shorter wavelengths, higher frequencies) are called its harmonics.

 Putting together de Broglie's duality principle and the planetary
atomic model, we can think of the electron's orbit as a circular string
of length $L=2\pi r$. Plugging in de Broglie's equation, $\lambda=h/mv$,
and imposing the condition of having stable eigenfunctions or standing
waves, see Enge (1972) and Figure 12, we have
 \[
    2\pi r= n\lambda= \frac{nh}{m_e v} \ \ \mbox{or} \ \ \   
    m_evr = n \frac{h}{2\pi} = n\hbar \ . 
 \]
 Planck's constant equals $6.626E-34$ joule-seconds or $4.136E-15$
electron-volt-second, and the electron mass is $9.11E-28$ gram. Since
the right hand side of this equation is the angular momentum of the
orbiting electron, de Broglie wave-particle duality principle imposes
its quantization.

 Bohr's atomic model was also able to, for the first time, provide an
explanation for another intriguing phenomenon, namely:

  (a) Atoms only emit light at sharply defined frequencies, known as
spectral lines;  

  (b) The frequencies, $\nu$, or wavelengths, $\lambda$, of these
spectral lines are related by integer algebraic expressions, like the
Balmer-Rydberg-Ritz-Paschen empirical formula,
 \[ 
    \frac{\nu_{n,m}}{c} = \frac{1}{\lambda_{n,m}} = 
    R \left( 
    \frac{1}{n^2} -\frac{1}{m^2} \right) \ , 
 \] 
 where $R=1.0973731568525(73) E7 \ m^{-1}$ is Rydberg's constant.

 Distinct combinations of integer numbers, $0<n<m$, in BRRP formula give
distinct wavelengths of the spectrum, see Enge (1972). It so happens
that these frequencies are in precise correspondence with the
differences of energy levels of orbital eigen-solutions, see Figure 13.
These are the Hydrogen spectral series of Lyman, $n=1$, Balmer, $n=2$,
Paschen, $n=3$, and Brackett, $n=4$, for  $m=n+1,\ldots \infty$. 
 Similar spectral series have been known for other elements, and used by
chemists and astronomers to identify the composition of matter from the
light it radiates. 
 Rydberg's constant can be written as $R=m_e e^4 / (8 \epsilon_0^2 h^3
c)$, where $m_e$ is the rest mass of the electron, $e$ is the elementary
charge, $\epsilon_0$ is the permittivity of free space, $h$ is the
Planck's constant, and $c$ is the speed of light in vacuum.

 The importance attributed by Bohr to the emergence of these sharp (discrete)
eigen-solutions out of a higher dimensional continuum of possibilities
is emphasized in Bohr (2007):  
 \begin{quotation} 
 {\it ``Your physical aha-experience? 

 Wavelengths of a complete series of spectral lines in the hydrogen
spectrum can be expressed with the aid of integers. This information, he
[Bohr] said, left an indelible impression on him.''}   \end{quotation}

   Approximation methods of perturbation theory can be used to  compute 
probabilities of spontaneous and induced transitions between  the
different orbitals or energy states of an atom, and these transition
rates can be observed as intensities  of the respective spectral lines, 
see Enge (1972, ch.8), Landshoff (1998, ch.7) and 
McGervey (1995, ch.14).  
   Comparative analyses between the value and accuracy of these   
theoretical calculations and empirical observations are of obvious 
interest. 
   However, the natural interpretation of these analyses immediately 
generates statements about the uncertainty of transition rates, 
expressed as probabilities of  probabilities. 
 Hence, as explained in section 5.6.3, these statements collide with
the canons of the subjectivist epistemological framework and are
therefore unaceptable in orthodox Bayesian statistics.

 \subsection{Quantization Entails Objective Probability}

 An objective form of probability is at the core of quantum
mechanics theory, as seen in previous sections. 
 However, probabilistic explanations or probabilistic causation have  
been, at least from a historical perspective, very controversial
concepts. This has been so since the earliest times. Aristotle (Physics,
II,4,195b-196a) discusses events resulting from coincidences or
incidental circumstances. If such an event serves a conscious human
purpose, it is called $\tau \upsilon \chi \eta$, translated as luck or
fortune. If it serves the ``unconscious purposiveness of nature'', it is
called $\alpha \upsilon \tau o \mu \alpha \tau o \nu$, translated as
chance or accident. 
 \begin{quotation} 
 {\it ``We must inquire therefore in what manner luck and chance are
present among the causes enumerated, and whether they are the same or
different, and generally what luck and chance are.
 
 Thus, we must inquire what luck and chance are, whether they are the
same or different, and how they fit into our division of causes.

 Some people even question whether they are real or not. They say
thatnothing happens by chance, but that everything which we ascribe to
luck or chance has some definite cause.  

 Others there are who, indeed, believe that chance is a cause, but that
it is inscrutable to human intelligence, as being a divine thing and
full of mystery.''}  
 \end{quotation} 

 Aristotle (Physics, II,4,195b-196a) also reports some older
philosophical traditions that made positive use of probabilistic
causation, such as a stochastic development or evolution theory due to
Empedocles: 
 \begin{quotation} 
 {\it ``Wherever then all the parts came about just what they would have
been if they had come be for an end, such things survived, being
organized spontaneously in a fitting way; whereas those which grew
otherwise perished and continue to perish, as Empedocles says...''} 
 \end{quotation} 

 Many other ancient cultures accepted probabilistic arguments and/or did
make use of randomized procedures, see  Davis (1969), 
Kaptchuk and Kerr (2004) and Rabinovitch (1973). 
 Even the biblical narrative, so averse to magic of any sort, presents 
the idea that destiny is ultimately inscrutable to human understanding, 
see for  example Exodus (XXXIII, 18-23):      
  
 Moses, who is always willing to speak his mind, 
 asks God for perfect knowledge:
 \begin{quotation}
 \noindent {\it   And Moses said: I pray You, show me Your glory!} 
 \end{quotation} 
  In response God, Who is always ready to explain to Moses Who makes the
 rules, tells him that perfect knowledge can not be achieved by a living
 creature.  
  This verse may also allegorically indicate that temporal 
 irreversibility is a necessary consequence of such veil of uncertainty: 
 \begin{quotation}
 \noindent {\it 
 And the Lord said: You cannot see My face, 
    for no man can see Me and live!...\\  
 I will enclose and confine you, and protect you in My manner... (so that)\\ 
 You shall see My back, but My face shall not be seen.} 
 \end{quotation} 

 Nevertheless, the concepts of stochastic evolution and probabilistic
causation lost prestige along the centuries. From the comments of Gould
and Lenoir in section 7.1, we may conclude that at the XVIII and early
XIX century its status reached the lowest level ever. It is ironic than
that stochastic evolution is the concept at the eye of the storm of some
of the most important scientific revolutions of the late XIX and XX
century. 

 As seen in section 6, Quantum Mechanics entails Heisenberg's
uncertainty principle, stating that we can not measure (in practice or
in theory) the classical variables describing the motion of a particle
with a precision beyond a hard threshold given by Planck's constant.
Hence, the available information about a physical system is, in quantum
mechanics, governed by {\it laws that are in nature essentially
probabilistic}, or, as stated in Ruhla (1992, p.162), 
 \begin{quotation} 
 {\it ``No longer is it chance as a matter of ignorance or of
incompetence: it is chance quintessential and unavoidable.''} 
 \end{quotation} 

  The path leading to an essentially stochastic world-view (or, in more 
 detail, a Weltanschauung including random systemic interactions) was 
 first foreseen by people far ahead of their time, like C.S.Peirce and
 L.Bozmann, a path that was than advanced by reluctant revolutionaries
 like M. Planck, A. Einstein, and E. Schr\"{o}dinger, who had a major
 participation in forging the new concept of probability, but that were
 at the same time, still emotionally attached to classical concepts.
  Finally, a third generation, including N.Bohr, W.Heisenberg and M.Born
 fully embraced the new concept of objective probability. Of course, as
 with all truly innovative concepts, it will take mankind at least a few
 generations to truly assimilate and incorporate the new idea.

 \section{Final Remarks} 

 The ``objectification of probability'' and the consequent raise of the
ontological status of stochastic evolution and/or probabilistic
causation was arguably one of the two greatest innovations of modern
physics. The other great innovation is the ``geometrization of
space-time'' in Einstein's theories of special and general relativity,
see French (1968) and Martin (1988) for intuitive introductions, 
Sachs and Wu (1977)  for a rigorous treatment, and Misner et al. (1973) 
for an encyclopedic treatise.

 The manifestation of physical quantization and (special) relativistic
geometry  is regulated by Planck's constant and the speed of light. 
 The value of these constants in standard (international) metric units,
$h=6.6E-34\,Js$ and $c=3.0E+8\,m/s$, have, respectively, a tiny and huge
order of magnitude, making it easy to understand why most of the effects
of modern physics are not immediately perceptible in our ordinary life
experience and, therefore, why classical physics can offer acceptable
approximations in many circumstances of common statistical practice.
 However, modern physics has forever changed some of our most basic
concepts related to space, time, causality and probability. Moreover, we
have seen in this chapter how some of these concepts, like modularity
and probabilistic causation, are essential to our theories and to
understand phenomena in many other fields. We have also seen how
quantization or stochastic evolution have a direct or indirect baring on
areas much closer to our daily life, like Biology and Engineering.
 Hence, it is of vital importance to incorporate these new concepts to a
contemporary epistemology or, at least, to use an epistemological
framework that is not incompatible with these new ideas.


%% file: CAPE6.TEX
 

 \chapter{The Living and Intelligent Universe}

 \mbox{} 

 {\flushright

 
 {\it 
  ``Cybernetics is the science of defensible metaphors.'' 
  }
  
 
   Gordon Pask (1928-1996).   

 \mbox{} 

 {\it 
  ``You, with all these words....'' 
  }
  
 
   Marisa Bassi Stern (my wife, when I speak too much).   

 \mbox{} 

 {\it 
  ``Yes I think to myself: What a wonderful world!'' 
  }
  
 
   B.Thiele and G.D.Weiss, in the voice of L.Armstrong.

 }

 \mbox{} \\

 In the article Mirror Neurons, Mirror Houses, and the Algebraic
Structure of the Self, by Ben Goertzel, Onar Aam, F. Tony Smith and Kent
Palmer (2008)  and the companion article of Goertzel (2007), the authors
provide an intuitive explanation for the logic of mirror houses, that is,
they study symmetry conditions for specular systems entailing the
generation of kaleidoscopic images.   
 In these articles, the authors share (in my opinion) several 
important  insights on autopoietic systems and constructivist
philosophy.     
 A more prosaic kind of mirror house used to be a popular
attraction in funfairs and amusement parks. 
 The entertainment then came from misperceptions about
oneself or other objects. More precisely, from the misleading ways
in which a subject sees how or where are the objects inside the 
mirrorhouse, or how or where himself stands in relation to other
objects.

 The main objective of this chapter is to show how similar misperceptions
 in  science can lead to ill-posed problems, paradoxical situations
 and even misconceived philosophical dilemmas.
 The epistemological framework of this discussion will be that of
 cognitive constructivism, as presented in previous chapters. 
 In this framework, objects within a scientific theory are tokens for
 eigen-solutions which Heinz von Foerster 
 characterized by four essential attributes, namely those of being discrete  
 (precise, sharp or exact), stable, separable and composable.
  The Full Bayesian Significance Test (FBST) is a  possibilistic
 belief calculus based on a (posterior) probabilistic measure originally 
 conceived as a statistical significance test to assess the 
 objectivity of such eigen-solutions, that is, to measure how well a 
 given object manifests or conforms to von Foerster's four essential 
 attributes. 

  The FBST belief or credal value of hypothesis $H$ given 
 the observed data $X$ is the {\it e-value}, $\mbox{ev}(H \g X)$,  
 interpreted as the {\it epistemic value} of hypothes $H$ (given $X$), 
 or the {\it evidence value} of data $X$ (supporting $H$). 
  A formal definition of the FBST and several of its implementations for 
 specific problems can be found in the author's previous articles,  
 and summarized in appendix A. 
 From now on, we will refer to Cognitive Constructivism accompanied by  
 Bayesian statistical theory and its tool boxes, as laid down in the 
 aforementioned articles, as the Cog-Con epistemological framework.      
   
 Instead of reviewing the formal definitions of the essential attributes
 of eigen-solutions, we analyze instead the Origami example,
 a didactic case presented by Richard Dawkins. This is the done in section 1.
 The origami example is so simple that it may look trivial and,
 in some sense, it is so. 
 In subsequent sections we analyze in which ways the eigen-solutions found 
 in the practice of science can be characterized as non-trivial,
 and also highlight some (in my view) common misconceptions about
 the nature of these non-trivial objects, just like distinct forms of illusion
 in a mirror-house.

 In section 2 we contrast the control, precision and stability of 
morphogenic folding processes in autopoietic and allopoietic systems. 
 In section 3 we concentrate in object orientation and code reuse, 
inter-modular adaptation and resonance, and also analyze the 
yoyo diagnostic problem.  
 In section 4 we explore auto-catalytic and hypercyclic networks, as
well as some related bootstrapping paradoxes. This section is heavily
influenced by the work  of Manfred Eigen. 
 Section 5 focus on explanations of specific components, single links or 
partial chains in long cyclic networks, including the meaning of 
some forms of directional (such as upward or downward) causation.     
 In section 6 we study the emergence of asymptotic eigen-solutions
such as thermodynamic variables or market prices,  
 and in section 7 we analyze the ontological status of such entities.
 In section 8 we study the limitations in the role and scope of conceptual 
distinctions used in science, and the importance of probabilistic
causation as a mechanism to overcome, in a constructive way, some of
the resulting dilemmas. 
 In short, section 2 to 8 discus autopoiesis, modularity, hypercycles,  
emergence, and probability as sources of complexity and forms of 
non-trivial organization.  
 Our final remarks are presented in section 9.

 In this chapter we have made a conscious effort to use examples that can
be easily visualized in space and time scales directly perceptible to our
senses, or at least as close as possible to it. We have also presented our arguments  
using, whenever possible, very simple (high school level) mathematics.
 We did so in order to make the examples intuitive and easy to 
understand, so that we could concentrate our attention on the
epistemological aspects and difficulties of the problems at hand. 
 Several interesting figures and images that illustrate 
some of the concepts discussed in this chapter are contained in the website 
 \verb#www.ime.usp.br/~jstern/pub/gallery2.pdf/# .

 \section{The Origami Example}

 The Origami example, from the following text in 
 Blackmore (1999, p.x-xii, emphasis are ours) 
 was given by Richards Dawkins to present the notion of
 reliable replication mechanisms in the context of
 evolutionary systems. 
 Dawkins' example contrasts two versions of the Chinese Whispers game
 using distinct copy mechanisms.

 \begin{quotation}
 Suppose we assemble a line of children.
 A picture, say, a Chinese junk, is shown to the first child, who is asked
 to draw it. The drawing, but not the original picture, is then shown to the
 second child, who is asked to make her own drawing of it. The second child's
 drawing is shown to the third child, who draws it again, and so the series
 proceeds until the twentieth child, whose drawing is revealed to everyone
 and compared with the first. Without even doing the experiment, we know
 what the result will be. The twentieth drawing will be so unlike the
 first as to be unrecognizable. Presumably, if we lay the drawings out
 in order, we shall note some resemblance between each one and its immediate
 predecessor and successor, but the mutation rate will be so high as to
 destroy all semblance after a few generations. A trend will be visible as we
 walk from one end of the  series of drawings to the other, and the
 direction of the trend will be degeneration...

 High fidelity is not necessarily synonymous with digital.
 Suppose we set up our Chinese Whispers Chinese Junk game again,
 but this time with a crucial difference. Instead of asking the
 first child to copy a drawing of the junk, we teach her, by
 demonstration, to make an origami model of a junk.
 When she has mastered the skill, and made her own junk,
 the first child is asked to turn around to the second child and teach
 him how to make one. So the skill passes down the line to the
 twentieth child. What will be the result of this experiment?
 What will the twentieth child produce, and what shall we observe if
 we lay the twenty efforts out in order along the ground? ...

 In several of the experiments, a child somewhere along the line will
 forget some crucial step in the skill taught him by the previous child,
 and the line of phenotypes will suffer an abrupt macromutation which will
 presumably then be copied to the end of the line, or until another
 discrete mistake is made. The end result of such mutated lines will not
 bear any resemblance to a Chinese junk at all.
 But in a good number of experiments the skill will correctly pass all
 along the line, and the twentieth junk will be no worse and no better,
 on average, than the first junk.
 If we lay then lay the twenty junks out in order, some will be more
 perfect than others, but imperfections will not be copied on down
 the line...

 Here are the first five instructions... for making a Chinese junk: 

 1. Take a {\em square} sheet of paper and fold  all four
    corners {\em exactly} into the {\em middle}.

 2. Take the reduced {\em square} so formed, and fold one side
    into the {\em middle}.

 3. Fold the opposite side into the {\em middle,  symmetrically}.

 4. In the same way, take the {\em rectangle} so formed, and fold its two
    ends into the {\em middle}.

 5. Take the small {\em square} so formed, and fold it backwards,
    {\em exactly} along the {\em straight line} where you last
    two folds met...

 These instructions, though I would not wish to call them digital,
 are potentially of very high fidelity, just as if they were digital.
 This is because they all make reference to idealized tasks like
 `fold the four corners exactly into the middle'...
 The instructions are self-normalizing.
 The code is error-correcting...

 \end{quotation}

 Dawkins recognizes that instructions for constructing an origami have  
 remarkable properties, providing the long term survival of the subjacent 
 {\it meme}, i.e. specific model or single idea, expressed as an origami.   
 Nevertheless, Dawkins is not sure how he ``wishes to call'' 
 these properties (digital? high fidelity?).  
 What adjectives should we use to appropriately describe the desirable 
 characteristics that Dawkins perceives in these instructions? 
 I claim that von Foerster's four essential attributes  
 of eigen-solutions offer an accurate description of the 
 properties relevant to the process in study.

 The instructions and the corresponding (instructed) operations 
 are precise, stable, separable and composable.
 A simple interpretation of the meaning of these four attributes in the 
 origami example is the following:

 {\it Precision}: An instruction like ``fold a paper joining  two opposite  
 corners of the square'' implies that the folding must 
 be done along a diagonal of the square. 
 A diagonal is a specific line, a 1-dimensional object in the 
 2-dimensional sheet of paper. 
 In this sense the instruction is precise or exact.   

 {\it Stability}: By interactively adjusting and correcting the position of 
the paper (before making a crease) it is easy to come very close to what
the instruction specifies. Even if the resulting fold is not
absolutely perfect (in practice it actually never is), it will probably
still work as intended.  

 {\it Composability and Separability}: We can compose or superpose  
multiple creases in the same sheet of paper. 
 Moreover, adding a new crease will not change or destroy   
the existing ones. 
 Hence, we can fold them one at a time, that is, separately.

 These four essential attributes are of fundamental importance 
 in order to  understand scientific activity in the 
 Cog-Con framework. 
 Moreover, Dawkins' origami example illustrates these attributes with 
 striking clarity and simplicity.     
  
 In the following sections we will examine other examples, which are
less simple, not so clear or non-trivial in a distinct and 
characteristic way. 
 We will also draw attention to some confusions and mistakes 
often made when analyzing systems with similar characteristics.

 \section{Autopoietic Control, Precision, Stability}

 The origami folding is performed and controlled by an external agent, 
the person folding the paper. 
  In contrast, organic development processes are self-organized.  
  These processes are not driven by an external agent, 
 do not require external supervision, 
 and usualy are not even amenable to external corrections.
 While artifacts and machines manufactured like an origami are called 
 {\it allopoietic}, from 
 $\alpha \lambda \lambda o$-$\pi o \iota \eta \sigma \iota \varsigma$ 
 - external production, 
 living organisms are called {\it autopoietic}, from 
 $\alpha \upsilon \tau o$-$\pi o \iota \eta \sigma \iota \varsigma$ 
 - self production. 

 Autopoiesis is a non-trivial process, in many interesting ways. 
 For example, the inexistence of external supervision or correction 
 mechanism requires an autopoietic process to be stable.  
 Moreover, typical biological processes occur in environments with high 
levels of noise and have large (extra) variability. 
 Hence the process must be intrinsically self-correcting and redundant  
so that its noisy implementation does not compromise the viability of
the final product.

 \subsection{Organic Morphogenesis: (Un)Folding Symmetries}

 In this section we make some considerations about morphogenic biological  
processes, namely, we study examples of tissue folding in early embryonic
development. 
 This process naturally invites not only strong analogies,  but also sharp
contrasts with  the origami example.  
 At a macroscopic (supra cellular) level, the organisms' organs and
structures are built by tissue movements, as described in 
 Forgacs and Newman (2005, p.109), and Saltzman (2004, p.38).

 The main types of tissue movements in morphogenic process are: 

 - Epiboly: spreading of a sheet of cells over deeper layers. 

 - Emboly: inward movement of cells which is of various types as: 

 - Invagination: infolding or insinking of a layer, 

 - Involution: inturning, inside rotation or inward movement of a tissue. 

 - Delamination: splitting of a tissue into 2 or more parallel layers. 

 - Convergent/Divergent Extension: stretching together/apart of two 
    distinct tissues.

 The blastula is an early stage in the embryonic development of most animals. 
 It is produced by cleavage of a fertilized ovum and consists of a 
 hollow sphere of around 128 cells surrounding a central cavity. 
 From this point on, morphogenesis unfolds by successive tissue movements. 
 The very first of such moves is known as gastrulation, a deep invagination 
 producing a tube, the archenteron or primitive digestive tract. 
 This tube may extend all the way to the pole opposing the 
 invagination point producing a second opening. 
 The opening(s) of the archenteron become mouth and anus of the 
 developing embryo.  
 
 Gastrulation produces three distinct (germ) layers, that will further 
differentiate into several body tissues. 
 Ectoderm, the exterior layer, will further differentiate into 
 skin and nervous systems. 
 Endoderm, the innermost layer at the archenteron, generates the 
digestive system.  
 Mesoderm, between the ectoderm and endoderm, differentiates into
 muscles, connective tissues, skeleton, kidneys, circulatory and
 reproductive organs. 
 We will use this example to highlight some important topics, 
some of which will be explored more thoroughly in further sections.

 \subsubsection*{Discrete vs. Exact or Precise Symmetries}  

 Notice that origami instructions, that implicitly rely on the 
 {\it symmetries} characterizing the shape of the paper, 
 require foldings at sharp edges or cresses. 
 Hence, a profile of the folded paper sheet may look 
 like it breaks (is non-differentiable) at a discrete or singular point.    
   
 Organic tissue foldings have no sharp edges. 
 Nevertheless, the (idealized) symmetries of the folded tissues,  
  like the spherical symmetry of the blastula, 
  or the cylindrical symmetry of the gastrula,     
 can be described by equations just as exact or precise, 
 see Beloussov (2008),  Nagpal (2002), Odel et al. (1980),  
 Tarasov (1986), and Weliky and Oster (1990).   
 This is why we usually prefer the adjectives {\it precise} or 
 {\it exact} to  the adjective {\it discrete} used by von Foester 
in his original definition  of the four essential properties of an
eigen-solution. \\

 \subsubsection*{Centralized vs. Decentralized Control} 

 In morphogenesis, there is no agent acting like a central controller,   
dispatching messages ordering every cell what to do. 
 Quite the opposite, the complex forms and tissue movements at 
a global or macroscopic (supra cellular) scale are the result of 
collective cellular behavior patterns based on distributed control.  
 The control mechanisms rely on simple local interaction between 
neighboring cells, see Keller et al. (2003), Koehl (1990), and Newman
and Comper (1990). 
 Some aspects of this process are further analyzed in sections 3 and 6.

 \section{Object Orientation and Code Reuse}

 At the microscopic level, cells at the several organic tissues studied
in the last section are differentiated by distinct metabolic reaction 
patterns. 
 However, the genetic code of any individual cell in a organism is
identical (as always in biology, there are exceptions, but they are not
relevant  to this analysis), and cellular differentiation at distinct
tissues are the result of differentiated (genetic) expressions of this
sophisticated program. 

 As studied in Chapter 5,  
 complex systems usually have a  modular hierarchical structure or, 
 in computer science jargon, an object oriented design. 
 In allopoietic systems object orientation is achieved by explicit
design, that is, it has to be introduced  by a knowledgeable and
disciplined programmer, see Budd (1999). 
 In autopoietic systems modularity is an implicit and emergent property, 
 as analyzed in Angeline (1996), Banzaff (1998), Iba (1992),  
 Lauretto at al. (2009)  and Chapter 5. 

 Object oriented design entails the reuse, over and over, of the  same
modules (genes, functions or sub-routines) as control mechanisms for
different processes. The ability to easily implement this kind of
feature was actively pursued in computer science and software
engineering. 
 Object orientation was also discovered, with some  surprise, to be 
naturally occurring in developmental biology, see  Carrol (2005).  
 
 However, like any abused feature, code reuse can
also become a burden in some circumstances.  
 The difficulty of locating the source of a functionality (or a bug) in
an intricate inheritance hierarchy, represented by a complex 
dependency graph, is known in computer science as the yoyo problem.   
 According to the glossary in Budd (1999, p.408), \   
 ``Yoyo problem: Repeated movements up and down the class hierarchy
that  may be required when the execution of a particular method
invocation  is traced.''  

  Systems undergoing many changes or modifications, under repeated
adaptation or expansion, or on rapid evolution are specially vulnerable
to yoyo effects. 
 Unfortunately, the design of the human brain and its mental abilities
are under all of the above conditions.  
 In the next subsection we study some examples in this area, related to
biological neural networks and language. These examples also include 
some mental dissociative phenomena that can be considered as
manifestations of the yoyo problem.

 \subsection{Doing, Listening and Answering}

 In this section we study some human  capabilities related to doing (acting), 
 listening (linguistic understanding) and answering (dialogue). 
 The capabilities we have chosen to study are related to the phylogenetic 
 acquisition and the ontogenetic development of: 
  \\ 
 - Mechanisms for precision manipulation, production of speech and 
empathic feeling; 
 \\ 
 - Syntax for complex manipulation procedures, language articulation and 
behavioral simulation; 
 \\ 
 - Semantics for action, communication and dialogue; and the learning of   
 \\ 
 - Technological know-how, social awareness and  self-awareness.

 When considering an action in a modern democratic society, we usually 
deliberate what to do (unless there is already a tacit agreement).  
 We then communicate with other agents involved to coordinate this 
action, so that we are finally able to do what has to be done. 
 Evolution, it seems, took exactly the other way around. 
 Phylogenetically, the path taken by our species follows a stepwise
development of several mechanisms (that were neither independent nor
strictly sequential), including:

 1. A mechanism for 3-dimensional vision and precision measurement, 
fine motor control of hands and mouth, and visual-motor coordination 
for the complex procedures of {\it precision manipulation}.    

 2. Mechanisms for {\it imitating}, {\it learning} and {\it simulating} 
the former procedures or actions. 

 3. Mechanisms for {\it  simulating} (possible) actions taken by other 
individuals, their consequences and motivations, that is, mechanisms for 
awareness and (behavioral) understanding of other individuals.   

 4. A mechanism for {\it communicating} (possible) actions, used for commanding, 
 controling and coordinating group actions. 
 The use of such a mechanism implies a degree of awareness of others, that is, 
 some ability to communicate, explain, listen and learn what you do,    
 {\it you} - an agent like {\it me}.

 5. Mechanisms for {\it dialoging} and {\it deliberating}, that is, 
 for negotiating, goal selecting and non-trivial social planning.   
 The use of such mechanisms implies some 
 {\it self-awareness} or {\it consciousness}, that is, 
 the conceptualization of an ego, an abstract 
 {\it I} - an agent like {\it you}.

 In a living individual, all of these mechanisms must be well integrated. 
 Consequently, it is natural that they work using coherent  
implicit grammars, reflecting compatible subjacent rules of composition 
for action, language and inter-individual interaction.   
 Indeed, resent research in neuro-science confirm the coherence  of
these mechanisms. Moreover, this research shows that this  coherence is
based not just on compatible designs of separate systems, but on
intricate schemes of use and reuse of the same structures, namely, the 
firmware code or circuits implemented as biological neural
networks.  

 {\it Mirror neuron} is a concept of neuroscience that highlights 
the reuse of the same circuits for distinct functions. 
 A mirror neuron is part of a circuit which is activated (fires) when
an individual executes an action, and also when the individual observes
another individual executing the same action as if he, the observer, were
performing the action himself.   
 The following passages, from important contemporary neuro-scientists, 
give some hints on how the mechanisms mentioned in the past paragraph
are structured.

 The first group of quotes, from Hesslow (2002, p.245), states the mirror
neuron {\it simulation hypothesis}, according to which, the same
circuits used to control our actions are  used to learn, simulate, and 
finally ``understand'' possible actions taken by other individuals.  
 According to the simulation hypothesis, we are then naturally 
endowed with the capability of observing, listening, and ``reading the
mind'' of (that is - understanding, by simulation, the meaning or intent
of the possible actions taken by) our fellow human beings.

  \begin{quote} 

  ...the simulation hypothesis states that thinking consists of simulated
 interaction with the environment and rests on the following three core
 assumptions: \\ 
  (1) simulation of actions: we can activate motor structures of the
 brain in a way that resembles activity during a normal action but does
 not cause any overt movement; \\ 
  (2) simulation of perception: imagining perceiving something is
 essentially the same as actually perceiving it, only the perceptual
 activity is generated by the brain itself rather than by external
 stimuli; \\ 
  (3) anticipation: there exist associative mechanisms that enable both
 behavioral and perceptual activity to elicit other perceptual activity
 in the sensory areas of the brain. 
  Most importantly, a simulated action can elicit perceptual activity
 that resembles the activity that would have occurred if the action had
 actually been performed. (p.5).  

  In order to understand the mental state of another
 when observing the other acting, the individual imagines
 herself/himself performing the same action, a covert simulation
 that does not lead to an overt behavior. (p.5). 


 \end{quote}

 The second group of quotes, from Rizzolatti and Arbib (1998), states
the mirror neuron  {\it linguistic hypothesis}, according to which, the
same structures  used for action simulation, are reused to support human
language.

 \begin{quote} 

 Our proposal is that the development of the human
lateral speech circuit is a consequence of the fact that
the precursor of Broca's area was endowed, before
speech appearance, with a mechanism for recognizing
actions made by others. This mechanism was the
neural prerequisite for the development of inter-individual
communication and finally of speech. We
thus view language in a more general setting than one
that sees speech as its complete basis. (Rizzo.p.190). 

 ...a `pre-linguistic grammar'
can be assigned to the control and observation of
actions. If this is so, the notion that evolution could
yield a language system 'atop' of the action system
becomes much more plausible. (p.191). 


 In conclusion, the discovery of the mirror system
suggests a strong link between speech and action representation.
 `One sees a distinctly linguistic way of
doing things down among the nuts and bolts of action
and perception, for it is there, not in the remote
recesses of cognitive machinery, that the specifically
linguistic constituents make their first appearance'. 
 (p.193-194). 
 \end{quote}

 Finally, a third group of quotes, from Ramachandran (2007), states the 
mirror neuron {\it self-awareness hypothesis}, according to which, the
same structures used for action simulation  are reused, over again, to
support abstract concepts related to consciousness and self-awareness. 
 According to this perspective, perhaps the most important of such 
concepts, that of an abstract self-identity or ego, is built upon one's
already developed simulation capability for looking at oneself as if
looking at another individual.

 \begin{quote} 

 I suggest that `other awareness' may have evolved first
and then counter-intuitively, as often happens in evolution, the same
ability was exploited to model one's own mind - what one calls self
awareness. 

 How does all this lead to self awareness? 
 I suggest that self awareness is simply using mirror neurons for
 `looking at myself as if someone else is look at me' (the word `me'
encompassing some of my brain processes, as well). 

 The mirror neuron mechanism - the same algorithm - that originally
evolved to help you adopt another's point of view was turned inward to
look at your own self. This, in essence, is the basis of things like
`introspection'.  

 This in turn may have paved the way for more conceptual types of
abstraction; such as metaphor (`get a grip on yourself').

  \end{quote}

 \subsubsection*{Yoyo Effects and the Human Mind}

 From our analyses in the preceding sections, one should expect, 
as a consequence of the heavy reuse of code under fast development and 
steady evolution, the sporadic occurrence of some mental yoyo problems. 
  Such yoyo effects break the harmonious way in which the same code is  
 (or circuits are) supposed to work as an integral part with several  
 functions used to {\it do}, {\it listen} and {\it answer}, that is,  
 to control action  performance, language communication, and self or
 other kind of awareness. 
  In psychology, many of such effects are known as 
 {\it dissociative phenomena.}  
 For carefully controlled studies of low level dissociative phenomena 
related to corporal action-perception, 
 see Schooler (2002) and Johansson et al. (2008).

 In the following paragraphs we give a glimpse on possible neuroscience
perspectives of some high level  dissociative phenomena.  
 Simulation mechanisms are (re)used to simulate one's actions, 
as well as other agents' actions. Contextualized action 
simulation is the basis for intentional and motivational inference. 
 From there, one can assess even higher abstraction levels such as  
tactical and strategic thinking, or even ethics and morality.     
 But these capabilities must rely on some principle of decomposition, 
that is, the ability to separate, to some meaningful degree, one's own 
mental state from the mental state of those whose
behavior is being simulated.  
 This premise is clearly stated in Decety and Gr\`{e}zes (2005, p.5):

 \begin{quote} 

 One critical aspect of the simulation theory of mind is the idea that in
trying to impute mental states to others, an attributor has to set aside
her own current mental states and substitute those of the target.   

 \end{quote}

 Unfortunately, as seen in the preceding section, the same low level
circuits used to support simulation are also used to support language. 
 This can lead to conflicting requests to use the same resources. 
 For example, verbalization requires introspection, a process that
conflicts with the need to set aside one's own current mental state.
 This conflict leads to  {\it verbal overshadowing} -   
 the phenomenon by which verbally describing or explaining an 
 experienced or simulated situation somehow modifies or impairs
 its correct identification (like recognition or recollection), 
 or distorts its understanding (like contextualization or meaning). 
 Some causes and consequences of this kind of conflict are addressed by 
 Iacoboni (2008, p.270):   

 \begin{quote} 

 Mirror neurons are pre-motor neurons, remember, and thus are cells not 
really concerned with our reflective behavior. 
 Indeed, mirroring behaviors such as the chameleon effect seem implicit, 
automatic, and pre-reflexive. 
 Meanwhile, society is obviously built on explicit, deliberate, reflexive 
discourse. Implicit and explicit mental processes rarely interact; 
indeed, they can even dissociate. (p.270). 

 \end{quote}

 Psychoanalysis can teach us a lot about high level dissociations such as  
 emotional / rational psychological mismatches and 
 individual / social  behavioral misjudgments. 
 For a constructivist perspective of psychotherapy see  
 Efran et al. (1990), and further comments on section 7.

 We end up this section by posing a tricky question capable of inducing 
the most spectacular yoyo bouncings. 
 This provocative question is related to the role played by division
algebras; Goertzel's articles mentioned at the introduction provide a good
source of references. 
 Division algebras capture the structure of eigen-solutions entailed by 
symmetry conditions for the recursively generated systems of 
specular images in a mirror house. 
 The same division algebras are of fundamental importance in many 
physical theories, see Dion et al. (1995), Dixon (1994) and 
Lounesto (2001).
 Finally, division algebras capture the structure of 2-dimensional (complex
numbers) and 3-dimensional (quaternion numbers) rotations and translations
governing human manipulation of objects, see  Hanson (2006).  
 We can thus ask: 
 Do we keep finding division algebras everywhere out there when trying to 
understand the physical universe because we already have the appropriate
hardware  to see them, or is it the other way around?   
 We can only suspect that any trivial choice in the dilemma posed by
this trick question, will only result in an inappropriate answer. 
 We shall revisit this theme at sections 7 and 8.

 \subsection{Mnemes, Memes, Mimes, and all that.} 

 We can make the ladder of hierarchical complexity in the systems analyzed in
the last sections go even further up, as if it did not climb high enough, 
by including new steps in the socio-cultural realms that stand 
above the level of simple or direct inter-individual interaction, 
such as art, law, religion, science, etc.   
 The origami example of section 1 is used by Richard Dawkins as a 
 prototypical {\it meme} or a unit of imitation. 
 The term {\it mneme}, derived from 
 $\mu \nu \eta \mu \eta$, the muse of memory, 
 was used by Richard Semon as a unit of retrievable memory.  
 Yet another variant of this term, {\it mime}, is derived from 
 $\mu \iota \mu \eta \sigma \iota \varsigma$ or imitation.   
 All these terms have been used to suggest a basic model, a single concept, 
 an elementary idea, a memory trace or unit, or to convey related meanings,  
 see Blackmore and Dawkins (1999), Dawkins (1976), van Driem (2007), 
 Schacter (2001), Schacter et al. (1978), and 
 Semon (1904, 1909, 1921, 1923). 

 Richard Semon's theory  was able to capture many important
characteristics  concerning the storage or memorization, retrieval, 
propagation, reproduction and survival of mnemes.   
 Semon was also able to foresee many important details and interconnections, 
at a time when there were no experimental techniques suitable for an 
empirical investigation of the relevant neural processes. 
 Unfortunately, Semon's analysis also suffers from the yoyo effect in some 
aspects. That is not surprising at all given the complexity of the systems 
he was studying and the lack of suitable experimental tools. 
 These yoyo problems were related to some mechanisms,
postulated by Semon, for mnemetic propagation across generations, or
mnemetic  hereditarity. 
 Such mechanisms had a Lamackian character, since they implied the 
possibility of hereditary transmission of learned or acquired
characteristics.   

 In modern Computer Science, the term {\it memetic algorithm} is used 
to describe evolutive programming based on populational evolution by code
(genetic) propagation that combines a Darwinian or selection phase, and
a local optimization or Lamackian learning phase, see Moscato (1989). 
 Such algorithms were inspired by the evolution of ideas and culture in 
human societies, and they proved to be very efficient for solving some
complex combinatorial problems, see Ong et al (2007) and Smith (2007). 
 Consequently, even knowing now, based on contemporary neural science, that
some of the concepts developed by Semon are not appropriate to explain
specific phenomena among those he was studying, he was definitely
postulating, far ahead of his time, some very interesting  and useful
ideas. 

 Nevertheless, for Semon's misfortune, he published his theory at the 
aftermath of the great Darwinian victory over the competing Lamarckian
view in the field of biological evolution. 
 At that time, any perceived contamination by Lamackian ideas was a 
kiss of death for a new theory, even if postulated within a
clearly distinct context. 
 As a regrettable consequence, the mneme concept was rejected and cast
into oblivion  for half a century, until its revival as Dawkin's
meme. 
 Such a drama is by no means unusual in the history of science. 
 It seems that some ideas, postulated ahead of their time, 
 have to be incubated and remain dormant 
 until the world is ready for them.  
 Another example of this kind, related to the concept of statistical
 randomization,  is analyzed in great detail in 
 Chapter 3. 

 \section{Hypercyclic Bootstrapping}

 On march 1st 2009, the Wikipedia definition for bootstrapping read: 

 \begin{quote} 

 Bootstrapping or booting refers to a group of metaphors that share a
common meaning, a {\em self-sustaining} process that proceeds without
external help. The term is often attributed to Rudolf Erich Raspe's
story The Adventures of Baron M\"{u}nchausen, where the main character
pulls himself out of a swamp, though it's disputed whether it was done
by his hair or by his bootstraps. 

 \end{quote}

 The attributed origin of this metaphor, the (literally) incredible  
adventures of Baron M\"{u}nchhausen, well known as a compulsive liar,
makes  us suspect that there may be something wrong with some of its uses.  
 There are, however, many examples where bootstrapping explanations can 
be rightfully applied. Let us analyze a few examples:

 1. The {\it Tostines mystery: 
 Does Tostines sell more because it is always fresh and crunchy, 
 or is it always fresh and crunchy because it sells more?}  

 This slogan was used at a very successful marketing campaign, that 
launched the relatively unknown brand Tostines, from Nestl\'{e}, to a
leading position in the Brazilian market of biscuits, crackers and
cookies. 
 The expression {\it Tostines mystery} became idiomatic in Brazilian
Portuguese, playing a role similar to that of the expression
bootstrapping in English.

 2. The C computer language and the UNIX operating system:  
 Perhaps the most successful and influential computer language ever 
designed, C was conceived having bootstrapping in mind. 
 The core language is powerful but spartan. 
 Many capabilities that are an integral part of other programming 
 languages are provided by functions in external standard libraries, 
 including all device dependent operations such as input-output,  
 string and file manipulation, mathematical computations, etc. 
 C was part of a larger project to write UNIX as a portable operating
 system. 
 In order to have UNIX and all of its goodies into a new machine (device
drivers should already be there), we only have to translate the assembly 
code for a core C compiler, compile a full C compiler, compile the
entire UNIX system, compile all the application programs we want, and
voil\`{a}, we are done. 
 Bootstrapping, as a technological approach, is of fundamental
importance for the computer industry as it allows the development of
evermore powerful software and the rapid substitution of hardware.

 3. The Virtuous cycle of open source software: 
 An initial or starting code contribution is made available at 
 an {\it open source code repository}. 
 {\it Developer communities} can use the resources at the repository
according to the established open source license.  
  Developers create software or application programs
according to their respective business models, affected by the open
source license agreements and the repository governance policy.  
 The use of existing {\it software} motivates new applications or
extensions to the existing ones, generating the development of new
programs and new contributions to the open source repository. 
 Code contributions to the repository are filtered by a  
 {\it controlling committee} according to a governance model. 
 The full development cycle works using the highlighted elements as 
catalysts, and is fuelled by the work of self-interested individuals 
acting according to their own motivations, see Heiss (2007).

  4. The Bethe-Weizs\"{a}cker main catalytic cycle (CNO-I):     
  
  \noindent 
  $\cado +\hy \rightarrow \nitr +\gamma +\mbox{1.95MeV}$;  \ \  
  $\nitr \rightarrow \catr +e^+ +\nu +\mbox{2.22MeV}$;  \\
  $\catr +\hy \rightarrow \niqa +\gamma +\mbox{7.54MeV}$;  \ \ 
  $\niqa +\hy \rightarrow \oxqi +\gamma +\mbox{7.35MeV}$; \\   
  $\oxqi \rightarrow \niqi +e^+ +\nu +\mbox{2.75MeV}$;  \ \  
  $\niqi +\hy \rightarrow \cado +\hequ +\mbox{4.96MeV}$. 

  \noindent    
  This example presents the nuclear synthesis of one atom of Helium 
 from four atoms of Hydrogen. Carbon, Nitrogen and Oxygen act as catalysts 
 in this cyclic reaction, that also produces gamma rays, positrons and 
 neutrinos. Note that the Carbon-12 atom used in the first reaction is 
 regenerated at the last one. 
 The CNO nuclear fusion cycle is the main source of energy in stars 
 with mass twice as large or more than that of the sun.  
 We have included this example from nuclear physics in order to stress
 the fact that catalytic cycles play an important role in phenomena occurring in  
 spatial and temporal scales which are much smaller than those typical of chemistry 
 or biology, where some of the readers may find them more familiar.

 5. RNA and DNA replication:   
 DNA and RNA duplication, translation, and copying may, in general, be 
considered the core cycle of life, since it is the central cycle of
biological reproduction. Although even a simple description 
of this process is far too complex to be included in this book,  
its worth noting that RNA and DNA copy mechanisms rely on many enzymes and
auxiliary structures, which are only available because they themselves are
synthesized or regenerated in the living cell of other, 
also very complex, cyclical networks.

 Examples 4 and 5 are taken from Eigen (1977).  
 Examples 3 and 5 are, in Manfred Eigen's nomenclature, hypercycles. 
 Eigen defines an {\it autocatalytic cycle} as a (chemical) reaction
cycle that, using  additional resources available in its environment,
produces an excess of one or more of its own reactants. 
 A {\it hypercycle} is an autocatalytic reaction of second or higher
order, that is, an autocatalytic cycle connecting autocatalytic units.  
 In a more general context, a hypercycle indicates self-reproduction of 
second or higher order, that is, a second or higher order cyclic
production network including lower order self-replicative units.  
 In the prototypical hypercycle architecture, a lower order 
self-replicative unit plays then a dual catalytic role: 
 First, it has an auto-catalytic function in its own reproduction. 
 Second, it acts like a catalyst promoting an intermediate step of the 
higher order cycle.

 \subsection{Bootstrapping Paradoxes} 

 Let us now examine some ways in which the bootstrapping metaphor is 
wrongfully applied, that is, it is used to generate incongruent or
inconsistent arguments, supposed to accommodate contradictory situations
or to explain the existence of impossible processes. 
 We will focus on four cases of historical interest and great 
epistemological importance.

 \subsubsection*{Perpetua Mobile}  

 Perhaps the best known paradox related to the bootstraping metaphor is
connected to a class of examples known as Perpetuum Mobile machines.  
 These machines are supposed to operate forever without any external help 
or even to produce some useful energy output. 
 Unfortunately, perpetual mobiles are only wishful thinking, since the
existence of such a machine would violate the first, second and third
laws of thermodynamics. 
 These are essentially ``no free lunch'' principles, formulated as 
inequalities for the flow (balance or transfer) of matter, energy and
information in a general system, see Atkins (1984), Dugdale (1996) 
and Tarasov (1988). 
  
 Hypercyclical processes are not magical and must rely on energy, 
information (order or neg-entropy) and raw materials available at their
environment.  
 In fact, the use of external sources of energy and information is so
important, that it entails the definition of metabolism used in 
 Eigen (1977):

 \begin{quote}

 Metabolism: 
 (The process) can become effective only for intermediate states which
are formed from energy-rich precursors and which are degraded to some
energy-deficient waste. The ability of the system to utilize the free
energy and the matter required for this purpose is called metabolism.
 The necessity of maintaining the system far enough from equilibrium by a
steady compensation of entropy production has been first clearly
recognized by Erwin Schr\"{o}dinger (1945). 

 \end{quote}

 The need for metabolism may come as a disappointment to professional
wishful thinkers, engineers of perpetuum mobile machines, narcissistic
philosophers and other anorexic designers. Nevertheless, it is important
to realize that metabolic chains are in fact an integral part of the
hypercycle concept.
 Hypercycles are built upon the possibility that       
 the raw material that is supposed to be freely available in
the environment for one autocatalytic reaction, may very well be the 
product of another catalytic cycle. 
 Moreover, the same thermodynamic laws that prevent the existence of
a perpeuum mobile, are fully compatible with a truly wonderful
property of hypercycles, namely, their almost miraculous efficiency, 
as stated in Eigen (1977):

 \begin{quote}
 Under the stated conditions, the product of the plain catalytic process
will grow linearly with time, while the autocatalytic system will show
exponential growth.
 \end{quote}

 \subsubsection*{Evolutionary View}

 The exponential or hyperbolic (super-exponential) growth of
processes based on auto-catalytic cycles and hypercycles have profound 
evolutionary implications. 
 Populations growing exponentially in environments with limited 
resources, or even with resources growing at a linear or polynomial
rate, find  themselves in the Maltusian conundrum of ever increasing
individual or group competition for evermore scarce resources. 
 In this setting, selection rules applied to a population of individuals
struggling to survive and reproduce inexorably leads to an evolutive
process. This qualitative argument goes back to Thomas Robert Malthus, 
Alfred Russel Wallace, and Charles Darwin, 
 see Ingraham (1982) and Richards (1989).

 Several alternative mathematical models for evolutive processes only
confirm the soundness of the original Malthus-Wallace-Darwin argument.  
 Eigen (1977, 1978a,b) analyses evolutionary processes on the basis of
dynamical systems models using the language of ordinary differential
equations. 
 Stern (2008, ch.5) takes 
 In Chapter 5 we take 
 a completely different approach,
 analyzing evolutionary processes on the basis of stochastic optimization
algorithms using the language of inhomogeneous Markov chains. 
 For other possible approaches see Jantsch and Waddington (1976) 
 and Jantsch (1980, 1981).  
 It is remarkable however, that the qualitative conclusions of 
 these distinct alternative analyses are in complete agreement.

 The evolutionary view replaces a static scenario by a dynamic context.  
 This replacement has the side effect of enhancing or amplifying most 
of the mirror-house illusions studied in this chapter. 
 No wonder then, that the adoption of an evolutionary view requires from
the observer a solid background on well founded scientific theories
together with the  firm domain of a logical and coherent epistemological
framework in order to keep his or her balance and maintain straight
judgment.

 \subsubsection*{Building Blocks and Modularity}

 Another consequence of the analysis of evolutionary processes, 
 using either the dynamical systems approach, 
 see Eigen (1977, 1978a,b), 
 or the stochastic optimization approach, 
 see Chapter 5,  
 is the spontaneous emergence of modular 
 structures and hierarchical organization of complex systems.

 A classic illustration of the need for modular organization is given by
the Hora and Tempus parable of Simon (1996), see also Growney (1982). 
 This is a parable about two watch makers, named Hora and Tempus, both
of whom are respected manufacturers and, under ideal conditions, produce
watches of similar quality and price. Each watch requires the assemblage
of $n=1000$ elementary pieces. However, while Hora uses a hierarchical modular
design, Tempus does not. Hora builds each watch with 10 large
blocks, each made of 10 small modules of 10 single parts each. Consequently, in
order to make a watch, Hora needs to assemble $m=111$ modules with
$r=10$ parts each, while Tempus needs to assemble only $m=1$ module with
$r=1000$ parts. It takes either Hora or Tempus one minute to put a part
in its proper place. Hence, while Tempus can assemble a watch in 1000
minutes, Hora can only do it in 1110 minutes. 
 However both work in a noisy environment, being subject to interruptions 
 (like receiving a telephone call).
 While placing a part an interruption occurrs with probability of $p=0.01$. 
 Partially assembled modules are unstable, braking down at an interruption. 
 Under these conditions, the expected time to assemble a watch is  
 \[ 
   \frac{m}{p}\left( \frac{1}{(1-p)^r} -1 \right) \ .  
 \] 
 Replacing $p$, $m$ and $r$ for the values in the parable, one finds
that Hora's manufacturing process is a few thousand times more efficient
then Tempus'.  
  After this analysis, it is not difficult to understand why Tempus
struggles while Hora prospers.

 Closing yet another cycle, we thus came to the conclusion
that the evolution of complex structures requires modular design. 
 The need for modular organization is captured by the following
dicta of Herbert Simon:  
 \begin{quotation} 
  {\it ``Hierarchy, I shall argue, is one of the central structural
 schemes that the architect of complexity uses.''} \ Simon (1996, p.184).
 

 {\it ``The time required for the evolution of a complex form from
 simple elements depends critically on the number and distribution of
 potential intermediate stable subassemblies.''} \ Simon (1996, p.190).  
 
 
  {\it ``The claim is that the potential for rapid evolution exists in 
any complex system that consists of a set of subsystems, each operating 
nearly independently of the detailed process going on within the other 
subsystems, hence influenced mainly by the net inputs and outputs of 
the other subsystems. If the near-decomposability condition is met, the
efficiency of one component (hence its contribution to organism 
fitness) does not depend on the detailed structure of other
components.''} \ Simon (1996, p.193).
 \end{quotation}

  \subsubsection*{Standards and Once-Forever Choices}

 An important consequence of emerging modularity in evolutive 
processes is the recurrent commitment to once-forever choices and 
the spontaneous establishment of standards.  
 This organizational side effect is responsible for mirror-house effects 
related to many misleading questions leading to philosophical dead-ends.  
 Why do (almost all) nations use the French {\it meter, m,} as the
standard unit of length, instead of the older Portuguese {\it vara}
($\approx 1.1m$) or the British {\it yard} ($\approx 0.9m$)? 
 Why did the automotive industry select 87 octane as ``regular'' 
gasoline and settled for 12V as the standard voltage for vehicles? 
 Why do we have chiral symmetry breaks, that is, why do we find only
one specific type among two or more possible isomeric molecular forms in
organic life?  
 What is so special about the DNA - RNA genetic code that it is shared 
by all living organisms on planet earth?

 In this mirror house we must accept that the deepest truth is often
pretty shallow. Refusing to do so, insisting on extraction by forceps of 
more elaborate explanations, can take us seriously astray into foggy
illusions, far away from clear reason and real understanding.    
 Eigen (1977, p.541-542) makes the following comments:

 \begin{quotation}  

 The Paradigm of Unity and Diversity in Evolution: 
 Why do millions of species, plants and animals, exist,
 while there is only one basic molecular machinery
 of the cell, one universal genetic code and unique
 chiralities of the macromolecules?   

 This code became finally established, not because it
 was the only alternative, but rather due to a peculiar
 `once-forever' selection mechanism, which could
 start from any random assignment. Once-forever selection
 is a consequence of hypercyclic organization. 

 \end{quotation}

 \section{Squaring the Cycle}  

 {\it Ouroboros} is a Greek name, 
 $Ou\rho o\beta o\rho \varsigma \  o \varphi i \varsigma$,    
 meaning the tail-devouring snake, see Eleazar (1760) and Franz (1981).  
 It is also an ancient alchemical symbol of self-reflexive or 
cyclic processes, of something perpetually re-creating itself. 
 In modern cybernetics it is used as a representation of autopoiesis. 
 The ourobouros is represented as a single, integral organism, 
the snake, whose head bites its own tail. 
 This pictorial representation would not make much sense if the 
snake were cut into several pieces, yet, that is what may happen, 
if we are not careful, when trying to explain a cyclic process.

 Let us illustrate this discussion with a schematic representation   
 of the  fiscal cycle of an idealized republic. 
 This cycle is represented by a diagram similar to the one presented 
 in section 7. This square diagram has four arrows pointing, 
 respectively, 
 \\ 
 - Down: Citizens pay taxes to fulfill their duties; 
 \\ 
 - Left: Citizens elect a senate or a house of representatives; 
 \\ 
 - Up: The senate legislates fiscal policies; and 
 \\ 
 - Right: A revenue service enforces fiscal legislation.

 Focusing on each one of the arrows we can speak, respectively, of 
 \\ 
 - Downward causation, whereby individuals comply with established 
 social constraints; 
 \\ 
 - Upward causation, whereby the systems constraints are established and 
 renewed; 
 \\ 
 - Leftward causation, whereby individuals (re)present new demands to 
 the republic; 
 \\ 
 - Rightward causation, whereby the status quo is maintained, 
 stabilized and enforced.

 Each one of these causal relations is indeed helpful to understand 
 the dynamic of our idealized republic. 
 On the other hand, the omission of any single one of these relations 
 breaks the cycle, and such an incomplete version of the schematic 
 diagram would no longer explain a dynamical system.

 The adjectives up and down capture our feelings as an individual living 
under social constraints (like costumes, moral rules, laws and
regulations) that may (seem to)  be overwhelming, 
 while the adjectives left and right are late echoes of the seating
arrangement in the French legislative assembly of 1791, with the 
conservatives,  protecting aristocratic privileges of the ancien
r\'{e}gime, seating on the right and the liberals, voicing the
 laissez-faire-laissez-passer slogans for free market 
capitalism, seating on the left. 
 How to assign intuitive and meaningful positional or directional
adjectives to  links in  a complex network is in general not so obvious. 
 In fact, insisting on similar labeling practices is a common source of 
unnecessary confusion and misunderstanding. 
 A practice that easily generates inappropriate interpretations is
{\it polysemy}, the reuse of the same tags in different contexts. 
 This is due to semantic contamination or spill over, that is, unwanted
or unforeseen transfers of meaning, induced by polysemic overloading.

 We can ask several questions concerning the relative importance 
of specific links in causal networks. For example:   
 Can we or should we by any means establish precedences between the 
links in our diagram? 
 Upward causes precede or have higher status then downward causes 
 or vice versa? 
 Rightward causes explain or have preponderance over leftward causes
 or vice versa? 
 Do any of the possible answers imply a progressive or revolutionary  
view? 
 Do the opposite answers  imply a conservative or reactionary view? 
 The same questions can be asked with respect to a similar diagram for scientific 
production presented in section 7. 
 Do any of the possible answers imply an empiricist or Aristotelic view? 
 Do the opposite answers imply an idealistic or Platonic view?

 To some degree these can be legitimate questions and consequently, 
 to the same degree, motivate appropriate answers.  
 Nevertheless, following the main goal of this chapter, namely, the
exploration  of mirror-house illusions, we want to stress that extreme
forms of  these questions often lead to ill posed problems.  
 Therefore, extreme answers to the same questions often give an  
over simplified, one sided, biased, or distorted view of reality.   
 The dangerous consequences of acceding to the temptation of having
an appetizing ourobourus' slice for supper are depicted, in the field of
psychology, by the  following quotations from Efran (1990, p.99,47):

 \begin{quotation}  

 Using language, any cycle can be broken into causes and purposes... 
 Note that inventing purposes - and they are invented - is usually an 
exercise in creating tautologies. 
 A description is turned into a purpose  that is then asked to account
for the description. 
 [A typical example] starts with the defining  characteristic of
life, self-perpetuation, and states that it is the purpose for which 
the characteristic exists. 
 Such circular renamings are not illegal, but they do not advance the
cause (no pun intended). (p.99)

 For a living system there is a unity between product and process: 
 In other words, the major line of work for a living system is creating 
more of itself. 

 Autopoiesis in neither a promise nor a purpose - it is an organizational 
characteristic. This means that life lasts as long as it lasts. It doesn't 
come with guarantees. In contrast to what we are tempted to believe, 
people do not stay alive because of their strong survival instincts or 
because they have an important job to complete. They stay alive because 
their autopoietic organization happens to permit it. When the essentials 
of that organization are lost, a person's career comes to an end - he or 
she disintegrates. (p.47) 

 \end{quotation}


 \section{Emergence and Asymptotics} 

  Asymptotic entities emerge in a model as a 
 {\it law of large numbers}, that is, as a stable behavior of a
quantity in the limiting case of model parameters corresponding to   
a system with very many (asymptotically infinite) components.   
  The familiar mathematical notation used in these cases takes the form  
  $\lim_{n \rightarrow \infty}   g(n)$ or 
  $\lim_{\epsilon \rightarrow 0} f(\epsilon)$.  
 Typically, the underlying model describes a local interaction in a
small or microscopic scale, while the resulting limit correspond  to a
global behavior in a large or macroscopic scale.

  The paradigmatic examples in this class express the behavior of
thermodynamic variables describing a system, such as volume, pressure
and temperature of a gas, as asymptotic limits in statistical mechanics
models for (infinitely) many  interacting particles, like atoms or
molecules, 
 see Atkins (1984), Tarasov (1988). 
 Other well known examples explain the behavior of macro-econometric
relations among descriptive variables of efficient markets, like
aggregated supply, demand, price  and production, form micro-economic
models for the interaction of individual agents,  
 see Ingrao and Israel (1990). 
 Even organic tissue movements in morphogenesis can be understood as the 
asymptotic limit of local cellular interactions at microscopic scale, 
as already mentioned in section 2. 
  In this section we have chosen to examine some aspects of the collective 
behavior of flocks, schools and swarms, that can be
easily visualized  in a space and time scale directly assessible to our
senses. 

  Large flocks of birds or schools of fish exhibit coordinated 
flight or swimming patterns and manifest collective reaction movements 
that give the impression that the collective entity has 
``a mind of its own''. 
 There are many explanations for why these animals swarm together.   
 For example, they may do so in order to achieve: 
 \\ 
 - Better aerodynamic or hydrodynamic performance  by flying or swimming 
in tight formation, 
 \\ 
 - More efficient detection of needed resources or dangerous threats  
by the pooling of many sensors;  
 \\ 
 - Increased reproductive and evolutive success by social selection  
rules; etc. 
 \\  
 In this section, however, we will focus on another advantage: 
 \\ 
 - Reducing the risk of predation by evasive maneuvers.

  The first point in the analysis of this example is to explain why it
is  a {\it valid example} of emergence,  that is, to describe a
possible local interaction model from which  the global behavior
emerges when the flock has a large number  of individuals. 
 We use the model programmed by  Craig Reynolds (1987).

 \begin{quotation} 

 In 1986 I made a computer model of coordinated animal motion such as
bird flocks and fish schools. It was based on three dimensional
computational geometry of the sort normally used in computer animation
or computer aided design. I called the generic simulated flocking
creatures {\it boids}. The basic flocking model consists of three simple
steering behaviors which describe how an individual boid maneuvers based
on the positions and velocities its nearby flockmates:  
 \\ \indent  
 {\it Separation:} steer to avoid crowding local flockmates  
 \\ \indent    
 {\it Alignment:} steer towards the average heading of local flockmates  
 \\ \indent    
 {\it Cohesion:} steer to move toward the average position of local flockmates  

 Each boid has direct access to the whole scene`s geometric description,
 but flocking requires that it reacts only to flockmates within a certain
 small neighborhood around itself. 

 \end{quotation}

 The second point is to explain why
being part of a flock can reduce the risk of predation: 
 Many predators, like a falcon hunting a sparrow, need to single out 
and focus on a chosen individual in order to strike accurately. 
 However, the rapid change of relative positions of individuals in 
the flock makes it difficult to isolate a single individual as the 
designated target and follow it inside the moving flock. 
 Computer simulation models show that this confusion effect greatly 
reduces the killing (success) rate in this kind of hunt.

 The third point in our analysis is to contrast the hunting of single
individuals, as analyzed in the previous paragraph, with other forms 
of predation based on the capture  of the entire flock, or a large chunk 
of it.   
 The focus of such alternative hunting techniques is, in the relative 
topology of the flock, not on local but on global variables describing 
the collective entity.  
 For example, as explained in Diachok (2006) and Leighton et al. 
 (2004, 2007), humpback whales collaborate using  sophisticated
 strategies for hunting herring, including specific tactics for:  

 {\it Detection:} Whales use active sonar detection techniques, using 
specific frequencies that resonates with and are attenuated by the swim
bladders of the herring. 
 In this way, the whales can detect schools over long distances, and
also measure its  pertinent characteristics.  

 {\it Steering:} Some whales broadcast loud sounds below the herring
school, driving them to the surface. 
 Other whales blow a bubble-net around the school, spiraling in as 
the school rises.  
 The herring is afraid of the loud sounds at the bottom, and also afraid
of swimming through the bubble-net, and is thus forced  into a dense pack
at a compact killing zone near the surface. 

 {\it Capture:} Finally, the whales take turns at the killing zone,
raising to the surface with their mouths wide open, catching hundreds of
fish at a time or, so to speak, ``biting of'' large chunks of the
school.

 Finally, let us propose two short statements that can be distilled 
from our examples. They are going to carry us to the next section.
 
 - Flocking makes it difficult for a predator to use {\it local tactics} 
 tracking the  trajectory of a single individual, consequently, 
 for a hunter that focus on local variables it is 
 {\it hard to know} what exactly is going on.  

 - On the other hand, the same collective behavior creates the
opportunity for {\it global strategies} that track and manipulate 
the entire flock. These hunting technique may be very efficient, 
in which case, we can say that the hunters {\it know very well} what
they are doing.

 \section{Constructive Ontologies}

 From the several examples mentioned in sections 2, 4 and 6, we can
suspect that the emergence of properties,  behaviors, organizational
forms and other entities are the rule rather than the exception for 
many non-trivial systems.  
 Hence it is natural to ask about the ontological status of such
entities. 
 Ontology is a term used in philosophy referring to a systematic
account of {\it existence} or {\it reality}. 
 In this section we analyze the ontological status of emergent 
entities according to the Cog-Con epistemological framework. 
 The following paragraphs give a brief summary of this perspective,  
as well as some specific epistemological terms as they are used in the
Cog-Con framework.

 The interpretation of scientific knowledge as an eigensolution of a
research process is part of a Cog-Con approach to epistemology.
 Figure 1 presents an idealized structure and dynamics of knowledge
production. 
 This diagram represents, on the Experiment side (left column) the  
laboratory or field operations of an empirical science, where experiments
are designed and built, observable effects are generated and
measured, and the experimental data bank is assembled. 
 On the Theory side (right column), the diagram represents the
theoretical work of statistical analysis, interpretation and (hopefully)
understanding according to accepted patterns. If necessary, new
hypotheses (including whole new theories) are formulated, motivating the
design of new experiments. 
 Theory and experiment constitute a double feed-back cycle making it clear
that the design of experiments is guided by the existing theory and its
interpretation, which, in turn, must be constantly checked, adapted or
modified in order to cope with the observed experiments. The whole
system constitutes an autopoietic unit.


 \begin{table}[bt] 
 \begin{center}
 \begin{tabular}{c c c c c} 

 Experiment & &     & & Theory \\  \\   

 Operation-  & $\Leftarrow$ & Experiment  

                       &  $\Leftarrow$ &  Hypotheses   \\ 
 alization & & design   & &  formulation                \\ 
   $\Downarrow$   & &               & & $\Uparrow$        \\  
 Effects       &  \multicolumn{3}{c}{True/False} & Creative  \\ 
 observation    &  \multicolumn{3}{c}{eigen-solution} & interpretation  \\     

   $\Downarrow$   & &               & & $\Uparrow$         \\ 
 Data    & & Mnemetic & &  Statistical   \\ 
 acquisition  & $\Rightarrow$ & explanation  
 & $\Rightarrow$ & analysis \\  \\ 
 
 \multicolumn{2}{l}{Sample space} & & 
 \multicolumn{2}{r}{Parameter space} 
 \end{tabular} 
 \mbox{} \\ \mbox{} \\ 
 \centerline{Figure 1: Scientific production diagram.}
 \end{center}    
 \end{table}

 The Cog-Con framework also includes the following definition of
reality and some related terms: 

 \begin{quotation} 
 \noindent {\it 1. Known (knowable) Object:} An actual (potential)
eigen-solution of a given system's interaction with its environment. 
 In the sequel, we may use a somewhat more friendly terminology by 
simply using the term Object.
 
 \noindent {\it 2. Objective (how, less, more):} Degree of conformance
of an object to the essential attributes of an eigen-solution 
 (to be precise, stable, separable and composable). 

 \noindent {\it 3.  Reality:} A (maximal) set of objects, as
recognized by a given system, when interacting with single objects or
with compositions of objects in that set. 
 \end{quotation} 

 The Cog-Con framework assumes that an object is 
always observed by an observer, just like a living organism or a more 
abstract system, interacting with its environment. 
 Therefore, this framework asserts that the manifestation of the
corresponding eigen-solution and the properties of
the object are respectively driven and specified by both the system and its environment. 
 More concisely, Cog-Con sustains: 

 \begin{quotation}
 \noindent {\it 4. Idealism:} The belief that a system's knowledge of an 
object is always dependent on the systems' autopoietic relations. 

 \noindent {\it 5. Realism:} The belief that a system's knowledge of an 
object is always dependent on the environment's constraints. 
 \end{quotation}

  Consequently, the Cog-Con perspective requires a fine equilibrium, 
called  {\it Realistic or Objective Idealism}. 
 {\it Solipsism or Skepticism} are symptoms of an epistemological
analyses that looses the proper balance by putting too much weight on 
the idealistic side. Conversely, 
 {\it Dogmatic Realism} is a symptom of an epistemological
analyses that looses the proper balance by putting too much weight on 
the realistic side.  
 Dogmatic realism has been, from the Cog-Con perspective, 
a very common (but mistaken) position in modern epistemology. 
 Therefore, it is useful to have a specific expression, namely, 
 {\it something in itself} to be used as a marker or label for such ill posed dogmatic statements. 
 The method used to access   
 something in itself is often described as: 
 - Something that an observer would observe if the (same)
observer did not exist, or   
 - Something that an observer could observe if he made no
observations, or  
 - Something that an observer should observe in the environment
without interacting with it (or disturbing it in any way),      
 and many other equally senseless variations.

 From the preceding considerations, it should become clear that, from the 
Cog-Con perspective, the ontological status of emergent entities can be 
perfectly fine, as long as these objects correspond to precise, stable,
separable and composable eigen-solutions. 
 However there is a long list of historical objections and complaints 
concerning such entities. 
 The following quotations from Pihlstrom and El-Hani (2002)  
 elaborate on this point. 

 \begin{quotation} 
 Emergent properties are not metaphysically real independently of our
practices of inquiry but gain their ontological status from the
practice-laden ontological commitments we make.

 [concerning] the issue of the ontological epistemological status of
emergents ... we simply need to be careful in our recognition of
emergent phenomena and continually ask the question of whether the
pattern we see is more in our eye than the pattern we are claiming to
see.

 Related to the supposed provisionality of emergents is the issue of
their ontological status: Are emergent phenomena part of the real,
authentic ``furniture of the world'', or are they merely a function of
our epistemological, cognitive apparatus with its ever-ready mechanism
of projecting patterns on to the world? 
 \end{quotation}

 From the summary of the Cog-Con epistemological
framework  presented above we conclude that, from this perspective,  
we have to agree with the first observations, and consider the last
question as an ill posed problem.

 Another set of historical issues concerning the ontological status of 
emergents relates to our ways of understanding them. 
 For some authors, ``real'' emergent entities must be genuinely ``new'', 
in the sense of being unanalyzable or unexplainable. 
 For such authors, understanding is a mortal sin that  threatens the very
existence on an entity, that is, understanding  undermines their
ontological status.  
 Hence, according to these authors, the most real of entities should
always be somewhat mysterious. 
  Vieira and El-Hani (2009, p.105), analyze this position:

 \begin{quotation} 

 A systemic property P of a system S will be irreducible if it does not 
follow, even in principle, from the behavior of the system's parts that 
S has property P.  

 If a phenomenon is emergent by reasons of being unanalyzable, it will be 
an unexplainable, brute fact, or, to use Alexander's (1920/1979) words, 
something to be accepted with natural piety. We will not be able to 
predict or explain it, even if we know its basal conditions. 

In our view, if the understanding of the irreducibility of emergent 
properties is limited to this rather strong sense, we may lose from 
sight the usefulness of the concept... 
 Indeed, claims about emergence turn out to be so strong, if interpreted  
exclusively in accordance with this mode of irreducibility, that they 
are likely to be false, at least in the domain of natural science (with 
are our primary interest in this chapter). 

 \end{quotation}

 We fully agree with Vieira and El-Hani in rejecting unanalyzability or 
unexplainability as conditions for the ``real existence'' of emergent
entities.  
 As expected, the Cog-Con framework does not punish
understanding, far from it. 
 In Chapter 4, 
 the Cog-Con perspective for the meaning of objects in a specific
reality is given by their interrelation in a network of causal nexus,
explaining  {\it why} the corresponding  eigen-solutions manifest
themselves the way they do. 
 Such explanations include, specially in modern science, the symbolic
derivation of scientific hypotheses from general scientific laws, the 
formulation of new laws in an existing theory, and even the conception
of new theories, as well as their general understanding based on
accepted metaphysical principles.    
 In the Cog-Con perspective, the understanding of an entity can
only strengthen its ontological status,   
embedding it even deeper in the system's life, 
endowing it with even wider connections in the web of concepts,  
revealing more of its links with the great chain of being!

 \section{Distinctions and Probability}

 In the last two sections we have analyzed emergent objects and their  
properties. In many of the examples used in our discussions, 
probability mechanisms where at the core of the emergence process. 
 In this section, other ways in which probability mechanisms 
can generate complex or non-trivial structures will be presented. 
 This section is also dedicated to the study of  the ontological status
of probability, and the role played by  explanations given by
probabilistic mechanisms and  stochastic causal relations.    
 We begin our discussion examining the concept of mixed strategies 
in game theory, due to von Neumann and Morgenstern.  
 
  Let us consider the {\it matching pennies} game, played by Odd and Even. 
  Each of the players has to show, simultaneously, a bit (0 or 1). 
  If both bits agree (i.e., 00 or 11), Odd wins. 
  If both bits disagree (i.e., 01 or 10), Even wins. 
  Both players only have two pure or deterministic strategies available 
 from which to choose: 
 $s_0$ - show a 0, or $s_1$ - show a 1. 

  A {\it solution, equilibrium} or {\it saddlepoint} of a game 
is a set of strategies that leaves each player at a local optimum, 
 that is, a point at which each player, having full knowledge of all
the other players' strategies at that equilibrium point, has nothing
to gain by unilaterally changing his own strategy. 
 It is easy to see that, considering only the two deterministic
strategies, the game of matching pennies has no equilibrium  point. 
 If Odd knows the strategy chosen by Even, he can just take the 
same strategy and win the game. In the same way, Even can take the 
opposite choice of Odd's, and win the game. 

 Let us now expand the set of strategies available to each player 
considering {\it mixed} or {\it randomized} strategies, where each 
player picks among the pure strategies according to a set of
probabilities he specifies. 
 We assume that a proper randomization device, like a dice, a  roulette
or a computer with a random number generator program, is available. 
 In the example at hand, Even and Odd can each specify a probability,   
respectively, $pe$ and $po$, for showing a $1$, and 
 $qe= 1-pe$ and $qo= 1-po$, for showing a $0$.  
 It is easy to check that $pe=po=1/2$ is a solution to this game.

 Oskar Morgenstern (2008, p.270) makes the following comments about the
philosophical significance of mixed strategies:

 \begin{quotation} 

 It is necessary to examine the significance of the use of mixed
strategies since they involve probabilities in situations in which
`rational' behavior is looked for. It seems difficult, at first, to
accept the idea that `rationality' - which appears to demand a clear,
definite plan, a deterministic resolution - should be achieved by the
use of probabilistic devices. Yet precisely such is the case.

 In games of chance the task is to determine and then to evaluate
probabilities inherent in the game; in games of strategy we introduce
probability in order to obtain the optimal choice of strategy. This is
philosophically of some interest. 

 \end{quotation}

 The role played by mixed strategies can be explained, at least in part, 
by convex geometry. 
 A {\it convex combination} of two points, $p_0$ and $p_1$, 
 is a point lying on the line segment joining them, 
 that is, a point of the form   
 $p(\lambda)= (1-\lambda)p_0 +\lambda p_1$, $0\leq \lambda \leq 1$. 
 A {\it convex set} is a set that contains all convex combinations 
of its points.  
 The {\it extreme points} of a convex set are those that can not be 
expressed as (non-trivial) convex combinations of other points in the set. 
 A function $f(x)$ is convex if its epigraph, $\mbox{epi}(f)$ - the set
of all point above the graph of $f(x)$, is convex. 
 A {\it convex optimization problem} consists of minimizing a convex 
function over a convex region. 
 The properties of convex geometry warrant that a convex optimization
problem  has an optimal solution, i.e. a minimum, $f(x^*)$. 
 Moreover, the minimum argument, $x^*$, is easy to compute using a
procedure such as the {\it steepest descent algorithm}, that can be
informally stated as  follows:
 Place  a particle at  some point over the graph of $f(x)$, and let it
``roll down the hill'' to the bottom of the valley, until it finds its
lowest point at $x^*$,   
 see Luenberger (1984) and Minoux (1986).  

 In the matching pennies game, let us consider a convex combination of 
the two pure strategies, that is, a strategy of the form  
 $s(\lambda)= (1-\lambda)s_0 +\lambda s_1$, $0\leq \lambda \leq 1$. 
 Since the pure strategies form a discrete set, such continuous combination 
 of pure strategies is not even
well defined, except for the trivial extreme cases, 
 $\lambda=0$ or $\lambda=1$. 
 The introduction of randomization gives a coherent definition for 
convex combinations of existing strategies and, in so doing, it expands
the set of available (mixed) strategies to a convex set where pure
strategies become extreme points. 
 In this setting, a game equilibrium point can be characterized as
the solution of a convex optimization problem.  
 Therefore, such an equilibrium point exists and is easy to compute.   
 This is one way of having a geometric understanding of von Neumann and 
Morgenstein theorems, as well as to subsequent extensions in game
theory due to John F. Nash, 
 see Bonassi et al. (2009), Mesterton-Gibbons (1992) and Thomas (1986).

 The matching pennies example poses a  
 $\delta \iota \lambda \eta \mu \mu \alpha$, dilemma - 
 a problem offering two possibilities, none of which is acceptable. 
 The conceptual dichotomy created by constraining the players to only
two deterministic strategies creates an ambush.  
 Caught in this ambush, both players would be trapped,
forever changing their minds between extreme options. 
 Randomization expands the universe of available possibilities and, in
so doing, allows the players to escape the perpetual flip-flopping at
this discrete logic decision trap. 
 In section 8.2, we extrapolate this example and generalize
these conclusions.
 However, before proceeding in this direction, we shall analyze, in the
next section, some objections to the  concepts of probability, statistics
and randomization posed by George Spencer-Brown, a philosopher of great
influence in the  field of radical constructivism.

 \subsection{Spencer-Brown, Probability and Statistics}

  Spencer-Brown (1953, 1957) analyzed some apparent paradoxes involving
the concept  of randomness, and concluded that the language of probability 
and statistics is inappropriate  for the practice of scientific 
inference. 
 In subsequent work, Spencer Brown (1969) reformulates classical logic
 using  only a generalized \textit{nor} operator 
 (marked \textit{not-and} unmarked \textit{or}), 
 that he represents \`{a} la mode of Charles Saunders Peirce or John Venn, 
 using a graphical boundary or  distinction mark,   
 see Edwards (2004), Kauffmann (2001, 2003), Meguire (2003), 
 Peirce (1880), Sheffer (1913).  
 Making distinctions is, according to Spencer-Brown, the basic (if not
the only) operation of human knowledge, an idea that has either influenced or 
been directly explored by several authors in the radical constructivist
movement.  
 Some typical arguments used by Spencer-Brown in his rejection of
probability and statistics are given in the next  quotations from 
Spencer-Brown (1957, p.66,105,113):

 \begin{quotation} 

  We have found so far that the concept of probability used in
statistical science is meaningless in its own terms; but we have found
also that, however meaningful it might have been, its meaningfulness
would nevertheless  have remained fruitless because of the impossibility
of gaining information from experimental results, however significant
 This final paradox, in  some ways the most beautiful, 
 I shall call the Experimental Paradox (p.66).    

 The essence of randomness has been taken to be absence of pattern.  
 But has not hitherto been faced is that the absence of one pattern 
logically demands the presence of another. It is a mathematical
contradiction to say that a series has no pattern; the most we can say
is that it has no pattern that anyone is likely to look for. 
 The concept  of randomness bears meaning only in relation to the
observer: If two  observers habitually look for different kinds of
pattern they are bound to disagree upon the series which they call
random. (p.105). 



 
 \end{quotation}

  In Section G.1 
 I carefully explain why I disagree with 
 Spencer-Brown's analysis of probability and statistics. 
  In some of my arguments I dissent from Spencer-Brown's  
 interpretation of measures of order-disorder in sequential signals.     
  These arguments are based on information theory and the notion of 
 entropy. 
  Atkins (1984), Attneave (1959), Dugdale (1996), Krippendorff (1986) 
 and Tarasov (1988) review some of the basic concepts 
 in this area using only elementary mathematics. 
  For more advanced works see  
   Kapur (1989), Rissanen (1989) and Wallace (2005). 
  %
  Several authors concur, at least in part, with my opinion  
 about Sencer-Brown's analysis of probability and statistics, 
 see Flew (1959), Falk and Konold (1997), Good (1958) 
 and Mundle (1959).

  I also  disapprove some of Spencer Brown's proposed 
 methodologies to detect ``relevant'' event sequences, that is, 
 his criteria to ``mark distinct patterns'' in empirical observations. 
  My objections have a lot in common with the standard caveats against 
 {\it ex post facto} ``fishing expeditions'' for interesting outcomes, 
 or simple {\it post hoc} ``sub-group analysis'' in experimental 
 data banks. 
  This kind of retroactive or retrospective data analyses is considered a 
 questionable statistical practice, and pointed as the culprit of 
 many misconceived studies, misleading arguments and mistaken conclusions.  
  The literature of statistical methodology for clinical trials has been 
 particularly active in warning against this kind of practice,  
 see Tribble (2008) and Wang (2007) for two interesting papers addressing  
 this specific issue and published in high impact medicine journals  
 less than a year before I began writing this chapter.  
  When consulting for pharmaceutical companies or advising in the design 
 of statistical experiments, I often find it useful to quote 
 Conan Doyle's Sherlock Holmes, in The Adventure of Wisteria Lodge: 
 \begin{quote} 
  Still, it is an error to argue in front of your data. 
  You find yourself insensibly twisting them around to fit your theories.
 \end{quote}

  Finally, I am suspicious or skeptical about some of the intended 
 applications of Spencer-Brown's research program, including the use of  
 extrasensory empathic perception for coded message communication,   
 exercises on object manipulation using paranormal powers, etc.    
  Unable to reconcile his psychic research program with statistical  
 science, Spencer-Brown had no regrets in disqualifying the later, 
 as he clearly stated at the prestigious scientific journal 
 {\it Nature}, Spence-Brown (1953b, p.594-595): 
 \begin{quote} 
  [On telepathy:] Taking the psychical research data (that is, the
residuum when fraud and incompetence are excluded), I tried to show that
these now threw  more doubt upon existing pre-suppositions in the theory
of probability  than in the theory of communication.

 [On psychokinesis:] If such an `agency' could thus `upset' a process of
randomizing, then all our conclusions drawn through the statistical
tests of  significance would be equally affected, including the the
conclusions about  the `psychokinesis' experiments themselves. (How are
the target numbers for  the die throws to be randomly chosen? By more
die throws?) 
 To speak of an `agency' which can `upset' any process of randomization
in  an uncontrollable manner is logically equivalent to speaking of an 
inadequacy in the theoretical model for empirical randomness, like  the
luminiferous ether of an earlier controversy, becomes, with the 
obsolescence of the calculus in which it occurs, a superfluous term. 
 \end{quote}

  Sencer-Brown's (1953, 1957) conclusions, including his 
 analysis of probability, were considered to be controversial 
 (if not unreasonable or extravagant) even by his own colleagues at 
 the Society of Psychical Research, see  Scott (1958), and Soal (1953).  
  It seems that current research in this area, even if not free 
 (or afraid) of  criticism, has abandoned the path of  na\"{\i}ve
 confrontation with statistical science, 
 see Atmanspacher (2005) and Ehm (2005).     
  For additional comments, see Henning (2006), Kaptchuk and Kerr (2004), 
 Utts (1991), and Wassermann (1955).

   Curiously, Charles Saunders Peirce and his student Joseph Jastrow, 
 who introduced the idea of randomization in statistical trials, 
 struggled with some of the very same dilemmas faced by Spencer-Brown,  
 namely, the eventual detection of distinct patterns or seemingly 
 ordered (sub)strings in a long random sequence. 
  Peirce and Jastrow did not have at their disposal the heavy 
 mathematical artillery I have cited in the previous paragraphs. 
  Nevertheless, like experienced explorers that when traveling in the 
 desert are not lured by the mirage of a misplaced oasis, 
 these intrepid pioneers were able to avoid the conceptual pitfalls 
 that lead Spencer-Brown so far astray. 
  For more details see Bonassi et al. (2008), Dehue (1997), Hacking (1988), 
 and Peirce and Jastrow (1885). 

 As stated in the introduction, the Cog-Con framework is supported by 
the FBST, a formalism based on a non-decision theoretic form of Bayesian
statistics. 
 The FBST was conceived as a tool for validating objective knowledge
and, in this role, it can be easily integrated to the Cog-Con
epistemological framework in the practice of scientific research. 
 Contrasting our distinct views of cognitive constructivism, it is not
at all surprising that I have come to conclusions  concerning the use of
probability and statistics, and also to the relation between probability and
logic, that are fundamentally different from those of Spencer-Brown.

 \subsection{Overcoming Dilemmas and Conceptual Dichotomies}

 As stated by William James, our ways of understanding require us to 
split reality with conceptual distinctions. 
 The non-trivial consequences of the resulting dichotomies are captured,
almost poetically, by James (1909, Lecture VI) in the following passage from  
 {\it A Pluralistic Universe}:

 \begin{quotation} 
 The essence of life is its continuously changing character; but our
concepts are all discontinuous and fixed, and the only mode of making
them coincide with life is by arbitrarily supposing positions of
arrest therein. With such arrests our concepts may be made congruent.
But these concepts are not parts of reality, not real positions
taken by it, but suppositions rather, notes taken by ourselves, and
you can no more dip up the substance of reality with them than you can
dip up water with a net, however finely meshed.

 When we conceptualize, we cut out and fix, and exclude everything
but what we have fixed. A concept means a that-and-no-other.
Conceptually, time excludes space; motion and rest exclude each other;  
approach excludes contact; presence excludes absence; unity excludes
plurality; independence excludes relativity; `mine` excludes `yours`; 
this connection excludes that connection - and so on indefinitely; 
whereas in the real concrete sensible flux of life experiences
compenetrate each other so that it is not easy to know just what is
excluded and what not... 


 The conception of the first half of the interval between Achilles and
the tortoise excludes that of the last half, and the mathematical
necessity of traversing it separately before the last half is traversed
stands permanently in the way of the last half ever being traversed.
Meanwhile the living Achilles... asks no leave of logic.



 \end{quotation}

 Sure enough, our way of understanding requires us to make those conceptual 
distinctions that are most adequate (or adequate 
enough) for a given reality domain. 
 However, the concepts that are appropriate to analyze reality at a given 
level, scale or granularity, may not be adequate at the next level, 
that may be lower or higher, larger or smaller, coarser or finer. 
 How then can we avoid being trapped by such distinctions? 
 How can we overcome the distinctions made at one level in order to be 
able to reach the next, and still maintain a coherent or congruent view
of the universe?  

 The Cog-Con endeavor requires languages and mechanisms  to
overcome  the limitations of conceptual distinctions and,  at the same
time, enable us to coherently build new concepts that can be
used at the next or new domains. 
 Of course, as in all scientific research, the goal of the new
conceptual  constructs is to entail theories and hypotheses providing
objective  knowledge (in its proper domain), and the success of the new
theories must  be judged pragmatically according to this goal. 
 I claim that statistical models and their corresponding probabilistic
mechanisms, have been, in the history of modern science, among the most
successful tools for accomplishing the task at hand. 
 In Chapter 5, 
 for example, we have shown in some detail how
 probabilistic reasoning can be used: 

 - In quantum mechanics, using the language of Fourier series and
transforms, to overcome the dilemmas posed by a physical theory using
concepts and laws coming from two distinct and seemingly incompatible
categories:  The mechanics of discrete particles and wave propagation in
continuous media or fields.  

 - In stochastic optimization, using the language of inhomogeneous
Markov chains, to overcome the dilemmas generated by dynamic populations
of individuals with the need of reliable reproduction, hierarchical
organization, and stable building blocks versus the need of creative
evolution with innovative change or mutation.

 In an empirical science, from a pragmatical perspective, probability
reasoning seems to be an efficient tool for overcoming artificial 
dichotomies, allowing us to bridge the gaps created by our own
conceptual distinctions.  
 Such probabilistic models have been able to generate new 
eigen-solutions with very good characteristics, that is, eigen-solutions
that are very objective (precise, stable, separable and composable). 
 These new objects can then be used as stepping stones or building
blocks for the construction of new, higher order theories. 
 In this context, we thus assign, coherently with the Cog-Con epistemological
framework, a high ontological status to 
probabilistic concepts and causation mechanisms, that is, 
we use a notion of probability that has a distinctively
 {\it objective} character.

 \section{Final Remarks and Future Research}

 The objective of this chapter was to use the Cog-Con framework  
 for the understanding of massively complex and non-trivial systems. 
 We have analyzed several forms of system complexity, several ways in
which systems become non-trivial, and some interesting  consequences,
side effects and paradoxes generated by such non-triviality.  
 How can we call the massive non-triviality found in nature? 
 I call it {\it The Living and Intelligent Universe.}  
 I could also call it {\it Deus sive natura} or,  
 according to Einstein, 
 \begin{quote}    
 {\it Spinoza's God, a God who reveals himself in the orderly  
  harmony of what exists...}  
 \end{quote}

 In future research we would like to extend the use of the same
Cog-Con framework to the analysis of the ethical conduct of 
agents that are conscious and (to some degree) self-aware. 
 The  definition of ethics given by Russell (1999, p.67), 
 reads:  
  \begin{quote}    
 {\it The problem of Ethics is to produce a harmony and 
  self-consistency in conduct, 
   but mere self-consistency within the limits of the individual 
   might be attained in many ways. There must therefore, to make 
   the solution definite, be a universal harmony; my conduct must 
   bring satisfaction not merely to myself, but to all whom it 
   affects, so far as that is possible. 
  } 
  \end{quote}      
 Hence, in this setting, such a research program should be concerned with 
the understanding and evaluation of choices and decisions made by 
agents, acting in a system in which they belong. 
 Such an analysis should provide criteria for addressing the coherence
and  consistency of the behavior of such agents, including the  direct,
indirect and reflexive consequences of their actions.     
 Moreover, since we consider conscious agents, their values, beliefs and 
ideas should also be included in the proposed models.  
 The importance of pursuing this line of research, and also the  inherent
difficulties of this task, are summarized by 
 Eigen (1992, p.126):  
 
 \begin{quote} 
 But long and difficult will be our ascent from the lowest landing up 
to the topmost level of life, the level of self-awareness: 
our continued ascent from man to humanity. 
 \end{quote}

 Goertzel (2008) points to generalizations of standard probabilistic and
logical formalisms,  and urges us to explore further connections between 
them, see for example Borges and Stern (2007), Caticha (2008), 
Costa (1986, 1993), Jaynes (1990), Stern (2004) and Youssef (1994, 1995).   
 I am fully convinced that this path of cross fertilization between 
probability and logic is another important field for future research.

%% file: CAPEPI.TEX
 \chapter*{Epilog}
 \addcontentsline{toc}{chapter}{Epilog}        
 \markboth{EPILOG}{EPILOG}

  In six chapters and ten appendices, we have presented our case in
defense of a constructivist epistemological framework and the use of
compatible statistical theory and inference tools. In this final remarks, 
we shall try to wrap up, as concisely as possible, the reasons for 
adopting the constructivist world-view.

 The basic metaphor of decision theory is the maximization of a
gambler's expected fortune, according to his own subjective utility,
prior beliefs an learned experiences. This metaphor has proven to be
very useful, leading the development of Bayesian statistics since its
XX-th century revival, rooted on the work of de Finetti, Savage and
others.      
   
 The basic metaphor presented in this text, as a foundation for
cognitive constructivism, is that of an eigen-solution, and the
verification of its objective epistemic status. The FBST is the
cornerstone of a set of statistical tolls conceived to assess the
epistemic value of such eigen-solutions, according to their four
essential attributes, namely, sharpness, stability, separability and
composability. We believe that this alternative perspective,
complementary to the one offered by decision theory, can provide
powerful insights and make pertinent contributions in the context of
scientific research.  

 To fulfill our promise of concision, we finish here this summer course
/ tutorial. We sincerelly thank the readers for their attention and
welcome their constructive comments. May the blessings of the three holy
knights in Figure J.2-4 protect and guide you in your way. Fair well and
goodbye!

  \vfill 

  \pagebreak 

  \mbox{} 

  \pagebreak

  \mbox{} \\   \mbox{} \\  \mbox{} \\  \mbox{} \\

  {\flushright

 {\it ``E aquela era a hora do mais tarde. \\ 
 O c\'{e}u vem abaixando. Narrei ao senhor. \\ 
 No que narrei, o senhor talvez at\'{e} ache, \\ 
 mais do que eu, a minha verdade. \\ 
 Fim que foi.'' 

 And it was already the time of later on, \\  
 the time of sun-down. My story I have told, \\  
 my lord, so that you may find, perhaps even \\ 
 better than me, the truth I wanted to tell. \\ 
 The End (that already was).

 \mbox{} \\ 
 
 ``Vivendo, se aprende; mas o que se aprende,  \\ 
 mais, \'{e} s\'{o} a fazer outras maiores perguntas.''} 

 Living one learns, but what one learns, \\ 
 is only how to ask even bigger questions. 
 
 \mbox{} \\ 

 Jo\~{a}o Guimar\~{a}es Rosa (1908-1967). \\ 
 Grande Sert\~{a}o: Veredas. 

 }

%% file: CAPREF.TEX
 

\chapter*{References}
\addcontentsline{toc}{chapter}{References}        
\markboth{REFERENCES}{REFERENCES} 

 \renewcommand{\baselinestretch}{0.92}
 \parskip 0.09cm
 \begin{small} 

 \rr E. Aarts, J. Korst (1989).
 {\it Simulated Annealing and Boltzmann Machines}.
 Chichester: John Wiley. 

 \rr J.Abadie, J.Carpentier (1969). Generalization of Wolfe Reduced 
 Gradient Method to the Case of Nonlinear Constraints. 
 p.37-47 in R.Flecher (ed) {\it Optimization}. London: Academic Press. 

 \rr K.M.Abadir, J.R.Magnus (2005).  
  {\it Matrix Algebra.}  Cambridge University Press. 

 \rr J.M.Abe, B.C.Avila, J.P.A.Prado (1998). 
 {\it Multi-Agents and Inconsistence.} 
 ICCIMA'98. 
 2nd International Conference on Computational Intelligence and 
 Multimidia Applications. Traralgon, Australia.

 \rr S.Abe, Y.Okamoto (2001). {\it Nonextensive Statistical Mechanics 
 and Its Applications.} NY: Springer. 

 \rr R.P.Abelson (1995). {\it Statistics as Principled Argument.} LEA. 

 \rr Abraham Eleazar (1760). {\it Uraltes chymisches Werk.} 
  2nd ed. Leipzig. 
 
 \rr P.Achinstein (1965). Theoretical Models. {\it British Journal
  for the Philosophy of Science,} 16, 102-20. 

  \rr P.Achinstein (1968). {\it Concepts of Science. A Philosophical 
  Analysis.} Baltimore. 

 \rr D.H. Ackley (1987).
 {\it A Connectionist Machine for Genetic Hillclimbing}.
 Boston: Kluwer.

 \rr J.Acz\'{e}l (1966). 
 {\it Lectures on Functional Equations and their Applications.} 
 NY:  Academic Press. 

 \rr P.Aczel (1988). {\it Non-Well-Founded Sets.} 
 Stanford, CA: CSLI - Center for the Study of language and Information.   

 \rr P.S.Addison (1997). {\it Fractals and Chaos: 
  An Illustrated Course.} Philadelphia: Institute of Physics. 

 \rr D.Aigner, K.Lovel, P.Schnidt (1977). Formulation and Estimation 
 of Stachastic Frontier Production Function Models. {\it Journal of 
 Econometrics}, 6, 21--37. 

 \rr J.Aitchison (2003). {\it The Statistical Analysis for Compositional 
 Data} (2nd edition). Caldwell: Blackburn Press. 

 \rr J.Aitchison,  S.M.Shen (1980). Logistic-Normal Distributions: Some 
Properties and Uses. {\it Biometrika}, 67, 261-72. 

 \rr H.Akaike (1969). Fitting Autoregressive Models for Prediction. 
 {\it Ann. Inst. Stat. Math,} 21, 243--247. 

 \rr H.Akaike (1970). Statistical Prediction Identification. 
 {\it Ann. Inst. Stat. Math,} 22, 203--217.  

 \rr H.Akaike (1974). A New Look at the Statistical Model 
 Identification. {\it IEEE Trans. Autom. Control.} 19, 716--723.  

 \rr J.H.Albert (1985). Bayesian Estimation Methods for Incomplete
Two-Way Contingency Tables using Prior Belief of Association, in
Bayesian Statistics 2:589-602, Bernardo, JM; DeGroot, MH; Lindley, DV;
Smith, AFM eds. Amsterdam, North Holland.

 \rr J.H.Albert, A.K.Gupta (1983). Bayesian Estimation Methods for 2x2
Contingency Tables using Mixtures of Dirichlet Distributions. 
 {\it JASA} 78, 831-41.

 \rr D.Z.Albert (1993). {\it Quantum Mechanics and Experience.}  
  Harvard University Press.  

 \rr R.Albright, J.Cox, D.Dulling, A.N.Langville, C.D.Meyer (2006). 
 Algorithms, Initializations, and Convergence for the Nonnegative 
 Matrix Factorization. 

 \rr J.Alcantara, C.V.Damasio, L.M.Pereira (2002).
 Paraconsistent Logic Programs.  
 JELIA-02. 8th European Conference on Logics in Artificial Intelligence. 
 {\it Lecture Notes in Computer Science,} 2424, 345--356. 

 \rr G.J.Alexander, J.C.Francis (1986). {\it Portfolio Analysis.} 
  Englewood Cliffs, NJ: Prentice-Hall.  

 \rr G.W.Allport, H.S.Odbert (1936). Trait Names: A Psycho-Lexical
  Study. {\it Psychological Monographs}, 47, No.211. 

  \rr S.I.Amari, O.E.Barndorff-Nielsen, R.E.Kass, S.L.Lauritzen, 
  C.R.Rao (1987). {\it Differential Geometry in Statistical
  Inference.}  IMS Lecture Notes Monograph, v.10. 
  Inst. Math. Statist., Hayward, CA.

 \rr S.I.Amari (2007). {\it Methods of Information Geometry.} 
 American Mathematical Society. 

 \rr  E.Anderson (1935). The Irises of the Gasp\'{e} Peninsula. 
 {\it Bulletin of the American Iris Society,} 59, 2-5. 

 \rr T.W.Anderson (1969). {\it Statistical Inference for Covariance 
 Matrices with Linear Structure.} in Krishnaiah, P. Multivariate 
 Analysis II, NY: Academic Press. 

 \rr P.Angeline. Two Self-Adaptive Crossover Operators for Genetic 
 Programming. ch.5, p.89-110 in Angeline and Kinnear (1996).

 \rr P.J.Angeline, K.E.Kinnear (1996). {\it Advances in Genetic Programming.  
 Vol.2, Complex Adaptive Systems.}. MIT. 

 \rr M.Anthony, N.Biggs (1992). {\it Computational Learning Theory.} 
 Cambridge Univ. Press.

 \rr  M.Aoyagi, A.Namatame (2005). Massive Individual Based Simulation: 
  Forming and Reforming of Flocking Behaviors. 
  {\it Complexity International,} 11. \\ 
  \verb#www.complexity.org.au:a\\vol11\\aoyagi01\# 

 \rr M.A.Arbib, E.J.Conklin, J.C.Hill (1987).  
 {\it From Schemata Theory to Language.} Oxford University Press.  

 \rr M.A.Arbib, Mary B. Hesse (1986). {\it The Construction of 
   Reality.} Cambridge University Press. 

 \rr O.Arieli, A.Avron (1996).
 Reasoning with Logical Bilattices.
 {\it Journal of Logic, Language and Information}, 5, 25--63. 

 \rr S.Assmann, S.Pocock, L.Enos, L.Kasten (2000). Subgroup analysis
  and other (mis)uses of baseline data in clinical trials. 
  {\it The Lancet,} 355, 9209, 1064-1069. 

 \rr P.W.Atkins (1984). {\it The Second Law.} 
  NY: The Scientific American Books. 

 \rr A.C.Atkinson (1970). A Method for Discriminating Between 
 Models. {\it J. Royal Statistical Soc. B}, 32, 323-354. 

 \rr  H.Atmanspacher (2005). Non-Physicalist Physical Aproaches. 
  Guest Editorial. {\it Mind and Matter,} 3, 2, 3-6.   

 \rr E.Attneave (1959). {\it Applications of Information Theory to 
 Psychology: A summary of basic concepts, methods, and results.} 
 New York: Holt, Rinehart and Winston.

 \rr A.Aykac, C.Brumat, eds. (1977). {\it New Developments in the 
 Application of Bayesian Methods.} Amterdam: North Holland.

 \rr J.Baggott (1992). {\it The Meaning of Quantum Theory.} 
  Oxford University Press. 

 \rr   L. H. Bailey (1894). Neo-Lamarckism and Neo-Darwinism.  
 {\it The American Naturalist}, 28, 332, 661-678. 

 \rr  T.Bakken, T.Hernes (2002). 
 {\it Autopoietic Organization Theory. 
  Drawing on Niklas Luhmann's Social  Systems Perspective.}  
 Copenhagen Business School. 

 \rr   G.van Balen (1988). The Darwinian Systhesis: A Critique of the 
   Rosenberg / Williams Argument. {\it British Journal of the 
   Philosophy of Science}, 39, 4, 441-448. 

 \rr J.D.Banfield,A.E.Raftery (1993). Model Based Gaussian and 
 nonGaussian Clustering. {\it Biometrics},803-21. 

 \rr W.Banzahf, P.Nordin, R.E.Keller, F.D.Francone (1998). 
 {\it Genetic Algorithms.}  

 \rr D.Barbieri (1992). Is Genetic Epistemology of Any Interest 
 for Semiotics? {\it Scripta Semiotica 1, 1-6.} 

 \rr R.E.Barlow, F.Prochan (1981). {\it Statistical Theory of 
  Reliability and Life Testing Probability Models.} 
 Silver Spring: To Begin With.  

 \rr G.A.Barnard (1947). The Meaning of Significance Level. 
 {\it Biometrika}, 34, 179--182. 

 \rr G.A.Barnard (1949). Statistical Inference. 
 {J. Roy. Statist. Soc. B}, 11, 115--149. 

 \rr A.R.Barron (1984) Predicted Squared Error: A Criterion for 
 Automatic Model Selection. in  Farlow (1984). 

 \rr V.Bryant, H.Perfect (1980). {\it Independence Theory in 
 Combinatorics: An Introductiory Account with Applications to Graphs 
 and Transversals.} London: Chapman and Hall. 

 \rr D.Basu (1988). Statistical Information and Likelihood.  
 Edited by J.K.Ghosh. {\it Lect. Notes in Statistics}, 45.  

 \rr D.Basu, J.K.Ghosh (1988). Statistical Information and Likelihood.  
  {\it Lecture Notes in Statistics,} 45.  

 \rr D.Basu, C.A.B.Pereira (1982). On the Bayesian Analysis of
 Categorical Data: The Problem of Nonresponse. 
 {\it JSPI} 6, 345-62. 

 \rr D.Basu, C.A.B.Pereira (1983). A Note on Blackwell Sufficiency and a
 Shibinsky Characterization of Distributions. 
 {\it Sankhya A}, 45,1, 99-104.

 \rr M.S.Bazaraa, H.D.Sherali, C.M.Shetty (1993). {\it Nonlinear 
 Programming: Theory and Algorithms.} NY: Wiley. 


 \rr J.L.Bell (1998). {\it A Primer of Infinitesimal Analysis.} 
    Cambridge Univ. Press. 

 \rr J.L.Bell (2005). {\it The Continuous and the Infinitesimal 
    in Mathematics and Philosophy.} Milano: Polimetrica.  

 \rr L.V.Beloussov (2008). Mechanically Based Generative Laws of 
  Morphogenesis. {\it Phys. Biol.}, 5, 1-19. 

 \rr A.H. Benade (1992). {\it Horns, Strings, and Harmony.} 
   Mineola: Dover.   

 \rr C.H.Bennett (1976). Efficient Estimation of Free Energy 
 Differences from Monte Carlo Data. 
 {\it Journal of Computational Physics} 22, 245-268. 

 \rr J.Beran (1994). {\it Statistics of Long-Memory Processes.} 
 London: Chapman and Hall. 

 \rr H.C.Berg (1993). {\it Random Walks in Biology.} 
 Princeton Univ. Press.  

 \rr J.O.Berger (1993). {\it Statistical Decision Theory and Bayesian 
  Analysis,} 2nd ed.  NY: Springer. 

 \rr J.O.Berger, J.M.Bernardo (1992). {\it On the Development of 
  Reference Priors.} Bayesian Statistics 4 
 (J. M. Bernardo, J. O. Berger, D. V. Lindley and A. F. M. Smith, eds). 
 Oxford: Oxford University Press, 35-60.  

 \rr J.O.Berger, R.L.Wolpert (1988). 
  {\it The Likelihood Principle,} 2nd ed. 
  Hayward, CA, Inst of Mathematical Statistic. 

 \rr C.A.Bernaards, R.I.Jennrich (2005). Gradient Projection
 Algorithms and Software for Arbitrary Rotation Criteria in Factor
 Analysis. {\it Educational and Psychological Measurement}, 65 (5), 676-696. 

 \rr J.M.Bernardo, A.F.M.Smith (2000).  
 {\it Bayesian Theory.} NY: Wiley.  

 \rr L.von Bertalanffy (1969). {\it General System Theory.} 
    NY: George Braziller. 

 \rr A.Bertoni, M.Dorigo (1993). Implicit Parallelism in Genetic 
 Algorithms. {\it Artificial Intelligence}, 61, 2, 307-314.

 \rr D.P.Bertsekas, J.N.Tsitsiklis (1989). 
 {\it Parallel and Distributed Computation, Numerical Methods.} 
 Englewood Cliffs: Prentice Hall.  

 \rr D.P.Bertsekas (1996). Thevelin Decomposition and Large Scale 
 Optimization. {\it JOTA}, 89, 1-15.

 \rr P.J.Bickel, K.A.Doksum (2001).  
 {\it Mathematical Statistics, 2nd ed.} USA: Prentice Hall. 

 \rr C.Biernacki G.Govaert (1998). Choosing Models in Model-based 
 Clustering and Discriminant Analysis. 
 Technical Report INRIA-3509-1998. 

 \rr K.Binder (1986).
 {\it Monte Carlo methods in Statistical Physics}. 
 Topics in current Physics 7. Berlin: Springer.

 \rr K.Binder, D.W.Heermann (2002). {\it Monte Carlo Simulation in 
 Statistical Physics, 4th ed.} NY: Springer.   

 \rr E.G.Birgin, R.Castillo, J.M.Martinez (2004). Numerical comparison of 
 Augmented Lagrangian algorithms for nonconvex problems. to appear in 
 {\it Computational Optimization and Applications.} 

 \rr A.Birnbaum (1962). On the Foundations of Statistical Inference. 
 {\it J. Amer. Statist. Assoc.} 57, 269--326.  

 \rr A.Birnbaum (1972). More on Concepts of Statistical Evidence. 
 {\it J. Amer. Statist. Assoc.} 67, 858--861. 
 
 \rr Z.W.Birnbaum, J.D.Esary, S.C.Saunders (1961). 
 Multicomponent Systems and Structures, and their Reliability. 
 {Technometrics,} 3, 55-77. 

 \rr B.Bj\"{o}rkholm, M.Sj\"{o}lund,, P.G.Falk, O.G.Berg, L.Engstrand, 
 D.I.Andersson (2001).  
 Mutation Frequency and Biological Cost of Antibiotic Resistance in 
 Helicobacter Pylori.  {\it PNAS}, 98,4, 14607-14612. 

 \rr S.J.Blackmore (1999). {\it The Meme Machine.} 
  Oxford University Press.  

 \rr D.Blackwell, M.A.Girshick (1954). {\it Theory of Games and Statistical 
 Decisions}. NY: Dover reprint (1976). 

 \rr J.R.S.Blair, B.Peyton (1993). An Introduction to Chordal Graphs 
    and Clique Trees. In George et al. (1993). 

 \rr C.R.Blyth (1972). On Simpson's Paradox and the Sure-Thing Principle. 
 {\it Journal of the American Statistical Association}, 67, p. 364. 

 \rr R.D.Bock, R.E.Bargnann (1966). Analysis of Covariance 
 Structure. {\it Psycometrica}, 31, 507--534. 

 \rr N.Bohr (1935). {\it Space-Time Continuity and Atomic Physics.}  
  H.H.Wills Memorial Lecture, Univ. of Bristol, Oct. 5, 1931. 
  In Niels Bohr Collected Works, 6, 363-370. 
  Complementarity, p.369-370. 

 \rr N.H.D.Bohr (1987a).  
    {\it The Philosophical Writings of Niels Bohr.} 
    V.I- Atomic Theory and the Description of Nature. 
    Woodbridge, Connecticut: Ox Bow Press. 

 \rr N.H.D.Bohr (1987b). 
    {\it The Philosophical Writings of Niels Bohr.} 
    V.II- Essays 1932-1957 on Atomic Physics and Human Knowledge. 
    Woodbridge, Connecticut: Ox Bow Press. 

 \rr N.H.D.Bohr (1987c). 
    {\it The Philosophical Writings of Niels Bohr.} 
    V.III- Essays 1958-1962 on Atomic Physics and Human Knowledge . 
    Woodbridge, Connecticut: Ox Bow Press.

 \rr N.H.D.Bohr (1999), J.Faye, H.J.Folse, eds. 
    {\it The Philosophical Writings of Niels Bohr.} 
    V.IV- Causality and Complementarity: Supplementary Papers. 
    Woodbridge, Connecticut: Ox Bow Press.

 \rr N.H.D.Bohr (1985), J.Kalckar ed. {\it Collected Works.}  
    V.6- Foundations of Quantum Physics I, (1926-1932). 
    Elsevier Scientific. 

 \rr N.H.D.Bohr (1996), J.Kalckar ed. {\it Collected Works.}  
    V.7- Foundations of Quantum Physics II, (1933-1958).  
    Elsevier Scientific. 

 \rr N.H.D.Bohr (2000), D.Favrholdt ed. {\it Collected Works.}  
    V.10- Complementarity beyond Physics, (1928-1962).  
    Elsevier Scientific. 
 
 \rr N.H.D.Bohr (2007). Questions Answered by Niels Bohr (1885-1962).  
  {\it Physikalisch-Technische Bundesanstalt.}   
 {\tt http://www.ptb.de/en/publikationen/blickpunkt/interviews/bohr.html.}

 \rr L.Boltzmann (1890). \"{U}ber die Bedeutung von Theorien. 
  Translated and edited by B.McGuinness (1974). 
  {\it Theoretical Physics and Philosophical Problems: Selected Writings}.  
  Dordretcht: Reidel.   

 \rr E.Bonabeau, M.Dorigo, G.Theraulaz (1999). {\it Swarm Intelligence: 
  From Natural to Artificial Systems.} Oxford University Press. 

 \rr J.A.Bonaccini (2000). 
 {\it Kant e o Problema da Coisa Em Si No Idealismo Alem\~{a}o.} 
 SP: Relume Dumar\'{a}.  

 \rr F.V.Bonassi, R.B.Stern, S.Wechsler (2008). The Gambler's Fallacy: 
  A Bayesian Approach. {\it MaxEnt 2008, AIP Conference Proceedings,} 
  v. 1073, 8-15. 

 \rr F.V.Bonassi, R.Nishimura, R.B.Stern (2009). 
  In Defense of Randomization: A Subjectivist Bayesian Approach. 
  To appear in {\it MaxEnt 2009, AIP Conference Proceedings.} 

 \rr W.Boothby (2002). 
 {\it An Introduction to Differential Manifolds and 
   Riemannian Geometry.} NY: Academic Press.

 \rr J.Bopry (2002). Semiotics, Epistemology, and Inquiry. 
 {\it Teaching \& Learning,} 17, 1, 5--18. 

 \rr K.C. Border (1989). {\it Fixed Point Theorems with Applications 
   to Economics and Game Theory.}  Cambridge University Press. 
 
 \rr W.Borges, J.M.Stern (2005). 
 On the Truth Value of Complex Hypothesis.  
 {\it CIMCA-2005 - International Conference on Computational 
 Intelligence for Modelling Control and Automation.} USA: IEEE.  


 \rr W.Borges, J.M.Stern (2007). The Rules of Logic Composition for the 
  Bayesian Epistemic e-Values. {\it Logic Journal of the IGPL}, 
  15, 5-6,  401-420.  doi:10.1093/jigpal/jzm032 .

 \rr G.E.P.Box, W.G.Hunter, J.S.Hunter (1978).
  {\it Statistics for Experimenters. An Introduction to Design, Data
  Analysis and Model Building.} NY: Wiley.

 \rr G.E.Box, G.M.Jenkins (1976). {\it Time Series Analysis, 
  Forcasting and Control.} Oakland: Holden-Day. 

 \rr G.E.P.Box and G.C.Tiao (1973). 
 {\it Bayesian Inference in Statistical Analysis.} 
 London: Addison-Wesley. 

 \rr   P.J.Bowler (1974). Darwin's Concept of Variation. {\it Journal 
  of the History of Medicine and Allied Sciences}, 29, 196-212.  

 \rr J.Boyar (1989). Inferring Sequences Produced by Pseudo-Random Number
 Generators. {\it Journal of the ACM}, 36, 1, 129-141. 

 \rr  R.Boyd, P.Gasper, J.D.Trout, (1991).  
 {\it The Philosophy of Science.} MIT Press.  

 \rr L.M.Bregman  (1967). The Relaxation Method for Finding the 
 Common Point Convex Sets and its Application to the Solution of 
 Problems in Convex Programming. {\it USSR Computational Mathematics 
 and Mathematical Physics,} 7, 200-217.   

 \rr L.Breiman, J.H.Friedman, C.J.Stone (1993). 
 {\it Classification and Regression Trees.} Chapman and Hall. 

 \rr R.Brent, J.Bruck (2006). Can Computers Help to Explain Biology?  
  {\it Nature}, 440/23, 416--417.  

 \rr S.Brier (1995) Cyber-Semiotics: 
 On autopoiesis, code-duality and sign games in bio-semiotics. 
 Cybernetics and Human Knowing, 3, 1, 3--14.   

 \rr S.Brier (2001). Cybersemiotics and Umweltlehre. {\it Semiotica},  
 Special issue on Jakob von Uexk\"{u}ll's Umweltsbiologie, 
 134 (1/4), 779-814.

 \rr S.Brier (2005). The Construction of Information and Communication: 
 A Cyber-Semiotic Re-Entry into Heinz von Foerster's Metaphysical 
 Construction of Second Order Cybernetics. 
 {\it Semiotica,} 154, 1, 355--399.   

 \rr P.J.Brockwell, R.A.Davis (1991). 
 {\it Time Series: Theory and Methods.}  NY: Springer. 

 \rr L.de Broglie (1946). {\it Matter and Light.} NY:Dover. 

 \rr M.W.Browne (1974). Gradient Methods for Analytical Rotation. 
 {\it British J.of Mathematical and Statistical Psychology}, 27, 115-121. 

 \rr M.W.Browne (2001). An Overview of Analytic Rotation in Exploratory
 Factor Analysis. {\it Multivariate Behavioral Research}, 36, 111-150.

 \rr P.Brunet (1938). {\it \'{E}tude Historique sur le Principe de la 
 Moindre Action.} Paris: Hermann.   

 \rr S.G.Brush (1961). Functional Integrals in Statistical Physics. 
 {\it Review of Modern Physics}, 33, 79-79. 

 \rr S.Brush (1968). A History of Random Processes: Brownian Movement from
 Brown to Perrin. {\it Arch. Hist. Exact Sci.} 5, 1-36. 

 \rr  T. Budd (1999). {\it Understanding Object-Oriented Programming 
 With Java.} Addison Wesley. (1999, Glossary, p.408):  

 \rr A.M.S.Bueno, C.A.B.Pereira, M.N.Rabelo-Gay, J.M.Stern (2002). 
 Environmental Genotoxicity Evaluation: Bayesian Approach for a 
 Mixture Statistical Model. {\it Stochastic Environmental Research 
 and Risk Assessment,} 16, 267-278.  

 \rr J.R.Bunch, D.J.Rose (1976). {\it Sparse Matrix Computations}.
    NY: Academic Press.   

 \rr A.Bunde, S.Havlin (1994). {\it Fractals in Science.} 
  NY: Springer.  

 \rr L.W.Buss (2007). {\it The Evolution of Individuality.} 
 Princeton University Press.  
 
 \rr E.Butkov (1968). {\it Mathematical Physics.} Addison-Wesley. 

 \rr F.W.Byron Jr., R.W.Fuller (1969). 
  Reading, MA: Addison-Wesley.    

 \rr H.B.Callen (1960). {\it Thermodynamics: An Introduction to the 
  PhysicalTheories of Equilibrium Thermostatics and Irreversible 
 Thermodynamics}. NY: John Wiley.   

 \rr C.A.Callaghan (2006). {\it Kinetics and Catalysis of the 
  Water-Gas-Shift Reaction: A Microkinetic and Graph Theoretic 
  Approach}. Dissertation, Worcester Polytechnic Institute, 
  Dept. of Chemical Engineering. 

 \rr T.Y.Cao (2003). Structural Realism and the Interpretation of 
  Quantum Field Theory. {\it Synthese}, 136, 1, 3-24. 

  \rr T.Y.Cao (2003). Ontological Relativity and Fundamentality - 
  Is Quantum Field Theory the Fundamental Theory?  
  {\it Synthese}, 136, 1, 25-30.  

  \rr T.Y.Cao (2003). Can We Dissolve Physical Entities into 
  Mathematical Structures? {\it Synthese}, 136, 1, 57-71. 

  \rr T.Y.Cao (2003). What is Ontological Synthesis? A Reply to Simon 
  Saunders. {\it Synthese}, 136, 1, 107-126. 

  \rr T.Y.Cao (2004). Ontology and Scientific Explanation. 
  In Conwell (2004).  

  \rr M.Carmeli, S.M.Malin (1976). {\it Representation of the Rotation 
  and Lorentz Groups}. Basel: Marcel Dekker.  

 \rr M.P.do Carmo (1976). 
  {\it Differential Geometry of Curves and Surfaces.} 
  NY: Prentice Hall.   

 \rr S.B.Carrol (2005). {\it Endless Forms Most Beautiful. 
 The New Science of Evo Devo.} NY: W.W.Norton. 

 \rr A.Caticha, A.Giffin (2007). 
 Updating Probabilities with Data and Moments. 
 27th International Workshop on Bayesian Inference and Maximum Entropy 
 Methods in Science and Engineering. 
 AIP Conf. Proc. 872, 74-84.  

 \rr A. Caticha (2007). Information and Entropy. 
 27th International Workshop on Bayesian Inference and Maximum Entropy 
 Methods in Science and Engineering. 
 AIP Conf. Proc. 872, 11-22.  

 \rr A. Caticha (2008). {\it Lectures on Probability, Entropy and 
  Statistical Physics.} Tutorial book for MaxEnt 2008, 
  The 28th International Workshop on Bayesian Inference 
  and Maximum Entropy Methods in Science and Engineering. 
  July 6-11 of 2008, Borac\'{e}ia, S\~{a}o Paulo, Brazil. 

 \rr  H.Caygill (1995).  
 {\it A Kant Dictionary.} Oxford: Blackwell. 

 \rr G.Celeux, G.Govaert (1995). Gaussian Parsimonious Clustering 
 Models. {\it Pattern Recog.} 28, 781-793.

 \rr G.Celeux, D.Chauveau, J.Diebolt (1996). 
 On Stochastic Versions of the EM Algorithm. 
 An Experimental Study in the mixture Case. 
 {\it Journal of Statistical Computation and Simulation,} 55, 287--314. 

 \rr Y.Censor, S.Zenios (1994). {\it Introduction to Methods of 
 Parallel Optimization.} IMPA, Rio de janeiro. 

 \rr Y.Censor, S.A.Zenios (1997). {\it Parallel Optimization: 
  Theory, Algorithms, and Applications.} NY: Oxford. 

 \rr C.Cercignani (1998). {\it Ludwig Boltzmann, 
 The Man who Trusted Atoms.}  Oxford Univ. 

 \rr F.V.Cerezetti, J.M.Stern (2012). Non-arbitrage in Financial 
  Markets: A Bayesian Approach for Verification. 
  {AIP Conf.Proc.}, 1490, 87-96. 

  \rr M.Ceruti (1989). {\it La Danza che Crea.} Milano: Feltrinelli.  

 \rr G.Chaitin (2004). On the Intelligibility of the Universe and the
 Notions of Simplicity, Complexity and Irreducibility.   
 pp. 517-534 in {\it Grenzen und Grenzberschreitungen, XIX.} 
 Berlin: Akademie Verlag.    

 \rr L.Chang (2005). Generalized Constraint-Based Inference. 
 M.S.Thesis, Univ.of British Columbia. 

 \rr V.Cherkaasky, F.Mulier (1998). {\it Learning from Data.} 
 NY: Wiley.   

 \rr M.Chester (1987). {\it Primer of Quantum Mechanics.}  
  John Wiley. 

 \rr U.Cherubini, E.Luciano, W.Vecchiato (2004). 
 {\it Copula Methods in Finance.} NY: Wiley. 

 \rr J.Y.Ching, A.K.C.Wong, K.C.C.Chan. ``Class-Dependent
 Discretization for Inductive Learning from Continuous and Mixed-Mode
 Data.'' {\it IEEE Transactions on Pattern Analysis and Machine
 Intelligence}, 17 n.7,  pp.641-651, 1995. 

 \rr  J.Christis (2001). Luhmann's Theory of Knowledge: 
 Beyond Realism and Constructivism?  {\it Soziale Systeme,} 
 7, 328--349. 

 \rr G.C.Chow (1983). {\it Econometrics.} Singapore: McGraw-Hill. 

 \rr A.Cichockt, R.Zdunek, S.I.Amari (0000). 
    New Algorithms for Non-Negative Matrix Factorization in 
    Applications to Bild Source Separaton. 

 \rr A.Cichockt, S.I.Amari, R.Zdunek, R.Kompass, G.Hori, 
    Z.He (0000). Extended SMART Algorithms for Non-Negative 
    Matrix Factorization. 
 
 \rr A.Cichockt, R.Zdunek, S.I.Amari (0000). 
    Csisz\'{a}r's Divergences for Non-Negative Matrix Factorization: 
    Family of New Algorithms. 

 \rr G.W.Cobb (1998). {\it Introduction to Design and Analysis of 
    Experiments.} NY: Springer. 

 \rr C.Cockburn (1996). The Interaction of Social Issues and Software 
 Architecture. {\it Communications of the ACM,} 39, 10, 40-46. 

 \rr D.W.Cohen (1989). {\it An Introduction to Hilbert Space and 
  Quantum Logic.} NY: Springer. 

 \rr R.W.Colby (1988). {\it The Encyclopedia of Technical Market Indicators.} 
  Homewood: Dow Jones - Irwin. 

 \rr E.C.Colla (2007). {\it Aplica\c{c}\~{a}o de T\'{e}cnicas de 
  Fatora\c{c}\~{a}o de Matrizes Esparsas para Infer\^{e}ncia em 
  Redes Bayesianas}. Ms.S. Thesis, 
  Institute of Mathematics and Statistics, University of S\~{a}o Paulo.   

 \rr E.C.Colla, J.M.Stern (2009). Factorization of Bayesian Networks.
 {\it Studies in Computational Intelligence}, 199, 275-285. 

 \rr N.E.Collins, R.W. Eglese, B.L. Golden (1988).
 {\it Simulated Annealing, An Annotated Bibliography}.
 In Johnson (1988).

 \rr M.L.L.Conde (1998). 
 {\it Wittgenstein: Linguagem e Mundo.}  
 SP: Annablume. 

 \rr P.C.Consul (1989). {\it Generalized Poisson Distributions.} 
  Basel: Marcel Dekker. 

 \rr W.J.Cook, W.H.Cunningham, W.R.Pulleyblank, A.Schrijver (1997).  
  {\it Combinatorial Optimization.} NY: Wiley. 

 \rr Cornwell (2004). {\it  Explanations: Styles of Explanation in 
    Science.}  

 \rr N.C.A.Costa (1963). Calculs Propositionnels pour les
  Systemes Formales Incosistants.
  {\it Compte Rendu Acad. des Scienes,} 257, 3790--3792.

 \rr N.C.A.da Costa (1986). {\it Pragmatic Probability.}  
 {\it Erkenntnis,}  25, 141-162.

 \rr N.C.A.da Costa (1993). {\it L\'{o}gica Indutiva e Probabilidade.}  
  S\~{a}o Paulo: Hucitec-EdUSP. 

 \rr N.C.A.da Costa, D. Krause (2004). 
    Complementarity and Paraconsistency. 
    In Rahman (2004, 557-568). 

 \rr N.C.A.Costa, V.S.Subrahmanian (1989).
  Paraconsistent Logics as a Formalism for Reasoning about
  Inconsistent Knowledge Bases.
  {\it Artificial Inteligence in Medicine}, 1, 167--174.

 \rr N.C.A.Costa, C.A.Vago, V.S.Subrahmanian (1991).
  Paraconsistent Logics Ptt.
  {\it Zeitschrift f\"{u}r Mathematische Logik und Grundlagen
  der Mathematik}, 37, 139-148.

 \rr N.C.A.Costa, J.M.Abe, V.S.Subrahmanian (1991).
  Remarks on Annotated Logic.
  {\it Zeitschrift f\"{u}r Mathematische Logik und Grundlagen 
  der Mathematik}, 37, 561--570.

 \rr N.C.A.Costa, J.M.Abe, A.C.Murolo, J.I.da Silva, C.F.S.Casemiro
  (1999). L\'{o}gica Paraconsistente Aplicada. S\~{a}o Paulo: Atlas.

 \rr F.G.Cozman (2000). Generalizing Variable Elimination in Bayesian 
  Networks. Proceedings of the Workshop in Probabilistic Reasoning in 
  Artificial Inteligence. Atibaia.  

 \rr J.F.Crow (1988). The Importance of Recombination. ch4, p.57-75 in 
  Michod and Levin (1988). 
 
 \rr I.Csiszar (1974). Information Measures. 
 {\it 7th Prage Conf.of Information Theory,} 2, 73-86. 

 \rr T.van Cutsem. ``Decision Trees for Detecting Emergency Voltage
 Conditions.'' {\it Proc. Second International Workshop on Bulk Power 
 System Voltage Phenomena}, pp.229-240, McHenry, USA, 1991. 

 \rr A. Damodaran (2003). {\it Investment Philosophies: Successful
  Investment Philosophies and the Greatest Investors Who Made Them Work.} 
  NY: Wiley. 

 \rr  A.Y.Darwiche, M.L.Ginsberg (1992). 
  A Symbolic Generalization of Probability Theory. 
  {AAAI-92.}  
  {10-th Conf. American Association for Artificial Intelligence.}   

 \rr A.Y.Darwiche (1993). {\it A Symbolic Generalization 
 of Probability Theory.} Ph.D. Thesis, Stanford Univ.

 \rr C.Darwin (1860). Letter to Asa Gray, dated 3 April 1860. 
 in F.Darwin ed. (1911). The Life and Letters of Charles Darwin, 
 London: John Murray. 

 \rr C.Darwin (1883). {\it The Variation of Animals and Plants under 
 Domestication.} V.2, Portland, OR: Book News Inc. 
 Reprint by Kissinger Press, 2004. 

 \rr C.Darwin (1859). {\it On the Origin of Species by Means of Natural 
 Selection.} 
 Reprinted as Great Books of the Western World V.49,   
 Chicago: Encyclopaedia Britanica Inc. 1952. 

 \rr F.N.David (1969). {\it Games, Gods and Gambling. A History of 
 Probability and Statistical Ideas.} London: Charles Griffin. 

 \rr L.Davis ed. (1987). Genetic Algorithms and Simulated Annealing. 
 Pittman, 1987.
 
 \rr L.Davis, M.Steenstrup (1987). Genetic Algorithms and Simulated 
 Annealing: An Overview. p.1-11 in Davis (1987). 

 \rr M.Davis (1977). {\it Applied Nonstandard Analysis.} 
    NY: Dover. 
 
 \rr R.Dawkins (1989). {\it The Selfish Gene}. 2nd ed.   
  Oxford University Press.

 \rr I.De\'{a}k (1990). {\it Random Number Generators and Simulation.} 
 Budapest: Akad\'{e}miai Kiad\'{o}. 

 \rr J.Decety, J.Gr\`{e}zes (2005). 
 The power of simulation: Imagining one`s own and other`s behavior. 
 {\it Brain Research,} 1079, 4-14. 

 \rr J.Decety, D.H.Ingvar (1990). Brain structures participating in
  mental simulation of motor behavior: a neuropsychological
  interpretation. {\it Acta Psychol,} 73, 13-24.

 \rr M.H.DeGroot (1970). {\it Optimal Statistical Decisions}.
  NY: McGraw-Hill.

 \rr T. Dehue (1997). Deception, Efficiency, and Random Groups: 
  Psychology and the Gradual Origination of the Random Group Design.  
  {\it Isis}, 88, 4, p.653-673 

 \rr A.Deitmar (2005). {\it A First Course in Harmonic Analysis}, 
    2nd ed. NY: Springer.  

 \rr B.P.Demidovich, I.A.Maron (1976). {\it Computational Mathematics.} 
  Moskow: MIR. 

 \rr M.Delgado, S.Moral (1987). 
   On the Concept of Possibility-Probability Consistency. 
   {\it  Fuzzy Sets and Systems,} 21, 3, 311-318. 
 
 \rr A.P.Dempster, N.M.Laird, D.B.Rubin (1977). 
 Maximum Likelihood from Incomplete Data via the EM Algorithm. 
 {\it J. of the Royal Statistical Society B.} 39, 1-38.   

 \rr D.G.T.Denison, C.C.Holmes, B.K.Mallick, A.F.M.Smith (2002). 
  {\it Bayesian Methods for Nonlinear Classification and Regression.} 
  John Wiley.  

 \rr I.S.Dhillon, S.Sra (0000). Generalized Nonnegative Matrix 
    Approximations with Bregman Divergences. 

 \rr O.Diachok (2006). {\it Do Humpback Whales Detect and Classify Fish by 
 Transmitting Sound Through Schools?} 
 151st ASA Meeting. Acoustical Society of America, Providence, RI. 

 \rr P.Diaconis (1978). Statistical Problems in ESP Research. 
    {\it Science}, 201, 131-136. 

 \rr P.Diaconis, D.Freeman (1987). A Dozen de Finetti Style Results 
 in Search of a Theory. {\it Ann. Inst. Poincar\'{e} Probab. Stat.}, 
 23, 397--423.  

 \rr P.Diaconis (1988). 
 {\it Group Representation in Probability and Statistics.}  
 Hayward: IMA. 

 \rr J.M.Dickey (1983). Multiple Hypergeometric Functions: Probabilistic
  Interpretations and Statistical Uses. {\it JASA}, 78, 628-37.

 \rr J.M.Dickey, T.J.Jiang, J.B.Kadane (1987). Bayesian Methods for
   Categorical Data. {\it JASA} 82, 773-81.

 \rr P.DiMaggio, W.W.Powell (1991).  
 {\it New Institutionalism in Organizational Analysis.} 
 Chicago Univ. 


  \rr M.Diniz, C.A.B.Pereira, J.M.Stern (2008). 
   FBST for Cointegration Problems. {\it AIP Conference Proceedings,} 
   v. 1073, p. 157-164.

 \rr M.Diniz, C.A.B.Pereira, J.M.Stern (2011). 
  Unit Roots: Bayesian Significance Test. 
  {\it Communications in Statistics - Theory and Methods}, 
  40, 4200-4213. 

  \rr M.Diniz, C.A.B.Pereira, J.M.Stern (2012). 
  Cointegration: Bayesian Significance Test. 
  {\it Communications in Statistics - Theory and Methods}, 
  41, 3562-3574.

 \rr M.Diniz, C.A.B.Pereira, A.Polpo, J.M.Stern, S.Wechsler (2012). 
 Relationship Between Bayesian and Frequentist Significance Indices 
 {\it International Journal for Uncertainty Quantification}, 
 2, 2, 161-172.

 \rr S.M.Dion, J.L.A.Pacca, N.J.machado (1995). 
  Quaternions: Sucessos e Insucessos de um Projeto de Pesquisa. 
  {\it Estudos Avan\,{c}ados,} 9, 25, 251-262. 

 \rr  G.Dixon (1994). {\it Division Algebras: Octonions, Quaternions, 
  Complex Numbers and the Algebraic Design of Physics.}  

 \rr B.Dodson (1994). {\it Weibull Analysis.} Milwaukee: ASQC Quality Press.

 \rr C.S.Dodson, M.K.Johnson, J.W.Schooler (1997). 
 The verbal overshadowing effect: 
 Why descriptions impair face recognition. 
 {\it Memory and Cognition,} 25 (2), 129-139

 \rr  M.G.Doncel, A.Hermann, L.Michel, A.Pais (1987). 
 {\it Symmetries in Physics (1600-1980).} 
 Seminari d'Hist\`{o}ria des les Ci\`{e}nces. 
 Universitat Aut\`{o}noma de Barcelona. 

 \rr I.M.L. D'Otaviano, M.E.Q.Gonzales (2000). 
 {\it Auto-Organiza\c{c}\~{a}o, Estudos Interdisciplinares.} 
 Campinas, Brazil: CLE-UNICAMP.  

 \rr G.van Driem  (2007). 
  Symbiosism, Symbiomism and the Leiden definition of the Meme. 
  Keynote lecture delivered at the pluridisciplinary symposium on 
 Imitation Memory and Cultural Change: Probing the Meme Hypothesis, 
 hosted by the Toronto Semiotic Circle at the University of Toronto, 
 4 May 2007. Retrieved from \\ 
 \verb#www.semioticon.com/virtuals/imitation/van_driem_paper.pdf#    

 \rr  L.E.Dubins L.J.Savage (1965). 
 {\it How to Gamble If You Must. 
 Inequalities for Stochastic Processes.}  
 NY: McGraw-Hill.   


 \rr D.Dubois, H.Prade, S.Sandri (1993). 
   On Possibility-Probability Transformations. 
   p.103-112 in Proceedings of Fourth IFSA Conference,
   Kluwer Academic Publ.

 \rr I.S.Duff (1986).
 {\it Direct methods for sparse matrices}.
 Oxford: Clarendon Press.
 
 \rr R.Dugas (1988). {\it A History of Mechanics.} Dover.  

 \rr J.S.Dugdale (1996). 
  {\it Entropy and Its Physical Meaning.} 
  London: Taylor and Francis. 

 \rr J.Dugundji (1966). {\it Topology.} Boston: Allyn and Bacon. 

 \rr M.L.Eaton (1989). 
 {\it Group Invariance Applications in Statistics.} 
 Hayward: IMA. 

  \rr   G.T.Eble (1999). On the Dual Nature of Chance in Evolutionary 
  Biology and Paleobiology. {\it Paleobilogy}, 25, 75-87. 

 \rr A.W.F.Edwards (2004). {\it Cogwheels of the Mind. The Story of Venn 
 Diagrams.} Baltimore: The Johns Hopkins University Press. 
 
 \rr J.S.Efran, M.D.Lukens, R.J.Lukens (1990). {\it Language, Structure 
  and Change: Frameworks of Meaning in Psychotherapy.}  
  NY: W.W.Norton. 

 \rr I.Eibel-Eibesfeldt (1970). {\it Ethology, The Biology of Behavior.} 
     NY: Holt, Rinehart and Winston.

 \rr  M.Eigen (1992). {\it Steps Towards Life.} 
  Oxford University Press. 


 \rr M.Eigen, P.Schuster (1977). The Hypercyde: 
 A Principle of Natural Self-Organization. 
 Part A: Emergence of the Hypercycle. 
 {\it Die Naturwissenschaften,} 64, 11, 541-565. 

 \rr M.Eigen, P.Schuster (1978a). The Hypercyde: 
 A Principle of Natural Self-Organization. 
 Part B: The Abstract Hypercycle. 
 {\it Die Naturwissenschaften,} 65, 1, 7-41. 

 \rr M.Eigen, P.Schuster (1978b). The Hypercyle: 
 A Principle of Natural Self-Organization. 
 Part C: The Realistic Hypercycle. 
 {\it Die Naturwissenschaften,} 65, 7, 341-369. 

 \rr C.Eisele edt. (1976). 
    {\it The New Elements of Mathematics of Charles S. Peirce.} 
    The Hague: Mouton.

 \rr A.Einstein (1905a). 
 \"{U}ber einen die Erzeugung und Verwandlung des Lichtes betreffenden
 heuristischen Gesichtspunkt. 
 (On a heuristic viewpoint concerning the production and
 transformation of light).  
 {\it Annalen der Physik}, 17, 132-148. 

 \rr A.Einstein (1905b). 
 \"{U}ber die von der molekularkinetischen Theorie der W\"{a}rme
 geforderte Bewegung von in ruhenden Fl\"{u}ssigkeiten suspendierten
 Teilchen. 
 (On the motion of small particles suspended in liquids at rest
 required by the molecular-kinetic theory of heat). 
 {\it Annalen der Physik}, 17, 549-560. 

 \rr A.Einstein (1905c). 
 Zur Elektrodynamik bewegter K\"{o}rper. 
 (On the Electrodynamics of Moving Bodies). 
 {\it Annalen der Physik}, 17, 891-921. 

 \rr A.Einstein (1905d). 
 Ist die T\"{a}rgheit eines Kr\"{o}pers von seinem Energiegehalt 
 abh\"{a}ngig?
 (Does the Inertia of a Body Depend Upon Its Energy Content?). 
 {\it Annalen der Physik}, 18, 639-641. 

 \rr A.Einstein (1905, 1956).  
 {\it Investigations on the Theory of the Brownian Movement.} 
 Dover. 

 \rr A.Einstein (1950). On the Generalized Theory of Gravitation. 
  {Scientific American,} 182, 4, 13-17. 
  Reprinted in Einstein (1950, 341-355). 

 \rr A.Einstein (1954). {\it Ideas and Opinions}. Wings Books.  

 \rr A.Einstein (1991). {\it Autobiographical Notes: A Centennial 
    Edition.} Open Court Publishing Company.

 \rr W.Ehm (2005). Meta-Analysis of Mind-Matter Experiments: 
  A Statistical Modeling Perspective. {\it Mind and Matter,} 3, 1, 85-132.

 \rr P.Embrechts (2002). Selfsimilar Processes
 Princeton University Press. 

 \rr C.Emmeche, J.Hoffmeyer (1991). From Language to Nature:  
    The Semiotic Metaphor in Biology. {\it Semiotica,}  84, 1/2, 1-42.   

 \rr H.A.Enge, M.R.Wehr, J.A.Richards (1972). 
  {\it Introduction to Atomic Physics.}  NY: Addison-Wesley. 

 \rr T.Elfving (1980). On Some Methods for Entropy maximization and 
 Matrix Scaling. {\it Linear algebra and its applications,} 34, 321-339. 

 \rr A.G. Exp\'{o}sito, L.G. Franquelo (1987).
 A New Contribution to the Cluster Problem. 
 {\it IEEE Transactions on Circuits and Systems}, 34, 546-552.

 \rr M.Evans (1997). Bayesian Inference Procedures Derived via
 the Concept of Relative Surprise.
 {\it Communications in Statistics}, 26, 1125--1143. 

 \rr M.Evans, T.Swartz  (2000).  {\it Approximating Integrals via Monte
 Carlo and Deterministic Methods.} Oxford University Press.

 \rr B.S.Everitt (1984). {\it Latent Variable Models.} 
  London: Chapman and Hall. 

 \rr R.Falk, C.Konold (1997). Making Sense of Randomness: 
 Implicit Encoding as a Basis for Judgment.  
 {\it Psychological Review,} 104, 2, 301-318. 

 \rr R.Falk, C.Konold (2005). Subjective Randomness. 
 {\it Encyclopedia of Statistical Sciences,} 13, 8397-8403.

 \rr S.C.Fang, J.R.Rajasekera, H.S.J.Tsao (1997). {\it Entropy Optimization 
 and Mathematical Programming.} Kluwer, Dordrecht. 

 \rr S.J. Farlow (1984) {\it Self-Organizing Methods in Modeling: 
 GMDH-type Algorithms.} Marcel Dekker, Basel.  

 \rr A. Faulstich-Brady (1993).    
 A Taxonomy of Inheritance Semantics
 {\it Proceedings of the Seventh International Workshop on
 Software Specification and Design,} 194-203. 
  
 \rr J.Feder (1988). {\it Fractals.} NY: Plenum.

  \rr J.D.Fehribach (2009). Vector-Space Methods and Kirchhoff 
  Graphs for REaction Networks. {\it SIAM J.on Applied Mathematics}, 
  70, 2, 543-562. 

 \rr W.Feller (1957). {\it An Introduction to Probability Theory and Its
 Applications} (2nd ed.), V.I. NY: Wiley.

 \rr W.Feller (1966). {\it An Introduction to Probability Theory and Its
 Applications} (2nd ed.), V.II. NY: Wiley.


 \rr  T.S.Ferguson (1996). {\it A Course in Large Sample Theory.} 
 NY: Chapman \& Hall.

 \rr P.J.Fernandes, J.M.Stern, M.S.Lauretto (2007). A New Media Optimizer 
  Based on the Mean-Variance Model. Presented at ARF'05 - Advertising 
  Research Foundation Conference. {\it Pesquisa Operacional}, 27, 427-456.

 \rr J.Ferreira (2006). {\it Semiotic Explorations in Computer Interface 
    Design.} Wellington: Victoria University.    

 \rr R.P.Feynman, A. R. Hibbs (1965). 
 {\it Quantum Mechanics and Path Integrals.} NY: McGraw-Hill.  

 \rr Feyerabend,P. (1993). {\it Against Method.} Verso Books.  

 \rr C.M. Fiduccia, R.M. Mattheyses (1982).
 A Linear Time Heuristic for Improving Network Partitions.
 {\it IEEE Design Automation Conferences}, 19, 175-181.

 \rr E.C.Fieller (1954). Some Problems in Interval Estimation. 
 {\it Journal of the Royal Statistical Society B,} 16, 175-185.   

 \rr B.de Finetti (1947). La pr\'{e}vision: Des lois logiques, ses sourses
 subjectives. Annalles de l'Institut Henri Poincar\'{e} 7,1-68. English
 translation:  Foresight: Its logical laws, its subjective sources, in
 Kiburg and Smoker Eds. (1963), Studies in Subjective Probability, 
 p.93-158, NY: Wiley.

 \rr B.de Finetti (1972). {\it Probability, Induction and Statistics.} 
 NY: Wiley. 

 \rr B.de Finetti (1974). {\it Theory of Probability,} 
 V1 and V2. London: Wiley.  

 \rr B.de Finetti (1975). {\it Theory of Probability. A Critical 
 Introductory Treatment.} London: Wiley. 

 \rr B.de Finetti (1977). Probabilities of Probabilities: 
  A Real Problem or a Misunderstanding? in 
   A.Aykac and C.Brumat (1977).

 \rr B.de Finetti (1980). Probability: Beware of Falsifications. 
    p. 193-224 in: H.Kyburg, H.E.Smokler (1980). {\it Studies in 
    Subjective Probability.} NY: Krieger. 

 \rr B.de Finetti (1981). 
 {\it Scritti.} V1: 1926-1930.  Padova: CEDAM

 \rr B. de Finetti (1991). 
 {\it Scritti.} V2: 1931-1936. Padova: CEDAM

 \rr B.de Finetti (1993). {\it Probabilit\'{a} e Induzione.} 
 Bologna: CLUEB. 

 \rr  D.Finkelstein (1993). Thinking Quantum. 
 {\it Cybernetics and Systems}, 24, 139-149.  

 \rr M.A.Finocchiaro (1991). {\it The Galileo Affair: A Documented 
 History.} NY: The Notable Trials Library. 

 \rr R.A.Fisher (1935). {\it The Design of Experiments}. 
    8ed.(1966). London: Oliver and Boyd,  
 
 \rr R.A.Fisher (1936). The Use of Multiple Measurements in 
 Taxonomic Problems. {\it Annals of Eugenics},7,179--188.   

 \rr R.A.Fisher (1926). The arrangement of Field Experiments. 
  {\it Journal of the Ministry of Agriculture}, 33, 503-513.  

 \rr R.A.Fisher (1934), Randomisation, and an Old Enigma of Card Play. 
  {\it Mathematical Gazette} 18, 294-297.

 \rr G.Fishman (1996). {\it Monte Carlo. Concepts, Algorithms and 
 Applications.} NY: Springer.

 \rr I.Fishtik, C.A.Callaghan, R.Datta (2004). 
 Reaction Route Graphs I: Theory and Algorithm.  
 {\it J. Phys. Chem. B}, 108, 5671-5682. 

 \rr I.Fishtik, C.A.Callaghan, R.Datta (2006). 
 Wiring Diagrams for Complex Reaction Networks. 
 {\it Ind. Eng. Chem. Res.}, 45, 6468-6476. 

 \rr H.Flanders (1989). 
 {\it Differential Forms with Applications to the Physical Sciences.} 
 NY: Dover. 

 \rr H.Fleming (1979). As Simetrias como Instrumento de Obten\c{c}\~{a}o 
 de Conhecimento. {\it Ci\^{e}ncia e Filosofia,} 1, 99--110. 

 \rr R.M.T.Fleming, C.M.Maes, M.A.Saunders, Y.Ye, B.O.Palsson (2012). 
 A Variational Principle for Computing Nonequilibrium Fluxes and 
 Potentials in Genome-Scale Biochemical Networks. 
 {\it Journal of Theoretical Biology}, 292, 71-77. 

 \rr M.Fleming (1962). {\it Domestic Financial Policies under Fixed and 
 Under Floating Exchange Rates.} International Monetary Fund Staff 
 Papers 9, 1962, 369-79. 

 \rr A.Flew (1959). Probability and Statistical Inference by 
 G.Spencer-Brown (review). {\it The Philosophical Quarterly,} 
 9, 37, 380-381.

 \rr  H.von Foerster (2003).
 {\it Understanding Understanding: 
 Essays on Cybernetics and Cognition.}   
 NY: Springer Verlag. 
 The following articles in this anthology are of special interest: \\ 
 (a) On Self-Organizing Systems and their Environments; p.1--19. \\ 
 (b) On Constructing a Reality; p.211--227. \\ 
 (c) Objects: Tokens for Eigen-Behaviors; p.261--271. \\  
 (d) For Niklas Luhmann: How Recursive is Communication? p.305--323. \\  
 (e) Introduction to Natural Magic. p.339--338. 

 \rr J.L.Folks (1984). Use of Randomization in Experimental Research. 
  p.17--32 in Hinkelmann (1984). 

 \rr H.Folse (1985). {\it The Philosophy of Niels Bohr.} 
    Elsevier. 

 \rr G.Forgacs, S.A.Newman (2005). {\it Biological Physics of the 
  Developing Embryo.}  Cambridge University Press.  

 \rr C.Fraley, A.E.Raftery (1999). Mclust: Software for Model-Based
  Cluster Analysis. {\it J. Classif.},16,297-306.

 \rr J.N.Franklin (1968). {\it Matrix Theory.} 
  Englewood-Cliffs: Prentice-Hall. 

 \rr M.L.von Franz (1981). {\it Alchemy: An Introduction to the Symbolism 
  and the Psychology.} Studies in Jungian Psychology, Inner City Books.   

 \rr A.P.French (1968). {\it Special Relativity.} 
 NY: Chapman and Hall.  

 \rr A.P.French (1974). {\it Vibrations and Waves.} 
    M.I.T. Introductory Physics Series. 

 \rr R.Frigg (2005). {\it Models and Representation: Why Structures 
    Are Not Enough.} Tech.Rep. 25/02, Center for Philosophy of Natural 
    and Social Science,  

 \rr  S.Fuchs (1996). The new Wars of Truth: 
 Conflicts over science studies as differential modes of observation. 
 {\it Social Science Information,} 307--326. 

 \rr M.C.Galavotti (1996). Probabilism and Beyond. 
    {\it Erkenntnis}, 45, 253-265.  

 \rr M.V.P.Garcia, C.Humes, J.M.Stern (2002). 
 Generalized Line Criterion for Gauss Seidel Method. 
 {\it Journal of Computational and Applied Mathematics}, 22, 1, 91-97. 

 \rr M.R. Garey, D.S. Johnson (1979).
 {\it Computers and Intractability, A Guide to the Theory of NP-Completeness}.
 NY: Freeman and Co. 

 \rr R.H.Gaskins (1992). 
 {\it Burdens of Proof in Modern Discourse.} 
 Yale Univ. Press.

 \rr  L.A.Gavrilov and N.S.Gavrilova (1991).
 {\it The Biology of Life Span: A Quantitative Approach}. 
 New York: Harwood Academic Publisher.

 \rr  L.A.Gavrilov and N.S.Gavrilova (2001). The Reliability Theory of
 Aging and  Longevity. {\it J. Theor. Biol.} 213, 527--545.

 \rr R.Karawatzki, J.Leydold, K.P\"{o}tzelberger (2005). Automatic
 Markov Chain  Monte Carlo Procedures for Sampling from Multivariate
 Distributions. Department of Statistics and Mathematics
 Wirtschaftsuniversit\"{a}t Wien Research Report Series. Report 27,
 December 2005.  Software available at
 {\it http://statistik.wu-wien.ac.at/arvag/software.html}. 
 
 \rr M.Gell'Mann (1994). {\it The Quark and the Jaguar: Adventures 
  in the Simple and the Complex.} New York: Freeman. 


 \rr A.Gelman, J.B.Carlin, H.S.Stern, D.B.Rubin (2003). 
   {\it Bayesian Data Analysis}, 2nd ed. NY: Chapman and Hall / CRC.    

 \rr S.Geman, D.Geman, (1984). Stochastic Relaxation, Gibbs Distribution 
  and Bayesian Restoration of Images. {\it IEE Transactions on Pattern 
  Analysis and Machine Intelligence,} 6, 721-741.

 \rr J.E.Gentle (1998). {\it Random Number Generator and Monte 
  Carlo Methods.} NY: Springer.

 \rr A.M.Geoffrion ed. (1972). {\it Perspectives on Optimization: 
    A Collection of Expository Articles.} NY: Addison-Wesley. 

 \rr A.George, J.W.H.Liu (1978). A Quotient Graph Model for Symmetric 
    Factorization. p.154-175 in: I.S.Duff, G.W.Stewart (1978) 
    {\it Spase Matrix Proceedings.} Philadelphia: SIAM. 
 
 \rr A.George, J.W.H.Liu, E.Ng (1989). Solution of Sparse Positive 
    Definite Systems on a Hypercube, in: 
    Vorst and van Dooren (1990). 

 \rr A.George, J.R.Gilbert, J.W.H.Liu (ed.) (1993). {\it Graph Theory and 
    Sparse Matrix Computation.} NY: Springer. 

 \rr A.George, J.W.H.Liu (1981). {\it Computer Solution of Large Sparse 
 Positive-Definite Systems.} NY: Prentice-Hall. 

 \rr C.J.Gerhardt (1890). 
 {\it Die philosophischen Schriften von Gottfried Wilhelm Leibniz.}  
 Berlin: Weidmannsche Buchhandlung.  

 \rr D.T.Gillespie (1992). 
 A Rigorous Derivation of the Chemical Master Equation. 
 {\it Physica A}, 188, 404-425. 

 \rr W.R.Gilks, S.Richardson, D.J.Spiegelhalter (1996). 
 {\it Markov Chain Monte Carlo in Practice.} NY: CRC Press. 
 
 \rr M.Ginsberg (1986). Multivalued Logics.
 {\it AAAI-86, 6th National Conference on Artificial Intelligence.}
  243--247.

 \rr Z.Ghaharamani, G.E.Hilton (1997). {\it The EM Algorithm for 
  Mixtures of Factor Analyzers.} Tech.Rep. CRG-TR-96-1. 
  Dept. of Computer Science, Univ. of Toronto.  
   
 \rr G.J.Chaitin (1975). Randomness and Mathematical Proof. 
  {\it Scientific American}, 232, 47-52.

 \rr G.J.Chaitin (1988). Randomness in Arithmetic. 
  {\it Scientific American}, 259, 80-85.

 \rr B.Goertzel, O.Aam, F.T.Smith, K.Palmer (2008). 
 Mirror Neurons, Mirrorhouses, and the Algebraic Structure of the Self,
 {\it Cybernetics and Human Knowing,} 15, 1, 9-28. 

 \rr B.Goertzel (2007). Multiboundary Algebra as Pregeometry. 
  {\it Electronic Journal of Theoretical Physics,} 16, 11, 173-186. 

 \rr D.E.Goldberg (1989). 
 {\it Genetic Algorithms in Search, Optimization, and Machine Learning}. 
 Reading, MA: Addison-Wesley. 

 \rr R.Goldblatt (1998). {\it Lectures on the Hyperreals: 
 An Introduction to Nonstandard Analysis.} NY: Springer.   

 \rr L. Goldstein, M. Waterman (1988).
 {\it Neighborhood Size in the Simulated Annealing Algorithm}.
 In Johnson (1988). 

 \rr H.H.Goldstine (1980). {\it A History of the Calculus of 
 Variations from the Seventeenth Through the Nineteenth Century}. 
 Studies in the History of Mathematics and the Physical Sciences. 
 NY: Springer. 

 \rr M.C.Golumbic (1980). 
 {\it Algorithmic Graph Theory and Perfect Graphs.} NY: Academic Press. 

 \rr D.V.Gokhale (1975). Maximum Entropy Characterization of some 
 Distributions. In Patil,G.P., Kotz,G.P., Ord,J.K. {\it Statistical 
 Distributions in Scientific Work.} V-3, 299-304. 

 \rr  G.H.Golub, C.F.van Loan (1989). {\it Matrix 
 Computations.} Baltimore: Johns Hopkins.

 \rr I.J.Good (1958). Probability and Statistical Inference by 
 G.Spencer-Brown (review). {\it The British Journal for the 
 Philosophy of Science,} 9, 35, 251-255.  

 \rr I.J.Good (ed.) (1962). {\it The Scientist Speculates. An Anthology 
    of Partly-Baked Ideas.} NY: Basic Books.  
 
 \rr  I.J.Good (1983). {\it Good thinking: The foundations of 
 probability and its applications}. 
 Minneapolis: University of Minnesota Press.  
 
 \rr I.J.Good (1988). The Interface Between Statistics and Philosophy of 
    Science. {\it Statistical Science}, 3, 4, 386-397. 

 \rr I.J.Good, Y.Mittal (1987). The Amalgamation and Geometry of
    Two-by-Two Contingency Tables. {\it Annals of Statistics}, 
    15, p. 695. 

 \rr P.C.Gotzsche (2002). Assessment of Bias. In S.Kotz, ed. (2006). 
  {\it The Encyclopedia of Statistics,} 1, 237-240.   

 \rr A.L.Goudsmit (1988).  
 Towards a Negative Understanding of Psychotherapy. 
 Ph.D. Thesis, Groningen University.  

  \rr M.Goupil (1991). {\it Du Flou au Clair? Histoire de 
   l'Affinit\'{e} Chimique de Cardan \`{a} Prigogine}. 
     Paris: CTHS. 

 \rr A.N.Gorban M.Shahzad (2011). The Michaelis-Menten-Stueckelberg 
  Theorem. {\it Entropy}, 13, 966-1019.

 \rr S.Greenland, J.Pearl. J.M.Robins (1999). 
 Confounding and Collapsibility in Causal Inference. 
 {\it Statistical Science}  14, 1, 29-46.

 \rr S.Greenland, H.Morgenstern1 (2001). Confounding in Health Research. 
  {\it Annual Review of Public Health}, 22, 189-212.  

 \rr J.S.Growney (1998). Planning for Interruptions. 
  {\it Mathematics Magazine}, 55, 4, 213-219. 

 \rr  B.Gruber et al. edit. (1980--98). 
 {\it Symmetries in Science, I--X.} NY: Plenum. 

 \rr E.Gunel (1984). A Bayesian Analysis of the Multinomial Model for a
 Dichotomous Response with Non-Respondents. {\it Communications in 
 Statistics - Theory and Methods}, 13, 737-51.

 \rr M.G\"{u}nther A.J\"{u}ngel (2003, p.117). 
 {\it Finanzderivate mit MATLAB. Mathematische Modellierung und 
  numerische Simulation .} Wiesbaden: Vieweg Verlag. 

 \rr I.Hacking (1988). Telepathy: Origins of Randomization in Experimental 
    Design. {\it Isis}, 79, 3, 427-451.  

 \rr G.Hadley (1964). {\it Nonlinear and Dynamic Programming.} 
 NY: Addison-Wesley.    

 \rr O.H\"{a}ggstr\"{o}m (2002). 
 {\it Finite Markov Chains and Algorithmic Applications.} 
 Cambridge Univ.

 \rr P.R.Halmos (1998). {\it Naive Set Theory.} 
    NY: Springer. 

 \rr J.H.Halton (1970). A Retrospective and Prospective Survey of 
 the Monte Carlo Method. {\it SIAM Review,} 12, 1, 1-63. 

 \rr H.D.Hamilton (1971). Geometry of the Selfish Herd. 
 {\it J. Theoretical Biology,} 31, 295--311.  

 \rr J.M.Hammersley, D.C.Handscomb (1964). {\it Monte Carlo Methods.} 
 London: Chapman and Hall. 

 \rr  J.Hanc, S.Tuleja, M.Hancova (2004). Symmetries and Conservation 
 Laws: Consequenses of Noether's Theorem.  
 {\it American Journal of Physics}, 72, 428--435.   

 \rr A.J.Hanson (2006). {\it Visualizing Quaternions.}  
  San Francisco, CA: Morgan Kaufmann - Elsevier. 
      
 \rr I.Hargittai (1992). {\it Fivefold Symmetry.} 
 Singapore: World Scientific. 

 \rr J.A.Hartigan (1983). {\it Bayes Theory.}   
 NY: Springer.  

 \rr C.Hartshorne, P.Weiss, A.Burks, edts. (1992).  
  {\it Collected Papers of Charles Sanders Peirce.} 
    Charlottesville: InteLex Corp. 

 \rr E.J.Haupt (1998). G.E.M\"{u}ller as a Source of American Psychology. 
 In R.W.Rieber, K.Salzinger, eds. (1998). 
 {\it Psychology: Theoretical-Historical Perspectives.}  
 American Psychological Association. 

 \rr D.A.Harville (2000). {\it Matrix Algebra From a Statistician's 
 Perspective}. NY: Springer. 

 \rr  L.L.Harlow, S.A.Mulaik, J.H.Steiger (1997). 
 {\it What If There Were No Significance Tests?}  
 London: LEA - Lawrence Erlbaum Associates. 

 \rr C.Hartshorne, P.Weiss, A.Burks, edts. (1992).  
    {\it Collected Papers of Charles Sanders Peirce.} 
    Charlottesville: InteLex Corp. 

 \rr Harville,D.A. (1997). {\it Matrix Algebra from a 
 Statistician's Perspective}. NY: Springer.  

 \rr W.K.Hastings (1970). Monte Carlo Sampling Methods Using Markov 
 Chains and their Applications. {\it Biometrika,} 57, 97-109.

 \rr M.Haw (2002). Colloidal Suspensions, Brownian Motion, Molecular
  Reality: A Short History. {\it J. Phys. Condens. Matter}. 
  14, 7769-7779.  

 \rr W.Heisenberg (1958). {\it Physics and Philosophy.} 
 London: Pinguin Classics reprint (2000). 

 \rr J.J.Heiss (2007). {\it The Meanings and Motivations of Open-Source 
 Communities.} Sun Developer Network, August 2007. 

 \rr W.Heitler (1956).  {\it Elementary Wave Mechanics with Applications 
   to Quantum Chemistry.}  Oxford University Press.   
 
 \rr E.Hellerman, D.C.Rarick (1971).
 Reinversion with the Preassigned Pivot Procedure. 
 {\it Mathematical Programming}, 1, 195-216.

 \rr Helmholtz (1887a). \"{U}ber die physikalische Bedeutung des Princips 
  der keinsten Wirkung. 
 {\it Journal f\"{u}r reine und angewandte Mathematik}, 
 100, 137-166, 213-222.  

 \rr Helmholtz (1887b). Zur Geschichte des Princips der kleinsten Action. 
 {\it Sitzungsberichte der K\"{o}niglich Preussichen Akademie der 
 Wissenschaften zu Berlin}, I, 225-236.  

 \rr N.D.Hemkumar, J.R.Cavallo (1994). Redundant and On-Line CORDIC 
 for Unitary Transformations. IEEE Transactions on Computers, 
 43, 8, 941--954. 

 \rr C.Henning (2006). {\it Falsification of Propensity Models by 
 Statistical Tests and the Goodness-of-Fit Paradox.} Technical Report, 
 Department of Statistical Science, University College, London.   

 \rr R.J.Hernstein, E.G.Boring (1966). 
 {\it A Source Book in Psychology.} Harvard Univ. 

 \rr M.B.Hesse (1966). {\it Models and Analogies in Science.}  
  University of Notre Dame Press. 

 \rr G.Hesslow (2002). Conscious thought as simulation of behavior
 and perception. {\it Trends Cogn. Sci.} 6, 242-247.

 \rr M.Heydtmann (2002). The nature of truth: Simpson's Paradox and the 
  Limits of Statistical Data. {\it QJM: An International Journal of 
  Medicine}. 95, 4, 247-249. 

 \rr by D.M.Himmelblau (1972). {\it Applied Nonlinear Programming.} 
  NY: McGraw-Hill. 

 \rr K.Hinkelmann (ed.) (1984). {\it Experimental Design, Statistical 
  Models and Genetic Statistics. Essays in Honor of Oscar Kempthorne.} 
  Basel: Marcel Dekker. 

 \rr Hitzer (2003). {\it Geometric Algebra - Leibnitz` Dream.    
  Innovative Teaching of Mathematics with Geometric Algebra.}  
  20-22 Nov. 2003, RIMS, Kyoto, Japan. 

 \rr E.M.S.Hitzer, L.Redaelli (2003). 
  Geometric algebra illustrated by Cinderella. 
  {\it Advances in Applied Clifford Algebras,} 13, 157-181. 

 \rr J.S.U.Hjorth (1984). {\it Computer Intensive Statistical Methods.} 
 Chapman and Hall, London. 

 \rr R.R.Hocking (1985). {\it The Analysis of Linear Models.} 
 Monterey: Brooks Cole. 

 \rr J.H.Holland (1975). 
 {\it Adaptation in Natural and Artificial Systems.}
 Ann Arbor: University of Michigan Press.

 \rr J.Honerkamp (1993).  {\it Stochastic Dynamical Systems: Concepts, 
 Numerical Methods, Data Analysis.} Wiley-VCH.  

 \rr F.H.C.Hotchkiss (1998). A ``Rays-as-Appendages'' Model for the Origin 
  of Pentamerism in Echinoderms.  {\it Paleobiology}, 24,2, 200-214.

 \rr R.Houtappel, H.van Dam, E.P.Wigner (1965). The Conceptual Basis 
 and Use of the Geometric Invariance Principles. {\it Reviews of 
 Modern Physics,} 37, 595--632.  

 \rr P.O.Hoyer (2004). Non-Negative Matrix Factorizations with Sparseness 
    Constrains. {\it J.of Machine Learning Research}, 5, 1457-1469.  

 \rr P.Hoyningen-Huene (1993). 
 {\it Reconstructing Scientific Revolutions. 
  Thomas S. Kuhn's Philosophy of Science.}  
 University of Chicago Press.  

 \rr C.Huang, A.Darwiche (1994). Inference in Belief Networks: 
 A Procedural Guide. Int.J.of Approximate Reasoning, 11, 1-58.  

 \rr M.D. Huang, F. Roameo, A. Sangiovanni-Vincentelli (1986).
 An Efficient General Cooling Schedule for Simulated Annealing. 
 {\it IEEE International Conference on Computer-Aided Design}, 
 381-384.

 \rr P.C.Hubert, M.Lauretto, J.M.Stern (2009). 
 FBST for a Generalized Poisson Distribution. 
 AIP Conference Proceedings, accepted. 

 \rr R.I.G.Hughes (1992). {\it The Structure and Interpretation of 
  Quantum Mechanics.} Harvard University Press.   

 \rr C.Humes, M.S.Lauretto, F.Nakano, C.A.B.Pereira, G.F.G.Rafare, 
  J.M.Stern (2012). TORC3: Token-ring Clearing Heuristic for 
  Currency Circulation. {AIP Conf.Proc.}, 1490, 179-188.  

 \rr T.P.Hutchinson (1991). {\it The engineering statistician's guide to
 continuous bivariate distributions.} Sydney: Rumsby Scientific Pub. 

 \rr M.Iacoboni (2008). {\it Mirroring People.} NY: FSG. 

 \rr H.Iba, T.Sato (1992). Meta-Level Strategy for Genetic Algorithms 
 Based on Structured Representation. p.548-554 in {\it Proc. of the 
 Second Pacific Rim International Conference on Artificial Intelligence.} 

 \rr I.A.Ibri (1992). {\it Kosmos Noetos. A Arquitetura Metaf\'{\i}sica 
 de Charles S. Peirce.} S\~{a}o Paulo: Prespectiva. 

 \rr R.Ingraham ed. (1982). {\it Evolution: A Century after Darwin.} 
 Special issue of San Jose Studies, VIII, 3. 

 \rr B.Ingrao, G.Israel (1990). {\it The Invisible Hand. Economic 
  Equilibrium in the History of Science.} Cambridge, MA: MIT Press. 

 \rr R.Inhasz, J.M.Stern (2010). Emergent Semiotics in Genetic
  Programming and the Self-Adaptive Semantic Crossover. 
  {\it Studies in Computational Intelligence}, 314, 381-392.
 
 \rr  T.Z.Irony, M.Lauretto, C.A.B.Pereira, and J.M.Stern (2002). 
 A Weibull Wearout Test: Full Bayesian Approach.
 In: Y.Hayakawa, T.Irony, M.Xie, edit.    
 {Systems and Bayesian Reliability}, 287--300. 
 {\it Quality, Reliability \& Engineering Statistics}, 5, 
 Singapore: World Scientific. 

 \rr T.Z.Irony, C.A.B.Pereira (1994). Motivation for the Use of Discrete
 Distributions in Quality Assurance. {\it Test}, 3,2, 181-93.

 \rr T.Z.Irony, C.A.B.Pereira (1995), Bayesian Hypothesis Test: Using 
 Surface Integrals To Distribute Prior Information Among The Hypotheses,
 Resenhas, Sao Paulo 2(1): 27-46.

 \rr T.Z.Irony, C.A.B.Pereira, R.C.Tiwari (2000). Analysis of Opinion 
 Swing: Comparison of Two Correlated Proportions. {\it The American
 Statistician}, 54, 57-62.

 \rr A.N.Iusem, A.R.De Pierro (1986). Convergence Results for an 
 Accelerated Nonlinear Cimmino Algorithm. {\it Numerische Matematik,}  
 46, 367-378. 

 \rr A.N.Iusem (1995). {\it Proximal Point Methods in Optimization.} 
 Rio de Janeiro: IMPA. 

 \rr A.J.Izzo (1992). A Functional Analysis Proof of the Existence of Haar 
 Measure on Locally Compact Abelian Groups
 {\it Proceedings of the American Mathematical Society}, 115, 2, 581-583. 

 \rr  B.Jaffe (1960). {\it Michelson and the Speed of Light.} 
 NY: Anchor.  

 \rr W.James (1909, 2004). {\it A Pluralistic Universe.}. 
 The Project Gutenberg, E-Book 11984, Released April 10, 2004.  
 
 \rr L.J\'{a}nossy, A.R\'{e}nyi, J.Acz\'{e}l (1950). 
 On Composed Poisson Distributions. 
 {\it Acta Math. Hungarica}, 1, 209-224.  

 \rr E.Jantsch (1980). {\it Self Organizing Universe: Scientific and 
  Human Implications.} Pergamon. 

 \rr E.Jantsch ed. (1981). {\it The Evolutionary Vision. Toward a 
  Unifying Paradigm of Physical, Biological and Sociocultural Evolution.}  
  Washington DC, AAA - American Association for the Advancement of Science. 

 \rr E.Jantsch, C.H.Waddington, eds. (1976). {\it Evolution and Consciousness. 
  Human Systems in Transition.} London: Addison-Wesley.  

 \rr J.Jastrow (1899). The mind's eye. 
 {\it Popular Science Monthly}, 54, 299-312. 
 Reprinted in Jastrow (1900). 
 
 \rr J.Jastrow (1900). {\it Fact and Fable in Psychology.} 
  Boston: Houghton Mifflin.

 \rr J.Jastrow (1988). A Critique of Psycho-Physic Methods. 
 {\it American Journal of of Psychology}, 1, 271-309.   

 \rr E.T.Jaynes (1980). The Minimum Entropy Production Principle. 
 {\it Ann. Rev. Phys. Chem.} 31, 579-601.  

 \rr E.T.Jaynes (1990). Probability Theory as Logic. 
  {\it Maximum-Entropy and Bayesian Methods,} 
  ed. P.F.Fougere, Kluwer. 

 \rr E.T.Jaynes (2003). {\it Probability Theory: The Logic of Science.} 
 Cambridge University Press. 

 \rr H.Jeffreys (1961). {\it Theory of Probability.} 
 Oxford: Clarendon Press. (First ed. 1939).

 \rr R.I.Jennrich (2001). A Simple General Method for Orthogonal Rotation. 
  {\it Psychometrica}, 66, 289-306. 

 \rr R.I.Jennrich (2002). A Simple Method for Oblique Rotation. 
 {\it Psychometrika}, 67,1,7-20.  

 \rr R.I.Jennrich (2004). Rotation to Simple Loadings using Component
 Loss Functions: The Orthogonal Case. {\it Psychometrika}, 69, 257-274. 

 \rr J.M.Jeschke, R.Tollrian (2007). Prey swarming: Which predators
 become confused and why. 
 {\it Animal Behaviour,} 74, 387--393. 

 \rr G.Jetschke (1989). On the Convergence of Simulated Annealing. 
 pp. 208-215 in Voigt et al. (1989). 

 \rr T.J.Jiang, J.B.Kadane,  J.M.Dickey (1992). Computation of Carsons
 Multiple Hipergeometric Function R for Bayesian Applications.  
 {\it Journal of Computational and Graphical Statistics}, 1, 231-51.

 \rr Jiang,G., Sarkar,S. (1998). Some Asymptotic Tests for the 
 Equality of Covariance Matrices of Two Dependent Bivariate Normals. 
 {\it Biometrical Journal}, 40, 205--225. 

 \rr  Jiang,G., Sarkar,S., Hsuan,F. (1999). A Likelihood Ratio Test  
 and its Modifications for the Homogeneity of the Covariance Matrices of 
 Dependent Multivariate Normals. {\it J. Stat. Plan. Infer.}, 81, 95-111. 

 \rr  Jiang,G., Sarkar,S. (2000a). Some Combination Tests for the 
 Equality of Covariance Matrices of Two Dependent Bivariate Normals. 
 Proc. {\it ISAS-2000, Information Systems Analysis and Synthesis.} 

 \rr  Jiang,G., Sarkar,S. (2000b). The Likelihood Ratio Test for 
 Homogeneity of the Variances in a Covariance Matrix with Block 
 Compound Symmetry. {\it Commun. Statist. Theory Meth.} 29, 1155-1178.  

 \rr D.S.Johnson, C.R.Aragon, L.A.McGeoch, C.Schevon (1989).
 Optimization by Simulated Annealing: An experimental evaluation, part 1.
 {\it Operations Research},  37, 865-892.

 \rr M.E.Johnson (1987). Multivariate Statistical Simulation. 
 NY: Wiley. 

 \rr M.E. Johnson (ed.) (1988).
 {\it Simulated Annealing \& Optimization}.
 Syracuse: American Science Press. This book is also the volume 8 
 of the {\it American Journal of Mathematical and Management Sciences.} 

 \rr P.Johansson, L.Hall, S.Silksr\"{o}m, A.Olsson (2008). 
  Failure to Detect Mismatches Between Intention and Outcome in 
  Simple Decision Task. {\it Science,} 310, 116-119.  

 \rr  M.C.Jones (1985). Generating Inverse Wishart Matrices. 
 {\it Commun. Statist. Simula. Computa.} 14, 511--514. 
  
 \rr  J\"{o}reskog,K.G. (1970). A General Method for Analysis of 
 Covariance Structures. {\it Biometrika}, 57, 239--251.  

 \rr C.G.Jung (1968). {\it Man and His Symbols.}  Laurel.

 \rr M.Kac (1983). What is Random? {\it American Scientist},
 71, 405-406.

 \rr J.B.Kadane (1985). Is Victimization Chronic? A Bayesian Analysis of
 Multinomial Missing Data. {\it Journal of Econometrics}, 29, 47-67.

 \rr J.Kadane, T.Seidenfeld (1990). Randomization in a Bayesian Perspective. 
    {\it J.of Statistical Planning and Inference}, 25, 329-345. 

 \rr J.B.Kadane, R.L.Winkler (1987). De Finetti's Methods of 
 Elicitation. In Viertl (1987).

 \rr I.Kant (1790). Critique of Teleological Judgment. 
 In Kant's Critique of Judgement, Oxford: Clarendon Press, 1980. 

 \rr I.Kant. {\it The critique of pure reason; The critique 
             of practical reason; The critique of judgment.} 
 Encyclopaedia Britannica Great books of the Western World, v.42, 1952.

 \rr S.Kaplan, C.Lin (1987). An Improved Condensation Procedure in 
 Discrete Probability Distribution Calculations. 
 {\it Risk Analysis,} 7, 15-19.  

 \rr   T.J.Kaptchuk, C.E.Kerr (2004). Commentary: Unbiased Divination, 
  Unbiased Evidence, and the Patulin Cliniacal Trial. 
  {\it International Journal of Epidemiology}, 33, 247-251.  

 \rr J.N.Kapur (1989). {\it Maximum Entropy Models in Science and 
 Engineering.}  New Delhi: John Wiley.    

 \rr J.N.Kapur, H.K.Kesevan (1992). {\it Entropy Optimization Principles
  with Applications.} Boston: Academic Press. 

 \rr T.R.Karlowski, T.C.Chalmers, T.C.Frankel, L.D.Kapikian, 
    T.L.Lewis, J.M.Lynch (1975). Ascorbic acid for the common cold: 
    a prophylactic and therapeutic trial. {\it JAMA}, 231, 1038-1042. 

 \rr A.Kaufmann, D.Grouchko, R.Cruon (1977). {\it Mathematical Models 
 for the Study of the Reliability of Systems.} NY: Academic Press. 

 \rr L.H.Kauffman (2001). The Mathematics of Charles Sanders Peirce. 
 {\it Cybernetics and Human Knowing,} 8, 79-110. 

 \rr  L.H.Kauffmann (2006). {\it Laws of Form: 
 An Exploration in Mathematics and Foundations.} \  
 \verb#http://www.math.uic.edu/~kauffman/Laws.pdf#  

 \rr M.J.Kearns, U.V.Vazirani (1994). {\it Computational Learning 
 Theory.} Cambridge: MIT Press.  

 \rr R. Keller, L.A. Davidson and D.R. Shook (2003). 
 How we are Shaped: The Biomechanics of Gastrulation.  
 {\it Differentiation,} 71, 171-205.

 \rr O.Kempthorne, L.Folks (1971). {\it Probability, Statistics and 
 Data Analysis.} Ames: Iowa State Univ. Press. 

 \rr O.Kempthorne (1976). Of what Use are Tests of Significance and 
 Tests of Hypothesis. {\it Comm. Statist.} A5, 763--777. 

 \rr O.Kempthorne (1977). Why Randomize? {\it J. of Statistical Planning
    and Inference,} 1, 1-25 

 \rr Kempthorne,O. (1980). Foundations of Statistical Thinking and 
 Reasoning. {\it Australian CSIRO-DMS Newsletter.} 68, 1--5; 69, 3--7.

 \rr M.G.Kendall (2004). {\it A Course in the Geometry of 
   $n$-Dimensions}. Mineola: Dover.  

 \rr B.W. Kernighan, S. Lin (1970).
 An Efficient Heuristic Procedure for Partitioning Graphs. 
 {\it The Bell System Technical Journal}, 49, 291-307.

 \rr A.I.Khinchin (1953). {\it Mathematical Foundations of Information 
 Theory.} NY: Dover.  

 \rr J.F.Kihlstrom (2006). {\it Joseph Jastrow and His Duck - Or Is It 
  a Rabbit?} On line document, University of California at Berkeley. 

 \rr D.A.Klein and G.C.Rota (1997). {\it Introduction to Geometric 
 Probability.} Cambridge Univ. Press.  

 \rr  G.J.Klir, T.A.Folger (1988). 
 {\it Fuzzy Sets, Uncertainty and Information.} NY: Prentice Hall.    

 \rr C.J.W.Kloesel (1993). {\it Writings of Charles S. Peirce. 
    A Chronological Edition}.   

 \rr S.Kocherlakota, KKocherlakota (1992). {\it Bivariate Discrete 
 Distributions.} 
 Basel: Marcel Dekker. 

 \rr M.A.R.Koehl (1990). Biomechanical Approaches to Morphogenesis. 
  {\it Sem. Dev. Biol.} 1, 367-378. 

 \rr J.Kokott (1998). {\it The Burden of Proof in Comparative and
  International Human Rights Law.}  Hague: Kluwer.

 \rr V.B.Kolmanovskii, V.R.Nosov (1986). {\it Stability of Functional 
 Differential Equations.} London: Academic Press. 

 \rr A.N.Kolmogorov ( 1965 ). Three Approaches to the Quantitative 
  Definition of Information. 
 {\it Problems in Information Transmission,} 1, 1-7.

  \rr A.N.Kolmogorov, S.V.Fomin (1982, Portuguese translation). 
 {\it Elements of the Theory of Functions and Functional Analysis.}    
 Moscow: MIR. 

 \rr B.O.Koopman (1940a). Axioms and Algebra of Intuitive 
 Probability. {\it Annals of Mathematics}, 41, 269--292. 

 \rr B.O.Koopman (1940b). Bases of Probability. {\it 
 Bulletin of the Ammerican Mathematical Society}, 46, 763--774. 

 \rr F.H.H.Kortlandt (1985). A Parasitological View of Non-Constructible 
  Sets. {\it Studia Linguistica Diachronica et Synchronica,} 477-483. 

 \rr S.Kotz, N.Balakrishnan, C.B.Read, B.Vidakovic, edts. (2005). 
    {\it Encyclopedia of Statistical Sciences,} 2nd ed. 
    Wiley-Interscience. 

 \rr J.R.Koza (1989). Hierarchical Genetic Algorithms Operating on 
  Populations of Computer Programs. 
  In Proceedings of the Eleventh International Joint Conference on
  Artificial Intelligence, IJCAI-89, N.S.Sridharan (ed.), 
  vol.1, p.768--774, Morgan Kaufmann. 
 
 \rr  K.L.Krasnov, G.I.Makarenko, A.I.Kiseliov (1973, 1984). 
 {\it C\'{a}lculo Variacional}. MIR, Moskow. 

 \rr K.Krippendorff (1986). {\it Information Theory: Structural Models 
   for Qualitative Data.} Quantitative Applications in the Social 
   Sciences V.62.  Beverly Hills: Sage. 

 \rr  W.Krohn, G.K\"{u}ppers, H.Nowotny (1990). 
 {\it Selforganization. Portrait of a Scientific Revolution.}  
 Dordrecht: Kluwer. 

 \rr  W.Krohn, G. K\"{u}ppers (1990). The Selforganization of  
 Science - Outline of a Theoretical Model.  
 in Krohn (1990), 208--222. 

 \rr P.Krugman (1999). O Canada: A Neglected Nation Gets its Nobel.
    Slate, October 19, 1999. 

 \rr A.Krzysztof Kw\'{a}sniewski (2008). {\it Glimpses of the Octonions 
  and Quaternions History and Today's Applications in Quantum Physics.}  
  eprint arXiv:0803.0119. 

 \rr O.S.Ksenzhek, A.G.Volkov (1998). {\it Plant Energetics.} 
  NY: Academic Press. 

 \rr T.S.Kuhn (1977). {\it The Essential Tension: Selected Studies 
    in Scientific Tradition and Change}. University of Chicago Press. 

 \rr T.S.Kuhn (1996). {\it The Structure of Scientific Revolutions.} 
 University of Chicago Press. 

 \rr H.Kunz, T.Z\"{u}blin, C.K.Hemelrijk (1000). 
 On Prey Grouping and Predator Confusion in Artificial Fish Schools.  

 \rr P.J.M. van Laarhoven, E.H.L. Aarts (1987).
 {\it Simulated Annealing: Theory and Applications}.
 Dordrecht: Reidel Publishing Co. 

 \rr C.L.Lanczos (1986). {\it The Variational Principles of Mechanics.}   
  Mineola: Dover. 
  Noether's Invariant Variational Problems, Appendix II, p.401-405. 

 \rr D.Landau, K.Binder (2000). {\it A Guide to Monte Carlo Simulations 
 in Statistical Physics.} Cambridge University Press.  

 \rr L.D.Landau, E. M. Lifchitz (1966). 
 {\it Cours de Physique Th\'{e}orique.} Moscou: MIR. 

 \rr P.V.Landshoff, A.Metherell, W.G.Rees (1998). 
 {\it Essential Quantum Physics.}  Cambridge University Press.   

 \rr K.Lange (2000). {\it Numerical Analysis for Statisticians.} 
 NY: Springer. 

 \rr I.Lakatos (1978a). {\it The Methodology of Scientific Research.} 
 Cambridge Univ. Press. 

 \rr I.Lakatos (1978b). {\it Mathematics, Science and Epistemology.} 
 Canbridge Univ. Press. 

 \rr G.Lakoff, M.Johnson (2003). {\it Metaphors We Live By.} 
 University of Chicago Press. 

 \rr L.S.Lasdon (1970). {\it Optimization Theory for Large Systems.} 
    NY: MacMillan. 

 \rr Laurent, John. 1999. A note on the origin of memes / mnemes, 
  {\it Journal of Memetics,} 3, 1, 20-21.

 \rr M.Lauretto, F.Nakano, C.A.B.Pereira, J.M.Stern (2009). 
 Hierarchical Forecasting with Polynomial Nets. 
 {\it Studies in Computational Intelligence}, 199, 305-315. 

 \rr M.Lauretto, F.Nakano, S.R.Faria, C.A.B.Pereira, J.M.Stern (2009). 
  A Straightforward Multiallelic Signicance Test for the Hardy-Weinberg 
  Equilibrium Law. {\it Genetics and Molecular Biology}, 32, 3, 619-625. 

 \rr M.S.Lauretto, F.Nakano, C.A.B.Pereira, J.M.Stern (2012). 
  Intentional Sampling by Goal Optimization with Decoupling by 
  Stochastic Perturbation. {AIP Conf.Proc.}, 1490, 189-201.   

 \rr   M.Lauretto, C.A.B.Pereira, J.M.Stern, S.Zacks (2003).
 Full Bayesian Significance Test Applied to Multivariate Normal 
 Structure Models.  
 {\it Brazilian Journal of Probability and Statistics,} 17, 147-168. 

 \rr M.Lauretto, F.Nakano, C.O.Ribeiro, J.M.Stern (1998). 
 REAL: Real Attribute Learning Algorithm. 
 ISAS-98 Proceedings, 2, 315-321.   

 \rr M.Lauretto, J.M.Stern (2005). FBST for Mixture Model Selection. 
 MaxEnt 2005, 24th International Workshop on Bayesian Inference 
 and Maximum Entropy Methods in Science and Engineering. 
 {\it American Institute of Physics Conference Proceedings}, 
 803, 121--128.  

  \rr M.Lauretto, S.R. de Faria Jr., B.B.Pereira, 
   C.A.B.Perreira, J.M.Stern (2007).  
  The Problem of Separate Hypotheses via Mixture Models. To appear, 
  {\it American Institute of Physics Conference Proceedings}.  

 \rr M.S.Lauretto, C.A.B.Pereira, J.M.Stern (2008). MaxEnt 2008 - 
  {\it Bayesian Inference and Maximum Entropy Methods in Science and 
  Engineering.} July 6-11, Borac\'{e}ia, S\~{a}o paulo, Brazil. 
  American Institute of Physics Conference Proceedings, v.1073.  

 \rr S.L.Lauritzen (2006). {\it Fundamentals of Graphical Models.} 
    Saint Flour Summer-school.   

 \rr J.W.Leech (1963). {\it Classical Mechanics.} London: Methuen. 

 \rr Lehmann,E.L. (1959). {\it Testing Statistical Hypothesis.} 
 NY: Wiley.  

 \rr T.G.Leighton, S.D.Richards, P.R.White (2004). 
 Trapped within a `Wall of Sound'
 A Possible Mechanism for the Bubble Nets of Humpback Whales. 
 {\it Acoustics Bulletin,} 29, 1, 24-29. 

 \rr T.Leighton, D.Finfer, E.Grover, P.White (2007). 
 An Acoustical Hypothesis for the Spiral Bubble Nets of
 Humpback Whales, and the Implications for Whale Feeding. 
 {\it Acoustics Bulletin,} 32, 1, 17-21. 

 \rr D.S.Lemons (2002).  {\it An Introduction to Stochastic Processes in 
  Physics.} Baltimore: John Hopkins Univ. Press. 

 \rr    T.Lenoir (1982). {\it The Strategy of Life. Teleology and Mechanics 
 in Nineteenth-Century German Biology.} Univ.of Chicago Press. 

 \rr  I.Levi (1974). {\it Gambling with Truth: 
 An Essay on Induction and the Aims of Science.}  
 MIT Press. 

 \rr  K.Lewin (1951). {\it Field Theory m Social Science: 
  Selected Theoretical Papers.} New York: Harper and Row. 

 \rr A.M.Liberman (1993). Haskins Laboratories Status Report on
  {\it Speech Research,} 113, 1-32

 \rr  D.V.Lindley (1957). A Statistical Paradox. 
 {\it Biometrika} 44, 187--192.  

 \rr D.V.Lindley (1991). {\it Making Decisions.} NY: John Wiley.  

 \rr D.V.Lindley, M.R.Novick (1981). The Role of Exchangeability in 
    Inference. {it The Annals of Statistics}, 9, 1, 45-58. 

 \rr R.J.A.Little, D.B.Rubin (1987). {\it Statistical Analysis with 
 Missing Data.} New York: Wiley.

 \rr J.L.Liu (2001). {\it Monte Carlo Strategies in Scientific Computing.} 
 NY: Springer. 

 \rr D.Loemker (1969). 
  {\it G.W.Leibniz Philosophical Papers and Letters.}  
  Reidel.

 \rr L.L.Lopes (1982). Doing the Impossible: A Note on Induction and
 the Experience of Randomness. {\it Journal of Experimental Psychology:
 Learning, Memory, and Cognition}, 8, 626-636.

 \rr L.L.Lopes,  G.C.Oden (1987). Distinguishing Between Random
 and Nonrandom Events. {\it Journal of Experimental Psychology: 
 Learning, Memory, and Cognition}, 13, 392-400.

 \rr  H.A.Lorentz, A.Einstein, H.Minkowski and H.Weyl (1952). 
 {\it The Principle of Relativity: A Collection of Original Memoirs 
 on the Special and General Theory of Relativity.} 
 NY: Dover.

 \rr R.H.Loschi, S.Wechsler (2002). {\it Coherence, Bayes's 
 Theorem and Posterior Distributions}, Brazilian Journal of 
 Probability and Statistics, 16, 169--185. 

 \rr LosDoggies web page (2010). Retrieved from  
  \verb#http://www.losdoggies.com/archives/858# 

 \rr P.Lounesto (2001). {\it Clifford Algebras and Spinors.} 
  2nd ed. Cambridge University Press.  

 \rr D.G.Luenberger (1984). {\it Linear and Nonlinear Programming.} 
  Reading: Addison-Wesley. 

 \rr  N.Luhmann (1989). 
 {\it Ecological Communication.}  Chicago Univ. Press. 

 \rr   N.Luhmann (1990a). {\it The Cognitive Program of 
 Constructivism and a Reality that Remains Unknown.} 
 in Krohn (1990), 64--86. 

 \rr  N.Luhmann (1990b). {\it Essays on Self-Reference.}  
 NY: Columbia Univ. Press. 

 \rr  N.Luhmann (1995). 
 {\it Social Systems.} Stanford Univ. Press. 

 \rr M. Lundy, A. Mees (1986). Convergence of an Annealing Algorithm.
 {\it Mathematical Programming}, 34, 111-124.

 \rr I.J.Lustig (1987).
 {\it An Analysis of an Available Set of Linear Programming Test Problems}.
 Tech. Rep. SOL-87-11, Dept. Operations Research, Stanford University.

 \rr D.K.C.MacDonald (1962). 
 {\it Noise and Fluctuations: An Introduction.} NY: Wiley.   

 \rr H.R.Madala, A.G.Ivakhnenko (1994). {\it Inductive Learning Algorithms
  for Complex Systems Modeling.} Boca Raton: CRC Press.

 \rr  M.R.Madruga, L.G.Esteves, S.Wechsler (2001).
 On the Bayesianity of Pereira-Stern Tests.
 {\it Test}, 10, 291--299.
 
 \rr   M.R.Madruga, C.A.B.Pereira, J.M.Stern (2003).
 Bayesian Evidence Test for Precise Hypotheses.
 {\it Journal of Statistical Planning and Inference,}
  117, 185--198.

 \rr  P.Maher, B.Skyrms, E.W.Adams, K.Binmore, J.Butterfield, 
 P.Diaconis, W.L. Harper (1993). {\it Betting on Theories.}  
 Cambridge Univ. Press. 

 \rr M.Maimonides (2001). {\it  Mishne Torah: Yad hachazakah.}
 NY: Yeshivath Beth Moshe. 

 \rr N.I.Mann, K.A.Dingess, P.J.B.Slater (2006). 
 Antiphonal four-part synchronized chorusing in a Neotropical Wren. 
 {\it Biol. Lett.,} 2, 1-4.  

 \rr V.T.L.Maranh\~{a}o, M.S.Lauretto, J.M.Stern (2012). 
  FBST for Covariance Structures of Generalized Gompertz Models. 
  {AIP Conf.Proc.}, 1490, 202-211. 

 \rr L.Margulis (1999). 
 {\it Symbiotic Planet: A New Look At Evolution.}  Basic Books.   

 \rr L.Margulis, D.Sagan (2003). 
 {\it Acquiring Genomes: The Theory of the Origins of the Species.} 
 Basic Books. 

 \rr D.D.Mari, S.Kotz (2001). {\it Correlation and Dependence.} 
    Singapore: World Scientific. 

 \rr J.B.Marion (1970). {\it Classical Dynamics of Particles and Systems.} 
 NY:  Academic Press. 

 \rr J.B.Marion (1975). {\it Classical Dynamics of Particles and 
    Systems.} NY: Academic Press.

 \rr H.M.Markowitz (1952).  Portfolio Selection.  {\it The Jounal of
  Finance}, 7(1), pp-77-91. 

 \rr H.M.Markowitz (1956). The optimization of a Quadratic Function 
 Subject to Linear Constraints. 
 {\it Naval Research Logistics Quarterly,} 3, 111-133. 
 
 \rr H.M.Markowitz (1987). {\it Mean-variance Analisys in Portfolio 
 Choice and Capital Markets}.  
 Cambridge, MA: Basil Blackwell.

 \rr G.Marsaglia (1968). Random Numbers Fall Mainly in the Planes. 
  {\it Proceedings of the National Academy of Sciences}, 61, 25-28. 

 \rr J.J.Martin (1975). 
 {\it Bayesian decision and probelms and Markov Chains.}  

 \rr J.L.Martin (1988). {\it Genearl Relativity. A Guide to its 
  Consequences for Gravity and Cosmology.} 
  Chichester: Ellis Horwood - John Willey.  

 \rr E.Martin-L\"{o}f(1966). The Definition of Random Sequences. 
  {\it Information and Control}, 9, 602-619.

 \rr E.Martin-L\"{o}f (1969). Algorithms and Randomness. 
 {\it Review of the Intern. Statistical Institute}, 37, 3, 265-272. 

 \rr J.M.Martinez, J.M. (1999). A Direct Search Method for Nonlinear 
  Programming. {\it ZAMM,} 79, 267-276. 

 \rr J.M.Martinez (2000). BOX-QUACAN and the Implementation of 
 Augmented Lagrangian Algorithms for Minimization with Inequality 
 Constraints. {\it Computational and Applied Mathematics.} 19, 31-56.  

 \rr J.Matou\u{s}ek (1991). {\it Geometric Discrepancy.} 
 Berlin: Springer. 

 \rr M.Matsumoto, T.Nishimura (1998) Mersenne Twister: A 623-dimensionally
 Equidistributed Uniform Pseudorandom Number Generator. 
 {\it ACM Trans. Model. Comput. Simul.}, 8, 3-30.   

 \rr M.Matsumoto, Y.Kurita (1992,1994). Twisted GFSR Generators. 
 {\it ACM Trans. Model. Comput. Simul.}, I:2,179-194; II:4,254-266.

  \rr  H.R.Maturana, F.J.Varela (1980). 
 {\it Autopoiesis and Cognition. The Realization of the Living.}  
 Dordrecht: Reidel. 

 \rr H.R.Maturana (1988). Ontology of Observing. The Biological 
 Foundations of Self Consciousness and the Physical Domain of Existence.  
 pp 18--23 in {\it Conference Workbook: Texts in Cybernetics.} 
 Felton, CA: American Society for Cybernetics.  

 \rr H.R.Maturana (1991). Science and Reality in Daily Life: 
    The Ontology of Scientific Explanations. In Steier (1991). 
  
 \rr H.R.Maturana, B.Poerksen (2004). Varieties of Objectivity. 
 {\it Cybernetics and Human Knowing.} 11, 4, 63--71.

 \rr P.L.M.de Maupertuis (1965), {\it Oeuvres, I-IV.} 
  Hildesheim: Georg Olms Verlagsbuchhandlung.  

 \rr G.P. McCormick  (1983).
 {\it Nonlinear Programming: Theory, Algorithms and Applications}.
 Chichester: John Wiley.

 \rr D.K.C.MacDonald (1962). {\it Noise and Fluctuations.} NY: Dover. 

 \rr  R.P.McDonald (1962). A Note on the Derivation of the General 
 Latent Class Model. {\it Psychometrika} 27, 203--206. 

 \rr R.P.McDonald (1974). Testing Pattern Hypotheses for Covariance  
 Matrices. {\it Psychometrika}, 39, 189--201. 

 \rr R.P.McDonald (1975). Testing Pattern Hypotheses for Correlation 
 Matrices. {\it Psychometrika}, 40, 253--255. 

 \rr R.P.McDonald, H.Swaminathan (1973). A Simple Matrix Calculus 
 with Applications to Multivariate Analysis. 
 {\it General Systems}, 18, 37--54 

  \rr A.L.McLean (1998), The Forecasting Voice: A Unified Approach to 
 Teaching Statistics. In {\it Proceedings of the Fifth International 
 Conference on Teaching of Statistics}, (eds L. Pereira-Mendoza, et al.), 
 1193-1199. Singapore: Nanjing University.

 \rr G.McLachlan, D.Peel (2000). {\it Finite Mixture Models.} NY: Wiley. 

 \rr J.D.McGervey (1995). {\it Quantum Mechanics: Concepts and 
  Applications.} San Diego: Academic Press.

 \rr W.H.McRea (1954). {\it Relativity Physics}. London: Methuen.  

 \rr E.J.McShane. The Calculus of Variations. Ch.7, p.125-130 in: 
  J.W.Brewer, M.K.Smith (1981). {\it Emmy Noether}.  

 \rr P.Meguire (2003). Discovering Boundary Algebra: A Simple Notation 
  for Boolean Algebra and the Truth Functions. 
  {\it Int. J. General Systems,} 32, 25-87. 

 \rr   J.G.Mendel (1866). Versuche \"{u}ber Plflanzenhybriden Verhandlungen 
 des naturforschenden Vereines in Br\"{u}nn, Bd. IV f\"{u}r das Jahr, 1865
 Abhandlungen: 3-47. For the English translation, see: Druery, C.T and
 William Bateson (1901). Experiments in Plant Hybridization. 
 {\it Journal of the Royal Horticultural Society}, 26, 1-32.  

 \rr M.B.Mendel (1989). {\it Development of Bayesian Parametric Theory 
 with Application in Control.} PhD Thesis, MIT, Cambridge: MA.

 \rr X.L.Meng, W.H.Wong (1996). Simulating Ratios of Normalizing 
 Constants via a Simple Identity: A Theoretical Exploration. 
 {\it Statistica Sinica}, 6, 831-860. 

 \rr R.Merkel (2005). 
 {\it Analysis and Enhancements of Adaptive Random Testing.} 
 Ph.D. Thesis. Swinburne University of Technology in Melbourne. 
 Melburne: Australia. 

 \rr M. Mesterton-Gibbons (1992). Redwood, CA: {\it An Introduction 
  to Game-Theoretic Modelling.} Addison-Wesley. 

 \rr N.Metropolis, S.Ulam (1949). The Monte Carlo method. 
  {\it J. Amer. Statist. Assoc.,} 44, 335-341.

 \rr N.Metropolis, A.W.Rosenbluth, M.N.Rosenbluth, A.H.Teller,  
    E.Teller (1953). Equations of State Calculations by Fast Computing 
    Machines. {\it Journal of Chemical Physics}, 21, 6, 1087-1092. 

 \rr  A.A.Michelson, E.W.Morley (1887). On the Relative Motion 
 of the Earth and the Luminiferous Ether. 
 {\it American Journal of Physics,} 34, 333--345. 

 \rr D.Michie, D.J.Spiegelhalter, C.C.Taylor (1994). {\it Machine Learning, 
  Neural and Statistical Classification.} Ellis Horwood. 

 \rr R.E.Michod, B.R.Levin (1988). The Evolution of Sex: An Examination 
 of Current Ideas. Sunderland, MA: Sinauer Associates. 

 \rr W.Millar (1951). Some General Theorems for Non-Linear Systems 
 Possessing Resistance. {\it Philosophical Magazine}, 
 7, 42 (333), 1150-1160. 

 \rr G.Miller (2000). Mental traits as fitness indicators - expanding
 evolutionary psychology's adaptationism. 
 Evolutionary Perspectives on Human Reproductive Behaviour. 
 {\it Annals of the New York Academy of Sciences}, 907, 62-74.

 \rr G.F.Miller (2001). {\it The Mating Mind: How Sexual Choice Shaped 
 the Evolution of Human Nature}. London: Vintage.

 \rr G.F.Miller, P.M.Todd (1995). The role of mate choice in
 biocomputation: Sexual selection as a process of search, optimization,
 and diversification. In: 
 W. Banzhaf, F.H. Eeckman (Eds.) Evolution and biocomputation: 
 Computational models of evolution (pp. 169-204). Berlin: Springer.

 \rr J.Miller (2006). Earliest Known Uses of Some of the Words of 
    Mathematics.\\ {\it http://members.aol.com/jeff570/mathword.html}

 \rr  J.Mingers (1995) {\it Self-Producing Systems: 
 Implications and Applications of Autopoiesis.} 
 NY: Plenum. 

 \rr  M.Minoux, S.Vajda (1986). {\it Mathematical Programming.} 
 John Wiley. 

 \rr C.W.Misner, K.S.Thorne, J.A.Wheeler, J.Wheeler (1973).  
  {\it Gravitation.} W.H.Freeman. 

 \rr O.Morgenstern (2008). Entry {\it Game Theory} at the
 {\it Dictionary of the History of Ideas} (v.2, p.264-275). 
 Retrieved from \\   
 \verb# http//etext.virginia.edu/cgi-local/DHI/dhi.cgi?id=dv2-32#  

 \rr O.Morgenstern, J.von Neumann (1947).  
 {\it The Theory of Games and Economic Behavior.}  
 Princeton University Press. 

 \rr W.J.Morokoff (1998). Generating Quasi-Random Paths for Stochastic 
 Processes. {\it SIAM Review,} 40, 4, 765-788.  

 \rr P.Moscato (1989). {\it On Evolution, Search, Optimization, 
  Genetic Algorithms and Martial Arts: Towards Memetic Algorithms.} 
  Caltech Concurrent Computation Program, Tech.Repport 826.  

 \rr A.Mosleh, V,M,Bier (1996). Uncertainty about Probability: 
 A Reconciliation with the Subjectivist Viewpoint. 
 {\it IEEE Transactions on Systems, Man and Cybernetics,} 
 A, 26, 3, 303-311. 

 \rr W.Mueller, F.Wysotzki (1994). Automatic Construction of 
 Decision Trees for Classification. {\it Ann. Oper. Res.} 52, 231-247.   

 \rr S.H.Muggleton (2006). Exceeding Human Limits. 
    {\it Nature}, 440/23, 409--410. 

 \rr P.Muir (1907). 
 {\it A History of Chemical Theories and Laws}. 
  NY: John Wiley.  Reprint, NY: Arno Press, 1975. 

 \rr R. Mundell (1963). Capital Mobility and Stabilization Policy under
  Fixed and Flexible Exchange Rates. {\it Canadian Journal of Economic 
 and Political Science}, 29, 475-85. 

 \rr C.W.K.Mundle (1959). Probability and Statistical Inference by 
 G.Spencer-Brown (review). {\it Philosophy,} 34, 129, 150-154.

 \rr I.L. Muntean (2006). {\it Beyond Mechanics: 
  Principle of Least Action in Maupertuis and Euler.} 
  On line doc., University of California at San Diego.    

 \rr J.J.Murphy (1986). {\it Technical Analysis of the Future Markets: 
 A Comprehensive Guide to Trading Methods ans Applications.} 
 NY: New York Institute of Finance.

 \rr B.A.Murtagh (1981). {\it Advanced Linear Programming.} 
  NY: McGraw Hill. 

 \rr T.Mikosch(1998). {\it Elementary Stochastic Calculus with 
  Finance in View.} Singapore: World Scientific. 

 \rr L.Nachbin (1965). {\it The Haar Integral.}  
    Van Nostrand. 
 
 \rr R.Nagpal (2002). Self-assembling Global Shape using Concepts from 
 Origami. p. 219-231 in T.C.Hull (2002). Origami3 
 Proceedings of the 3rd International Meeting of
 Origami Mathematics, Science, and Education. 
 Natick Massachusetts: A.K.Peters Ltd.

 \rr J.Nash (1951). Non-Cooperative Games. 
 {\it The Annals of Mathematics}, 54,2, 286-295. 

 \rr L.K.Nash (1974). {\it Elements of Statistical Thermodynamics.} 
  NY: Dover. 

 \rr R.B.Nelsen (2006, 2nd ed.). {\it An Introduction to Copulas.}
    NY: Springer. 

 \rr E.Nelson (1987). {\it Radically Elementary Probability Theory.} 
    AM-117. Princeton University Press.  

 \rr  W.Nernst (1909). {\it Theoretische Chemie vom Standpunkte der 
 Avogadroschen Regel und der Thermodynamik.} Stuttgart: F.Enke.  

 \rr J.von Neumann (1928). Zur Theories der Gesellschaftsspiele. 
 {\it Mathematische Annalen}, 100, 295-320. English translation in
 R.D.Luce, A.W.Tucker eds. (1959). {\it Contributions to the Theory of 
 Games IV}. pp.13-42. Princeton University Press.

 \rr M.C.Newman, C.Strojan (1998). 
 {\it Risk Assessment: Logic and Measurement.} CRC. 

 \rr S.A.Newman, W.D.Comper (1990). Generic Physical Mechanisms of
  Morphogenesis and Pattern Formation. {\it Development,} 110, 1-18.

 \rr N.Y.Nikolaev, H.Iba (2001). Regularization Approach to Inductive 
  Genetic Programming. {\it IEEE Transactions on Evolutionary Computation}, 
  5, 4, 359-375. Recombinative Guidance. 

 \rr N.Y.Nikolaev, H.Iba (2003). Learning Polynomial Fedforward Neural 
  Networks by Genetic Programming and Backpropagation.  
  {\it IEEE Transactions on Neural Networks,} 14, 2, 337-350.  

 \rr N.Y.Nikolaev, H.Iba (2006). {\it Adaptive Learning of Polynomial 
 Networks.} Genetic and Evolutionary Computation. NY: Springer.  

 \rr W.Noeth (1995). {\it Handbook of Semiotics.} 
 Indiana University Press.  

 \rr  E.Noether (1918). Invariante Varlationsprobleme. 
 {\it Nachrichten der K\"{o}nighche Gesellschaft der Wissenschaften 
 zu G\"{o}ttingen.} 235--257. 
 Translated: {\it Transport Theory and Statistical Physics,} 
 1971, 1, 183--207.  

 \rr J.H.Noseworthy, G.C.Ebers, M.K. Vandervoort, R.E.Farquhar, 
    E.Yetisir, R.Roberts (1994). The impact of blinding on 
    the results of a randomized, placebo-controlled multiple 
    sclerosis clinical trial. {\it Neurology,} 44, 16-20. 

 \rr M.F.Ochs, R.S.Stoyanova, F.Arias-Mendoza, T.R.Brown (1999). 
    A New Methods for Spectral Decomposition Using  a Bilinear 
    Bayesian Approach. {\it J.of Magnetic Resonance }, 137, 161-176. 

 \rr G.M.Odel, G.Oster, P.Alberch, B.Burnside (1980). The Mechanical
  Basis of Morphogenesis. I - Epithelial Folding and Invagination. 
  {\it Dev. Biol.} 85, 446-462.

 \rr G.\"{O}kten (1999). {\it Contributions to the Theory of Monte Carlo 
 and Quasi monte Carlo Methods.} Ph.D. Thesis, 
 Clearmont University. Clearmont, CA: USA. 

 \rr K.Olitzky edt. (2000). {\it Shemonah Perakim: 
 A Treatise on the Soul by Moshe ben Maimon.} URJ Press.  

 \rr Y.S.Ong,  N.Krasnogor,  H.Ishibuchi (2007). 
 Special Issue on Memetic Algorithms. 
 {\it IEEE Transactions on Systems, Man, and Cybernetics,} 
 part B, 37, 1, 2-5.  

 \rr D.Ormoneit, V.Tresp (1995). Improved Gaussian Mixtures 
 Density Estimates Using Bayesian Penalty Terms and Network Averaging. 
 {\it Advances in Neural Information Processing Systems 8}, 
 542--548. MIT. 

 \rr J.Ortega y Gasset (1914). Ensayo de Estetica a Manera de 
 Prologo. in {\it El Pasajero} by J.Moreno Villa. Reprinted in 
 p.152-174 of J.Ortega y Gasset (2006). 
 {\it La Deshumanizacion del Arte.} 
 Madrid: Revista de Occidente en Alianza Editorial.

 \rr R.H.J.M. Otten, L.P.P.P. van Ginneken (1989).
 {\it The Annealing Algorithm}.  Boston: Kluwer.

 \rr C.C.Paige, M.A.Saunders (1977). Least Squares Estimation of 
   Discrete Linear Dynamic Systems using Orthogonal Transformations. 
   {\it Siam J. Numer. Anal.} 14,2, 180-193.  

 \rr A.Pais (1988). {\it Inward Bound: Of Matter and Forces in the 
    Physical World.} Oxford University Press. 

 \rr C.D.M.Paulino,  C.A.B.Pereira (1992). 
 Bayesian Analysis of Categorical Data Informatively Censored. 
 {\it Communications in Statistics - Theory and Methods}, 21, 2689-705.

 \rr C.D.M.Paulino, C.A.B.Pereira (1995). Bayesian Methods for 
 Categorical Data under Informative General Censoring. 
 {\it Biometrika}, 82,2, 439-446.

 \rr Y.Pawitan (2001). {\it In All Likelihood: Statistical Modelling 
 and Inference Using Likelihood.} Oxford University Press. 

 \rr J.Pearl (2000). 
  {\it Caysality: Models, Reasoning, and Inference.''}
  Cambridge University Press.

 \rr J.Pearl (2004). {\it Simpson's Paradox: An Anatomy.} 
    Rech.Rep. Cognitive Systems Lab., Computer Science Dept., 
    Univ.of California at Los Angeles.  

 \rr  C.S.Peirce (1880). A Boolean Algebra with One Constant. 
 In Hartshorne et al. (1992), 4, 12-20. 

 \rr C.S.Peirce (1883). {\it The John Hopkins Studies in Logic.} 
  Boston: Little, Brown and Co. 

 \rr C.S.Peirce, J.Jastrow (1885). On small Differences of Sensation.
  {\it Memoirs of the National Academy of Sciences}, 3 (1884), 75-83. 
  Also in  \rr C.S.Peirce, J.Jastrow (1885). On small Differences of Sensation.
  {\it Memoirs of the National Academy of Sciences}, 3 (1884), 75-83. 
  Also in Kloesel (1993), v.5 (1884-1886), p.122-135.  (1993), v.5 (1884-1886), p.122-135. 

 \rr  P.Penfield, R.Spence, S. Duinker (1970a). 
  A Generalized Form of Tellegens Theorem. 
  {\it IEEE Transactions on Circuit Theory}, CT-17, 3, 302-305

 \rr  P.Penfield, R.Spence, S. Duinker (1970b). 
  {\it Tellegen's Theorem and Electrical Networks.} 
  Cambrige, MA: MIT Press.  
  
 \rr C.A.B.Pereira, D.V.Lindley (1987). Examples Questioning the use 
 of Partial Likelihood. {\it The Statistician}, 36, 15--20. 

 \rr  C.A.B.Pereira, J.M.Stern (1999a).  
 A Dynamic Software Certification and Verification Procedure. 
 Proc. {\it ISAS-99, Int.Conf.on Systems Analysis and Synthesis,} 
 2, 426--435. 

 \rr  C.A.B.Pereira, J.M.Stern, (1999b). Evidence and Credibility:
 Full Bayesian Significance Test for Precise Hypotheses.
 {\it Entropy Journal}, 1, 69--80.

 \rr C.A.B.Pereira, J.M.Stern (2001a). Full Bayesian Significance
 Tests for Coefficients of Variation. In: George, E.I. (Editor). 
 Bayesian Methods with Applications to Statistics, 391-400. 
 Monographs of Official Statistics, EUROSTAT.    

 \rr C.A.B.Pereira,  J.M.Stern (2001b). Model Selection: Full
 Bayesian Approach. {\it Environmetrics} 12, (6), 559-568. 

 \rr C.A.B.Pereira, J.M.Stern (2005). 
 {\it Infer\^{e}ncia Indutiva com Dados Discretos: 
 Uma Vis\~{a}o Genuinamente Bayesiana.} 
 COMCA-2005. Chile: Universidad de Antofagasta. 

 \rr C.A.B.Pereira, J.M.Stern (2008). 
 Special Characterizations of Standard Discrete Models.        
 {\it REVSTAT Statistical Journal}, 6, 3, 199-230.  

 \rr C.A.B.Pereira, S.Wechsler (1993). On the Concept 
 of $p$-value. {\it Brazilian Journal of Probability and 
 Statistics}, 7, 159--177.  

 \rr C.A.B.Pereira, M.A.G.Viana (1982). 
 {\it Elementos de Infer\^{e}ncia Bayesiana.} 5o Sinape, S\~{a}o Paulo. 

 \rr C.A.B.Pereira, S.Wechsler, J.M.Stern (2008). 
 Can a Significance Test be Genuinely Bayesian? 
 {\it Bayesian Analysis,} 3, 1, 79-100. 

 \rr P.Perny, A.Tsoukias (1998).
 {\it On the Continuous Extension of a Four Valued Logic for
 Preference Modelling.} IPMU-98, 302--309.
 7th Conf. on Information Processing and Management of Uncertainty 
 in Knowledge Based Systems. Paris, France. 
 
 \rr J.Perrin (1903). {\it Trait\'{e} de Chimie Physique}. 
    Paris: Gauthier-Villars. 

 \rr J.Perrin (1906). La discontinuit\'{e} de la Mati\`{e}re. 
  {\it Revue de Mois,} 1, 323-343.  
 
 \rr J. B. Perrin (1909). Mouvement Brownien et R\'{e}alit\'{e} 
 Mol\'{e}culaire. {\it Annales de Chimie et de Physiqe}, VIII 18, 5-114.  
 also in p.171-239 of Perrin (1950). 
 Translation: Brownian Movement and Molecular Reality, 
 London: Taylor and Francis. 

 \rr J.B.Perrin (1913). {\it Les Atomes}. Paris: Alcan. 
 Translation: {\it Atoms.} NY: Van Nostrand. 

 \rr J.Perrin (1950). {\it Oeuvres Scientifiques}. Paris: CNRS. 

 \rr L.Peusner (1986). {\it Studies in Network Thermo-Dynamics.} 
  Amsterdam: Elsevier. 

 \rr D.Pfeffermann, A.M.Krieger, Y.Rinott (1998). 
 Parametric Distributions of Complex Survey 
 Data under Informative Probability Sampling. 
 {\it Statistica Sinica} 8, 1087-1114

 \rr D.Pfeffermann, M.Sverchkov (2003). 
  Fitting Generalized Linear Models Under informative Sampling. 
  In C.Skinner, R.Chambers (2003), 175-195.

 \rr G.C.Pflug (1996). {\it Optimization of Stochastic Models: The
 Interface Between Simulation and Optimization}. Boston: Kluwer.  

 \rr L.Phlips (1995). 
 {\it Competition Policy: A Game-Theoretic Perspective.}
 Cambridge University Press.

 \rr J.Piaget (1975). {\it L'\'{e}quilibration des Structures Cognitives: 
    Probl\`{e}me Central du D\'{e}veloppement.} Paris: PUF.

 \rr J.Piaget (1985). {\it Equilibration of Cognitive Structures: 
 The Central Problem of Intellectual Development.} 
 Univ.of Chicago.  

 \rr J.Piaget, B.Inhelder (1951). {\it The Origin of the Idea of Chance 
  in Children.} Translated by L.Leake,  E.Burrell, H.D.Fishbein (1975),  
  New York: Norton. 

 \rr S.D.Pietra, V.Pietra, J.Lafferty (2001). 
 {\it Duality and Auxiliary Functions for Bregman Distances.}  
 Tec.Rep. CMU-CS-01-109R, Carnegie Mellon. 

 \rr  S.Pihlstrom, C.N.El-Hani (2002). 
    Emergence Theories and Pragmatic Realism.  
    {\it  Essays in Philosophy.} 
    Arcata, CA, USA: Humboldt State University. \\   
    \verb#www.humboldt.edu/~essays/pihlstrom.html#  

 \rr S.Pissanetzky (1984). {\it Sparse Matrix Technology.} 
  NY: Academic Press. 

 \rr M.Planck (1915). Das Prinzip der kleinsten Wirkung. 
 {\it Kultur der Gegenwart}. Also in p.25-41 of Planck (1944). 

 \rr M.Planck (1944) Wege zur physikalischen Erkenntnis. 
 Reden und Vortr\"{a}ge. Leipzig: S.Hirzel.  

 \rr M.Planck (1937). Religion and Natural Science. 
 Also in Planck (1950).  

 \rr M.Planck (1950). Scientific Autobiography and other Papers. 
  London: Williams and Norgate. 

 \rr R.J.Plemmons, R.E.White (1990).  Substructuring Methods for Computing 
 the Nullspace of Equilibrium Matrices. 
 {\it SIAM Journal on Matrix Analysis and Applications}, 11, 1-22.


 \rr  K.R.Popper (1959). 
 {\it The Logic of Scientific Discovery.}  NY: Routledge. 

 \rr  K.R.Popper (1963). 
 {\it Conjectures and Refutations: The Growth of Scientific Knowledge.}   
 NY: Routledge. 

 \rr I.Prigoine (1961). {\it Introduction to the Thermodynamics of 
  Irreversible Processes}, 2nd ed. NY: Interscience. 

 \rr H.Pulte (1989). 
  Das Prinzip der kleinsten Wirkung und die Kraftkonzeptionen der 
  rationalen Mechanik: Eine Untersuchung zur Grundlegungsproblemematik 
  bei Leonhard Euler, Pierre Louis Moreau de Maupertuis und Joseph Louis 
  Lagrage. {\it Studia Leibnitiana}, sonderheft 19.   

 \rr J.R.Quinlan (1986). Induction of Decision Trees.  
  {\it Machine Learning} 1, 221-234.  

 \rr N.L.Rabinovitch (1973). {\it Probability and Statistical Inference 
  in Ancient and Medieval Jewish Literature.}   
  University of Toronto Press. 

 \rr H.Rackham (1926). {\it Aristotle, Nicomachean Ethics.} 
 Harvard University Press. 

 \rr S.Rahman, J.Symons, D.M.Gabbay J.P. van Bendegem, eds. (2004).  
  {\it Logic, Epistemology, and the Unity of Science}. NY: Springer. 

 \rr V.S.Ramachandran (2007). The Neurology of Self-Awareness. 
 The Edge 10-th Anniversary Essay. 

 \rr  W.Rasch (1998). Luhmann's Widerlegung des Idealismus: 
 Constructivism as a two-front war. 
 {\it Soziale Systeme,} 4, 151--161. 

 \rr W.Rasch (2000) Niklas Luhmanns Modernity. Paradoxes of
 Differentiation. Stanford Univ.Press. 
 Specially chapter 3 and 4, also published as:  
 W.Rasch (1998). Luhmanns Widerlegung des Idealismus:
 Constructivism as a  Two-Front War. 
 {\it Soziale Systeme,} 4, 151-159; and     
 W.Rasch (1994). In Search of Lyotard Archipelago, or: 
 How to Live a Paradox and Learn to Like It. 
 {\it New German Critique,} 61, 55-75.   

 \rr A.Recski (1989). {\it Matroid Theory and its Applications 
   in Electrical Network Theory and in Statics.} 
   Budapest: Akad\'{e}miai Kiad\'{o}.  

 \rr C.R.Reeves (1993). 
 {\it Modern Heuristics for Combinatorial Problems.} 
 Blackwell Scientific.  

 \rr C.R.Reeves, J.E.Rowe (2002). {\it Genetic Algorithms - 
 Principles and Perspectives: A Guide to GA Theory.} 
 Berlin: Springer. 

 \rr F.Reif (1965). {\it Statistical Physics.}  NY: McGraw-Hill. 

 \rr R.Reintjes, A.de Boer, W.van Pelt, J.M.de Groot (2000).  
  Simpson's Paradox: An Example from Hospital Epidemiology. 
  {\it Epidemiology}, 11, 1, 81-83. 

 \rr A.Renyi (1970). {\it Probability Theory.} 
  Amsterdam: North-Holland. 

 \rr A.Renyi (1961). On Measures of Entropy and Information. 
 {\it Proc. 4-th Berkeley Symp. on Math Sats. and Prob.} V-I, 547-561.  

 \rr  H.L.Resnikoff, R.O.Wells (2002).  {\it Wavelet Analysis: 
  The Scalable Structure of Information.} Springer Verlag.

 \rr P.Ressel (1985). DeFinetti Type Theorems: Analytical approach. 
 {\it Annals Probability,} 13, 898--900. 
  
 \rr P.Ressel (1987). 
 A Very General De Finetti Type Theorem. In: Viertl (1987). 

 \rr P.Ressel (1988). 
 Integral Representations for Distributions of Symmetric Processes. 
 {\it Probability Theory and Related Fields,} 79, 451--467. 

 \rr C.Reynolds (1987). Flocks Herds and Schools: A Distributed 
  Behavioral Model. {\it Computer Graphics,} 21, 25-34. 
  Updated version at  \\  
  \verb#www.red3d.com/cwr/boids/#    

 \rr R.J.Richards (1989). {\it Darwin and the Emergence of Evolutionary 
 Theories of Mind and Behavior.} University Of Chicago Press.  

 \rr C.J.van Rijsbergen (2004). {\it The Geometry of Information 
    Retrieval.} Cambridge University Press. 

 \rr B.D.Ripley (1987). {\it Stochastic Simulation.} NY: Wiley. 

 \rr B.D.Ripley (1996). {\it Pattern Recognition and Neural Networks.} 
 Cambridge University Press. 

 \rr J.Rissanen (1978). Modeling by Shortest Data Description. 
 {\it Automatica,} 14, 465--471. 

 \rr J.Rissanen (1989). {\it Stochastic Complexity in Statistical 
  Inquiry.} NY: World Scientific. 

 \rr G.Rizzolatti, M.A.Arbib (1998). Language within our grasp. 
    {\it TINS,} 21, 5, 188-194. 

 \rr G.Rizzolatti, C.Sinigalia (2006). {\it Mirrors in the Brain. 
  How our Minds Share Actions and Emotions.} Oxford University Press. 

 \rr A.M.Robert (2003). {\it Nonstandard Analysis.} Mineola: Dover. 

 \rr C.P.Robert (1996). Mixture of Distributions: Inference and 
 Estimation. in Gilks et al. (1996). 

 \rr R.Robertson (2001). One, Two, Three, Continuity. C.S.Peirce 
    and the Nature of the Continuum. {\it Cybernetics \& Human 
    Knowing,} 8, 7-24. 


 \rr V.Ronchi  (1970). {\it Nature of Light: An Historical Survey.} 
  Harvard Univ. Press.  

 \rr  H.Rouanet, J.M.Bernard, M.C.Bert, B.Lecoutre, M.P.Lecoutre,
 B.Le Roux (1998). {\it New Ways in Statistical Methodology. From
 Significance Tests to Bayesian Inference.} Berne: Peter Lang. 

 \rr D.J.Rose (1972) {\it Sparse Matrices and Their Applications.} 
    NY: Springer. 

 \rr D.J.Rose, R.A.Willoughby (1972). {\it Sparse Matrices.} 
    NY: Plenum Press. 

 \rr L.Rosenfeld (2005). {\it Classical Statistical Mechanics.} 
  S\~{a}o Paulo: CBPF - Livraria da F\'{\i}sica. 

 \rr J.Ross, S.R.Berry (2008). {\it Thermodynamics and Fluctuations 
  far from Equilibrium}. NY: Springer.  

 \rr R.Royall (1997). {\it Statistical Evidence:
 A Likelihood Paradigm}. London: Chapman \& Hall.

 \rr E.Rubin (1915). {\it Visuell wahrgenommene Figuren}. 
    Copenhagen: Cyldenalske Boghandel. 

 \rr D.B.Rubin (1978). Bayesian Inference for Causal Effects: 
  The Role of Randomization. {\it The Annals of Statistics,} 
  6, 34-58. 

 \rr D.Rubin, D.Thayer (1982). EM Algorithm for ML Factor Analysis. 
  {\it Psychometrika,} 47, 1, 69-76.   

 \rr H.Rubin (1987). A Weak System of Axioms for ``Rational'' 
 Behaviour and the Non-Separability of Utility from Prior. 
 {\it Statistics and Decisions}, 5, 47--58. 

 \rr R.Y.Rubinstein, D.P.Kroese (2004). {\it The Cross-Entropy Method: 
  A Unified Approach to Combinatorial Optimization, Monte-Carlo
  Simulation and Machine Learning.} NY: Springer. 

 \rr C.Ruhla (1992). {\it The Physics of Chance: From Blaise Pascal to 
    Niels Bohr.} Oxford University Press. 

 \rr  B.Russell (1894). Cleopatra or Maggie Tulliver? 
 Lecture at the Cambridge Conversazione Society. 
 Reprinted as Ch.8, p.57-67, in C.R.Pigden, ed. (1999). 
 {\it Russell on Ethics.} London: Routledge.  

 \rr S.Russel (1988). Machine Learning: The EM Algorithm. 
 Unpublished note. 

 \rr S.Russell (1998). {\it The EM Algorithm.} 
 On line doc, Univ. of California at Berkeley.   

 \rr Ruta v. Breckenridge-Remy Co., USA, 1982.

 \rr A.I.Sabra (1981). {\it Theories of Light: From Descartes to Newton.} 
  Cambridge University Press. 

 \rr R.K.Sacks, H.Wu (1977). 
  {\it Genearl Relativity for Mathematicians.} NY: Springer. 

 \rr L.Sadun (2001). 
  {\it Applied Linear Algebra: The Decoupling Principle.} 
  NY: Prentice Hall. 

 \rr Sakamoto,Y. Ishiguro,M. Kitagawa,G. (1986). {\it Akaike Information 
 Criterion Statistics.} Dordrecht: Reidel - Kluwer. 

 \rr V.H.S.Salinas-Torres, C.A.B.Pereira, R.C.Tiwari (1997). 
 Convergence of Dirichlet Measures Arising in Context of Bayesian Analysis 
 of Competing Risks Models. {\it J. Multivariate Analysis}, 62,1, 24-35.

 \rr V.H.S.Salinas-Torres, C.A.B.Pereira, R.C.Tiwari (2002). Bayesian
 Nonparametric Estimation in a Series System or a Competing-Risks Model.
 {\it J.of Nonparametric Statistics}, 14,4, 449-58.

 \rr M.Saltzman (2004). {\it Tissue Engineering.} Oxford University Press.   

 \rr A. Sangiovanni-Vincentelli, L.O. Chua (1977).
 An Efficient Heuristic Cluster Algorithm for Tearing Large-Scale Networks. 
 {\it IEEE Transactions on Circuits and Systems}, 24, 709-717.

 \rr L.Santaella (2002). {\it Semi\'{o}tica Aplicada.} 
    S\~{a}o Paulo: Thomson Learning. 

 \rr L.A.Santal\'{o} (1973). {\it Vectores y Tensores.} 
  Buenos Aires: Eudeba. 

 \rr G.de Santillana (1955). {\it The Crime of Galileo}. 
  University of Chicago Press.  

 \rr L.A.Santalo (1976). {\it Integral Geometry and Geometric 
 Probability.} 
 London: Addison-Wesley.  

 \rr  J.Sapp, F.Carrapio, M.Zolotonosov (2002). Symbiogenesis: The
  Hidden Face of Constantin Merezhkowsky. {\it History and Philosophy 
  of the Life Sciences,} 24, 3-4, 413-440. 

 \rr S.Sarkar (1988). Natural Selection, Hypercycles and the Origin 
 of Life.   Proceedings of the Biennial Meeting of the Philosophy of 
 Science Association, Vol.1, 197-206. 
 The University of Chicago Press. 

 \rr  L.J.Savage (1954): {\it The Foundations of Statistics.}  
 Reprint 1972. NY: Dover.

 \rr L.J.Savage (1981). {\it The writings of Leonard Jimmie Savage: 
 A memorial selection.} Institute of Mathematical Statistics.

 \rr D.Schacter (2001). {\it Forgotten Ideas, Neglected Pioneers: 
  Richard Semon and the Story of Memory.} Philadelphia: Psychology Press. 

 \rr D.L.Schacter, J.E.Eich, E.Tulving, (1978). Richard Semon's 
  Theory of Memory. {\it Journal of Verbal Learning and Verbal Behavior,}  
  17, 721-743.

 \rr J.D.Schaffer (1987). Some Effects of Selection Procedures on 
 Hyperplane Sampling by Genetic Algorithms. 
 p. 89-103 in L.Davis (1987). 

 \rr J.M.Schervich (1995). {\it Theory of Statistics.} Berlin, Springer. 

 \rr  M.Schlick (1920). Naturphilosophische Betrachtungen \"{u}ber 
 das Kausalprintzip. {\it Die Naturw issenschaften}, 8, 461-474. 
 Translated as, Philosophical Reflections on the Causal Principle, 
 ch.12, p.295-321, v.1 in M.Schlick (1979).    

 \rr M.Schlick (1979). Philosophical Papers. Dordrecht: Reidel. 

 \rr H.Scholl (1998). 
 Shannon optimal priors on independent identically distributed
 statistical experiments converge weakly to Jeffreys' prior. 
 {\it Test}, 7,1, 75-94.  

 \rr J.W.Schooler (2002). Re-Representing Consciousness: 
 Dissociations between Experience and Metaconsciousness. 
 {\it Trends in Cognitive Sciences,} 6, 8, 339-344. 

 \rr A.Schopenhauer (1818, 1966). 
  {\it The World as Will and Representation.} NY: Dover. 

 \rr E.Schr\"{o}dinger (1926). 
  Quantisierung als Eigenwertproblem.  
  (Quantisation as an Eigenvalue Problem). 
  {\it Annalen der Physic}, 489, 79-79. 
  {\it Physical Review}, 28, 1049-1049. 

 \rr E.Schr\"{o}dinger (1945). {\it What Is Life?} 
  Cambridge University Press.  
 
 \rr G.Schwarz (1978). Estimating the Dimension of a Model. 
 {\it Ann. Stat.}, 6, 461-464. 

 \rr C.Scott (1958). G.Spencer-Brown and Probability: A Critique. 
 {\it J.Soc. Psychical Research,} 39, 217-234. 

 \rr C. Sechen, K. Lee (1987).  An Improved Simulated Annealing Algorithm 
 for Row-Based Placement. {\it Proc. IEEE International Conference on 
 Computer-Aided Design}, 478-481.

 \rr  L.Segal (2001). {\it The Dream of Reality. 
 Heintz von Foerster's Constructivism.}  NY: Springer. 
 
 \rr R.W.Semon (1904). {\it Die Mneme}. Leipzig: W. Engelmann.
  Translated (1921), {\it The Mneme}. London: Allen and Unwin.

 \rr R.W.Semon (1909). {\it Die Mnemischen Empfindungen}. 
  Leipzig: Leipzig: W.Engelmann. 
  Translated (1923), {\it Mnemic psychology.} London: Allen and Unwin.

 \rr S.K.Sen, T.Samanta, A.Reese (2006). Quasi Versus Pseudo Random 
 Generators: Discrepancy, Complexity and Integration-Error Based 
 Comparisson.  {\it Int.J. of Innovative Computing, Information and 
 Control,} 2, 3, 621-651. 

 \rr S.Senn (1994). Fisher's game with the devil. 
 {\it Statistics in Medicine,} 13, 3, 217-230.  

 \rr G.Shafer (1982),  Lindley's Paradox. 
 {\it J. American Statistical Assoc.}, 77, 325--51. 

 \rr  G.Shafer, V.Vovk (2001). {\it Probability and Finance, 
  It's Only a Game!} NY: Wiley. 

 \rr B.V.Shah, R.J.Buehler, O.Kempthorne (1964). 
 Some Algorithms for Minimizing a Function of Several Variables. 
 {\it J. Soc. Indust. Appl. Math.} 12, 74--92.  

 \rr J.Shedler, D.Westen (2004). Dimensions of Personality Pathology: 
 An Alternative to the Five-Factor Model. 
 {\it American Journal of Psychiatry}, 161, 1743-1754. 

 \rr J.Shedler, D.Westen (2005). Reply to T.A.Widiger, T.J.Trull.      
 A Simplistic Understanding of the Five-Factor Model
 {\it American J.of Psychiatry}, 162,8, 1550-1551.  

 \rr H.M.Sheffer (1913). A Set of Five Independent Postulates for 
  Boolean Algebras, with Application to Logical Constants. 
  {\it Trans. Amer. Math. Soc.}, 14, 481-488.  

 \rr Y.Shi (2001). {\it Swarm Intelligence.} Morgan Kaufmann. 

 \rr R.Shibata (1981). An Optimal Selection of Regression Variables. 
 {\it Biometrika,} 68, 45--54.

 \rr G.Shwartz (1978). Estimating the Dimension of a Model. 
 {\it Annals of Statistics,} 6, 461--464. 

 \rr  B.Simon (1996). {\it Representations of Finite and Compact Groups.}  
  AMS Graduate Studies in Mathematics, v.10. 

 \rr H.A.Simon (1996). {\it The Sciences of the Artificial}. 
  MIT Press. 

 \rr E.H.Simpson (1951). The Interpretation of Interaction in
  Contingency Tables. {\it Journal of the Royal Statistical Society}, 
  Ser.B, 13, 238-241. 

 \rr S.Singh, M.K.Singh (2007). 
 `Impossible Trinity' is all about Problems of Choice:  
 Of the Three Options of a Fixed Exchange Rate, Free Capital Movement,
 and an Independent Monetary Policy: One can choose only two at a time. 
 LiveMint.com, The Wall Street Journal. 
 Posted: Mon, Nov 5 2007. 12:30 AM IST \\ 
 \verb#www.livemint.com2007/11/05003003/Ask-Mint--8216Impossible-t.html#  

 \rr J.Skilling (1988). The Axioms of MaximumEntropy. 
  Maximum-Entropy and BayesianMethods in Science and Engineering, 
  G. J. Erickson and C. R. Smith (eds.) Dordrecht: Kluwer.  

 \rr J.E.Smith (2007). Coevolving Memetic Algorithms: 
  A Review and Progress Report. 
  {\it IEEE Transactions on Systems Man and Cybernetics,} 
  part B, 37, 1, 6-17.

 \rr P.J.Smith, E.Gunel (1984). Practical Bayesian Approaches to the 
 Analysis of 2x2 Contingency Table with Incompletely Categorized Data. 
 {\it Communication of Statistics - Theory and Methods}, 13, 1941-63.

 \rr C.Skinner, R.Chambers (2003). 
 {\it Analysis of survey Data}, 
 New York: Wiley, 175-195.

 \rr S.G.Soal, F.J.Stratton, R.H.Thouless (1953). Statistical Significance 
  in Psychical Research. {\it Nature,} 172, 594. 

 \rr G.M.Souza, A.G.Manzatto (2000). {\it Hierarchquia Auto-Organizada 
  em Sistemas Biol\'{o}gicos.} p.153-173 in D'Otaviano and Gonzales (2000).  

 \rr J.C.Spall (2003). {\it Introduction to Stochastic Search and 
 Optimization.} Hoboken: Wiley.

 \rr G.Spencer-Brown (1953a). Statistical Significance in Psychical 
 Research. {\it Nature,} 172, 154-156. 
 
 \rr Spencer-Brown (1953b). Answer to Soal et al. (1953). 
  {\it Nature,} 172, 594-595.

 p.594-595)

 \rr G.Spencer-Brown (1957). {\it Probability and Scientific Inference.} 
  London: Longmans Green.  

 \rr G.Spencer-Brown (1969). {\it Laws of Form.} Allen and Unwin.

 \rr M.D.Springer (1979). 
 {\it The Algebra of Random Variables.} NY: Wiley. 

 \rr F.Steier, edt. (1991) {\it Research and Reflexivity.} 
    SAGE Publications. 

 \rr M.Stephens (1997). {\it Bayesian Methods for Mixtures of 
 Normal Distributions.} 
 Oxford University.   

 \rr J.Stenmark, C.S.P.Wu (2004). Simpsons Paradox, Confounding Variables 
  and Insurance Ratemaking. 

 \rr   C.Stern (1959). Variation and Hereditary Transmission.  
 {\it Proceedings of the American Philosophical Society}, 103, 2, 183-189. 

 \rr J.M.Stern  (1992). Simulated Annealing with a Temperature 
 Dependent Penalty Function. {\it ORSA Journal on Computing,} 
 4, 311-319. 

 \rr J.M.Stern (1994). {\it Esparsidade, Estrutura, Estabilidade 
    e Escalonamento em \'{A}lgebra Linear Computacional.}  
    Recife: UFPE, IX Escola de Computa\c{c}\~{a}o. 
 
 \rr J.M.Stern (2001) The Full Bayesian Significant Test for the 
  Covariance Structure Problem. 
 Proc. {\it ISAS-01, Int.Conf.on Systems Analysis and Synthesis,} 
 7, 60-65. 

 \rr J.M.Stern (2003a). 
  Significance Tests, Belief Calculi, and Burden of Proof in 
  Legal and Scientific Discourse. Laptec-2003, 
  {\it Frontiers in Artificial Intelligence and its Applications,}    
  101, 139--147.     

 \rr J.M.Stern (2004a). Paraconsistent Sensitivity Analysis for 
 Bayesian Significance  Tests. SBIA'04,   
 {\it Lecture Notes Artificial Intelligence,} 
 3171, 134--143. 

 \rr J.M.Stern (2004b). Uninformative Reference Sensitivity in 
 Possibilistic Sharp Hypotheses Tests. MaxEnt 2004, 
 {\it American Institute of Physics Proceedings}, 735, 581--588.

 \rr J.M.Stern (2006a). 
 Decoupling, Sparsity, Randomization, and Objective Bayesian Inference. 
 Tech.Rep. MAC-IME-USP-2006-07.  

 \rr J.M.Stern (2006b). Language, Metaphor and Metaphysics: 
  The Subjective  Side of Science. Tech.Rep. MAC-IME-USP-2006-09.  

 \rr J.M.Stern (2007a).   
  Cognitive Constructivism, Eigen-Solutions, and Sharp
  Statistical Hypotheses. {\it Cybernetics and Human Knowing}, 
  14, 1, 9-36.  
  Early version in 
  Proceedings of FIS-2005, 61, 1--23. Basel: MDPI. 

 \rr J.M.Stern (2007b).  Language and the Self-Reference Paradox. 
  {\it Cybernetics and Human Knowing}, 
  14, 4, 71-92. 

 \rr J.M.Stern (2007c). Complex Structures, Modularity and Stochastic 
  Evolution. Tech.Rep. IME-USP MAP-0701.  

 \rr J.M.Stern (2008a).  Decoupling, Sparsity, Randomization, and 
  Objective Bayesian Inference. 
  {\it Cybernetics and Human Knowing}, 
  15, 2, 49-68. 

 \rr J.M.Stern (2008b). {\it Cognitive Constructivism and the Epistemic 
  Significance of Sharp Statistical Hypotheses.} Tutorial book for 
  MaxEnt 2008, The 28th International Workshop on Bayesian Inference 
  and Maximum Entropy Methods in Science and Engineering. 
  July 6-11 of 2008, Borac\'{e}ia, S\~{a}o Paulo, Brazil.

 \rr J.M.Stern (2011). Spencer-Brown vs. Probability and Statistics: 
  Entropys Testimony on Subjective and Objective Randomness. 
  {\it Information}, 2, 2, 277-301. 

 \rr J.M.Stern (2011). Constructive Verification, Empirical Induction, 
  and Falibilist Deduction: A Threefold Contrast. 
  {\it Information}, 2, 635-650 

 \rr J.M.Stern (2011). Symmetry, Invariance and Ontology in Physics 
  and Statistics. {\it Symmetry}, 3, 3, 611-635. 

 \rr J.M.Stern, C.Dunder, M.S.Laureto, F.Nakano, C.A.B.Pereira,  
  C.O.Ribeiro (2006). {\it Otimiza\c{c}\~ao e Processos Estoc\'{a}sticos 
  Aplicados \`{a} Economia e Finan\c{c}as.} S\~{a}o Paulo: IME-USP.  

 \rr J.M.Stern, M.S.Lauretto, A.Polpo, M.A.Diniz (2012). 
  EBEB 2012 - XI Brazilian Meeting on Bayesian Statistics. 
  AIP Conference Proceedings v.1490. 
  Melville, NY: American Institute of Physics.  

 \rr J.M.Stern, C.O.Ribeiro, M.S.Lauretto, F.Nakano (1998). REAL:
 Real Attribute Learning Algorithm. {\it Proc. ISAS/SCI-98} 2, 315--321.

 \rr J.M.Stern, S.A.Vavasis (1994). 
  Active Set Algorithms for Problems in Block Angular Form. 
  {\it Computational and Applied Mathemathics}, 12, 3, 199-226. 

 \rr J.M.Stern, S.A.Vavasis (1993).  
  Nested Dissection for Sparse Nullspace Bases. 
  {\it SIAM Journal on Matrix Analysis and Applications,} 14, 3, 766-775. 

 \rr  J.M.Stern, S.Zacks (2002). Testing  Independence 
 of Poisson Variates under the Holgate Bivariate Distribution. 
 The Power of a New Evidence Test. {\it Statistical and
 Probability Letters}, 60, 313--320. 

 \rr J.M.Stern, S.Zacks (2003). 
  {\it Sequential Estimation of Ratios, with Applications to 
  Bayesian Analysis.} Tech. Rep. RT-MAC-2003-10.  

 \rr R.B.Stern (2007). {\it An\'{a}lise da Responsabilidade Civil do
  Estado com base nos Princ\'{\i}pios da Igualdade e da Legalidade.} 
  Graduation Thesis. 
  Faculdade de Direito da Pontif\'{\i}cia Universidade Cat\'{o}lica
  de S\~{a}o Paulo.  

 \rr R.B.Stern, C.A.B.Pereira (2008). 
 A Possible Foundation for Blackwell's Equivalence.  
 {\it AIP Conference Proceedings,} v. 1073, 90-95. 

 \rr S.M.Stigler (1978). Mathematical Statistics in the Early States. 
 {\it The Annals of Statistics,} 6, 239-265. 

 \rr S.M.Stigler (1986). {\it The History of Statistics: The Measurement
    of Uncertainty before 1900}.  Harvard Univ.Press. 

 \rr M.St\"{o}ltzner (2003). The Principle of Least Action as the
 Logical Empiricist's Shibboleth. 
 {\it Studies in History and Philosophy of Modern Physics}, 34, 285-318.   

 \rr R.D.Stuart (1966). {\it An Introduction to Fourier Analysis}. 
 London: Methuen.  

 \rr M.N.S.Swamy, K.Thulasiraman (1981). {\it Graphs, Networks and 
  Algorithms}. NY: Wiley. 

 \rr L Szilard (1929). \"{U}ber die Entropieverminderung in einem
  Thermodynamischen System bei Eingriffen Intelligenter Wesen.  
  {\it Zeitschrift f\"{u}r Physik}, 53, 840.

 \rr Taenzer, Ganti, and Podar (1989). Object-Oriented Software Reuse:
  The Yoyo Problem. {\it Journal of Object-Oriented Programming,} 
  2, 3, 30-35. 

 \rr H.Takayasu (1992). {\it Fractals in Physical Science.} 
 NY: Wiley. 

 \rr L.Tarasov (1988). {\it The World is Built on Probability.} 
 Moscow: MIR. 

 \rr L.Tarasov (1986). {\it This Amazingly Symmetrical World.} 
 Moscow: MIR. 

 \rr M.Teboulle (1992). Entropic Proximal Mappings with Applications 
 to Nonlinear Programming. {\it Mathematics of operations Research,}  
 17, 670-690. 

 \rr L.C.Thomas (1986). {\it Games, Theory and Applications.} 
 Chichester, England: Ellis Horwood.   

 \rr C.J.Thompson (1972). {\it Mathematical Statistical Mechanics.} 
  Princeton University Press. 
 
 \rr G.L.Tian, K.W.Ng, Z.Geng (2003). Bayesian Computation for
 Contingency Tables with Incomplete Cells-Counts. 
 {\it Statistica Sinica}, 13, 189-206.

 \rr W.Tobin (1993). Toothed Wheels and Rotating Mirrors. 
 {\it Vistas in Astronomy}, 36, 253-294. 

 \rr S.Tomonaga (1962). {\it Quantum Mechanics.  
  V.1 Old Quantum Theory; V.2, New quantum theory.} 
  North Holland and Interscience Publishers.

 \rr C.A. Tovey (1988). {\it Simulated Simulated Annealing}.
 In Johnson (1988).

 \rr C.G.Tribble (2008). Industry-Sponsored Negative Trials and the
 Potential Pitfalls of Post Hoc Analysis. {\it Arch Surg,} 143, 933-934.  

 \rr M.Tribus (1961). {\it Thermostatics and Thermodynamics: 
  An Introduction to Energy, Information and States of Matter, 
  with Engineering Applications}. Princeton: van Nostrand. 

 \rr M.Tribus, E.C.McIrvine (1971). Energy and Information.  
 {\it Scientific American}, 224, 178-184.  

 \rr P.K.Trivedi, D.M.Zimmer (2005). {\it Copula Modeling: 
 An Introduction for Practitioners.} Boston: NOW.

 \rr C.Tsallis (2001). Nonextensive Statistical Mechanics and 
 Termodynamics: Historical Background and Present Status. 
 p. 3-98 in Abe and Okamoto (2001) 

 \rr S.M.Ulam (1943). What is a Measure? 
  {\it The American Mathematical Monthly.} 50, 10, 597-602. 

 \rr S.Unger, F.Wysotzki (1981). 
  {\it Lernfachige Klassifizirungssysteme.} Berlin: Akademie Verlag. 

 \rr J.Uffink (1995). The Constraint Rule of the Maximum Entropy 
 Principle. {\it Studies in the History and Philosophy of Modern 
 Physics,} 26B, 223-261. 

 \rr J.Uffink (1996). Can the Maximum Entropy principle be Explained as 
 a Consistency Requirement?. {\it Studies in the History and Philosophy of 
 Modern Physics,} 27, 47-79.  

 \rr J.Utts (1991). Replication and Meta-Analysis in Parapsychology. 
 {\it Statistical Science,} 6, 4, 363-403, with comments by 
 M.J.Bayarri, J.Berger, R.Dawson, P.Diaconis, J.B.Greenhouse, 
 R.Hayman. R.L.Morris and F.Mosteller.   

 \rr I.V\'{a}g\'{o} (1985). {\it Graph Theory: Applications to the 
  Calculation of Electrical Networks.} Amsterdam: Elsevier. 
 
 \rr V.N.Vapnik (1995). {\it The Nature of Statistical Learning Theory.} 
 NY: Springer.   

 \rr V.N.Vapnik (1998). {\it Statistical Learning Theory: Inference 
 for Small Samples.} NY: Wiley.  

 \rr F.Varela (1978). {\it Principles of Biological Autonomy.} 
  North-Holland.

 \rr A.M.Vasilyev (1980). {\it An Introduction to Statistical Physics.} 
  Moscow: MIR. 

 \rr M.Vega Rod\'{\i}guez, (1998). La Actividad Metaf\'{o}rica: 
  Entre la Raz\'{o}n Calculante y la Raz\'{o}n Intuitiva. 
  {\it Es\'{e}culo, Revista de estudios literarios}. 
  Madrid: Universidad Complutense.   

 \rr E.S.Ventsel (1980). {\it Elements of Game Theory.} Moscow: MIR.  

 \rr M.Viana (2003). {\it Symmetry Studies, An Introduction.} 
 Rio de Janeiro: IMPA.  

 \rr B.Vidakovic (1999). {\it Statistical Modeling by Wavelets.} 
 Wiley-Interscience.  

 \rr F.S.Vieira, C.N.El-Hani (2009). Emergence and Downward Determination 
 in the Natural Sciences. {\it Cybernetics and Human Knowing,} 15, 101-134. 

 \rr R.Viertl (1987). 
 {\it Probability and Bayesian Statistics.} NY: Plenum. 

 \rr M.Vidyasagar (1997). {\it A Theory of Learning and Generalization.} 
 Springer, London. 

 \rr H.M.Voigt, H.Muehlenbein, H.P.Schwefel (1989). 
 {\it Evolution and Optimization.} 
 Berlin: Akademie Verlag. 
 
 \rr  H.A.van der Vorst, P.van Dooren, eds. (1990). 
    {\it Parallel Algorithms for Numerical Linear Algebra.} 
    Amsterdam: North-Holland.   

 \rr Hugo de Vries (1889). {\it Intracellular Pangenesis Including 
 a paper on Fertilization and Hybridization}.   
 Translated by C.S.Gager (1910). Chicago: The Open Court Publishing Co.   

 \rr H.de Vries (1900). 
 Sur la loi de disjonction des hybrides. 
 {\it Comptes Rendus de l'Academie des Sciences}, 130, 845-847. 
 Translated as  Concerning the law of segregation of hybrids. 
 {\it Genetics}, (1950), 35, 30-32. 

 \rr S.Walker(1986). A Bayesian Maximum Posteriori Algorithm for
 Categorical Data under Informative General Censoring. 
 {\it The Statistician}, 45, 293-8.

 \rr C.S.Wallace, D.M.Boulton (1968), An Information Measure for
 Classification. {\it Computer Journal}, 11,2, 185-194. 

 \rr C.S.Wallace, D.Dowe (1999). Minimum Message Length and 
 Kolmogorov Complexity. {\it Computer Journal}, 42,4, 270-283. 

 \rr C.S.Wallace (2005). {\it Statistical and Inductive Inference by 
 Minimum Message Lenght.} NY: Springer. 

 \rr W.A.Wallis (1980). The Statistical Research Group, 1942-1945.   
 {\it Journal of the American Statistical Association}, 75, 370, 320-330. 

 \rr R.Wang, S.W.Lagakos,J.H.Ware, D.J.Hunter, J.M.Drazen (2007). 
 Statistics in Medicine - Reporting of Subgroup Analyses in Clinical Trials
 {\it The New England Journal of Medicine,} 357, 2189-2194.
 
 \rr G.D.Wassermann (1955). Some Comments on the Methods and Statements 
 in Parapsychology and Other Sciences. {\it The British Journal for the 
 Philosophy of Science,} 6, 22, 122-140.

 \rr L.Wasserman (2004). {\it All of Statistics: A Concise Course in 
  Statistical Inference.}  NY: Springer. 

 \rr L.Wasserman (2005). {\it All of Nonparametric Statistics.} 
  NY: Springer.

 \rr  S.Wechsler, L.G.Esteves, A.Simonis, C.Peixoto (2005). 
 Indiference, Neutrality and Informativeness: 
 Generalizing the Three Prisioners Paradox. 
 {\it Synthese}, 143, 255-272. 

 \rr S.Wechsler, C.A.B.Pereira, P.C.Marques (2008). 
  Birnbaum's Theorem Redux. 
 {\it AIP Conference Proceedings,} 1073, 96-100. 


 
 \rr  R.L.Weil, P.C.Kettler (1971).
 Rearranging Matrices to Block Angular Form for Decomposition Algorithms. 
 {\it Management Science}, 18, 1, 98-108.

 \rr M.Weliky, G.Oster (1990). The Mechanical Basis of Cell 
  Rearrangement. {\it Development,} 109, 373-386. 

 \rr  E.Wenger, R.Pea, J.S.Brown, C.Heath (1999). 
 {\it Communities of Practice: Learning, Meaning, and Identity.} 
 Cambridge Univ. Press. 

 \rr S.R. White (1984). Concepts of Scale in Simulated Annealing.
 {\it American Institute of Physics Conference Proceedings}, 122, 261-270.

 \rr T.A.Widiger, E.Simonsen (2005). Alternative Dimensional Models of
  Personality Disorder: Finding a Common Ground. 
  {\it J.of Personality Disorders}, 19, 110-130

 \rr F.W.Wiegel (1986). {\it Introduction to Path-Integral Methods in 
 Physics and Polymer Science.} Singapore: World Scientific. 

 \rr E.P.Wigner (1960).  
 The Unreasonable Effectiveness of Mathematics in the Natural Sciences.  
 {\it Communications in Pure and Applied mathematics,} 13,1. 
 Also ch.17, 222-237 of Wigner (1967). 

 \rr E.P.Wigner (1967). {\it Symmetries and Reflections.} 
 Bloomington: Indiana University Press.  

 \rr D.Williams (2001) {\it Weighing the Odds.} 
 Cambridge Univ. Press.  

 \rr R.C.Williamson (1989). {\it Probabilistic Arithmetic.} 
 Univ. of Queensland.  

 \rr S.S.Wilks (1962). {\it Mathematical Statistics.}  NY: Wiley. 

 \rr R.L.Winkler (1975). 
 {\it Statistics: Probability, Inference, and Decision.}  
 Harcourt School. 

 \rr  T.Winograd, F.Flores (1987).             
  {\it Understanding Computers and Cognition: 
  A New Foundation for Design} 
  NY: Addison-Wesley. 
 \rr H.Weyl (1952). {\it Symmetry.} Princeton Univ. Press. 

 \rr   R.G.Winther (2000). Darwin on Variation and Hereditarity. 
    {\it Journal of the History of Biology}, 33, 425-455. 

 \rr A.Wirfs-Brock, B.Wilkerson (1989). Variables Limit Reusability,
  {\it Journal of Object-Oriented Programming,} 2, 1, 34-40. 

 \rr D.A.Wismer, ed. (1971). {\it Optimization Methods for Large-Scale 
    Systems with Applications.} NY: McGaw-Hill. 

 \rr S.Wi\'{s}niewski, B.Staniszewski, R.Szymanik (1976). 
    Thermodynamics of Nonequilibrium Processes. 
    Dirdrecht: Reidel. 

 \rr  L.Wittgenstein (1921). 
 {\it Tractatus Logico Philosophicus}  
 (Logisch-Philosophische Abhandlung). 
 (Ed.1999) NY: Dover.  

 \rr L.Wittgenstein (1953). 
  {\it Philosophische Untersuchungen.} 
  Philosophical Investigations, english transl. by 
  G.E.M.Anscombe. Oxford: Blackwell. 

 \rr P.Wolfe (1959). The Simplex Method for Quadratic Programming. 
  {\it Econometrica}, 27, 383--398.  

 \rr W.Yourgrau S.Mandelstam (1979). {\it Variational Principles in 
 Dynamics and Quantum Theory.} NY: Dover. 

 \rr S.Youssef (1994). Quantum Mechanics as Complex Probability
  Theory. {\it Mod. Physics Lett.} A, 9, 2571-2586.

 \rr S.Youssef (1995). Quantum Mechanics as an Exotic Probability Theory.  
 Proceedings of the Fifteenth International Workshop on Maximum Entropy 
 and Bayesian Methods, ed. K.M.Hanson and R.N.Silver, Santa Fe.

 \rr S.L. Zabell (1992). The Quest for Randomness and its Statistical 
 Applications. In E.Gordon, S.Gordon (Eds.), Statistics for the 
 Twenty-First Century (pp. 139-150). 
 Washington, DC: Mathematical Association of America. 

 \rr L.A.Zadeh (1987). {\it Fuzzy Sets and Applications}.
 NY: Wiley. 

 \rr A.Zahavi (1975).  Mate selection: A selection for a handicap.  
 {\it Journal of Theoretical Biology}, 53, 205-214.

 \rr W.I.Zangwill (1969). {\it Nonlinear Programming: A Unified Approach.} 
  NY: Prentice-Hall.  

 \rr  W.I.Zangwill, C.B.Garcia (1981). 
  {\it Pathways to Solutions, Fixed Points, and Equilibria.} 
  NY: Prentice-Hall.  
 
 \rr  M.Zelleny (1980). {\it Autopoiesis, Dissipative Structures, 
 and Spontaneous Social Orders.}  
 Washington: 
 American Association for the Advancement of Science. 

 \rr A.Zellner (1971). {\it Introduction to
  Bayesian Inference in Econometrics.} NY:Wiley.

 \rr A.Zellner (1982). Is Jeffreys a Necessarist? 
 {\it American Statistician}, 36, 1, 28-30.  

 \rr H.Zhu (1998). {\it Information Geometry, Bayesian Inference,
 Ideal Estimates and Error Decomposition.} 
 Santa Fe Institute, 1399 Hyde Park Road, Santa Fe, NM 87501.

 \rr V.I.Zubov (1983). {\it Analytical Dynamics of Systems of 
  Bodies.} Leningrad Univ.  

 \rr M.A.Zupan (1991). Paradigms and Cultures: Some Economic Reasons for 
  Their Stickness. {\it The American Journal of Economics and Sociology},  
  50, 99-104.

\end{small} 
\renewcommand{\baselinestretch}{1.1}
\parskip 0.2cm








%% file: CAPFBST.TEX
 

 \chapter{FBST Review} 

 {\flushright 

  {\it  
  ``(A) man's logical method should be loved and reverenced as  
   \mbox{} \\
    his bride, whom he has chosen from all the world. He need not 
   \mbox{}  \\  
    contemn the others; on the contrary, he may honor them deeply, 
   \mbox{}  \\ 
    and in doing so he honors her more. But she is the one that he  
   \mbox{} \mbox{}  \\ 
    has chosen, and he knows  that he was right in making that choice.''
   }
   
    C.S.Peirce (1839 - 1914), \ \ \mbox{} \\  
    The Fixation of Belief (1877). \ \  \mbox{} \\ 
   
  \mbox{} \\ 

  {\it  
  ``Make everything as simple as possible, but not simpler.'' 
  }

    Albert Einstein (1879 - 1955). \ \  \mbox{} \\ 
   
 }

 \section{Introduction}

 The FBST was specially designed to give a measure of the 
 {\it epistemic value} of a sharp statistical hypothesis $H$, 
 given the observations, that is, to give a measure of the 
 {\it value of evidence} in support of $H$ given by the observations. 
 This measure is given by the support function $\ev(H)$, 
 the FBST {\it e-value}. 
 Furthermore the e-value has many necessary or desirable properties 
for a statistical support function, such as: 

 (I) Give an intuitive and simple measure of significance for the 
hypothesis in test, ideally, a {\it probability} defined directly
in the  original or {\it natural parameter space}. 

 (II) Have an intrinsically geometric definition, independent of any
non-geometric aspect, like the particular parameterization of the
(manifold representing the) null hypothesis being tested, or the
particular coordinate system chosen for the parameter space, i.e., be an
 {\it invariant} procedure. 

 (III) Give a measure of significance that is smooth, i.e. 
{\it continuous and differentiable}, on the hypothesis parameters and 
sample statistics, under appropriate regularity conditions for the model.    

 (IV) Obey the {\it likelihood principle} , i.e., the information
gathered from observations should be represented by, and only by, the
likelihood function, 
 see Berger and Wolpert (1988), Pawitan (2001, ch.7)  and 
 Wechsler et al. (2008).

 (V) Require {\it no ad hoc artifice} like assigning a positive prior
probability to zero measure sets, or setting an arbitrary initial
belief ratio between hypotheses. 

 (VI) Be a {\it possibilistic} support function, where 
the support of a logical disjunction is the maximum support 
among the support of the disjuncts. 

 (VII) Be able to provide a {\it consistent} 
test for a given sharp hypothesis.

 (VIII) Be able to provide {\it compositionality} operations in  
complex models. 

 (IX) Be an {\it exact} procedure, i.e., make no use of ``large sample'' 
asymptotic approximations when computing the $e$-value.

 (X) Allow the incorporation of previous experience or expert's opinion 
via (subjective) {\it prior distributions}.

 The objective of this section is to provide a very short review  of the
FBST theoretical framework,  summarizing the most important
statistical properties of its support function, the $e$-value. 
 It also summarizes the logical (algebraic) properties of the $e$-value, 
and its relations to other classical support calculi, including 
possibilistic calculus and logic, paraconsistent and classical.     
 Further details, demonstrations of theoretical properties, comparison
with other statistical tests for sharp hypotheses, and an extensive
list of references can be found in the author's previous papers. 
   
 \section{Bayesian Statistical Models}

   A standard model of (parametric) Bayesian statistics concerns  
  an observed (vector) random variable, $x$, that has a 
  {\it sampling} distribution with a specified functional form, 
  $p(x \g \theta)$, indexed by the (vector) parameter $\theta$.   
   This same functional form, regarded as a function of the free 
  variable $\theta$ with a fixed argument $x$, is the model's 
  {\it likelihood} function. 
    In {\it frequentist} or classical statistics, one is allowed 
  to use probability calculus in the sample space, but strictly 
  forbidden to do so in the parameter space, that is, $x$ is to be 
  considered as a random variable, while $\theta$ is not to be 
  regarded as random in any way. 
   In frequentist statistics,  $\theta$ should be taken as a 
  `fixed but unknown quantity' (whatever that means).

   In the Bayesian context, the parameter $\theta$ is regarded as a 
  latent (non-observed) random variable. 
   Hence, the same formalism  used to express credibility or 
  (un)certainty, namely, probability theory, is used in 
  both the sample and the parameter space.   
   Accordingly, the joint probability distribution, $p(x,\theta)$ 
  should summarize all the information available in a statistical model.  
   Following the rules of probability calculus, 
   the model's joint distribution of $x$ and $\theta$ can be 
  factorized either as the likelihood function of the parameter 
  given the observation times the {\it prior} distribution on $\theta$, 
  or as the {\it posterior} density of the parameter times the 
  observation's marginal density,   
  \[ 
     p(x,\theta) = p(x \g \theta) p(\theta) 
                 = p(\theta \g x) p(x) \ . 
  \]

    The {\it prior} probability distribution $p_0(\theta)$  
  represents the initial information available about the parameter.  
   In this setting, a {\it predictive} distribution for 
  the observed random variable, $x$, is represented by a mixture 
  (or superposition) of stochastic processes, all of them 
  with the functional form of the sampling distribution,
  according to the prior mixing (or weights) distribution,      
   \[ 
      p(x) = \int_{\theta} p(x \g \theta) p_0(\theta) d \theta \ . 
   \]

    If we now observe a single event, $x$, it follows from the 
  factorizations of the joint distribution above that the 
  {\it posterior}  probability distribution of $\theta$, representing 
  the available information about the parameter after the observation, 
  is given by  
   \[ 
      p_1(\theta) \propto p(x \g \theta) p_0(\theta) \ . 
   \]  

    In order to replace the `proportional to' symbol, $\propto$,  
   by an equality, it is necessary to divide the right hand site 
   by the normalization constant,  
    $c_1 = \int_\theta p(x \g \theta) p_0(\theta) d\theta$. 
    This is the {\it Bayes rule}, giving the (inverse) probability 
   of the parameter given the data. That is the basic learning 
   mechanism of Bayesian statistics.  
    Computing normalization constants is often difficult or 
   cumbersome. 
    Hence, especially in large models, it is customary to work with 
   unormalized densities or {\it potentials} as long as possible 
   in the intermediate calculations, computing only the final 
   normalization constants.    
    It is interesting to observe that the joint distribution function, 
   taken with fixed $x$ and free argument $\theta$, is a potential 
   for the posterior distribution.

    Bayesian learning is a recursive process, where the 
   posterior distribution after a learning step becomes the 
   prior distribution for the next step. 
    Assuming that the observations are i.i.d. (independent and identically 
   distributed) the posterior distribution after $n$ observations, 
   $x^{(1)},\ldots x^{(n)}$, becomes,             
   \[ 
      p_n(\theta) 
      \ \propto \ 
      p(x^{(n)} \g \theta) p_{n-1}(\theta) 
      \ \propto \  \pprod_{i=i}^n 
        p(x^{(i)} \g \theta) p_0(\theta) \ . 
   \]  
    If possible, it is very convenient to use a {\it conjugate prior}, 
  that is, a mixing distribution whose functional form is invariant 
  by the Bayes operation in the statistical model at hand.   
   For example, the conjugate priors for the Normal and Multivariate 
  models are, respectively, Wishart and the Dirichlet distributions. 
  The explicit form of these distributions is given in 
  the next sections.   
 
    The `beginings and the endings' of the Bayesian learning process 
  really need  further discussion, that is, we should present 
  some rationale for choosing the prior distribution used to start the 
  learning process, and some convergence theorems for the posterior 
  as the number observations increases.    
   In order to do so, we must access and measure the information 
  content of a (posterior) distribution.  
   Appendix E is dedicated to the concept of entropy, the key that 
  unlocks many of the mysteries related to the problems at hand. 
   In particular, Sections E.5 and E.6 discuss some fine details about 
  criteria for prior selection and posterior convergence properties.

 \section{The Epistemic $e$-values} 

  Let $\theta \in \Theta \subseteq  \Re^p$ be a vector parameter 
 of interest, and $p(x \g \theta)$ be the likelihood associated to 
 the observed data $x$, as in the standard statistical model. 
  Under the Bayesian paradigm the posterior density, 
 $p_n(\theta)$, is proportional to the product of the 
 likelihood and a prior density,  
 \[ 
    p_n(\theta) \propto p(x \g \theta) \, p_0(\theta). 
 \]  
 
  The (null) hypothesis $H$ states that the parameter lies in the 
 null set, defined by inequality and equality constraints given by 
 vector functions $g$ and $h$ in the parameter space.     
  \[ 
         \Theta_H = \{ \theta \in \Theta \g 
         g(\theta) \leq \zero \wedge h(\theta) = \zero \} 
  \] 
  From now on, we use a relaxed notation, writing 
 $H$ instead of $\Theta_H$.    
  We are particularly interested in sharp (precise) hypotheses, i.e., 
 those in which there is at least one equality constraint and hence, 
 $\dim(H) < \dim(\Theta)$.

 The FBST defines $\ev(H)$, the $e$-value supporting (in favor of) 
 the hypothesis $H$, and $\evb(H)$, the $e$-value against $H$, as    
  \[  
     s(\theta) =   \frac{p_n(\theta)}{r(\theta)} \ , \ \   
     s^* = s(\theta^*) = 
     \sup\nolimits_{\theta \in H}  s(\theta) \ , \ \        
     \sh = s(\tho) = 
     \sup\nolimits_{\theta \in \Theta}  s(\theta) \ ,         
  \]   
  \[  
     T(v) =  \{ \theta \in \Theta \g s(\theta) \leq v \}   
      \ , \ \  
     W(v) = \int_{T(v)} p_n\left(\theta \right) d\theta 
      \ , \ \ 
     \ev(H) = W(s^*)  \ , 
  \] 
  \[  
     \Tb(v) = \Theta - T(v) \ , \ \ 
     \Wb(v) = 1-W(v)  \ , \ \  
     \evb(H) = \Wb(s^*) = 1-\ev(H) \ .  
  \] 

  The function $s(\theta)$ is known as the posterior surprise 
 relative to a given reference density, $r(\theta)$. 
  $W(v)$ is the cumulative surprise distribution.  
  The surprise function was used, among other statisticians,  
 by Good [23], Evans [16]  and Royall [48]. 
  Its role in the FBST is to make $\ev(H)$ explicitly invariant 
 under suitable transformations on the coordinate system of the 
 parameter space, see next section. 

  The tangential (to the hypothesis) set $\Tb=\Tb(s^*)$, is a 
 Highest Relative Surprise Set (HRSS). 
  It contains the points of the parameter space with higher 
 surprise, relative to the  reference density, than any point in the 
 null set $H$. 
  When $r(\theta)\propto 1$, the possibly improper uniform density, 
 $\Tb$ is the Posterior's Highest Density Probability Set (HDPS) 
 tangential to the null set $H$. 
  Small values of $\evb(H)$ indicate that the hypothesis traverses high 
 density regions, favoring the hypothesis.

  Notice that, in the FBST definition, there is an optimization step  
 and an integration step. 
  The optimization step follows a typical {\it maximum probability} 
 argument, according to which, 
 ``a system is best represented by its highest probability 
 realization''. 
  The integration step extracts information from the system as a   
 probability weighted average. 
  Many inference procedures of classical statistics rely basically on 
 maximization operations, while many inference procedures of Bayesian 
 statistics rely on integration (or marginalization) operations. 
  In order to achieve all its desired properies, the FBST procedure 
 has to use both, as explained in this appendix.

  The  evidence value, defined above, has a simple and intuitive 
 geometric characterization. We now illustrate the above definitions 
 with two  simple but non-trivial examples. These two exemples are 
 easy to visualize, since they have a two dimensional parameter space, 
 and are also non-trivial, in the sense that they have a non-linear 
 hypothesis.

 \subsection*{Coefficient of Variation}

  The Coefficient of Variation (CV) of a random variable $X$ is defined 
 as the ratio
 $CV(X)=\sigma(X)/E(X)$, 
 i.e. the ratio of its standard deviation to its mean. 
  Let $X$ be a normal random variable, with unknown mean and  variance.  
  We want to compute the evidence value supporting the hypothesis that 
  the coefficient of variation of $X$ is equal to a given constant,   
 $$X \sim N(\beta,\sigma)\ \ , \ \ \  H :\ \sigma / \beta = c$$

 The conjugate family for this problem is the
family  of bivariate distributions, where the conditional distribution
of the  mean $\beta$, for a fixed precision $\rho=1/\sigma^2$, is
normal, and the  marginal distribution of the precision $\rho$ is gamma,
DeGroot (1970). 
 Using the standard improper priors, uniform on $]-\infty,+\infty[$ for
$\beta$, and $1/\rho$ on  $]0,+\infty[$ for $\rho$, we get the posterior
 joint distribution for $\beta$ and $\rho$:  
 \[ p_n(\beta ,\rho \g x) \propto 
    \sqrt{\rho}\; exp(-n\rho(\beta-\bar{x})^2/2) \;  
    \rho^{\frac{n-2}{2}} exp(-\rho s n/2 ) 
 \] 
 \[  
   x=[x_1\ldots x_n]\ , 
    \bar{x}=\frac{1}{n}\sum_{i=1}^n{x_i}\ ,\  
    s= \sum_{i=1}^n{(x_i-\bar{x})^2} 
 \] 
   
 \begin{figure}[hbt] 
 \centerline{\includegraphics*[height=4.0in, width=5.5in, 
                                      angle=0]{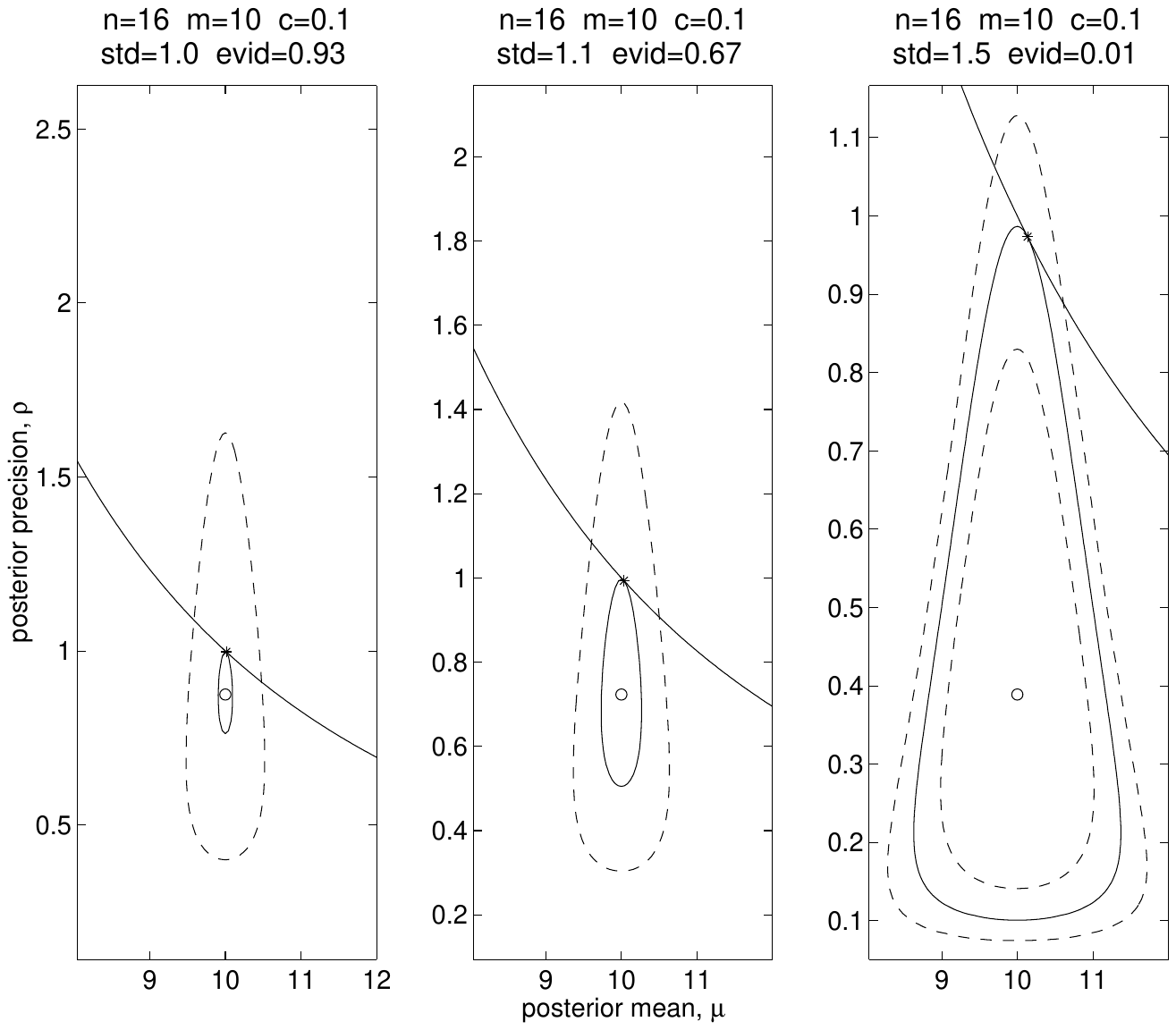}} 
 \centerline{Figure A.1: FBST for H: CV=0.1}
 \end{figure}

 Figure A.1 shows the null set $H$, 
 the tangential HRSS $\Tb$, and the points of constrained and 
 unconstrained maxima, $\theta^*$ and $\tho$,  for testing the 
 hypothesis at hand with the following numerical example:  
 $CV=0.1$ with 3 samples of size $n=16$,  mean $\bar{x}=10$ and 
 standard deviations $std=1.0$, $std=1.1$ and $std=1.5$.
  We can see the tangent set expanding as the sample 
 standard deviation over mean ratio gets farther away from the 
 coefficient of variation being tested, $CV(X)=\sigma(X)/E(X)=0.1$. 
  In this example we use the standard improper prior density and 
 the uniform reference density. 
  In the first plot, the sample standard deviation over mean ratio equals 
 the coefficient of variation tested. 
  Nevertheless, the evidence against the null hypothesis is not zero; 
 this is because of the non uniform prior. 
 In order to test other hypotheses we only have to change the
constraint(s)  passed to the optimizer. Constraints for the hypothesis
$\beta=c$ and  $\sigma=c$ would be represented by, respectively,
vertical and horizontal lines.  All the details for these and other
simple examples, as well as comparisons with standard frequentist and
Bayesian tests, can be found in Irony et al. (2001),  
Pereira and Stern (1999b, 2000a,b) and Pereira and Wechsler (1993).

  \subsection*{Hardy-Weinberg equilibrium}

 \begin{figure}[hbt] 
  \centerline{\includegraphics*[height=2.0in, width=4.0in, 
  angle=0]{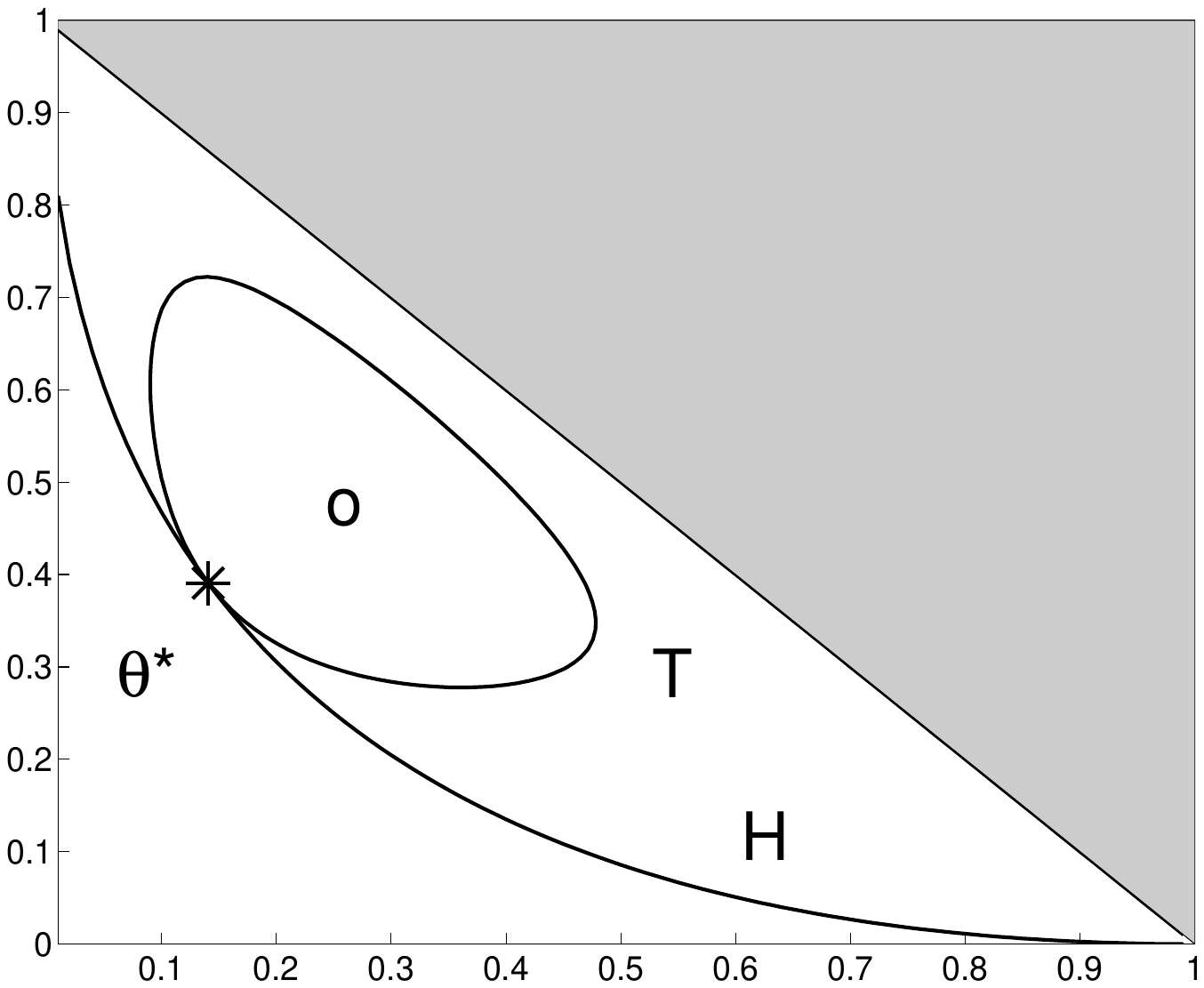}}
 \centerline{Figure A.2: H-W Hypothesis and Tangential Set}
 \end{figure}

 Figure A.2 shows the null set $H$, 
 the tangential HRSS $\Tb$, and the points of constrained and 
 unconstrained maxima, $\theta^*$ and $\tho$,  for testing 
 Hardy-Weinberg equilibrium law in a population genetics problem, 
 as discussed in Pereira and Stern (1999). 
  In this biological application 
 $n$ is the sample size,  
 $x_1$ and $x_3$ are the two homozygote sample counts 
 and $x_2 = n -x_1 -x_3$ is heterozygote sample count.    
 $\theta= [ \theta_1, \theta_2, \theta_3 ]$ is the parameter vector.   
  The posterior and maximum entropy reference densities for this 
 trinomial model, the parameter space and the null set are:  
  \[  
   p_n(\theta \mid x) \propto  
    \theta_1^{x_1+y_1-1} \theta_2^{x_2+y_2-1} \theta_3^{x_3+y_3-1} \ , \ \   
     r(\theta)  \propto  
    \theta_1^{y_1-1} \theta_2^{y_2-1} \theta_3^{y_3-1} \ , \ \ 
    y = [0,0,0] \ , 
  \] 
  \[  
   \Theta =   
    \{ \theta \geq 0 \mid \theta_1 + \theta_2 +\theta_3 = 1 \} \ , \ \  
    H =  \{ \theta \in \Theta \mid \theta_3 
                  = ( 1 -\sqrt{\theta_1} \  )^2  \} \ .  
  \]

  \subsection*{Nuisance Parameters}

  Let us consider the situation where the hypothesis constraint, 
 $H:\ h(\theta)=h(\delta)=0\ , \theta=[\delta,\lambda]$ is not a function
of some of the parameters, $\lambda$. 
 This situation is described by D.Basu in Ghosh (1988):   

 \begin{quotation} 
 {\it ``If the inference problem at hand relates only to $\delta$, 
 and if information gained on $\lambda$ is of no direct relevance to the 
problem, then we classify $\lambda$ as the Nuisance Parameter. 
 The big question in statistics is: How can we eliminate the 
nuisance parameter from the argument?''}  
 \end{quotation} 

 Basu goes on listing at least 10 categories of procedures to  achieve
this goal, like using $max_\lambda$ or $\int \ d\lambda$,  the
maximization or integration  operators, in order to obtain a projected
profile or marginal  posterior function, $p(\delta \g x)$. 
 The FBST does not follow the nuisance parameters elimination paradigm, 
 working in the original parameter space, in its full dimension.

 \section{Reference, Invariance and Consistency}

 In the FBST the role of the reference density, $r(\theta)$ is to make
 $\evb(H)$ explicitly invariant under suitable transformations of
 the coordinate system. 
 The natural choice of reference density is an uninformative prior,
 interpreted as a representation of no information in the parameter
 space, or the limit prior for no observations, or the neutral
 ground state for the Bayesian operation.
 Standard (possibly improper) uninformative priors include the
 uniform and maximum entropy densities, 
 see Dugdale (1996) and Kapur (1989)   
 for a detailed discussion.  
 Invariance, as used in statistics, is a metric concept.
 The reference density can be interpreted as induced by 
 the information  metric in the parameter space, 
 $dl^2=d\theta'G(\theta)d\theta$. 
 Jeffreys' invariant prior is given by  
 $p(\theta)= \sqrt{\mbox{det} G(\theta)}$, see Section E.5.

 In the H-W example, using the notation above,   
 the uniform density can be represented by 
 $y=[1,1,1]$ observation counts, and 
 the standard maximum entropy density  
 can be represented by 
 $y=[0,0,0]$ observation counts.

 Let us consider the cumulative distribution 
 of the evidence value against the hypothesis,  
 $\Vb(c)= \Pr(\evb \leq c)$,   
 given $\theta^0$, the true value of the parameter.  
 Under appropriate regularity conditions, for increasing sample size, 
 $n\rightarrow \infty$, we can say the following:

 - If $H$ is false, $\theta^0\notin H$, then  
 $\evb$ converges (in probability) to 1, that is,  
 $\Vb(0\leq c <1)\rightarrow 0$.

 - If $H$ is true, $\theta^0\in H$,  then $\Vb(c)$, 
   the confidence level,  is approximated  by the function   
 \[   
    QQ(t,h,c) = 
    \Chi2\left(t-h, \Chi2^{-1}\left(t,c\right) \right) \ , \ \ 
    \wwhere 
 \] 
 \[ 
   \Chi2(k,x) = 
    \frac{\Gamma(k/2, x/2)}{\Gamma(k/2, \infty)} \ , \  \  
    \Gamma(k,x) = \int_0^x y^{k-1}e^{-y} dy \ ,  
 \] 
 $t=\dim(\Theta)$, $h=\dim(H)$ and $\Chi2(k,x)$ is the 
 cumulative chi-square distribution with $k$ degrees of freedom. 
 Figure A.3 portrays 
 $QQ(t,h,c)$ $\Chi2(t-h, \Chi2^{-1}(t,c))$ for 
 $t=2\ldots 4$ and $h=0\ldots t-1$.  

  Under the same regularity conditions, an appropriate choice of 
 threshold or critical level, $c(n)$, provides a consistent test, 
 $\tau_c\ $, that rejects the hypothesis if $\evb(H) > c$.    
  The empirical power analysis developed in Stern and Zacks (2002) and 
 Lauretto et al. (2003), provides critical levels that are consistent 
 and also effective for small samples.

 \begin{figure}[htb] 
  \centerline{\includegraphics*[height=1.5in, width=5.5in, 
  angle=0]{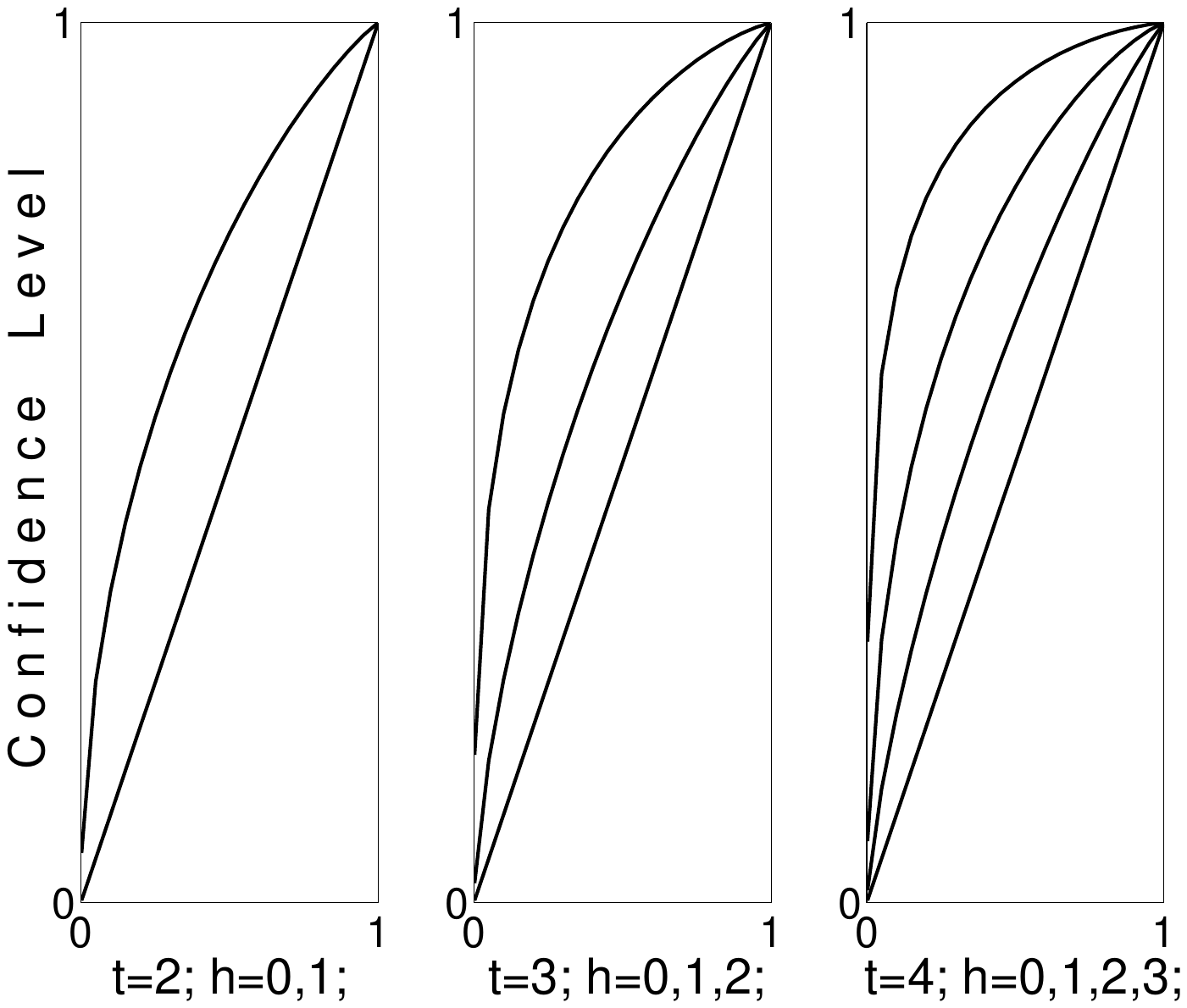}}
 \centerline{Figure A.3: Test $\tau_c$ critical level vs. confidence level}
 \end{figure}

 \subsection*{Proof of invariance:} 

 Consider a proper 
 (bijective, integrable, and almost surely continuously 
 differentiable) reparameterization 
 $\omega=\phi(\theta)$. 
  Under the reparameterization, the Jacobian, surprise, 
 posterior and reference functions are:  
  \[      
  J(\omega)  =   
  \left[ \frac{\del\theta}{\del\omega} \right] =  
  \left[ \frac{\del \phi^{-1}(\omega)}{\del \omega} \right] =  
   \left[ \begin{array}{ccc} 
   \frac{\del\theta_1}{\del\omega_1} & \ldots & 
   \frac{\del\theta_1}{\del\omega_n} \\ 
   \vdots & \ddots & \vdots \\ 
   \frac{\del\theta_n}{\del\omega_1} & \ldots & 
   \frac{\del\theta_n}{\del\omega_n}  
   \end{array} \right]    
   \]  
   \[ 
  \widetilde{s}(\omega)   =    
   \frac{ \widetilde{p}_n(\omega) }{ \widetilde{r}(\omega) } = 
   \frac{ p_n(\phi^{-1}(\omega)) \left| J(\omega) \right| } 
        { r(\phi^{-1}(\omega))   \left| J(\omega) \right| } 
   \] 

 Let $\Omega_H = \phi(\Theta_H)$. 
 It follows that 
 \[ \widetilde{s}^* = 
 \sup_{\omega \in \Omega_H} \widetilde{s}(\omega)  = 
 \sup_{\theta \in \Theta_H}  s(\theta)  =  s^*  
 \] 
 hence, the tangential set, 
 $\Tb \mapsto \phi(\Tb) = \widetilde{\Tb}$, and  
 \[ 
 \widetilde{\mbox{ev}}(H) = 
 \int_{\widetilde{\Tb}} \widetilde{p}_n(\omega) d \omega = 
 \int_{\Tb} p_n(\theta) d \theta = \evb(H) .  
 \]

 \subsection*{Proof of consistency:}

 Let  $\Vb(c)= \Pr(\evb \leq c)$  be the cumulative distribution of the
evidence value against the  hypothesis, given $\theta$. 
 We stated that, under appropriate regularity conditions, for increasing
sample size, $n\rightarrow \infty$, if $H$ is true, i.e. $\theta\in H$,
then $\Vb(c)$, is approximated  by the function   
 \[ 
    QQ(t,h,c) = 
    \Chi2\left(t-h, \Chi2^{-1}\left(t,c\right) \right) \ . 
 \]

 Let $\theta^0$, $\tho$ and $\theta^*$ be the true value, 
 the unconstrained MAP (Maximum A Posteriori), and constrained 
 (to $H$) MAP estimators of the parameter $\theta$.  

 Since the FBST is invariant, we can chose a coordinate system  where,
the (likelihood function) Fisher information matrix at the true
parameter value  is the identity, i.e., $J(\theta^0)=I$. 
 From the posterior Normal approximation theorem, 
 see Section 5 of Appendix E, we know that the standarized total
 difference between $\tho$ and $\theta^0$  converges in distribution  to 
 a standard Normal distribution, i.e. 
 \[ 
    \sqrt{n}(\tho -\theta^0) \rightarrow 
    N\left( 0, J(\theta^0)^{-1} J(\theta^0) J(\theta^0)^{-1} \right) = 
    N\left( 0, J(\theta^0)^{-1} \right) = 
    N\left( 0, I \right) 
 \] 

 This standarized total difference can be decomposed into tangent
(to the  hypothesis manifold) and transversal orthogonal components, i.e. 
 \[ 
    d_t = d_h + d_{t-h} \ , \  
    dt = \sqrt{n}(\tho -\theta^0) \ , \    
    d_h = \sqrt{n}(\theta^* -\theta^0) \ , \     
    d_{t-h} = \sqrt{n}(\tho -\theta^*)  \ .  
 \] 
 Hence, the total, tangent and transversal distances ($L^2$ norms), 
 $||d_t||$, $||d_h||$ and $||d_{t-h}||$, 
 converge in distribution to chi-square variates with, respectively,
 $t$, $h$ and $t-h$ degrees of freedom.  

 Also from, the MAP consistency, we know that the MAP 
estimate of the Fisher information matrix, $\widehat{J}$,  
converges in probability to true value, $J(\theta^0)$. 

 Now, if $X_n$ converges in distribution to $X$, and $Y_n$ converges 
in probability to $Y$, we know that the pair $[X_n, Y_n]$ converges 
in distribution to $[X, Y]$. 
 Hence, the pair $[||d_{t-h}||, \widehat{J}]$ converges in 
distribution to $[x, J(\theta^0)]$, where $x$ is a chi-square 
variate with $t-h$ degrees of freedom. 
 So, from the continuous mapping theorem, the evidence value 
against $H$, $\evb(H)$, converges in distribution to 
 $\overline{e}=\Chi2(t,x)$, where $x$ is a chi-square variate with 
 $t-h$ degrees of freedom. 
 
 Since the cumulative chi-square distribution is an increasing function, 
we can invert the last formula, i.e., 
 $\overline{e}=\Chi2(t,x)\leq c \Leftrightarrow x \leq \Chi2^{-1}(t,c)$. 
 But, since $x$ in a chi-square variate with 
 $t-h$ degrees of freedom, 
 \[ 
   \Pr(\overline{e} \leq c) =  
    QQ(t,h,c) = 
   \mbox{Q.E.D.} 
 \] 
 A similar argument, using a non-central chi-square distribution, 
proves the other asymptotic statement. 

 If a random variable, $x$, has a continuous and increasing cumulative 
 distribution function, $F(x)$, the random variable $u=F(x)$ has uniform 
 distribution.     
  Hence, the tranformation $\sevb =QQ(t,h,\evb)$, defines 
 a ``standarized $e$-value'', $\sev =1-\sevb$, that can be used 
 somewhat in the same way as a $p$-value of classical statistics. 
  This standarized $e$-value may be a convenient form to report,  
 since its asymptotically uniform distribution provides a large-sample 
 limit interpretation, and many researchers will feel already familiar 
 with consequent diagnostic procedures for scientific hypotheses 
 based on abundant empirical data-sets.

 \section{Loss Functions}

 In orthodox decision theoretic Bayesian statistics, a significance  test
is legitimate if and only if it can be characterized as an Acceptance
(A) or  Rejection (R) decision procedure defined by the minimization of 
the posterior expectation of a loss function, $\Lambda$.  
 Madruga (2001) gives the following family of loss functions 
characterizing the FBST. This loss function is based on indicator 
functions of $\theta$ being or not in the tangential set $\Tb$:    
 \[ 
   \Lambda(R,\theta)= a\, I(\theta \notin \Tb )   
   \ \ , \ \  
   \Lambda(A,\theta)= b +d\, I(\theta \in \Tb ) 
 \] 
 The interpretation of this loss function is as follows: 
 If $\theta \in \Tb$ we want to reject $H$,  
 for $\theta$ is more probable than anywhere on $H$; 
 If $\theta \in T$ we want to accept $H$, 
 for $\theta$ is less probable than anywhere on $H$.  
 The minimization of this loss function gives the optimal test: 
 \[   
 \mbox{Accept \ } H \mbox{\ iff \ } 
   \ev(H) \geq \varphi= (b+c)/(a+c) \ . 
 \] 
 Note that this loss function is dependent on the observed sample
(via the likelihood  function), on the prior, and on the reference
density, stressing the important point of non-separability of 
utility and probability, 
 see Kadane and Winkler (1987) and Rubin (1987). 
 
 This type of loss function can be easily adapted in order to provide  
an asymptotic indicator checking if the true parameter belongs to the 
hypothesis set, $I(\theta^0 \in H)$. 
 Consider the {\it tangential reference mass}, 
 \[ 
 \mb = \left[ \int_{\Tb(s^*)} r(\theta ) d\theta \right]^\gamma   
 \] 
 If $\gamma=1$, $\mb$ is the reference density mass of the tangencial set.    
 If $\gamma=1/t$, $\mb$ is a pseudo-distance from $\tho$ to $\theta^*$ . 
 Consider also a threshold of form 
 $\varphi_1 = b \mb$ or $\varphi_2 = b \mb / (a +\mb)$, $a,b>0$,  
 in the expression of the optimal test above.  

 If $\theta^0 \notin H$, then 
 $\tho \rightarrow \theta^0$ and 
 $\theta^* \rightarrow \theta^{0*}$, where 
 $\theta^{0*} \neq \theta^0$,  therefore 
 $||\tho - \theta^* || \rightarrow c_1 >0$. \\ 
 But the standarized posterior, $p_n$, converges to 
 a normal distribution centered on $\theta^0$. \\  
 Hence, $\mb \rightarrow c_2>0$ and 
 $\varphi \rightarrow c_3>0$. \ Finally, 
 since $\ev(H)\rightarrow 0$,  
 $\Pr( \ev(H) > \varphi ) \rightarrow 0$.

 If $\theta^0 \in H$, then 
 $\tho \rightarrow \theta^0$ and $\theta^* \rightarrow \theta^0$, 
 therefore $||\tho -\theta^*|| \rightarrow 0$. 
 Hence, $\mb \rightarrow 0$ and $\varphi \rightarrow 0$.  
 But $\ev(H)$ converges to a propper distribution, see section A.3, 
 and, therefore, $\Pr(\ev(H) > \varphi ) \rightarrow 1$.

 \section{Belief Calculi and Support Structures}

 Many standard Belief Calculi can be formalized in the context of 
 Abstract Belief Calculus, ABC, 
 see Darwiche and Ginsberg (1992), Darwiche (1993) and Stern (2003).    
 In a Support Structure, 
 $\langle \Phi, \oplus, \oslash \rangle$,     
 the first element is a Support Function, 
 $\Phi$, on a universe of statements, $\uu$. 
 Null and full support values are represented by $\zero$ and $\uno$. 
 The second element is a support Summation operator, $\oplus$, 
and the third is a support Scaling or Conditionalization operator, 
$\oslash$. 
 A Partial Support Structure, $\langle \Phi, \oplus \rangle$, 
lacks the scaling operation. 

 The Support Summation operator, $\oplus$,  gives the support value 
of the disjunction of any two logically disjoint statements from their 
individual support values, i.e.,  
  \[  
      \neg (A \wedge B) \Rightarrow  
      \Phi(A \vee B) = \Phi(A) \oplus \Phi(B) \ .  
  \]

  The support {\it scaling} operator updates an old state of belief 
 to the new state of belief resulting from making  an observation. 
  Hence it can be interpreted as predicting or propagating changes  
 of belief after a possible observation.
  Formally, the support scaling operator, $\oslash$, gives the 
 conditional support value of $B$ given $A$ from the unconditional 
 support values of $A$ and the conjunction $C=A\wedge B$, i.e.,
 \[   
   \Phi_A(B) = \Phi(A\wedge B) \oslash \Phi(A) \ . 
 \]

  The support {\it unscaling} operator reconstitutes the old state of 
 belief from a new state of belief and the observation that has led to it. 
  Hence it can be interpreted as explaining or back-propagating changes  
 of belief for a given observation.
  If $\Phi$ does not reject $A$, the support unscaling operatior, 
 $\otimes$, gives the inverse of the scaling operator, i.e.,  
 \[ 
   \Phi(A \wedge B) = \Phi_A(B) \otimes \Phi(A) \ . 
 \]

 Support structures for some standard belief calculi
 are given in Table A.1, where   
 the support value of two statements their conjunction 
 are given by  
 $a=\Phi(A)$, $b=\Phi(B)$, $c=\Phi(C=A\wedge B)$. 
 In Table A.1, the relation $a \preceq b$ indicates that the value 
$a$ represents a stringer support than the value $b$.       
  Darwiche and Ginsberg (1992) and Darwiche (1993) also give a set o
axioms defining  the essential functional properties of a (partial)
support function. Stern (2003) shows that the support 
 $\Phi(H)=\ev(H)$ complies with all these axioms.

 \begin{table}[bth]  
 \begin{center}
 {Table A.1: Support structures for some belief calculi, \\ 
    $a=\Phi(A)$, \ $b=\Phi(B)$, \ $c=\Phi(C=A\wedge B)$.} \\ 
 \mbox{} \\ 
 \resizebox{6.2in}{0.8in}{  
 { \large 
 \begin{tabular}{c l c c c c c | l } 
  \hline
   $\Phi(\uu)$ & $a \oplus b$ & $\zero$ & $\uno$ & $a \preceq b$ & 
  $c \oslash a$ & $a \otimes b$ &  Calculus \\  
 \hline 
   $[0,1]$   & $a + b$ & $0$ & $1$ & $a\leq b$ & 
  $c/a$ & $a \times b$ & Probability \\   
   $[0,1]$   & $\max(a,b)$ & $0$ & $1$ & $a\leq b$ &  
  $c/a$ & $a \times b$ & Possibility  \\ 
   $\{0,1\}$ & $\max(a,b)$ & $0$ & $1$ & $a\leq b$ & 
  $\min(c,a)$ & $\min(a,b)$ & Classic.Logic \\   
   $[0,1]$ & $a+b-1$  & $1$  & 0 & $b\leq a$ & 
  $(c-a)/(1-a)$ & $a+b-ab$ & Improbablty \\    
   $\{0..\infty \}$ & $\min(a,b)$  & $\infty$  & 0 & $b\leq a$ & 
  $c-a$ & $a + b$ & Disbelief \\    
 \hline 
 \end{tabular} 
 }
 }
 \end{center} 
 \end{table}

 In the FBST, the support values, $\Phi(H)=\ev(H)$, are computed using 
standard probability calculus on $\Theta$ which has an intrinsic  
conditionalization operator. 
 The computed evidences, on the other hand, have a possibilistic 
summation, i.e., the value of evidence in favor of a
composite  hypothesis $H=A \vee B$, is the most favorable value of
evidence in favor of each of its terms, i.e.,
 $\ev(H)=\max\{ \ev(A), \ev(B) \}$. 
  It is impossible however to define a simple scaling operator for 
 this possibilistic support that is compatible with the 
FBST's evidence, $\ev$, as it is defined.  

 Hence, two belief calculi are in simultaneous use in the FBST setup:  
 $\ev(H)$ constitutes a possibilistic partial support structure 
coexisting in harmony with the probabilistic support structure given by 
the posterior probability measure, $p_n(\theta)$, in the parameter space, 
 see Dubois et al. (1993), Delgado and Moral (1987).

  Requirements (V) and (VI), i.e. {\it no ad hoc artifice} and 
 {\it possibilistic support}, find a rich interpretation in the
 juridical or legal context, where they correspond to the some of 
 the most basic juridical principles, see Stern (2003). 

  {\it Onus Probandi} is a basic principle of legal reasoning, also
known as Burden of Proof, see Gaskins (1992) and Kokott (1998). 
  It also manifests itself in accounting through the 
 Safe Harbor Liability Rule:  

  \begin{quotation} 
 {\it ``There is no liability as long as there is a reasonable basis for
 belief, effectively placing the burden of proof (Onus Probandi) on the
 plaintiff, who, in a lawsuit, must prove false a defendant's
 misstatement, without making any assumption not  explicitly stated by
 the defendant, or tacitly implied by an existing law or regulatory 
 requirement.''}    
  \end{quotation} 
  The Most Favorable Interpretation principle, which, depending
 on the context, is also known as Benefit of the Doubt, 
 {\it In Dubito Pro Reo}, or Presumption of Innocence, is a consequence 
 of the Onus Probandi principle, and requires the court to consider 
 the evidence in the light of what is most favorable  to the
 defendant. 
 
  \begin{quotation} 
 {\it ``Moreover, the party against whom the motion is directed is
 entitled to have the trial court construe the evidence in support of
 its claim as truthful, giving it its most favorable interpretation, as
 well as having the benefit of all reasonable inferences drawn from that
 evidence.''}   
  \end{quotation}


 \section{Sensitivity and Inconsistency}

   For a given prior, likelihood and reference density, let 
  $\eta = \ev(H;p_0,L_x,r)$ 
  denote the $e$-value supporting $H$.  
   Let $\eta', \eta'' \ldots$ denote the $e$-value with respect to 
  references $r', r'' \ldots$. 
   The degree of inconsistency of the $e$-value supporting $H$, 
  induced by a set of references, 
  $\{r, r', r''\ldots \}$ is defined by the index 
   \[ 
       I \left\{ \eta, \eta', \eta'' \ldots  \right\} \ = \ 
     \max \left\{ \eta, \eta', \eta'' \ldots \right\}  
    -\min \left\{ \eta, \eta', \eta'' \ldots \right\}  
   \]  
   
   The same index can be used to study the degree of inconsistency 
  of the $e$-value induced by a set of priors, 
  $\{p_0, p_0', p_0''\ldots \}$.      
   One could also study the sensitivity of the $e$-value 
  to a set of vitual sample sizes, 
  $\{1n, \gamma' n, \gamma''n \ldots \}$, $\gamma\in [0,1]$,  
  corresponding to scalled likelihoods, 
  $\{L, L^{\gamma'}, L^{\gamma''} \ldots \}$. 
   This intuitive measure of inconsistency can be made rigorous in 
  the context of paraconsistent logic and bilattice structures,   
  see Abe et al. (1998), Alcantara et al. (2002), Arieli and Avron (1996), 
  Costa (1963), Costa and Subrahmanian (1989) and 
  Costa et al. (1991), (1999).

 The bilattice
 $B(C,D)=\left< C\times D, \leqk, \leqt \right>$, 
 given two complete lattices, 
 $\left< C, \leqc \right>$, and
 $\left< D, \leqd \right>$, 
 has two orders,
 the knowledge order, $\leqk$,
 and the truth order, $\leqt$, given by:
 \begin{eqnarray*}
 \left< c_1, d_1 \right> \leqk \left< c_2, d_2 \right>
 &\Leftrightarrow&
 c_1 \leqc c_2 \aand d_1 \leqd d_2 \\
 \left< c_1, d_1 \right> \leqt \left< c_2, d_2 \right>
 &\Leftrightarrow&
 c_1 \leqc c_2 \aand d_2 \leqd d_1
 \end{eqnarray*}
 The standard interpretation is that $C$ provides the ``credibility''
 or value in favor of a hypothesis (or statement) $H$,
 and $D$ provides the ``doubt'' or value against $H$.
   If $\left< c_1, d_1 \right> \leqk \left< c_2, d_2 \right>$,
 then we have more information (even if inconsistent) about
 situation 2 than  1. Analogously,  if
 $\left< c_1, d_1 \right> \leqt \left< c_2, d_2 \right>$,
 then we have more reason to trust (or believe)
 situation 2 than  1 (even if with less information).

 For each of the bilattice orders we define a 
 join and a meet operator, based on the join and the meet operators 
 of the single lattices orders. More precisely, 
 $\sqcupk$ and $\sqcapk$, 
 for the knowledge order, and  
 $\sqcupt$ and $\sqcapt$, 
 for the truth order, 
 are defined by the folowing equations:
 \[
 \left< c_1, d_1 \right> \sqcupk \left< c_2, d_2 \right>  =  
 \left< c_1 \sqcupc c_2 , d_1 \sqcupd d_2 \right>         ,  
 \left< c_1, d_1 \right> \sqcapk \left< c_2, d_2 \right>  =  
 \left< c_1 \sqcapc c_2 , d_1 \sqcapd d_2 \right>
 \]
 \[ 
 \left< c_1, d_1 \right> \sqcupt \left< c_2, d_2 \right>  =   
 \left< c_1 \sqcupc c_2 , d_1 \sqcapd d_2 \right>         ,   
 \left< c_1, d_1 \right> \sqcapt \left< c_2, d_2 \right>  = 
 \left< c_1 \sqcapc c_2 , d_1 \sqcupd d_2 \right> 
 \]

 The ``unit square'' bilattice,
 $\left< [0,1] \times [0,1], \leq, \leq \right>$
 has been routinely used to represent fuzzy or rough pertinence
 relations, logical probabilistic annotations, etc.
  The lattice $\left< [0,1], \leq \right>$ is the standard unit
 interval, where the join and meet  
 coincide with the max and min operators, 
 $\sqcup=\max$ and $\sqcap=\min$.

   In the unit square bilattice the ``truth'', ``false'',
 ``inconsistency'' and ``indetermination'' extremes are
 $t$, $f$, $\top$, $\bot$,
 whose coordinates are given in Figure A.4.
  As a simple example, let region $R$ be the convex hull of the
 four vertices $n$, $s$, $e$ and $w$, given in Figure A.4.
  Points $kj$, $km$, $tj$ and $tm$ are the 
  knowledge and truth join and meet, over $r\in R$.

  In the unit square bilattice, the degree of trust and degree of 
inconsistency for a point $x=\left< c,d \right>$ are given by 
 $\bt \left (\left< c,d \right> \right) = c-d$, and 
 $\bi \left (\left< c,d \right> \right) = c+d-1$, 
 a convenient linear reparameterization of 
 $[0,1]^2$, to $[-1,+1]^2$. 
  Figure A.4 also compares the credibility-doubt and 
 trust-inconsistency coordinates. 

   Let $\eta=\ev(H)$, and $\etab=\evb(H)= 1-\ev(H)$. 
 The point $x=\left< \eta , \etab \right>$ 
 in the unit square bilattice, represents herein a single evidence. 
  Since $\bi(x)=0$, such a point is consistent.
  It is also easy to verify that for the multiple $e$-values, the
 definition of degree of  inconsistency given above,  is the degree of
 inconsistency of the knowledge join of all the single evidence points, 
 i.e.,  
 %
 \[
        I(\eta,\eta',\eta''\ldots ) \ = \
        \bi \left(     \left< \eta , \etab   \right>
   \sqcupk  \left< \eta'  , \etab'  \right>
   \sqcupk  \left< \eta'' , \etab'' \right> \ldots \right) \ .
 \]

 Negation type operators are not an integral part of the bilattice
 structure but, in the unit square, one can define negation as 
 $\neg \left< c, d \right> = \left< d, c \right>$,  and conflation as
 $ - \left< c, d \right> = \left< 1-c, 1-d \right>$,  
 so that negation reverses trust, but preserves knowledge, 
 and conflation reverses knowledge, but preserves trust.

 \begin{figure}[h]
  \centerline{\includegraphics*[height=2.4in, width=6.0in, 
              angle=0]{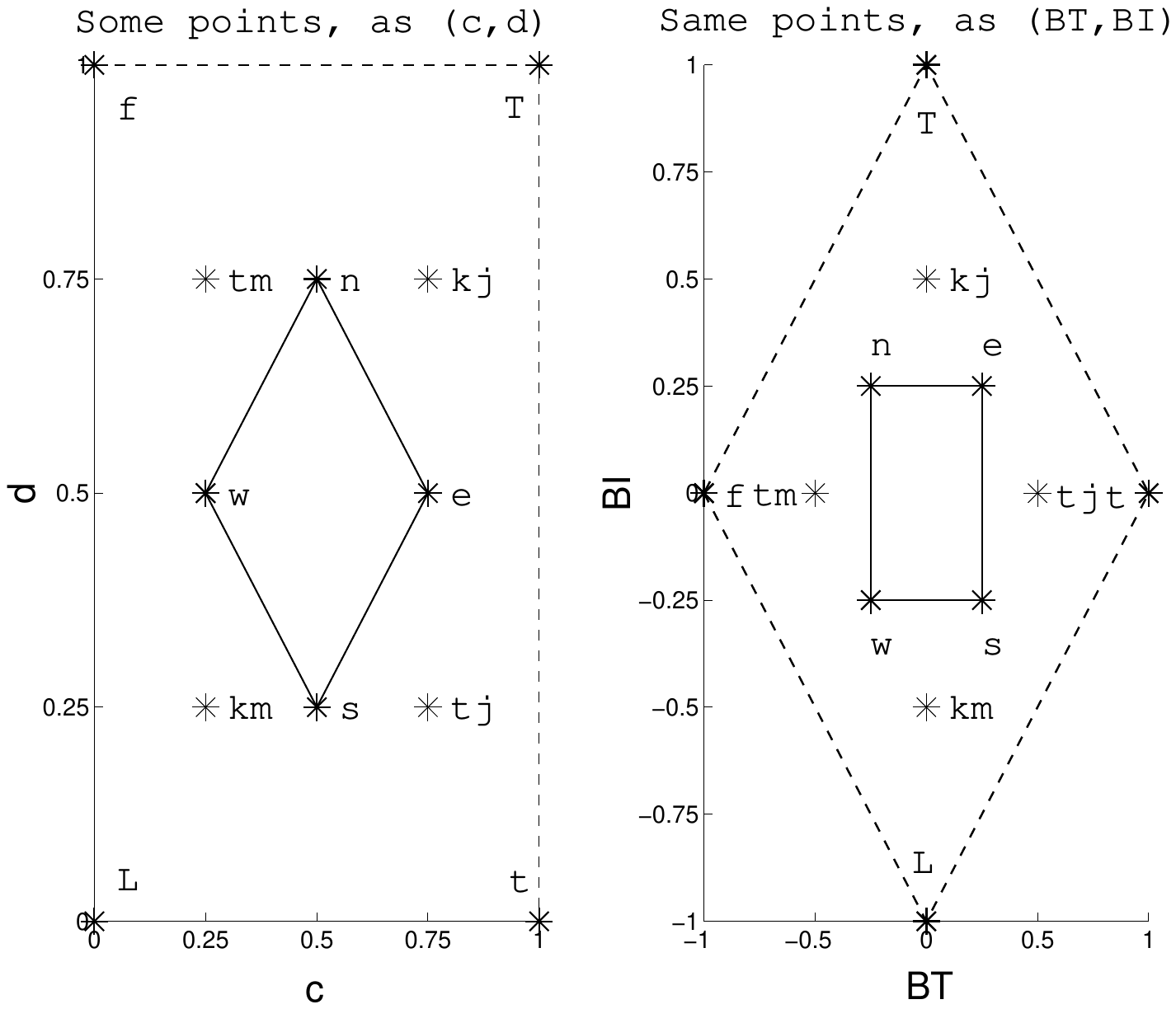}}
  \centerline{Figure A.4: credibility-doubt and 
              trust-inconsistency coordinates}
 \end{figure}

  As an example of sensitivity analysis we use the HW model with the
standard uniformative references, the uniform and the  maximum entropy
densities, represented by $[1,1,1]$ and $[0,0,0]$  observation
counts. For a motivation for this particular analysis, 
see the observations at the end of section E.5.    
 Between these two uninformative references, we also consider perturbation
references corresponding to $[0,1,1]$, $[1,0,1]$ and  $[1,1,0]$ 
observation counts. Each of these references can be interpreted as the
exclusion of a single observation of the corresponding type from the
observed data set.

 \begin{figure}[bth]
  \centerline{\includegraphics*[height=3.1in, width=6.0in, 
              angle=0]{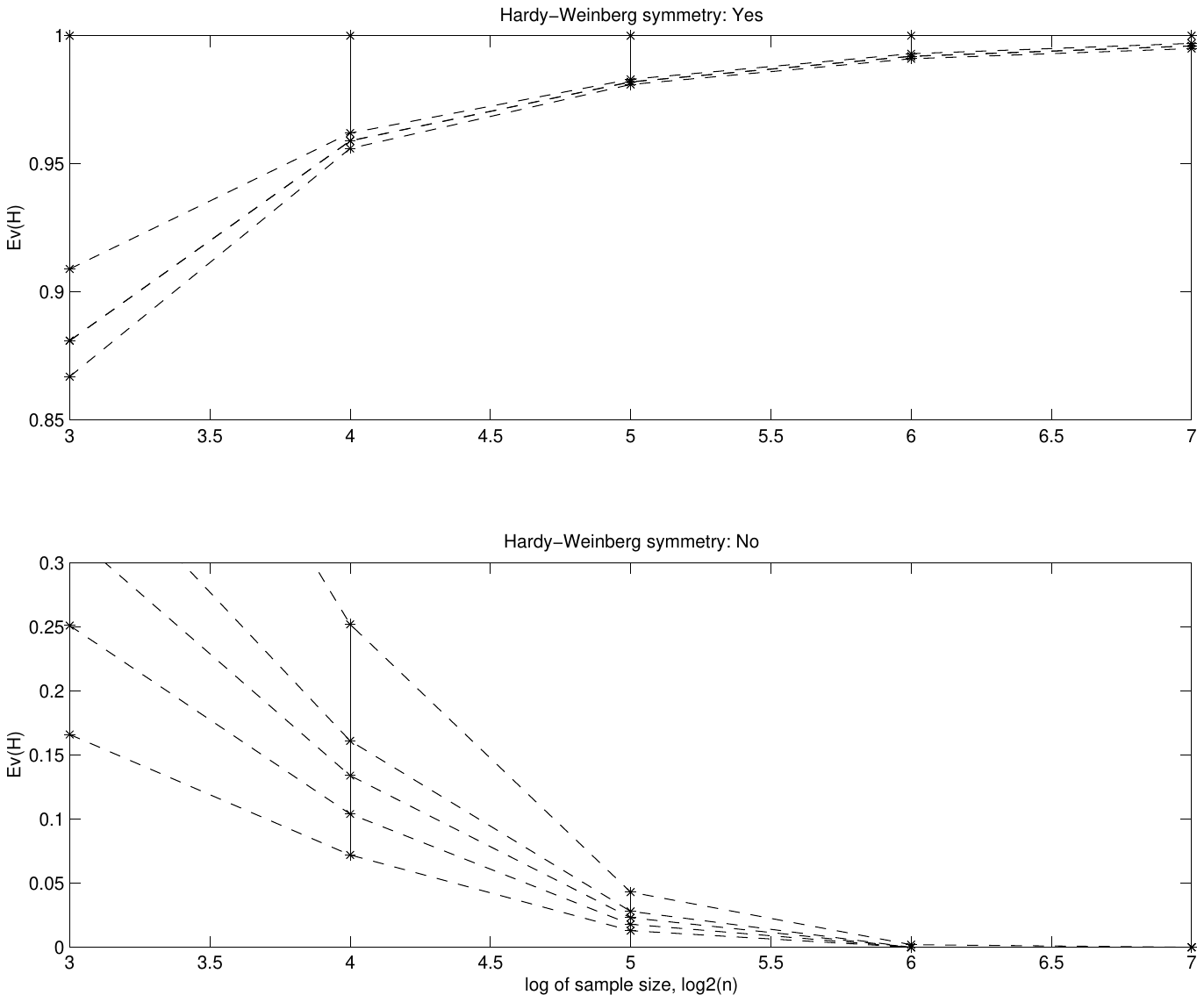}}
  \centerline{Figure A.5: Sensitivity analysis}
 \end{figure}

 The $e$-values in the example are calculated using two sample
proportions, 
 $[x_1,x_2,x_3]$ $=n[1/4, 1/4, 1/2]$ and $=n[1/4, 1/2, 1/4]$. 
 The first exhibits the HW hypothesis symmetry, the second does not.
 The log 2 of sample size, $\log_2(n)$, ranged from $3$ to $7$.  
 In Figure A.5, the $e$-values corresponding to each choice of reference, 
are given by an interpolated dashed line. 
 The interpretation of the vertical interval (solid bars) between the
dashed lines  is similar to that of the usual statistical error bars. 
 However, the uncertainty represented by these bars does not have a 
probabilistic nature, being rather a possibilistic measure of 
inconsistency, defined in the partial support structure given by the 
FBST evidence value, see Stern (2004).

 \section{Complex Models and Compositionality}

  The relationship between the credibility of a complex hypothesis, $H$, 
and those of its constituent elementary hypothesis, $H^{(i,j)}$,  
in the independent setup, can be analyzed under the FBST, 
see Borges and Stern (2006) for precise definitions, and 
detailed interpretation.

 Let us consider elementary hypotheses, $H^{(i,j)}$, in $k$ independent 
constituent models, $M^j$, and the complex or composit hypothesis $H$,
equivalent to a (homogeneous) logical  composition 
 (disjunction of conjunctions) of  elementary hypotheses, in the
composit product model, $M$. 

 The possibilistic nature of the $e$-value measure makes it easy to 
compute the support for disjunctive complex hypotheses. 
 Conjunction of elementary hypotheses require a more sophisticated 
analysis.  
 First we must observe that knowing the $e$-values of the elementary 
 hypotheses is not enough to know the $e$-value of the conjunction;   
 Elementary $e$-values can give only lower and upper bounds to the 
 support for the conjunction.  
 Figure A.6 illustrates these bounds, and also the following results, 
for further details see Borges and Stern (2006). 
 For conjunctive compositions, the models' truth functions, $W^j$, 
are the key element for the required algebraic manipulation, 
as stated in the next result.

 If $H$ is expressed in HDNF or Homogeneous Disjunctive Normal Form,     
 \[
  H \ = \ \bbigvee_{i=1}^q \bbigwedge_{j=1}^{k}  H^{(i,j)} 
  \ , \ \    
  M^{(i,j)} = \{ \Theta^j, H^{(i,j)}, p_0^j, p_n^j, r^j \} \ ,   
 \] 
 \[ 
    M = \{ \Theta, H, p_0, p_n, r \} \ , \ \  
    \Theta = \pprod_{j=1}^k \Theta^j \ , \ \   
    p_n =    \pprod_{j=1}^k p_n^j \ , \ \  
    r =    \pprod_{j=1}^k r^j \ ;   
 \] 
 then the e-value supporting $H$ is   
 \[ 
    \ev(H)  = \ 
    \ev \left( \bbigvee_{i=1}^{q} \bbigwedge_{j=1}^{k} H^{(i,j)} \right) 
     = \ 
    W \left(  \max_{i=1}^{q} 
              \pprod_{j=1}^{k} s^{*(i,j)} \right)  
     = \  
 \] 
 \[
    W \left(  \max_{i=1}^{q} s^{*i} \, \right)  
    = \ 
    \max_{i=1}^{q}  W \left(  s^{*i} \, \right)     
    = \  
    \max_{i=1}^{q} 
       \ev \left( \bbigwedge_{j=1}^{k} H^{(i,j)} \right)  
    = \  
    \max_{i=1}^{q} 
       \ev \left( H^{i} \right) \ ;  
 \] 
 where the cumulative surprise distribution of the composite model, 
 $W(v)$, is given by the Mellin convolution operation,   
 see Springer (1979), defined as  
 \[ 
   W  =  \bigotimes_{1\le j\le k} \; W^j  
   \ \ , \ \ \    
   W^1 \otimes W^2(v) \ = \ \int_0^\infty W^1(v/y)W^2(dy) \ .
 \]

 The probability distribution of the product of two independent positive
random variables is the Mellin convolution of each of their
distributions. 
 From this interpretation, the we immediately see that $\otimes$  
is a commutative and associative operator. 

 Mirroring Wittgenstein, in the FBST context, we can call
the e-value, $\ev(H)$, the cumulative surprise distribution, 
$W(v)$, and the Mellin convolution operation, $\otimes$,  
respectively, truth value, truth function, and truth operation.

 Finally, we observe that, in the extreme case of    
 null-or-full support, that is, when, for 
 $1 \le i \le q$ and $1 \le j \le k$, 
 $s^{*(i,j)}=0 \ \oor \ s^{*(i,j)}= \sh^j$, 
 the evidence values (or, in this context, truth values) of the 
 constituent elementary hypotheses are either 0 or 1, 
 and the conjunction 
 and disjunction composition rules of classical logic hold.

 \subsubsection*{Numerical Aspects} 

 In appendix G we detail an efficient Monte Carlo algorithm for 
 computing $\ev(H;p_n,r)$. 
 In this algorithm, the bulk of the work consists in generating 
 random points in the parameter space, $\theta^j \in \Theta$, 
 and evaluating the surprise function, $s^j=s(\theta^j)$. 
 The Monte Carlo algorithm proceeds updating several accumulators 
 based on the tangential set ``hit indicator'',  
 \[ 
    I^*(\theta^j;p_n,r) = \uno(\theta^j \in \Tb) 
                        = \uno(s(\theta^j) > s^* ) \ . 
 \] 

 In order to compute a $k$-step function approximation of $W(v)$, 
 we only have to split the surprise range interval, 
 $[0,\sh]$ with a vector of $k$ intermediate points, 
 $0<s^1<s^2<\ldots s^h<s^*<s^{h+1}<\ldots s^k<\sh$, 
 and set up a set of vector accumulators based on the 
 vector threshold indicator,  
 $I^k(\theta^j;p_n,r) = \uno(s(\theta^j) > s^k )$. 
 Updating the vector accumulators usualy imposes only a small 
 overhead on the Monte Carlo algorithm.

 Numerical convolutions of step functions can be easily computed 
 with the help of good condensation procedures, 
 see Kaplan and Lin (1987). For alternative approaches to numerical  
 convolution see Springer (1979) and Williamson (1989).     
 In the case of dependent models, the composite truth function 
 can be solved with the help of analytical and numerical copulas, see 
 Cherubini (2004), Mari and Kotz (2001) and Nelsen (2006).

 \mbox{} \vfill \pagebreak

 \centerline{\includegraphics*[height=8.5in, width=6.5in, angle=0]{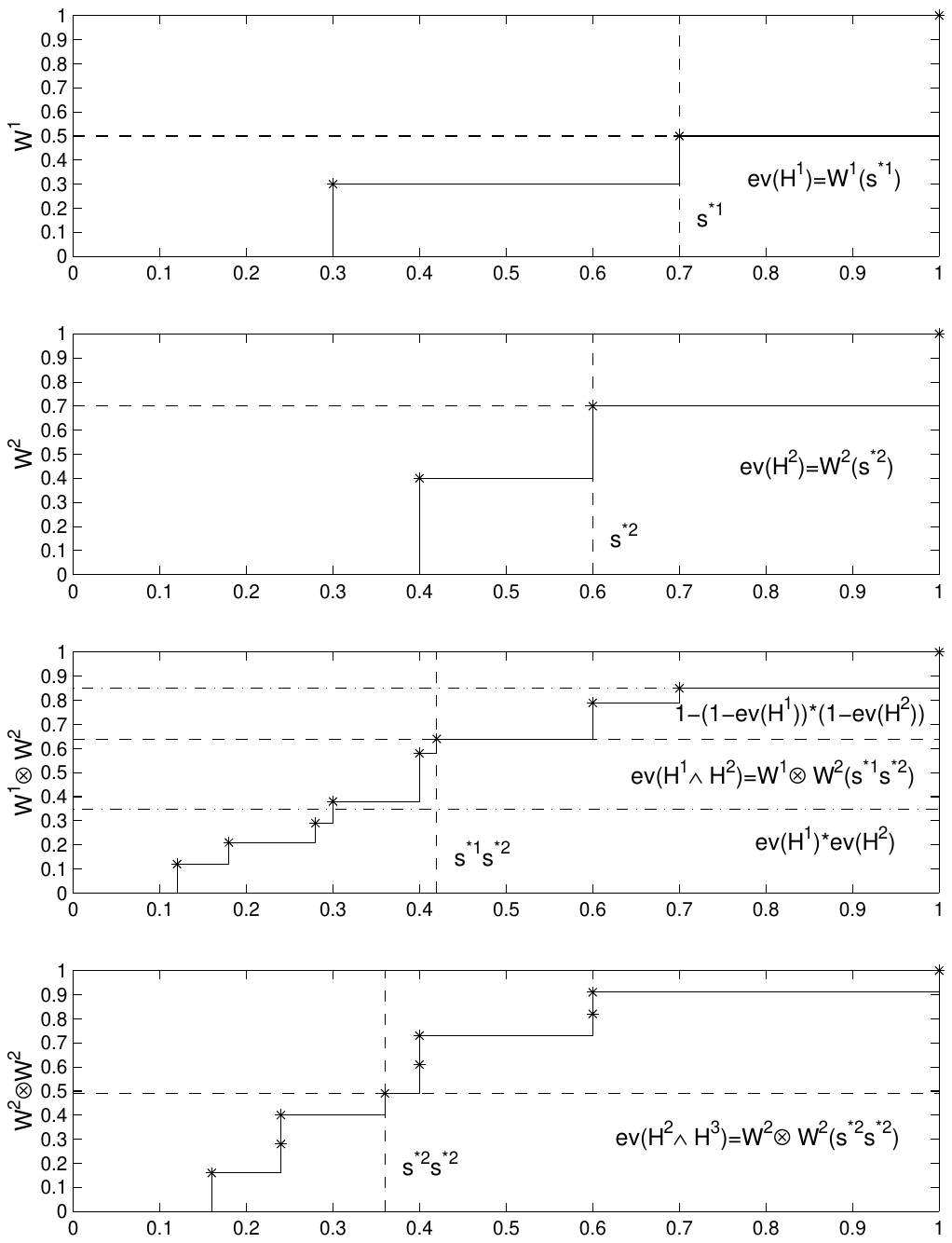}} 
 \vspace{-2.0cm} 
 \begin{center} 
 { 
  Fig A.6. Subplots 1,2: $W^j$, $s^{*j}$, and $\ev(H^j)$, for $j=1,2$; \\ 
  Subplot 3: $W^1\otimes W^2$, $s^{*1}s^{*2}$, $\ev(H^1 \wedge H^2)$ 
             and bounds;  \\ 
  Subplot 4: 
 Structure $M^3$ is an independent replica of $M^2$, \\   
 $\ev(H^1)<\ev(H^2)$, but   
 $\ev(H^1\wedge H^3)>\ev(H^2\wedge H^3)$.    
 } 
 \end{center}  

%% file: CAPBIN.TEX
     
\newcommand{\calX}{\mathcal{X}}
\newcommand{\vsp}{\mbox{\\}}
 

 \chapter{Binomial, Dirichlet, Poisson and Related Distributions}


 This essay has been published as Pereira and Stern (2008). 

 \noindent 
 The matrix notation used in this section is defined in section F.1.

 \section{Introduction and Notation}
 \markboth{APPENDIX B: BINOMIAL, DIRICHLET AND RELATED DISTRIBUTIONS}
  {B.1 \ INTRODUCTION AND NOTATION}

 This essay presents important properties of the distributions used  for
categorical data analysis.  
 Regardless of the population size being known or unknown, or the 
specific observational stopping rule, the Bernoulli Processes generates
the  sampling distributions considered. 
 On the other hand, the Gamma  distribution generates the prior and
posterior  distributions obtained: Gamma, Gamma-Poisson, Dirichlet, and 
Dirichlet-Multinomial.  
 The Poisson Processes as generator of sampling distributions is also 
considered.

 The generation form of the discrete sampling distributions presented in
Section 2 is, in fact, a characterization method of such distributions. 
 If one recalls that all the distribution classes being 
mixed are complete classes and are Blackwell sufficient for the 
Bernoulli processes, the mixing distributions are unique.  
 This characterization method is completely described in Basu and
Pereira  (1983). 

 Section 9 describes the Reny-Aczel characterization of the Poisson
distribution. 
 Although it could be thought as a de Finetti type characterization 
this characterization is based on alternative requirements. 
 While de Finetti chaparcterization is based on a permutable infinite 
0-1 process, Reny-Aczek characterization is based on a homogeneous
Markov process in a finite interval, generating finite discrete Markov
Chains. 
 Using Reny-Aczel characterization, together with Theorem 4, one can
obtain a  characterization of Multinomial distributions.

 Section 7 describes the Dirichlet of Second Kind.  
 In this section we also show how to use a multivariate normal
approximation to the logarithm of a random vector distributed  as
Dirichlet of Second Kind, and a log-normal approximation to a Gamma  
distribution, see Aitchison and Shen (1980). 
 In many examples of the authors' consulting practice these approximations 
proved to be a powerful modeling tool, leading to efficient computational 
procedures.  

 The development of the theory in this essay is self contained, seeking 
a unified treatment of a large variety of problems, including finite and
infinite populations, contingency tables of arbitrary dimension,
deficiently categorized data, logistic regressions, etc. 
 These models also present a way of introducing non parametric solutions. 

 The singular representation adopted is unusual in statistical  texts.
 This singular representation makes it simpler to extend  and
generalize the results and greatly facilitates numerical and
computational implementation.  
 In this essay, corollaries, lemmas, propositions and theorems are
numbered  sequentially.

 We introduce the following notation for observation matrices, 
and respective summation vectors: 
 \[ U=[u^1,u^2,\ldots] \ , \ 
    U^{1 \ate n} = [ u^1,u^{2},\ldots u^n] \ , \ 
    x^{n} = U^{1 \ate n} \uno = \ssum_{j=1}^n u^j \ .  
 \] 
 The tilde accent indicates some form of normalization like,  
 for example, $\widetilde{x}=(1/\uno'x)x$.

 \noindent 
 {\bf Lemma 1:} 
 If $u^1,\ldots u^n$ are i.i.d random vectors,  
 \[ 
 x= U^{1 \ate n} \uno \Rightarrow 
 \E(x) = n\E(u^1) \And 
 \Cov(x) = n\Cov(u^1) \ . 
 \] 
 The first result is trivial. 
 For the second result, we only have to remember the 
 transformation properties of for the expectation and 
 covariance operators by a linear operation on their argument, 
 \[
 E(AY+b) = AE(Y)+b \ \ , \ \  
 \Cov(AY+b)= A\Cov(Y)A' \ , 
 \]  
 and write  
 \begin{eqnarray*} 
 \lefteqn{  \Cov(x) = \Cov(U^{1 \ate n}\uno) }  \\ 
 &=&  
 \Cov \left( (\uno' \kron I) \Vec(U^{1 \ate n}) \right) 
 \ = \ 
 \left( \uno' \kron I \right) 
 \left( I \kron \Cov(u^1) \right)   
 \left( \uno \kron I \right) \\ 
 &=& 
 \left( \uno' \kron \Cov(u^1) \right)    
 \left( \uno \kron I \right) 
 \ = \ n \Cov(u^1) \ . 
 \end{eqnarray*}

 \section{The Bernoulli Process}
 \markboth{APPENDIX B: BINOMIAL, DIRICHLET AND RELATED DISTRIBUTIONS}
  {D.2 \ BERNOULLI PROCESS}

 Let us consider a sequence of random vectors  
 $u^1, u^2, ...$ where, $\forall u^i$  can assume only two 
 values 
 \[ I^1 = \left[ 
     \begin{array}{c}1 \\ 0 \end{array} \right]  \Or 
    I^2 = \left[ 
     \begin{array}{c}0 \\ 1 \end{array} \right]   \mbox{ where } 
    I = \left[ \begin{array}{cc} 1 & 0 \\ 0 & 1 \end{array} \right]    
 \]  
 representing success or failure. 
 That is, $u^i$ can assume the value of any column of the identity 
 matrix, I. 
 We say that $u^i$ is of class $k$, $c(u^i)=k$, 
 iff $u^i=I^k$, $k\in [1,2]$.

 Also assume that (in your opinion), this sequence is  
 exchangeable, that is, if  
 $p=[p(1), p(2), \ldots p(n)]$  
 is a permutation of $[1,2, \ldots n]$, 
 than, $\forall n, p$,  
 \[ 
  \Pr \left(u^1,... u^n\right) = 
  \Pr \left(u^{p(1)},... u^{p(n)}\right) \ . 
 \]  
 Just from this exchangeability constraint, that can be interpreted as 
saying that the index labels are non informative, de Finetti Theorem 
establishes the existence of an unknown vector 
 \[ 
    \theta \in \Theta = 
    \{ \zero \leq \theta =  
    \left[ \begin{array}{cc} \theta_1 \\ \theta_2 \end{array} \right] 
    \leq \uno \g \ \uno'\theta=1 \} 
 \]   
 such that,
 conditionally on $\theta$, 
 $ \ u^1, u^2, \ldots$ are mutually independent, 
 and the conditional probability of  
 $\Pr(u^i=I^k \g \theta)$ is $\theta_k$, i.e.  
 \[ 
    (u^1 \amalg u^2 \amalg \ldots) \g \theta  \Or 
    \coprod_{i=1}^{\infty} u_i \g \theta \ , \And   
    \Pr(u^i=I^k \g \theta) = \theta_k \ .   
  \] 
 Vector $\theta$ is characterized as the limit of proportions 
 \[ 
    \theta = \lim_{n\rightarrow \infty} \frac{1}{n} x^{n} 
    \ \ , \ \ \  
    x^n = U^{1 \ate n} \uno = \ssum_{j=1}^n u^j \ .   
 \] 

 Conditionally on $\theta$ , the sequence  
 $u^1,u^2, \ldots$ receives the name of Bernoulli process.  
 As we shall see, many well known discrete distributions can be 
 obtained from transformations of this process. 

   The expectation and covariance (conditionally on $\theta$)  
 of any vector in the sequence are: 
 \begin{itemize} 
 \item $\E(u^i) = \theta$ , 
 \item $\Cov(u^i) 
       = \E \left( u^i \kron (u^i)' \right) 
       -\E \left( u^i \right) \kron \E \left( (u^i)' \right)
       = \diag(\theta) -\theta \kron \theta'$ .  
 \end{itemize} 

 When the summation domain $1 \ate n$, is understood, 
 we may use the relaxed notation $x$ instead of $x^{n}$. 
 We also define the Delta operator, or  
 ``pointwise power product'' between two vectors of same dimension:  
 Given $\theta$, and $x$, $n\times 1$,   
 \[ 
    \theta\tria x \equiv \prod_{i=1}^n (\theta_i)^{x_i} \ . 
 \]

 A stopping rule, $\delta$, establishes, for every   
 $n=1,2,\ldots$, a decision  of observing (or not) $u^{n+1}$, 
 after the observations $u^1,\ldots u^n$.  

 For a good understanding of this text, it is necessary to have a 
 clear interpretation of conditional expressions like 
 $x^n \g n$ or $x^n_2 \g x^n_1$. 
 In both cases we are referring to a unknown vector, $x^n$, 
 but with a different partial information. 
 In the first case, we know $n$, and therefore we know the sum of 
 components, $x^n_1+x^n_2=n$; however, we know neither component 
 $x^n_1$ nor $x^n_2$. 
 In the second case we only know the first component, of $x^n$, $x^n_1$, 
 and do not know the second component, $x^n_2$, 
 obviously we also do not know the sum, $n=x^n_1+x^n_2$. 
 Just pay attention: We list what we know to the right of the bar and, 
 (unless we have some additional information) everything that can not 
 be deduced from this list is unknown. 
 
 The first distribution we are going to discuss is the Binomial. 
 Let $\delta(n)$ be the stopping rule where $n$ is the pre-established 
number of observations. 
 The (conditional) probability of the observation sequence 
 $U^{1 \ate n}$ is     
 \[
    \Pr(U^{1 \ate n} \g \theta) = \theta\tria x^n  \ . 
  \]
 
 The summation vector, $x^n$,  has Binomial distribution with parameters 
 $n$ and $\theta$, and we write  
 $x^n \g [n,\theta] \sim \Bin(n,\theta)$.  
 When $n$ (or $\delta(n)$) is implicit in the context we may write 
 $x \g \theta$ instead of $x^n \g [n,\theta ]$. 
 The Binomial distribution has the following expression: 
 \[
 \Pr(x^n \g n,\theta)=      
 \left( \begin{array}{c} n \\ x^n \end{array} \right) 
 (\theta \tria x^n)   
 \]
 where  
 \[ 
    \left( \begin{array}{c} n \\ x \end{array} \right) \equiv 
   \frac{\Gamma(n+1)}{\Gamma(x_1+1) \, \Gamma(x_2+1)}       =  
   \frac{n!}{x_1! \, x_2!}  \And n=\uno'x \ .  
 \]
 
 A good exercise for the reader is to check that expectation vector 
and the covariance matrix of $x^n \g [n,\theta]$ 
have the following expressions: 
 \[  
  \E(x^n) = n\theta \ \And \  
  \Cov(x^n) = n \, (\theta \tria \uno) \,  
  \left[ \begin{array}{cc} 1 & -1 \\ -1 & 1 \end{array} \right] \ . 
 \]

 The second distribution we discuss is the Negative Binomial. 
 Let $\delta(x^n_1)$ be the rule establishing to stop at observation 
 $u^{n}$ when obtaining a pre-established number of $x^n_1$ successes.  
 The random variable $x^n_2$, the number of failures he have when 
 we obtain the required $x^n_1$ successes, is called a Negative Binomial 
 with parameters $x^n_1$ and $\theta$.    
 It is not hard to prove that the Negative Binomial distribution 
 $x^n_2 \g [x^n_1 , \theta] \sim \Bne(x^n_1 , \theta)$,  
 has expression, $\forall \ x^n_2 \in \Na$, 
 \[
 \Pr(x^n \g x^n_1 ,\theta)     
 \ = \ 
 \frac{x^n_1}{n} \,    
 \left( \begin{array}{c} n \\ x^n \end{array} \right) 
 ( \theta \tria x^n ) 
 \ = \ \theta_1 
 \Pr\left( ( x^n - I^1 ) \g (n-1), \theta ) \right) \ . 
 \]

 Note that, from the definition this distribution, $x^n_1$ is a positive
integer number. Nevertheless, we can extend the definition  above for
any real positive value $a$, and still obtain a probability function. 
 For this, we use   
 \[ 
  \sum_{j=0}^{\infty}
  \frac{\Gamma(a + j)}{\Gamma(a )j!}(1-\pi)^j=
  \pi^{-a}   
  \ , \ \forall \, a \in [0,\infty[ \And \pi \in ]0,1[ \ . 
 \] 
 The reader is asked to check the last equation, as well as the 
following expressions for the expectation and variance of $x^n_2$:  
 \[  
  \E\left(x^n_2\g x^n_1,\theta \right) = 
 \frac{x^n_1 \theta_2 }{\theta_1 } 
 \ \And \ 
 \Var\left(x^n_2\g x^n_1 , \theta \right) = 
 \frac{x^n_1 \theta_2 }{ (\theta_1)^2 } \ . 
 \] 
 
 In the special case of $\delta(x^n_1=1)$, the Negative Binomial 
distribution is also known as the Geometric distribution with 
parameter $\theta$. 
 If $a$ random variables are independent and identically distributed 
(i.i.d.) as a geometric distribution with parameter $\theta$, 
then the sum of these variables has Negative Binomial distribution 
with parameters $a$ and $\theta$.

 The third distribution studied in this essay is the Hypergeometric. 
 Going back to the original sequence, $u^1,u^2,...$, 
assume that a first observer knows the first $N$ observations, 
while a second observer knows only a subsequence of $n<N$ 
of these observations. 
 Since the original sequence, $u^1,u^2,\ldots$, is exchangeable, 
we can assume, without loss of generality, that the subsequence known to 
the second observer is the subsequence of the first $n$ observations, 
 $u^1,\ldots u^n$ .
 Using de Finetti theorem, we have that 
 $x^n$ and $x^N-x^n = U^{n+1 \ate N}\uno$ 
 are conditionally independent, given $\theta$. 
 That is, $x^n \amalg (x^N -x^n) \g \theta$. 
 Moreover, we can write 
 \[ 
 x^n \g [n,\theta] \sim \Bin(n,\theta) \ , \  
 x^N \g [N,\theta] \sim \Bin(N,\theta) \ , \ \And  
 \] 
 \[ 
 (x^N -x^n) \g [(N-n),\theta] \sim \Bin(N-n,\theta) \ .  
 \] 

 Our goal is to find the distribution function of $x^n \g x^N $.  
 Note that   
 $x^N$ is sufficient for $U^{1 \ate N}$ given $\theta$, and   
 $x^n$ is sufficient for $U^{1 \ate n}$.  
 Moreover 
 $x^n \g [n,x^N]$ has the same distribution of  
 $x^n \g [n,x^N,\theta]$. 
 Using the basic rules of probability calculus and the properties above, 
we have that
 \begin{eqnarray*}
 \lefteqn{ \Pr(x^n\g n,x^N, \theta) } \\ 
   & = &  
   \frac{\Pr(x^n,x^N \g n,N,\theta)}{\Pr(x^N \g n,N,\theta)} 
   \ = \ 
   \frac{\Pr(x^n, (x^N -x^n) \g n,N,\theta)}{\Pr(x^N \g n,N,\theta)} \\  
   & = & 
   \frac{\Pr(x^n \g n,N,\theta) \; 
   \Pr( x^N -x^n \g n,N,\theta)}{\Pr(x^N \g n,N,\theta)} \ . 
 \end{eqnarray*}

 Hence, $x^n \g [n,x^N]$ has distribution function 
 \[ 
   \Pr(x^n \g n,x^N) = 
    \frac { \left( \begin{array}{c}
                   n \\ x^n \end{array}
            \right) \;
            \left( \begin{array}{c}
                    N-n \\ x^N -x^n \end{array} 
            \right) 
          }
          { \left( \begin{array}{c}
                    N \\ x^N \end{array} 
            \right)
          } 
       \] 
       \[  
         \Where 
         \zero \leq x^n \leq x^N \leq N \uno 
         \ , \  \uno'x^n=n \ , \ \uno'x^N=N \ .  
       \]
  This is the vector representation of the Hypergeometric 
probability distribution. 
 \[ 
     x^n \g [n,x^N] \sim \Hip(n,N,x^N) \ .  
 \]

 The reader is asked to check the following expressions for the 
expectation and (conditional) covariance of $x^n \g [n,N,x^N]$, 
and covariance of $u^i$ and $u^j$, $i,j \leq n$:
 \[  
  \E(x^n) = \frac{n}{N}\,x^N \ \And \  
  \Cov(x^n) = \frac{n(N-n)}{(N-1)}\, 
  ( x^N \tria \uno ) \,   
  \left[ \begin{array}{cc} 1 & -1 \\ -1 & 1 \end{array} \right] 
 \] 
 \[ 
  \Cov(u^i,u^j \g x^N) = \frac{1}{(N-1)N^2}\,
  ( x^N \tria \uno ) \,    
  \left[ \begin{array}{cc} -1 & 1 \\ 1 & -1 \end{array} \right] \ . 
 \]

 We finish this section presenting the derivation of the 
Beta-Binomial distribution. 
 Let us assume that the first observer observed $x^n_2$ failures, 
until observing a pre-established number of $x^n_1$ successes. 
 A second observer makes more observations, observing $x^N_2$ failures 
until completing the pre-established number of 
$x^N_1$ successes, $x^n_1 < x^N_1$. 
 
 Since $x^n_1$ and $x^N_1$ are pre-established, we can write 
 \[ 
   x^N_2 \g \theta \sim \Bne(x^N_1 , \theta) \ , \ \ 
   x^n_2 \g \theta \sim \Bne(x^n_1 , \theta) 
 \]
 \[ 
   (x^N_2 -x^n_2) \g \theta \sim    
   \Bne(x^N_1 -x^n_1 , \theta) \ \And  
    \ x^n_2 \amalg (x^N_2 -x^n_2) \g \theta \ . 
 \]             
 As before, our goal is to describe the distribution of 
 $x^n_2 \g [x^n_1, x^N]$. 
 If one notices that $[x^n_1, x^N]$ is sufficient for $[x^n, (x^N -x^n)]$, 
with respect to $\theta$, the problem becomes similar to the Hypergeometric 
case, and one can obtain 
 \[   \Pr (x^n_2 \g x^n_1, x^N )=\frac{x^N_2 ! \;   
      \Gamma(x^N_1)}{\Gamma(x^N_2  +x^N_1 )} \;  
 \frac{\Gamma(x^n_2 +x^n_1)}{x^n_2 ! \; \Gamma(x^n_1)} \;
 \frac{\Gamma(x^N_2 -x^n_2 +x^N_1 -x^n_1)}{(x^N_2 -x^n_2)! \; 
       \Gamma(x^N_1 -x^n_1)} \ , 
 \] 
 \[   
 x^n_2 \in \{0,1,...,x^N_2 \} \ . 
 \]
 This is the distribution function of a random variable called 
 Beta Binomial with parameters  $x^n_1$ and $x^N$. 
 \[
    x^n_2 \g(x^n_1, x^N) \sim \Bbi(x^n_1, x^N) \ . 
 \] 
 The properties of this distribution will be studied in the general case 
of the Dirichlet-Multinomial, in the following sections.

 Generalized categories for $k>2$  can be represented 
by the orthonormal base $I^1, I^2,\ldots I^k$, 
i.e., the columns of the $k$-dimensional identity matrix. 
 The Multinomial and Hypergeometric multivariate distributions, 
presented in the next sections, are distributions derived of this basic
generalization.

 \section{Multinomial Distribution}
 \markboth{APPENDIX B: BINOMIAL, DIRICHLET AND RELATED DISTRIBUTIONS}
  {B.3 \ MULTINOMIAL DISTRIBUTION}

 Let $u^i$, $i= 1,2,\ldots$ be random vectors with possible  
results in the set of columns of the $m$-dimensional identity matrix, 
 $I^k, k\in 1\ate m$.   
 We say that  $u^i$ is of class $k$, $c(u^i)=k$, 
 iff $u^i=I^k$.  

 Let $\theta \in [0,1]^m$ be the vector of probabilities for an observation 
of class $k$ in a $m$-variate Bernoulli process, i.e., 
 \[ 
    \Pr(u^i=I^k \g \theta)=\theta_k \ , \ 
     \zero \leq \theta \leq \uno \ , \ \uno'\theta=1 \ . 
 \]  
 Like in the last section, let $U$  
 \[ 
  U = [u^1,u^2,\ldots ]  \And 
  x^n = U^{1 \ate n} \uno  \ .    
 \]   

 \noindent
 {\bf Definition: } If the knowledge of $\theta$ makes the vectors 
 $u^i$ independent, then the (conditional) distribution of $x^n$ given 
$\theta$ is the Multinomial distribution of order $m$ with 
parameters $n$ and $\theta$, given by 
 \[
 \Pr(x^n \g n,\theta)=
 \left( \begin{array}{c} n \\ x^n \end{array} \right) 
 (\theta \tria x^n)      
 \]
 where 
 \[ 
    \left( \begin{array}{c} n \\ x \end{array} \right) \equiv 
   \frac{\Gamma(n+1)}{\Gamma(x_1+1) \ldots \Gamma(x_m+1)}       =  
   \frac{n!}{x_1! \ldots x_m!}  \And n=\uno'x \ .  
 \]

 We represent the $m$-Multinomial distribution writing 
 \[ 
    x^n \g [n,\theta] \sim \Mno_m(n,\theta) \ . 
 \]
 When $m=2$, we have the binomial case.

 Let us now examine some properties of the Multinomial distribution. 

 \noindent 
 {\bf Lemma 2:} 
  If $x \g \theta \sim \Mno_m(n,\theta)$ 
 then the (conditional) expectation and covariance of $x$ are 
 \[  
    \E(x) = n\theta  \And 
    \Cov(x) = n ( \diag(\theta) - \theta \kron \theta' ) \ . 
 \] 
 
 \noindent 
 {\bf Proof:} 
 Analogous to the binomial case.  

 The next result presents a characterization of the Multinomial 
in terms of the Poisson distribution.  
 
 \noindent 
 {\bf Lemma 3:} Reproductive property of the Poisson distribution.   
 \[ x_i \sim \Poi(\lambda_i) \Rightarrow \ 
    \uno'x \g \lambda \ \sim \Poi( \uno'\lambda ) \ . 
 \]   
 that is, the sum of (independent) Poisson variates is also Poisson.

 \mbox{}

 \noindent 
 {\bf Theorem 4:} 
 Characterization of the Multinomial by the Poisson. \\  
 Let $x= [x_1, ..., x_m]'$ be a vector with independent Poisson 
 distributed components with parameters in the known vector 
 $\lambda=[\lambda_1,...,\lambda_m]'>0$. 
 Let $n$ be a positive integer. 
 Then, given $\lambda$,
 \[ 
    x  \g [n=\uno'x, \lambda]   \ \ 
       \sim \Mno_m(n,\theta) \Where 
       \theta = \frac{1}{\uno'\lambda} \lambda \ . 
 \]

 \noindent
 {\bf Proof:}  
 The joint distribution of $x$, given $\lambda$ is   
 \[
    \Pr(x \g \lambda)  = 
  \prod_{k=1}^m \frac{e^{-\lambda_k} \lambda_i^{x_k}}{x_k!} \ . 
 \]
 Using the Poisson reproductive property, 
 \begin{eqnarray*} 
 \lefteqn{ \Pr\left( x \g \uno'x=n , \lambda \right) } \\ 
  & = & 
  \frac{ \Pr\left( \uno'x=n \wedge x \g \lambda \right) }
       { \Pr\left( \uno'x=n \g \lambda \right) }  
  \ = \ 
   \delta( n=\uno'x ) 
   \frac{ \Pr(x \g \lambda)}{ \Pr( \uno'x=n \g \lambda) } \ .  
 \end{eqnarray*}

 The following results state important properties of the Multinomial 
distribution. 
 The proof of these properties is simple, using the characterization of 
the Multinomial by the Poisson, and the Poisson reproductive property.

 \mbox{}

 \noindent 
 {\bf Theorem 5:} Multinomial Class Partition \\  
 Let $1\ate m$ be the index domain for the classes of a order $m$ 
Multinomial distribution. 
 Let $\olP$ be a partition matrix breaking the $m$-classes 
into $s$-super-classes.  
 Let $x \sim \Mno_m(n,\theta)$,  
 then  $y= \olP x \sim \Mno_s(n, \olP \theta)$. 
   
 \mbox{}

 \noindent
 {\bf Theorem 6:} Multinomial Conditioning on the Partial Sum. \\ 
 If $x \sim \Mno_m(n,\theta)$, then the distribution of part of the vector 
 $x$ conditioned on its sum has Multinomial distribution, 
 having as parameter the corresponding part of the original 
 (normalized) parameters. 
 In more detail, conditioning on the $t$ first components, we have: 
 \[ 
    x_{1 \ate t}  \g ( \uno'x_{1 \ate t}=j ) \sim 
    \Mno_t \left( j , \frac{1}{\uno'\theta_{1 \ate t}} 
                \theta_{1 \ate t} \right)  
    \Where 0 \leq j \leq n  \ . 
 \]  
      
 \mbox{}

 \noindent 
 {\bf Theorem 7:} Multinomial--Binomial Decomposition. \\  
 Using the last two theorems, if  
 $x \sim \Mno_m(n,\theta)$, 
 \begin{eqnarray*} 
  \Pr(x\g n, \theta)  
  &=& 
   \sum_{j=0}^n 
   \Pr \left( x_{1 \ate t} \g j,    
        \frac{1}{\uno'\theta_{1 \ate t}}\theta_{1 \ate t} \right) \\  
  & & \  \ \ \ \  
   \Pr \left( x_{t+1 \ate m} \g (n-j), 
         \frac{1}{\uno'\theta_{t+1 \ate m}}\theta_{t+1 \ate m} \right) \\  
  & & \ \ \ \ \ 
   \Pr \left(                    
     \left[ \begin{array}{c} j \\ (n-j)  \end{array} \right]  \g   n , 
     \left[ \begin{array}{c} 
               \uno'\theta_{1 \ate t} \\  \uno'\theta_{t+1 \ate m}  
     \end{array} \right]   \right)  \ . 
  \end{eqnarray*} 

 Analogously, we could write the Multinomial-Trinomial decomposition 
 for a three-partition of the class indices in three super-classes. 
 More generally, we could also write the 
 $m$-nomial-$s$-nomial decomposition for the partition of the 
 $m$ class indices into $s$ super-classes.

 \section{Multivariate Hypergeometric Distribution}
 \markboth{APPENDIX B: BINOMIAL, DIRICHLET AND RELATED DISTRIBUTIONS}
  {B.4 \ MULTIVARIATE HYPERGEOMETRIC}

 In the first section we have shown how an Hypergeometric variate  
can be generated from a Bernoulli process. The natural generalization 
of this result is obtained considering a Multinomial process. 
 As in the last section, 
 we say that $u^i$ is of class $k$, $c(u^i)=k$, iff $u^i=I^k$.  

 We take a sample of size $n$ from a finite population of size $N (>n)$, 
that is partitioned into $m$ classes. 
 The population frequencies (number of elements in each category) 
are represented by $[\psi_1,\ldots \psi_m]$, hence $N=\uno'\psi$. 
 Based on the sample, we want to make an inference on $\psi$. 
 $x_k$ \'{e} is the sample frequency of class $k$. 

 One way of describing this problem is to consider an urn with $N$ balls 
of $m$ different colors, indexed by $1,\ldots m$. 
 $\psi_k$ is the number of balls of color $k$. 
 Assume that the $N$ balls are separated into two smaller boxes, so that 
 box 1 has $n$ balls and box 2 has the remaining $N-n$ balls.  
  The statistician can observe the composition of box 1, represented by 
 vector $x$ of sample frequencies. 
 The quantity of interest for the statistician is the vector 
 $\psi-x$ representing the composition of box 2. 

 As in the bivariate case, we assume that $U^{1 \ate N}$ is a finite 
sub-sequence in an exchangeable process and, therefore, 
any sub-sequence extracted from $U^{1 \ate N}$ has the same 
distribution of  $U^{1 \ate n}$.
 Hence, $x= U^{1 \ate n}\uno$ has the same distribution of the 
frequency vector for a sample of size $n$.

 As in the bivariate case, our objective is to find the distribution of 
 $x \g \psi$.  
 Again, using de Finetti theorem, there is a vector 
 $\zero \leq \theta \leq \uno \ , \ \uno'\theta=1$,  such that 
 $\coprod_{j=0}^{N}u^j \g \theta$ \ and \  
 $\Pr\left( c(u^j)=k \right) = \theta_k$ .   

 \noindent 
 {\bf Theorem 8:}
 As in the Multinomial case, the following results follow: 
 \begin{itemize}
 \item $\psi \g \theta \sim \Mno_m(N,\theta)$  ; 
 \item $x \g \theta \sim \Mno_m(n,\theta)$  ; 
 \item $(\psi-x) \g \theta \sim \Mno_m((N-n),\theta)$  ; 
 \item $(\psi-x)\amalg x \g \theta$  . 
 \end{itemize}

 Using the results of the last section and following the same steps 
as in the $\Hip_2$ case in the first section, we obtain the following 
expression for $m$-variate Hypergeometric distribution,  
 $x^n \g [n,N,\psi] \sim \Hip_m(n,N,\psi)$ :   
  \[ 
   \Pr(x^n \g n, \psi) = 
    \frac { \left( \begin{array}{c}
                   n \\ x^n \end{array}
            \right) \;
            \left( \begin{array}{c}
                    N-n \\ \psi -x^n \end{array} 
            \right) 
          }
          { \left( \begin{array}{c}
                    N \\ \psi \end{array} 
            \right)
          } 
       \] 
       \[  
         \Where 
         \zero \leq x^n \leq \psi \leq N \uno 
         \ , \  \uno'x^n=n \ , \ \uno'\psi=N \ .  
       \]
  This is the vector representation of the Hypergeometric 
probability distribution. 
 \[ 
     x^n \g [n,x^N] \sim \Hip(n,N,x^N) \ .  
 \]

  Alternatively, we can write the more usual formula,   
 \[ 
 \Pr(x \g \psi) =  \frac{ 
 \left(\begin{array}{c} \psi_1 \\ x_1 \end{array} \right)
 \left(\begin{array}{c} \psi_2 \\ x_2 \end{array} \right)
 \cdots 
 \left(\begin{array}{c} \psi_m \\ x_m \end{array} \right)
 }{\left(\begin{array}{c} N \\ n \end{array} \right)} \ . 
 \]

 \noindent 
 {\bf Theorem 9:} The expectation and covariance of a random vector 
with Hypergeometric distribution, 
 $x \sim \Hip_m(n,N,\psi)$, are: 
 \[ 
    \E(x)   =   n \widetilde{\psi} \ , \ 
    \Cov(x) =   n \frac{N-n}{N-1} 
               \left(  \diag(\widetilde{\psi}) 
            -\widetilde{\psi} \kron \widetilde{\psi}' \right) 
    \Where  \widetilde{\psi} = \frac{1}{N}\psi \ . 
 \]     
 
 \noindent 
 {\bf Proof:} Use that 
 \begin{eqnarray*}  
  \Cov(x^n) 
    &=& n \Cov(u^1) +n(n-1) \Cov(u^1,u^2) \\   
  \Cov(u^1) 
    &=& \E \left( u^1 \kron (u^1)' \right) -\E(u^1) \kron \E(u^1)' 
   \ = \ \diag(\widetilde{\psi}) -\widetilde{\psi} \kron \widetilde{\psi}'  \\  
  \Cov(u^1,u^2) 
    &=& \E \left( u^1 \kron (u^2)' \right) -\E(u^1) \kron \E(u^2)' \ . 
 \end{eqnarray*} 
  The second term of the last two equations are equal, and the 
  first term of the last equation is 
 \[ 
    \E \left( u^1_i u^2_j \right) = \left\{ 
    \begin{array}{l} 
    \frac{\psi_i}{N} \frac{\psi_i -1}{N-1} \If i= j \\  
    \frac{\psi_i}{N} \frac{\psi_j}{N-1}    \If i\neq j   
    \end{array}                     \right. 
 \] 
  Algebraic manipulation yields the result. 

 Note that, as in the order 2 case, the diagonal elements of 
 $\Cov(u^1)$ are positive, while the diagonal elements of  
 $\Cov(u^1,u^2)$ are negative. 
 In the off diagonal elements, the signs are reversed.

 \section{Dirichlet Distribution}
 \markboth{APPENDIX B: BINOMIAL, DIRICHLET AND RELATED DISTRIBUTIONS}
  {B.5 \ DIRICHLET DISTRIBUTION}

 In the second section we presented the multinomial distribution, 
 $\Mno_m(n,\theta)$. In this section we present the Dirichlet 
distribution for the parameter $\theta$. 
 Let us first recall the univariate Poisson and Gamma distributions. 

 A random variable has Gamma distribution, 
 $x \g [\aalpha,\bbeta] \sim G(\aalpha,\bbeta), \aalpha,\bbeta>0$, 
 if its distribution is continuous with density 
 \[ 
    f(x\g \aalpha,\bbeta) = 
    \frac{\bbeta^\aalpha}{\Gamma(\aalpha)} x^{\aalpha-1} \exp(-\bbeta x) 
    \ , \ x>0 \ . 
 \] 
 The expectation and variance of this variate are 
 \[ 
    E(x) = \frac{\aalpha}{\bbeta} \And \Var(x) = 
    \frac{\aalpha}{\bbeta^2} \ . 
 \] 

 \noindent 
 {\bf Lemma 10:} Reproductive property for the Gamma distribution. \\ 
 If $n$ independent random variables  
 $x_i \g \aalpha_i,\bbeta \sim G(\aalpha_i,\bbeta)$,  
 then 
 \[ 
   \uno'x \sim G( \uno'\aalpha,\bbeta) \ . 
 \]

 \noindent  
 {\bf Lemma 11:} The Gamma distribution is conjugate to the Poisson 
distribution.

 \noindent 
 {\bf Proof:} \\ 
 If $y \g \lambda \sim \Poi(\lambda)$ and $\lambda$ has prior  
 $\lambda \g \aalpha,\bbeta \sim G(\aalpha,\bbeta)$, then 
 \begin{eqnarray*}
 \lefteqn{ f(\lambda \g y,\aalpha,\bbeta) 
  \ \propto \ L(\lambda \g y) f(\lambda) } \\   
 &=& 
 \exp(-\lambda)\frac{\lambda^y}{y!}  \   
 \frac{\bbeta^\aalpha}{\Gamma(\aalpha)} \lambda^{\aalpha-1} 
  \exp(-\bbeta\lambda) \ \propto \  
 \lambda^{y+\aalpha-1} exp(-(\bbeta+1)\lambda ) \ . 
 \end{eqnarray*} 
 That is, the posterior distribution of $\lambda$ is  
 Gamma with parameters $[\aalpha+y,\bbeta+1]$.

 \mbox{} 
 
 \noindent 
 {\bf Definition:} Dirichlet distribution. \\ 
 A random vector 
 \[ 
    y \in \mathcal{S}_{m-1} \equiv 
    \{ y \in \Re^m \g \zero \leq y \leq \uno \wedge \uno'y=1 \} 
 \]  
 has Dirichlet distribution of order $m$ with positive 
 $\aalpha\in \Re^m$ if its density is  
 \[ 
    \Pr(y \g \aalpha ) = \frac{y \tria (\aalpha -\uno)}{ B(\aalpha) } \ . 
 \]

 Note that $\mathcal{S}_{m-1}$, the $m-1$ dimensional Simplex, 
 is the region of $\Re^m$ subject to the ``constraint'', 
 $\uno'y=1$. Hence, a point in the Simplex has only 
 $m-1$ ``degrees of freedom''. 
 In this sense we say that the Direchlet distribution has a ``singular'' 
 representation. 
 It is possible to give a non-singular representation to the 
 distribution  $[y_1,\ldots y_{m-1}]'$, known as the Multivariate Beta 
 distribution, but at the cost of obtaining a convoluted algebraic 
 formulation that also loses the natural geometric interpretation of 
 the singular form. 

 The normalization factor for the Dirichlet distribution is
 \[ 
     B(\aalpha) \equiv \int_{ y\in \mathcal{S}_{m-1} }  
              \left( y \tria (\aalpha-1) \right) dy  \ . 
 \] 
         
 \mbox{} 
 
 \noindent 
 {\bf Lemma 12:} Beta function.  \\ 
 The normalization factor for the Dirichlet distribution 
 defined above is the Beta function, defined as 
 \[  
    B(\aalpha) = \frac{ \prod_{k=1}^{m} \Gamma(\aalpha_k)} 
                     { \Gamma( \uno'\aalpha ) } \ . 
 \] 
 The proof is given at the end of this section. 

 \mbox{}

 \noindent
 {\bf Theorem 13:} Dirichlet as Conjugate of the Multinomial: \\ 
 If $\theta \sim \Dir_m(\aalpha)$ and 
 $x\g \theta \sim \Mno_m(n,\theta)$ then  
 \[
    \theta\g x \sim \Dir_m(\aalpha+x) \ . 
 \]

 \mbox{}

 \noindent
 {\bf Proof:} 

 We only have to remember that the Multinomial likelihood 
is proportional to $\theta \tria x$, and that a Dirichlet prior is 
proportional to $\theta \tria (\aalpha-\uno)$. 
 Hence, the posterior is proportional to $\theta \tria (x+\aalpha-1)$. 
 At the other hand, $B(\aalpha+x)$ is the normalization factor, i.e., 
equal to the integral on $\theta$ of $\theta \tria (x+\aalpha-1)$,  
and so we have a Dirichlet density function, as defined above.

 \mbox{}

 \noindent
 {\bf Lemma 14:} Dirichlet Moments. \\
 If $\theta \sim \Dir_m(\aalpha)$ and $p\in \Na^m$, then  
 \[ \E \left( \theta \tria p \right) \ = \ 
    \frac{B(\aalpha+p)}{B(\aalpha)} \ . 
 \] 

 \noindent 
 {\bf Proof:} \\ 
 \[  
   \int_\Theta (\theta \tria p) f(\theta \g \aalpha) d\theta  =    
   \frac{1}{B(\aalpha)} \, 
   \int_\Theta \left( \theta \tria p \right) 
            \left( \theta \tria (\aalpha-1) \right) d\theta = 
 \] 
 \[ 
  \frac{1}{B(\aalpha)} \,  
  \int_\Theta \left( \theta \tria (\aalpha+p-1) \right) d\theta
  \ = \ \frac{B(\aalpha+p)}{B(\aalpha)} \ . 
 \]   

 Choosing the exponents, $p$, appropriately, we have 
 
 \noindent
 {\bf Corollary 15:} If $\theta \sim \Dir_m(\aalpha)$ , 
 then 
 \begin{eqnarray*}
 \E(\theta) &=& \widetilde{\aalpha} \equiv \frac{1}{\uno'\aalpha}\aalpha  \\  
 \Cov(\theta) &=& \frac{1}{\uno'\aalpha +1} \left(   
  \diag(\widetilde{\aalpha}) 
  - \widetilde{\aalpha} \kron \widetilde{\aalpha}' \right) \ . 
 \end{eqnarray*}

 \noindent 
 {\bf Theorem 16:} Characterization of the Dirichlet by the Gamma: \\ 
 Let the components of the random vector $x\in \Re^m$ be 
 independent variables with distribution   
 $G(\aalpha_k,\bbeta)$. 
 Then, the normalized vector 
 \[ 
    y = \frac{1}{\uno'x} x \sim \Dir_m(\aalpha) 
    \ , \  \uno'x \sim \Gam(\uno'\aalpha)  
    \And  y \amalg \uno'x   \ . 
 \]

 \mbox{} 

 \noindent 
 {\bf Proof:} \\ 
  
 Consider the normalization, 
 \[ 
    y   \ = \  \frac{1}{t} x  \ , \ 
    t \ = \  \uno'x  \ , \ 
    x   \ = \  t y \ ,  
 \]    
 as a transformation of variables. 
 Note that one of the new variables, say \\ 
 $y_m \equiv t(1 -y_1 \ldots -y_{m-1})$, becomes redundant.  

 The Jacobian matrix of this transformation is     
  \[ 
    J \ = \ 
  \frac{ \del ( x_1,x_2,\ldots x_{m-1},x_m ) }
       { \del ( y_1,y_2,\ldots y_{m-1},t ) } 
  \ = \ \left[ \begin{array}{ccccc} 
    t     &  0     & \cdots &  0      & y_1              \\ 
    0     &  t     & \cdots &  0      & y_2              \\ 
   \vdots & \vdots & \ddots & \vdots  & \vdots           \\ 
    0     &   0    & \cdots &  t      & y_{m-1}          \\ 
   -t     &  -t    & \cdots & -t      & 1-y_1\cdots -y_{m-1}    
  \end{array} \right] \ . 
  \] 
  By elementary operations (see appendix F) that add all rows 
 to the last one, we obtain the LU factorization the Jacobian matrix, 
 $J=LU$, where  
 \[ 
    L = \left[ \begin{array}{ccccc} 
    1      & 0      & \cdots & 0      & 0      \\ 
    0      & 1      & \cdots & 0      & 0      \\  
    \vdots & \vdots & \ddots & \vdots & \vdots \\ 
    0      & 0      & \cdots & 1      & 0      \\ 
    -1     & -1     & \cdots & -1     & 1      
    \end{array} \right] 
    \ \And \ 
    U = \left[ \begin{array}{ccccc} 
    t      & 0      & \cdots & 0      & y_1      \\ 
    0      & t      & \cdots & 0      & y_2      \\  
    \vdots & \vdots & \ddots & \vdots & \vdots   \\ 
    0      & 0      & \cdots & t      & y_{m-1}  \\ 
    0      & 0      & \cdots & 0      & 1      
    \end{array} \right] \ . 
 \] 
 A triangular matrix determinant is equal to the product of the 
 elements in its main diagonal, hence   
 $|J|= |L|\,|U|= 1\, t^{m-1}$. 

 At the other hand, the joint distribution of $x$ is 
 \[ 
   f(x) \ = \ \prod_{k=1}^{m} \Gam(x_k \g \aalpha_k,\bbeta) \ = \   
   \prod_{k=1}^m \frac{\bbeta^{\aalpha_k}}{\Gamma(\aalpha_k)}  
   e^{-\bbeta x_k} (x_k)^{\aalpha_k-1}  \ . 
 \]
 and the joint distribution in the new system of coordinates is  
 \begin{eqnarray*}
 \lefteqn{ g([y,t]) = |J|\, f\left( x^{-1}([y,t]) \right) } \\
 & = &  
  t^{m-1} \, \prod_{k=1}^{m} 
  \frac{\bbeta^{\aalpha_k}}{\Gamma(\aalpha_k)} 
   e^{-\bbeta x_k} (x_k)^{\aalpha_k-1} 
 \ = \ 
   t^{m-1} 
   \prod_{k=1}^m \frac{\bbeta^{\aalpha_k}}{\Gamma(\aalpha_k)}  
   e^{-\bbeta t y_k} (t y_k)^{\aalpha_k-1}   \\ 
 & = & 
   \left( \prod_{k=1}^m \frac{(y_k)^{\aalpha_k-1}}
   {\Gamma(\aalpha_k)} \right) \, 
    \bbeta^{\uno'\aalpha} e^{-\bbeta t} t^{\uno'\aalpha-m} t^{m-1} 
 \ = \  
    \left( \prod_{k=1}^m \frac{(y_k)^{\aalpha_k-1}}
    {\Gamma(\aalpha_k)} \right) \, 
    \bbeta^{\uno'\aalpha} e^{-\bbeta t} t^{\uno'\aalpha-1} \ .  
 \end{eqnarray*}    
 Hence, the marginal distribution of  
 $y=[y_1,\ldots y_k]'$ is 
 \begin{eqnarray*} 
  \lefteqn{ g(y) = \int_{t=0}^\infty g([y,t]) dt } \\ 
  & = &  
  \left( \prod_{k=1}^m \frac{(y_k)^{\aalpha_k-1}}
   {\Gamma(\aalpha_k)} \right) \,  
  \int_{t=0}^\infty \bbeta^{\uno'\aalpha} 
      e^{-\bbeta t} t^{\uno'\aalpha-1} dt  \\ 
  & = & 
  \left( \prod_{k=1}^m \frac{(y_k)^{\aalpha_k-1}}{\Gamma(\aalpha_k)} \right) \, 
  \Gamma(\uno'\aalpha) 
  \ = \  
  \frac{ y \tria (\aalpha-1) }{ B(\aalpha) } \ . 
 \end{eqnarray*} 
 In the last passage, we have replaced the integral by the normalization 
factor of a Gamma density, $\Gam(\uno'\aalpha,\bbeta)$.  
 Hence, we obtain a density proportional to $y \tria (\aalpha-1)$, 
 i.e., a Dirichlet, Q.E.D. 

 In the last passage we also obtain the Dirichlet normalization factor, 
prooving the Beta function lemma.

 \mbox{} 

 \noindent 
 {\bf Lemma 17:} Bipartition of Indices for the Dirichlet. \\ 
 Let $1 \ate t$, $t+1\ate m$ be a bipartition of the 
 class index domain, $1\ate m$, of an order $m$ Dirichlet, 
 in two super-classes. 
 Let $y \sim \Dir_m(\aalpha)$, and   
 \[ 
    z^1 = \frac{1}{\uno'y_{1 \ate t}} y_{1 \ate t} \ , \ 
    z^2 = \frac{1}{\uno'y_{t+1 \ate m}} y_{t+1 \ate m} \ , \ 
    w = \left[ \begin{array}{c} 
        \uno'y_{1 \ate t}  \\  
    \uno'y_{t+1 \ate m} \end{array} \right] \ .  
 \]   
 We than have,  \ 
 $z^1 \amalg z^2 \amalg w$ \ and  
 \[   
    z^1 \sim \Dir_{t}(\aalpha_{1 \ate t})  \ , \    
    z^2 \sim \Dir_{m-t}(\aalpha_{t+1 \ate m})                \And 
    w \sim \Dir_2\left( \left[ \begin{array}{c} 
       \uno'\aalpha_{1 \ate t}  \\  \uno'\aalpha_{t+1 \ate m} 
      \end{array} \right] \right) \ . 
 \]

 \mbox{} 

 \noindent 
 {\bf Proof:} \\ 
 From the Dirichlet characterization by the Gamma we can imagine 
 that the vector $y$ is built by normalizing of a vector $x$, 
 as follows,  
 \[ 
            y = \frac{1}{\uno'x} x  
    \ , \   x_k \sim \Gam(\aalpha_k ,\bbeta) 
    \ , \  \coprod_{k=1}^{m} x_k  \ .  
 \] 
 Considering isolatetly each one of the super-classes, 
 we build the vectors $z^1$ and $z^2$ that are distributed as   
   \begin{eqnarray*}   
    z^1 
    & = & 
    \frac{1}{\uno'y_{1 \ate t}} y_{1 \ate t} 
    \ = \ 
    \frac{1}{\uno'x_{1 \ate t}} x_{1 \ate t} 
    \ \sim \ 
    \Dir_{t}(\aalpha_{1 \ate t})   \\ 
    z^2 
    & = & 
    \frac{1}{\uno'y_{t+1 \ate m}} y_{t+1 \ate m} 
    \ = \ 
    \frac{1}{\uno'x_{t+1 \ate m}} x_{t+1 \ate m} 
    \ \sim \ 
    \Dir_{m-t}(\aalpha_{t+1 \ate m}) \ .   
    \end{eqnarray*} 
  $z^1 \amalg z^2$, that are in turn independent of the 
  partial sums  
 \[ 
  \uno'x_{1 \ate t}    \sim \Gam(\uno'\aalpha_{1 \ate t} ,\bbeta)  \And 
  \uno'x_{t+1 \ate m}  \sim \Gam(\uno'\aalpha_{t+1 \ate m} ,\bbeta) \ . 
 \] 

 Using again the theorem characterizing the Dirichlet by the Gamma 
 distribution for these two Gamma variates, we obtain the result,
 Q.E.D. 
 
 We can generalize this result for any partition of the set of classes, 
 as follows. 
 If $y \sim \Dir_m(\aalpha)$ and $\olP$ \'{e} is a $s$-partition 
 of the $m$ classes,  the intra and extra super-class distributions 
 are independent Dirichlets, as follows 
 \begin{eqnarray*} 
    z^r 
    & = & \frac{1}{ {\olP_r} y }\, {_rP} y 
    \ \sim \ \Dir_{{\olP_r}1}
    \left( {_rP} \aalpha \right) \\ 
    w 
    & = &  \olP y 
    \ \sim \ \Dir_s (\olP \aalpha) \ .    
  \end{eqnarray*}

 \section{Dirichlet-Multinomial}
 \markboth{APPENDIX B: BINOMIAL, DIRICHLET AND RELATED DISTRIBUTIONS}
  {B.6 \ DIRICHLET-MULTINOMIAL}

 We say that a random vector $x\in \Na^n \g \uno'x=n$ 
 has Dirichlet-Multinomial (DM) distribution with parameters 
 $n$ and $\aalpha \in \Re^m$, iff 
  \[  
     \Pr(x \g n,\aalpha) 
     \ = \ 
      \frac{B(\aalpha+x)}{B(\aalpha)} 
      \left( \begin{array}{c} n \\ x \end{array} \right)  
     \ = \ 
     \frac{B(\aalpha+x)}{B(\aalpha)\,B(x)} \, \frac{1}{ x \tria \uno } \ . 
  \] 

  \mbox{} 

   \noindent 
  {\bf Theorem 18:} Characterization of the DM as a 
Dirichlet mixture of Multinomials. \\  
  \[ \mbox{Se}\  
     \theta \sim \Dir_m(\aalpha) \And   
     x \g \theta \sim \Mno(n,\theta)  \Then 
     x \g [n,\aalpha] \sim \DM_m(n,\aalpha) \ . 
  \]

  \noindent 
  {\bf Proof:} 
 
 The joint distribution of $\theta,x$ is  
 proportional to $\theta \tria (\aalpha+x-1)$, which integrated on 
 $\theta$ is $B(\aalpha+x)$. 
 Hence, multiplying by the joint distribution constants, we have the 
 marginal for $x$, Q.E.D. 
 Therefore, we have also proved that the function $\DM$ 
 is normalized, that is 
 \begin{eqnarray*} 
  \lefteqn{ 
    \Pr(x) 
    \ = \   
    \int_{\theta \in \mathcal{S}_{m-1}} 
      \left( \begin{array}{c} n \\ x \end{array} \right) 
      ( \theta \tria x ) \, 
      \frac{1}{B(\aalpha)} \theta \tria (\aalpha-\uno) \, d\theta     } \\  
   & = &  
    \frac{1}{B(\aalpha)} 
    \left( \begin{array}{c} n \\ x \end{array} \right) 
    \int_{\theta \in \mathcal{S}_{m-1}} 
      \left( \theta \tria (x+\aalpha-\uno) \right) \, d\theta 
    \ = \ 
        \frac{B(x+\aalpha)}{B(\aalpha)} 
    \left( \begin{array}{c} n \\ x \end{array} \right) \ . 
  \end{eqnarray*}

  \noindent 
 {\bf Theorem 19:} Characterization of the DM by $m$ Negative Binomials. \\  
 Let $\aalpha \in \Na^m_+$, and $x \in \Na_m$, be a vector whose 
 components are independent random variables,  
 $\aalpha_k \sim \Bne(\aalpha_k,\theta)$. Then  
 \[ 
    x \g [ \uno'x=n , \aalpha ] \sim \DM_m(n,\aalpha) \ . 
 \]      

 \noindent 
 {\bf Proof:}
 \[ 
  \Pr(x \g \theta, \aalpha) 
  \ = \ 
  \prod_{k=1}^{m} 
  \left( \begin{array}{c} \aalpha_k + x_k -1 \\ x_k \end{array} \right)  
   \theta^{\aalpha_k} (1-\theta)^{x_k}  
  \] 
  \[  
   \Pr(\uno'x \g \theta, \aalpha) 
   \ = \ 
   \left( \begin{array}{c} \uno'\aalpha + \uno'x -1 \\ \uno'x 
     \end{array} \right)  
    \theta^{\uno'\aalpha} (1-\theta)^{\uno'\aalpha}  \ . 
   \] 
   Then,  
  \[ 
      \Pr(x \g \uno'x=n , \theta , \aalpha) 
      \ = \ 
      \frac{ \Pr(x \g \aalpha , \theta ) }{ \Pr( \uno'x=n \g \theta ) } 
      \ = \ 
   \frac{    \prod_{k=1}^{m} 
       \left( \begin{array}{c} \aalpha_k + x_k -1 \\ x_k \end{array} \right) }  
   { \left( \begin{array}{c} \uno'\aalpha +\uno'x -1 \\ \uno'x 
     \end{array} \right) } \ . 
    \] 
   Hence, 
  \begin{eqnarray*} 
   \lefteqn{       
    \Pr( x \g \uno'x=n, \theta, \aalpha) \ = \ \Pr(x \g \uno'x=n , \aalpha ) } \\ 
   & = & 
    \prod_{k=1}^m 
    \frac{ \Gamma(\aalpha_k +x_k) }{ x! \Gamma(\aalpha_k) }  \ \slash \  
    \frac{ \Gamma(\uno'\aalpha +n) }{ \Gamma(\uno'\aalpha) n! }   
   \ = \  
   \frac{ B(\aalpha+x) }{ B(\aalpha) } 
   \left( \begin{array}{c} n \\ x \end{array} \right) \ . 
   \end{eqnarray*}

   \mbox{} 

   \noindent 
   {\bf Theorem 20:} The DM as Pseudo-Conjugate for the 
                   Hypergeometric\\  
   \[ 
    \mbox{Se\ } 
     x \sim \Hip_m(n,N,\psi) \And \psi \sim \DM_m(N,\aalpha) \Then 
     (\psi -x) \g x \sim \DM_m( N-n, \aalpha ) \ . 
   \] 
 
   \noindent 
   {\bf Proof:}
   Using the properties of the Hypergeometric already presented, 
   we have the independence relation, 
   $(\psi-x) \amalg x \g \theta$. 
   We can therefore use the Multinomial sample $x \g \theta$ for  
   updating the prior and obtain the posterior 
   \[ 
       \theta \g x \sim \Dir_m (\aalpha+x) \ . 
   \] 
    Hence, the distribution of the non sampled pat of the population, 
    $\psi-x$, given the sample $x$, is a mixture of 
    $(\psi-x) \theta$ buy the posterior for $\theta$. 
    By the characterization of the DM as a mixture of Multinomials 
    by a Dirichlet, the theorem follows, i.e.,  
    \[ 
      \left. 
      \begin{array}{c} 
      (\psi-x) \g [\theta, x] \sim (\psi-x) \g \theta \sim  
      \Mno_m(N-n,\theta)    \\  
      \theta \g x \sim \Dir_m(\aalpha+x) 
      \end{array} \right\} 
       \Rightarrow 
    \] 
    \[ 
       \Rightarrow 
       (\psi-x) \g x \sim \Dir_m(N-n,\aalpha+x) \ . 
    \] 
     
  \noindent 
  {\bf Lemma 21:} DM Expectation and Covariance. \\ 
  If \ $x \sim \DM_m(n,\aalpha)$ \ then  
  \begin{eqnarray*}    
   \E(x) & = & n \widetilde{\aalpha} \equiv \frac{1}{\uno'\aalpha} \aalpha \\   
   \Cov(x) & = & \frac{n(n +\uno'\aalpha)}{\uno'\aalpha +1}         \left(   
    \diag(\widetilde{\aalpha}) 
    - \widetilde{\aalpha} \kron \widetilde{\aalpha}' \right) \ . 
  \end{eqnarray*} 
   
  \noindent 
  {\bf Proof:} 
  \begin{eqnarray*} 
   \E(x) 
   & = &  \E_\theta\left( \E_x (x\g \theta) \right)  
   \ = \  \E_\theta (n \theta) \ = \ n \widetilde{\aalpha} \\ 
  \E(x \kron x')  
   & = &  
    \E_\theta \left( \E_x (x \kron x' \g \theta ) \right) \\ 
   & = & 
    \E_\theta \left(  \E(x\g \theta) \kron \E(x\g \theta)' 
                     +\Cov(x\g \theta ) \right) \\ 
   & = & 
    \E_\theta \left( 
         n\left( \diag(\theta) -\theta \kron \theta' \right) 
        +n^2 \theta \kron \theta' \right)  \\ 
   & = &  
     n \E_\theta \left( \diag(\theta) \right) 
                       +n(n-1) \E_\theta (\theta \kron \theta') \\ 
   & = &  
     n \diag(\widetilde{\aalpha}) 
    +n(n-1) \left( \E(\theta) \kron \E(\theta)' +\Cov(\theta) \right) \\      
   & = &  
     n \diag(\widetilde{\aalpha}) +n(n-1) 
       \left( \widetilde{\aalpha} \kron \widetilde{\aalpha}'  
       +\frac{1}{\uno'\aalpha +1} 
     \left( \diag(\widetilde{\aalpha}) 
       -\widetilde{\aalpha} \kron \widetilde{\aalpha}' \right)          
                                    \right) \\      
   & = &          
     n \diag(\widetilde{\aalpha})  +n(n-1) \left(  
        \frac{1}{\uno'\aalpha +1} \diag(\widetilde{\aalpha})
       +\frac{\uno'\aalpha}{\uno'\aalpha +1} 
         \widetilde{\aalpha} \kron \widetilde{\aalpha}'          
                                    \right) \\      
  \Cov(x) 
   & = &  
    \E(x \kron x') -\E(x) \kron \E(x)'    
    \ = \ 
    \E(x \kron x') -n^2 \widetilde{\aalpha} \kron \widetilde{\aalpha}' \\    
    & = & 
     \left( n +\frac{n(n-1)}{\uno'\aalpha+1} \right) 
      \diag(\widetilde{\aalpha}) 
    +\left( n(n-1)\frac{\uno'\aalpha}{\uno'\aalpha+1} -n^2 \right) 
      \widetilde{\aalpha} \kron \widetilde{\aalpha}' \\ 
     & = & 
    \frac{n(n +\uno'\aalpha)}{\uno'\aalpha +1}         \left(   
    \diag(\widetilde{\aalpha}) 
     - \widetilde{\aalpha} \kron \widetilde{\aalpha}' \right)
     \ \ \mbox{Q.E.D.} 
   \end{eqnarray*} 

 \noindent
 {\bf Theorem 22:} DM Class Bipartition \\
 Let $1\ate t$, $t+1\ate m$ a bipartition of the index domain 
 for the classes of an order $m$ DM,  $1\ate m$, in two super-classes. 
 Then, the following conditions (i) to (iii) are equivalent to 
 condition (iv): 
 \begin{description} 
 \item{i: \ \ \ \ \ }  
   $x_{1:t} \amalg x_{t+1:m} \g n_1= \uno'x_{1:t}$ ; 
 \item{ii-1: \ }  
   $x_{1:t} \g n_1=\uno'x_{1;t} \sim \DM_t(n_1, \aalpha_{1:t})$ ; 
 \item{ii-2: \ }  
   $x_{t+1:m} \g n_2=\uno'x_{t+1:m} \sim \DM_{m-t}(n_2, \aalpha_{t+1:m})$ ;  
 \item{iii: \ \ }  
   $\left[ \begin{array}{c} n_1 \\ n_2 \end{array} \right] 
         \sim \DM_2\left( n, 
           \left[ \begin{array}{c} 
              \uno'\aalpha_{1:t} \\ \uno'\aalpha_{t+1:m} 
           \end{array} \right] \right)$ ; 
 \item{iv: \ \ \ }   $x  \sim \DM_m(n,\aalpha)$ . 
 \end{description} 

 \noindent 
 {\bf Proof:} 
 We only have to show that the joint distribution can be 
factored in this form. By the DM characterization as a mixture, 
we can write it as Dirichlet mixture of Multinomials. 
 By the bipartition theorems, we can factor both, the 
Multinomials and the  Dirichlet, so the theorem follows.

 \section{Dirichlet of the Second Kind}
 \markboth{APPENDIX B: BINOMIAL, DIRICHLET AND RELATED DISTRIBUTIONS}
  {B.7 \ DIRICHLET OF THE SECOND KIND}

 Consider $y \sim \Dir_{m+1}(\aalpha)$. The vector 
 $z= (1/y_{m+1}) y_{1\ate m}$ 
 has Dirichlet of the Second Kind (D2K) distribution.   
 
 \noindent 
 {\bf Theorem 23:} Characterization of D2K by the Gamma distribution. \\  
 Using the characterization of the Dirichlet by the Gamma, 
 we can write the D2K variate as a function of $m+1$ 
 independent Gamma variates, 
 \[ 
    z_{1\ate m} \sim (1/x_{m+1}) x_{1\ate m}  
    \ \ \mbox{where} \ \ 
    x_k \sim Ga(a_k,b)  \ .  
 \]

 Similar to what we did for the Dirichlet (of the first kind), 
 we can write the D2K distribution and its moments as: 
 \[ 
    f(z \g a) = \frac{ z \tria (a_{1\ate m} -1) } 
                     { (1+\uno'z)^{\uno'a} \ B(a) } \ , 
 \]  
 \[ 
   E(z) = e = (1/a_{m+1}) a_{1\ate m} \ ,   
 \] 
 \[ 
   \Cov(z) = \frac{1}{a_{m+1}-2} \left( \diag(e) + e \kron e' \right) \ . 
 \]   

 The logarithm of a Gamma variate is well approximated by a Normal 
variate, see  Aitchison \& Shen (1980).  
 This approximation is the key to several efficient computational 
procedures, and motivates the computation of the first two moments 
of the log-D2K distribution.
 For that, we use the Digamma, $\psi(\ )$,  
 and Trigamma function, $\psi'(\ )$, defined as:   
 \[ 
  \psi(a) = \frac{d}{da} \ln \Gamma(a) = \frac{\Gamma'(a)}{\Gamma(a)} 
  \  \ , \ \ 
  \psi'(a) = \frac{d}{da} \psi(a) \ . 
 \] 
 
 \noindent 
 {\bf Lemma 24:} 
 The expectation and covariance of a log-D2K variate are: 
 \[ 
   E(\log(z)) = \psi(a_{1\ate m}) -\psi(a_{m+1}) \uno  \ , 
 \] 
 \[  
   \Cov(\log(z)) = \diag \left( \psi'(a_{1\ate m}) \right)  
                   +\psi'(a_{m+1}) \uno \kron \uno' \ . 
 \] 

 \noindent     
 {\bf Proof:} 
 Consider a Gamma variate, $x\sim G(a,1)$ : 
 \[ 
   1 = \int_0^\infty f(x) dx = \int_0^\infty  
    \frac{1}{\Gamma(a)} x^{a-1} \exp(-x) dx  \ . 
 \] 
 Taking the derivative with respect to parameter $a$, we have   
 \[ 
  0 = \int_0^\infty \ln(x) x^{a-1} \frac{\exp(-x)}{\Gamma(a)} dx 
      -\frac{\Gamma'(a)}{\Gamma^2(a)}\Gamma(a) = 
  E(\ln(x)) -\psi(a) \ . 
 \] 
 Taking the derivative with respect to parameter $a$ a second time, 
 \begin{eqnarray*} 
   \psi'(a) &=& \frac{d}{da}E(\ln(x)) \ = \ \frac{d}{da}\int_0^\infty  
    \frac{\ln(x)}{\Gamma(a)} x^{a-1} \exp(-x) dx   \\ 
  &=& \int_0^\infty \ln(x)^2 x^{a-1} \frac{\exp(-x)}{\Gamma(a)} dx 
     - \frac{\Gamma'(a)}{\Gamma(a)} E(\ln(x)) \\ 
  &=& E(\ln(x)^2) -E(\ln(x))^2 = \Var(\ln(x)) \ . 
 \end{eqnarray*} 
 The lemma follows from the D2K characterization by the Gamma.

 \section{Examples}
 \markboth{APPENDIX B: BINOMIAL, DIRICHLET AND RELATED DISTRIBUTIONS}
  {B.8 \ EXAMPLES}

 {\bf Example 1:}  Let $A$, $B$ be two attributes, each one of them 
present or absent in the elements of a population. Then each element of
this population  can be classified in exactly one of $2^2 = 4$
categories 
 \begin{center}
 \begin{tabular}{cccc}
 A & B & $k$ & $I^k$ \\
 \hline
 present & present & 1 & $[1,0,0,0]'$ \\
 present & absent  & 2 & $[0,1,0,0]'$ \\
 absent & present  & 3 & $[0,0,1,0]'$ \\
 absent & absent   & 4 & $[0,0,0,1]'$ \\
 \end{tabular}
 \end{center}
 According to the notation above, we can write 
 $x \g n, \theta \sim \Mno_4(n,\theta)$.

 If $\theta=[0.35,\;0.20,\;0.30,\;0.15]$ and $n=10$, then 
 \[ 
 \Pr(x^{10} \g n,\theta)=
 \left( \begin{array}{c} 10 \\ x^{10} \end{array} \right) 
 (\theta \tria x^{10})  \ . 
 \]
 Hence, in order to compute the probability of $x=[1,2,3,4]'$ given 
$\theta$, we use the expression above, obtaining  
 \[ 
  \Pr \left( 
 \left[ \begin{array}{c} 1 \\ 2 \\ 3 \\ 4 \end{array} \right]  \g 
 \left[ \begin{array}{c} 0.35 \\ 0.20 \\ 0.30 \\ 0.15 \end{array} \right] 
 \right) = 0.000888  \ . 
 \]

 \noindent 
 {\bf Example 2:} If $X\g\theta \sim \Mno_3(10,\theta)$, 
 $\theta=[0.20,\;0.30,\;0.15]$, 
 one can conclude, using the result above, that 
 \[
   \E(X)=(2,3,1.5) \ , 
 \]
 while the covariance matrix is 
 \[\Sigma=\left[
 \begin{array}{ccc}
 1.6 & -0.6 & -0.3 \\
 -0.6 & 2.1 & -0.45 \\
 -0.3 & -0.45 & 1.28
 \end{array}
 \right] \ . 
 \]

\mbox{}

 \noindent
 {\bf Example 3:} Assume that $X \g \theta \sim \Mno_3(10,\theta)$, 
 with $\theta = [0.20,0.30,0.15]$, as in Example 2. 
 Let us take $A_0 = \{0,1\}$, $A_1 = \{2,3\}$. Then,  
 \[
   \sum_{A_1} X_i\g\theta = X_2 + X_3 \g \theta \sim 
   \Mno_1(10, \theta_2 + \theta_3) \ , 
 \]
 or 
 \[ 
    X_2 + X_3 \g \theta \sim \Mno_1(10, 0.45) \ . 
 \]
 Analogously, 
 \begin{eqnarray*}
 X_0 + X_1 \g \theta & \sim & \Mno_1(10,0.55) \ , \\
 X_1 + X_3 \g \theta & \sim & \Mno_1(10,0.35) \ , \\
 X_2 \g \theta & \sim & \Mno_1(10,0.30) \ .
 \end{eqnarray*}

 Note that, in general, if $X\g\theta \sim \Mno_k(n,\theta)$ then  
 $X_i\g\theta \sim \Mno_1(n,\theta_i)$, $i=1,...,k$.

 \noindent
 {\bf Example 4:} 3x3 Contingency Tables. \\
 Assume that $X\g\theta \sim \Mno_8(n,\theta)$, 
 as in a 3x3 Contingency Tables: 
 \begin{center}
 \begin{tabular}{|c|c|c|c}
 \cline{1-3}
 $x_{11}$ & $x_{12}$ & $x_{13}$ & $x_{1\bullet}$ \\
 \cline{1-3}

 $x_{21}$ & $x_{22}$ & $x_{23}$ & $x_{2\bullet}$ \\
 \cline{1-3}
 $x_{31}$ & $x_{32}$ & $x_{33}$ & $x_{3\bullet}$ \\
 \cline{1-3}
 \multicolumn{1}{c}{$x_{\bullet 1}$} & 
 \multicolumn{1}{c}{$x_{\bullet 2}$} & 
 \multicolumn{1}{c}{$x_{\bullet 3}$} & 
 \multicolumn{1}{c}{$n$} \\
 \end{tabular}
 \end{center}

 \noindent
 Applying Theorem 5 we get   
 \[(X_{1\bullet}, X_{2\bullet})\g\theta \sim \Mno_2(n,\theta'), 
 \theta'=(\theta_{1\bullet},\theta_{2\bullet}), \theta'_0 = \theta_{3} \ . 
 \]
 This result tell us that  
 \[
   (X_{i1}, X_{i2}, X_{i3})\g \theta \sim \Mno_3(n,\theta'_i) \ , 
 \]
 with 
 \[
   \theta'_i=(\theta_{i1},\theta_{i2}, \theta_{i3}) \ , \ 
   \theta'_{0i} = 1-\theta_{i\bullet} \ , \ \ i=1,2,3 \ . 
 \]
 We can now apply Theorem 6 to obtain the probability
 distribution of each row of the contingency table, conditioned on its
 sum, or conditioned on the sum of the other rows.  
 We have  
 \[
   (X_{i1},X_{i2})\g x_{i\bullet} \; , 
   \theta \sim \Mno_2(x_{i\bullet}, \theta'_i) 
 \]
 with 
 \[
  \theta'_i = \frac{(\theta_{il},\theta_{i2})}{\theta_{i\bullet}} \ , \ 
  \theta'_{0i} = \frac{\theta_{i3}}{\theta_{i\bullet}} \ . 
 \]

 \mbox{}
 
 The next result expresses the distribution of $X\g\theta$ in  
 term of the conditional distributions, of each row of the table, 
 in its sum, and in term of the distribution of these sums. 

\mbox{}

 \noindent
 {\bf Proposition 25:} If $X\g\theta \sim \Mno_{r^2 - 1}(n,\theta)$, as 
in an $r \times r$, contingency table, then $P(X\g\theta)$
can be written as 

 \[ 
   P(X\g\theta) = 
  \left[ \prod_{i=1}^r P(X_{i1},...,X_{i,r-1} \g x_{i\bullet} , \theta)\right] 
    P(X_{1\bullet},...,X_{r-1\bullet} \g \theta) \ .  
 \]

 \mbox{} 

 \noindent
 {\bf Proof:} We have:
 \begin{eqnarray*}
  P(X\g\theta) & = & n! \prod_{i=1}^{r} \frac{\theta_i^{x_i}}{x_i!} 
   = n! \ \frac{\theta_{11}^{x_{11}} \ ... \ 
   \theta_{rr}^{x_{rr}}}{x_{11}! \  ... \ x_{rr}!} \\
   & = & \left[ \prod_{i=1}^r \frac{x_{i\bullet}!}{x_{i1}! \ ... \ x_{ir}!} 
    \left( \frac{\theta_{i1}}{\theta_{i\bullet}} \right)^{x_{i1}} ... \
    \left( \frac{\theta_{ir}}{\theta_{i\bullet}} \right)^{x_{ir}} \right] 
         \  \frac{n!}{x_{i\bullet}! \ ... \ x_{r\bullet}!} \ 
        \theta_{1\bullet}^{x_{1\bullet}} \ ... \ 
    \theta_{r\bullet}^{x_{r\bullet}} \ . 
 \end{eqnarray*}
 From Theorems 5 and 6, as in the last example, we recognize
each  of the first $r$ factors above as  the probabilities of each row
in the  table , conditioned on its sum, and recognize the last factor as
the  joint probability distribution of sum of these $r$ rows. 

 \mbox{}

 \noindent
 {\bf Corollary 26:} If $X\g\theta \sim \Mno_{r^2 - 1}(n,\theta)$, 
as in Theorems 5 and 6, then 
 \[
  P(X \g x_{1\bullet},...,x_{r-1\bullet},\theta) = 
  \prod_{i=1}^r P(X_{i1},...,X_{i,r-1} \g x_{i\bullet}, \theta) 
 \]
 and, knowing $\theta, x_{1\bullet},... ,x_{r-1\bullet}$,
 \[
    (X_{11},...,X_{1,r-1})\amalg ... \amalg(X_{r1},...,X_{r,r-1}) \ . 
 \]

 \mbox{}

 \noindent
 {\bf Proof:} Since 
 \[ 
  P(X\g\theta) = P(X\g x_{1\bullet},...,x_{r-1\bullet},\theta) 
  P(X_{1\bullet}, X_{2\bullet},..., X_{r-1\bullet}\g  \theta) \ , 
 \]
 from Theorems 5 and 6 we get the proposed equality. 

 \mbox{}

 The following result will be used next to express Theorem 7    
 as a canonical representation for $P(X\g \theta)$.

 \mbox{}

 \noindent
 {\bf Proposition 27:} If $X\g \theta \sim \Mno_{r^2 - 1}(n,\theta)$,
as in Proposition, then a transformation 
 \[ 
    T : (\theta_{11},...,\theta_{1r},...,\theta_{r1},...,
   \theta_{r,r-1})\rightarrow 
   (\lambda_{11},...,\lambda_{1,r-1},...,\lambda_{r1},...,\lambda_{r,r-1},
    \eta_1,...,\eta_{r-1})
 \]
 given by 
 \[ \begin{array}{ccccc}
 \lambda_{11} = \frac{\theta_{11}}{\theta_{1\bullet}} & , & ... & , & 
 \lambda_{1,r-1} = \frac{\theta_{1,r-1}}{\theta_{1\bullet}} \\
 \vdots & & & & \\
 \lambda_{r1} = \frac{\theta_{r1}}{\theta_{r\bullet}} & , & ... & , & 
 \lambda_{r,r-1} = \frac{\theta_{r,r-1}}{\theta_{r\bullet}}
 \end{array} \]
 \[ \eta_1 = \theta_{1\bullet}, \ \eta_2 = \theta_{2\bullet}, ..., 
   \eta_{r-1} = \theta_{(r-1)\bullet} \]
 is a onto transformation defined in 
 $\{0<\theta_{11}+...+\theta_{r,r-1}<1 \; ; \ \ 
 0<\theta{ij}<1\}$ over the unitary cube of dimension $r^2-1$. 
 Moreover, the Jacobian of this transformation, $t$, is 
 \[
   J = \eta^{r-1} \; \eta_1^{r-1} \; ... \; 
   \eta_{r-1}^{r-1} \; (1-\eta_1-...-\eta_{r-1})^{r-1} \ . 
 \]

 \mbox{}

 The proof is left as an exercise. 

 \mbox{}

 \noindent
 {\bf Example 5:} Let us examine the case of a $2 \times 2$ 
 contingency table: 
 \begin{center}
 \begin{tabular}{|c|c|c}
 \cline{1-2}
 $x_{11}$ & $x_{12}$ &  \\
 \cline{1-2}
 $x_{21}$ & $x_{22}$ &  \\
 \cline{1-2}
 \multicolumn{2}{c}{ } & \multicolumn{1}{c}{$n$} 
 \end{tabular}
 \hspace{0.5cm}
 \begin{tabular}{|c|c|c}
 \cline{1-2}
 $\theta_{11}$ & $\theta_{12}$ &  \\
 \cline{1-2}
 $\theta_{21}$ & $\theta_{22}$ &  \\
 \cline{1-2}
 \multicolumn{2}{c}{ } & \multicolumn{1}{c}{1} 
 \end{tabular}
 \end{center}

 \noindent
 In order to obtain the canonical representation of $P(X\g \theta)$
 we use the transformation  $T$ in the case $r=2$:
 \begin{eqnarray*} 
     \lambda_{11} & = & \frac{\theta_{11}}{\theta_{11}+\theta_{12}} \ , \\
     \lambda_{21} & = & \frac{\theta_{11}}{\theta_{21}+\theta_{22}} \ , \\
     \eta_1 & = & \theta_{11}+\theta_{12} \ ,
 \end{eqnarray*} 
 hence, 
 \begin{eqnarray*} 
 \lefteqn{  P(X \g  \theta) = }  \\ 
 &=&  
  \left(\begin{array}{c} x_{1\bullet} \\ x_{11} 
   \end{array} \right) \lambda_{11}^{x_{11}} (1 - \lambda_{11})^{x_{12}} 
  \left(\begin{array}{c} x_{2\bullet} \\ x_{21} \end{array} \right) 
   \lambda_{21}^{x_{21}} (1 - \lambda_{21})^{x_{22}} 
  \left(\begin{array}{c} n \\ x_{1\bullet} \end{array} \right) 
    \eta_1^{x_{1\bullet}} (1 - \eta_1)^{x_{2\bullet}} \ , \\ 
   & \  \  \  \  \  &  
    0 < \theta_{11} < 1  \ , \ \ 0 < \theta_{21} < 1 \ , \ \ 
    0 < \eta_1 < 1 \ . 
  \end{eqnarray*} 
 %


 \section{Functional Characterizations}
 \markboth{APPENDIX B: BINOMIAL, DIRICHLET AND RELATED DISTRIBUTIONS}
  {B.9 \ FUNCTIONAL CHARACTERIZATIONS}

 The objective of this section is to derive the general form of a 
 homogeneous Markov random process. 
 Theorem 28, by Reny and Aczel, states that such a process is 
 described by a mixture of Poisson distributions. 
 Our presentation follows Aczel (1966, sec. 2.1 and 2.3) and 
 Janossy, Reny and Aczel (1950). 
 It follows from the characterization of the Multinomial by the 
 Poisson distribution given in theorem 4, that 
 Reny-Aczel characterization of a homogeneous and local time point 
 process is analogous to de Finetti characterization of 
 an infinite exchangeable 0-1 process as a mixture of  Bernoulli 
 distributions, 
 see for example Feller (1971, v.2, ch.VII, sec. 4). 
 %

 \subsection*{Cauchy's Functional Equations}      

 Cauchy's additive functional equation has the form 
 \[ 
    f(x+y) = f(x) +f(y) \ . 
  \]   
 The following argument from Cauchy (1821) shows that a continuous 
 solution of this functional equation must have the form 
 \[ 
    f(x) = c x \ . 
 \]

 Repeating the sum of the same argument, $x$, $n$ times, we must have  
 $f(nx) = n f(x)$.   
 If $x=(m/n)t$, then $nx=mt$ and 
 \[ 
   n f(x) = f(nx) = f(mt) = m f(t)  \ \mbox{hence} , 
 \] 
 \[ 
    f\left( \frac{m}{n} t \right) = \frac{m}{n} f(t) \ , 
 \] 
 taking $c=f(1)$, and $x=m/n$, it follows that $f(x)=c x$, 
 over the rationals,  $x\in \Qe$. 
 From the continuity condition for $f(x)$, the last result must also  
 be valid over the reals, $x\in \Re$. Q.E.D.

 Cauchy's multiplicative functional equation has the form 
 \[ 
    f(x+y) = f(x) f(y)  \ , \ \ \forall x, y > 0 \ , f(x) \geq 0 \ .  
 \]   
 The trivial solution of this equation is $f(x)\equiv 0$. 
 Assuming $f(x)>0$, we take the logarithm, reducing the   
 multiplicative equation to the additive equation,  
 \[ 
    \ln f( x_y) = \ln f(x) +\ln f(y) \ , \ \ \mbox{hence} , 
 \] 
 \[ 
    \ln f(x) = c x   \ , \mbox{or} \ \ 
     f(x) = \exp( cx ) \ . 
 \]

 \subsection*{Homogeneous Discrete Markov Processes} 

 We seek the general form of a homogeneous discrete Markov process. 
 Let $w_k(t)$, for $t\geq 0$, be the probability of occurrence of  
 exactly $k$ events. Let us also assume the following hypotheses: 

 Time Locality: If $t_1 \leq t_2 \leq t_3 \leq t_4$ then, 
 the number of events in $[t_1,t_2[$ is independents of  
 the number of events in $[t_3,t_4[$. 

 Time Homogeneity: The distribution for the number of events 
 occurring in $[t_1,t_2[$ depends only on the interval length, 
 $t=t_2 -t_1$. 

 From time locality and homogeneity, 
 we can decompose the occurrence of no (zero) 
 events in $[0,t+u[$ as , 
 \[ 
    w_0(t+u) = w_0(t)w_0(u) \ . 
 \] 
 Hence, $w_0(t)$ must obey Cauchy's functional equation, and 
 \[ 
    w_0(t) = \exp(ct) = \exp(-\lambda t) \ . 
 \] 
 Since $w_0(t)$ is a probability distribution, 
 $w_0(t)\leq 1$, and $\lambda > 0$. 

 Hence, $v(t)= 1-w_0(t)= 1-\exp(-\lambda t)$, 
 the probability of one or more events occurring before $t>0$, 
 must be the familiar exponential distribution. 

 For $k\geq 1$ occurrences before $t+u$, 
 the general decomposition relation is  
 \[ 
    w_n(t+u)= \sum_{k=0}^{n} w_k(t) w_{n-k}(u) \ . 
 \]

 {\bf Theorem 28:} (Reny-Aczel)
 The general (non trivial) solution of this 
 this system of functional equations has the form: 
 \[ 
    w_k(t)=  e^{-\lambda t} 
  \sum_{<r,k>} 
  \prod_{j=1}^{k} \frac{\left(c_j t\right)^{r_j}}{r_j !} 
  \ , \ \ 
  \lambda = \sum_{j=1}^{\infty} c_j \ .   
 \] 
 where the index set $<r,k,n>$ is defined as 
 \[ 
      <r,k,n> = 
    \{ r_1, r_2, \ldots r_k \g r_1+2r_2\ldots +kr_k=n \}  \ . 
 \]   
 and $<r,k>$ is a shorthand for $<r,k,k>$.

 {\bf Proof.} By induction: 
 The theorem is true for $k=0$. 
 Let us assume, as induction hypothesis, that it is true to $k<n$. 
 The last equation in the recursive system is 
 \[ 
    w_n(t+u)= \sum_{k=0}^n  w_k(t) w_{n-k}(u) = 
 \] 
 \[
    w_n(t) e^{-\lambda u}  +w_n(u) e^{-\lambda t} 
    +e^{-\lambda(t+u)} \;  
    \sum_{k=1}^{n-1} \; \sum_{<r,k>} \; \sum_{<s,n-k>} \;  
    \prod_{i=1}^k \frac{(c_i t)^{r_i}}{r_i !} \;        
    \prod_{j=1}^k \frac{(c_j u)^{s_j}}{s_j !} \ . 
 \]        

 Defining 
 \[ 
    f_n(t) = e^{\lambda t} w_n(t) 
            -\sum_{<r,n-1,n>} \; \prod_{j=1}^{n-1} 
             \frac{(c_j t)^{r_j}}{r_j !} \ ,  
 \] 
 the recursive equation takes the form 
 \[ 
   f_n(t+u)= f_n(t) +f_n(u) \ , 
 \] 
 and can be solved as a general Cauchy's equation, that is, 
 \[ 
    f_n(t)= c_n t \ .    
 \] 
 From the last equation and the definition of $f_n(t)$, 
 we get the expression of $w_n(t)$ as in theorem 28. 
 The constant $\lambda$ is chosen so that the distribution 
 is normalized.

  The general solution given by theorem 28 represents a composition 
 (mixture) of Poisson processes, where an event in the $j$-the process 
 in the composition corresponds to the simultaneous occurrence of 
 $j$ single events in the original homogeneous Markov process.   
 If we impose the following rarity condition, the general solution is 
 reduced to a mixture of ordinary Poisson processes.  
  
 Rarity Condition: The probability that an event occurs in a short time 
 at least once is approximately equal to the probability that it occurs 
 exactly once, that is, 
 the probability of simultaneous occurrences is zero.

 \section{Final Remarks}
 \markboth{APPENDIX B: BINOMIAL, DIRICHLET AND RELATED DISTRIBUTIONS}
  {B.10 \ FINAL REMARKS}

 This work is in memory of Professor D Basu who was the supervisor of
the first author PhD dissertation, the starting point for the research
in Bayesian analysis of categorical data presented here.  A long list of
papers follows Basu and Pereira (1982).  We have chosen a few that we
recommend for additional reading:  Albert (1985), Gunel (1984), Irony,
Pereira and Tiwari (2000), Paulino and Pereira (1992, 1995) and Walker
(1996).  To make the analysis more realistic, extensions and mixtures of
Dirichlet also were considered. For instance see Albert and Gupta (1983),
Carlson (1977), Dickey (1983), Dickey, Jiang and Kadane (1987), and
Jiang, Kadane and Dickey (1992).

 Usually the more complex distributions are used to realistic represent
situations for which the strong properties of Dirichlet seems to be not
realistic.  For instance, in a 2 x 2 contingency table, the first line
to be conditional independent of the second line given the marginal
seems to be unrealistic in some situations.  Mixtures of Dirichlet in
some cases take care of the situation as shown by Albert and Gupta
(1983).  

 The properties presented here are also important in non-parametric
Bayesian statistics in order to understand the Dirichlet process for the
competitive risk survival problem.  See for instance Salinas-Torres,
Pereira and Tiwari (1997, 2002).  
 In order to be historically correct we cannot forget the important book
of Wilks, published in 1962, where one can find the definition of
Dirichlet distribution. 

  The material presented in this essay adopts a singular representation 
 for several distributions, as in Pereira and Stern (2005). 
  This representation  is unusual in the statistical literature, 
 but the singular representation makes it simpler to extend  and
 generalize the results and greatly facilitates numerical and
 computational implementations.

 We end this essay presenting the Reny-Aczel characterization of 
 the Poisson mixture. 
 This result can be interpreted as an alternative to de Finetti
characterization theorem introduced in Finetti (1937). Using the
characterization of binomial distributions by Poisson processes
conditional arguments, as given by Theorem 4, and Blackwell (minimal)
sufficiency properties discussed in Basu and Pereira (1983), Section 9
leads in fact to a De Finetti characterization for Binomial
distributions. Also, if one recall the indifference principle (Mendel,
1989) the finite version of Finetti argument can simply be obtained. 
See also Irony and Pereira (1994) for the motivation of these arguments. 
The consideration of Section 9 could be viewed as a very simple
formulation of the binomial distribution finite characterization.

%% file: CAPMIX.TEX
 

 \chapter{Model Miscellanea} 

 {\flushright

 
 {\it 
  ``Das Werdende, das ewig wirkt und lebt, \\
  Umfass euch mit der Liebe holden Schranken, \\
  Und was in schwankender Erscheinung schwebt, \\ 
  Befestiget mit dauernden Gedanken!'' 
  }
 
   The becoming, which  forever works and lives, \\ 
  Holds you in love's gracious bonds, \\ 
  And what fluctuates in apparent oscillations, \\ 
  Fix it in place with enduring thoughts!  
 
 
   Johann Wolfgang von Goethe (1749-1832), \\ 
   The Lord, in Faust, prologue in heaven.   

 \mbox{} 

 {\it
  ``Randomness and order do not contradict each \\  
  other; more or less both may be true at once. \\ 
  The randomness controls the world and due to \\ 
  this in the world there is order and law, which \\ 
  can be expressed in measures of random events \\ 
  that follow the laws of probability theory.'' 
  }


  Alfr\'{e}d R\'{e}nyi (1921 - 1970).

 }

 \mbox{}

 This appendix collects the material in some slide presentations on a
miscellanea of  statistical models used during the curse to illustrate
several  aspects of the FBST use and implementation. 
 This appendix is not intended to be a self sufficient reading material, 
but rather a guide or further study.     
 Section 1, on contingency table models, is (I hope)  fully supplemented
by the  material on the Multinomial-Dirichlet distribution presented in 
appendix B. 
 These models are of great practical importance, and also relatively
simple to implement and easy interpret. These characteristics  make  
them ideal for the several statistical ``experiments'' required in the 
home works. 
  Section 2, on a Wibull model, should require only minor additional 
reading, for further details see Barlow and Prochan (1981) and 
Ironi et al. (2002). 
 This model highlights the importance of being able to incorporate 
expert opinion as prior information. 

 Sections 3 to 5, presenting several models based on the 
Normal-Wishard distribution, may require extensive additional readings. 
 Some epistemological aspects of these models are discussed in chapters 
4 and 5. The material in these sections is presented for completude, 
but its reading is optional, and only recommended for those students with 
a degree in statistics or equivalent knowledge. 
  Of course, it is also possible to combine Normal-Wishad and 
 Multinomial-Dirichlet models, in the form of mixture models,  
 see section 6 and Lauretto and Stern (2005).  
 Section 7 presents an overview of the REAL classification tree
algorithm, for further details see Lauretto et al. (1998).

 \section{Contingency Table Models} 

 \subsection*{Homogeneity test in $2 \times 2$ contingency table} 
 
 This model is useful in many applications, like comparison of 
two communities with relation to a disease incidence, consumer behavior, 
electoral preference, etc.    
 Two samples are taken from two binomial populations, and the objective 
is to test whether the success ratios are equal. Let $x$ and $y$ be the 
number of successes of two independent binomial experiments of sample
sizes  $m$ and $n$, respectively. 
 The posterior density for this multinomial model is, 
 $$f(\theta \mid x,y,n,m) \propto 
    \theta_1^{x} \theta_2^{n-x} \theta_3^{y} \theta_4^{m-y}$$    
 The parameter space and the null hypothesis set are:  
  $$\Theta = \{ 0 \leq \theta \leq 1 \mid 
    \theta_1 + \theta_2 =1 \wedge  \theta_3 + \theta_4 = 1 \}$$  
  $$\Theta_0 = \{ \theta \in \Theta \mid \theta_1 = \theta_3 \}$$  
       
 The Bayes Factor considering a priori 
 $Pr\{H \} = Pr\{\theta_1=\theta_3\}=0.5$
and uniform densities over $\Theta_0$ and $\Theta-\Theta_0$ is given 
in the equation below. See Irony (1994) and Irony (1995) for
details and discussion about properties.
 $$BF= \frac{\left( \begin{array}{c} m \\ x \end{array} \right) 
      \left( \begin{array}{c} n \\ y \end{array} \right)} 
      {\left( \begin{array}{c} m+n \\ x+y \end{array} \right)}\  
      \frac{(m+1)(n+1)}{m+n+1}$$

 \subsection*{Independence test in a $2 \times 2$ contingency table}  

 Suppose that laboratory test is used to help in the diagnostic of a
disease.  It should be interesting to check if the test results are
really related to the health conditions of a patient.  A patient chosen
from a clinic is classified as one of the four states of the set 
 $$\{(h,t) \mid h,t=0\ or\ 1\}$$ in such a way that $h$ is the indicator
of the occurrence or not of the disease and $t$ is the indicator for the
laboratory test being positive or negative.  For a sample of size $n$ we
record $(x_{00},x_{01},x_{10},x_{11})$, the vector whose components are
the sample frequency of each the possibilities of $(t,h)$.  The
parameter space is the simplex 
 $$\Theta = \{(\theta_{00},\theta_{01},\theta_{10},\theta_{11}) \mid
   \theta_{ij} \geq 0 \ \wedge \  \sum_{i,j}\theta_{ij}=1\}$$ 
 and the null hypothesis, h and t are independent, is defined by
 $$\Theta_0= \{\theta \in \Theta \mid
\theta_{00}=\theta_{0\bullet}\theta_{\bullet 0},\ 
\theta_{0\bullet}=\theta_{00}+\theta_{01},\  \theta_{\bullet
0}=\theta_{00}+\theta_{10}\}.$$ 
 The Bayes Factor for this case is discussed by [Iro 95] and has the
following expression:
 $$BF= \frac{\left( \begin{array}{c} x_{0\bullet} \\ x_{00} \end{array}
 \right) \left( \begin{array}{c} x_{1\bullet} \\ x_{11} \end{array}
 \right)}{\left( \begin{array}{c} n \\ x_{\bullet 0}\end{array}
 \right)}\left\{\frac{(n+2)\left\{(n+3)-(n+2)[P(1-P)+Q(1-Q)]\right\}}
                 {4(n+1)}\right\}$$
 where $x_{i\bullet}=x_{i0}+x_{i1}, x_{\bullet j}=x_{0j}+x_{1j}$,
 $P=\frac{x_{0\bullet}}{n+2}$, and $Q=\frac{x_{\bullet 0}}{n+2}$.

 \section{Weibull Wearout Model} 


 We where faced with the problem of testing the wearout of a lot of 
used display panels. A panel displays 12 to 18 characters. 
 Each character is displayed as a $5\times 8$ matrix of pixels, and each
 pixel is made of 2 (RG) or 3 (RGB) individual color elements, (like a
light emitting diode  or gas plasma device). 
 A panel fails when the first individual color element fails.
 The construction characteristics of a display panel makes the weibull
distribution specially well suited to model its life time. 
 The color elements are ``burned in'' at the production process, so we
assume they are not at the infant mortality region, i.e. we assume the 
Weibull's shape parameter to be greater than one, with wearout or
increasing hazard rates. 
 
 The panels in question were purchased as used components, taken from
surplus machines. The dealer informed the machines had been operated for
a given time, and also informed the mean life of the panels at those
machines. Only working panels were acquired. The acquired panels were
installed as components on machines of a different type. 
 The use intensity of the panels at each type of machine corresponds to
a different time scale, so mean lifes are not directly comparable.  
 The shape parameter however is an intrinsic characteristic of the
panel. The used time over mean life ratio, $\rho=\alpha/\mu$, is
adimensional, and can therefore be used as an intrinsic measure of 
wearout. 
 We have recorded the time to failure, or times of withdrawal with no
failure, of the panels at the new machines, and want to use this data to
corroborate (or not) the wearout information provided by the surplus
equipment dealer.

 \subsection*{Weibull Distribution} 

 The two parameter Weibull probability density, reliability (or survival
probability) and hazard functions, for a failure time $t\geq 0$,
given the shape, and characteristic life (or scale) parameters, 
 $\beta>0$, and $\gamma>0$, are:    
 \begin{eqnarray*} 
 w(t\g \beta,\gamma) &=& 
  (\beta\,t^{\beta-1}/\gamma^\beta)\,exp(-(t/\gamma)^\beta) \\ 
 r(t\g \beta,\gamma) &=& exp(-(t/\gamma)^\beta) \\ 
 z(t\g \beta,\gamma) &\equiv& w(\ )/r(\ ) =                     
  \beta\,t^{\beta-1}/\gamma^\beta      
 \end{eqnarray*} 
 The mean and variance of a Weibull variate are given by: 
 \begin{eqnarray*} 
 \mu &=& \gamma \, \Gamma(1 +1/\beta) \\ 
 \sigma^2 &=& \gamma^2 ( \Gamma(1+2/\beta) +\Gamma^2(1+1/\beta) ) \\ 
 \end{eqnarray*} 

 By altering the parameter, $\beta$, $W(t\g \beta,\gamma)$ takes a
variety of shapes, Dodson(1994). Some values of shape parameter are 
important special cases: 
 for $\beta=1$,   $W$ is the exponential distribution; 
 for $\beta=2$,   $W$ is the Rayleigh distribution;   
 for $\beta=2.5$, $W$ approximates the lognormal distribution;   
 for $\beta=3.6$, $W$ approximates the normal distribution; and    
 for $\beta=5.0$, $W$ approximates the peaked normal distribution.   
 The flexibility of the Weibull distribution makes it very useful  for
empirical modeling, specially in quality control and reliability.    
 The regions $\beta<1$, $\beta=1$, and $\beta>1$ correspond to
decreasing, constant and increasing hazard rates. 
 These three regions are also known as infant mortality, memoryless, 
and wearout failures.
  $\gamma$ is approximately the 63rd percentile of the life time,
regardless of the shape parameter.     

 The Weibull also has important theoretical properties. 
 If $n$ i.i.d. random variables have Weibull distribution, 
 $X_i \sim w(t\g \beta,\gamma)$, then the first failure is a 
 Weibull variate with characteristic life $\gamma/n^{1/\beta}$, i.e.  
 $X_{[1,n]} \sim w(t\g \beta,\gamma/n^{1/\beta})$. 
 This kind of property allows a characterization of the Weibull as a
 limiting life distribution in the context of extreme value theory,
 Barlow and Prochan (1975).    

 The affine transformation $t=t'+\alpha$ leads to the three parameter 
truncated Weibull distribution. 
 A location (or threshold) parameter, $\alpha>0$ represents beginning
observation of  a (truncated) Weibull variate at $t=0$, after it has
already survived the period $[-\alpha,0[$.  
 The three parameter truncated Weibull is given by: 
 \begin{eqnarray*} 
 w(t\g \alpha,\beta,\gamma) &=& 
  (\beta\,(t+\alpha)^{\beta-1}/\gamma^\beta)
  \,exp(-((t+\alpha)/\gamma)^\beta)/r(\alpha\g \beta,\gamma) \\ 
 r(t\g \alpha,\beta,\gamma) &=& exp(-((t+\alpha)/\gamma)^\beta)
  /r(\alpha\g \beta,\gamma) \\ 
 \end{eqnarray*}

 \subsection*{Wearout Model} 

 The problem described at the preceding sections can be tested using the 
 FBST, with parameter space, hypothesis and posterior joint density: 
 \begin{eqnarray*} 
 \Theta &=& \{ (\alpha,\beta,\gamma) \; \in \; \:    
   ]0,\infty]\times [1,\infty] \times [0,\infty[ \; \} \\  
 \Theta_0 &=& \{ (\alpha,\beta,\gamma) \in \Theta 
   \g \alpha = \rho \mu(\beta,\gamma) \: \}\\ 
 f(\alpha,\beta,\gamma\g D) &\propto& 
   \prod_{i=1}^{n} w(t_i\g \alpha,\beta,\gamma) 
   \prod_{j=1}^{m} r(t_j\g \alpha,\beta,\gamma)    
 \end{eqnarray*} 
 where the data $D$ are all the recorded failure times, $t_i>0$,  
 and the times of withdrawal with no failure, $t_j>0$. 

 At the optimization step it is better, for numerical stability, to 
 maximize the log-likelihood, $fl(\ )$. 
 Given a sample with $n$ recorded failures and $m$ withdrawals,    
 \begin{eqnarray*}  
  wl_i &=& \log(\beta) +(\beta-1)\log(t_i +\alpha) 
    -\beta\log(\gamma) -((t_i +\alpha)/\gamma)^\beta 
    +(\alpha/\gamma)^\beta \\ 
  rl_j &=& -((t_j +\alpha)/\gamma)^\beta +(\alpha/\gamma)^\beta \\ 
  fl &=& \sum_{i=1}^{n} wl_i +\sum_{j=1}^{m} rl_j \\   
 \end{eqnarray*}   
 the hypothesis being represented by the constraint 
 $$h(\alpha,\beta,\gamma) = 
   \rho\, \gamma\, \Gamma(1+1/\beta) -\alpha = 0$$   

 The gradients of $fl(\ )$ and $h(\ )$ analytical 
 expressions, to be given to the optimizer, are:  
 \begin{eqnarray*} 
  \lefteqn{dwl =} \\       
  & & [\;   (\beta-1)/(t+\alpha)   
           -((t+\alpha)/\gamma)^\beta\beta/(t+\alpha)   
           +(\alpha/\gamma)^\beta\beta/\alpha \, , \\ 
  & & \;\;  1/\beta+\log(t+\alpha) -\log(\gamma)  
           -((t+\alpha)/\gamma)^\beta\log((t+\alpha)/\gamma)  
           +(\alpha/\gamma)^\beta\log(\alpha/\gamma) \, , \\ 
  & & \;\; -\beta/\gamma +((t+\alpha)/\gamma)^\beta\beta/\gamma  
           -(\alpha/\gamma)^\beta\beta/\gamma \; ] \\ 
  \lefteqn{drl =} \\ 
  & & [\;  -((t+\alpha)/\gamma)^\beta\beta/(t+\alpha) 
           +(\alpha/\gamma)^\beta\beta/\alpha \, , \\ 
  & & \;\; -((t+\alpha)/\gamma)^\beta\log((t+\alpha)/\gamma)  
           +(\alpha/\gamma)^\beta\log(\alpha/\gamma) \, , \\ 
  & & \;\;  ((t+\alpha)/\gamma)^\beta\beta/\gamma  \, , \,  
           -(\alpha/\gamma)^\beta\beta/\gamma \; ]\\  
  \lefteqn{dh =} \\  
  & & [\;  -1 \, , \,    
     -\rho\,\gamma\,\Gamma'(1+1/\beta)\,\Gamma(1+1/\beta)/\beta^2 \, ,\,  
      \rho\,\Gamma(1+1/\beta) \; ]\\  
 \end{eqnarray*} 
 For gamma and digamma functions efficient algorithms  
 see Spanier and Oldham (1987). 


  In this model, some prior distribution of the shape parameter is
needed  to stabilize the model. Knowing  color elements'  life time to
be approximately normal, we consider $\beta\in [3.0,4.0]$.

 \section{The Normal-Wishart Distribution} 

 The matrix notation used in this section is defined in section F.1. 

 The Bayesian research group at IME-USP has developed several 
applications based on multidimensional normal models, including 
structure models, mixture models and factor analysis models. 
 In this appendix we review the core theory of some of these models, 
since they are used in some of the illustrative examples in chapters 
4 and 5. For implementation details, practical applications, 
case studies, and further comments, see Lauretto et al. (2003).

 The conjugate family of priors for multivariate normal distributions 
 is the Normal-Wishart family of distributions, DeGroot (1970).  
 Consider the random matrix $X$ with elements 
 $X_i^j\ , i=1\ldots k\ , \ j=1\ldots n\ , \ n>k$,  
 where each column, $x^j$, contains  a sample vector from a 
 $k$-multivariate normal distribution with parameters $\beta$ 
(mean vector) and $V$ (covariance matrix), or $R=V^{-1}$ 
(precision matrix).  

 Let $\bar{x}$ and $W$ denote, respectively, the statistics: 
 \begin{eqnarray*} 
  \bar{x} &=& \frac{1}{n}\, \sum_{j=1}^n x^j  
          \ = \   \frac{1}{n} X{\bf 1}  \\   
   W &=& \sum_{j=1}^n (x^j -\beta)\, (x^j -\beta)' 
     \ = \  (X-\beta) (X-\beta)'  
 \end{eqnarray*} 
 The random matrix $W$ has Wishart distribution with $n$ degrees of
freedom and precision matrix $R$. The Normal and Wishart pdfs have
the  expressions:  
 \begin{eqnarray*} 
  f(\bar{x}\g n,\beta,R) &=&  
   (\frac{n}{2\pi})^{k/2} {|R|}^{1/2} \, 
       \exp( \, -\frac{n}{2}(\bar{x}-\beta)'R(\bar{x}-\beta) \, ) \\ 
 f(W\g n,\beta,R) &=&    
    c\, {|W|}^{(n-k-1)/2} \, \exp(\, -\frac{1}{2}\tr(W\, R) \, ) \\  
 c^{-1} &=&   
      {|R|}^{-n/2}\, 2^{nk/2}\, \pi^{k(k-1)/4}\,  
       \prod_{j=1}^k \Gamma ( \frac{n+1-j}{2} ) 
 \end{eqnarray*} 

 Now consider the matrix $X$ as above, with unknown mean $\beta$ and unknown 
precision matrix $R$, and the statistic  
 $$S = \sum_{j=1}^n (x^j -\bar{x})\, (x^j -\bar{x})' 
     = (X -\bar{x}) (X -\bar{x})'  
 $$ 
 Taking as prior distribution for the precision matrix $R$ the wishart
distribution with $a>k-1$ degrees of freedom and precision matrix
$\dot{S}$ and, given $R$, taking as prior for $\beta$ a multivariate
normal with mean $\dot{\beta}$ and precision $\dot{n}R$, i.e. 
  \begin{eqnarray*} 
 p(\beta,R) &=&  p(R)\, p(\beta\g R) \\ 
 p(R) &\propto&  
   {|R|}^{(a-k-1)/2} \, \exp(\, -\frac{1}{2}\tr(R\,\dot{S}) \, ) \\  
 p(\beta\g R) &\propto&   {|R|}^{1/2}\, 
  \exp(\, -\frac{\dot{n}}{2}(\beta-\dot{\beta})'R(\beta-\dot{\beta}) \, ) \\ 
 \end{eqnarray*} 
 The posterior distribution for the parameters $\beta$ and $R$ has the form:   
  \begin{eqnarray*} 
 p_n(\beta,R\g n,\bar{x},S) &=& 
    p_n(R\g n,\bar{x},S)\, p_n(\beta\g R,n,\bar{x},S) \\ 
 p_n(R\g n,\bar{x},S) &\propto&  
   {|R|}^{(a+n-k-1)/2} \, \exp(\, -\frac{1}{2}\tr(R\,\ddot{S}) \, ) \\  
 p_n(\beta\g R,n,\bar{x},S) &\propto& {|R|}^{1/2}\, 
 \exp(\, -\frac{\ddot{n}}{2}(\beta-\ddot{\beta})'R(\beta-\ddot{\beta}) \, ) \\  
 \ddot{\beta} &=& (n\bar{x} +\dot{n}\dot{\beta})/\ddot{n} \ \ ,\ \  
 \ddot{n}  =  n +\dot{n} \\   
 \ddot{S} &=& S +\dot{S} 
     +\frac{n\dot{n}}{n+\dot{n}}(\dot{\beta}-\bar{x})(\dot{\beta}-\bar{x})' 
 \end{eqnarray*} 
 Hence, the posterior distribution for $R$ is a Wishart distribution with 
$a+n$ degrees of freedom and precision $\ddot{S}$, and the conditional 
distribution for $\beta$, given $R$, is $k$-Normal with mean 
$\ddot{\beta}$ and precision $\ddot{n}R$. 
 All covariance and precision matrices are supposed to be positive 
definite, $n>k$, $a>k-1$, and $\dot{n}>0$.       
  
 Non-informative improper priors are given by $\dot{n}=0$,
$\dot{\beta}=0$, $a=0$, $\dot{S}=0$, i.e. we take a Wishart with $0$ 
degrees of freedom as prior for $R$, and a constant prior for $\beta$,
Box and Tiao (1973), DeGroot (1970), Zellner (1971). 
 Then, the posterior for $R$ is a Wishart with $n$ degrees of freedom
and precision $S$, and the posterior for $\beta$, given $R$, is
$k$-Normal with mean $\bar{x}$ and precision $nR$.   

 We can now write the simplified log-posterior kernels:  
 \begin{eqnarray*} 
 fl(\beta,R\g n,\bar{x},S) &=&   
   fl(R\g n,\bar{x},S)\, +fl(\beta\g R,n,\bar{x},S) \\ 
 fl(R\g n,\bar{x},S) = flr &=&  
  \frac{a+n-k-1}{2} \log(|R|) \, -\frac{1}{2}\tr(R\,\ddot{S}) \\  
 fl(\beta\g R,n,\bar{x},S) = flb &=&  \frac{1}{2}\log(|R|) \, 
      -\frac{\ddot{n}}{2}(\beta-\ddot{\beta})'R(\beta-\ddot{\beta}) \\ 
 \end{eqnarray*}   
 For the surprise kernel, relative to the uninformative prior, we 
 only have to replace the factor $(a+n-k-1)/2$ by $(a+n)/2$.

 \section{Structural Models}

 In this section we study the dose-equivalence hypothesis. 

 The dose-equivalence hypothesis, $H$, asserts a proportional 
response of a pair of response measurements to two different stimuli.
 The hypothesis also asserts proportional standard deviations, 
and equivalent correlations for each response pair. 
 The proportionality coefficient, $\delta$, is interpreted as 
the second stimulus dose equivalent to one unit of the first. 

 This can be seen as a simultaneous generalization of the linear mean
structure, the linear covariance structure, and the Behrens-Fisher
problems. The test proved to be useful when comparing levels of genetic 
expression, as well as to calibrate micro array equipment at 
BIOINFO, the genetic research task force at University of Sao Paulo.  
 The application of the dose-equivalence model is similar to the 
much simpler bio-equivalence model used in pharmacology, and closely
related by several other classic covariance structure models used in
biology, psychology,  and social sciences, as described in
 Anderson (1969), Bock and Bargnann (1966), 
 Jiang and Sarkar (1998, 1999, 2000a,b), J\"{o}reskog (1970), and 
 McDonald (1962, 1974, 1975).
 We are not aware of any alternative test for the dose-equivalence 
hypothesis published in the literature.

 \subsection{Mean and Covariance Structure}

  As it is usual in the covariance structure literature,
 we will write $V(\gamma)=\sum \gamma_h G\{h\}$, where the matrices 
 $G\{h\}$, $h=1,\ldots k(k+1)/2$ form a basis for the space of 
 $k\times k$ symmetric matrices; in our case, $k=4$.     
  The matrix notation is presented at Section F.1.  
 \begin{eqnarray*}
 V(\gamma) &=& \sum_{h=1}^{10} \gamma_h\, G\{h\} \, = \,  
 \left[ \begin{array}{cccc} 
 \gamma_1 & \gamma_5    & \gamma_7 & \gamma_8 \\  
 \gamma_5 & \gamma_2    & \gamma_9 & \gamma_{10} \\  
 \gamma_7 & \gamma_9    & \gamma_3 & \gamma_6 \\  
 \gamma_8 & \gamma_{10} & \gamma_6 & \gamma_4 \\    
 \end{array} \right]  \ , \ \ \mbox{where} \\    
 G\{h\} &=&  
 \left[ \begin{array}{cccc} 
 \delta_h^1 & \delta_h^5  & \delta_h^7 & \delta_h^8 \\  
 \delta_h^5 & \delta_h^2  & \delta_h^9 & \delta_h^{10} \\  
 \delta_h^7 & \delta_h^9  & \delta_h^3 & \delta_h^6 \\  
 \delta_h^8 & \delta_h^{10} & \delta_h^6 & \delta_h^4 \\    
 \end{array} \right] \ , \ \ 
 \end{eqnarray*}           
 and the Kronecker-delta is  
 $\delta_h^j=1$ if $h=j$ and $\delta_h^j=0$ if $h\neq j$.

 The dose-equivalence hypothesis, $H$, asserts a proportional 
response of a pair of response measurements to two different stimuli.
 Each pair of response measurements is supposed to be a bivariate 
normal variate.  
 $H$ also asserts proportional standard deviations, and 
equivalent correlations for each pair of response measurements. 
 The proportionality coefficient, $\delta$, is interpreted as 
the dose, calibration or proportionality coefficient. 

 In order to get simpler expressions for the log-likelihood, the 
constraints and its gradients, we use in the numerical procedures 
an extended parameter space including the coefficient $\delta$, 
and state the dose-equivalence optimization problem on the extended 
15-dimentional space, with a 5-dimentional constraint: 
 \begin{eqnarray*}  
 \Theta &=& \{ \theta=[\gamma',\beta',\delta]' \in R^{10+4+1} \ , \ 
            V(\gamma)>0 \} \\   
 \Theta_0 &=& \{ \theta \in \Theta \g h(\theta)=0 \} \\ 
 h(\theta) &=& \left[ \begin{array}{c}  
  \delta^2 \gamma_1 -\gamma_3 \\  
  \delta^2 \gamma_2 -\gamma_4 \\ 
  \delta^2 \gamma_5 -\gamma_6 \\ 
  \delta   \beta_1  -\beta_3      \\ 
  \delta \beta_2    -\beta_4      \\ 
  \end{array} \right] 
  \end{eqnarray*} 

  In order to be able to compute some gradients needed in 
 the next section, we recall some matrix derivative identities, 
 see Anderson (1969), Harville (1997), 
 McDonald and Swaminathan (1973), Rogers (1980). 
  We use $V=V(\gamma)$, $R=V^{-1}$, and $C$ for a constant matrix. 
 \[    
  \frac{\del V}{\del \gamma_h} = G\{h\}  \ ,  
  \ \ \ \ \ \ 
  \frac{\del R}{\del \gamma_h}  =   -R\, G\{h\}\, R \ , 
 \] 
 \[  
  \frac{\del \beta'\, C\, \beta}{\del \beta} =  2\, C\, \beta \ ,      
  \ \ \ \ \  
  \frac{\del \log(|V|)}{\del \gamma_h}  =  \tr (R\, G\{h\} ) \ , 
  \] 
  \[     
  \frac{\del \  \mbox{frob2}(V-C)}{\del \gamma_h}  =    
     2 \sum_{i,j} (V-C)\odot G\{h\}  \ .  
  \] 
  We also define the auxiliary matrices: 
  \[ 
  P\{h\} = R\,G\{h\} \ , \ \ \ \ \   
  Q\{h\} = P\{h\}\,R \ . 
  \]

 \subsection{Numerical Optimization} 

  To find $\theta^*$ we use an objective function, 
to be minimized on the extended parameter space, given by  
a centralization term minus the log-posterior kernel,     
 \begin{eqnarray*} 
 f(\theta \g n,\bar{x},S) &=&  
    c\,n\, \mbox{frob2}(V -C)\, -flr\, -flb  \\  
   &=&  c\,n\, \mbox{frob2}(V -C)\,  -\frac{a+n-k}{2} \log(|R|) \\ 
   & &  +\frac{1}{2}\tr(R\,\ddot{S})\,   
        +\frac{\ddot{n}}{2}(\beta-\ddot{\beta})'R(\beta-\ddot{\beta}) 
 \end{eqnarray*}    
 Large enough centralization factors, $c$, times the squared Frobenius
norm of $(V-C)$, where $C$ are intermediate approximations of the 
constrained minimum, make the first points of the
optimization sequence remain in the neighborhood of the empirical
covariance (the initial $C$). 
 As the optimization proceeds, we relax the centralization
factor, i.e. make $c\rightarrow 0$, and  maximize the pure posterior
function. 
 This is a standard optimization procedure following the regularization 
strategy of Proximal-Point algorithms, see  
Bertzekas and Tsitsiklis (1989), Iusem (1995), Censor and Zenios (1997). 
 In practice this strategy let us avoid handling explicitly 
the difficult constraint $V(\gamma)>0$.   

  Using the matrix derivatives given in the last section, we find 
 the objective function's gradient, $\del f /\del \theta$, 
 \begin{eqnarray*} 
  \frac{\del f}{\del \gamma_h} &=&  
  \frac{a+n-k}{2} \tr(P\{h\})\: -\frac{1}{2}\tr(Q\{h\}\,\ddot{S})\: \\ 
 & & -\frac{\ddot{n}}{2}(\beta-\ddot{\beta})'\,Q\{h\}\,(\beta-\ddot{\beta}) \\    
 & &   +2 c\,n\, \sum_{i,j=1}^n (V-C)\odot G\{h\}    \\ 
 \frac{\del f}{\del \beta} &=&  
     \ddot{n}\,R\, (\beta -\ddot{\beta}) \\   
 \end{eqnarray*}                    
 For the surprise kernel and its gradient, relative to the uninformative 
 prior, we only have to replace the factor $(a+n-k)/2$ by $(a+n+1)/2$.   

 The Jacobian matrix of the constraints, 
 ${\del h}/{\del \theta}$, is:   
 \[ 
 \left[ \begin{array}{ccccccccccccccc} 
 \delta^2 & 0 & -1 & 0 & 0 & 0 & 0 & 0 & 0 & 0 & 
            0 & 0 & 0 & 0 & 2 \delta \gamma_1 \\   
 0 & \delta^2 & 0 & -1 & 0 & 0 & 0 & 0 & 0 & 0 & 
            0 & 0 & 0 & 0 & 2 \delta \gamma_2 \\   
 0 & 0 & 0 & 0 & \delta^2 & -1 & 0 & 0 & 0 & 0 & 
            0 & 0 & 0 & 0 & 2 \delta \gamma_5 \\   
 0 & 0 & 0 & 0 & 0 & 0 & 0 & 0 & 0 & 0 & 
        \delta & 0 & -1 & 0 & \beta_1 \\   
 0 & 0 & 0 & 0 & 0 & 0 & 0 & 0 & 0 & 0 & 
        0 & \delta & 0 & -1 & \beta_2 \\   
 \end{array} \right]  
 \]  

 At the optimization step, Variable-Metric Proximal-Point algorithms,
working with the explicit analytical derivatives given above,  
proved to be very stable, in contrast with the often unpredictable  
behavior of some methods found in most statistical software, 
like  Newton-Raphson or ``Scoring''. 
 Optimization problems of small dimension, like above, allow us 
to use dense matrix representation without significant loss, 
Stern (1994). 

 In order to handle several other structural hypotheses, we only have 
to replace the constraint, and its Jacobian, passed to the optimizer. 
 Hence, many different hypothesis about the mean and covariance or 
correlation structure can be treated in a coherent, efficient, exact,
robust, simple, and unified way. 

 The derivation of the Monte Carlo procedure for the numerical
integrations required to implement the FBST in this model is 
presented in appendix G.

 \section{Factor Analysis} 

 This section reviews the most basic facts about FA models.  
 For a synthetic introduction to factor analysis, see 
 Ghaharamani and Hilton (1997) and Everitt (1984). 
 For some of the matrix analytic and algorithmic details, 
 see Abadir and Magnus (2005), Golub and Loan (1989), Harville (2000), 
 Rubin and Thayer (1982), and Russel (1998).   
 For the technical issue of factor rotation, see Browne (1974, 2001),  
 Jennrich (2001, 2002, 2004) and Bernaards and Jennrich (2005).

 The generative model for Factor Analysis (FA) is 
 $x= \Lambda z +u$, where 
 $x$ is a $p\times 1$ vector of observed random variables, 
 $z$ is a $k\times 1$ vector or latent (unobserved) random variables, 
 known as {\it factors} and 
 $\Lambda$ is the $p\times k$ matrix of factor {\it loadings}, or weights. 
 FA is used as a dimensionality reduction technique, so $k<p$.  

 The vector variates $z$ and $u$ are assumed to be distributed as 
 $\mathcal{N}(0,I)$ and $\mathcal{N}(0,\Psi)$, where $\Psi$ is diagonal. 
 Hence, the observed and latent variables joint distribution is 
 \[ 
    \left[ \begin{array}{c} x \\ z \end{array} \right] \sim 
    \mathcal{N} \left( 
    \left[ \begin{array}{c} 0 \\ 0 \end{array} \right] \ , \  
    \left[ \begin{array}{cc} 
      \Lambda \Lambda' +\Psi & \Lambda \\ \Lambda' & I 
    \end{array} \right] \right) \ .  
 \]

 For two jointly distributed Gaussian (vector) variates, 
 \[ 
    \left[ \begin{array}{c} x \\ z \end{array} \right] \sim 
    \mathcal{N} \left( 
    \left[ \begin{array}{c} a \\ b \end{array} \right] \ , \  
    \left[ \begin{array}{cc} 

      A  & C \\ C' & D  
    \end{array} \right] \right) \ ,   
 \] 
 the distribution of $z$ given $x$ is given by, 
 see Zellner (1971),    
 \[ 
    z \g x \sim   \mathcal{N} \left( 
    b + C'A^{-1}(x-a) , \ D-C'A^{-1}C \right) \ . 
 \] 
 Hence, in the FA model, 
 \[ 
    z \g x \sim   \mathcal{N} \left( 
    Bx , \ I-B\Lambda \right) \ , \ \mbox{where} \ 
 \] 
 \[ 
    B= \Lambda'(\Lambda\Lambda' +\Psi)^{-1}   
    =\Lambda' \left(\Psi^{-1} 
     -\Psi^{-1}\Lambda 
        \left(I+\Lambda'\Psi^{-1}\Lambda \right)^{-1} 
      \Lambda'\Psi^{-1} \right) 
 \]

 \subsection{EM Algorithm} 

 In order to obtain the Maximum Likelihood (ML) estimator of 
 the parameters, one can use the EM-Algorithm, 
 see Rubin and Thayer (1982) and Russel (1998). 
 The E-step for the FA model computes the expected first and second 
 moments of the latent variables, for each observation, $x$.  
 \[ 
    E(z\g x)= B x \ , \ \mbox{and} 
 \] 
 \[ 
    E(zz'\g x)= \Cov(z\g x) +E(z\g x) E(z\g x)' = 
    I+ B\Lambda +Bx\; x'B'   
 \] 

 The M-step optimizes the parameters $\Lambda$ and $\Psi$, 
 of the expected log likelihood for the FA (completed data) model, 
 \[ 
   q(\Lambda, \Psi) = 
   E \left( \log \pprod_{j=1}^n f(x \g z, \Lambda, \Psi) \right) 
 \] 
 \[ 
   = E \left( \log \pprod_{j=1}^n (2\pi)^{p/2} \left| \Psi \right|^{-1/2} 
   \exp\left( -\frac{1}{2}\left( x^j-\Lambda z\right)' 
              \Psi^{-1} \left( x^j-\Lambda z\right) \right) \right) 
 \]  
 \[ 
   =c -\frac{n}{2}\log \left| \Psi \right| -\ssum_{j=1}^n E \left( 
    \frac{1}{2} {x^j}'\Psi^{-1}x^j -{x^j}'\Psi^{-1}\Lambda z 
    +\frac{1}{2} z'\Lambda' \Psi^{-1} \Lambda z \right) 
 \] 
 Using the results computed in the E-step, 
 the last summation can be written as 
 \[ 
   \ssum_{j=1}^n \left( \frac{1}{2} {x^j}'\Psi^{-1}x^j 
    -{x^j}'\Psi^{-1}\Lambda E(z\g x^j)  
    +\frac{1}{2} \tr \left( \Lambda' \Psi^{-1} \Lambda E(zz'\g x^j) 
    \right) \right) 
 \] 
 The ML estimator, $(\Lambda^*,\Psi^*)$, 
 is a stationary point in $\Lambda^*$, therefore  
 \[ 
    \frac{\del q}{\del \Lambda}= 
    -\ssum_{j=1}^n  \Psi^{-1}x^j E(z\g x^j)' 
    +\ssum_{j=1}^n  \Psi^{-1} \Lambda E(zz'\g x^j) =0 
    \ , \ \ \mbox{hence} 
 \] 
 \[ 
    \Lambda^* =  \left( \ssum_{j=1}^n E(zz'\g x^j)' \right)^{-1} 
               \ssum_{j=1}^n x^j E(z\g x^j)' 
 \]    
 and also a stationary point in $\Psi^*$, or in its inverse, therefore,  
 substituting the stationary value of $\Lambda^*$ computed in the last 
 equation,  \[ 
   \frac{\del q}{\del \Psi^{-1}}= \frac{n}{2}\Psi 
   -\ssum_{j=1}^n \left( \frac{1}{2} x^j {x^j}'
      -\Lambda^* E(z\g x^j) {x^j}'  
      +\frac{1}{2}\Lambda^* E(zz'\g x^j) {\Lambda^*}' \right) =0 
 \] 
 Solving for $\Psi$, and using the diagonality constraint, 
 \[ 
   \Psi^* = \frac{1}{n} \diag^2 \left( 
     \ssum_{j=1}^n  x^j {x^j}'  
    -\Lambda^* \ssum_{j=1}^n  E(z\g x^j) {x^j}' \right)  
 \] 

 The equation for $\Lambda^*$, in the M-step of the EM algorithm 
 for FA, formally resembles the equation giving the LS estimation in a 
 Linear Regression model, 
 $\beta' = y'X(X'X)^{-1}$. 
 This is why, in the FA literature, the matrix $\Lambda^*$ 
 is sometimes interpreted as 
 ``the linear regression coefficients of the $z$'s on the $x$'s''.

 \subsection{Orthogonal and Oblique Rotations} 

 Given a FA model and a non-singular coordinate transform, $T$,  
 it is possible to obtain transformed factors together with transformed 
 loadings giving an equivalent FA model. 
 Both, a {\it direct}, and an {\it inverse}, form of the factor loadings   
 transform are common in the literature.    

 In the direct form, 
 \[ 
   \til{z}=T^{-1}z \ \mbox{and} \ \  \til{\Lambda} =\Lambda T,
 \] 
 hence, in the new model, 
 \[ 
   x= \Lambda z +u = \Lambda T T^{-1} z +u = \til{\Lambda}\til{z} +u 
   \ \ \mbox{and} 
 \] 
 \[ 
   \Cov(x)= \til{\Lambda}\til{\Lambda}'+\Psi 
          = \Lambda T (T^{-1} I T^{-t}) T' \Lambda' +\Psi 
          = \Lambda \Lambda'+\Psi \ .   
 \] 
 In the inverse form, 
 \[ 
   \til{z}=T'z \ \mbox{and} \ \  \til{\Lambda} =\Lambda T^{-t},
 \] 
 hence, in the new model, 
 \[ 
   x= \Lambda z +u = \Lambda T^{-t} T'z +u = \til{\Lambda}\til{z} +u 
   \ \ \mbox{and} 
 \] 
 \[ 
   \Cov(x)= \til{\Lambda}\til{\Lambda}'+\Psi 
          = \Lambda T^{-t} (T' I T) T^{-1} \Lambda' +\Psi 
          = \Lambda \Lambda'+\Psi \ . 
 \]

 This shows that the FA model is only determined by the $k$ dimensional 
 subspace of $\mathcal{R}^p$ spanned by the factors. 
 Any change of coordinates in this (sub) space, given by $T$, 
 leads to an equivalent model. 

  An operator $T$ is an {\it orthogonal rotation} iff $T'T=I$.   
  Hence, orthogonal transformed factors are still normalized and  
 uncorrelated.   

  An operator $T$ is an {\it oblique rotation} (in the inverse form) 
 iff $\diag^2(T'T)=I$. 
  Hence, oblique transformed factors are still normalized, 
 but correlated.   

 We want to chose either an orthogonal or an oblique rotation, $T$, 
 so to minimize a complexity criterion function of $\til{\Lambda}$. 
  Before discussing appropriate criteria and how to use them, 
 we examine some technical details concerning matrix norms and 
 projections in the following subsection.

 \subsection{Frobenius Norm and Projections} 

 The matrix Frobenius product and the matrix Frobenius norm 
 are defined as follows: 
 \[ 
   \left< \, A \g B \, \right>_F = \tr(A'B) 
   = \uno' \left( A \odot B \right) \uno \ , 
 \] 
 \[ 
   \left| \left| A \right| \right|_F^2 = \left< \, A \g A \, \right>_F 
    = \ssum_j  \left|\left| A^j \right|\right|_2^2   
    = \ssum_i  \left|\left| A_i \right|\right|_2^2  \ .  
 \]

 Lemma 1: 
 The projection, $T$, with respect to the Frobenius norm, 
 into the algebraic sub-manifold of the oblique rotation matrices 

 of a square matrix, $A$, is given as follows,    
 \[ 
    T= A \diag^2(A'A)^{1/2} \ . 
 \]

 A matrix $T$ represents an {\it oblique rotation} iff it has 
 normalized columns, that is, iff  

 $\diag(T'T)=\uno$.  
 We want to minimize    
 \[ 
   \left|\left| A-T \right|\right|_F^2 = 
   \ssum_j  \left|\left| A^j -T^j \right|\right|_2^2   
 \]  
 But, in the 2-norm, the normal vector $T^j$ that is closest to 
 the $A^j$ is the one that has the same direction of vector $A^j$, 
 that is, 

 \[ 
    T^j= \frac{1}{||A^j||_2} A^j = 
          \frac{1}{({A^j}'A^j)^{1/2}} A^j \ , 
 \] 
  hence, the lemma.

 Lemma 2: 
 The projection, $Q$, with respect to the Frobenius norm, 
 into the algebraic sub-manifold of the orthogonal rotation matrices 
 of a square matrix, $A$, is given by its SVD factorization, 
 as follows,    
 \[ 
    Q= UV' \ \ \mbox{where} \ \ U'(A)V= \diag(s) \ . 
 \]

 In order to prove the second lemma, we will consider the following problem.    
 The {\it orthogonal Procrustes problem} seeks the orthogonal rotation, 
 $Q\g Q'Q=I$, that minimizes the Frobenius norm of the difference between 
 a given matrix $A$, $m\times p$, and the rotation of a second matrix, $B$, 
 Formally, the problem is stated as 
 \[ 
    \min_{Q\g Q'Q=I} \left|\left| A-BQ \right|\right|_F^2 
 \]  

 The norm function being minimized can be restated as 
 \[ 
    \left|\left| A-BQ \right|\right|_F^2 = 
    \tr(A'A) +\tr(B'B) -2\tr(Q'B'A) 
 \] 
 Hence the problem asks for the maximum of the last term. 
 Let $Z$ be an orthogonal matrix defined by $Q$ and the SVD factorization 
 of $B'A$ as follows,  
 \[ 
    U'(B'A)V = S = \diag(s) \ , \ \  Z= V'Q'U \ . 
 \] 
 We have, 
 \[ 
   \tr(Q'B'A)= \tr(Q'USV')= \tr(ZS)= 
    s'\diag(Z) \leq s'\uno  \ . 
 \] 
 But the last inequality is tight if $Z=I$, 
 hence the optimal solution for the orthogonal Procrustes problem is 
 \[ 
    Q= UV' \ \ \mbox{where} \ \ U'(B'A)V= \diag(s) \ . 
 \] 
 In order to prove lemma 2, just consider the case $B=I$.

 \subsection{Sparsity Optimization}

  In the FA literature, minimizing the complexity of the factor 
 loadings, $\til{\Lambda}$, is accomplished by maximizing a measure 
 of its sparsity, $\sp(\til{\Lambda})$. 

 A natural sparsity measure in engineering applications is the 
 Minimum Entropy measure. 
 This measure and its (matrix) derivative  are given by 
 \[ 
    \sp_{me}(\Lambda) = 
     -\left< \, \Lambda 2 \g \log(\Lambda 2) \, \right>_F \ , \ \ 
    \left(\Lambda 2\right)_i^j= \left(\Lambda_i^j\right)^2 \ . 
 \] 
 \[ 
    \frac{d \sp_{me}(\Lambda)}{d \Lambda} =  
     -\Lambda \odot \log(\Lambda 2) -\Lambda \ . 
 \] 
 Several variations of the entropy sparsity measure are used in the 
 literature, see Bernaards and Jennrich (2005).

 Hoyer (2004) proposes the following sparsity measure for a vector 
 $x\in \mathcal{R}^k_+$, based on the difference of two $p$-norms, 
 namely $p=1$ and $p=2$, 
 \[ 
   \sp_{ho}(x) = \frac{1}{\sqrt{k} -1} \left( \sqrt{k}  
     -\frac{\left| \left| x \right| \right|_1}
           {\left| \left| x \right| \right|_2} \right) \ . 
 \] 
 From Cauchy-Schwartz inequality, 
 we have the bounds, 
 \[ 
    \frac{1}{\sqrt{n}} \left| \left| x \right| \right|_1 \leq 
    \left| \left| x \right| \right|_2 \leq 
    \left| \left| x \right| \right|_1  
    \ , \ \mbox{hence} \ 
    0 \leq \sp_{ho}(x) \leq 1 \ . 
 \]

 Similar interpretations can be given to the 
 Carroll's Oblimin, on the parameter $\gamma$, and 
 Crawford-Ferguson, on the parameter $\kappa$, 
 families of sparsity measures.  
 These measures, for $\Lambda \ p\times k$, and its (matrix) derivative  
 are given by, 
 \[ 
    \sp_\gamma (\Lambda) = \frac{1}{4}  
    \left< \, \Lambda 2 \g B(\gamma) \, \right>_F 
     \ , \ \mbox{where} 
 \] 
 \[ 
    B(\gamma) = (I-\gamma C) \Lambda 2 \, N \ \ , 
 \] 
 \[  
    \left(\Lambda 2\right)_i^j= \left(\Lambda_i^j\right)^2 \ , \ \ 
    C_i^j = \frac{1}{p} \ , \ \ 
    N_i^j = 1- \delta_i^j \ , \ \ 
    \delta_i^j = \mathcal{I}(i=j) \ .    
 \] 
 \[ 
    \frac{d \sp_\gamma(\Lambda)}{d \Lambda}=  
     \Lambda \odot B(\gamma)  \ . 
 \] 
 \[ 
    \sp_\kappa (\Lambda) = \frac{1}{4} 
    \left< \, \Lambda 2 \g B(\kappa) \, \right>_F 
     \ , \ \mbox{where} 
 \] 
 \[ 
    B(\kappa) = 
     (1-\kappa) \Lambda 2 \, N +\kappa M \, \Lambda 2  \ \ , 
 \] 
 \[  
    \left(\Lambda 2\right)_i^j= \left(\Lambda_i^j\right)^2 \ , \ \ 
    M_i^j = 1- \delta_i^j \ , p\times p \ , \ \ 
    N_i^j = 1- \delta_i^j \ , k\times k \ . 
 \] 
 \[ 
    \frac{d \sp_\kappa(\Lambda)}{d \Lambda} =  
     \Lambda \odot  B(\kappa) \ . 
 \] 
 These parametric families include many sparsity measures,  
 or simplicity criteria, traditionally used in psychometric studies, 
 for example, setting 
 $\gamma$ to $0$, $1/2$, or $1$, we have the 
 Quartmin, Biquartmin or Covarimin criterium, also, setting 
 $\kappa$ to $0$, $1/p$, $k/(2p)$ or $(k-1)/(p+k-2)$, we have the 
 Quartimax, Varimax, Equamax or Parsimax criterion. 
 
 In order to search for an optimal transformation, $T^*$,  
 we need to express the  sparsity function and its matrix derivative  
 as functions of $T$.   
 In the direct form, 
 \[ 
    \frac{d \sp(\til{\Lambda})}{ d T} =  
    \frac{d \sp(\Lambda T)}{ d T} =    
    -\left( \Lambda' \frac{d \sp(\Lambda)}{d \Lambda}  \right)' \ .  
 \] 
 In de inverse form,  
 \[ 
    \frac{d \sp(\til{\Lambda})}{ d T} =  
    \frac{d \sp(\Lambda T^{-t})}{ d T} =    
    -\left( \Lambda' \frac{d \sp(\Lambda)}{d \Lambda} T^{-1} \right)' \ . 
 \] 

 This expressions, together with the projectors obtained in the last section,
can be used in standard  gradient projection optimization algorithms, 
like the Generalized Reduced Gradient (GRG) or other standard primal 
optimization algorithms, see Bernaards and Jennrich (2005), Jennrich (2002), 
Luenberger (1984), Minoux and Vajda (1986), Shah et al. (1964), and 
Stern et al. (2006).   
 
 The projection operation for oblique rotations only requires inexpensive 
matrix operations, like a matrix inversion, performed numerically as a 
LU or QR factorization. 
 The projection operation for orthogonal rotations, on the other hand, 
requires a SVD factorization, an operation that requires much more 
computational work. 
 Therefore, a constraint free representation of an orthogonal matrix can
be very useful in designing optimization  algorithms, 
 see Browne (1974, 2001). 
 The Cayley transform establishes one-to-one correspondence
between  skew-symmetric operators, $K$, and the orthogonal operators,
$Q$, that do not have $-1$ as a characteristic value, 
 see  Gantmacher (1959, I, 288-289). 
 Although extreme reversal operators, like a coordinate reflection  or 
permutation can not be represented in this form, there is a Cayley 
representation for any local, that is, not too far from the identity,
orthogonal operator. 
 \[ 
    J=K+I \ , \ \  K_i^j = -K_j^i \ .   
 \]  
 \[   
    K= (I-Q)(I+Q)^{-1} = 2(I+Q)^{-1} -I \ ,  
 \] 
 \[ 
    Q= (I-K)(I+K)^{-1} = 2J^{-1} -I \ . 
 \] 

 The spasity measure derivatives
 of the direct orthogonal rotation of the factor loadings, 
 using the Cayley representation,  
 are given by,   
 \[ 
  \sp(\til{\Lambda}) = \sp(\Lambda T) 
   \ , \ \  T= J^{-1} -I 
 \] 
 \[ 
   \frac{\del \sp(\til{\Lambda}) }{\del J_i^j} =   
   \tr\left( \frac{\del \sp(\til{\Lambda}) }{\del T'} 
             \frac{\del T}{\del J_i^j} \right) = 
   -2 \tr \left( \frac{\del \sp(\til{\Lambda})' }{\del T} 
                J^{-1} \frac{\del J}{\del J_i^j} J^{-1} \right) =        
 \] 
 \[ 
    2 (Y_i^j -Y_j^i) \ , \ \mbox{where} \ \ 
    Y= J^{-1} \frac{\del \sp(\til{\Lambda})'}{\del T} J^{-1}  \ .   
 \]

 \section{Mixture Models} 

 The matrix notation used in this section is defined in section F.1. 
 In this section,  
 $h,i$ are indices in the range $1\fto d$,  
 $k$ is in $1\fto m$, and $j$ is in $1\fto n$.

 In a $d$-dimensional multivariate finite mixture model with 
 $m$ components (or classes), and  sample size $n$, 
 any given sample $x^j$ is of class $k$ with probability 
 $w_k$; the weights, $w_k$, give the probability that a 
 new observation is of class $k$.  
 A sample $j$ of class $k=c(j)$ is distributed with density 
 $f(x^j\g \psi_k)$.

 The classifications $z_k^j$ are boolean variables indicating whether 
 or not $x^j$ is  of class $k$, i.e. $z_k^j=1$ iff $c(j)=k$. 
 $Z$ is not observed, being therefore named latent variable or 
 missing  data.  
 Conditioning on the missing data, we get:     
 \begin{eqnarray*}  
  f(x^j\g \theta) &=& 
  \ssum_{k=1}^m f(x^j \g \theta,z^j_k) f(z_k^j\g \theta) \ \ = \ \   
  \ssum_{k=1}^m w_k f(x^j \g \psi_k) \\     
  f(X\g \theta )  &=& 
   \pprod_{j=1}^n  f(x^j\g \theta) \ \ = \ \    
   \pprod_{j=1}^n \ssum_{k=1}^m  w_k  f(x^j\g \psi_k) 
 \end{eqnarray*} 
   
 Given the mixture parameters, $\theta$, and the observed data, 
 $X$, the conditional classification probabilities, 
 $P=f(Z\g X, \theta)$, are:   
 \begin{eqnarray*}   
  p_k^j &=& f(z_k^j\g x^j, \theta) \ \ = \ \  
  \frac{ f(z_k^j, x^j \g \theta) }{ f( x^j \g \theta ) } \ \ = \ \      
  \frac{w_k f(x^j\g \psi_k)}{\ssum_{k=1}^m w_k f(x^j\g \psi_k)}   
 \end{eqnarray*}

 We use $y_k$ for the number of samples of class $k$, i.e. 
 $y_k= \sum_j z_k^j$, or $y= Z{\bf 1}$.     
 The likelihood for the ``completed'' data, $X, Z$, is: 
  \begin{eqnarray*} 
  f(X, Z\g \theta ) &=& 
   \pprod_{j=1}^n f(x^j \g \psi_{c(j)} ) f(z_k^j \g \theta )  \ \ = \ \  
   \pprod_{k=1}^m \bigl(  {w_k}^{y_k}  
             \pprod_{j \g c(j)=k} f(x^j \g \psi_k ) \bigr)   
 \end{eqnarray*}

  We will see in the following sections that considering the missing 
 data $Z$, and the conditional classification probabilities 
 $P$, is the key for successfully solving the numerical integration 
 and optimization steps of the FBST. 
  In this article we will focus on Gaussian  finite mixture models,
 where $f(x^j\g \psi_k)= N(x^j\g b^k, R\{k\})$, 
 a normal density with mean  $b^k$ and variance matrix $V\{k\}$, 
 or precision $R\{k\}=(V\{k\})^{-1}$. 
  Next we specialize the theory of general mixture models 
 to the Dirichlet-Normal-Wishart case.

 \subsection{Dirichlet-Normal-Wishart Mixtures}

 Consider the random matrix 
 $X_i^j$, $i$ in $1\fto d$, $j$ in $1\fto n$, $n>d$,  
where each column contains  a sample element from a $d$-multivariate normal
distribution with parameters $b$ (mean) and $V$ (covariance), 
or $R=V^{-1}$ (precision).  
 Let $u$ and $S$ denote the statistics: 
   %
 \[  
  u = (1/n)\, \ssum_{j=1}^n x^j = (1/n)\, X{\bf 1}  
  \ \ , \ \   
  S = \ssum_{j=1}^n (x^j -b)\kron (x^j -b)' 
    = (X-b) (X-b)'  
 \] 

 The random vector $u$ has normal distribution with mean $b$ 
and precision $nR$. 
 The random matrix $S$ has Wishart distribution with $n$ degrees of
freedom and precision matrix $R$. 
 The Normal, Wishart and Normal-Wishart pdfs have expressions:  
 \[ 
   N(u\g n,b,R) =   
   ( {\textstyle\frac{n}{2\pi}})^{d/2} {|R|}^{1/2} \, 
       \exp \left( \, -(n/2) (u-b)'R(u-b) \, \right) 
 \] 
 \[ 
   W(S\g e,R)  =  c^{-1}\, 
                {|S|}^{(e-d-1)/2} \, 
        \exp \left(\, -(1/2) \tr(S\, R) \, \right)  
 \]
 with normalization constant \   
 \(  c = {|R|}^{-e/2} \, 2^{ed/2}\, \pi^{d(d-1)/4}\,  
           \prod_{i=1}^d \Gamma ( (e-i+1)/2 ) 
 \) .  

 Now consider the matrix $X$ as above, with unknown mean $b$ and unknown 
precision matrix $R$, and the statistic   
  \begin{eqnarray*} 
   S = \ssum_{j=1}^n (x^j -u)\kron (x^j -u)' 
     = (X -u) (X -u)'  
  \end{eqnarray*} 
  
 The conjugate family of priors for multivariate normal distributions 
is the Normal-Wishart, 
 see DeGroot (1970).  
 Take as prior distribution for the precision matrix $R$ the wishart
distribution with $\dot{e}>d-1$ degrees of freedom and precision matrix
$\dot{S}$ and, given $R$, take as prior for $b$ a multivariate
normal with mean $\dot{u}$ and precision $\dot{n}R$, i.e. let us take   
 the Normal-Wishart prior 
 $NW(b,R\g \dot{n},\dot{e},\dot{u},\dot{S})$.   
 Then, the posterior distribution for $R$ is a Wishart distribution
with  $\ddot{e}$ degrees of freedom and precision $\ddot{S}$, and the
posterior for  $b$, given $R$, is $k$-Normal with mean $\ddot{u}$ and
precision  $\ddot{n}R$, i.e., we have the Normal-Wishart posterior: 
  \begin{eqnarray*} 
 \lefteqn{ NW(b,R\g \ddot{n},\ddot{e},\ddot{u},\ddot{S}) \ \ = \ \ 
  W(R\g \ddot{e},\ddot{S})\ N(b\g \ddot{n},\ddot{u},R)        } \\    
  \ddot{n}  &=&  \dot{n} +n  \ \ , \ \ \ddot{e}  = \dot{e} +n   
 \ \ , \ \  \ddot{u} = (nu +\dot{n}\dot{u})/\ddot{n} \\   
 \ddot{S} &=& S +\dot{S} 
    + (n\dot{n} / \ddot{n})   
   (u-\dot{u})\kron (u-\dot{u})'  
 \end{eqnarray*}  

 All covariance and precision matrices are supposed to be positive 
definite, and proper priors have $\dot{e}\geq d$, and $\dot{n}\geq 1$.         
 Non-informative Normal-Wishart improper priors are given by $\dot{n}=0$,
$\dot{u}=0$, $\dot{e}=0$, $\dot{S}=0$, i.e. we take a Wishart with $0$ 
degrees of freedom as prior for $R$, and a constant prior for $b$,
 see DeGroot (1970). 
 Then, the posterior for $R$ is a Wishart with $n$ degrees of freedom
and precision $S$, and the posterior for $b$, given $R$, is
$d$-Normal with mean $u$ and precision $nR$.   

 The conjugate prior for a multinomial distribution  is a 
 Dirichlet distribution:  
 \begin{eqnarray*} 
  M(y\g n,w) &=& \left( n! \big/ y_1!\ldots y_m! \right) \, 
                  {w_1}^{y_1}\ldots {w_m}^{y_m} \\    
  D(w\g y) &=& \left( \Gamma(y_1+\ldots +y_k) \big/  
   \Gamma(y_1)\ldots\Gamma(y_k)  \right) \, 
   \pprod_{k=1}^m {w_k}^{y_k-1} 
 \end{eqnarray*} 
 with $w>{\bf 0}$ and  $w{\bf 1}=1$.
 Prior information given by $\dot{y}$, 
 and observation $y$, result in the 
 posterior parameter $\ddot{y} =\dot{y} +y$. 
 A non-informative prior is given by $\dot{y}={\bf 1}$.   
      

 Finally, we can write the posterior and completed posterior
 for the model as: 
 \[  
   f(\theta\g X,\dot{\theta}) =    
   f(X\g \theta) f(\theta \g \dot{\theta})        
 \] 
 \[   
   f(X\g \theta) =   
    \pprod_{j=1}^n \ssum_{k=1}^m p_k^j w_k N(x^j\g b^k, R\{k\})  
 \] 
 \[  
  f(\theta\g \dot{\theta}) = D(w\g \dot{y}) \pprod_{k=1}^m 
   NW(b^k,R\{k\}\g \dot{n}_k,\dot{e}_k,\dot{u}^k,\dot{S}\{k\} )  
 \] 
 \[ 
   p_k^j  =   {w_k N(x^j\g b^k,R\{k\})} \   \big/   \  
                    {\ssum_{k=1}^m w_k N(x^j\g b^k,R\{k\})}   
 \]  
 \[  
    f(\theta\g X,Z,\dot{\theta}) =  
   f(\theta\g X,Z) f(\theta \g \dot{\theta}) =    
    D(w\g \ddot{y}) \,  \pprod_{k=1}^m 
    NW( b^k, R\{k\}\g \ddot{n}_k,\ddot{e}_k,\ddot{u}^k,\ddot{S}\{k\} )  
 \] 
 \[ 
   y = Z{\bf 1} \ \ ,\ \ \ddot{y} =  \dot{y} +y \ \ , \ \  
  \ddot{n} = \dot{n} +y \ \ , \ \ \ddot{e} = \dot{e} +y  
 \] 
 \[ 
  u^k = (1/y_k) \ssum_{j=1}^n z_k^j x^j \ \ , \ \   
  S\{k\} = \ssum_{j=1}^n z_k^j (x^j -u^k)\kron (x^j -u^k)'  
 \] 
 \[        
 \ddot{u}^k = (1/\ddot{y}_k) ( \dot{n}_k\dot{u}^k +y_k u^k )   
  \ \ , \ \ 
 \ddot{S}\{k\} =  S\{k\} +\dot{S}\{k\} 
   +( \dot{n}_k y_k \big/ \ddot{n}_k )   
    (u^k -\dot{u}^k)\kron (u^k -\dot{u}^k)'   
 \]

 \subsection{Gibbs Sampling and Integration} 

 In order to integrate a function over the posterior measure, 
 we use an ergodic Markov Chain. The form of the Chain below is 
 known as Gibbs sampling, and its use for numerical integration 
 is known as Markov Chain Monte Carlo, or MCMC.    
  
 Given $\theta$, we can compute $P$. Given $P$, 
 $f(z^j\g p^j)$ is a simple multinomial distribution.   
 Given the latent variables, $Z$, we have simple conditional 
 posterior density expressions for the mixture parameters:   
 %
 \[ 
 f(w \g Z,\dot{y})  =  D(w\g \ddot{y} ) \ \  , \ \    
 f(R\{k\} \g X,Z,\dot{e}_k,\dot{S}\{k\}) =  
 W(R\g \ddot{e}_k, \ddot{S}\{k\}) 
 \] 
 \[ 
 f(b^k \g X,Z,R\{k\},\dot{n}_k,\dot{u}^k) 
  = N(b \g \ddot{n}_k, \ddot{u}^k, R\{k\})  
 \]

 Gibbs sampling is nothing but the MCMC generated by cyclically 
 updating variables $Z$, $\theta$, and $P$, by drawing $\theta$ and $Z$ 
 from the above distributions, 
 see Gilks aet al. (1996) and H\"{a}ggstr\"{o}m (2002). 
 A uniform generator is all what is  needed to the multinomial variate.   
 A Dirichlet variate $w$ can be drawn using a gamma generator  
 with shape and scale parameters $\alpha$ and $\beta$, 
 $G(\alpha,\beta)$, see Gentle (1998). 
 Johnson (1987) describes a simple procedure to generate the Cholesky 
 factor of a Wishart variate $W=U'U$ with $n$ degrees of freedom, from 
 the Cholesky factorization of the covariance 
 $V=R^{-1}=C'C\; ,$ and a chi-square generator: \  
  a) $g_k = G(y_k,1)\; ;$      
  b) $w_k = g_k \; / \; \sum_{k=1}^m g_k\; ;$  
  c) for $i<j\; ,$  $B_{i,j} = N(0,1)\; ;$        
  d) $B_{i,i} = \sqrt{ \chi^2(n-i+1) }\; ;$  and   
  e) $U = BC\; .$
 %
 All subsequent matrix computations proceed  
 directly from the Cholesky factors, see Jones (1985). 

 \subsubsection{Label Switching and Forbidden States}

 Given a mixture model, we obtain an equivalent model 
 renumbering the components $1\fto m$ by a permutation 
 $\sigma([1\fto m])$. 
 This symmetry must be broken in order to have an identifiable model, 
 see Stephens (1997). 
 Let us assume there is an order criterion that can be used when 
 numbering the components. 
 If the components are not in the correct order, Label Switching is 
 the operation of finding permutation $\sigma([1\fto m])$ and 
 renumbering the components, so that the order criterion is satisfied. 
 If we want to look consistently at the classifications produced 
 during a MCMC run, we must enforce a label switching to break 
 all non-identifiability symmetries. 
 For example, in the Dirichlet-Normal-Mixture model, we could   
 choose to order the components (switch labels) according to the
 the rank given by:   
 1) A given linear combination of the vector means, $c'*b^k$; 
 2) The variance determinant $|V\{k\}|$.    
 The choice of a good label switching criterion should consider not 
 only the model structure and the data, but also the semantics and 
 interpretation of the model.

 The semantics and interpretation of the model may also dictate that 
 some states, like certain configurations of the latent variables $Z$, 
 are either meaningless or invalid, and shall not be considered as 
 possible solutions.   
  The MCMC can be adapted to deal with forbidden states by implementing 
 rejection rules, that prevent the chain from entering the forbidden 
 regions of the complete and/or incomplete state space, 
 see Bennett (1976) and Meng (1996).

 \subsection{EM Algorithm for ML and MAP Estimation}

 The EM algorithm optimizes the log-posterior function 
 $fl(X\g \theta)+fl(\theta\g \dot{\theta})$, 
 see Dempster (1977), Ormoneit (1995) and Russel (1988).   
 The EM is derived from the conditional 
 log-likelihood, and the Jensen inequality:   
 If $w,y>{\bf 0}, w'{\bf 1}=1$ then $\log w'y \geq w' \log y$.    
 Let $\theta$ and $\tilde{\theta}$  be our current and next 
 estimate of  the MAP (Maximum a Posteriori), and 
 $p_k^j=f(z_k^j\g x^j,\theta)$  
 the conditional classification probabilities.  
 At each iteration, the log-posterior improvement is:

 \[ 
  \delta(\tilde{\theta},\theta\g X, \dot{\theta} ) \ = \  
           fl(\tilde{\theta}\g X,\dot{\theta}) 
           -fl(\theta\g X,\dot{\theta})            \ = \   
    \delta(\tilde{\theta},\theta\g X) 
   +\delta(\tilde{\theta},\theta\g \dot{\theta})       
 \]  
 \[  
   \delta(\tilde{\theta},\theta\g \dot{\theta}) \ = \  
   fl(\tilde{\theta}\g \dot{\theta}) -fl(\theta\g \dot{\theta})  
 \] 
 \[   
   \delta(\tilde{\theta},\theta \g X)  \ = \   
     fl(X\g \tilde{\theta}) -fl(X\g \theta)   \ = \       
    \ssum_j \delta(\tilde{\theta},\theta \g x^j)  
 \] 
 \[ 
   \delta(\tilde{\theta},\theta \g x^j)  \ = \  
     fl(x^j \g \tilde{\theta}) -fl(x^j \g \theta)    \ = \      
  \log  \ssum_k \tilde{w}_k f(x^j \g \tilde{\psi}_k )  
             \   -fl(x^j \g \theta)                       \ = \      
 \] 
 \[ 
    = \  
      \log \ssum_k 
                     \frac{ p_k^j }{ p_k^j } 
    \frac{ \tilde{w}_k f(x^j \g \tilde{\psi}_k )}{ f(x^j \g \theta)} 
                 \  \  \geq \ \  
  \Delta(\tilde{\theta},\theta \g x^j )  \ = \      
  \ssum_{k} p_k^j \log \frac{ \tilde{w}_k f(x^j \g \tilde{\psi}_k )}
                       { p_k^j f(x^j \g \theta)}    
 \]
 
 Hence,  $\Delta(\tilde{\theta},\theta\g X, \dot{\theta}) = 
          \Delta(\tilde{\theta},\theta\g X) 
          +\delta(\tilde{\theta},\theta\g \dot{\theta})$, 
 is a lower bound to     
         $\delta(\tilde{\theta},\theta\g X, \dot{\theta})$. 
 Also  $\Delta(\theta,\theta\g X, \dot{\theta}) =  
       \delta(\theta,\theta\g X, \dot{\theta}) = 0$.  
 So, under mild differentiability conditions, both surfaces 
 are tangent, assuring convergence of EM to the nearest local maximum.  
 But maximizing $\Delta(\tilde{\theta},\theta\g X, \dot{\theta})$ 
 over $\tilde{\theta}$ is the same as maximizing  
 \begin{eqnarray*} 
 Q(\tilde{\theta},\theta) &=& \ssum_{k,j} p_k^j \log 
      \left( \tilde{w}_k f(x^j \g \tilde{\psi}_k ) \right)  
        +fl(\tilde{\theta}\g \dot{\theta})   
 \end{eqnarray*} 
 and each iteration of the EM algorithm breaks down in two steps: \\ 
 E-step: Compute $P= E(Z \g X,\theta)$ . \\   
 M-step: Optimize  $Q(\tilde{\theta},\theta)$ , given $P$. \\  
      
 For the Gaussian mixture model, with a Dirichlet-Normal-Wishart prior,  
 \begin{eqnarray*}  
  Q(\tilde{\theta},\theta) &=& 
    \ssum_{k=1}^m \ssum_{j=1}^n  p_k^j \bigl( 
    \log \tilde{w}_k +\log N(x^j\g \tilde{b}^k, \tilde{R}\{k\}) \bigr)   
    +fl(\tilde{\theta} \g \dot{\theta}) \\ 
  fl(\tilde{\theta} \g \dot{\theta}) &=& 
    \log D(\tilde{w}\g \dot{y}) \, 
   +\ssum_{k=1}^m  \log NW(\tilde{b}^k, \tilde{R}\{k\} \g 
                           \dot{n}_k,\dot{e}_k,\dot{u}^k,\dot{S}\{k\}) 
 \end{eqnarray*} 
 %
 Lagrange optimality conditions give 
 a simple analytical solutions for the M-step: 
 \[ 
 y = P{\bf 1}  \ \ , \ \  
 \tilde{w}_k  =  \left( y_k +\dot{y}_k -1 \right) \big/ 
                 \left( n -m +\ssum_{k=1}^m \dot{y}_k \right)     
 \]  
 \[ 
  u^k = {\textstyle \frac{1}{y_k}} 
        \ssum_{j=1}^n p_k^j x^j \ \ , \ \  
  S\{k\} = \ssum_{j=1}^n p_k^j 
         (x^j -\tilde{b}^k)\kron (x^j -\tilde{b}^k)'   
 \] 
 \[         
 \tilde{b}^k =  \frac{ \dot{n}_k \dot{u}^k +y_k u^k } 
                            { \dot{n}_k +y_k }   \ \ , \ \ 
 \tilde{V}^k \ \ = \ \  \frac{ S\{k\}
  +\dot{n}_k (\tilde{b}^k -\dot{u}^k)\kron (\tilde{b}^k -\dot{u}^k)' 
   +\dot{S}\{k\} }{ y_k +\dot{e}_k -d }  
 \]

 \subsubsection{Global Optimization} 

  In more general (non-Gaussian) mixture models, if an analytical 
 solution for the M-step is not available, a robust local 
 optimization algorithm can be used, for example Martinez (2000). 
  The EM is only a local optimizer, but the MCMC provides plenty
 of good starting points, so we have the basic  elements 
 for a global optimizer.  
  To avoid using many starting points going to a 
same local maximum, we can filter the (ranked by the  posteriori) top
portion of the MCMC output using a clustering  algorithm, and select 
a starting point from each cluster. 
  For better efficiency, or more complex problems, the Stochastic  EM or
SEM algorithm can be used to provide starting points near each
important local maximum,   
 see Celeux (1995), Pflug (1996) and Spall (2003).

 \subsection{Experimental Tests and Final Remarks}

 The test case used in this study is given by a sample $X$ assumed to 
follow a mixture of bivariate normal distributions with unknown 
parameters, including the number of components. 
 $X$ is the {\it Iris virginica} data set, with sepal and petal length 
of 50 specimens (1 discarded outlier). 
 The botanical problem consists of determining whether or not 
 there are two distinct subspecies in the population, 
 see Anderson (1935), Fisher (1936) and McLachlan (2000). 
 Figure 1 presents the dataset and 
 posterior density level curves 
 for the parameters, $\theta^*$ and $\widehat{\theta}$, 
 optimized for the 1 and 2 component models.

 \begin{center} 
 \centerline
 {\includegraphics*[height=3.0in, width=7.0in, angle=0]{FIGIRIS.PDF}} 
 {Figure1: Iris virginica data and models with 
           one (left) and two (right) components}
 \end{center} 

 In the FBST formulation of the problem, the 2 components is the base  
model, and the hypothesis to be tested is the constraint of having only
1 component.   
 When implementing the FBST one has to be careful with trapping states
on the MCMC. These typically are states where one component has a small
number of  sample points, that become (nearly) collinear, resulting in a
singular posterior. This problem is particularly serious with the Iris
dataset because of the small precision,  only 2 significant digits, of 
the measurements. 
 A standard way to avoid this inconvenience is to use flat or minimally
informative priors, instead of non-informative priors, 
 see Robert (1996). 
 
 We used as flat prior parameters: 
 $\dot{y}={\bf 1}$, $\dot{n}=1$, $\dot{u}=u$, 
 $\dot{e}=3$, $\dot{S}=(1/n)S$.   
 Robert (1996) 
 uses, with similar effects,    
 $\dot{e}=6$, $\dot{S}=(1.5/n)S$.

 The FBST selects the 2 component model, rejecting $H$, if the evidence
against the hypothesis  is above a given threshold, $\ev(H)> \tau$, and
selects the 1 component model, accepting $H$, otherwise.  
 The threshold $\tau$ is chosen by empirical power analysis, 
 see Stern and Zacks (2002) and Lauretto et al. (2003). 
 Let $\theta^*$ and $\widehat{\theta}$ represent the  constrained 
 (1 component) and unconstrained (2 component)  maximum a posteriori 
(MAP) parameters optimized to the Iris dataset. 
 Next, generate two collections of $t$ simulated datasets of size $n$, 
the first collection at $\theta^*$, and the second at $\widehat{\theta}$. 
 $\alpha(\tau)$ and $\beta(\tau)$, the empirical type 1 and type 2
statistical errors, are the rejection rate in the first collection and 
the acceptance  rate in the second collection. 
A small, $t=500$, calibration run sets the threshold $\tau$ so to 
minimize the total error, $(\alpha(\tau)+\beta(\tau))/2$. 
 Other methods like sensitivity analysis, 
 see Stern (2004a,b), 
 and loss functions, 
 see Madruga (2001), 
 could also be used. 



 Biernacki and Govaert (1998) 
 studied similar mixture problems and
 compared several selection criteria, pointing as the best overall 
 performers:  
 AIC - Akaike Information Criterion,  
 AIC3 - Bozdogan's modified AIC, and 
 BIC - Schwartz' Bayesian Information Criterion. 
 These are regularization criteria, weighting the model fit
against the number of parameters, 
 see  Pereira and Stern (2001).  
 If $\lambda$ is the model log-likelihood, $\kappa$ its number of 
parameters, and $n$ the sample size, then, 
 \[ 
   AIC= -2\lambda +2\kappa \ , \ \  
   AIC3= -2\lambda +3\kappa \ \mbox{and} \ \ 
   BIC= -2\lambda +\kappa \log(n) \ . 
 \] 

 Figure 2 show $\alpha$, $\beta$, and the total error 
 $(\alpha+\beta)/2$. 
 The FBST outperforms all the regularization criteria. 
 For small samples, BIC is very biased, always selecting the 1 component
model.  
 AIC is the second best criterion, caching up with the FBST for sample 
sizes larger than $n=150$.

 \begin{center} 
 \centerline
 {\includegraphics*[height=3.0in, width=7.0in, angle=0]{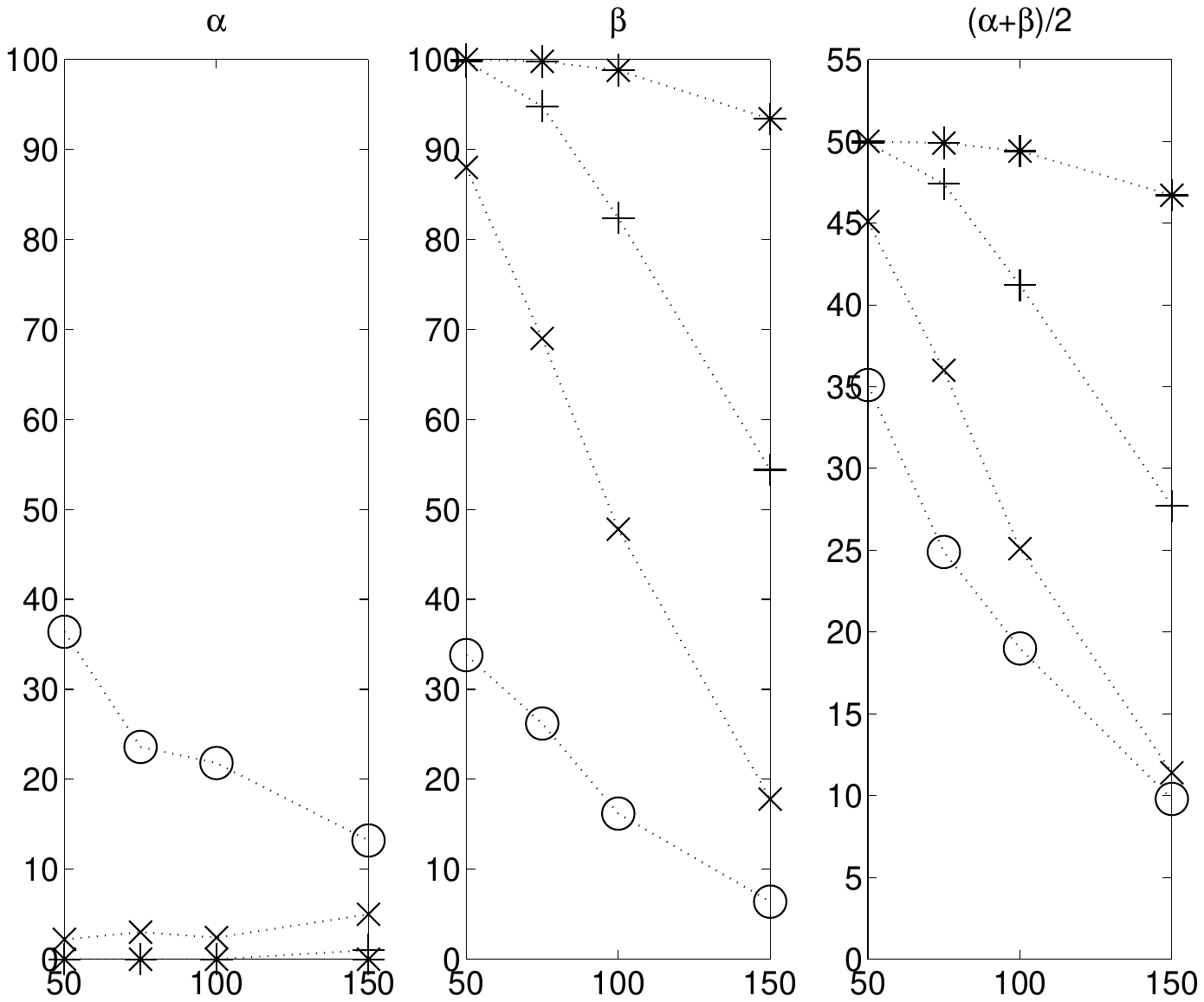}} 
 {Figure 2: Criteria  O= FBST, X= AIC, += AIC3, *= BIC,  \\
  Type 1, 2 and total error rates for different sample sizes}
 \end{center} 

 Finally, let us point out a related topic for further research: 
 The problem of discriminating between models consists of determining
which of $m$  alternative models, $f_k(x,\psi_k)$, more adequately fits
or  describes a given dataset. 
 In general the parameters $\psi_k$ have distinct dimensions, 
 and the models $f_k$ have distinct (unrelated) functional forms. 
 In this case it is usual to call them ``separate'' models 
(or hypotheses).   
  Atkinson (1970), 
 although in a very different theoretical  
framework, was the first to analyse this problem using a  mixture
formulation, 
 \[ f(x\g \theta)= \ssum_{k=1}^m w_k f_k(x,\psi_k) \ .\]

 The general theory for mixture models presented in this article can be
adapted to analyse the problem of discriminating  between separate
hypotheses. This is the subject of the authors'  ongoing research with
Carlos Alberto de Bragan\c{c}a Pereira and  Bas\'{\i}lio de Bragan\c{c}a
Pereira, to be presented in forthcoming  articles.  

 The authors are grateful for the support of  CAPES -
Coordena\c{c}\~{a}o de Aperfei\c{c}oamento de Pessoal de N\'{\i}vel
Superior,  CNPq - Conselho Nacional de Desenvolvimento Cient\'{\i}fico e
Tecnol\'{o}gico,  and FAPESP - Funda\c{c}\~{a}o de  Apoio \`{a} Pesquisa
do Estado de S\~{a}o Paulo.

\section{REAL Classification Trees}

This section presents an overview of REAL, The 
Real Attribute Learning Algorithm for automatic 
construction of classification trees. 
 The REAL project  started as an application to be used at the Brazilian
BOVESPA and BM\&F financial markets, trying to provide  a good algorithm
for predicting the adequacy of operation strategies.  In this context,
the success or failure of a given operation strategy corresponds to
different classes,  and the attributes are real-valued technical
indicators. The users demands for a decision  support tool also explain
several of the algorithm's unique features.

 The classification problems are stated as an $n \times (m+1)$ matrix
$A$. Each row, $A(i,:)$,  represents a different example, and each
column, $A(:,j)$, a  different attribute. The first $m$ columns in each
row are real-valued attributes, and  the last column , $A(i,m+1)$ is the
example's class. Part of these samples, the training set, is used by the
algorithm to generate a  classification tree, which is then tested with
the remaining examples. The error  rate in the classification of the
examples in the test set is a simple way of evaluating the
classification tree.

 A market operation strategy is a predefined set of rules determining an
operator's actions in the market. The strategy shall have a predefined
criterion for classifying a strategy application as success  or failure.
 
 As a simple example, let us define the strategy $buysell(t,d,l,u,c)$: 
 \begin{itemize} 
 \item At time $t$ buy a given asset {\it A}, at its price $p(t)$. 
 \item Sell {\it A} as soon as: 
 \begin{enumerate} 
  \item $t' = t+d$ , or 
  \item $p(t') = p(t)*(1+u/100)$ , or  
  \item $p(t') = p(t)*(1-l/100)$ .  
 \end{enumerate} 
 \item The strategy application is successful if  
       $c \leq 100*p(t')/(p(t) \leq u$ 
 \end{itemize} 
 The parameters $u$, $l$, $c$ and $d$ can be interpreted as the desired
and  worst accepted returns (low and upper bound), the strategy
application cost,  and a time limit.  


 \subsubsection{Tree Construction}  

 Each main iteration of the REAL algorithm corresponds to the branching
of a  terminal node in the tree. The examples at that node are
classified according to  the value of a selected attribute, and new
branches generated to each specific interval. The partition of a
real-valued attribute's domain in adjacent non-overlapping (sub)
intervals is the discretization process. Each main iteration of  REAL
includes: 
 \begin{enumerate}  
 \item The discretization of each attribute, and its evaluation by a
loss function. 
 \item Selecting the best attribute, and branching the node accordingly.
 \item Merging adjacent intervals that fail to reach a minimum
conviction threshold.   
 \end{enumerate}

 \subsection{Conviction, Loss and Discretization}  

Given a node of class $c$  with $n$ examples, $k$ of which are
misclassified and $(n-k)$ of  which are correctly classified, we needed
a single scalar parameter, $cm$, to measure the probability of
misclassification and its confidence level. Such a simplified conviction
(or trust) measure was a demand of REAL users operating at the  stock
market.    

Let $q$ be the misclassification probability for an example at a given
node, let $p=(1-q)$ be the probability of correct classification,  and
assume we have a Bayesian  distribution for $q$ , namely 
 $$D(c) = Pr(q \leq c) = Pr(p \geq 1-c)$$  

We define the conviction measure: $100*(1-cm)\%$, 
where 
 $$cm =  \mbox{min}\ c \ \ | \ \  Pr( q \leq c ) \geq 1 -g(c)$$ 

 \noindent 
 and $g(\ )$ is a monotonically increasing bijection of $[0,1]$ onto
itself. From our  experience in the stock market application we learned
to be extra cautious about  making strong statements, so we make 
 $g(\ )$ a convex function.  

In this paper $D(c)$ is the posterior distribution for a sample taken
from the  Bernoulli distribution, with a uniform prior for $q$:  
 \begin{eqnarray*}
  B(n,k,q) &=& comb(n,k) * q^k * p^{n-k} \\  
  D(c,n,k) &=& \int_{q=0}^c B(n,k,q) \ \ \  / \ \ \int_{q=0}^1 B(n,k,q) \\    
           &=& \mbox{betainc}(c,k+1,n-k+1)  
 \end{eqnarray*}  

Also in this paper, we focus our attention on  
 $$g(c)  =  g(c,r)  =  c^r  , \ \  r  \geq 1.0 $$ 
 we call $r$ the convexity parameter. 

With these choices, the posterior is the easily computed incomplete beta 
function, and $cm$ is the root of the monotonically decreasing function:  
 \begin{eqnarray*} 
  cm(n,k,r) &=& c \ \ |  \ \ f(c) = 0 \\ 
  f(c)      &=& 1 -g(c) -D(c,n,k) \\   
            &=& 1 -c^r -\mbox{betainc}(c,k+1,n-k+1)  
 \end{eqnarray*}   
 \noindent

 Finally, we want a loss function for the discretizations, based on the
conviction  measure. In this paper we use the overall sum of each
example classification  conviction, that is, the sum over all intervals
of the interval's conviction measure  times the number of examples in
the interval.   
 $$loss = \sum_i  n_i*cm_i$$


 Given an attribute, the first step of the discretization procedure is to
order  the examples in the node by the attribute's value, and then to
join together the  neighboring examples of the same class. So, at the
end of this first step, we  have the best ordered discretization for the
selected attribute with uniform class clusters. 

In the subsequent steps, we join intervals together, in order to
decrease the overall loss  function of the discretization. The gain of
joining $J$ adjacent  intervals,   $I_{h+1}$, $I_{h+2}$, \ldots
$I_{h+J}$ ,  is the relative decrease in the loss function 
  $$gain(h,j) = \sum_j loss(n_j,k_j,r)\ \  - loss(n,k,r) $$ 
 where $n= \sum_j n_j$ and $k$ counts the minorities' examples in the
new cluster (at  the second step $k_j=0$, because we begin with uniform
class clusters).  

At each step we perform the cluster joining operation with maximum gain.
The discretization procedure stops when  there are no more joining
operations with positive gain.  

The next examples show some clusters that would be joined together at
the first  step of the discretization procedure. The notation
$(n,k,m,r,\pm )$ means the we have  two uniform clusters of the same
class, of size $n$ and $m$, separated by a uniform  cluster of size $k$
of a different class; $r$ is the convexity parameter, and  $+$ ($-$) 
means we would (not) join the clusters together.

\begin{verbatim} 
 ( 2,1, 2,2,+) 
 ( 6,2, 7,2,-) ( 6,2, 8,2,+) ( 6,2,23,2,+) ( 6,2,24,2,-) 
 ( 7,2, 6,2,-) ( 7,2, 7,2,+) ( 7,2,42,2,+) ( 7,2,43,2,-) 
 (23,3,23,2,-) (23,3,43,2,-) (23,3,44,2,+) 
 (11,3,13,3,-) (11,3,14,3,+) (11,3,39,3,+) (11,3,40,3,-) 
 (12,3,12,3,-) (12,3,13,3,+) (12,3,54,3,+) (12,3,55,3,-) 
\end{verbatim}

In these examples we see that it takes extreme clusters of a balanced
and large  enough size, $n$ and $m$, to ``absorb'' the noise or impurity
in the middle cluster of  size $k$. A larger convexity parameter, $r$,
implies a larger loss at small clusters,  and therefore makes it  easier
for sparse impurities to be absorbed.

 \subsection{Branching and Merging}  

For each terminal node in the tree, we 
 \begin{enumerate} 
 \item perform the discretization procedure for each available attribute, 
 \item measure the loss function of the final discretization,  
 \item select the minimum loss attribute, and 
 \item branch the node according this  attribute discretization. 
 \end{enumerate} 
 If no attribute discretization decreases the loss function  by a
numerical precision threshold $\epsilon > 0$, no branching takes place.  

 A premature discretization by a parameter selected at a given level may
preclude further improvement of the classification tree by the branching
process.  For this reason we establish a conviction threshold, $ct$, and
after each branching  step we merge all adjacent intervals that do not 
achieve $cm < ct$. To prevent an  infinite loop, the loss function value
assigned to the merged interval is sum of the losses of the merging 
intervals. At the final leaves, this merging is undone.  
 The conviction threshold naturally stops the branching process,
so there is no need for an external pruning procedure, like in 
most TDIDT algorithms.   


 In the straightforward implementation, REAL spends most of the execution
time  computing the function $cm(n,k,r)$.  We can greatly accelerate the
algorithm by  using precomputed tables of $cm(n,k,r)$ values for small
$n$, and precomputed tables of $cm(n,k,r)$ polynomial interpolation
coefficients  for larger $n$. To speed up the algorithm we can also
restrict the search for join operations at the discretization step to
small neighborhoods, i.e. to  join only $3 \leq J \leq Jmax$ clusters:
Doing so will expedite the algorithm without any noticeable consistent
degradation.

 For further details on the numerical implementation, benchmarks, 
and the specific market application, see Lauretto et al. (1998).

%% file: CAPDOPT.TEX
\chapter{Deterministic Evolution and Optimization}  

 This chapter presents some methods of deterministic optimization. 
 Section 1 presents the fundamentals of Linear Programming (LP), its
duality theory, and  some variations of the Simplex algorithm.  
 Section 2 presents some basic facts of constrained and unconstrained
Non-Linear Programming (NLP), the Generalized Reduced Gradient (GRG)
algorithm for constrained NLP problems, the ParTan method for
unconstrained NLP problems, and some simple line search algorithms 
for uni-dimensional problems. 
 Sections 1 and 2 also presents some results about these algorithms 
local and global convergence properties. 
 Section 3 is a very short introduction to variational problems and 
the Euler-Lagrange equation.

 The algorithms presented in sections 1 and 2 are within the class of 
{\it active set} or active constraint algorithms. 
 The choice of concentrating on this class is motivated by
some properties of active set algorithms, that makes them specially
useful in the applications concerning the statistics, namely:

 - Active set algorithms maintain viability throughout in the search 
path for the optimal solution. 
 This is important if the objective function can only be computed  at
(nearly) feasible arguments, as it is often the case in statistics 
or simulation problems. 
 This feature also makes active set algorithms relatively easy to 
expalain and implement. 

 - The general convergence theory of active set algorithms  and the
analysis of specific problems may offer a constructive proof of the 
existence or the verification of stability conditions for an 
equilibrium or fixed point, representing a systemic eigen-solution  
 see, for example,  Border (1989), Ingrao and Israel (1990)   
 and Zangwill (1964). 

 - Active set algorithms are particularly efficient for small or medium
size re-optimization  problems, that is, for optimization problems
where the initial solution  or staring point for the optimization
procedure is (nearly) feasible and already close to the optimal
solution, so that the optimization algorithm is only used to 
finetune the solution. 
 In FBST applications, such good staring points can be obtained  from an
exploratory search performed by the Monte Carlo or Markov Chain Monte
Carlo procedures used to numerically integrate the FBST $e$-value,  
$\ev(H)$, or truth function $W(v)$, see appendices A and G.

 \section{Convex Sets and Polyhedra}

 The matrix notation used in this book is defined at section F.1.

 \subsubsection{Convex Sets}

 A point $y(l)$ is a {\em convex combination} of $m$ points of $\Re ^n$, 
given by the columns of matrix $X,\ n\times m$, iff 
 \[ \forall i \ , \ \  
    y(l)_i = \sum_{j=1}^{m} l_j * X_i^j \ , \ \ \  
    l_j\geq 0 \ \mid \ \sum_{j=1}^{m} l_j = 1 \ , 
 \] 
 or, equivalently, in matrix notation, iff 
 \[  y(l) = \sum_{i=1}^m l_i * X^{j} \ , \ \ \ 
     l_j\geq 0 \ \mid \ \sum_{j=1}^{m} l_j = 1 \ , 
 \] 
 or, in yet another equivalent form, 
 replacing the summations by inner products, 
 \[ 
   y(l) = Xl \ ,\ \ l\geq 0 \mid {\bf 1}'l=1 \ . 
 \]

 In particular, the point $y(\lambda )$ is a convex combination of 
two points, $z$ e $w$, if  
 \[ y(\lambda ) = (1-\lambda )z +\lambda w \ , \ \ 
    \lambda \in [0,\ 1] \ . 
 \] 
 Geometrically, these are the points in the line segment from $z$ to $w$. 
 
 A set, $C\in \Re ^n$, is {\em convex} iff it contains all convex 
combinations of any two of its points.  
 A set, $C\in \Re ^n$, is {\em bounded} iff the distance between 
any two of its points is bounded: 
 \[ \exists \delta \mid \forall x1 ,\ x2 \in C \ , \ \ 
    || x1 - x2 || \leq \delta 
 \]
 Figure D.1 presents some sets exemplifying the definitions above.

 \begin{figure}[hbt] 
 \centerline{\includegraphics*[height=4.4in, width=6.4in, angle=0]{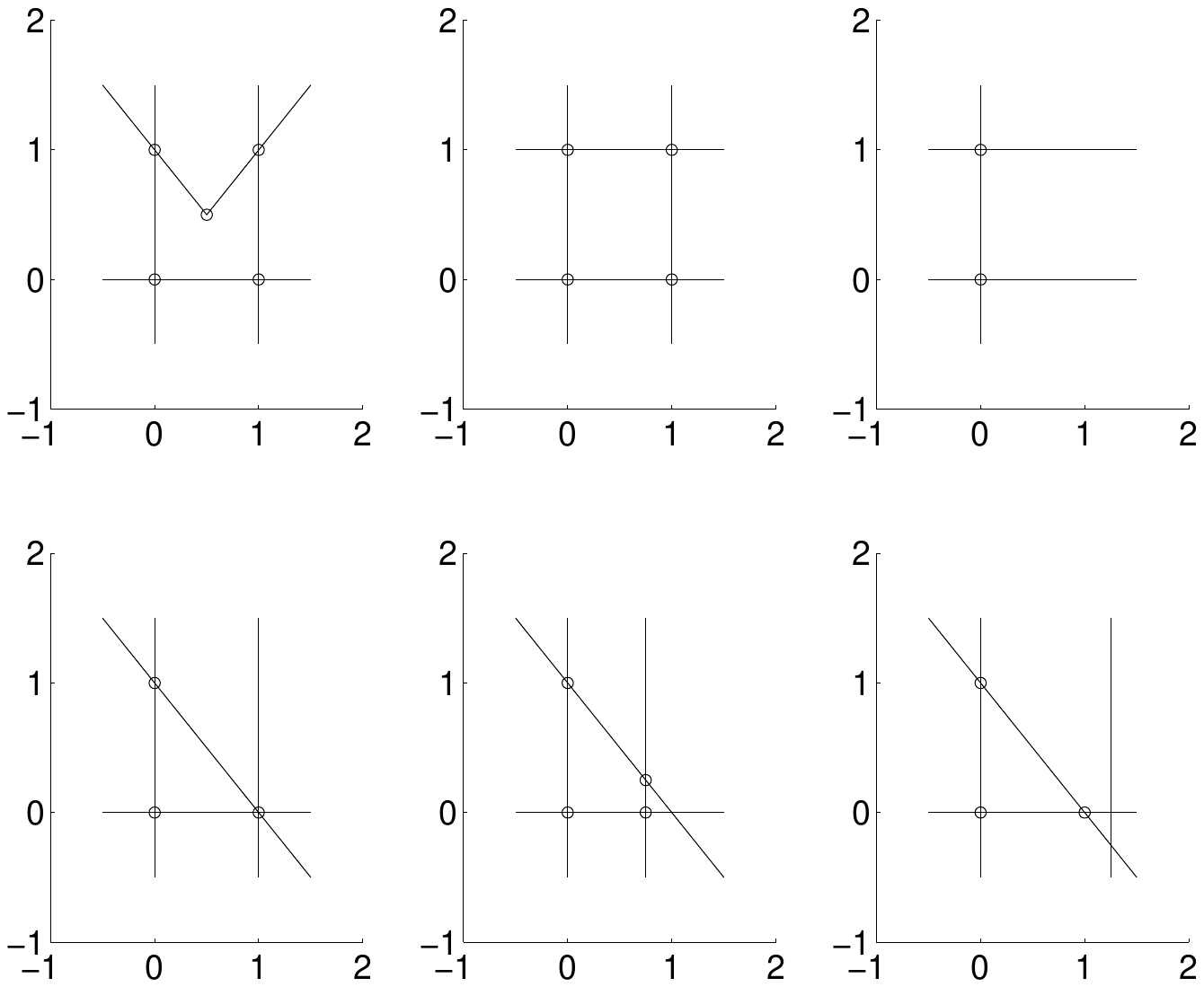}}
 \centerline{Figure D.1: (a) non-convex set, 
              (b,c) bounded and unbounded polyhedron,} 
 \centerline{(d-f) degenerate vertex perturbed to a single 
             or two nondegenerate ones.} 
  \end{figure}

 An {\it extreme point} of a convex set $C$ is a point $x\in C$ 
 that can not be represented as a convex combination of  two other
points of $C$. 
 The {\it profile} of a convex set $C$, $\mbox{ext}(C)$, 
 is the set of its extreme points. 
 The {\it Convex hull}  and the {\it closed convex hull} of a set $C$, 
 $\mbox{ch}(C)$ and $\mbox{cch}(C)$, are the intersection of all convex 
 sets, and closed convex sets, containing $C$. 

 {\bf Theorem:} A compact (closed and bounded) convex set is equal 
 to the closed convex hull of its profile, that is, 
 $C=\mbox{cch}(\mbox{ext}(C))$.

 The {\em epigraph} of a curve in $\Re ^2$, $y=f(x)$, $x\in [a,b]$,
 is the set defined as  
  $ \mbox{epig}(f) \equiv \{ (x,y)\mid x\in [a,b] \wedge y\geq f(x) \}$.  
 A curve is said to be convex iff its epigraph is convex. 
 A curve is said to be {\it concave} iff $-f(x)$ is convex.

 {\bf Theorem:} A curve, $y=f(x)$, $\Re \mapsto \Re$, 
 that is continuously differentiable and has monotonically 
 increasing first derivative is convex.

 {\bf Theorem:} The convex hull of a finite set of points, $V$, 
 is the set of all convex combinations of points of $V$, that is, if 
 $V=\{x^i, i=1\ldots n\}$, then  
 $\mbox{ch}(V)=\{x \mid x=[x^1,\ldots x^n]l, l\geq 0, {\bf 1}'l=1\}$.

  A (non-linear) {\it constraint}, in $\Re ^n$, is an inequality 
 of the form $g(x)\leq 0$, $g:\ \Re ^n \mapsto \Re $.  
  The {\it feasible region} defined by $m$ constraints,   
 $g(x)\leq 0$, $g:\ \Re ^n \mapsto \Re ^m$, 
 is the set of feasible (or viable) points 
 $\{ x \mid g(x)\leq 0 \}$.  
 At the feasible point $x$, the constraint $g_i(x)$ is said to be 
 {\it active} or {\it tight} if the equality, $g_i(x)=0$, holds, and 
 it is said to be {\it inactive} or {\it slack} if the strict 
 inequality, $g_i(x)<0$, holds.

 \subsubsection{Polyhedra}

 A {\em polyhedron} in $\Re ^n$ is a feasible region  defined by 
 {\it linear constraints}:  $Ax\leq d$.  
 We can always compose an equality constraint, $a'x= \delta$ with two 
 inequality constraints,  $a'x\leq \delta$ e $a'x\geq \delta$.

 {\bf Theorem:} Polyhedra are convex, but not necessarily bounded.

 A {\em face} of dimension $k$, of a polyhedron in $\Re^n$ with $m$
 equality constraints, is a feasible region that obeys tightly to 
 $n-m-k$ of the polyhedron's inequality constraints. 
 Equivalently, a point that obeys to $r$ active inequality constraints 
 is at a face of dimension $k=n-m-r$.   
 A  {\em vertex} is a face of dimension $0$. 
 An {\em edge} is a face of dimension $1$, 
 An {\em interior point} of the polyhedron has all inequality 
 constraints slack or inactive, that is, $k=n-m$. 
 A {\it facet} is a face of dimension $k=n-m-1$. 

 It is possible to have a point in a face of negative dimension.  
 For example, Figure D.1 shows a point where $n-m+1$ inequality 
 constraints are active.  
 This point is ``super determined'', since it is a point in $\Re^n$ 
 that obeys to $n+1$ equations, $m$ equality constraints and 
 $n-m+1$ active inequality constraints. 
 Such a point is said to be {\it degenerate.}    
 From now on we assume the {\it non-degenerescence hypothesis},  
 stating that such points do not exist in the optimization problem
 at hand. 
 This hypothesis is very reasonable, since the slightest 
 perturbation to a degenerate problem transforms a degenerate point 
 into one or more vertices, see Figure D.1.

 A polyhedron in {\em standard form}, $P_{A,d} \subset \Re^n$, 
 is defined by $n$ {\em signal constraints}, $x_i\geq 0$, and 
 $m<n$ {\em equality constraints}, that is,  
 \[ 
    P_{A,d} = \left\{ x\geq 0 \ \mid \ Ax=d \right\} 
    \ \ ,\ A\ m\times n 
 \]     
 
 We can always rewrite a polyhedron in standard form  
 (at a higher dimensional space) using the following artifices: 
 \begin{itemize}  
 \item[1.] Replace an unconstrained variable, $x_i$ by the difference 
  of two positive ones, $x_i^{+} - x_i^{-}$ where 
    $x_i^{+} = max \{0,x_i \}$ e $x_i^{-} = max \{0,-x_i \}$.
     
 \item[2.] Add a {\em slack variable}, $\chi \geq 0$ to each 

 inequality  
    \[ a'x\leq \delta \ \Leftrightarrow \  
       \left[ \begin{array}{cc} a & 1 \end{array} \right] 
       \left[ \begin{array}{c} x \\ \chi \end{array} \right] 
       = \delta \ \ .  
    \]  
 \end{itemize}

 From the definition of vertex we can see that, in a polyhedron
in standard form, $P_{A,d}$, a vertex is a feasible point where $n-m$ 
constraints are active. 
 Hence, $n-m$ variables are null; these are the 
 {\em residual variables} of this vertex. 
 Let us permute the vector $x$ so to place the residual variables 
at the last $n-m$ positions. 
 Hence, the remaining (non-null)  variables, the {\em basic variables}
will be at the first $m$ positions. 
 Applying the same permutation to the columns of matrix $A$, 
 the block of the first $m$ columns is called the {\em basis}, $B$,   
 of this vertex, while the block of the remaining $n-m$ columns of $A$ 
 is called the {\em residual matrix}, $R$. 
 That is, given vectors $b$ and $r$ with the basic and residual indices, 
 the permuted matrix $A$ can be partitioned as 
 \[  
 \left[ \begin{array}{cc} A^b & A^r \end{array} \right] = 
 \left[ \begin{array}{cc} B & R \end{array} \right] 
 \] 
 In this form, it is easy to write the non-null variables explicitely, 
 \[ 
 \left[ \begin{array}{c} x_b \\ x_r \end{array} \right] 
 \geq 0 \mid 
 \left[ \begin{array}{cc} B & R \end{array} \right] 
 \left[ \begin{array}{c} x_b \\ x_r \end{array} \right] = d 
 \ \ \mbox{hence,} 
 \] 
 \[ 
    x_b = B^{-1} \left[ d- Rx_r \right] \ \ 
 \]  
 Equating the residual variables to zero, it follows that 
 \[  
    x_b = B^{-1}d \ \ . 
 \] 

 From the definition of degenerescence we see that the vertex of a 
polyhedron in standard form is degenerate iff it has a null basic variable.

 \section{Linear Programming} 

 This section presents Linear Programming, the simplest optimization
problem studied in  multi-dimensional mathematical programming. 
 The simple structure of LP allows the formal development of relatively 
simple solution algorithms, namely, the primal and dual simplex. 
 This section also presents some decomposition techniques used for
solving LP problems in special forms.

 \subsection{Primal and Dual Simplex Algorithms} 

 A LP problem in standard form asks for the minimum of a linear 
function inside a polyhedron in standard form, that is, 
 \[ 
    \mbox{min}\ cx, \ \ x\geq 0 \mid Ax=d \ \ . 
 \] 

 Assume we know which are the residual (zero) variables of a given 
vertex. In this case we can form basic and residual index vectors, 
$b$ and $r$, and obtain the basic (non-zero) variables of this vertex. 
 Permuting and partitioning all objects of the LP
problem according  to the order established by the basic and residual
index vectors,  the LP problem is written as      
 \[ 
   \mbox{min}\ \ 
 \left[ \begin{array}{cc} c^b & c^r \end{array} \right] 
 \left[ \begin{array}{c} x_b \\ x_r \end{array} \right] \ , \ \ 
 x \geq 0 \mid 
 \left[ \begin{array}{cc} B & R \end{array} \right] 
 \left[ \begin{array}{c} x_b \\ x_r \end{array} \right] = d \ \ . 
 \] 
 using the notation 
 \[ \tilde d \equiv B^{-1}d \ \mbox{and} \ 
  \tilde R \equiv B^{-1}R \ , 
 \] 
 the basic solution corresponding to this vertex is \ 
 $x_b = \tilde d$ (and $x_r=0$).

 Let us now proceed with an analysis of the sensitivity of this basic 
solution by a perturbation of a single residual variable. 
 If we change a single residual variable, say the $j$-th element 
of $x_r$, allowing it to become positive, that is, making $x_{r(j)} > 0$, 
the basic solution, $x_b$ becomes   
 \begin{eqnarray*} 
  x_b 
  &=& \tilde d - \tilde R x_r \\ 
  &=& \tilde d - x_{r(j)} \tilde R^j  
 \end{eqnarray*} 

 This solution remains feasible as long as it remains non-negative. 
 Using the non-degeneregescence hypothesis, $\tilde d >0$, and 
 we know that it is possible to increase the value of $x_{r(j)}$, 
 while keeping the basic solution feasible, up to a threshold 
 $\epsilon >0$, when some basic variable becomes null. 

 The value of this prturbed basic solution  is 
 \begin{eqnarray*} 
   cx 
   &=&  c^b x_b + c^r x_r \\
   &=&  c^b B^{-1} [d- Rx_r] + c^r x_r \\  
   &=&  c^b \tilde d + (c^r - c^b \tilde R) x_r \\ 
   &\equiv &  \varphi - z x_r \\ 
   &=& \varphi - z^j x_{r(j)} 
 \end{eqnarray*}   
 Vector  $z$ is called the {\em reduced cost} of this basis. 

 The sensitivity analysis suggests the following algorithm used to 
generate a sequence of vertices of decreasing values, 
starting from an initial vertex, $[x_b\g x_r]$. 
 
\vspace{0.5cm}

\noindent
{\bf Simplex Algorithm:} 

\vspace{0.5cm}

 \begin{enumerate}  
 \item Find a residual index $j$, such that $z^j >0$. 
 \item Compute, for $k \in K\equiv \{l \mid \tilde R _l^j >0\}$\ , 
      \ \ $\epsilon _k= \tilde d _k / \tilde R _k^j$ \ ,  \\     
      and take \ $i\in \mbox{Argmin} _{k \in K} \ \  \epsilon_k$ \ ,  
      \ \ i.e. $\epsilon (i) = \min_k \epsilon_k$\ .  
             
 \item Make the variable $x_{r(j)}$ basic, and $x_{b(i)}$ residual. 
 \item Go back to step 1. 
 \end{enumerate} 

 The simplex can not proceed if $z\leq 0$ at the first step, 
 or if, at the second step, the mimimum is taken over the empty set.  
 In the second case the LP problem is unbounded. 
 In the first case the current vertex is an optimal solution!   

 Changing the status basic / residual of a pair of variables is,  
 in the LP jargon, {\it to pivot}. 
 After each pivoting operation  the basis inverse needs to be
 recomputed, that is, the basis needs to be {\it reinverted}. 
  Numerically efficient implementation of the Simplex do not actually
 keep the basis inverse, instead, the basis inverse is represented  
 by a numerical factorization, like $B=LU$ or $B=QR$.  
  At each pivot operation the basis is changed by a single column, 
 and there are efficient numerical algorithms used to update the 
 numerical factorization representing the basis inverse, 
 see Murtagh (1981) and Stern (1994).     
  
 Example 1: Let us illustrate the Simplex algorithm 
 solving the following simple example.

 Let us consider the LP problem  
 $\mbox{min} [-1,-1] x , \ 0 \leq x \leq 1$. \\ 
 This problem can be restated in standard form:  
 \[ 
  c= \left[ \begin{array}{cccc} -1 & -1 & 0 & 0 \end{array} \right] \ \ \ 
  A= \left[ \begin{array}{cccc} 1 & 0 & 1 & 0 \\ 
                                0 & 1 & 0 & 1 \end{array} \right]    \ \ \ 
  d= \left[ \begin{array}{c} 1 \\ 1 \end{array} \right]
 \] 
 The initial vertex $x=[0,0]$ is assumed to be known. \\ 
 Step 1:  
 $r=[1,2]$, $b=[3,4]$, $B=A(:,b)=I$, $R=A(:,r)=I$, \\  
 $-z= c^r-c^b\tilde{R}= [-1,-1] -[0,0] \Rightarrow z= [1,1]$, 
 $j=1$, $r(j)=1$, \\ 
 $x_b= \tilde{d} -\epsilon \tilde{R}^j = 
  \left[ \begin{array}{c} 1 \\ 1 \end{array} \right]
  -\epsilon \left[ \begin{array}{c} 1 \\ 0 \end{array} \right] 
 \Rightarrow \epsilon^*=1 , \  i=1 , \ b(i)=3$ \\  
 Step 2:  
 $r=[3,2]$, $b=[1,4]$, $B=A(:,b)=I$, $R=A(:,r)=I$, \\  
 $-z= c^r-c^b\tilde{R}= [0,-1] -[-1,0] \Rightarrow z= [-1,1]$, 
 $j=2$, $r(j)=2$, \\ 
 $x_b= \tilde{d} -\epsilon \tilde{R}^j = 
  \left[ \begin{array}{c} 1 \\ 1 \end{array} \right]
  -\epsilon \left[ \begin{array}{c} 0 \\ 1 \end{array} \right] 
 \Rightarrow \epsilon^*=1 , \  i=2 , \ b(i)=4$ \\  
 Step 3:  
 $r=[3,4]$, $b=[1,2]$, $B=A(:,b)=I$, $R=A(:,r)=I$, \\  
 $-z= c^r-c^b\tilde{R}= [0,0] -[-1,-1] \Rightarrow z= [-1,-1] <0$

 \subsubsection*{Obtaining the initial vertex} 

 In order to obtain an initial vertex, used to start the simplex, 
 we can use the auxiliary LP problem, 
 \[ 
 \mbox{min} \ 
 \left[ \begin{array}{cc} 0 & 1 \end{array} \right] 
 \left[ \begin{array}{c} x \\ y \end{array} \right] 
 \ \ \mid \ \ 
 \left[ \begin{array}{c} x \\ y \end{array} \right] 
 \geq 0 \ \wedge \  
 \left[ \begin{array}{cc} A & 
        \mbox{diag}(\mbox{sign}(d))\end{array} \right] 
 \left[ \begin{array}{c} x \\ y \end{array} \right] 
 = d 
 \]  

 An initial vertex for the auxiliary problem is given by 
 $ \left[ \begin{array}{cc} 0 & \mbox{abs}(d') \end{array} \right]'$.  
 If the auxiliary problem has an optimal solution of value zero, 
 the optimal solution gives a feasible vertex for the original problem; 

 if not, the original problem is unfeasible.

 \subsubsection{Duality} 

 Given an LP problem, called the {\em primal} LP problem, we define 
a second problem, the {\em dual} problem (of the primal problem).  
 Duality theory establishes important relations between the solution 
of the primal LP and the solution of its dual.  
 
 Given a LP problem in {\em canonic} form, 
 \[ 
     \mbox{min}\ \ cx \mid x\geq 0 \wedge Ax\geq d \ , 
 \] 
 its dual LP problem is defined as 
 \[ 
    \mbox{max}\ \ y'd \mid y\geq 0 \wedge y'A\leq c \ .  
 \]

 The primal and dual problems in canonic form have an intuitive 
economical interpretation. 
 The primal problem can be interpreted as the classic ration problem: 
 $A_i^j$ is the quantity of nutrient of type $j$ found in one unit of 
aliment of type $i$. 
 $c^i$ is the cost of one unit of aliment $i$, and $d_j$ the 
minimum daily need of nutrient $j$. 
 The primal optimal solution is a nutritionally feasible ration of 
minimum cost. 
 The dual problem can be interpreted as a manufacturer of synthetic 
nutrients, looking for the ``market value'' for its nutrients line. 
 The manufacturer income per synthetic ration is the objective function 
to be maximized. In order to keep its line of synthetic nutrients 
competitive, no natural aliment should provide nutrients cheaper than 
the corresponding synthetic mixture, these are the dual problem's 
constraints.      
 The optimal prices for the synthetic nutrients, $y^*$ can also be 
interpreted as marginal prices, giving the differential price increment  
of aliment $i$ by differential increase of its content of nutrient $j$. 
 The correctness of these interpretations are demonstrated by the duality 
properties discussed next.

 {\bf Lemma 1:} The dual of the dual is the primal PL problem. 

 Proof: Just observe that the dual of the primal LP in canonic form 
is equivalent to 
 \[ 
     \mbox{min}\ \ -y'd \mid y\geq 0 \wedge -y'A\geq -c \ \ .  
 \] 
 This problem is again in canonic form, and can be immediately dualized, 
yielding a problem equivalent to the original LP problem. 

\vspace{0.5cm}

 {\bf Weak Duality Theorem:} 
 If $x$ and $y$ are, respectively, feasible solutions of the primal and 
dual problems, then there is a non-negative gap between their values 
as solutions of these problems, that is,   
 \[ 
     cx \geq y'd \ \ . 
 \]   

 Proof: 
 By feasibility, $Ax\geq d$ and $y\geq 0$. Hence, $y'Ax\geq y'd$. 
 In the same way, $y'A\leq c$ and $x\geq 0$. Hence $y'Ax\leq cx$.  
 Therefore, $cx\geq y'd$.  QED. 

\vspace{0.5cm}

 {\bf Corollary 1:} 
 If we have a pair of feasible solutions, $x$ for the primal LP problem 
and $y$ for its dual, and their values as primal and dual solutions 
coincide, that is, $cx^* =(y^*)'d$, then both solutions are optimal. 

\vspace{0.5cm}
 
 {\bf Corollary 2:} 
 If the primal LP problem is unbounded, its dual is unfeasible. 

\vspace{0.5cm}

 As we could re-write any LP problem in standard for, we can re-write 
any LP problem in canonical form. 
 Hence, from Lemma 1, we know that the duality relation is defined 
between pairs of LP problems, whatever the form they have been writen. 

 \vspace{0.5cm}

 {\bf Lemma 2:}  
 Given a primal in standard form, 
 \[ 
    \mbox{min}\ \ cx \mid x\geq 0 \wedge Ax = d \ \ , 
 \]
 its dual is  
 \[ 
     \mbox{max}\ \ y'd \mid y\in \Re ^m \wedge y'A\leq c \ \ . 
 \] 

\vspace{0.5cm}
  
 {\bf Theorem} (Simplex proof of correctness):  
  
 We shall prove that the Simplex stops at an optimal vertex. 
 At the Simplex halting point we have 
 $z = -( c^r - c^b B^{-1}R ) \leq 0$. 
 Let us consider $y'= c^bB^{-1}$ 
 as a candidate solution for the dual.  
 \begin{eqnarray*} 
  \lefteqn{ 
     \left[ \begin{array}{cc} c^b & c^r \end{array} \right] 
      - y' \left[ \begin{array}{cc} B & R \end{array} \right] } \\  
    &=& \left[ \begin{array}{cc} c^b & c^r \end{array} \right] 
     - c^b B^{-1}  \left[ \begin{array}{cc} B & R \end{array} \right] \\ 
    &=& \left[ \begin{array}{cc} c^b & c^r \end{array} \right] 
     - c^b  \left[ \begin{array}{cc} I & \tilde R \end{array} \right] \\ 
    &=&  \left[ \begin{array}{cc} c^b & c^r \end{array} \right] 
     - \left[ \begin{array}{cc} c^b & c^b \tilde R \end{array} \right] \\  
    &=& \left[ \begin{array}{cc} 0 & -z \end{array} \right] \geq 0 
 \end{eqnarray*}   
 Hence, $y$ \'{e} is a feasible dual solution.  
 Moreover, its value (as a dual solution) is 
 $y'd = c^b B^{-1} d = c^b \tilde d = \varphi$, 
 and, by corollary 1, both solutions are optimal. 

\vspace{0.5cm}

 {\bf Theorem} (Strong Duality):  
 If the primal problem is feasible and bounded, so is its dual. 
 Moreover, the value of the primal and dual solutions coincide. 

 Proff: Constructive, by the Simplex algorithm. 

\vspace{0.5cm}
 
 {\bf Theorem} (Complementary Slackness): 
 Let $x$ and $y'$ be feasible solutions to a LP in standard form and 
its dual. These solutions are optimal iff 
 $w'x=0$, where $w=(c-y'A)$. 
 The vectors $x$ and $w$ represent the slackness in the inequality 
constraints of the primal and dual LP problems. 
 Since $x\geq 0$ and $w\geq 0$, the scalar product $w'x$ is null iff  
each of its terms, $w_j x_j$,  is null;  
 or equivalently, if for each slack inequality constraint in the primal,
the corresponding inequality constraint in the dual is tight, and
vice-versa. Hence the name complementary slackness. 

 Proof: If the solutions are optimal, we could have obtained them 
by the Simplex algorithm. 
 As in the Simplex proof of correctness, let 
 \[ 
   (c-y'A)x =   
   \left[ \begin{array}{cc} 0 & z \end{array} \right] 
   \left[ \begin{array}{c} x_b \\ 0 \end{array} \right] 
   = 0 
 \]  
 If $(c-y'A)x=0$, then $y'(Ax)=cx$, or $y'd=cx$, 

 and by the first corollary of the weak duality theorem, 
 both solutions are optimal.

 \subsubsection*{General Form of Duality} 

 The following are LP problem of the most general form and its dual.   
 An asterisk, $*$, indicates an unconstrained sub-vector 

 The general Primal LP problem, 
 \[  
 \min 
 \left[ \begin{array}{ccc} 
   \map{c}{1} & \map{c}{2} & \map{c}{3} \end{array} \right]
 \left[ \begin{array}{c} 
    \map{x}{1} \\ \map{x}{2} \\ \map{x}{3} \end{array} \right] \ \ \  
 \left. 
 \begin{array}{ccc} 
  \map{x}{1} & \geq & 0 \\ 
  \map{x}{2} & * &  \\ \map{x}{3} & \leq & 0  
 \end{array} \right.
 \ \ \mid \   
 \] 
 \[ 
 \left[ \begin{array}{ccc} 
  \mapq{A}{1}{1} & \mapq{A}{1}{2} & \mapq{A}{1}{3} \\  
  \mapq{A}{2}{1} & \mapq{A}{2}{2} & \mapq{A}{2}{3} \\ 
  \mapq{A}{3}{1} & \mapq{A}{3}{2} & \mapq{A}{3}{3} \\\end{array} \right]
 \left[ \begin{array}{c} 
    \map{x}{1} \\ \map{x}{2} \\ \map{x}{3} \end{array} \right]
 \left. \begin{array}{cc} 
    \leq & \map{d}{1} \\ = & \map{d}{2} \\ \geq & \map{d}{3} 
  \end{array} 
 \right.  
 \] 
 
 and its Dual LP problem: 
 \[ 
 \max 
 \left[ \begin{array}{ccc} 
   \map{d}{1}' & \map{d}{2}' & \map{d}{3}' \end{array} \right]
 \left[ \begin{array}{c} 
   \map{y}{1} \\ \map{y}{2} \\ \map{y}{3} \end{array} \right] \ \ \  
 \left. \begin{array}{ccc} 
  \map{y}{1} & \leq & 0 \\ \map{y}{2} & * &  \\ \map{y}{3} & \geq & 0  
 \end{array} \right.
 \ \ \mid \   
 \] 
 \[ 
 \left[ \begin{array}{ccc} 
  \mapq{A}{1}{1} & \mapq{A}{1}{2} & \mapq{A}{1}{3} \\  
  \mapq{A}{2}{1} & \mapq{A}{2}{2} & \mapq{A}{2}{3} \\ 
  \mapq{A}{3}{1} & \mapq{A}{3}{2} & \mapq{A}{3}{3}  
 \end{array} \right]'
 \left[ \begin{array}{c} 
   \map{y}{1} \\ \map{y}{2} \\ \map{y}{3} 
 \end{array} \right]
 \left. \begin{array}{cc} 
    \leq & \map{c}{1}' \\ 
    = & \map{c}{2}' \\ 
  \geq & \map{c}{3}' 
 \end{array} \right.
 \]

 The following interesting special case is known as the 
 standard linear programming problem with box constraints:   

 \[ 
 \mbox{Primal:} \ \min c'x \ \mid \ 
     Ax = d \ \wedge \ l \leq x \leq u  \ , 
 \] 
 \[  
 \mbox{Dual:} \ \max d' y +l' p  -u' q   \ \mid \ 
   A' y +p -q = c \ \wedge \ p, q \geq 0    \ .  
 \]

 \subsubsection{Dual Simplex Algorithm} 

 The Dual Simplex algorithm is analogous to the standard Simplex, 
 but it works caring a basis that is dual feasible, and works to 
 achieve primal feasibility. 
 The Dual Simplex is very useful in several situations in which we 
 solve a LP problem, and subsequently have to alter some constraints, 
 loosing primal feasibility. 
 We work with a standard LP program and its dual, 
 \[ 
   P:\; \min cx\; , \; x\geq 0 \; Ax=d \; \; \; \; \mbox{and} \; \; \; \; 
   D:\; \max y'd\; , \; y'A\leq c 
 \] 

 In dual feasible basis, $y=c^bB^{-1}$ 
 is a dual feasible solution, that is 
 \begin{eqnarray*} 
  \lefteqn{ 
     \left[ \begin{array}{cc} c^b & c^r \end{array} \right] 
      - y' \left[ \begin{array}{cc} B & R \end{array} \right] } \\  
    &=& \left[ \begin{array}{cc} c^b & c^r \end{array} \right] 
     - c^b B^{-1}  \left[ \begin{array}{cc} B & R \end{array} \right] \\ 
    &=& \left[ \begin{array}{cc} c^b & c^r \end{array} \right] 
     - c^b  \left[ \begin{array}{cc} I & \tilde R \end{array} \right] \\ 
    &=&  \left[ \begin{array}{cc} c^b & c^r \end{array} \right] 
     - \left[ \begin{array}{cc} c^b & c^b \tilde R \end{array} \right] \\  
    &=& \left[ \begin{array}{cc} 0 & -z \end{array} \right] \geq 0 
 \end{eqnarray*}

 Now, let us rewrite the dual in a form that is analogous to the 
 standard form, adding slack variables and using a partition 
 of the coefficient matrix, $[B,R]$, as follows: 
 \begin{eqnarray*} 
  \max d'y  & &  A'y\leq c'  \\ 
  \max d'y  & & 
       \left[ \begin{array}{c} B' \\ R' \end{array} \right] y \leq 
       \left[ \begin{array}{c} {c^b}' \\ {c^r}' \end{array} \right] \\ 
  \max d'y  & & 
  \left[ \begin{array}{ccc} B' & I & 0 \\ R' & 0 & I \end{array} \right] 
  \left[ \begin{array}{c} y \\ w_b \\ w_r \end{array} \right] = 
  \left[ \begin{array}{c} {c^b}' \\ {c^r}' \end{array} \right] 
  \; , \; w\geq 0  
 \end{eqnarray*}  
 In this form, the dual basis, its inverse, and corresponding 
basic solution are given by, 
 \[
  \left[ \begin{array}{cc} B' & 0 \\ R' & I \end{array} \right] \; , \;    
  \left[ \begin{array}{cc} B^{-t} & 0 \\ -R'B^{-t} & I \end{array} \right] 
 \] 
 \[  
  \left[ \begin{array}{c} y \\ w_r \end{array} \right] =     
  \left[ \begin{array}{cc} B^{-t} & 0 \\ -R'B^{-t} & I \end{array} \right] 
  \left[ \begin{array}{c} {c^b}' \\ {c^r}' \end{array} \right]     
  -\left[ \begin{array}{cc} B^{-t} & 0 \\ -R'B^{-t} & I \end{array} \right] 
  \left[ \begin{array}{c} I \\ 0 \end{array} \right] w_b \; \; \mbox{i.e.} 
 \] 
 \[     
   y= B^{-t}{c^b}' -B^{-t}w_b \; \; \; \; e \; \; \; \; 
   w_r= {c^r}' -R'B^{-t}{c^b}' +R'B^{-t}w_b 
 \] 

 Note that the indices in $b$ and $r$ correspond to basic and residual 
indices in the primal, the situation being reversed in the dual. 
 As in the standard Simplex, we can increase a zero element of the 
residual vector in order to have a better dual solution, 
 \[ 
   d'y = d'B^{-t}({c^b}' -w_b) = \mbox{const} -{\tilde{d}}' w_b 
 \] 
 If $\tilde{d}\geq 0$ the primal basic solution is feasible, 
 and we have the optimal solutions for both the primal and the dual 
 problems. 
 If there is an element $\tilde{d}_i < 0$, 
 we can increase the value of the dual solution increasing $w_{b(i)}$. 
 We can increase $w_{b(i)}=\nu$, without loosing dual feasibility, 
 as long as we maintain 
 \[ 
  w_r= {c^r}'-R'B^{-t}{c^b}' +\nu R'B^{-t}I^i 
   \geq 0\;\;\mbox{transposing} 
 \] 
 \[ 
  c^r -\tilde{c^b}R +\nu B^{-1}_i R \geq 0 \; \; \mbox{i.e.} 
 \]  
 \[ 
    -z +\nu \tilde{R}_i \geq 0  
 \] 

 Making   
 $j= \mbox{arg} \min \{\nu(j)=\: z^j / \tilde{R}_i^j 
      \; , \; j \mid \; \tilde{R}_i^j < 0 \}$, 
 we have the index that leaves the dual basis. 
  Hence, in the new list of indices $b$, that are primal basic, 
 we can exclude $b(i)$, include $r(j)$, update the basis' inverse 
 and proceed to a new dual simplex iteration, 
 until we reach dual optimality or, equivalently, primal feasibility.

 \subsection{Decomposition Methods}


 Suppose we have a LP problem in the form 
 \[ 
    \min cx, \ x \geq 0 \g Ax=b, 
     \ \ \ \mbox{where the matrix}  \  \ 
  A=\left[ \begin{array}{c} \dot{A} \\ \ddot{A} \end{array} \right] \ , 
 \] 
 and the polyhedron described by $\ddot{A}x=\ddot{b}$ 
 has a very ``simple'' structure, while $\dot{A}x=\dot{b}$ 
 implies only a ``few'' additional constraints that, unfortunately, 
 greatly complicate  the problem.

 For example, let $\ddot{A}x=\ddot{b}$ describe a set of separate 
 LP problems, while $\dot{A}x=\dot{b}$ imposes global constrains 
 coupling the variables of the several LP problems. 
 This structure is known as Row Block Angular Form (RBAF), 
 see section 5.2. 
  
 We now study the Danzig-Wolf method, that allow us to solve the 
 original LP problem, by successive iterations between a ``small''  
 {\it main} or {\it master} problem, and a large but ``simple'' 
 {\it subproblem} or {\it slave} problem. 
 We assume that the simple polyhedron is bounded, hence being the 
 convex hull of its vertices  
 \[ 
   \ddot{X}=\{x\geq 0 \mid \ddot{A}x=\ddot{b}\} = ch(V) 
   = Vl \; , \; l\geq 0 \mid {\bf 1}'l=1 \ . 
 \] 

 The origianl LP problem is equivalent to the following 
 master problem: 
 \[ 
  M:\;\; \min cVl \; , \; l\geq 0 \mid 
 \left[ \begin{array}{c} \dot{A}V \\ {\bf 1}' \end{array} \right] l = 
 \left[ \begin{array}{c} \dot{b} \\  1 \end{array} \right] 
 \] 
 obviously this representation has only theoretical interest, 
 for it is not practical to find the many vertices of $V$. 
 A given basis $B$ is optimal iff 
 \[  
  -z=[cV]_R -([cV]_B B^{-1})R \equiv [cV]_R - [y,\gamma]R \geq 0 \ . 
 \] 
 This condition  is equivalent of having, for any residual index, $j$,   
 \[ 
   cV^j - [y,\gamma] 
   \left[ \begin{array}{c} \dot{A}V^j \\ 1 \end{array} \right] \geq 0
   \; ,\; \mbox{or} 
 \] 
 \[   
   \gamma \leq cV^j - y \dot{A}V^j = (c - y \dot{A} )V^j \; , \; \mbox{or} 
 \] 
 \[  
  \gamma \leq \min (c - y \dot{A})v \; , \; v \in \ddot{X} 
 \]  

 Hence, we define the o sub-problem  
 \[ 
   S:\;\; \min (c - y \dot{A})v\; , \; 
   v\geq 0 \; \mid \; \ddot{A}v= \ddot{b} 
 \] 
 If the optimal solution of $S$, $v^*$ has optimal value  
 $(c - y\dot{A})v^*\geq \gamma$, 
 the basis $B$ is optimal for $M$. 
 If not, $v^*$ give us the next column for entering the basis, 
 $\left[ \begin{array}{c} \dot{A}v^* \\ 1 \end{array} \right]$.  

 The optimal solution of the auxiliary problem also give us a lower 
 bound for the original problem. 
 Let $x$ be any feasible solution for the original problem, that is,  
 $x\in \ddot{X} \mid \dot{A}x=\dot{b}$. 
 Since $x$ is more constrained, 
 $(c-y\dot{A})x\geq (c-y\dot{A})v^*$, hence, 
 $cx \geq y\dot{b} +(c-y\dot{A})v^*$. 
 Note that $y\dot{b}$ is the current upper bound. 
 Also note that it is not necessary to have a monotonic 
 increase in the lower bound. Hence we must keep track of the 
 best lower bound found so far. 

 As we have seen, the Danzig-Wolf works very well for LP problems in 
 RBAF.  If we had a problem in CBAF - Column Block Angular Form, 
 we could use Danzig-Wolf decomposition method on the problem's dual. 
 This is essentially Benders decomposition method, that can be 
 efficiently implemented using the Dual Simplex algorithm.

 \subsection*{Exercises} 

 \begin{itemize}

 \item[1.] Geometry and simple lemmas: 

 a- Draw the simplex, $S_n$,  and the cube, $C_n$ of dimension  
 2 and 3. $S=\{ x\geq 0 \mid {\bf 1}'x \leq 1 \}$, 
 $C=\{ x\geq 0 \mid Ix \leq {\bf 1} \}$. 

 b- Rewrite $S_2$, $S_3$, $C_2$ and $C_3$ as standard form plyedra in  
 $R^n$, where $n= 3, 4, 4, 6$, respectively.   

 c- Prove that a polyhedron (in standard form) is convex. 

 d- Prove duality lemmas 1 and 2.  

 e- Prove that a bounded polyhedron is the set of convex combinations 
of its vertices.

 \item[2.] Write a program to solve a LP problem in standard form 
 by exhaustive enumeration of its vertices. 
 Suggestion: Write a function to enumerate all arrangements, 
 $b=[b(1),\ldots b(m)]$, of $m$ indices from $1:n$, 
 in increasing order that is, $b(j)>b(i)$ for $j>i$.    
 For each arrangement, $b$ form the basis $B=A^b$. 
 Check if $B$ is invertible and, if so, check if the 
 basic solution is a vertex, that is, if it is feasible, 
 $\tilde d = B^{-1}d >0$. 
 Compute the value of all feasible basic solutions, 
 and select the best one.

 \item[3.] Adapt the Simplex algorithm to use the QR factorization of 
the basis. Explain how to update the factorization after a pivoting 
operation. 

 \item[4.] Adapt and implement the Simplex for LP problems with 
 box constraints, that is 
 \[ 
    \min cx\; , \;  l\leq x\leq u \: \mid \; Ax=d \ . 
 \]

 Hint: Consider a given feasible basis, $B$, and a partition 
 $\left[ \begin{array}{ccc} B & R & S \end{array} \right]$, \ where \\  
 $l_b < x_b < u_b$, $x_r = l_r$, $x_s = u_s$, \ so that, \\ 
 $x_b = B^{-1}d -B^{-1}Rx_r -B^{-1}Sx_s$ \ and \\      
 $cx = c^b B^{-1} d +(c^r -c^b B^{-1} R)x_r +(c^s -c^b B^{-1} S)x_s 
     = \varphi +z^r x_r +z^s x_s$    
 
 If $z^{r(k)}<0$, we can improve the current solution 
 increasing this residual variable residual at its lower bound, 
  $x_{r(k)} = l_{r(k)} +\delta_{r(k)}$,  \ making \\     
  $x_b = B^{-1}d -B^{-1}R l_r -B^{-1}S u_s -\delta_{r(k)} B^{-1}R^k$. \\ 
 However, $\delta_{r(k)}$ shall respect the following bounds: \\ 
 1- $x_{r(k)} = l_{r(k)} +\delta_{r(k)} \leq u_{r(k)}$, \ \   
 2- $x_b \geq l_b$, \ \ 
 3- $x_b \leq u_b$. \\ 
 In a similar way, if $z^{s(k)}>0$, we can improve the current solution 
 decreasing this residual variable at its upper bound, 
 $x_{s(k)} = u_{s(k)} -\delta_{s(k)}$.

 \item[5.] Adapt and implement the Dual Simplex for LP problems with 
box constraints. 

 \item[6.] Implement Danzig-Wolf decomposition methods for RBAF problems.

\end{itemize}

 \section{Non-Linear Programming}

 \subsubsection{Optimality and Lagrange Multipliers}

 We start this section giving an intuitive explanation of 
 Lagrange's optimality conditions for 
 a Non-Linear Programming (NLP) problem, given as   
 \[ 
  \min f(x),\ x \mid g(x)\leq 0 \wedge h(x)=0\ , \  
   f:\Re ^n\mapsto \Re\ , \ g:\Re ^n\mapsto \Re ^m\ , \ 
   h:\Re ^n\mapsto \Re ^k \ . 
 \] 

 We can imagine the function $f$ as {\em potential}, or the ``height'' 
 of a surface. An {\em equipotential} is a manifold where the function 
 is constant, $f(x)=c$.  
 The gradient 
 \[ 
   \nabla f \equiv \partial f /\partial x = \left[ \begin{array}{cccc} 
   \partial f /\partial x_1, & \partial f /\partial x_2, & \ldots & 
   \partial f /\partial x_n 
   \end{array} \right] 
 \] 
 gives steepest ascent direction of the function at point $x$. 
 Hence, the gradient $\nabla f(x)$ is orthogonal to  
 the equipotential at this point. 

 Imagine a particle being ``pulled down'' by the force  
 $-\nabla f(x)$. 
 The optimal solution must be a point of equilibrium for the particle.  
 Hence, either the force pulling the particle down is null, 
 or else the force must be equilibrated by ``reaction'' forces 
 exercised by the constraints. 
 The reaction force exercised by an inequality constraint  
 $g_i(x)\leq 0$, must obey the following conditions:
 \begin{itemize}
 \item[a)] Be a force orthogonal to the equipotential curve of this 
 constraint (since only the value of $g_i(x)$ is relevant for this 
 constraint);   
 \item[b)] Be a force pulling the particle ``inwards'', that is, to the 
 inside of the feasible region;  
 \item[c)] Moreover, a inequality constraint can only exercise a 
 reaction force if it is tight, otherwise there is a slack allowing 
 the particle to move even closer to this constraint. 
 \end{itemize} 
 An equality constraint, $h_i(x)=0$,  can be seen as a pair of inequality 
constraints, $h_i(x)\leq 0$ and $h_i(x)\geq 0$, but unlike an inequality 
constraint, an equality constraint is always active. 

 Our intuitive discussion can be summarized analytically 
by the following conditions known as 
 {\it Lagrange's optimality conditions}: 

 If $x^* \in \Re ^n$ \'{e} is an optimal point, then 
 \[ 
   \exists u\in \Re ^m\ , \ v\in \Re ^k\ \mid 
   u\nabla g(x^*) +v\nabla h(x^*) -\nabla f(x^*) =0 
   \ ,\ \mbox{onde}\ \ u\leq 0 \wedge ug(x^*)=0 \ . 
 \]  
 The condition $u\leq 0$ implies that the inequality's reaction force
points to the inside  of the feasible region, while the  
 {\em complementarity condition}, $ug=0$, implies that only active
constraints can exercise reaction forces. 
 The vectors $u$ and $v$ are known as {\em Lagrange multipliers}.  

 These necessary conditions can also be presented 
 by means of the Lagrangen function,    
 \[ 
     L(x,\lambda) =  f(x) +u g(x) +v h(x) \ , 
 \] 
 where $\lambda=[u,v]$, 
 $u \in R_+^m$ and $v \in R^k$,    
 as a necessary condition for a saddle point:   
 \[ 
    \frac{\del L(x^*,\lambda^*)}{\del x} =0 \ , \ \ 
    \frac{\del L(x^*,\lambda^*)}{\del \lambda} =0 \ . 
 \] 
  The Lagrangean function can be used to define a duality theory 
 for non-linear optimization problems, 
  see Luenberger (1984) and  Minoux and Vajda (1986).

 \subsubsection{Quadratic and Linear Complementarity Problems} 

 Quadratic Programming (QP) is an important problem in its own right,
and is  also frequently used as a subproblem in methods designed to
solve  more general problems like, for example, Sequential Quadratic
Programming, see Luenberger (1984) and  Minoux and Vajda (1986).   

 The QP problem with linear constraints is stated as 
 \[ 
   min f(x)\equiv (1/2)x'Qx -\eta p'x  
 \mid x\geq 0 \wedge Te*x=te \wedge Tl*x\leq tl 
 \] 
 where the matrix dimensions are  
 $Te \ me\times n,\ me<n$, 
 $Tl \ ml\times n,\ ml<n$,  
 $Ml={1,2,\ldots ml},\ Me={1,2,\ldots me},\ N={1,2,\ldots n}$. 
 We assume that the quadratic form defining the problem is 
 symmetric and positive definite, that is, 
 $Q=Q',\ Q>0$. 
 
 In the QP problem above, the objective function's gradient is  
 \[ 
    \nabla f = x'Q -\eta p'\ , 
 \] 
 and the gradients of the constraint functions are 
 \[ 
    g_i(x) =  T_{i}x \leq t_{i} \Rightarrow \nabla g_{i} = T_{i}\ .  
 \] 
 Hence, the Lagrange optimality conditions are 
 \[ 
     x\in R_+^n, s\in R_+^n, l\in R_+^{ml}, e\in \Re ^{me},\   \ \mid \  
    -(x'Q -\eta p')  +s' -l'Tl +e'Te =0  
 \] 
 \[   
    \ \wedge \ \forall i\in N \ , \ x_{i}s_{i}=0  
    \ \wedge \ \forall k\in Ml \ , \  (Tl*x -tl)_{k}l_{k}=0  
    \ \ \mbox{or} 
 \]  
 \[ 
   x\in R_+^n, s\in R_+^n, l\in R_+^{ml}, e\in \Re ^{me}, y\in R_+^{ml}\   
    \ \mid \ Qx -s' +Tl'*l +Te'e  = \eta p  
 \]  
 \[ 
    \ \wedge \ \forall i\in N \ , \  x_{i}s_{i}=0  
    \ \wedge \ \forall k\in Ml \ , \  yl_{k}l_{k}=0 \   
    \mbox{onde}\ yl=(tl -Tl*x) 
 \] 
 
 The Complementarity Conditions (CC), $x's=0$ e $yl'l=0$,
 indicate that only active constraints can help to equilibrate 
 non-negative components of the objective function's gradient. 
 Using the change of variables $e=ep-em$, $ep,\ em \geq 0$, 
 the optimal solution is characterized by the 
 Viability and Optimality conditions (VOC):  
 \[  
 \left[ \begin{array}{c}  
  x\\ l\\ ep\\ en\\ s\\ yl \end{array} \right] \geq 0 
 \mid 
 \left[ \begin{array}{rrrrrr} 
  Tl & 0   & 0   & 0    & 0  & I \\ 
  Te & 0   & 0   & 0    & 0  & 0 \\ 
  Q  & Tl' & Te' & -Te' & -I & 0   
 \end{array} \right]
 \left[ \begin{array}{c}  
  x\\ l\\ ep\\ en\\ s\\ yl \end{array} \right]
 = \left[ \begin{array}{c} tl\\ te\\ \eta p \end{array} \right] 
 \] 
 \[  
   x's=0,\ yl'l=0. 
 \]

 \subsubsection{Tabu Simplex}

  Observe that: 
  (1) the VOC system  (viability and optimality conditions) 
 stated above formally resembles a LP (linear programming) problem 
 in standard form; 
  (2) all the non-linearity of the original QP (quadratic programming) 
 problem is encapsulated and locked inside the CC 
 (complementarity conditions); 
  (3) the CC take a logical form of mutual exclusion.   
  These observations are the key to adapt the Simplex to solve a 
 (positive definite) QP, see Hadley (1964), Stern et al. (2006) 
 and Wolfe (1959).  

  The VOC plus CC stated above imply that, 
 at an optimal solution, there must be many null elements, 
 more specifically: 
 There are at least  $n$ null elements among $x$ and $s$; and 
 there are at least $ml$ null elements among $yl$ e $l$.  
  Moreover, $ep$ and $en$ are, respectively, 
 the positive and negative parts of the unconstrained vector $e$, 
 so that, by construction, 
 there are at least $me$ null elements among $ep$ and $en$.  
  Hence, in the optimal solution, there are no more than $ml+me+n$ 
 non-zero variables, that can be written as a basic solution to 
 the VOC linear system.   
  This suggests using of the Simplex algorithm for solving QP.

   Let us assume, for convenience and without loss of generality, 
  that $tl\geq 0$. 
   The need to solve the VOC system, respecting the CCs, 
   is expressed by the following Linear Complementarity (LC) problem:     
 \[  
 \min 
 \left[ \begin{array}{cccccccc} 0 & 0 & 0 & 0 & 0 & 0 & 1 & 1 
 \end{array} \right]   
 \left[ \begin{array}{c}  
  x\\ l\\ ep\\ en\\ s\\ yl\\ ye\\ yq \end{array} \right]  
 \ \ \left[ \begin{array}{c}  
  x\\ l\\ ep\\ en\\ s\\ yl\\ ye\\ yq \end{array} \right] \geq 0  
 \mid 
 \] 
 \[   
 \left[ \begin{array}{rrrrrrrr} 
  Tl & 0 & 0 & 0 & 0 & I & 0 & 0 \\ 
  Te & 0 & 0 & 0 & 0 & 0 & De & 0 \\ 
  Q & Tl' & Te' & -Te' & -I & 0 & 0 & Dq  
 \end{array} \right]
 \left[ \begin{array}{c}  
  x\\ l\\ ep\\ en\\ s\\ yl\\ ye\\ yq \end{array} \right]
 = \left[ \begin{array}{c} tl\\ te\\ \eta p \end{array} \right] 
 \] 
 where the CCs, the new matrix blocks, and the initial vertex are 
 given by, respectively, \\    
 $ 
 x's=0,\ yl'l=0,\ \  
 Dq=diag( sign( \eta p )),\   
 De=diag( sign( te )),\ \  
 tl\geq 0,$ \\  
 \[ 
 { \left[ \begin{array}{c}  
  x\\ l\\ ep\\ en\\ s\\ yl\\ ye\\ yq \end{array} \right] }_0 =
 \left[ \begin{array}{c} 0\\ 0\\ 0\\ 0\\ 0\\ tl\\ 
         |\ te\ | \\ |\ p\ | \end{array} \right] \ . 
 \] 

  At the initial solution the CCs are satisfied, for $x's=0'0=0$,
 and $yl'l=tl'0=0$. 
  Assuming feasible equality and inequality constraints, one can 
 start driving the artificial variables $ye$ out of the basis, 
 as in a phase 0 Simplex for a standard LP problem.  
  The next phase of the algorithm consists of driving the remaining 
 artificial basis out of the basis, respecting however the CC.  
  In order to insure that the CC continue to be satisfied as the 
 Simplex progresses, we use the following prohibition or 
 {\it Tabu rules:} \\   
  - {\it Forbid} variable $x_i$ to enter the basis 
         if $s_i$ is currently basic, and vice-versa;  \\ 
  - {\it Forbid} variable $yl_i$ to enter the basis 
         if $l_i$ is currently basic, and vice-versa.

  The Tabu Simplex is an efficient algorithms for 
 {\em Parametric Quadratic Programming} (PQP).  
  The original version of the Tabu Simplex is presented in  
 Wolfe (1959); Hadley (1964) gives a simple proof of the algorithms 
 subject only to equality constraints, and Stern et al. (2006) 
 details this proof including inequality constraints. 

  A prototypical application of PQP is the computation of 
 {\em Efficient Frontiers}, see Alexander and Francis (1986) and 
 Markowitz (1952, 1956, 1987).  
  Many theoretical aspects of financial portfolio analysis can 
 be easily stated based on optimality conditions of the underlying 
 optimization problems, see Stern et al. (2006).

 \subsection{GRG: Generalized Reduced Gradient} 

 Let us consider a NLP problem with non-linear equality constraints, 
plus box constraints over the variables' range, 
 \[ 
     \min f(x) 
     \ \ , \ \ \  f: \Re^n \mapsto \Re 
 \] 
 \[ 
    l \leq x \leq u \ \g \ h(x)=0   
      \ \ , \ \ \ h: \Re^n \mapsto \Re^m  
 \] 

 The Generalized Reduced Gradient (GRG) method emulates the 
behaiviour of the Simplex method, for a local linearization of 
the NLP problem, see Abadie and Carpentier (1969) and 
Minoux and Vajda (1986),   
 for an intuitive presentation see Himmelblau (1972). 
 Let $x$ be an initial feasible point.   
 As for LP, we assume a non-degenerescence hypothesis, that is, 
we assume that, at a given feasible point,  a maximum of $(n-m)$ 
box constraints can be active. 
 Hence, we can take $m$ of the variables with slack box constraints 
as basic (or dependent) variables, and the remaining $n-m$ variables 
as residual (or independent) variables.   
 As in the Simplex algorithm, we permute and partition all 
vector and matrix objects to better display this distinction,  
 \[ 
    x=   
 \left[ \begin{array}{c} x_b \\ x_r \end{array} \right] \ , \ \ 
    l=   
 \left[ \begin{array}{c} l_b \\ l_r \end{array} \right] \ , \ \ 
    u=   
 \left[ \begin{array}{c} u_b \\ u_r \end{array} \right] \ , \ \ 
    \nabla f(x)=   
 \left[ \begin{array}{cc} 
   \nabla^b f(x) & \nabla^r f(x) 
 \end{array} \right] 
 \]    
 \[ 
   J(x) \ = \ \ 
 \left[ \begin{array}{cc} 
   J^B(x) &  J^R(x) 
 \end{array} \right]  \ = \ \   
 \left[ \begin{array}{cc} 
   \nabla^b h_1(x) & \nabla^r h_1(x) \\  
   \nabla^b h_2(x) & \nabla^r h_2(x) \\  
   \vdots          & \vdots          \\ 
   \nabla^b h_m(x) & \nabla^r h_m(x) 
 \end{array} \right] 
 \]

 Let us consider the effect of a small alteration to the 
current feasible point, $x+\delta$, assuming that the functions 
$f$ and $h$ are continuous and differentiable.   
 The corresponding alteration to the solution's value is 
 \[ 
    \Delta f \ = \ \ 
   f(x+\delta) -f(x) \ \ \approx \ \ 
   \nabla f(x) \; \delta \ \ = \ \ 
   \left[ \begin{array}{cc} 
     \nabla^b f(x) & \nabla^r f(x) 
   \end{array} \right] \ 
   \left[ \begin{array}{c} 
     \delta_b \\ \delta_r 
   \end{array} \right]       
 \] 
 We also want the altered solution, $x+\delta$, to remain \
 (approximately) feasible, that is,  
 \[  
    \Delta h \ = \ \ 
   h(x+\delta) -h(x) \ \ \approx \ \ 
    J(x) \; \delta \ \ = \ \ 
   \left[ \begin{array}{cc} 
     J^b(x) & J^r(x) 
   \end{array} \right] \ 
   \left[ \begin{array}{c} 
     \delta_b \\ \delta_r 
   \end{array} \right] \ \ = \ \ 0       
 \] 
  Isolating $\delta_b$, and assuming that the basis 
 $J^b(x)$ is invertible,  
 \begin{eqnarray*} 
   \delta_b  & = &  
       -\left( J^b(x) \right)^{-1} \; J^r(x) \; \delta_r \\         
   \Delta f & \approx & 
    \nabla^b f(x) \; \delta_b +\nabla^r f(x) \; \delta_r \\ 
            & = & 
  \left( \nabla^r f(x) -\nabla^b f(x) 
       \left( J^b(x) \right)^{-1} \; J^r(x)  \right) \; \delta_r 
    \ \ = \ \ \ z(x) \; \delta_r 
 \end{eqnarray*}  

 Since the problem is non-linear, we can not assure that an optimal 
solution has all residual variables with one active constraint, 
that is, are at one side of the box, as in a standard LP problem.  
 Therefore, there is no motivation to restrict $\delta_r$ 
to have only one non-zero component, as in the Simplex. 
 Instead, we suggest to move the current solution  
(in the space of residual variables) along the direction given by 
the vector $v_r$, opposed to the reduced gradient, as long as 
the corresponding box constraint is slack, that is, 
 \[  
    v_{r(i)} = \left\{ \begin{array}{cl} 
         -z^i & \mbox{if} \ z^i > 0 \ \mbox{and} \ x_{r(i)} > l_{r(i)} \\ 
         -z^i & \mbox{if} \ z^i < 0 \ \mbox{and} \ x_{r(i)} < u_{r(i)} \\ 
         0 & \mbox{otherwise}  
          \end{array} \right. 
 \] 

 In  subsection D.3.4 shall we give general conditions that assure 
 global convergence for NLP algorithms, and we will see that 
 the discontinuity of vector $v_r$ as a function of the box constraints 
 slacks is undesirable. 
 Hence, we shall use a continuous version 
 of the search direction like, for example, 
 \[  
    v_{r(i)} = \left\{ \begin{array}{cl} 
         -\gamma( x_{r(i)} -l_{r(i)} ) z^i & \mbox{if} \ z^i > 0 
              \ \mbox{and} \ x_{r(i)} > l_{r(i)} \\ 
         -\gamma( u_{r(i)} -x_{r(i)} ) z^i & \mbox{if} \ z^i < 0 
              \ \mbox{and} \ x_{r(i)} < u_{r(i)} \\ 
         0 & \mbox{otherwise}  
          \end{array} \right. 
 \] 
 \[ 
   \mbox{where} \ \gamma(x) = x/\epsilon, 
   \ \mbox{if} \  0\leq x \leq \epsilon ;  \ \mbox{and} \ 
   \gamma(x)=1 , \ \mbox{otherwise} \ . 
 \]

  The basic idea of one iteration of the GRG method is to move 
 the feasible point by a step 
 $x +\delta$  with  $\delta=\eta v$, where 
 $v_b= -\left( J^b(x) \right)^{-1}\;J^r(x)\;v_r$, 
 that is, a step (in the space of residual variables) of size 
 $\eta$ in the direction $v_r$. 
  In order to determine the step size, $\eta$, we need to perform 
 a line search, always respecting the box constraints. 

  Note that the direction, in the space of basic variables, $v_b$, 
 has been chosen so that  $x+ \eta v$, remains (approximately) 
 feasible, since we are moving inside a hyperplane that is tangent 
 to the algebraic manifold defined by $h(x)=0$.           
  The new nearly feasible point, $x$, shall then receive a correction  
 $\Delta x$ in order to regain exact feasibility for the non-linear 
 constraints, that is, so that $h(x+\Delta x)=0$. 
  The nearly feasible point $x$  can be used a the starting point 
 for a recursive method used to get exact feasiblity, like  
 the Newton-Raphson method, that uses the basic Jacobian, $J^b(x)$,  
 to compute the correction 
 \[ 
    \Delta x_b= -\left( J^b(x) \right)^{-1} \; h(x_b, x_r) \ . 
 \]

 \subsection{Line Search and Local Convergence}

  This section analyses the problem of minimizing an unidimensional 
function, $f(x)$. 
 First, let us consider the problem of finding 
the root (zero) of a differentiable function, approximated by its 
first order Taylor expansion,  
 $g(x) \approx q(x^k) +g'(x^k)(x-x^k)$. 
 This approximation implies that 
 $g(x^{k+1})\approx 0$, where   
 \[ 
    x^{k+1} = x^k -g'(x^k)^{-1} g(x^k) 
 \] 
 This is Newton's method, used to find the root of an unidimensional 
function. 
 
 If a function $f(x)$ is differentiable, its minimum is at a point 
where the function's first derivative is null. 
 Hence, we can use Newton's method for minimizing $f(x)$, 
 \[ 
    x^{k+1} = x^k -f''(x^k)^{-1} f'(x^k)  
 \] 

 Let us examine how fast the sequence generated by Newton's method 
approaches the optimal solution, $x^*$, assuming the starting point, 
$x^0$, is already close enough to $x^*$. 
 Assuming third order differentiability, we can write 
 \[ 
   0 = f'(x^*) = f'(x^k) + f''(x^k)(x^*-x^k) 
      +(1/2)f'''(y^k)(x^*-x^k)^2 \ , \ \mbox{or} 
 \] 
 \[ 
    x^* = x^k -f''(x^k)^{-1}f'(x^k) 
         -(1/2)f''(x^k)^{-1}f'''(y^k)(x^*-x^k)^2 
 \] 
 Subtracting the equation that defines Newton's method, we have 
 \[ 
    (x^{k+1}-x^*) = 
      (1/2)f''(x^k)^{-1}f'''(y^k) \, (x^k-x^*)^2  
 \] 
   
 As we shall see in the following, this result implies that 
Newton's method converges very fast (quadratically), if we are 
already close enough to the optimal solution. 
 However,  Newton's method needs a lot of differential information 
about the function, something that may be hard to obtain. 
 Moreover, far from the optimum, one can not be sure about the 
method's convergence.      
 The following methods overcome these difficulties. 

 Let us now examine the Golden Ratio search method, for minimizing 
a unidimensional and unimodal function, $ f(x)$, in the interval, 
 $[x^1, x^4]$. 
 Assume we know the function's value at four points, the extremes 
of the interval and two interior points, $x^1<x^2<x^3<x^4$. 
 From the unimodality hypothesis we can know that the point of 
minimum, $x^*$, is in one of the sub-itervals, that is  
 \[ 
    f(x^2)\leq f(x^3) \Rightarrow x^* \in [x^1, x^3] \ \ , \ \ 
    f(x^2) > f(x^3)   \Rightarrow x^* \in [x^2, x^4] \ \ . 
 \] 

 without loss of generality, let us consider the way to divide 
the interval $[0,1]$. A ratio $r$ defines a symmetric division 
in the form $0<1-r<r<1$. 
 Dividing the subinterval $[0,r]$ by the same ratio $r$, 
 we obtain the points $0<r(1-r)<r^2<r$. 
 We want the points $r^2$ and $1-r$ to coincide, so that it will 
 only be necessary to evaluate the function at one additional point, 
 taht is, we want $r^2+r-1=0$. Hence, $r=(\sqrt{5}-1)/2$, 
 this is the {\it golden ratio} $r\approx 0.6180340$.     
   
 The golden ratio search method is robust, working for 
any unimodal function, and using only the function's value 
at the search points. 
 However, the extremes of the size of the search interval decreases 
only linearly with the number of iterations.  

 Polynomial methods, studied next, try to conciliate the best 
 characteristics of the methods already presented.    
  Polynomial methods for minimizing an unidimensional function,  
 $\min f(x +\eta)$, on $\eta \geq 0$, rely on a polynomial, $p(x)$, 
 that locally approximates $f(x)$, and the subsequent minimization 
 of the adjusted polynomial. 
  The simplest of these methods is quadratic adjustment. 
  Assume we know at three points, $\eta_1, \eta_2, \eta_3$, 
 the respective function values, $f_i=f(x +\eta_i)$. 
 Considering the equations for the interpolating polynomial 
 \[ 
    q(\eta)= a \eta^2 +b \eta + c  
     \ \ , \ \ 
    q(\eta_i) = f_i  
 \] 
 we obtain the polynomial      
 \begin{eqnarray*}
  a &=& \frac{ 
  f_1(\eta_2 -\eta_3) +f_2(\eta_3 -\eta_1) +f_3(\eta_1 -\eta_2)
 }{ -(\eta_2 -\eta_1)(\eta_3 -\eta_2)(\eta_3 -\eta_1) } \\ 
  b &=& \frac{  
  f_1(\eta_3^2 -\eta_2^2) +f_2(\eta_1^2 -\eta_3^2) 
 +f_3(\eta_2^2 -\eta_1^2)
 }{ -(\eta_2 -\eta_1)(\eta_3 -\eta_2)(\eta_3 -\eta_1) } \\ 
  c &=& \frac{  
  f_1(\eta_2^2\eta_3 -\eta_3^2\eta_2) 
 +f_2(\eta_3^2\eta_1 -\eta_1^2\eta_3) 
 +f_3(\eta_1^2\eta_2 -\eta_2^2\eta_1)
 }{ -(\eta_2 -\eta_1)(\eta_3 -\eta_2)(\eta_3 -\eta_1) }   
 \end{eqnarray*} 

 Equating the first derivative of the interpolating polynomial to 
 zero, $q'(\eta_4)= 2a\eta +b$, we obtain its point of minimum, 
 $\eta_4= a/2b$ or, directly from the function's values, 
 \[ 
  \eta_4 = \frac{1}{2}  \frac{ 
  f_1(\eta_3^2 -\eta_2^2) +f_2(\eta_1^2 -\eta_3^2) 
 +f_3(\eta_2^2 -\eta_1^2) 
 }{ 
  f_1(\eta_3 -\eta_2) +f_2(\eta_1 -\eta_3) 
 +f_3(\eta_2 -\eta_1)
 } 
 \] 

 We should try to use the initial points in the 
 ``interpolating pattern'' 
 $\eta_1 <\eta_2 <\eta_3$ e $f_1 \geq f_2 \leq f_3$, 
 that is, three points where the intermediary point has 
 the smallest function's value. 
 So doing, we know that the minimum of the interpolating polynomial 
 is inside of the initial search interval, that is, 
 $\eta_4\in [\eta_1, \eta_3]$. 
 In this situation we are interpolating and not extrapolating the 
 function, favoring the numerical stability of the procedure. 

  Choosing $\eta_4$ and two more points from the initial three, 
 we have a new set of three points in the desired interpolating pattern, 
 and are ready to proceed for the next iteration. 
  Note that, in general, we can not guaranty that $\eta_4$   
 is the best point in the new set of three. 
  However, $\eta_4$ will always replace the worst point in the old set.  
  Hence, the sum  $z=f_1+f_2+f_3$ is monotonically decreasing. 
  In section D.3.4 we shall see that these properties assure 
 the global convergence of the quadratic adjustment line search algorithm. 

  Let us now consider the errors relative to the minimum argument, 
 $\epsilon_i= x^* -x_i$.  
  We can write  
 $\epsilon_4 = g(\epsilon_1,\epsilon_2,\epsilon_3)$, 
 where the function $g$ is a second order polynomial. 
 This is because $\eta_4$ is obtained by a quadratic adjustment, 
 that is also symmetric in its arguments, 
 since the order of the first three points is irrelevant. 
  Moreover, it is nor hard to check that $\epsilon_4$ is zero 
 if two of the three initial errors are zero. 
  Hence, close to the minimum, $x^*$, 
 we have the following approximation for the forth error: 
 \[ 
    \epsilon_4 = C \left( \epsilon_1 \epsilon_2 
     +\epsilon_1 \epsilon_3 +\epsilon_2 \epsilon_3 \right) 
 \]  

  Assuming that the process is converging, the 
  $k$-th error is approximately  
  $\epsilon_{k+4} = C \epsilon_{k+1} \epsilon_{k+2}$ 
  or 
  $C\epsilon_{k+3} = C\epsilon_{k+1} \; C\epsilon_{k+0}$.  
  Let us now assume a power-law convergence,   
  $C \epsilon_k \approx \delta^{r_k}$,  
  so that we have 
  $\delta^{r_{k+3}} = \delta^{r_{k+1}} \; \delta^{r_{k+0}}$ 
  or, taking the logarithm, 
  ${r_{k+3}} = {r_{k+1}} + {r_{k+0}}$. 
  The general solution of this finite difference equation 
  has the form 
  ${r_{k}} = A\lambda_1^k +B\lambda_2^k +C\lambda_3^k$, 
  where $\lambda_i$ are the roots  of the characteristic equation 
  $\lambda^3 -\lambda^1 -\lambda^0 = \lambda^3 -\lambda -1 = 0$. 
  The order of convergence of this method, as defined in the followig 
  paragraph, is the largest root of the characteristic equation,  
  $\lambda \approx 1.32$, 
  Notice that $1.32^3 \approx 2.30$, making three steps of this 
 method ``as good as'' one step of the quadratically convergent 
 Newton method, with the advantages of being globally convergent 
 and not requiring the computation of expensive derivatives.

  We say that a sequence of real numbers 
 $r^k\rightarrow r^*$ converges at least in order $p>0$ if  
 \[  
    0 \leq \lim_{k\rightarrow \infty} 
    \frac{ | r^{k+1}-r^* | }{ | r^k-r^* |^p }  = \beta < \infty  
 \]  
  The sequence {\em order of convergence} is the supremum of 
 constants $p>0$ in such conditions. 
  If $p=1$ and $\beta<1$, we say that the sequence has 
 {\em linear convergence} with rate $\beta$. 
  If $\beta=0$, we say that the sequence has {\em super linear} 
 convergence. 

  For example, for $c\geq 1$, $c$ is the order of convergence 
 of the sequence $a^{(c^k)}$. 
  We can also see that $1/k$ converges in order $1$, 
 although it is not linearly convergent, because 
 $r^{k+1}/r^k\rightarrow 1$.   
  Finally, $(1/k)^k$ converges in order $1$, 
 because for any $p>1$,   
 $r^{k+1}/(r^k)^p\rightarrow \infty$, 
 However, this convergence is super-linear, because         
 $r^{k+1}/r^k\rightarrow 0$.  
  
 

 \subsection{The Gradient ParTan Algorithm} 

  In this section we present the method of Parallel Tangents, ParTan, 
 developed by Shah, Buehler and Kempthorne (1964) 
 for solving the problem of minimizing an unconstrained convex function. 
  We present a particular case of the General ParTan algorithm, 
 the Gradient ParTan, following the presentation in Luenberger (1983). 

  The ParTan algorith was developed to solve exactly, after $n$ steps,
 a general  quadratic function $f(x) = x'Ax +b'x +c$. 
  If $A$ is real, symmetric and full rank matrix, it is possible to find 
 the  eigenvalue decomposition $V'AV = D = \diag(d)$, see section F.2. 
 If we had the eigen-vector matrix, $V$, we could consider the coordinate 
 transformation $y=V'x$, $x=Vy$,  $f(y)= y'V'AVy + b'Vy= y'Dy +e'y +c$. 
  The coordinate transformation given by (the orthogonal) matrix $V$ can 
 be interpreted as a  decoupling operator, see Chap.3, for it transforms 
 an $n$-vector optimization problem into $n$ independent scalar 
 optimization problems, $y_i \in \arg \min d_i (y_k)^2 +e_i y_i +c$.   
  However, finding the eigenvalue decomposition of $A$ is even 
 harder than solving the original optimization problem. 
  A set of vectors (or directions), $w^k$ is $A$-conjugate 
 iff, for $k\neq j$,  $(w^k)' A w^j =0$. 
  A (non-orthogonal) matrix of $n$ $A$-conjugate vectors,  
 $W=[w^1\ldots w^n]$ provides an alternative, and much cheaper 
 decoupling operator for the quadratic optimization problem. 
  The Partan algorithm finds, on the fly, a set of $n$   
 $A$-conjugate vectors $w^k$.

  To simplify the notation we assume, without loss of generality, a 
 quadratic function that is centered at the origin, $f(x) = x'Ax$. 
  Therefore, $\grad(x)=Ay$, so that $y'Ax=y'\grad(x)$, and   
  vectors $x$ and $y$ are $A$-conjugate 
  iff $y$ is ortogonal to $\grad(x)$.    
  The Partan algorithm is defined as follows, progressing through 
 points $x^0, x^1, y^1, x^2, \ldots x^{k-1}, y^{k-1}, x^k$, 
 see Figure D.2 (left). 
 The algorithm is initialized by choosing an arbitrary starting point,  
 $x^0$, by an initial Cauchy step to find $y^0$, 
 and by taking $x^1=y^0$.

 $N$-Dimensional (Gradient) ParTan Algorithm:

 - Cauchy step: 
 For $k=0,1,\ldots n$, 
 find  $y^k = x^k + \alpha_k g^k$ in an exact line search 
 along the $k$-th steepest descent direction,  $g^k = -\grad f(x^k)$.  

 - Acceleration step: 
 For $k=1,\ldots n-1$, 
 find $x^{k+1} = y^k +\beta_k (y^k - x^{k-1})$ 
 in an exact line search along the $k$-th acceleration direction, 
 $(y^k - x^{k-1})$.

 In order to prove the correctness of the ParTan algorithm, 
 we will prove, by induction, two statements: 

 (1) The directions $w^k = (x^{k+1}-x^k)$ are $A$-conjugate.  

 (2) Although the ParTan never performs the 
 {\it conjugate direction line search},   
 $x^{k+1} = x^k +\gamma_k w^k$, this is what implicitly happens, 
 that is, the point $x^{k+1}$, actually found at the acceleration step, 
 would also solve the (hypothetical) conjugate direction line search. 

  The basis for the induction, $k=1$, is trivially true. 
  Let us assume the statements are true up to $k-1$, 
 and prove the induction step for the index $k$, see Figure D.2 (right).

 \mbox{} 

 \centerline{\input{PARTAN1.PIC}}  

 \mbox{} \vspace{-13.0cm} \mbox{} 

 \centerline{Figure D.2: The Gradient ParTan Algorithm.}

 By the induction hypothesis, $x^k$ is the minimum of $f(x)$ on 
 the $k$-dimensional hyperplane through $x^0$ spanned by all 
 previous conjugate directions, $w^j$, $j<k$. 
 Hence, $g^k=-\grad f(x^k)$ is orthogonal to all $w^j$, $j<k$. 
 All previous search directions lie in the same $k$-hyperplane, 
 hence, $g^k$ is also orthogonal to them. 
 In particular, $g^k$ is orthogonal to 
 $g^{k-1}=-\grad f(x^{k-1})$.  
 Also, from the exact Cauchy step from $x^k$ to $y^k$, 
 we know that $g^k$ must be orthogonal to $\grad f(y^k)$.   
 Since $\grad f(x)$ is a linear function, it must be orthogonal 
 to $g^k$ at any point in the line search 
 $x^{k+1} = y^k +\beta_k (y^k - x^{k-1})$. 
 Since this line search is exact, 
 $\grad f(x^{k+1})$ is orthogonal to $(y^k - x^{k-1})$. 
 Hence $\grad f(x^{k+1})$  is orthogonal to any linear combination of  
 $g^k$ and $(y^k - x^{k-1})$, including $w^k$. 
  For all other products $(w^j)' A w^k$, $w^j$, $j<k-1$,  
 we only have to write $w^k$ as a linear combination of 
 $g^k$ and $w^{k-1}$ to see that they vanish.  
 This is enough to conclude the induction step of statements 
 (1) and (2). QED.

 Since a full rank matrix $A$ can have at most $n$ simultaneous 
 $A$-conjugate directions,  the Gradient ParTan must find the optimal 
 solution of a quadratic function in at most $n$ steps. 
 This fact can be used to show that, if the quadratic 
 model of the objective function is good, the ParTan algorithm 
 converges quadratically.  
 Nevertheless, even if the quadratic model for the objective function 
 is poor, the Cauchy (steepest descent) steps can make good progress.  
 This explains the Gradient ParTan robustness as an optimization
 algorithm, even if it starts far away from the optimal solution. 

 The ParTan needs two line searches in order to obtain each conjugate 
 direction.  
 Far away from the optimal solution a Cauchy method would use only 
 one line search. 
 Close to the optimal solution alternative versions of the ParTan 
 algorithm, known as Conjugate Gradient algorithms, achieve quadratic 
 convergence using only one line search per dimension. 
 Nevertheless, in order to use these algorithms one has to 
 devise a monitoring system that keeps track of how well the 
 quadratic model is doing, and use it to decide when to make the 
 transition from the Cauchy to the Conjugate Gradient algorithm. 
 Hence, the Partanization of search directions  provides a simple
 mechanism  to upgrade an algorithm based on Cauchy (steepest descent)
 line search steps, accelerating it to achieve quadratic convergence,
 while keeping the robustness  that is so characteristic of Cauchy
 methods.


 \subsection{Global Convergence} 

 In this section we give some conditions that assures global 
convergence for a NLP algorithm. 
 We follow the ideas of Zangwill (1964), 
 similar analyses are presented in Luenberger (1984) and   
 Minoux and Vajda (1986).

 We define an Algorithm as an iterative process generating a sequence 
of points, $x^0,x^1,x^2\ldots$, that oby a recursion equation of the form 
 $x^{k+1} \in A_k(x^k)$, where the {\em point-to-set map} $A_k(x^k)$ 
 defines the possible successors of $x^k$ in the sequence. 
 
 The idea of using an point-to-set map, instead of a ordinary function 
 or point-to-point map, allows us to study in a unified way a hole 
 class of algorithms, including alternative implementations of 
 several details, approximate or inexact computations, 
 randomized steps, etc.   
 The basic property we look for on the maps defining an algorithm  
 is {\em closure}, defined as follows. 

 A point-to-set map from space $X$ to space $Y$, is 
 {\em closed} at $x$ if the following condition holds: 
 If a sequence $x^k$ converges to $x\in X$, 
 and the sequence $y^k$ converges to $y\in Y$, 
 where $y^k \in A(x)$, then the also the limit 
 $y$ is in the image $A(x)$, that is,   
  \[ 
     x^k \rightarrow x \ , \ 
     y^k \rightarrow y \ , \ 
     y^k \in A(x^k) \ \Rightarrow \ 
     y \in A(x) \ . 
  \] 
 The map is closed in $C\subseteq X$ if it is closed at any point of $C$. 
 Note that if we replace, in the definition of closed map, the 
 inclusion relation by the equality relation, we get the definition 
 of continuity for point-to-point functions. 
 Therefore, the closure property is a generalization of continuity.  
 Indeed, a continuous function is closed, although the contrary is 
 not necessarily true. 

 The basic idea of Zangwill's global convergence theorem is to 
 find some characteristic that is continuously ``improved'' 
 at each iteration of the algorithm. 
 This characteristic is represented by the concept of 
 {\em descendence function}. 

 Let $A$ be an algorithm in $X$ for solving the problem $P$, 
 and let $S\subset X$ be the solution set for $P$. 
 A function $Z(x)$ \'{e} is a descendence function for $(X,A,S)$ 
 if the composition of $Z$ and $A$ is always decreasing outside 
 the solution set, and does not increase inside the solution set, 
 that is, 
 \[ 
   x\notin S \wedge y\in A(x) \Rightarrow Z(y) < Z(x) 
   \ \ \ \mbox{and} \ \ 
   x\in S \wedge y\in A(x) \Rightarrow Z(y) \leq Z(x)\ . 
 \]               
 
 In optimization problems, some times the very objective function 
 is a good descendence function. 
 Other times, more complex descendence functions have to be used, 
 for example, the objective function with auxiliary terms, 
 like  penalties for constraint violations. 
 
 Before we state Zangwill's theorem, let us review two basic concepts 
 of set topology: 
 An {\it accumulation point} os a sequence is a limit point for 
 one of its sub-sequences. 
 A set is {\em compact} iff any (infinite) sequence has an accumulation 
 point inside the set. 
 In $R^n$, a set is compact if it is closed and bounded.

 Zangwill's Global Convergence Theorem: 

 Let $Z$ be a descendence function for the algorithm $A$ defined in 
 $X$ with solution set $S$, and let  $x^0,x^1,x^2,\ldots$ be a 
 sequence generated by this algorithm such that: \\ 
 A) The map $A$ is closed in any point outside $S$, \\ 
 B) All points in the sequence remain inside a compact set 
    $C\subseteq X$, and \\ 
 C) $Z$ is continuous. \\ 
 Then, any accumulation point of the sequence is in the solution set. 

 Proof: From $C$ compacity, a sequence generated by the algorithm has 
 a limit point, $x\in C\subseteq X$, for a subsequence, $x^{s(k)}$. 
 From the continuity of $Z$ in $X$, the limit value of $Z$ in the 
 subsequence coincides withg the value of $Z$ at the limit point, 
 that is,  $Z(x^{s(k)})\rightarrow Z(x)$.  
  But the complete sequence, $Z(x^k)$ is monotonically decreasing, 
 hence, if $s(k) \leq j \leq s(k+1)$ then  
 $Z(x^{s(k)}) \geq Z(x^j) \geq Z(x^{s(k+1)})$, 
 and the value of $Z$ in the complete sequence also converges 
 to the value of $Z$ at the accumulation point, that is 
 $Z(x^k)\rightarrow Z(x)$.

 Let us now imagine, for a proof by contradiction, 
 that $Z(A(x))<Z(x)$. 
 Let us consider the sub-sequence of the successors of 
 the points in the first sub-sequence, $x^{s(k)+1}$. 
 This second sub-sequence, again by compacity, also has an 
 accumulation point, $x'$. 
 But from the result in the last paragraph, the value of 
 the descendence function in both sub-sequences converge to 
 the limit value of the hole sequence, that is,   
 $\lim Z(x^{s(k)+1}) = \lim Z(x^k) = \lim Z(x^{s(k)})$. 
 So we have prooved the impossibility of $x$ not being a solution.

 Several algorithms are formulated as a composition of several steps. 
 Hence, the map describing the hole algorithm is the composition 
 of several maps, one for each step. 
 A typical example would be a step for choosing a search direction, 
 followed by a step for a line search.  
 The following lemmas are useful in the construction of such 
 composite maps. 

 First Composition Lemma:  
 Let $A$ from $X$ to $Y$, and $B$ from $Y$ to $Z$, be point-to-set maps, 
 $A$ closed in $x\in X$, $B$ closed in $A(x)$. 
 If any sequence $x^k$ converging to  $x$, 
 $y^k \in A(x^k)$ has an accumulation point $y$, 
 then the composed map $B\circ A$ is closed in $x$. 

 Second Composition Lemma:  
 Let $A$ from $X$ to $Y$, and $B$ from $Y$ to $Z$, be point-to set maps,  
 $A$ closed in $x\in X$, $B$ closed in $A(x)$. 
 If $Y$ is compact, then the composed map,  
 $B\circ A$ is closed in $x$.

 Third Composition Lemma:  
 Let $A$ be a point-to point map from $X$ in $Y$, 
 and $B$ a point-to-set map from $Y$ to $Z$.  
 If $A$ is continuous in $x$, and $B$ is closed in $A(x)$. 
 then the composed map  
 $B\circ A$ is closed in $x$.

 \section{Variational Principles}

 The variational problem asks for the function $q(t)$ that minimizes a
global functional (function of a function), $J(q)$, with fixed boundary
conditions, $q(a)$ and $q(b)$, as shown in Figure D.3. 
 Its general form is given by a local functional, $F(t,q,q')$, and an
integral or global functional,  
 \[
    J(q)= \int_a^b F(t,q,q') dt \ ,
 \]
 where the prime indicates, as usual, the simple derivative with respect
to $t$, that is, $q'=dq/dt$.

 \begin{figure}[hbt] 
 \centerline{\includegraphics*[height=2.4in, width=6.4in, angle=0]{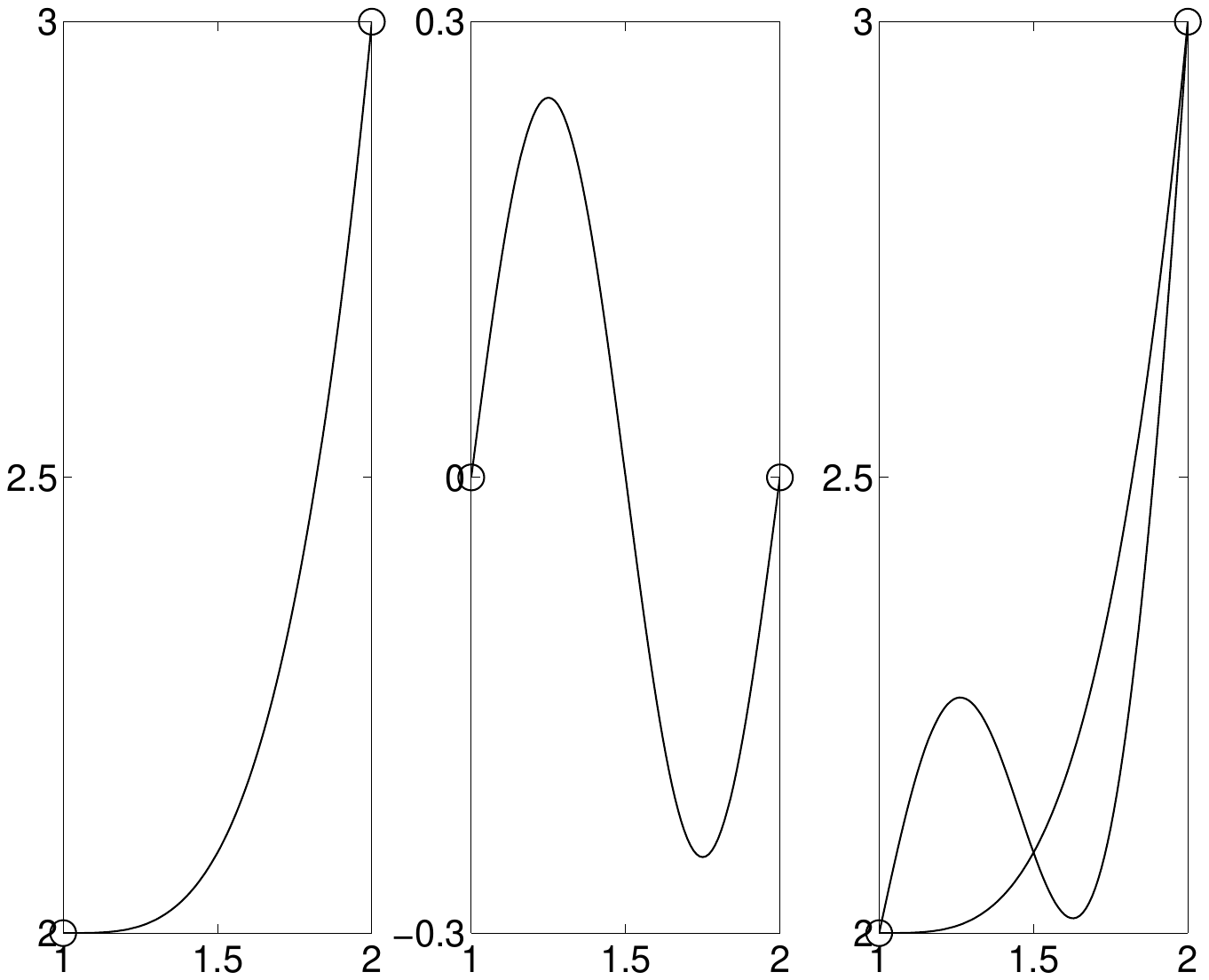}}
 \centerline{Figure D.3: Variational problem, 
             $q(x)$, $\eta(x)$, $q(x)+\eta(x)$.} 
  \end{figure}

 \subsection*{Euler-Lagrange Equation} 

 Consider a `variation' of $q(t)$ given by another curve, $\eta(t)$,
satisfying the fixed boundary conditions, $\eta(a)=\eta(b)=0$,  
 \[
   q= q(\epsilon,t)= q(t) +\epsilon \eta(t) \ \ \mbox{and} 
 \]
 \[
   J(\epsilon) = \int_a^b 
    F\left(t,q(\epsilon,t),q'(\epsilon,t) \right) dt \ . 
 \]

 A minimizing $q(t)$ must be stationary, that is, 
 \[
   \frac{\del J}{\del \epsilon} =  
   \frac{\del}{\del \epsilon} \int_a^b 
    F\left(t,q(\epsilon,t),q'(\epsilon,t) \right) dt =0 \ . 
 \] 
 Since the boundary conditions are fixed, the differential operator
affects only the integrand, hence
 \[ 
   \frac{\del J}{\del \epsilon} = \int_a^b \left( 
    \frac{\del F}{\del q} \frac{\del q}{\del \epsilon} 
   +\frac{\del F}{\del q'} \frac{\del q'}{\del \epsilon} 
    \right) dt  
 \] 

 From the definition of $q(\epsilon,t)$ we have
 \[
   \frac{\del q}{\del \epsilon} = \eta(t) \ , \ \ 
   \frac{\del q'}{\del \epsilon} = \eta'(t)   
   \ , \ \ \mbox{hence} \ ,  
 \] 
 \[
   \frac{\del J}{\del \epsilon} = \int_a^b \left( 
    \frac{\del F}{\del q} \eta(t)  
   +\frac{\del F}{\del q'} \eta'(t)   
    \right) dt \ .  
 \]
 Integrating the second term by parts, we get 
 \[
   \int_a^b \frac{\del F}{\del q'} \eta'(t) dt = 
    \left. \frac{\del F}{\del q'} \eta(t) \right|_a^b
   -\int_a^b  \frac{d}{dt} \left( \frac{\del F}{\del q'} \right) 
    \eta(t) dt \ ,  
 \]
 where the first term vanishes, since the extreme points, 
 $\eta(a)=\eta(b)=0$, are fixed. Hence
 \[
   \frac{\del J}{\del \epsilon} = \int_a^b \left( 
    \frac{\del F}{\del q}   
   -\frac{d}{dt} \frac{\del F}{\del q'}  
    \right) \eta(t) dt \ .  
 \]
 Since $\eta(t)$ is arbitrary and the integral must be zero, the parenthesis 
 in the integrand must be zero. This is the Euler-Lagrange equation:
 \[
   \frac{\del F}{\del q} - \frac{d}{dt} \frac{\del F}{\del q'} =0 \ . 
 \]

 \subsection*{Noether Theorems} 

 Nother theorems establishes very general conditions under which the
existence of a symmetry in the system, described by the invariance under
the action of a continuous group, implies the existence of a quantity
that remains constant in the system's evolution, that is, a conservation
law, see for example Byron  and Fuller (1969, V-I, Sec. 2.7).   

 For example, consider a functional $F(t,q,q')$ that does not depends
explicitly of $q$. This situation reveals a symmetry: The system is
invariant by a translation on the coordinate $q$. From Euler-Lagrange
equatuion, it follows that the quantity $p=\del F / \del q'$ is
conserved. In the language of classical mechanics, $q$ would be called a
``cyclic coordinate'', while $p$ would be called a ``generalized
moment''. 

 Let us consider the lifeguard's problem from section 5.5. Using the
variable $t$ instead of $x$, and $q$ instead of $y$, the length of an
infinitesimal arch is $ds^2=dt^2+dq^2$ and we can build the total travel
time using the functional
 \[ 
    F(t,q,q')= \nu(t) \sqrt{1+q'} 
 \] 
 Since the local functional is not a function of $q$, the Euler-Lagrange
equation reduces to $\del F / \del q'=K$, where $K$ is a constant.
Hence, the lifeguard's problem solution is
 \[
   \nu(t) \frac{q'}{\sqrt{1+q'}} = K  \ . 
 \]
 If the resistance index $\nu(t)$ is also independent of $t$, $q'$ must
be a constant, so that $q$ is a straight line, as we have guessed in our
very informal solution. In general, the solution to the lifeguard's
problem is given by
 \[
   \nu(t) \frac{\tan(\theta)}{\sqrt{1+\tan(\theta)}}=  
   \nu(t) \sin(\theta) = K  \ . 
 \]


%% file: CAPENT.TEX
 \chapter{Entropy and Asymptotics}

 \begin{flushright} 


 {\it 
 ``...we can identify that quantity which we commonly designate as \\ 
   (thermodynamic) entropy with the probability of the actual state.'' 
 }

 Ludwig Bolttzmann (1844 - 1906). \\ 
 W\"{a}rmetheorie und der Wahrscheinlischkeitrechnung, 1877. 


 \end{flushright}

 The origins of the entropy concept lay in the fields of Thermodynamics 
and Statistical Physics, but its applications have extended far and wide 
to many other phenomena, physical or not. 
 The entropy of a probability distribution, $H(p(x))$, is a measure of 
uncertainty (or impurity, confusion) in a system whose states,  
 $x\in \mathcal{X}$, have $p(x)$ as probability distribution. 
 We follow closely the presentation in the following references.  
 For the basic concepts: Csisz\'{a}r (1974), Dugdale (1996),  
 Kinchine (1957) and Renyi (1961, 1970).  
 For MaxEnt characterizations: Gokhale (1975), Kapur (1989), 
 and Kapur and Kesavan (1992). 
 For MaxEnt optimization: Bertsekas and Tsitsiklis (1989), 
 Censor and Zenios (1994, 1997), Elfving (1980), 
 Fang et al. (1997) and Iusem and Pierro (1987).  
 For posterior asymptotic convergence: Gelman (1995). 

 For a detailed analysis of the connection between MaxEnt optimization and
 Bayesian statistics' formalisms, that is, for a deeper view of the 
 relation between MaxEnt and Bayes' rule updates, 
 see Caticha and Giffin (2007) and Caticha (2007).

 \subsubsection*{Convexity} 
 
 We first introduce the concept of convexity, that is going to be 
important throughout this chapter.

 {\bf Definition:} A region $S \in R^n$ is Convex iff, 
 for any two points, $x^1, x^2 \in S$, and weights   
 $0\leq l_1, l_2 \leq 1 \g l_1+l_2=1$, the convex combination of 
 these two points remains in $S$,  i.e.     
 $l_1 x^1 +l_2 x^2 \in S$.

 {\bf Theorem } Finite Convex Combination: 
 A region $S\in R^n$ is Convex iff 
 any (finite) convex combination of its points 
 remains in the region, i.e.,
 $\forall \ $ $0\leq l\leq 1 \g 1'l=1$,  
 $X=[x^1,x^2,\ldots x^m]$, $x^j\in S$,    
 \[ 
    X\, l= 
    \left[ \begin{array}{cccc}  
           x_1^1 & x_1^2 & \ldots & x_1^m \\  
           x_2^1 & x_2^2 & \ldots & x_2^m \\  
           \vdots & \vdots & \ddots & \vdots \\  
           x_n^1 & x_n^2 & \ldots & x_n^m \\  
    \end{array} \right] 
    \left[ \begin{array}{c} 
           l_1 \\ l_2 \\ \ldots \\ l_m 
    \end{array} \right] 
    \in S  
 \] 

 {\bf Proof:} By induction in the number of points, $m$. 

 {\bf Definition:} The Epigraph of the function  
 $\varphi: R^n \rightarrow R$ 
 is the region of $X$ ``above the graph'' of $\varphi$, i.e. 
 \[ 
    \mbox{Epi }(\varphi) = 
    \left\{ x\in R^{n+1} \g x_{n+1} \geq \varphi \left( \left[ 
    x_1 , x_2 , \ldots , x_n  \right]' \right) \right\} 
 \] 
 
 {\bf Definition:} A function $\varphi$ is convex iff  
 its epigraph is convex. 
 A function $\varphi$ is concave iff $-\varphi$ is convex. 

 {\bf Theorem:} A differentiable function, $\varphi: R \rightarrow R$, 
 with non negative second derivative is convex. 

 {\bf Proof:} Consider $x^0= l_1 x^1 +l_2 x^2$, 
 and the Taylor expansion around $x^0$,  
 \[ 
    \varphi(x)= \varphi(x^0) +\varphi'(x^0)(x-x^0) 
                +(1/2)\varphi''(x^*)(x-x^0)^2 
 \] 
 where $x^*$ is an appropriate intermediate point. 
 If $\varphi''(x^*)>0$ the last term is positive.  
 Now, making $x=x^1$ and $x=x^2$ we have, respectively, that  
 $\varphi(x^1)\geq \varphi(x^0) +\varphi'(x^0)l_1(x^1-x^2)$ and 
 $\varphi(x^2)\geq \varphi(x^0) +\varphi'(x^0)l_2(x^2-x^1)$ 
 multipying the first inequality by $l_1$, the second by $l_2$,  
 and adding them, we obtain the desired result.

 {\bf Theorem } Jensen Inequality: 
 If $\varphi$ is a convex function, 
 \[ 
    \E\left( \varphi(x) \right) \geq \varphi\left( \E (X) \right) 
 \] 

 For discrete distributions the Jensen inequality is a special case 
of the finite convex combination theorem. 
 Arguments of Analysis allow us to extend the result to continuous 
distributions.

 \section{Boltzmann-Gibbs-Shannon Entropy} 
 
 If $H(p(x))$ is to be a measure of uncertainty, it is reasonable that 
it should satisfy the following list of requirements. 
 For the sake of simplicity, we present the theory for finite spaces.  

 1) If the system has $n$ possible states, $x_1,\ldots x_n$, 
the entropy of the system with a given  distribution,   
 $p_i \equiv p(x_i)$, is a function  
 \[ H = H_n (p_1,\ldots,p_n) \] 

 2) $H$ is a continuous function.     

 3) $H$ is a function symmetric in its arguments. 

 4) The entropy is unchanged if an impossible state is added 
to the system, i.e., 
    \[ H_n(p_1,\ldots p_n) = H_{n+1}(p_1,\ldots p_n,0) \] 

 5) The system's entropy in minimal and null when the system is 
fully determined, i.e., 
    \[ H_n(0,\ldots,0,1,0,\ldots 0) = 0 \] 

 6) The system's entropy is maximal when all states are equally 
probable, i.e.,
    \[ \frac{1}{n} \uno = \arg \max H_n  \]

 7) A system maximal entropy increases with the number of states, i.e. 
    \[ H_{n+1}\left(\frac{1}{n+1}\uno \right) 
       > H_{n}\left(\frac{1}{n}\uno \right)    \] 

 8) Entropy is an extensive quantity, i.e., 
    given two independent systems, with distributions  
    $p$ e $q$, the entropy of the composite system is additive, i.e., 
    \[ H_{nm}(r) = H_n(p) +H_m(q) \ \ , \  
       r_{i,j} = p_i\,q_j \]

 The Boltzmann-Gibbs-Shannon measure of entropy,  
 \[ 
    H_n(p) = -I_n(p) = -\ssum_{i=1}^n p_i \log(p_i) = -\E_i \log(p_i)  
    \ \ \ , \ \ 0\,\log(0)\equiv 0 
 \]    
 satisfies requirements (1) to (8), 
 and is the most usual measure of entropy. 
  In Physics it is usual to take the logarithm in Nepper base, 
 while in Computer Science it is usual to take base 2  
 and in Engineering it is usual to take base 10.  
  The opposite of the entropy, $I(p)=-H(p)$, the Neguentropy, 
 is a measure of Information available about the system. 

  For the Boltzmann-Gibbs-Shannon entropy we can extend requirement 8, 
and compute the composite Neguentopy even without independence: 
 \begin{eqnarray*} 
 \lefteqn{ I_{nm}(r) \ \ = \ \ 
   \ssum_{i=1,j=1}^{n,m} r_{i,j}\, \log(r_{i,j})   
 \ \ = \ \   
   \ssum_{i=1,j=1}^{n,m} p_i \Pr(j\g i)\, 
                   \log \left( p_i \Pr(j\g i) \right) } \\ 
 &=& 
   \ssum_{i=1}^{n} p_i \log(p_i) \ssum_{j=1}^m \Pr(j\g i) 
  +\ssum_{i=1}^{n} p_i \ssum_{j=1}^m \Pr(j\g i) \, 
          \log\left( \Pr(j\g i) \right)                \\ 
 &=& 
   I_n(p) +\ssum_{i=1}^n p_i \, I_m( q^i )            
 \ \ \Where \ 
                \ q^i_j = \Pr(j \g i) 
 \end{eqnarray*} 
 
 If we add this last identity as item number 9 in the list of 
requirements, we have a characterization of Boltzmann-Gibbs-Shannon 
entropy, see Kinchine (1957) and Renyi (1961, 1970).  

 Like many important concepts, this measure of entropy was discovered and 
re-discovered several times in different contexts, and sometimes the 
uniqueness and identity of the concept was not immediately recognized. 
 A well known anecdote refers the answer given by von Neumann, after  
Shannon asked him how to call a ``newly''  discovered concept in 
Information Theory. As reported by Shannon in 
 Tribus and McIrvine (1971, p.180):

 {\it ``My greatest concern was what to call it. 
  I thought of calling it information, but the word was overly used, so
 I decided to call it uncertainty. When I discussed it with John von
 Neumann, he had a better idea. Von Neumann told me, You should call it
 entropy, for two reasons. 
  In the first place your uncertainty function has been used in
statistical mechanics under that name, so it already has a name. 
 In the second place, and more important, nobody knows what entropy
really is, so in a debate you will always have the advantage.''}

  A simple proof that requirement (6) is satisfied can be 
 obtained directly from the convexity of $I_n(p,q)$ 
 as a function of $p$, see Kapur and Kesavan (1992, Sec.IV.2). 
  Convexity properties of $I_n(p,q)$, on either of its vector 
 arguments, can, in turn, be asserted from the gradient vectors 
 and positive definite Hessian matrices given by the following 
 derivatives:     
 \[      
    \frac{\del I_n(p,q)}{\del p_i} = 
     1 + \log \left( \frac{p_i}{q_i} \right)  \ , \ \ 
    \frac{\del I_n(p,q)}{\del q_i} = -\frac{p_i}{q_i} \ ,  
 \] 
 \[      
    \frac{\del^2 I_n(p,q)}{\del p_i \, \del p_j} = 
     \delta_i^j \; \frac{1}{p_i} \ , \ \ 
    \frac{\del^2 I_n(p,q)}{\del q_i \, \del q_j} = 
     \delta_i^j \; \frac{p_i}{q_i^2} \ .    
 \] 
  These Hessian matrices are not only positive definite, 
 but also diagonal. This observation is the basis for several 
 analogies between minimum divergence problems and generalized 
 network flow problems, see observations at Section D.3.1.

 \section{Csisz\'{a}r's $\varphi$-divergence}

 We present an alternative demonstration that requirement (6) is 
 satisfied (just use $q\propto 1$), based on the following lemma:  

 \noindent 
 {\bf Lemma:} Shannon Inequality \\ 
 If $p$ and $q$ are two distributions over a system with $n$ possible 
 states, and $q_i \neq 0$, then the {\it Information Divergence} of $p$ 
 relative to $q$, $I_n(p,q)$, is positive, except if $p=q$, 
 when it is null, 
 \[ 
    I_n(p,q) \equiv \ssum_{i=1}^n p_i \, 
            \log \left( \frac{p_i}{q_i} \right)  
    \ \ , \ I_n(p,q)\geq 0 
    \ \ , \ I_n(p,q) = 0 \Rightarrow p=q 
 \] 
 
 {\bf Proof:} 
 By Jensen inequality, if $\varphi$ is a convex function,  
 \[ \E\left( \varphi(x) \right) \geq \varphi\left( \E (X) \right) \] 
 Taking 
 \begin{eqnarray*} 
    \varphi(t) &=& t\ln(t)        \  \And \ 
     t_i \ \ = \ \frac{p_i}{q_i}    \\ 
    \E_q\, (t) &=& 
      \ssum_{i=1}^n q_i\;\frac{p_i}{q_i} \ = \ 1 \\   
    I_n(p,q) &=&  
    \ssum q_i\; t_i \log t_i \geq  1\, \log (1) \ = \ 0 
 \end{eqnarray*}

   Shannon's inequality motivates the use of the Information Divergence  
 as a measure of  (non symmetric) ``distance'' between distributions. 
  In Statistics this measure is known as the Kullback-Leibler distance.
  The denominations Directed Divergence, Cross Information and Cross 
 Neg-Entropy (XEnt) are used in Engineering. 
  The proof of Shannon inequality motivates the following 
 generalization of (directed) divergence: 
  
 \noindent 
 {\bf Definition:} 
 Csisz\'{a}r's $\varphi$-divergence. \\   
 Given a convex function $\varphi$,    
 \begin{eqnarray*}  
    d_{\varphi}(p,q) &=&  
   \ssum_{i=1}^n q_i \, \varphi \left( \frac{p_i}{q_i} \right) \\ 
   0\,\varphi \left( \frac{0}{0} \right) &=&  0  \ \ , \ \ 
   0\,\varphi \left( \frac{c}{0} \right) \ = \  
   c\, \lim_{t\rightarrow \infty} \, \frac{\varphi(t)}{t}       
 \end{eqnarray*} 

 For example, we can define the quadratic and the absolute 
 divergence as  
 \begin{eqnarray*} 
 \xi^2(p,q) &=& \ssum \frac{ (p_i -q_i)^2 }{q_i} \ \ , \ \ 
  \For \varphi(t) \ = \ (t-1)^2 \\ 
 Ab(p,q) &=& \ssum \frac{ | p_i -q_i | }{q_i} \ \ , \ \ 
  \For \varphi(t) \ = \ |t-1| 
 \end{eqnarray*}

 \section{Minimum Divergence under Constraints}

 Given a prior distribution, $q$, we would like to find a vector $p$ 
that minimizes the Information Divergence $I_n(p,q)$, where $p$ is under 
the constraint of being a probability distribution, and maybe also under 
additional constraints over the expectation of functions taking values 
on the system's states, that is, we want 
 \[ 
   p^* \in \arg \min I_n(p,q) \ \ , \ 
   p\geq 0\ \g \uno'p=1 \And Ap=b \ \ , \ A \ (m-1)\times n 
 \] 
 $p^*$ is the {\it Minimum Information Divergence}   
distribution, relative to $q$, given the constraints $\{A,b\}$. 
 We can write the probability normalization constraint as a 
generic linear constraint, including $\uno$ and $1$ as the 
 $m$-th (or $0$-th) rows of matrix $A$ and vector $b$.  
 So doing, we do not need to keep any distinction between the 
normalization and the other constraints. 
 In this chapter, the operators $\odot$ e $\oslash$ 
 indicate the point (element) wise product and division between matrices 
of same dimension. 

 The Lagrangean function of this optimization problem, 
 and its derivatives are: 
 \[  
  L(p,w) =  p' \log(p\oslash q) + w'(b -A p) \ , 
 \]  
 \[ 
  \frac{\del L}{\del p_i} =  
     \log(p_i / q_i) +1 -w'A^i  \ , \ \  
  \frac{\del L}{\del w_k} =    b_k -A_k p  \ .      
 \] 
 
 Equating the $n+m$ derivatives to zero, we have a system with  
 $n+m$ unknowns and equations, giving viability and optimality 
 conditions (VOCs) for the problem: 
 \begin{eqnarray*}  
  p_i &=&  q_i \exp \left( w' A^i -1 \right)  
  \ \ \mbox{ou} \ \  
  p \ = \ q \odot \exp \left( (w'\,A)' -\uno \right)  \\ 
  A_k p &=& b_k   \ \ , \ p \geq 0  
 \end{eqnarray*}

 We can further replace the unknown probabilities, $p_i$, 
 writing the VOCs only on $w$, the dual variables (Lagrange multipliers), 
 \[ 
    h_k(w) \equiv A_k 
   \left( q \odot \exp \left( (w'\,A)' -\uno \right) \right) -b_k = 0 
 \]  

 The last form of the VOCs motivates the use of iterative algorithms 
of Gauss-Seidel type, solving the problem by cyclic iteration. 
 In this type of algorithm, one cyclically ``fits'' one equation of 
the system, for the current value of the other variables. 
 For a detailed analysis of this type of algorithm, see 
 Bertsekas and Tsitsiklis (1989), Censor and Zenios (1994, 1997), 
 Elfving (1980), Garcia et al. (2002) and Iusem and Pierro (1987).  

 \noindent 
 {\bf Bregmann Algorithm:} 
 
 Initialization: Take $t=0$, $w^t \in \Re^m$, and  
 \[ 
    p^t_i = q_i \exp \left( {w^t}'A^i \, -1 \right)  
 \] 

 Iteration step: for $t=1,2,\ldots$, Take 
 \[ k= (t \ \mbox{mod}\ m)  \ \ \And \ \   
    \nu \g \varphi(\nu)=0 \ , \ \Where 
 \] 
 \begin{eqnarray*}  
   w^{t+1} &=& \left[ 
 w^t_1, \ldots w^t_{k-1}, w^t_k +\nu, w^t_{k+1}, \ldots w^t_m \right]' \\ 
   p^{t+1}_i &=&        q_i \exp( {w^{t+1}}'\,A^i -1) 
             \ \ = \ \  p^t_i \exp( \nu A^i_k )  \\    
   \varphi(\nu) &=&  A_k p^{t+1} -b_k 
  \end{eqnarray*}

  From our discussion of Entropy optimization under linear constraints, 
it should be clear that the minimum information divergence distribution  
for a system under constraints on the expectation of functions taking 
values on the system's states, 

 $E_{p(x)} a_k(x)=\int a_k(x)p(x)dx =b_k$, 
 (including the normalization constraint, $a_0=\uno , b_0=1$) 
 has the form  
 \[ 
    p(x) = q(x) \exp \left( -\theta_0 
   -\theta_1\, a_1(x) -\theta_{2} \, a_2(x) \ldots \right) 
 \]   
 Note that we took $\theta_0=-(w_0-1)$, $\theta_k=-w_k$, 
 and we have also indexed the state $i$ by variable $x$, 
 so to write the last equation in the standard form used in the 
 statistical literature. 
 
 Several distributions commonly used in Statistics can be interpreted 
 as minimum information (or MaxEnt) densities 
 (relative to the uniform distribution, if not otherwise stated) 
 given some constraints over the expected value of state functions. 
 For example: 

 The Normal distribution is characterized as the distribution of maximum
 entropy on $\Re^n$, given the expected values of its first and second 
 moments, i.e., mean vector and covariance matrix.

 The Wishart distribution: 
 \[ 
  f(S\g \nu,V) \equiv c(\nu,V) \exp \left( 
   \frac{\nu-d-1}{2} \log( \det (S) ) 
  -\ssum_{i,j} V_{i,j} S_{i,j} \right) 
 \] 
 is characterized as the distribution of maximum entropy in the support 
 $S>0$, given the expected value of the elements and log-determinant of 
 matrix $S$. 
 That is, writing $\Gamma'$ for the digamma function,  
 \[ 
    \E(S_{i,j}) = V_{i,j} \ \ , \ \ 
    \E(\log(\det(S))) = \ssum_{k=1}^d \Gamma' 
     \left( \frac{\nu-k+1}{2} \right) 
 \]

  The Dirichlet distribution 
 \[ 
   f(x\g \theta) = c(\theta) \exp \left( \ssum_{k=1}^m 
   (\theta_k -1) \log( x_k ) \right)  
 \] 
 is characterized as the distribution of maximum entropy in the support 
 $x\geq 0 \g \uno'x=1$, given the expected values of the 
 log-coordinates, $\E(\log(x_k))$.

 \noindent 
 {\bf Jeffrey's Rule:}

 Richard Jeffrey considered the problem of updating an old probability 
 distribution, $q$, to a new distribution, $p$, 
 given new constraints on the probabilities of a partition, that is, 
 \[ 
     \ssum_{i\in S_k} p_i =\alpha_k \ , \ \  
     \ssum_k \alpha_k =1 \ , \ \ 
     S_1 \cup \ldots \cup S_m = \{1,\ldots n\} \ , \ \     
     S_l \cap S_k = \emptyset , \ l\neq k \ . 
 \]  
 His solution to this problem, known as the {\it Jeffrey's rule},      
 coincides with the minimum information divergence  distribution, 
 relative to $q$, given the new constraints.  
 This solution can be expressed analytically as 
 \[ 
    p_i = \alpha_{k} q_i \slash \ssum_{j\in S_{k}} q_j  
    \ , \ \  k \g i \in S_k \ . 
 \]

 \section{Fisher's Metric and Jeffreys' Prior}

 The Fisher Information Matrix, $J(\theta)$, is defined as minus the 
 expected Hessian of the log-likelihood.  
 Under appropriate regularity conditions, the {\it information geometry} 
 is defined by the metric in the parameter space given by the 
 Fisher information matrix, that is, the geometric lenght of a curve 
 is computed integrating the form  
 $dl^2 = d\theta' J(\theta) d\theta$.

 Lemma: The Fisher information matrix can also be written as the 
 covariance matrix of for the gradient of the same likelihood, i.e.,   
 \[ 
    J(\theta) \equiv  -\E_{\mathcal{X}}  
    \frac{ \del^2 \log p(x \g \theta) }{\del \theta^2} 
    =  \E_{\mathcal{X}} \left(
    \frac{ \del \log p(x \g \theta) }{\del \theta} \;   
    \frac{ \del \log p(x \g \theta) }{\del \theta} \right)      
 \] 
 
 \noindent 
 {\bf Proof: } 
 \[ 
   \int_{\mathcal{X}} p(x \g \theta) dx =1 \Rightarrow 
   \int_{\mathcal{X}}  \frac{\del p(x \g \theta) }{\del \theta} dx=0 
   \Rightarrow      
 \] 
 \[ 
   \int_{\mathcal{X}}   
   \frac{\del p(x \g \theta) }{\del \theta} 
   \frac{p(x\g \theta)}{p(x\g \theta)} dx= 
   \frac{\del \log p(x \g \theta) }{\del \theta} 
   p(x\g \theta) dx = 0 
 \] 
 differentiating again relative to the parameter, 
 \[ 
   \int_{\mathcal{X}} \left( 
    \frac{ \del^2 \log p(x \g \theta) }{\del \theta^2} p(x\g \theta) 
    + \frac{ \del \log p(x \g \theta) }{\del \theta} \;  
    \frac{ \del p(x \g \theta) }{\del \theta} 
    \right) dx =0  
 \] 
 observing that the second term can be written as 
 \[ 
   \int_{\mathcal{X}} 
    \frac{ \del \log p(x \g \theta) }{\del \theta} \;  
    \frac{ \del p(x \g \theta) }{\del \theta} \; 
    \frac{ p(x \g \theta) }{ p(x \g \theta) }   dx = 
   \int_{\mathcal{X}} 
    \frac{ \del \log p(x \g \theta) }{\del \theta} \;  
    \frac{ \del \log p(x \g \theta) }{\del \theta} \;  
    p(x \g \theta)  dx 
 \] 
 we obtain the lemma.

 Harold Jeffreys used the Fisher metric to define a class of 
 prior distributions, proportional to the determinant of the 
 information matrix,   
 \[ 
    p(\theta) \propto 
   \left| J \left( \theta \right) \right|^{1/2} .  
 \] 

 Lemma: Jeffreys' priors are geometric objects in the sense of being 
 invariant by a continuous and differentiable change of coordinates 
 in the parameter space, $\eta = f(\theta)$. 
 The proof follows Zellner (1971, p.41-54): 
 
 \noindent 
 {\bf Proof: } 
 \[ 
   J(\theta) = \left[ \frac{\del \eta}{\del \theta} \right] J(\eta)  
             \left[ \frac{\del \eta}{\del \theta} \right]'  
   \ , \ \ \mbox{hence}    
 \] 
 \[ 
    \left| J \left( \theta \right) \right|^{1/2} = 
    \left| \frac{\del \eta}{\del \theta} \right| \ 
    \left| J \left( \eta \right) \right|^{1/2} \ , \ \ \mbox{and} 
 \] 
 \[ 
    \left| J \left( \theta \right) \right|^{1/2} \ d\theta = 
    \left| J \left( \eta \right) \right|^{1/2} \ d\eta \ .  
    \ \ \mbox{Q.E.D.}  
 \] 
 
 Example: For the multinomial distribution, 
 \[ p(y \g \theta) = n! \pprod_{i=1}^m \theta_i^{x_i}  
    \left/ \; \pprod_{i=1}^m x_i ! \right. \ , \ 
    \theta_m = 1 - \ssum_{i=1}^{m-1} \theta_i \ , \  
     x_m = n - \ssum_{i=1}^{m-1} x_i \ , 
 \] 
 \[    
    L= \log p(\theta \g x) = \ssum_{i=1}^m x_i \log \theta_i \ , 
 \]   
 \[ 
    \frac{\del^2 L}{(\del \theta_i)^2} = 
     -\frac{x_i}{\theta_i^2} +\frac{x_m}{\theta_m^2} \ , \ \ 
    \frac{\del^2 L}{\del \theta_i \del \theta_j} = 
     -\frac{x_m}{\theta_m^2}  \ , \ \ i,j=1\ldots m-1 \ , 
 \] 
 \[ 
   -\E_{\mathcal{X}} \frac{\del^2 L}{(\del \theta_i)^2} = 
     \frac{n}{\theta_i} +\frac{n}{\theta_m} \ , \ \ 
   -\E_{\mathcal{X}} \frac{\del^2 L}{\del \theta_i \del \theta_j} = 
     \frac{n}{\theta_m} \ , 
 \] 
 \[ 
    \left| J(\theta) \right| = 
      \left( \theta_1 \theta_2 \ldots \theta_m \right)^{-1} \ , \ \ 
    p(\theta) \propto  
      \left( \theta_1 \theta_2 \ldots \theta_m \right)^{-1/2} \ , 
 \] 
 \[ 
    p(\theta \g x) \propto  
    \theta_1^{x_1-1/2} \theta_2^{x_2-1/2} \ldots \theta_m^{x_m-1/2}  \ .    
 \] 

  Hence, in the multinomial exemple, Jeffreys' prior ``discounts'' half 
 an observation of each kind, while the maxent prior discounts one full 
 observation, and the flat prior discounts none.    
  Similarly, slightly different versions of uninformative priors for the 
 multivariate normal distribution are shown in section C.3. 
 This situation leads to the possible criticism stated in 
 Berger (1993, p.89): 
 \begin{quote} 
 {\it ``Perhaps the most embarassing feature of noninformative priors, 
 however, is simply that there are often so many of them.''}
 \end{quote} 
  One response to this this criticism, to which Berger (1993, p.90) 
  explicitly subscribes, is that 
 \begin{quote} 
 {\it ``it is rare for the choice of a noninformative prior to makedly 
 affect the answer... so that any reasonable noninformative prior can be 
 used. Indeed, if the choice of noninformative prior does have a pronouced 
 effect on the answer, then one is probably in a situation where it is 
 crucial to involve subjective prior information.''}
 \end{quote} 
    
  The robustness of the inference procedures to variations on the form of 
 the uninformative prior can tested using sensitivity analysis, as 
 discussed in section A.6. For alternative approaches, on robustness 
 and sensitivity analysis, see Berger (1993, sec.4.7).

 In general Jeffrey's priors are not minimally informative in any sense. 
 However, Zellner (1971, p.41-54, Appendix do chapter 2:   
 Prior Distributions Representing ``Knowing Little'') 
 gives the following argument (attributed to Lindley) to present 
 Jeffreys' priors as asymptotically minimally informative.  
 The information measure of $p(x \g \theta)$, $I(\theta)$; 
 The prior average information, $A$; 
 The information gain, $G$, that is, the prior average information 
 associated with an observation, $A$, minus the prior information measure;   
 and The asymptotic information gain, $G_a$, are defined as follows:  
 \[ 
    I(\theta) = \int p(x \g \theta) \log p(x \g \theta) dx \ ; 
 \]  
 \[ 
    A = \int I(\theta) p(\theta) d\theta \ ; 
 \] 
 \[ 
    G =  A - \int p(\theta) \log p(\theta) d\theta \ ;  
 \] 
 \[ 
    G_a = \int p(\theta) \sqrt{ n \left| J(\theta) \right| } d\theta     
        - \int p(\theta) \log p(\theta) d\theta \ . 
 \] 
 Although Jeffreys' priors does not in general maximize the information 
 gain, $G$, the asymptotic convergence results presented in the next 
 section imply that Jeffrey's priors maximize the 
 asymptotic information gain, $G_a$.  
 For further details and generalizations, see 
 Amari (2007), Amari et al. (1987), Berger and Bernardo (1992), 
 Berger (1993), Bernardo and Smith (2000), Hartigan (1983), 
 Jeffreys (1961), Scholl (1998), and Zhu (1998).

 \section{Posterior Asymptotic Convergence} 

  The Information Divergence, $I(p,q)$, can be used to proof several 
asymptotic results that are fundamental to Bayesian Statistics. 
 We present in this section two of these basic results, following 
 Gelman (1995, Ap.B).

 \noindent 
 {\bf Theorem } Posterior Consistency for Discrete Parameters: 

 Consider a model where $f(\theta)$ is the prior in a discrete parameter 
space, $\Theta=\{\theta^1, \theta^2, \ldots \}$, 
 $X=[x^1,\ldots x^n]$ is a series of observations, 
 and the posterior is given by 
 \[ 
    f(\theta^k \g X) \propto f(\theta^k )\, p(X \g \theta^k ) = 
      f(\theta^k )\, \pprod_{i=1}^n p(x^i \g \theta^k ) 
 \]  

 Further, assume that this model there is a single value for the 
vector parameter, $\theta^0$, that gives the best approximation for the 
``true'' predictive distribution $g(x)$, in the sense that it 
minimizes the information divergence 
 \begin{eqnarray*} 
   \{ \theta^0 \} 
   &=& 
   \arg \min_k I\left( g(x), p(x\g \theta^k) \right) \\ 
   I\left( g(x), p(x\g \theta^k) \right) 
   &=&  
   \int_{\mathcal{X}} g(x) \log \left( 
    \frac{g(x)}{p(x\g \theta^k)} \right) dx \ \ = \ \ 
   \E_{\mathcal{X}} \log \left( 
    \frac{g(x)}{p(x\g \theta^k)} \right)   
 \end{eqnarray*} 
 Then, 
 \[ 
    \lim_{n\rightarrow \infty} f(\theta^k \g X) = 
     \delta(\theta^k,\theta^0)  
 \] 

 \noindent 
 {\bf Heuristic Argument:} 
 Consider the logarithmic coefficient   
 \[ 
 \log \left( \frac{f(\theta^k \g X)}{f(\theta^0 \g X)} \right) 
 \ \ = \ \ 
 \log \left( \frac{f(\theta^k)}{f(\theta^0)} \right) 
 +\ssum_{i=1}^n \log \left( 
  \frac{p(x^i \g \theta^k)}{p(x^i \g \theta^0)}  \right)  
 \] 
 The first term is a constant, and the second term is a sum  
which terms have all negative expected (relative to $x$, for $k\neq0$) 
value since, by our hypotheses, $\theta^0$ is the unique 
argument that minimizes $I(g(x),p(x\g\theta^k))$. 
 Hence, (for $k\neq 0$), the right hand side goes to minus infinite as 
$n$ increases. Therefore, at the left hand side, 
 $f(\theta^k \g X)$ must go to zero. 
 Since the total probability adds to one, 
 $f(\theta^0 \g X)$ must go to one, QED. 

 We can extend this result to continuous parameter spaces, 
assuming several regularity conditions, like continuity, differentiability, 
and having the argument $\theta^0$ as a interior point of $\Theta$
with the appropriate topology. 
 In such a context, we can state that, given a pre-established 
small neighborhood around $\theta^0$, like $C(\theta^0,\epsilon)$ 
 the cube of side size $\epsilon$ centered at $\theta^0$,   
 this neighborhood concentrates almost all mass of $f(\theta \g X)$, 
 as the number of observations grows to infinite. 
 Under the same regularity conditions, we also have that Maximum a 
 Posteriori (MAP) estimator is a consistent estimator, i.e., 
 $\widehat{\theta} \rightarrow \theta^0$. 

 The next results show the convergence in distribution of the 
posterior to a Normal distribution. 
 For that, we need the Fisher information matrix identity 
 from the last section.

 \noindent 
 {\bf Theorem } Posterior Normal Approximation: \\ 
 The posterior distribution converges to a Normal distribution with 
 mean $\theta^0$ and precision $nJ(\theta^0)$. 

 \noindent 
 {\bf Proof} (heuristic):  
 We only have to write the second order log-posterior Taylor expansion 
 centered at $\widehat{\theta}$,  
 \begin{eqnarray*} 
 \lefteqn{ 
  \log f(\theta \g X) \ \ = \ \ 
  \log f(\widehat{\theta} \g X)          \    
  +\frac{ \del \log f(\widehat{\theta} \g X) }{\del \theta}  
   (\theta -\widehat{\theta})   
   }  \\     
    &  &  
  +\frac{1}{2}(\theta -\widehat{\theta})' 
   \frac{ \del^2 \log f(\widehat{\theta} \g X) }{\del \theta^2} 
   (\theta -\widehat{\theta})             \  
  +\mathcal{O}(\theta -\widehat{\theta})^3         
 \end{eqnarray*} 

 The term of order zero is a constant. 
 The linear term is null, for $\widehat{\theta}$ is the MAP 
 estimator at an interior point of $\Theta$. 
 The Hessian in the quadratic term is  
 \[   
   H(\widehat{\theta}) = 
   \frac{ \del^2 \log f(\widehat{\theta} \g X) }{\del \theta^2} 
   \ \ = \ \ 
   \frac{ \del^2 \log f(\widehat{\theta}) }{\del \theta^2}
   +\ssum_{i=1}^n 
   \frac{ \del^2 \log p(x^i\g \widehat{\theta}) }{\del \theta^2} 
 \] 
 The Hessian is negative definite, by the regularity conditions, 
 and because $\widehat{\theta}$ is the MAP estimator. 
 The first term is constant, and the second is the sum of $n$ 
 i.i.d. random variables. 
 At the other hand we have already shown that the MAP estimator, 
 and also that all the posterior mass concentrates around 
 $\theta^0$. 
 We also see that the Hessian grows (in average) linearly with $n$,  
 and that the higher order terms can not grow super-linearly. 
 Also for a given $n$ and 
 $\theta \rightarrow \widehat{\theta}$,  
 the quadratic term dominates all higher order terms.  
 Hence, the quadratic approximation of the log-posterior 
 in increasingly more precise, Q.E.D.

 Given the importance of this result, we present an alternative proof,
also giving the reader an alternative way to visualize the convergence 
process, see Figure 1.   
 
  \noindent 
 {\bf Theorem } MLE Normal Approximation: \\ 
 The Maximum Likelihood Estimator (MLE) is asymptotically 
 Normal, with mean $\theta^0$ and precision  $nJ(\theta^0)$.

 \noindent 
 {\bf Proof } (schematic):  
 Assuming all needed regularity conditions, 
 from the first order optimality conditions, 
 \[ 
    \frac{1}{n} \ssum_{i=1}^{n} 
    \frac{\del \log p(x^i \g \widehat{\theta}) }{\del \theta} = 0 
 \] 
 hence, by the mean value theorem, there is an intermediate point 
 $\widetilde{\theta}$ such that   
 \[ 
    \frac{1}{n} \ssum_{i=1}^{n} 
    \frac{\del \log p(x^i \g \theta^0 ) }{\del \theta} = 
    \frac{1}{n} \ssum_{i=1}^{n} 
    \frac{ \del^2 \log p(x^i\g \widetilde{\theta}) }{\del \theta^2}  
    (\theta^0 - \widehat{\theta}) 
 \] 
  or, equivalently,    
 \[ 
    \sqrt{n} (\widehat{\theta} - \theta^0 ) = 
    -\left[ \frac{1}{n} \ssum_{i=1}^{n} 
    \frac{ \del^2 \log p(x^i\g \widetilde{\theta}) }{\del \theta^2}  
    \right]^{-1}    \frac{1}{\sqrt{n}}  \ssum_{i=1}^{n} 
    \frac{\del \log p(x^i \g \theta^0 ) }{\del \theta}       
 \] 

 We assume the regularity conditions are enough to assure that 
 \[ 
    -\left[ \frac{1}{n} \ssum_{i=1}^{n} 
    \frac{ \del^2 \log p(x^i\g \widetilde{\theta}) }{\del \theta^2}  
    \right]^{-1} \rightarrow  J(\theta^0)^{-1}  
 \]  
 for the MLE is consistent, $\widehat{\theta}\rightarrow \theta^0$,  
 and hence so is the mean value point, 
 $\widetilde{\theta}\rightarrow \theta^0$; and  
 \[ 
     \frac{1}{\sqrt{n}} \ssum_{i=1}^{n} 
    \frac{ \del \log p(x^i\g \theta^0) }{\del \theta}  
    \rightarrow  N(0,J(\theta^0))  
 \]  
 because we have the sum of $n$ i.i.d. vectors with mean $0$ and, 
 by the Information Matrix Identity lemma covariance $J(\theta^0)$. 
  
 Hence, we finally have 
 \[ 
    \sqrt{n}(\widehat{\theta} -\theta^0) \rightarrow 
    N\left( 0, J(\theta^0)^{-1} J(\theta^0) J(\theta^0)^{-1} \right) = 
    N\left( 0, J(\theta^0)^{-1} \right)
 \] 
 Q.E.D. 


   \subsection*{Exercises:} 
    
   1) Implement Bregmann's algorithm. It may be more convenient to number 
   the rows of $A$ from $1$ to $m$, and take $k=(t \ \mbox{mod}\ m)+1$. 

   2) I was given a dice, that I assumed to be honest. 
   A friend of mine lent the dice and reported playing it 60 times, 
   obtaining  4 i's, 8 ii's, 11 iii's, 14 iv's, 13 v's and 10 vi's. \\  
   A) What is my Bayesian posterior? \\ 
   Bi) What was the mean face value? (3.9). \\ 
   Bii) What is the expected posterior value of this statistic? \\ 
   C) I called the dice manufacturer, and he told me that this dice 
   is made so that the expected value of this statistic is exactly 4.0. 
   Use Bregmamnn algorithm to obtain the ``entropic posterior'', 
   that is, the  distribution closest to the prior that obeys the 
   given constraints. 
   Use as prior: 1) the uniform; ii) the Bayesian posterior. 

   3) Discuss the difference between the Bayesian update and the 
   entropic update. What is the information given in each case? 
    Observations or constraints? 

   4) Discuss the possibility of using the FBST to make hierarchical 
   tests for complex hypotheses using these ideas. 

   5) Try to give MaxEnt characterizations and Jeffrey's priors 
   for all distributions you know.

%% file: CAPFAC.TEX
 \chapter{Matrix Factorizations} 


 \section{Matrix Notation}

 Let us first define some matrix notation. 
 The operator $f\fto s\fto t$, to be read  
 {\it from} $f$ {\it to} $t$ with {\it step} $s$,
 indicates the vector $[f,f+s,f+2s,\ldots t]$  
 or the corresponding index domain. 
 $f\fto t$ is a short hand for $f\fto 1\fto t$.  
 The element in the $i$-th row and  $j$-th column of matrix $A$
 is written as $A(i,j)$ or, with subscript row index 
 and superscript column index, as $A_i^j$. 
 Index vectors can be used to build a matrix by extracting from a 
 larger matrix a given sub-set of rows and columns. 
 For example, $A({1\ate m/2} , {n/2\ate n})$ or $\maij{A}{1:m/2}{n/2:n}$ 
 is the northeast block, i.e. the block with the first rows and last 
 columns, from $A$. 
   The next example shows a more general case of this notation, 
 \[
 A=\left[ \begin{array}{ccc} 11 & 12 & 13 \\ 21 & 22 & 23 \\
 31 & 32 & 33 \end{array} \right] \ , \  \
 r=\left[ \begin{array}{cc} 1 & 3 \end{array} \right] \ , \  \
 s=\left[ \begin{array}{ccc} 3 & 1 & 2 \end{array} \right] \ ,
 \]
 \[
 A_r^s= A(r,s)= 
 \left[ \begin{array}{ccc} 13 & 11 & 12 \\ 33 & 31 & 32
 \end{array} \right] \ .
 \]
 The suppression of an index vector indicates that the corresponding 
 index spans all values in its current context. 
 Hence, $A( i , \ate )$ or $\mai{A}{i}$ indicates the $i$-th row, 
 and  $A( \ate , j )$ or $\maj{A}{j}$
 indicates the $j$-th column of matrix $A$.

 A single or multiple list of matrices is referenced by one or more 
 indices in braces, like $A\{k\}$ or $A\{p,q\}$. 
 As for element indices, for double lists we may also use the 
 subscript - superscript alternative notation for $A\{p,q\}$,   
 namely, $\mapq{A}{p}{q}$. 
 This compact notation is specially usefull for building 
 block matrices, like in the following example, 
 \[ 
  A  = 
   \left[  \begin{array}{cccc} 
    \mapq{A}{1}{1} & \mapq{A}{1}{2} & \ldots & \mapq{A}{1}{s} \\ 
    \mapq{A}{2}{1} & \mapq{A}{2}{2} & \ldots & \mapq{A}{2}{s} \\ 
    \vdots & \vdots & \ddots & \vdots \\ 
    \mapq{A}{r}{1} & \mapq{A}{r}{2} & \ldots & \mapq{A}{r}{s} \\     
   \end{array} \right] \ . 
 \] 
 Hence, $A\{p,q\}(i,j)$ or $\mapqij{A}{p}{q}{i}{j}$ 
 indicates the element in the  
 $i$-th row and $j$-th column of the block situated at the 
 $p$-th block of rows and $q$-th block of columns of matrix $A$, 
 $A\{p,q\}(:,j)$ or $\mapqj{A}{p}{q}{j}$ indicates  
 the $j$-th column of the same block, and so on.

 An upper case letter usually stands for (or starts) a matrix name, 
 while lower case letters are used for vectors or scalars.   
 Whenever recommended by style or tradition, we may slightly
 abuse the notation using upper case for the name of a matrix and 
 lower case for some of its parts. 
 For example, we may write $x^j$, instead of $X^j$ for the $j$-th 
 column of matrix $X$.



  The vectors of zeros and ones, 
 with appropriate dimension given by the context, are 
 ${\bf 0}$ and ${\bf 1}$. 
  The transpose of matrix $M$ is $M'$, 
  and the transpose inverse, $M^{-t}$. 
  In $(M+v)$, where $v$ is a column (row) vector of compatible dimension, 
 $v$ is added to each column (row) of matrix $M$.

 A tilde accent, $\widetilde{A}$, indicates some simple transformation 
 of  matrix $A$. 
 For exemple, it may indicate a row and / or column permutation, 
 see next subsection.   
 A tilde accent may also indicate a normalization, like   
 $\widetilde{x}=(1/||x||)x$.

 The $p$-norm of a vector $x$ is given by  
 $||x||_p = \left( \sum |x_i|^p \right)^{-p}$. 
 Hence, for a non-negative vector  $x$, we can write its 1-norm as 
 $||x||_1 = \uno' x$. 
 $V>0$ is a positive definite matrix. 
  The Hadamard or pointwise product, $\odot$, is defined by   
 $M=A\odot B \Leftrightarrow M_i^j = A_i^j\, B_i^j$. 
 The squared Frobenius norm of a matrix is defined by  
 $\mbox{frob2}(M) = \sum_{i,j} (M_i^j)^2$.

 The Diagonal operator, $\diag$, 
 if applied to a square matrix, extracts the main diagonal as a vector, 
 and if applied to a vector, produces the corresponding diagonal matrix. 
 \[ 
   \diag(A) = 
   \left[  \begin{array}{c} 
    A_1^1 \\ A_2^2 \\ \vdots \\ A_n^n 
   \end{array} \right] 
   \ , \ \ 
   \diag(a) = 
   \left[  \begin{array}{cccc} 
    a_1 & 0  & \ldots & 0 \\ 
    0 & a_2 & \ldots & 0 \\ 
    \vdots & \vdots & \ddots & \vdots \\ 
    0 & 0 & \ldots & a_n    
   \end{array} \right] 
   \ , \ \ 
   \diag^2(A) = 
   \left[  \begin{array}{cccc} 
    A_1^1 & 0  & \ldots & 0 \\ 
    0 & A_2^2 & \ldots & 0 \\ 
    \vdots & \vdots & \ddots & \vdots \\ 
    0 & 0 & \ldots & A_n^n    
   \end{array} \right] \ . 
 \]

 The Kroneker product of two matrices is a block matrix where 
 block $\{i,j\}$ is the second matrix multiplied by element 
 $(i,j)$ of the first matrix:  
 \[ 
    A \otimes B \ = \ 
  \left[ \begin{array}{ccc} 
  A^1_1 B & A^2_1 B & \cdots \\ 
  A^1_2 B & A^2_2 B & \cdots \\ 
  \vdots & \vdots & \ddots 
  \end{array} \right] 
 \]   
 The following properties are easy to check: 
 \begin{itemize}  
 \item $(A\kron B)(C\kron D) = (AC)\kron(BD)$  
 \item $(A\kron B)' = A'\kron B'$  
 \item $(A\kron B)^{-1} = A^{-1}\kron B^{-1}$  
 \end{itemize}

 The $\Vec$ operator stacks the columns of a matrix into a single 
 column vector, that is, if  $A$ is $m\times n$, 
 \[ \Vec(A) = \left[ \begin{array}{c} 
    A^1 \\ \vdots \\ A^n \end{array} \right] 
 \]   
 The following properties are easy to ckeck: 
 \begin{itemize} 
 \item $\Vec(A+B) = \Vec(A) +\Vec(B)$ 
 \item $\Vec(AB) 
 =  \left[ \begin{array}{c} AB^1 \\ \vdots \\ AB^n \end{array} \right] 
 = (I\kron A) \Vec(B)$  
 \end{itemize}

 \subsubsection*{Permutations and Partitions}

 We now introduce some concepts and notations related to the permutation 
and partition of an $m \times n$ matrix $A$.  
 A permutation matrix is a matrix obtained by permuting rows and columns 
of the identity matrix, $I$. 
 To perform on $I$ a given row (column) permutation yields the 
corresponding row (column) permutation matrix. 

 Given row and column permutation matrices, $P$ and $Q$, 
the corresponding vectors of permuted row and column indices are 
 \[ p= (P \left[ \begin{array}{c} 1\\ 2\\ \vdots \\ m 
  \end{array} \right] )' 
 \] 
 \[ 
   q= \left[ \begin{array}{cccc} 1 & 2 & \ldots & n 
  \end{array} \right] Q 
 \] 

 To perform a row (column) permutation on a matrix $A$, 
 obtaining the permuted matrix $\tilde A$, is equivalent to 
 multiply it at the left (right) by the corresponding 
 row (column) permutation matrix.  
 Moreover, if $p$ ($q$) is the corresponding vector of 
permuted row (column) indices, 
 \[  
     A_p = P A =  I_p  A  \ \ , \ \   A^q = A Q = I^q  \ .  
 \] 

 Exemple: Given the martices 
 \[ 
 A=\left[ \begin{array}{ccc} 11 & 12 & 13 \\ 21 & 22 & 23 \\
 31 & 32 & 33 \end{array} \right] \ , \  \ 
 P=\left[ \begin{array}{ccc} 0 & 0 & 1 \\ 1 & 0 & 0 \\ 0 & 1 & 0 
 \end{array} \right] \ , \  \ 
 Q=\left[ \begin{array}{ccc}0 & 1 & 0 \\ 0 & 0 & 1 \\ 1 & 0 & 0 
 \end{array} \right] \ , 
 \]  
 \[ 
 p=q=\left[ \begin{array}{ccc} 3 & 1 & 2 \end{array} \right] \ , \ \ 
 PA=\left[ \begin{array}{ccc} 31 & 32 & 33 \\ 11 & 12 & 13 \\ 
 21 & 22 & 23 \end{array} \right] \ ,\ \ 
 AQ=\left[ \begin{array}{ccc} 13 & 11 & 12 \\ 23 & 21 & 22 \\ 
 33 & 31 & 32  \end{array} \right] \ . 
 \]

 A square matrix, $A$, is {\it symmetric} iff it is equal to its 
transpose, that is, iff $A=A'$.  
 A {\it symmetric permutation} of a square matrix $A$ is a permutation 
of form $\tilde A =PAP'$ or $\tilde A =Q'AQ$, where $P$ or $Q$ are 
(row or column) permutation matrices. 
 A square matrix, $A$, is {\it orthogonal} iff its inverse equals its 
transpose, that is, iff $A^{-1}=A'$. 
 The following statements are easy to check: \\
 (a) A permutation matrix is orthogonal. \\ 
 (b) A symmetric permutation of a symmetric matrix is still symmetric.

 A permutation vector, $p$, and a termination vector, $t$, 
 define a partition of $m$ original indices in $s$ classes: 
 \[ 
    \left[ \begin{array}{c}  
     p\left( 1 \right) \\ \vdots \\ p\left( t(1) \right)  
    \end{array} \right] \ , \ 
    \left[ \begin{array}{c}  
     p\left( t(1)+1 \right) \\ \vdots \\ p\left( t(2) \right)  
    \end{array} \right] \ \ldots \ 
    \left[ \begin{array}{c}  
     p\left( t(s-1)+1 \right) \\ \vdots \\ p\left( t(s) \right)  
    \end{array} \right] 
 \] 
 \[ 
    \Where 
    t(0)=0 < t(1) < \ldots < t(s-1) < t(s)=m \ . 
 \] 
 We define the corresponding permutation and partition matrices, 
 $P$ and $T$, as 
 \[   P = I_{ p\left(1 \ate m \right) }  = 
      \left[ \begin{array}{c} \map{P}{1} \\ \map{P}{2} \\ 
      \vdots \\ \map{P}{s} 
      \end{array} \right] 
   \ \ , \ \ 
    \map{P}{r} = I_{ p\left( t(r-1)+1 \ate t(r) \right) }  
   \ \ ,    
 \] 
 \[ 
     {\olP_r} = \uno' \left( {\map{P}{r}} \right) 
    \ \ \And \ \     
    \olP = \left[ \begin{array}{c} 
     {\olP_1} \\ \vdots \\ {\olP_s}   
      \end{array} \right] \ . 
 \]

  These matrices facilitate writing functions of a given partition, like 
  \begin{itemize} 
  \item The indices in class $r$ \\ 
   \[ 
      {\map{P}{r}}  \, (1\ate m) = 
      {\map{P}{r}} \left[ \begin{array}{c} 1 \\ \vdots \\ m 
          \end{array} \right]  = 
          \left[ \begin{array}{c} 
            p\left( t(r-1)+1 \right) \\ \vdots \\ p\left( t(r) \right)
           \end{array} \right] \ ; 
   \] 
  \item The number of indices in class $r$ \\ 
   \[ 
      {\olP_r} \, \uno = t(r) -t(r-1) \ ; 
   \] 
  \item A sub-matrix with the row indices in class $r$ \\ 
   \[ 
      {\map{P}{r}} \, A  = 
     \left[ \begin{array}{c} 
      A_{ p\left( t(r-1)+1 \right) } \\ \vdots \\ 
      A_{ p\left( t(r) \right) } 
     \end{array} \right] \ ; 
   \] 
  \item The summation of the rows of a submatrix with row indices  
        in class $r$  
   \[ 
     {\olP_r} \, A = \uno' \left( {\map{P}{r}} \, A \right) \ ; 
   \] 
  \item The rows of a matrix, added over each class  
    \[ \olP \, A =  \left[ \begin{array}{c} 
       {\olP_1} \, A \\ \vdots \\ {\olP_s} \, A  
       \end{array} \right] \ . 
    \] 
  \end{itemize}

 Note that a matrix $\olP$ represents a partition of $m$ idices into 
$s$ classes if $\olP$ has dimension $s\times m$,  
$\olP_h^j \in \{0,1\}$ and $\olP$ has orthogonal rows.   
 The element $\olP_h^j$ indicates if the index 
 $j\in 1\ate m$ is in class $h\in 1\ate s$.

 \section{Dense LU, QR and SVD Factorizations}

 \subsection*{Vector Spaces and Projectors}

 Given two vectors, $x,y \in {\Re }^n$, their {\it scalar product}  
 is defined as 
 \[  
      x'y = \sum_{i=1}^{n} x_{i}y^{i} \ . 
 \] 
 With this definition in mind, it is easy to check that the scalar product 
 satisfies the following properties of the {\it inner product} operator: 
 \begin{enumerate}
 \item $<x\mid y> = <y\mid x>$, symmetry. 
 \item $<\alpha x+\beta y\mid z> = 
      \alpha <x\mid z> + \beta <y\mid z>$, linearity. 
 \item $<x\mid x> \geq 0$ , semi-positivity.
 \item $<x\mid x>=0 \Leftrightarrow x=0$ , positivity.
 \end{enumerate}

 A given inner product defines the following norm, 
 \[ 
    \| x \| \equiv <x\mid x>^{1/2} \ ; 
 \]  
 that can in turn be used to define the angle between two vectors: 
 \[  
    \Theta(x,y) \equiv \arccos ( <x\mid y>  /  \| x\| \| y\| ) \ . 
 \]

 Let us consider the linear subspace generated by the columns of a matrix 
 $A$, $m$ by $n$, $m\geq n$: 
 \[ 
    C(A) = \{ y=Ax, x\in {\Re }^n \} \ .  
 \] 
 $C(A)$ is called the {\it image} of  $A$, and the complement of $C(A)$, $N(A)$,
 is called the {\it null space} of $A$, 
 \[  
    N(A) = \{ y \mid  A'y=0 \} \ .  
 \] 

 The projection of a vector  $b\in {\Re }^m$ in the column space of $A$ 
 is defined by the relations:  
 \[ 
     y = P_{C(A)}b \leftrightarrow y\in C(A) \wedge (b-y)\perp C(A) 
 \]  
 or, equivalently,  
 \[ 
    y= P_{C(A)}b \leftrightarrow y=Ax \wedge A'(b-y)=0 \ . 
 \] 

 In the sequel we assume that $A$ has full rank, i.e., that its 
columns are linearly independent. 
 It is easy to check that the projection of $b$ in $C(A)$ is given by 
the linear operator 
 \[ 
    P_A = A(A'A)^{-1}A' \ .  
 \]  
 If $y = A((A'A)^{-1}A'b)$, then  it is obvious that $y\in C(A)$.  
 At the other hand, 
 \[ 
     A'(b-y) = A'(I-A(A'A)^{-1}A')b = (A' - IA')b = 0 \ . 
 \]

\subsection*{Orthogonal Matrices}

 A real square matrix $Q$ is said to be {\it orthogonal} iff its inverse 
 is equal to its transpose, that is, $Q'Q=I$. 
 The columns of an orthogonal matrix $Q$ are a orthonormal basis for 
 ${\Re }^n$. 
 The quadratic norm of a vector $v$,  given by 
 \[ 
     {\| v \|}^{2} \equiv \sum (v_{i})^2 = v'v \ ,  
 \]  
 is not changed by an orthogonal transform, since 
 \[ 
    (Qv)'(Qv) = v'Q'Qv = v'Iv = v'v \ . 
 \] 
 
 Given a vector in ${\Re }^2$, 
 $\left[ \begin{array}{c} x_1 \\ x_2 \end{array} \right] $, 
 a rotation of this vector by an angle $\theta$ 
 is given by the linear transform 
 \[  
 G\{\theta \}x = \left[ \begin{array}{cc} 
 \cos (\theta ) & \sin (\theta ) \\ -\sin (\theta ) & \cos (\theta ) 
 \end{array} \right] 
 \left[ \begin{array}{c} x_1 \\ x_2 \end{array} \right] .
 \] 
 A rotation is an orthogonal transform, since 
 \[  
 G\{\theta \}' G\{\theta \}
 =
 \left[ \begin{array}{cc} {\cos (\theta )}^2 + {\sin (\theta 
 )}^2 & 0 \\ 
       0 & {\cos (\theta )}^2 + {\sin (\theta )}^2 
 \end{array} \right] 
 =
 \left[ \begin{array}{cc} 1 & 0 \\  0 & 1 \end{array} 
 \right] \ .
 \] 

 The Givens rotation is a linear operator whose matrix is the 
identity, except for the insertion of a bidimensional rotation matrix:  
  \[ 
 G\{i,j,\theta \} =
 \left[ \begin{array}{cccccccc} 
 1 & & & & & & & \\
   & \ddots & & & & & & \\
   & & \cos (\theta ) & & \sin (\theta ) & & & \\
   & & & \ddots & & & & \\
   & & -\sin (\theta ) & & \cos (\theta ) & & & \\
   &  & &  & &  & \ddots &  \\
   &  & &  & &  &  & 1 \\
 \end{array} \right] \ .
 \] 
 The left multiplication of matrix $A$ by a Givens transform, 
 $G'A$, rotates rows $i$ and $j$ of $A$ counterclockwise by an angle 
 $\theta$. 
 Since the product of orthogonal transforms is still orthogonal, we can 
 use a sequence of Givens rotations to build more complex orthogonal 
 transforms.

 We now define some simple bidimensional rotations that will be used as 
building blocks in the construction of several algorithms.   
 Let us consider, in $\Re^2$, a vector $v$,  a symmetric matrix $S$, 
and an asymmetric matrix $A$, 
 \[ 
 v= \left[ \begin{array}{c} x \\ y \end{array} \right]  
 \  , \ \  
 S= \left[ \begin{array}{cc} p & q \\ q & r \end{array} \right] 
 \ , \ \ 
 A= \left[ \begin{array}{cc} a & b \\ c & d \end{array} \right] 
 \] 

 In order to set to zero the second component of vector $v$ by means of 
a left rotation, $G\{\theta_v\}'\,v$, it is possible to use the angle 
 \[ 
    \theta_v=\arctan\left(\frac{y}{x}\right) \ . 
 \]  
 In order to diagonalize the symmetric matrix by a symmetric rotation, 
 $G\{ \theta_{diag} \}'\,S\,G\{ \theta_{diag} \}$, 
 it is possible to use the angle  
 \[ 
    \theta_{diag}=\frac{1}{2}\arctan\left(\frac{2q}{r-p}\right) \ . 
 \]  
 
 In order to symmetrize the asymmetric matrix by means of a left rotation, 
 $G\{\theta_{sym}\}'\,A$, 
 it is possible to use the angle  
  \[ 
       \theta_{sym}=\arctan\left(\frac{b-c}{a+d}\right) \ . 
  \]

 Hence, it is possible to diagonalize the asymmetric matrix by means of 
a symmetrization followed by a diagonalization operation. 
 Alternatively, it is possible to use the left and right of Jacobi 
rotations, $J\{\theta_{r}\}'\,A\,J\{\theta_{l}\}$,  defined as follows   
 \[ 
    \theta_{sum} = \theta_{r} +\theta_{l}  
                      = \arctan\left(\frac{c+b}{d-a}\right)  \ , \ \  
    \theta_{dif} = \theta_{r} -\theta_{l}  
                      = \arctan\left(\frac{c-b}{d+a}\right)  \ \Or        
 \]  
 \[ 
   J\{\theta_r\}' =  G\{\theta_{sum}/2\}'\,G\{-\theta_{dif}/2\}' \ , \ \  
   J\{\theta_l\}  =  G\{\theta_{dif}/2\} \,G\{\theta_{dif}/2\} \ .  
 \] 

 when computing the rotation matrices, there is no need to make explicit 
use of the rotation angles, nor is it necessary to use trigonometric 
functions, but only to compute  the factors 
 $c=\sin(\theta)$ \ and \ $s=\sin{\theta}$,  
 directly as 
 \[ 
    c = \frac{x}{\sqrt{x^2 +y^2}} \ , \ \ 
    s= \frac{-y}{\sqrt{x^2 +y^2}} \ . 
 \]   
 In order to avoid numerical overflow, one can use the procedure  
  \begin{itemize} 
  \item Se $y==0$\ , \ \Then $c=1 \ , \ s=0$ \ . 
  \item Se $y\geq x$ \ , \ \Then 
        $t=-x/y \ , \ s=1/\sqrt{1+t^2} \ , \ c=st$ \ . 
  \item Se $y<x$ \ , \ \Then 
        $t=-y/x \ , \ c=1/\sqrt{1+t^2} \ , \ s=ct$ \ . 
  \end{itemize}

\subsection*{QR Factorization}

 Given a full rank real matrix $A$, $m\times n$, $m\geq n$, 
 it is always possible to find an orthogonal matrix $Q$ such that 
 $A=Q \left[ \begin{array}{c} R \\ 0 \end{array} \right] $, 
 where $R$ is a square upper triangular matrix. 
 This is the QR factorization (or decomposition) of matrix $A$. 
 The orthogonal factor,  $Q=[C\mid N]$ gives an orthonormal basis for 
 ${\Re }^m$, where the first $n$ columns give an orthonormal base for 
 $C(A)$, and the last $m-n$ columns give an orthonormal base for $N(A)$, 
 as can be easily checked by the identity 
 $Q'A=\left[ \begin{array}{c} R\\ 0 \end{array} \right]$.  
 In the sequel a QR factorization algorithm is presented. 

 The following example illustrates a rotation sequence that takes a 
 $5\times 3$ matrix to upper triangular form. 
 Every index pair, $(i,j)$, indicates a rotation used to zero the 
 position at row $i$ column $j$. 
 We assume that the original matrix is dense, that is, that the matrix 
 has no zero elements, and illustrate the sparsity pattern in the matrix 
 as the algorithm progresses. 
 \[ 
 (1,5) * (1,4) (1,3) (1,2) * (2,5) (2,4) (2,3) * (3,5) (3,4) *
 \] 
 \[ 
 \left[ \begin{array}{ccc} 
  x & x & x \\ x & x & x \\ x & x & x \\ x & x & x \\ 0 & x & x
 \end{array} \right] \ \  
 \left[ \begin{array}{ccc} 
  x & x & x \\ 0 & x & x \\ 0 & x & x \\ 0 & x & x \\ 0 & x & x
 \end{array} \right] \ \  
 \left[ \begin{array}{ccc} 
  x & x & x \\ 0 & x & x \\ 0 & 0 & x \\ 0 & 0 & x \\ 0 & 0 & x
 \end{array} \right] \ \  
 \left[ \begin{array}{ccc} 
  x & x & x \\ 0 & x & x \\ 0 & 0 & x \\ 0 & 0 & 0 \\ 0 & 0 & 0
 \end{array} \right]  
 \] 

\subsubsection*{Least Squares}

 Given an over-determined system, $Ax=b$ where $A$ is $m\times n$, 
 $m>n$, vector $x^*$ is a least squares solution to the system iff 
 $x^*$ minimizes the quadratic norm of the residual, that is,  
 \[ 
    x^* = Arg \min_{x\in {\Re }^n} \| Ax - b {\|} \ , 
 \] 

 Since an orthogonal rotation does not change the square norm of a vector, 
one can seek the least square solution to this system minimizing the 
residual of the system transformed by the orthogonal factor of 
the QR factorization of $A$,   
 \[ 
 \| Q'(Ax-b) {\|}^2 = 
 \| \left[ \begin{array}{c} R \\ 0 \end{array} \right] x - 
    \left[ \begin{array}{c} c \\ d \end{array} \right] 
 {\|}^2 = 
 \| Rx-c {\|}^2  +  \| 0x-d {\|}^2 .
 \]

 From the last expression one can see that the solution and the residual 
of the original problem are given by 
 \[ 
 x^* = R^{-1}c \ , \ \ 
 y = Ax^* \mbox{\ \ \mbox{and} \ \ }
 z = Q \left[ \begin{array}{c} 0 \\ d \end{array} \right] .
 \] 
 Since the last $m-n$ columns of $Q$ are an orthonormal basis of $N(A)$, 
 we see that $z \perp C(A)$, and can therefore conclude that $y=P_{A}b$.

\subsection*{LU and Cholesky Factorizations}

 Given a matrix $A$, the {\it elementary operation} given by the  {\it
multiplier} $m^j_i$, is the operation of subtracting from row $i$ 
 the row $j$ multiplied by $m^j_i$.
 The elementary operation applied to the identity matrix generates the 
corresponding {\it elementary matrix},
 \[ 
 M\{ i,j \}=
 \left[ \begin{array}{ccccccc} 
   1 & & & & & &  \\
   & \ddots & & & & &  \\
   & & 1 & & & &  \\
   & & \vdots & \ddots & & &  \\
   & & -m_i^j & & 1 & &  \\
   & & \vdots & & & \ddots &  \\
   & & & & & & 1 \\
 \end{array} \right] \ 
 \begin{array}{c} \\ \\ j \\ \\ i \\ \\ \\ \end{array} \ . 
 \]  
 Applying a elementary operation to matrix $A$ is equivalent to 
multiplying $A$ from the left by the corresponding elementary matrix.

 In the {\it Gaussian elimination} algorithm we use a 
sequence of elementary operations to bring $A$ to upper triangular form, 
 \[ 
    MA=M\{n,n-1\}M\{n-1,n-2\}M\{n,n-2\}\cdots 
 \] 
 \[  
      M\{3,2\}\cdots M\{n-1,2\}M\{n,2\}M\{2,1\}\cdots 
      M\{n-1,1\}M\{n,1\}A=U \ . 
 \]  
 Multiplier $m^j_i$ is computed as the current matrix element at position 
$(i,j)$ divided by the {\it pivot} element at the diagonal position 
$(j,j)$.  
 Elementary operation $M\{i,j\}$ is used to {\it eliminate} (zero) the 
position $(i,j)$.  
 The elementary operations are performed in an order that prevents 
the zeros created at previous steps to be filled again.  

 The next example shows the steps of Gaussian elimination on a small matrix. 
The multipliers, in italic, are stored at the positions corresponding 
to the zeros they created. 
 \[   
 \left[ \begin{array}{rrr} 
  2 & 1 & 3 \\ 2 & 3 & 6 \\ 4 & 4 & 6 
 \end{array} \right] \ \  \rightarrow \ \  \left[ \begin{array}{rrrr} 
  2 & 1 & 3 \\  {\it 1} & 2 & 3 \\ {\it 2} & 2 & 0  
 \end{array} \right] \ \  \rightarrow \ \  \left[ \begin{array}{rrrr} 
  2 & 1 & 3 \\ {\it 1} & 2 & 3 \\ {\it 2} & {\it 1} & -3  
 \end{array} \right] \ \ 
 \] 

 The inverse of the product of this sequence of elementary matrices 
has the lower triangular form, that is, 
 \[ 
   M^{-1}=        
   M^{-1}\{n,1\}M^{-1}\{n-1,1\}\cdots
   M^{-1}\{2,1\}M^{-1}\{n,2\}M^{-1}\{n-1,2\} 
 \] 
 \[  
   \cdots M^{-1}\{3,2\}
   \cdots M^{-1}\{n,n-2\}M^{-1}\{n-1,n-2\}M^{-1}\{n,n-1\} \ . 
 \]  
 \[ 
 L= M^{-1}=
 \left[ \begin{array}{lllll}
 1 & & & & \\
 m^1_2 & 1 & & \\
 \vdots & \vdots & \ddots & & \\
 m^1_{n-1} & m^2_{n-1} &  & 1 & \\  
 m^1_n & m^2_n & \cdots & m^{n-1}_n & 1 \\
 \end{array} \right] \ .  
 \] 
 Therefore the algorithm finds the LU factorization, $A=LU$. 
 The lower and upper triangular form of $L$ and $U$ allow us to easily 
 compute $L^{-1}z$ and $U^{-1}z$ by simple forward and backward 
 substitution. 
 Hence, $A^{-1}z=U^{-1}(L^{-1}z)$ can be computed in just two 
 substitution steps.

 In case we factor a symmetric matrix $V=LU$, 
we can collect the diagonal elements of $U$ in a diagonal matrix $D$, 
and write $V=LDL'$. 
 If $S$ is positive definite we can take the square roots of the diagonal 
elements and write $D=D^{1/2}D^{1/2}$. 

 Defining $C=LD^{1/2}$, we have $V=CC'$, the {\it Cholesky factor} of $V$.
 For reasons of numerical stability, it is recommended to take the square 
roots of each diagonal elements just before we use it as a 
{\it pivot element}, and then eliminate the elements of its column, 
 see Pissanetzky (1984).

\subsubsection*{Quadratic Programming}

 The {\it quadratic programming} problem with equality constraints 
is the minimization of the objective function 
 \[  f(y) \equiv (1/2)y'Wy + c'y \ , \ \ W=W' \] 
 with the constraints 
 \[  g_{i}(y) \equiv N_{i}'y = d_i . \] 
The gradients of $f$ and $g_i$ are given by  
 \[  {\nabla}_{y}f = y'W + c' \ , \ \mbox{and} \  
     {\nabla}_{y}g_i = N_{i}' \ .
 \] 
 The Lagrange (first order) optimality conditions state that the 
constraints are in effect, and that objective function gradient 
equals a linear combination of gradients of the constraint functions, 
 Hence, the solution may be obtained from the {\it Lagrange multipliers}, 
i.e., the vector $l$ with the coefficients of the aformentioned linear 
combination.  
 \[ 
   N'y = d \ \wedge \ 
   y'W + c' = l' N' \ , 
 \]  
 or, in matrix form, 
 \[ 
 \left[ \begin{array}{cc} N' & 0 \\ W & N \end{array} \right] 
 \left[ \begin{array}{c} y \\ l \end{array} \right] = 
 \left[ \begin{array}{c} d \\ c \end{array} \right] \ .  
 \] 
 These equations are known as the {\it normal system}, with a symmetric 
coefficient matrix. If quadratic form $W$ is positive definite,  
 i.e. if $\forall x\ x'Wx \geq 0 \ \wedge \ x'Wx=0 \Leftrightarrow x=0$, 
 and the constraint matrix $N$ is full rank, the coefficient matrix of the 
normal system is also positive definite.

\subsection*{SVD Factorization} 

 The SVD factorization takes a real matrix $A$, 
 $m\times n,\, m\geq n$, to a diagonal matrix, $D$, by left and right 
 multiplication by orthogonal matrices  $D=U'AV$, 
 Let us first consider the case $m=n$, i.e. a square matrix. 

 The Jacobi algorithm is an iterative procedure that, at each iterations, 
 ``concentrates'' the matrix in the diagonal by a Jacobi rotation, 
 \[ 
    J\{i,j,\theta_r\}'\,A\{k\}\,J\{i,j,\theta_l\} = A\{k\ma1 \} = 
 \] 
 \[
    \left[ \begin{array}{ccccccc} 
    \mapij{A}{k\ma1 }{1}{1} & \cdots & \mapij{A}{k\ma1 }{1}{i} & \cdots & 
    \mapij{A}{k\ma1 }{1}{j} & \cdots & \mapij{A}{k\ma1 }{1}{n} \\  
    \vdots        & \ddots & \vdots        & \ddots & 
    \vdots        & \ddots  & \vdots \\ 
    \mapij{A}{k\ma1 }{i}{1} & \cdots & \mapij{A}{k\ma1 }{i}{i} & \cdots & 
    0             & \cdots & \mapij{A}{k\ma1 }{i}{n} \\  
    \vdots        & \ddots & \vdots        & \ddots & 
    \vdots        & \ddots  & \vdots \\ 
    \mapij{A}{k\ma1 }{j}{1} & \cdots & 0             & \cdots & 
    \mapij{A}{k\ma1 }{j}{j} & \cdots & \mapij{A}{k\ma1 }{j}{n} \\  
    \vdots        & \ddots & \vdots        & \ddots & 
    \vdots        & \ddots  & \vdots \\ 
    \mapij{A}{k\ma1 }{n}{1} & \cdots & \mapij{A}{k\ma1 }{n}{i} & \cdots & 
    \mapij{A}{k\ma1 }{n}{j} & \cdots & \mapij{A}{k\ma1 }{n}{n}   
    \end{array} \right] 
  \]   

  Let us consider the sum of squares of of-diagonal elements of $A$, 
  $\mbox{Off}_2(A)$.  We can see that 
  \[ 
     \mbox{Off}_2(A\{k\ma1 \}) = \mbox{Off}_2(A\{k\})     
         -(\mapij{A}{k}{i}{j})^2 -(\mapij{A}{k}{j}{i})^2 
  \] 
   Hence, choosing at each iteration the index pair that maximizes the 
   sum of squares of the corresponding elements, the algorithms 
   converges linearly to a diagonal matrix. 

   The Jacobi algorithm gives a constructive proof for the existence 
   of the SVD factorization, and is the basis of several efficient 
   numerical algorithms. 
 
   If $A$ is a rectangular matrix, one can first find its 
   QR factorization, and then apply Jacobi algorithm to the 
   upper triangular $R$ factor. 
   If $A$ is square and symmetric, the obtained factorization is 
   known as the {\it eigenvalue decomposition} of $A$. 

   The orthogonal matrices $U$ and $V$ can be interpreted as orthonormal 
 bases in the respective $m$ and $n$ dimensional spaces. 
  The values at the diagonal of $S$ are called the {\it singular values} 
 of matrix $A$, and can be interpreted geometrically as the scaling factors 
 of the map $A=UDV'$, taking each versor of the basis $V$ to a scaled 
 versor of the basis $U$.

 \subsection*{Complex Matrices} 
  
 Many techniques developed in this section for real matrices can be 
generalized to complex matrices. Practical and elegant methods of obtaining  
and describing such generalizations are the described by Hemkumar (1994) 
using {\it Cordic transforms} (COordinate Rotation Digital Computer). 
 Such a transform is applied to a  $2\times 2$ complex matrix $M$ 
 in the form of {\it internal and external rotations pairs}, 
 \[ 
 \left[ \begin{array}{cc} 
  c(\phi) & -s(\phi) \\ s(\phi) & c(\phi) 
 \end{array} \right]  
 \left[ \begin{array}{cc} 
  e(i\alpha) & 0 \\ 0 & e(i\beta)  
 \end{array} \right]  
 \left[ \begin{array}{cc} 
  Ae(ia) & Be(ib) \\ Ce(ic) & De(id)   
 \end{array} \right]  
 \left[ \begin{array}{cc} 
  e(i\gamma) & 0 \\ 0 & e(i\delta)  
 \end{array} \right]  
 \left[ \begin{array}{cc} 
  c(\psi) & -s(\psi) \\ s(\psi) & c(\psi)  
 \end{array} \right]   
 \] 

 The elegance of these Cordic transforms comes from the following 
 observations: 
 The internal transform affects only the imaginary exponents 
 of the matrix elements, while the external transform can be 
 independently applied to the real and the imaginary parts of the 
 matrix, that is,  
 \[ 
 \left[ \begin{array}{cc} 
  e(i\alpha) & 0 \\ 0 & e(i\beta)  
 \end{array} \right]  
 \left[ \begin{array}{cc} 
  Ae(ia) & Be(ib) \\ Ce(ic) & De(id)   
 \end{array} \right]  
 \left[ \begin{array}{cc} 
  e(i\gamma) & 0 \\ 0 & e(i\delta)  
 \end{array} \right]  = 
 \]  
 \[ 
 \left[ \begin{array}{cc} 
  Ae(ia') & Be(ib') \\ Ce(ic') & De(id')   
 \end{array} \right] =  
 \left[ \begin{array}{cc} 
  Ae(i(a+\alpha+\gamma)) & Be(i(b+\alpha+\delta)) \\ 
  Ce(i(c+\beta+\gamma)) & De(i(d+\beta+\gamma))   
 \end{array} \right] 
 \] 
 \[ 
 \left[ \begin{array}{cc} 
  c(\phi) & -s(\phi) \\ s(\phi) & c(\phi) 
 \end{array} \right]  
 \left[ \begin{array}{cc} 
  A'_r +iA'_i & B'_r +iB'_i \\ C'_r +iC'_i & D'_r +iD'_i   
 \end{array} \right]  
 \left[ \begin{array}{cc} 
  c(\psi) & -s(\psi) \\ s(\psi) & c(\psi)  
 \end{array} \right] =  
 \] 
 \[ 
 \left[ \begin{array}{cc} 
  c(\phi) & -s(\phi) \\ s(\phi) & c(\phi) 
 \end{array} \right]  
 \left[ \begin{array}{cc} 
  A'_r & B'_r \\ C'_r & D'_r    
 \end{array} \right]  
 \left[ \begin{array}{cc} 
  c(\psi) & -s(\psi) \\ s(\psi) & c(\psi)  
 \end{array} \right] 
 \] 
 \[ 
 + i \left(   
 \left[ \begin{array}{cc} 
  c(\phi) & -s(\phi) \\ s(\phi) & c(\phi)  
 \end{array} \right]  
 \left[ \begin{array}{cc} 
  A'_i & B'_i \\ C'_i & D'_i     
 \end{array} \right]  
 \left[ \begin{array}{cc} 
  c(\psi) & -s(\psi) \\ s(\psi) & c(\psi)  
 \end{array} \right] \right)   
 \] 

 The following table defines some useful internal and external transforms. 
 Type I transforms change the imaginary exponents of  the matrix elements 
 at one of the diagonals. Transforms of Type R, C and D  make real the 
 elements in a row, column or diagonal.

 \begin{center} 
 \begin{tabular}{|c|c|} 
 \hline 
 Type & Value \\ 
 \hline $I_{main}$ & 
 $\alpha = -\beta = \gamma = -\delta = (d-a)/2$ \\ 
 \hline $I_{off}$ & 
 $\alpha = -\beta = -\gamma = \delta = (c-b)/2$ \\ 
 \hline $R_{up}$ & 
 $\alpha = \beta = -(b+a)/2 \ ; \  \gamma = -\delta = (b-a)/2$ \\ 
 \hline $R_{low}$ & 
 $\alpha = \beta = -(d+c)/2 \ ; \  \gamma = -\delta = (d-c)/2$ \\ 
 \hline $C_{left}$ & 
 $\alpha = -\beta = (c-a)/2 \ ; \  \gamma = \delta = -(c+a)/2$ \\ 
 \hline $C_{right}$ & 
 $\alpha = -\beta = (d-b)/2 \ ; \  \gamma = \delta = -(d+b)/2$ \\ 
 \hline $D_{main}$ & 
 $\alpha = \beta = -(d+a)/2 \ ; \  \gamma = -\delta = (d-a)/2$ \\ 
 \hline $D_{off}$ & 
 $\alpha = \beta = -(b+c)/2 \ ; \  \gamma = -\delta = (b-c)/2$ \\ 
 \hline 
 \end{tabular} 
 \end{center} 

 It is easy to see that a sequence of internal transforms is equivalent 
 to a single internal transform whose parameters are the sum of the 
 coresponding parameters of the transforms in the sequence.  

 Combining internal and external transforms, it is possible to create 
 HT's for several interesting algorithms. 
 For example, the HT's of type I, II and III in the following table 
 can be used to obtain the SVD factorization of a complex matrix, 
 much like the Jacobi algorithm. 
 A type I transform applies $R_{low}$ followed by a rotation, making the 
 matrix upper triangular. 
 A type II transform applies $D_{main}$, $I_{off}$ followed by a 
 diagonalization. 
 For Hermitian (self-adjoint) matrices, the diagonalization is 
 obtained using only one transform of type III

 \begin{center} 
 \begin{tabular}{|c|c|c|} 
 \hline 
 Type & Internal & External \\ 
 \hline I & 
 $\alpha = \beta = -(d+c)/2 \ ; \  \gamma = -\delta = (d-c)/2$ &  
 $\phi=0 \ ; \ \psi = \arctan(C/D)$ \\ 
 \hline II & 
 $\alpha = -(a+b)/2 \ ; \ \beta = \gamma = -\delta = (b-a)/2$ &  
 $\phi \pm \psi = \arctan(B/(D \mp A))$ \\ 
 \hline III & 
 $\alpha = -\beta = -\gamma = \delta = -b/2$ &  
 $\phi = \psi = \arctan(2B/(D-A))/2$ \\ 
 \hline 
 \end{tabular} 
 \end{center}

\subsection*{Exercises}

 \begin{enumerate}
 \item Use the fundamental properties of the inner product to prove that: 
  \begin{enumerate}
  \item The Cauchy-Scwartz inequality:
   $|<x\mid y>| \leq \| x\| \| y\|$.
   Suggestion: Compute ${\| x-\alpha y \|}^2$ for  
   $\alpha =<x\mid y>^2 / \| y\|$.
  \item The triangular inequality:
   $\| x+y\| \leq \| x\| +\| y\| $.
  \item In which case do we have equality or strict Cauchy-Schwartz 
   inequality? Relate your answer to the definition of angle between two 
   vectors.  
  \end{enumerate}
\item  Use the definition of inner product in ${\Re }^n$ to prove  
 the parallelogram law: 
 ${\| x+y\|}^2 + {\| x-y\|}^2 = 2{\| x\|}^2 + 2{\| y\|}^2 $.
 \item A matrix is idempotent, or a non-orthogonal projector, iff 
 $P^{2} = P$. Prove that: 
\begin{enumerate}
 \item $R = (I-P)$ is idempotent.
 \item ${\Re }^n = C(P) + C(R)$.
 \item All eigenvalues of $P$ are $0$ or $+1$. Suggestion:
  Show that if $0$ is a root of the characteristic  polynomial 
  of $P$, ${\varphi}_{P}(\lambda )\equiv \det (P-\lambda I)$, than 
  $(1-\lambda)=1$ is a root of ${\varphi}_{R}(\lambda )$.
\end{enumerate}

\item Prove that $\forall P$ idempotent and symmetric, 
  $P = P_{C(P)}$. Suggestion: Show that $P'(I-P)=0$.

\item Prove that the projection operator into a given vector subspace,  
  $V$, $P_{V}$, is unique and symmetric.

\item Prove Pythagoras theorem: 
  $\forall b \in {\Re }^{m} , u \in V$ we have 
  ${\| b-u \| }^2 = {\| b-P_{V}b \| }^2 + {\| P_{V}b - u \| }^2$.

\item Assume we have the QR factorization of a matrix $A$. 
  Consider a new matrix, $\tilde A$, obtained from $A$ by the substitution 
 of a single column. How could we update our orthogonal factorization 
 using only $3n$ rotations? 
 Suggestion: (a) Remove the altered column of $A$ and update the 
 factorization using at most $n$ rotations. 
 (b) Rotated by the new column by the current orthogonal factor. 
  $\tilde a = Q'a = R^{-t}A'a$. 
 (c) Add $\tilde a$ as the last column of $\tilde A$, 
 and update the factorization using $2n$ rotations.

\item Compute the $LDL$ and Cholesky factorizations of matrix 
      \[ \left[ \begin{array}{rrrr}
         4 & 12 & 8 & 12 \\ 12 & 37 & 29 & 38 \\
         8 & 29 & 45 & 50 \\ 12 & 38 & 50 & 113 \\
         \end{array} \right] \ . 
      \] 
\item Prove that: 
  \begin{enumerate}
  \item $(AB)'=B'A'$.
  \item $(AB)^{-1}=B^{-1}A^{-1}$.
  \item $A^{-t}\equiv (A^{-1})' = (A')^{-1}$.
  \end{enumerate}
\item Describe four algorithms to compute $L^{-1}x$ and $L^{-t}x$,
   accessing the unit diagonal and lower triangular matrix $L$ 
   row by row or column by column. 
\end{enumerate}

 \section{Sparse Factorizations}

 As indicated in chapter 4, we present in this appendix 
some aspects related to the sparse factorization. 
 This material has strong connections with the issues discussed in 
chapter 4, but is more mathematical in its nature,  and can be omitted
by the reader interested mostly in the purely  epistemological aspects
of decoupling.

 Computing the Cholesky factorization of a $n\times n$ matrix involves 
 on the order of $n^3$ arithmetical operations. 
 Large models may have thousands of variables, so it seems that 
 decoupling large models requires a lot of work. 
 Nevertheless, in practice, matrices appearing in large models are 
 typically sparse and structured.   
 A matrix is called {\it sparse} if it has many zero elements, 
 otherwise it is called {\it dense}. 
 A sparse matrix is called {\it structured} if its non-zero-elements (NZEs) 
 are arranged in a ``nice'' pattern. 
 As we will see in the next sections, we may be able to obtain a 
 Cholesky factor, $L$, of a (permuted) sparse and structured matrix $V$, 
 that `preserves' some of its sparsity and structure, 
 hence decreasing the computational work.

 \subsection{Sparsity and Graphs}

 In the discussion of sparsity and structure, the language of graph 
theory is very helpful. This section gives a quick review of 
some of the basic concepts on directed and undirected graphs, 
and also defines the process of vertex elimination. 
 
 A Directed Graph, or DG, $\mathcal{G}=(\mathcal{V},\mathcal{A})$ has a
set of vertices or nodes, $\mathcal{V}$,  indexed  by natural numbers,
and a set or directed arcs, $\mathcal{A}$, where each arc joins two
vertices. 
 We say that arc $(i,j)\in \mathcal{A}$ goes from node $i$ to node $j$. 
 When drawing a graphical representation of a DG, it is usual to 
represent vertices by dots, and arcs by a arrows between the dots.  
 In a DG, we say that $i$ is a parent of $j$, $i\in \pa(j)$, or that $j$
is a  child of $i$, $j\in \ch(i)$, if there is an arc going from $i$
to $j$. The  children of $i$, the children of its children, and so on,
are the  descendents of $i$. 
 If $j$ is a descendent of $i$ we say that there is a {\it path} in 
$\mathcal{G}$ going from $i$ to $j$. A {\it cycle} is a path from a
given vertex to itself. 
 An arch from a vertex to itself, $(j,j)$ is called a {\it loop}. 
 In some situations we spare the effort of multiple definitions
of essentially the same objects by referring to the same graph with   
or without all possible loops.    

  There is yet another representation for a DG, $\mathcal{G}$, given by
 $(\mathcal{V},B)$, where the {\em adjacency matrix}, $B$, is the Boolean 
 matrix $B(i,j)=1$ if arc $(i,j)\in \mathcal{A}$, and $0$ otherwise. 
 The key element relating the topics presented in this and the previous 
section, is the Boolean matrix $B$ indicating the non-zero elements of 
the numerical matrix $A$, $B_i^j = I(A_i^j \neq 0)$. 
 In this way, the graph $\mathcal{G}=(\mathcal{V},B)$ is used to 
represent the sparsity pattern of a numerical matrix $A$.

 A Directed Acyclic Graph, DAG, has no cycles. 
 A {\it separator} $S\subset \mathcal{V}$ separates $i$ from
$j$ if any path  from $i$ to $j$ goes through a vertex in $S$. 
 A vertex $j$ is a {\em spouse} of vertex $i$, $j\in \sp(i)$, 
 if they have a child in common. 
 A {\em tree} is a DAG where each vertex has exactly one parent, except 
for the root vertex, that has no parent. The leafs of a tree are the 
vertices with no children.  
 A graph composed by several trees is a {\it forest}.  

 An Undirected Graph, or UG, is a DG where, if arc $(i,j)$ is in the
graph, so is its opposite, $(j,i)$. An UG can also be represented as 
 $\mathcal{G}=(\mathcal{V},\mathcal{E})$, where each undirected edge,
$\{i ,j\}\in \mathcal{E}$, stands for the  pair of opposite directed
arcs, $(i,j)$ and $(j,i)$. Obviously, the adjacency matrix of a UG 
is a symmetric matrix, and vice-versa.

 \[ 
 \begin{array}{ccccc} 
 1 & & 3 & \rightarrow & 5 \\ 
 \downarrow & \nearrow &  & \searrow &  \\ 
 2 & \rightarrow & 4 & \rightarrow & 6 \\ 
 \end{array} 
 \ \ \ , \ \ \ 
 \begin{array}{ccccc} 
 1 &  & 3 & - & 5 \\ 
 | & \slash & | & \backslash & \\ 
 2 & - & 4 & - & 6 \\ 
 \end{array} 
 \ \ \ , \ \ \ 
 \begin{array}{ccccc} 
  &  & 3  & & \\ 
  &  &  &  & \\ 
 2 &  & {\bf 4} &  & 6 \\ 
 \end{array} 
 \] 
 \centerline{Figure 2: A DAG and its Moral Graph.} 
 \\ \mbox{}

 The moral graph of the DAG $\mathcal{G}$, $\mathcal{M}(\mathcal{G})$,
is the undirected graph  with the same nodes as $\mathcal{G}$, and edges
joining nodes $i$ and $j$ if they are  immediate relatives in
$\mathcal{G}$. The immediate relatives of a node in $\mathcal{G}$ 
include its parents, children and spouses (but not brooders or sisters). 
 The set of immediate relatives of $i$ is also called the Markov blanket
of $i$, $\mb(i)$, hence, $j\in \mb(i)$ if $j$ is a neighbor of $i$ in
the moral graph. 
 Figure 2 represents a DAG, its moral graph,   
 and the Markov blanket of one of its vertices.

 Sometimes it is important to consider an order on the vertex set, 
 established by an `index vector' $q$, in (a subset of) 
 $\mathcal{V}=\{1,2,\ldots N\}$. For example, 
 we can consider the natural order $q=[1,2,\ldots N]$, 
 or the order given by a permutation, $q=[q(1),q(2),\ldots q(N)]$.  

 In order not to make language and notation too heavy, we may refer to 
 the vertex `set' $q$, meaning the set of elements in vector $q$.  
 Also, given two index vectors, $a=[a(1),\ldots a(A)]$ and  
 $b=[b(1),\ldots b(B)]$, the index vector   
 $c= a\cup b$, has all the indices in $a$ or $b$. 
 Similarly, $c= a\backslash b$ has all the indices in $a$ that are not
 in $b$. 
 These are essentially set operations but, since an index vector also 
 establishes an order of its elements, $c=[c(1),\ldots c(C)]$,  
 this order, if not otherwise indicated, has somehow to be chosen.  

 We define the elimination process in the UG, 
 $\mathcal{G}=(\mathcal{V},\mathcal{E})$, $\mathcal{V}=\{1,\ldots N\}$
 given an elimination order, $q=[q(1),\ldots q(N)]$, as the sequence of 
 {\it elimination graphs} 
 $\mathcal{G}_k=(\mathcal{V}_k,\mathcal{E}_k)$ where, for
 $k=1\ldots n$,
 \[
   \mathcal{V}_k=\{q(k), q(k+1), \ldots q(n)\}, \ \ \ 
   \mathcal{E}_1=\mathcal{E}, \ \ \mbox{and, for}\ k>1\ , 
 \] 
 \[ 
   \{i,j\}\in \mathcal{E}_k \ \Leftrightarrow \ 
   \left\{ \begin{array}{l} 
    \{i,j\} \in \mathcal{E}_{k-1} \ , \ \mbox{or} \\ 
    \{q(k-1),i\} \in \mathcal{E}_{k-1} \ \mbox{and} \  
     \{q(k-1),j\} \in \mathcal{E}_{k-1} \ . 
   \end{array} \right. 
 \] 
 that is, when eliminating vertex $q(k)$, we make its neighbors a 
{\it clique}, adding all missing edges between them.  

 The {\it filled graph} is the graph $(\mathcal{V},\mathcal{F})$, where  
 $\mathcal{F}= \cup_{k=1}^n \mathcal{E}_k$.  
 The original edges and the filled edges in $\mathcal{F}$ are,
 respectively, the edges in $\mathcal{E}$ and in 
 $\mathcal{F} \backslash \mathcal{E}$.

 Figure 3 shows a graph with 6 vertices, the elimination graphs,  and the
filled graph, considering the  elimination order $q=[1,3,6,2,4,5]$. 

 \[ 
 \begin{array}{cccc} 
   {\bf 1} & -      & 3 \\ 
         | & \times &  & \backslash \\ 
         2 &        & 6 & \mbox{}\; | \\ 
           &  /     &   & / \\ 
         5 &        & 4 
 \end{array} \ \ \  
 \begin{array}{cccc} 
     &        & {\bf 3} \\ 
     & /      & {\bf |} & \backslash \\ 
   2 & {\bf -} & 6 & \mbox{}\; | \\ 
     & /      &   & / \\ 
   5 &        & 4 
 \end{array} \ \ \  
 \begin{array}{ccc} 
   2 & -      & {\bf 6}  \\ 
     & \times & {\bf |}  \\ 
   5 &        & 4 
 \end{array} \ \ \  
 \begin{array}{ccc} 
   {\bf 2} &        &   \\ 
    {\bf |} & \backslash  &   \\ 
   5 & {\bf -} & 4  
 \end{array} \ \ \  
 \begin{array}{c} 
  {\bf 4} \\ | \\ 5    
 \end{array} \ \ \  
 \begin{array}{cccc} 
   1 &  -     & 3 \\ 
   | & \times & | & \backslash \\ 
   2 & -      & 6 & \mbox{}\; | \\ 
   | & \times & | & / \\ 
   5 & -      & 4 
 \end{array} 
 \]  
 \centerline{Figure 3: Elimination Graphs.}  
 \\ \mbox{}

 There is a computationally more efficient form of obtaining the 
filled graph, known as {\it simplified elimination}:  
 In the simplified version of the elimination graphs, $\mathcal{G}_k^*$, 
when eliminating vertex $q(k)$, we add only the clique edges 
incident to its neighbor, $q(l)$, that is next in the elimination 
order. 
 Figure 4 shows the simplified elimination graphs and the filled graph 
corresponding to the elimination process in Figure 3; 
 The vertex being eliminated is in boldface, and his next (in the 
elimination order) neighbor in italic.

 \[ 
 \begin{array}{cccc} 
   {\bf 1} & -      & {\it 3} \\ 
         | & \times &  & \backslash  \\ 
         2 &        & 6 & \mbox{}\; | \\ 
           & /      &   & / \\ 
         5 &        & 4 
 \end{array} \ \ 
 \begin{array}{cccc} 
     &        & {\bf 3} \\ 
     & /      & | & {\bf \backslash } \\ 
   2 &        & {\it 6} & \mbox{}\; | \\ 
     & /      &   & / \\ 
   5 &        & 4 
 \end{array} \ \ 
 \begin{array}{cccc} 
   {\it 2} & {\bf - } & {\bf 6}  \\ 
     & /      & {\bf |} \\ 
   5 &        & 4 
 \end{array} \ \ 
 \begin{array}{cccc} 
   {\bf 2} &        &   \\ 
   {\bf | } & {\bf \backslash } &   \\ 
   5 &        & {\it 4}  
 \end{array} \ \ 
 \begin{array}{c} 
  {\bf 4} \\ | \\ {\it 5}     
 \end{array} \ \ 
 \begin{array}{cccc} 
   1 &  -     & 3 \\ 
   | & \times & | & \backslash \\ 
   2 & -      & 6 & \mbox{}\; | \\ 
   | & \times & | & / \\ 
   5 & -      & 4 
 \end{array} 
 \]  
 \centerline{Figure 4: Simplified Elimination Graphs.}   
 \\ \mbox{}

 An elimination order is perfect if it generates no fill.   
 Perfect elimination is the key to relate the vertex elimination
process to the theory of chordal graphs, see Stern (1994). 
 Chordal graph theory provides a unified framework for similar
elimination processes in several other contexts, see 
Golumbic (1980) Stern (1994) and Lauritzen (2006). 
 Nevertheless, we will not explore this connection any further 
in this paper.

 The material presented in this section will be used in 
 the next two sections for the analysis of the sparsity structure
 in Cholesky  factorization and Bayesian networks. 
 This structure is the key for efficient decoupling, allowing the 
computation of large models, used in the analysis of large
systems. 
 These structural aspects have been an area of intense 
research by the designers of efficient numerical algorithms.    
 However, the same area has not been able to attract so much interest 
in statistical modeling. 
 From the epistemological considerations in the following chapters, 
we hope to convince the reader that this is a topic that deserves to
receive much more attention from the statistical modeler.

 \subsection{Sparse Cholesky Factorization}

 Let us begin with some matrix notation. 
 Given a matrix A, and index vectors $p$ and $q$, the equivalent 
 notations $A(p,q)$ or $A_p^q$ indicate the (sub) matrix of rows and 
 columns extracted from $A$ according to the indices in $p$ and $q$. 
 In particular, if $p$ and $q$ have single indices, $i$ and $j$, 
 $A(i,j)$ or $A_i^j$ indicate the element of $A$ in row $i$ and 
 column $j$. The next example shows a more general case:  
 \[ 
  p= \left[ \begin{array}{c} 2\\ 3\\ 1 \end{array} \right] 
   \ , \ \ 
  q= \left[ \begin{array}{c} 3\\ 2  \end{array} \right] 
   \ , \ \  
  A= \left[ \begin{array}{ccc} 
    11 & 12 & 13 \\ 21 & 22 & 13 \\ 31 & 32 & 33  \end{array} \right] 
   \ , \ \ 
  A_p^q= \left[ \begin{array}{cc} 
    23 & 22 \\ 33 & 32 \\ 13 & 12  \end{array} \right] 
   \ . 
 \]   
 If $q=[q(1),\ldots q(N)]$ is a permutation of $[1,\ldots N]$, 
 and $I$ is the identity matrix, $Q=I_q$ and $Q'=I^q$ are the 
 corresponding row and column permutation matrices. 
 Moreover, if $A$ a $N\times N$ matrix, $A_q=QA$ and $A^q=AQ'$. 
 The symmetric permutation of $A$ in order $q$ is 
 $A(q,q)=QAQ'$. 
     
 Let us consider the covariance structure model of section 3. 
 If we write the variables of the model in a permuted order, $q$, the
new covariance matrix is $V(q,q)$. The statistical model is of course
the same,  but the Cholesky factor of the two matrices may have a quite
a  different sparsity structure.

 Figure 5 shows the positions filled in the Cholesky 
factorization of a matrix $A$, and in the Cholesky factorization of 
two symmetric permutation of the same matrix, $A(q,q)$. 
 Initial Non Zero Elements, NZEs, are represented by $x$, initial zeros 
filled during the factorization are represented by $0$, and initial 
zeros left unfilled are represented by blank spaces. 
 \[  
 \begin{array}{c} 1\\ 2\\ 3\\ 4\\ 5\\ 6 \end{array} \ \
 \left[ \begin{array}{cccccc} 
    1 & x & x &   &   & x \\  
    x & 2 & x &   &   & 0 \\ 
    x & x & 3 & x &   & 0 \\ 
      &   & x & 4 &   & 0 \\ 
      &   &   &   & 5 & x \\ 
    x & 0 & 0 & 0 & x & 6 
    \end{array} \right] \ 
 \left[ \begin{array}{cccccc} 
    1 & x & x & x &   &   \\  
    x & 3 & 0 & x & x &   \\ 
    x & 0 & 6 & 0 & 0 & x \\ 
    x & x & 0 & 2 & 0 & 0 \\ 
      & x & 0 & 0 & 4 & 0 \\ 
      &   & x & 0 & 0 & 5 
    \end{array} \right] \ 
 \left[ \begin{array}{cccccc} 
    5 &   &   & x &   &   \\  
      & 4 &   &   & x &   \\ 
      &   & 2 &   & x & x \\ 
    x &   &   & 6 &   & x \\ 

      & x & x &   & 3 & x \\ 
      &   & x & x & x & 1 
    \end{array} \right] 
 \]  
 \centerline{Figure 5: Filled Positions in Cholesky Factorization.}  
 \\ \mbox{}

 The next lemma connects the numerical elimination process in the
Cholesky  factorization of a symmetric matrix $A$, to the vertex
elimination process in the UG having as  adjacency matrix, $B$, the
sparsity pattern of $A$.  

 Elimination Lemma: 
 When eliminating the $j$-th column in the Cholesky factorization of 
matrix $A(q,q)=LL'$, we fill the positions in $L$ corresponding  to the
filled edges in $\mathcal{F}$ at the elimination of vertex $q(j)$.

 Given a matrix $A$, $\mathcal{G}=(\mathcal{V},\mathcal{E})$, an
elimination order $q$,  and the respective filled graph, let us consider
the set of  row indices  of NZE's in $L^j$, 
 the $j-th$ column  of the Cholesky factor, 
 $L\mid QAQ'=LL'$: 
 \[ 
   \nze(L^j)=\{i\mid i>j \wedge \{q(i),q(j)\}\in \mathcal{F}\} + \{ j\}\ . 
 \]

 \[ 
 \begin{array}{ccccccccc} 
  6 & \rightarrow & 5 \\ 
   & \searrow  & 4 & \rightarrow & 3 & \rightarrow & 2 & \rightarrow & 1 
 \end{array} \ \ , \ \ \  
 \begin{array}{ccccc} 
  6 & \rightarrow & 5 & \rightarrow & 4 \\ 
    &             &   &             & \downarrow \\ 
  1 & \leftarrow & 2 & \leftarrow & 3 
 \end{array} \ \ , \ \ \ 
 \begin{array}{ccccc} 
    &             &   & \nearrow    & 2 \\  
  6 & \rightarrow & 5 & \rightarrow & 3 \\  
   & \searrow    & 4 & \rightarrow & 1 
 \end{array} \ \ .   
 \] 
 \centerline{Figure 6: Elimination Trees.}  
  \\ \mbox{}

 We define the {\it elimination tree}, $\mathcal{H}$, by 
 \[ 
  h(j)= \left\{ \begin{array}{l}
  j, \ \ \mbox{if}\ \ \nze(L^j) = \{ j\} , \ \ \mbox{or} \\
  \min \{i>j \mid i \in \nze(L^j) \} 
  \ , \ \ \mbox{otherwise} \ . \  
  \end{array} \right. 
 \] 
 %
 where $h(j)$, the parent of $j$ in $\mathcal{H}$,
 is the first (non diagonal) NZE in column $j$ of $L$. 
 Figure 6 shows the elimination trees corresponding to the examples 
 in Figure 5.  

 {\bf Elimination Tree Theorem:}   
 For any row index $i$ bellow the diagonal in column $j$ of $L$,    
 $j$ is a descendant of $i$ in the elimination tree, that is, 
 for any $i>j \g i\in \nze(L^j)$, the is a path in $\mathcal{H}$ 
 going from $i$ to $j$.

 Proof (see Figure 7): 
 If $i=h(j)$, the result is trivial. 
 Otherwise, (see Figure 7), let $k=h(j)$. 
 But $L_i^j\neq 0 \wedge L_k^j\neq 0 \Rightarrow L_i^k\neq 0$, 
 because 
 $\{q(j),q(i)\},\{q(j),q(k)\} \in \mathcal{E}_j 
  \Rightarrow \{q(k),q(i)\}\in \mathcal{E}_{j+1}$.  
 Now, either $i=h(k)$, or, applying the argument recursively, 
 we trace a branch of $\mathcal{H}$ 
 $(i,l,\ldots k,j)$, $i>l>\ldots >k>j$. QED.

 \[
 \begin{array}{ccccccccccc} 
 1 & \\ 
   & \ddots & \\ 
   &        & j \\ 
   &        & \vdots & \ddots \\ 
   &        &   x    & \ldots & k \\ 
   &        &        &        &   & \ddots \\ 
   &        &        &        &   &        &  l \\ 
   &        &        &        &   &        & \vdots & \ddots \\ 
   &        & \bullet & & \bullet &        &  x     & \ldots & i \\  
   &        &        &        &   &        &        &  &  & \ddots \\ 
   &        &        &        &   &        &        &  &  & & n \\ 
 \end{array} 
 \]    
 \centerline{Figure 7: A Branch in the Elimination Tree.} 
  \\ \mbox{} 

 From the proof of the last theorem we see that the elimination tree 
portrays the dependencies among the columns for the numeric 
factorization process. 
 More exactly, we can eliminate column $j$ of $A$. i.e. compute all  the
multipliers in column $j$, $M^j$, and update all the elements  affected
by these multipliers, if and only if we have already eliminated all the 
descendents of $j$ in the elimination tree.

 If we are able to perform parallel computations, we can simultaneously
eliminate all the columns at a given level of the elimination  tree,
beginning with the leaves, and finishing at the root. 
 Example 4 considers the elimination of a matrix with the same 
sparsity pattern of the last permutation in example 1. 
 Its elimination tree is the last one presented at Figure 6. 
 This elimination tree has three levels that, from the leaves to 
the root, are: $\{1,3,2\}$, $\{4,5\}$, e $\{6\}$. 

 Hence, we can perform a Cholesky factorization with this 
sparsity pattern in only 2 steps, as illustrated in the following 
numerical example:  
 \[  
 \left[ \begin{array}{cccccc} 
    {\bf 1} &   &   & 7 &   &   \\  
      & {\bf 2} &   &   & 8 &   \\ 
      &   & {\bf 3} &   & 6 & 9 \\ 
    7 &   &   & 53 &   & 2 \\ 
      & 8 & 6 &   & 49 & 23 \\ 
      &   & 9 & 2 & 23 & 39 
    \end{array} \right]  
 \left[ \begin{array}{cccccc}
    1 &   &   & 7 &   &   \\
      & 2 &   &   & 8 &   \\
      &   & 3 &   & 6 & 9 \\
    {\it 7} &   &   & {\bf 4} &   & 2 \\
      & {\it 4} & {\it 2} &   & {\bf 5} & 5 \\
      &   & {\it 3} & 2 & 5 & 12 
    \end{array} \right] 
 \left[ \begin{array}{cccccc} 
    1 &   &   & 7 &   &   \\
      & 2 &   &   & 8 &   \\
      &   & 3 &   & 6 & 9 \\
    {\it 7} &   &   & 4 &   & 2 \\
      & {\it 4} & {\it 2} &   & 5 & 5 \\
      &   & {\it 3} & {\it \frac{1}{2}} & {\it 1} & {\bf 6}   
    \end{array} \right] 
 \]

  The sparse matrix literature has many heuristics designed for finding  
 good elimination orders.  The example in Figures 8 and 9 show a
 good elimination  order for a $13\times 13$ sparse matrix.

 \[   
 \begin{array}{c} 
  1 \\ 2 \\ 3 \\ 4 \\ 5 \\ 6 \\ 7 \\ 8 \\ 9 \\ 10 \\ 11 \\ 12 \\ 13  
 \end{array} \ \ \ \   
 \begin{array}{ccccccccccccc} 
  3  & x &    &     &    &    &    &    &     &    &    &    &    \\ 
  x  & 8 &    &     &  x &    &    &    &     &    &    &    &    \\ 
     &   &  1 &     &  x &    &    &    &     &    &    &  x &  x \\ 
     &   &    & 10  &  x &    &    &    &     &    &    &  x &    \\ 
     & x &  x &  x  &{\bf9}&  &    &    &     &    &    &  0 &    \\ 
     &   &    &     &    & 12 &    &    &     &  x &    &    &  x \\ 
     &   &    &     &    &    & 13 &    &     &  x &    &    &    \\ 
     &   &    &     &    &    &    &  2 &   x &  x &    &    &    \\ 
     &   &    &     &    &    &    &  x &   7 &  0 &    &    &    \\ 
     &   &    &     &    &  x &  x &  x &   0 &{\bf 6}& &    &  0 \\ 
     &   &    &     &    &    &    &    &     &    & 11 &    &  x \\ 
     &   &  x &  x  &  0 &    &    &    &     &    &    &{\bf 4}& 0 \\ 
     &   &  x &     &  0 &  x &    &    &     &  0 &  x &  0 &{\bf 5}\\ 
 \end{array}  
  \]  
 \centerline{Figure 8: Gibbs Heuristic's Elimination Order.} 
  \\ \mbox{}

  The elimination order in Figure 8 was found using the Gibbs heuristic, 
 described in Stern (1994, ch.6) or Pissanetzky (1984, ch.x). 
 The intuitive idea of Gibbs heuristic, see Figure 9, 
 is as follows:  
 1- Starting from a `peripheral' vertex, in our example, vertex 3; 
 2- Grow a breath-first tree $\mathcal{T}$ in $\mathcal{G}$. 
 Notice that the vertices at a given level, $l$, of $\mathcal{T}$ form a  
 separator, $S_l$, in the graph $\mathcal{G}$.  
 3- Chose a separator, $S_l$, that is `small', i.e. with few vertices, 
    and `central', i.e. dividing $\mathcal{G}$ in `balanced' components.  
 4- Place in $q$, first the indices of each component separated by 
    $S_l$, and, at last, the vertices in $S_l$. 
 5- Proceed recursively, separating each large component into smaller ones.  
 In our example, we first use separator    
 $S_5=\{4,5\}$, dividing $\mathcal{G}$ in three components,  
 $C_1=\{3,8,1,10,9\}$ $C_2=\{12,13,2,7,6\}$ $C_3=\{11\}$.    
 Next, we use separators  
 $S_3=\{9\}$ in $C_1$, and $S_7=\{6\}$ in $C_2$.

 The main goal of the techniques studied in this and the last section 
is to find an elimination order filling as few positions as possible in 
the Cholesky factor. Once the elimination order has been chosen, 
simplified elimination can be used to prepare in advance all the data
structures holding the sparse matices, hence separating the
symbolic (combinatorial) and  numerical steps of the factorization.   
 This separation is important in the production of high performance 
computer programs.

 \[ 
 \begin{array}{ccccccccccc}
     &   &   & -  & -  & -  & 1  &            & 2  \\ 
     &   & \mbox{}\: / & &  &  / & |  &            & |  & \backslash \\ 
   3 &   & | &    & 4  &    & 5  &            & 6  &            & 7  \\ 
   | &   & | &    & |  &    & |  & \backslash & |  & \backslash \\ 
   8 & - & 9 & -  & 10 &    & 11 &            & 12 &            & 13  
 \end{array} \ \ \  
 \] 


 \[  
 \begin{array}{ccc} 
           &   5;4  & \\ 
       \mbox{} \ \ \ \ \swarrow & \downarrow & \searrow \ \ \ \ \mbox{} \\ 
     6    &  11   &   9  \\ 
  \swarrow \downarrow \searrow  &     &  \swarrow \downarrow \searrow   \\ 
  12 \ 13 \ \ 7;2  &  &  10 \ \ 1 \ \ 8;3      
 \end{array} 
 \]  


 \[  
 \begin{array}{cccccccccccccccccccc} 
   & &  &  &  &  &  &  & 10& \rightarrow & 4&  & 11&  &  &  & 13 \\   
   & &  &  &  &  &  & \nearrow &   &  &  & \nearrow &   &  &  & \nearrow & \\ 
 \mathcal{T}= 
   & & 3& \rightarrow & 8& \rightarrow & 9& \rightarrow &  1& \rightarrow & 
    5& \rightarrow & 12& \rightarrow & 6& \rightarrow & 2& \rightarrow & 7 \\ 
    \\ 
 l=& & 1&  & 2&  & 3&  &  4&  & 5&  & 6&  & 7&  &  8&  & 9  
 \end{array} 
 \]  
 \centerline{Figure 9: Nested Dissection by Gibbs Heuristic.} 
 \\ \mbox{}

 \section{Bayesian Networks}

 The objective of this section is to show that the sparsity techniques 
described in the last two section can be applied, almost immediately, 
to an other important statistical model, namely, Bayesian networks. 
 The presentation in this section follows very closely Cozman (2000). 
 A Bayesian network is represented by a DAG. 
 Each node, $i$, represents a random variable, $x_i$. 
 Using the notation established in section 9, we write $i \in n$ , 
 where $n$ is the index vector $n=[1,2,\ldots N]$.   
 The DAG representing the Bayesian network has an arc from node $i$  to
node $j$ if the probability distribution of variable $x_j$ is directly
dependent on variable $x_i$.

 In many statistical models that arc is interpreted as 
a direct influence or causal effect of $x_i$ on $x_j$. 
 Technically, we assume that the joint distribution of the vector 
 $x$ is given in the following product form. 
 \[ 
    p(x) = \pprod_{j\in n} p\left( x_j \g x_{\pa(j)} \right) \ .  
 \]

 The important property of Markov blankets in a Bayesian network is
that, given the variables in its Markov blanket, a variable $x_i$ is
conditionally independent of any  other variable, $x_j$, in the network,
that is, the Markov blanket of a variable `decouples' this variable 
from the rest of the network,    
 \[ 
   p(x_i \g x_{\mb(i)}, x_j) = p(x_i \g x_{\mb(i)} ) \ . 
 \]

 Inference in Bayesian networks is based on queries, where 
 the distribution of some `query' variables,  
 $x_q$, $q=[q(1),\ldots q(Q)]$, is computed,   
 given the observed values of some `evidence' variables, 
 $x_e$, $e=[e(1),\ldots e(E)]$.   
 Such queries are performed eliminating, that is marginalizing, 
 integrating, or summing out, all the remaining variables, $x_s$, that is, 
  \[ 
   p(x_q \g x_e) = \ssum_{x_s} p(x)  = \ssum_{x_s} 
   \pprod_{j\in r} p\left( x_j \g x_{\pa(j)} \right) \ . 
  \]   
  We place the indices of the variables to be eliminated in the
elimination  index vector, $s=r\backslash (q\cup e)$. 
 For now, let us consider the `requisite' index vector, $r$, as being
just a permutation (reordering) of the  original indices in
the network, that is, $r=[r(1),\ldots r(R)]$, $R=N$.
  The `elimination order' or `elimination sequence', 
 $s=[s(1),\ldots s(S)]$, will play an important role in what follows.

 Let us mention two technical points: 
 First, not all variables of the original network may be needed for a
given query. If so, the indices of the unnecessary ones can be removed
from the requisite index vector, and the query is performed involving
only a proper subset of the original variables, hence, $R<N$. 
 For example, if the network has disconnected components, all the
vertices  in components having no query variables are unnecessary. 
 Second, the normalization constant of distributions that appear in
intermediate computations are costly to obtain  and, more important, not
needed. Hence, we can perform this intermediate computations with
un-normalized distributions, also called `potentials'.

 Making explicit use of the elimination order,  
 $s=[s(1),\ldots s(S)]$, we can write the last equation as  
 \[ 
  p(x_q \g x_e) = \ssum_{x_{s(S)}} \cdots \ssum_{x_{s(1)}}  
    p(x_{r(1)} \g x_{\pa(r(1))}) \times \ldots \times  
    p(x_{r(R)} \g x_{\pa(r(R))}) \ . 
 \] 
  
 Because $x_{s(1)}$ can only appear in densities $p(x_j\g x_{\pa(j)})$ 
 for $j=s(1)$ or $j\in \ch(s(1))$, we can separate the first summation, 
 writing 
 \[ 
  p(x_q \g x_e) = 
   \ssum_{x_{s(S)}} \cdots \ssum_{x_{s(2)}} \left( 
    \pprod_{j\in r\backslash (\ch(s(1))\cup s(1))} 
      p(x_j \g x_{\pa(j)} ) \right) 
 \] 
 \[ 
    \times \left(  \ssum_{x_{s(1)}} 
    \pprod_{ j \in \ch(s(1))\cup s(1)}  
      p(x_j \g x_{\pa(j)} ) \right) \ . 
 \] 

 Eliminating, i.e. integrating out, the first variable in the 
 elimination order, $x_{s(1)}$, we create a new (joint) 
 potential of the children of the eliminated variable, 
 given its parents and spouses, that is,  
 \[ 
  p(x_{\ch(s(1)} \g x_{\pa(s(1)}, x_{\sp(s(1))} ) = 
  \ssum_{x_{s(1)}} 
    \pprod_{ j \in \ch(s(1))\cup s(1)}  p(x_j \g x_{\pa(j)} )
 \] 
 
 Next we eliminate $x_{s(2)}$, that is, we collect all potentials
 containing $x_{s(2)}$, form their joint product, and marginalize on 
 $x_{s(2)}$. We proceed in the elimination order eliminating 
 $x_{s(3)}, x_{s(4)} \ldots x_{s(S)}$, at which point the normalized 
 potentials left give us the distribution $p(x_q \g x_e)$.  

 We refer to the variables appearing in a joint potential as 
that potential's cluster.  
 Forming a joint potential is a computation of a complexity that is  
exponential in the size of its cluster. 
 Hence, it is vital to chose an elimination order that keeps the   
cluster sizes as small as possible. 
 But the clusters formed in the elimination process of a BN correspond 
to the cliques appearing in the elimination graphs, as defined in the 
last two sections. 
 Hence all techniques and heuristics used for finding a good elimination 
order for Cholesky factorization can be used to obtain a good elimination 
order for querying a BN. 
 Also, all the abstract combinatorial structures appearing in sparse
Cholesky factorization, like elimination trees, have their analogues 
for computation in BNs. 
 Cozman (2000) develops the complete theory of BNs in a very simple and
intuitive way, a way that naturally highlights this analogy.  
 Other authors have already commented on the similarities between several 
graph decomposition algorithms, see for example Lauritzen (2006, Lecture 4, 
Probability propagation and related algorithms) for a 
very general and abstract, but highly mathematical overview.


%% file: CAPMM.TEX
 \chapter{Monte Carlo Miscellanea} 

 Monte Carlo or, if necessary, Markov Chain Monte Carlo, is the basic
tool we use for numerical integration. 
  There are several excellent books on the subject. 
  Hammersley and Handscomb (1964) is a short and intuitive introduction, 
 including some important topics not usually covered at this level, 
 like pseudo-random and quasi-random generators, 
 importance sampling and other variance reduction techniques, 
 and the solution of linear systems. 
  This book is now out of print, but has the advantage of being freely 
 available for download at the internet. 
  Ripley (1987) is an other excellent text covering this material that  
 is still in print. 
  Gilks et al. (1996) gives several excellent and up-to-date  review
 papers on areas that are of interest for statistical modeling.  
  There is a vast literature on MC and MCMC written by physicists. 
  It contains many original, interesting and useful ideas, but sometimes
 it employs a terminology that is unfamiliar to statisticians.   
 The article of Meng and Wong (1996) can help to overcone this gap.

 \section{Pseudo, Quasi and Subjective Randomness}

  The implementation of Monte Carlo methods, as described in the 
 following sections, requires a random number generator of 
 i.i.d (independent and identically distributed) random variables 
 uniformly distributed in the unit interval, $[0,1[$.  
  From this basic uniform generator one gets a uniform generator in 
 the $d$-dimensional unit box, $[0,1[^d$ and, from there,  
 non-linear generators for many other multivariate distributions.

 \subsection*{Random and Pseudo-Random Generators}

  The concept of randomness is usually applied to a variable (to be) 
 generated or observed  process involving some uncertainty, as in the 
 definition presented by  Hammersley and Handscomb (1964, p.10):

 \begin{quotation} 
 {\it ``A {\it random event} is an event which has a chance of happening, 
and {\it probability} is a numerical measure of that chance.''}   
 \end{quotation}

  Monte Carlo, and several other applications, require a random 
 number generator. 
  With the last definition in mind, engineering devices based on 
 sophisticated physical processes have been built in the hope of 
 offering a source of ``true'' random numbers. 
  However, these special devices were cumbersome, expensive, 
 not portable nor universally available, and often unreliable.  
  Moreover, practitioners soon realized that simple deterministic 
 sequences could successfully be used to emulate a random generator,  
 as stated in the following quotes (our emphasis) by  
  Hammersley and Handscomb (1964, p.26) 
  and Ripley (1987, p.15):

 \begin{quotation} 
 {\it ``For electronic digital computer it is most convenient to calculate 
a sequence of numbers one at a time as required, by a completely 
specified rule which is, however, so devised that no {\bf reasonable}  
statistical test will detect any significant departure from randomness. 
 Such a sequence is called {\em pseudorandom}. The great advantage of 
a specified rule is that the sequence can be exactly reproduced for 
purposes of computational checking.''}  
 \end{quotation}

 \begin{quotation} 
 {\it ``A sequence of {\em pseudorandom} numbers ($U_i$) is a
 deterministic  sequence of  numbers in $[0,1]$ having the same 
 {\bf relevant} statistical  properties as a sequence of random
 numbers.''} 
 \end{quotation}

  Many deterministic random emulators used today are 
 Linear Congruential Pseudo-Random Generators (LCPRG),  
 as in the following example:  
 \[ 
   x_{i+1} = (a x_i +c) \mod m \ , 
 \]    
 where the multiplier $a$, the increment $c$ and the modulus $m$ 
 should obey the conditions: 
 (i)   $c$ and $m$ are relatively prime;  
 (ii)  $a-1$ is divisible by all prime factors of $m$;  
 (iii) $a-1$ is a multiple of $4$ if $m$ is a multiple of $4$. 
  LCPRG's are fast and easy to implement if $m$ is taken as 
 the computer's word range, $2^s$, where $s$ is the computer's 
 word size, typically $s=32$ or $s=64$. 
  The LCPRG's starting point, $x_0$, is called the seed. 
  Given the same seed the LCPG will reproduce the same sequence,  
 what is very important for tracing, debugging and verifying 
 application programs.

   However, LCPRG's are not an universal solution. 
   For example, it is trivial to devise some statistics 
  that will be far from random, see Marsaglia (1968). 
   There the importance of the words {\bf reasonable} and {\bf relevant} 
  in the last quotations becomes clear: 
   For most Monte Carlo applications these statistics are irrelevant. 
   LCPRG's can also exhibit very long range auto-correlations and,  
  unfortunately, these are more likely to affect long simulated 
  time series required in some special applications. 
   The composition of several LCPRG's by periodic seed refresh may   
  mitigate some of these difficulties, see Pereira and Stern (1999b). 
   LCPRG's are also not appropriate to some special applications in 
  cryptography, see Boyar (1989). 
   Current state of the art generators are given in 
  Matsumoto and Kurita (1992,1994) and Matsumoto and Nishimura (1998).

 \subsection*{Chance is Lumpy - Quasi-Random Generators}

 {\it ``Chance is Lumpy''} is Robert Abelson's First Law of Statistics, 
 see Abelson (1995, p.xv). 
  The probabilistic expectation is a linear operator, that is, 
 $E(Ax+b)=AE(x)+b$, where $x$ in random vector and $A$ and $b$ 
 are a determined matrix  and vector. 
  The Covariance operator is defined as 
 $\Cov(x)= E((x-E(x))\otimes (x-E(x)))$. 
 Hence, $\Cov(Ax+b)= A\Cov(x)A'$. 
  Therefore, given $n$ i.i.d. scalar variables,  
  $x_i \g \Var(x_i)=\sigma^2$, the variance of their mean,  
  $m=(1/n)\uno'x$,  is given by  
 \[ 
   \frac{1}{n}\uno' \; \diag(\sigma^2\uno) \; \frac{1}{n}\uno \ = \ \ 
   \left[ \begin{array}{cccc} 
    \frac{1}{n} & \frac{1}{n} & \ldots & \frac{1}{n} 
   \end{array} \right] \;  
   \left[ \begin{array}{cccc} 
    \sigma^2 & 0 & \ldots & 0 \\ 0 & \sigma^2 & \ldots 0 \\ 
    \vdots & \vdots & \ddots & \vdots \\ 0 & 0 & \ldots & \sigma^2   
   \end{array} \right] \;  
   \left[ \begin{array}{c} 
    \frac{1}{n} \\ \frac{1}{n} \\ \vdots \\ \frac{1}{n}  
   \end{array} \right] \ = \ \  \sigma^2/n \ . 
 \]  
  Hence, the mean's standard deviation is 
  $\mbox{std}(m)=\sigma/\sqrt(n)$.  
  So, mean values of iid random variables converge to their  
 expected values at a rate of $1/\sqrt(n)$. 
  
  Quasi-random sequences are deterministic sequences built not to 
 emulate random sequences, as pseudo-random sequences do, 
 but to achieve faster convergence rates. 
  For $d$-dimensional quasi-random sequences, an appropriate measure 
 of fluctuation, called discrepancy, only grows at a rate of 
 $\log(n)^d$, hence growing much slower than $\sqrt(n)$. 
  Therefore, the convergence rate corresponding to quasi-random
 sequences, $\log(n)^d/n$, is much faster than the one corresponding 
 to (pseudo) random sequences, $\sqrt(n)/n$.  
  Figure 1 allows the visual comparison of typical (pseudo)
 random (left) and quasi-random (right) sequences in $[0,1[^2$. 
  By visual inspection we see that the points of the quasi-random 
 sequence are more ``homogeneously scattered'' that is, they do not 
 ``clump together'', as the point of the (pseudo) random sequence 
 often do.

 Let us consider an axis-parallel rectangles in the unit box, 
 \[ 
    R = [a_1, b_1[ \, \times \, [a_2, b_2[ \, \times \, \ldots \, 
    [a_d, b_d[  \ \subseteq \ [0,1[^d \ . 
 \] 
 The discrepancy of the sequence $s_{1:n}$ in box $R$, 
 and the overall discrepancy of the sequence are defined as 
 \[ 
    D(s_{1:n},R) = n\mbox{Vol}(R) - \left| s_{1:n} \cap R \right| 
    \ , \ \  
    D(s_{1:n}) = \sup_{R\in [0,1[^d}  \left| D(s_{1:n},R) \right| 
    \ . 
 \] 
  It is possible to prove that the discrepancy of the Halton-Hammersley 
 sequence, defined next, is of order $O(\log(n)^{d-1})$, 
 see Matou\u{s}ek (1991, ch.2).

  Halton-Hammersley sets: 
  Given $d-1$ distinct prime numbers, $p(1), p(2), \ldots p(d-1)$, 
  the $i$-th point, $x^i$, in the 
  Halton-Hammersley set, $\{x^1, x^2,\ldots x^n \}$, is   
  \[ 
     x^i = \left[ i/n, r_{p(1)}(i), r_{p(2)}(i), 
      \ldots r_{p(d-1)}(i) \right]'    
     \ ,  \ \  \mbox{for} \, i=1:n-1 \ , \mbox{where}  
  \] 
  \[ 
     i = a_0 +p(k)a_1 +p(k)^2 a_2 +p(k)^3 a_3 +\ldots \ , \ \ 
     r_{p(k)}(1) = \frac{a_0}{p(k)} +\frac{a_1}{p(k)^2} 
        +\frac{a_2}{p(k)^3} +\ldots \ . 
  \]   
  That is, the $(k+1)$-th coordinate of $x^i$, 
  $x^i_{k+1}=r_{p(k)}(i)$, 
  is obtained by the bit reversal of $i$ written in 
  $p(k)$-ary  or base $p(k)$  notation.

   The Halton-Hammersley set is a generalization of van der Corput 
 set, built in the bidimensional unit square, $d=2$, using 
 the first prime number, $p=2$. 
 The following example, from Hammersley (1964, p.33) and 
 G\"{u}nther and J\"{u}ngel (2003, p.117) builds the 
 8-point van der Corput set, expressed in binary and decimal notation.

 \begin{minipage}{3in} 
 \begin{verbatim} 

 function x= corput(N,b)
 % size N base b v.d.corput set    
 m = fix(log(N)/log(b)); 
 D = [ ];   n = 1:N; 
 for i = 0:m 
    d = mod(n,b);
    n = (n-d)/b;
    D = [D; d];
 end
 x = ((1/b).^(1:(m+1)))*D;  
 \end{verbatim} 
 \end{minipage} 
 \mbox{} \mbox{} 
 \begin{minipage}{3in}  
 \begin{tabular}{|rl|rl|} 
  \hline   
  \multicolumn{2}{|c|}{Decimal} & \multicolumn{2}{c|}{Binary} \\ 
  \hline  
  $i$  & $r_2(i)$ & $i$  & $r_2(i)$  \\ 
  \hline  
  1 & 0.5    & 1    & 0.1    \\ 
  2 & 0.25   & 10   & 0.01   \\ 
  3 & 0.75   & 11   & 0.11   \\ 
  4 & 0.125  & 100  & 0.001  \\ 
  5 & 0.625  & 101  & 0.101  \\ 
  6 & 0.375  & 110  & 0.011  \\ 
  7 & 0.875  & 111  & 0.111  \\ 
  8 & 0.0625 & 1000 & 0.0001 \\ 
  \hline 
 \end{tabular} 
 \end{minipage}


  \centerline{
    \includegraphics*[height=3.0in, width=7.0in]{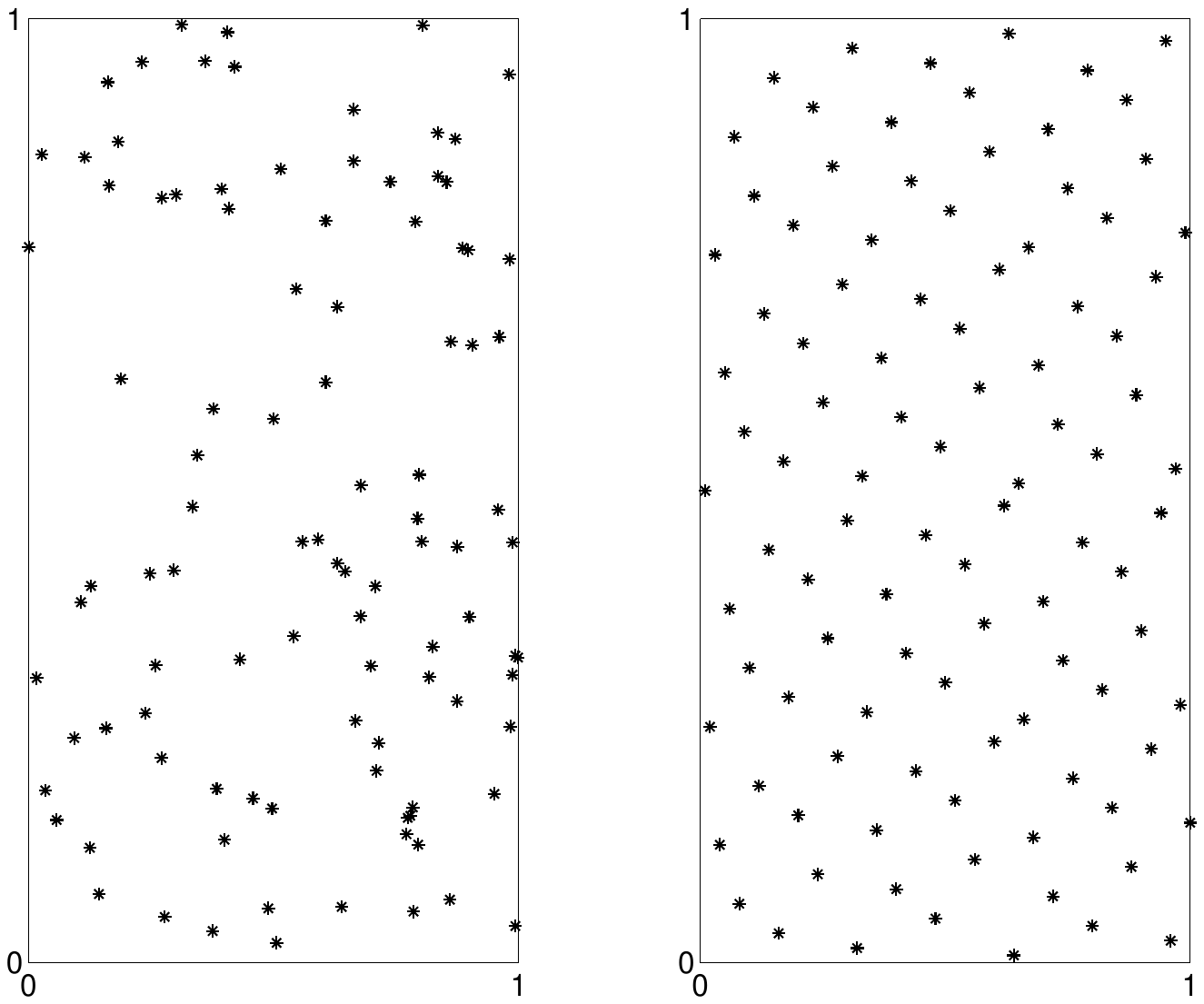}}

  \centerline{Figure G.1: (Pseudo)-random  and quasi-random 
              point sets on the unit box}

  Quasi-random sequences, also known as low-discrepancy sequences, 
 can substitute pseudo-random sequences in some applications of 
 Monte Carlo methods, achieving higher accuracy with less  
 computational effort, see Merkel (2005), \"{O}kten (1999) 
 and Sen, Samanta and Reese (2006).  
  Nevertheless, since by design the points of a quasi-random sequence 
 tend to avoid each other, strong (negative) correlations are expected 
 to appear. In this way, the very reason that can make quasi-random 
 sequences so helpful, can ultimately impose some limits to their 
 applicability. 
  Some of these problems are commented by Morokoff (1998, p.766): 

 \begin{quote} 
 {\it ``First, quasi-Monte Carlo methods are valid for integration
problems, but may not be directly applicable to simulations, due to the
correlations between the points of a quasi-random sequence. ...  
 A second limitation: the improved accuracy of quasi-Monte Carlo methods
is generally lost for problems of high dimension or problems in which
the integrand is not smooth.''} 
 \end{quote}

 \subsection*{Subjective Randomness and its Paradoxes}

  When asked to look at patterns like those in Figure 1, 
 many subjects perceive the quasi-random set as ``more random'' 
 than the (pseudo) random set. How can this paradox be explained? 
  This was the topic of many psychological studies in the field of 
 subjective randomness. 
  The quotation in the next paragraph is from one of these 
 studies, namely, Falk and Konold (1997, p.306, emphasis are ours):

 \begin{quotation} 
 {\it ``One major source of confusion is the fact that randomness
involves two distinct ideas: {\bf process} and {\bf pattern} 
(Zabell, 1992). 
 It is natural to think of randomness as a process that generates
unpredictable outcomes (stochastic process according to Gell'Mann,
1994). Randomness of a {\bf process} refers to the {\bf unpredictability} 
of the individual event in the series (Lopes, 1982). 
 This is what Spencer Brown (1957) calls {\bf primary randomness}. 
 However, one usually determines the randomness of the process by means
of its output, which is supposed to be {\bf patternless}. 
 This kind of randomness refers, by definition, to a sequence. 
 It is labeled {\bf secondary randomness} by Spencer Brown. 
 It requires that all symbol types, as well as all ordered pairs
(diagrams), ordered triplets (trigrams)... n-grams in the sequence
be equiprobable. 
 This definition could be valid for any n only in infinite sequences,
and it may be approximated in finite sequences only up to ns much
smaller than the sequence's length. 
 The entropy measure of randomness (Attneave, 1959, chaps. 1 and 2) 
is based on this definition. 

 These two aspects of randomness are closely related. We ordinarily
expect outcomes generated by a random process to be patternless. Most of
them are. Conversely, a sequence whose order is random supports the
hypothesis that it was generated by a random mechanism, whereas
sequences whose order is not random cast doubt on the random nature of
the generating process.''} 
 \end{quotation}

  Spencer-Brown was intrigued by the apparent incompatibility 
 of the notions of primary and secondary randomness.  
  The apparent collision of these two notions generates several 
 interesting paradoxes, taking Spencer-Brown to question the 
 applicability of the concept of randomness to probability and 
 statistical analysis, see Spencer-Brown (1953, 1957) and 
 Flew (1959), Good (1958) and Mundle (1959). 
 See also Henning (2006), Kaptchuk and Kerr (2004), 
 Utts (1991), and  Wassermann (1955).          
  In fact, several subsequent psychological studies were able to confirm 
 that, for many subjects, the intuitive or common-sense perception of 
 primary and secondary randomness are quite discrepant. 
  However, a careful mathematical analysis makes it 
 possible to reconcile the two notions of randomness. 
  These are the topics discussed in this section.

  The relation between the joint and conditional entropy 
 for a pair of random variables, see appendix E.2, 
 \[ 
    H(i,j) = H(j) +H(i\g j) = H(i) +H(j\g i) \ , 
 \] 
 motivates the definition of first, second and higher order 
 entropies, defined over the distribution of words of size $m$ 
 in a string of letters from an alphabet of size $a$.    
 \[ 
   H_1 = \sum_{j} p(j) \log p(j) \ , \ \ 
   H_2 = \sum_{i,j} p(i) p(j\g i) \log p(j \g i) \  , \ \  
 \] 
 \[ 
   H_3 = \sum_{i,j,k} p(i) p(j\g i) p(k\g i,j) \log p(k \g i,j) \ \ldots  
 \]

  It is possible to use these entropy measures to access the disorder 
 or lack of pattern in a given finite sequence, using the empirical 
 probability  distributions of single letters, pairs, triplets, etc. 
  However, in order to have a significant empirical distribution of 
 $m$-plets any possible $m$-plet must be well represented in the 
 sequence, that is, the word size, $m$, is required to be very short, 
 relative to the sequence log-size, $m << \log_a(n)$.

  In the article of Falk and Knold (1997), Figure 2 displays the 
 typical perceived or apparent randomness of Boolean (0-1) bit sequences 
 or black-and-white pixel grids versus the second order entropy of 
 the same strings and grids, see also Attneave (1959). 
  Clearly, there is a remarkable bias of the apparent randomness 
 relative to the entropic measure.


 \begin{figure}[!h]
  \centerline{\includegraphics*[height=2.5in,  width=4.5in]{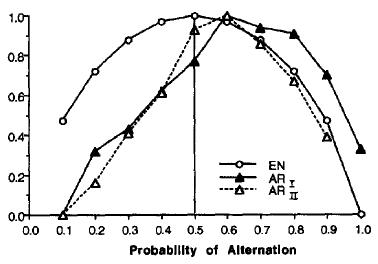}} 
  \centerline{ Figure 2: EN, $H_2$-entropy vs. AR, apparent randomness.} 
 \end{figure}

 \begin{quotation} 
 {\it `` When people invent superfluous explanations because they
perceive patterns in random phenomena, they commit what is known in
statistical parlance as Type I error The other way of going awry, known
as Type H error, occurs when one dismisses stimuli showing some
regularity as random. 
 The numerous randomization studies in which participants generated too
many alternations and viewed this output as random, as well as the
judgments of overalternating sets as maximally random in the perception
studies, were all instances of type II error in research results.''} 
 Falk and Konold (1997, p.303).  
 \end{quotation}

  This effect is also known as the {\it gambler's fallacy} when 
 betting on {\it cool spots}, expecting the random sequence to 
 ``compensate'' finite average fluctuations from  expected values.  
  Of course, some gamblers exhibit the opposite behavior,  
 preferring to bet on {\it hot spots}, expecting the same 
 fluctuations to reoccur.  
  These effects are the consequence of a perceived coupling,  
 by a negative or positive correlation or other measure of association, 
 between non overlapping segments that are in fact supposed to be 
 decoupled, uncorrelated or have no association,  that is, to be Markovian. 
 For a statistical analysis, see Bonassi et al. (2008).  
  A possible psychological explanation of the gambler's fallacy is given 
 by the constructivist theory of Jean Piaget, 
 see  Piaget and Inhelder (1951), in which any ``lump'' 
 in the sequence is (miss) perceived as non-random order: 

 \begin{quotation} 
 {\it ``In analogy to Piaget's operations, which are conceived as
internalized actions, perceived randomness might emerge from
hypothetical action, that is, from a thought experiment in which
one describes, predicts, or abbreviates the sequence. 
 The harder the task in such a thought experiment, the more random the
sequence is judged to be.''} 
 Falk and Konold (1997, p.316).  
 \end{quotation}

  The same hierarchical decomposition scheme used for higher order 
 conditional entropy measures can be adapted to measure the disorder or 
 patternless  of a sequence, relative to a given subject's model of  
 ``computer'' or generation mechanism. 
  In the case of a discrete string, this generation model could be, 
 for example, a deterministic or probabilistic Turing machine, 
 a fixed or variable length Markov chain, etc. 
  It is assumed that the model is regulated by a code, program or vector 
  parameter, $\theta$, and outputs a data vector or observed string, $x$. 
  The hierarchical complexity measure of such a model 
  emulates the Bayesian prior and conditional likelihood decomposition, 
  $H(p(\theta,x)) = H(p(\theta)) +H(p(x\g \theta))$,   
  that is, the total complexity is given by the complexity of the program 
  plus the complexity of the output given the program.      
   This is the starting point for several complexity models, like 
   Andrey Kolmogorov, Ray Solomonoff and Gregory Chaitin's 
  computational comlexity models, 
   Jorma Rissanen's Minimum Description Length (MDL), 
   and Chris Wallace and David Boulton's Minimum Message Length (MML). 
   All these alternative complexity models can also be used to 
  successfully reconcile the notions of primary and secondary randomness, 
  showing that they are asymptotically equivalent, 
   see  Chaitin (1975, 1988), Kac (1983), Kolmogorov (1965), 
   Martin-L\"{o}f(1966, 1969).

 \section{Integration and Variance Reduction} 

  This section presents the derivation of generic Monte Carlo procedures 
 for numerical integration. We follow the presentation of 
 Hammersley (1964). 
  Let us consider the integration of a bounded function, 
 $0\leq f(x) \leq 1$ in the unit interval, $x \in [0,1]$. 
  The {\it crud Monte Carlo} unbiased estimate of this integral is the 
 mean value of the function evaluated at uniformly distributed iid  
 random points, $x_i \in [0,1]$, $i=1:n$, with variance  
 \[ 
  \gamma = \int_0^1 f(x) dx  \ ,  \  \ 
  \haat{\gamma}_c = \frac{1}{n} \sum_1^n f(x_i) \ , \ \ 
  \sigma_c^2 = \frac{1}{n} \int_0^1 \left( f(x) -\gamma \right)^2 dx \ . 
 \] 

 An alternative unbiased estimator is the {\it hit-or-miss Monte Carlo}, 
 defined by the auxiliary hit indicator function, $h$,  
 \[ 
    h(x,y)= I(f(x)\geq y)  \ , \ \ 
    \gamma= \int_0^1 \int_0^1 h(x,y) dx dy \ , \ \ 
    \haat{\gamma}_h = \frac{1}{n} \sum_1^n h(x_i,y_i) = \frac{n^*}{n} \ . 
 \] 
 The variance of this method is that of a Bernoulli variate. 
 Simple manipulation shows that 
 \[ 
    \sigma_h^2= \frac{\gamma (1-\gamma )}{n} \ , \ \ 
    \sigma_h^2 -\sigma_c^2 = 
    \frac{1}{n} \int_0^1 f(x)(1-f(x)) dx >0 \ . 
 \] 
 Hence, hit-or-miss MC is worst than crude MC, as one could guess 
 from the fact that it is using far less information about $f$ at 
 any given point, $x_i$. 

 An other alternative is {\it importance sampling} MC, defined by an 
 auxiliary {\it sampling distribution}, $g$, in the integration interval, 
 \[      
    \gamma = \int f(x)  dx = \int \frac{f(x)}{g(x)} g(x) dx 
      = \int \frac{f(x)}{g(x)} dG(x) \ , \ \ 
 \] 
 \[ 
   \haat{\gamma}_s = \frac{1}{n} \sum_1^n \frac{f(x_i)}{g(x_i)} 
    \ , \ \ x_i \sim g \ , \ i=1:n \ ; \ \ 
    \sigma_s^2 = \frac{1}{n} \int
    \left( \frac{f(x)}{g(x)} -\gamma \right)^2 dG(x) \ . 
 \] 
  The importance sampling method can be used on an arbitrary integration 
 interval, as long as we know how to draw the points $x_i$ according 
 to the sampling distribution. 
  The variance of this method is minimized if $g\propto f$, that is if 
 the sampling distribution is (approximately) proportional to the 
 integrand. 
  In order to achieve a small variance and numerical stability, it is 
 important to keep the sampling ratio bounded, $f/g\leq c$. 
  In particular, if the integration interval is unbounded, the tails of 
 the sampling distribution should ``cover'' the tails of the integrand. 
 
 The formula for $\sigma^2_s$ suggests yet another strategy of variance 
reduction.  Let $\varphi(x)$ be a  function that closely emulates or
mimics $f(x)$, but is easy to integrate  analytically (or even
numerically). 
 Such a $\varphi(x)$ is known as a {\it control variate} 
for $f(x)$. 
 The desired integral can be computed as  
 \[ 
   \gamma = \ \int \varphi(x) dx +\int (f(x) -\varphi(x)) dx 
    \ = \   \gamma' +\int (f(x) -\varphi(x)) dx \ .  
 \] 

 Consider the following estimators and variances,  
 \[ 
   \haat{\gamma} = \frac{1}{n} \sum_1^n f(x_i) \ , \ \ 
   \haat{\gamma}' = \frac{1}{n} \sum_1^n \varphi(x_i) \ ,  
 \] 
 \[ 
   \Var(\haat{\gamma} -\haat{\gamma}' ) \ = \  
   \Var(\haat{\gamma}) +\Var(\haat{\gamma}' ) 
   -2 \Cov \left( \haat{\gamma},\haat{\gamma}' \right)  \ . 
 \] 
 That is, this the method is useful if the integration and the 
 control variates are  strongly (positively) correlated.

 \subsection*{Non-Uniform Random Generators}

  This section considers some elementary methods for producing 
 i.i.d. non-uniform variates, $x_i$, from a source of uniform variates 
 in the unit interval, $u_i \sim U_{]0,1[}$. 
  Perhaps the simplest example is to produce a Bernoulli variate:  

  (a) If $0\leq u_i \leq p$, then $x_i=1$, else ($p< u_i \leq 1$), $x_i=0$.

  If $F(x)$ is the cumulative distribution of $f(x)$, 
  and $x_i\sim f$, then $u=F(x_i)\sim U_{[0,1]}$. 
  Hence, if $F(x)$ is invertible, we can just 
 take $x_i = F^{-1}(u_i)$ as a mechanism for generating $f$ distributed 
 variates.  For example: 

  (b) The exponential distribution with mean $1/\lambda$ 
 is given by $f(t)= \lambda\exp(-\lambda t)$, and 
 $F(t)= 1-\exp(-\lambda t)$. 
  Hence, $t = (-1/\lambda) \ln (u)$ produces an exponential variate. 

  (c) The Cauchy  distribution with location and scale parameters, 
  $a,b$, is given by  
  $1/f(x)= \pi b(1+((x-a)/b)^2)$,  
  $F(x)= (1/2) +(1/\pi)\arctan((x-a)/b)$.  
  Hence, $x=a+b\tan(\pi(u-(1/2)))$ produces the corresponding 
  Cauchy variate.

  The characterizations of a distribution in terms of a second 
 distribution may offer an implicit generation mechanism. 
  For example: 

  (d) The Chi-squared distribution with $2$ degrees of freedom, 
 $\chi^2_2$, is a particular case of the exponential with mean 
 $(1/\lambda)=2$. 
  Hence, $x= -2\ln(u)$ generates a $\chi^2_2$ variate. 
 
  (e) A $\chi^2_d$ variate is characterized as a sum of squares of 
 $d$ normal variates. 
  Hence, if $d$ is even, we can generate a $\chi^2_d$ variate as   
 $x= -2\ln(u_1 u_2 \ldots u_{d/2})$.       

  (f) Counting consecutive $\lambda$-exponential arrivals 
 until the threshold $t_1 +t_2 \ldots +t_k \geq 1$ produces a 
 rate-$\lambda$ Poisson variate, 
 $f(k)= \exp(-\lambda)\lambda^k/k!$. 

  (g) Appendix B presents characterizations of many discrete 
 distributions by the Poisson, hence providing implicit 
 generation mechanisms for those distributions.

  (h) The following two dimensional transformation method generates 
 two i.i.d. standard Normal variates, see Ripley (1987).   
 \[ 
    u,v \sim U_{[0,1]} \ , \ \      
    \theta = 2\pi u \ , \ \ r = \sqrt{-2\ln(v)} \ , \ \  
    x = r \cos(\theta) \ , \ \  y = r \sin(\theta) \ . 
 \] 
  To check the method consider the transformation to polar coordinates, 
 $[r,\theta]$, of  a standard bivariate normal 
 $[x,y]\sim (1/2\pi)\exp((-1/2)(x^2+y^2))$.  
 \[ 
   [r,\theta] \sim 
   \frac{1}{2\pi}\exp\left(\frac{-r^2}{2}\right) \; 
    \left| \begin{array}{cc} \cos(\theta) & \sin(\theta) \\ 
     -r \sin(\theta) & r \cos(\theta) \end{array} \right| 
    \ = \ \ \frac{1}{2\pi} \; r \exp\left(\frac{-r^2}{2}\right) \ .   
  \] 
   Hence, $r$ and $\theta$ are independent, 
  $\theta$ is uniformly distributed in $[0,2\pi]$,   
  and $r^2$  is a $\chi^2_2$ variate. 
   Finally, we see that $r$ is produced by the transformation defined 
  in item (d) above to generate a $\chi^2_2$ variate.

  If the scaled density, $\kappa g$, can be used as an 
 {\it envelope} dominating the density $f$, that is, 
 $f\leq \kappa g$, the following {\it acceptance-rejection} method 
 due to von Neumann can be used: \\ 
 (1) Generate $[y_i,u_i] \sim g \times U_{[0,1]}$  
     \ until \  $\kappa u_i \leq f(y_i)/g(y_i)$. \  
 (2) Take $x_i = y_i$.

   The Gamma distribution with parameter $c$ is   
  $f(x)= x^{c-1} \exp(-x)/ \Gamma(c)$. 
  For $c=1$ this is the exponential distribution, 
  also, the sum of two gamma variates with parameters 
  $c_1, c_2$ is a gamma variate with parameter 
  $c_1+c_2$. 
  The following results given in De\'{a}k (1990, sec.4.5) 
  provide implicit acceptance rejection generation methods:

  (i) For $c<1$, $f(x)$ is dominated by the following 
  density $g(x)$ scaled by the factor  
  $\kappa = (c\Gamma(c))^{-1} +(e\Gamma(c))^{-1}$. 
  Moreover, $G^{-1}$ has an easy analytic form. 
  \[ 
     g(x)= \left\{ \begin{array}{ll} 
     \frac{ec}{(e+c)}x^{c-1} , \  \mbox{if} \ x\in [0,1] \\ 
     \frac{ec}{(e+c)}e^{-x} , \ \mbox{if} \ x\in \; [1,\infty[  
     \end{array} \right. 
     \ , \ 
     G(x)= \left\{ \begin{array}{ll} 
     \frac{e}{(e+c)}x^{c} , \  \mbox{if} \ x\in [0,1] \\ 
     \frac{e}{(e+c)} +\frac{c}{(e+c)}(1-e^{1-x}) 
      , \ \mbox{if} \ x\in \; [1,\infty[  
     \end{array} \right. 
  \] 

 (ii) For $c>1$, $f(x)$ is dominated by a Cauchy with parameters 
  $a=1/\sqrt{2c-1}$ and $b=c-1$, scaled by the factor 
  $\kappa= \pi \sqrt{2c-1} \exp(-c+1) (c-1)^{c-1} / \Gamma(c)$. 

  (iii) For $c>1$, $f(x)$ is dominated by the envelope density, $g_c(x)$, 
  and scale factor, $\kappa_c$, described as follows. 
  First, let us consider an auxiliar variate distributed as the 
  t-density with 2 degrees of freedom. The auxiliar  density, 
  $g(y)$, cumulative distribution, $G(y)$, and 
  generation method by direct inversion are as follows:  
  \[ 
     g(y) = \frac{1}{2\sqrt{2}} \left( 1+ 
        \frac{y^2}{2} \right)^{-\frac{3}{2}} 
   \ , \ \ 
     G(y) = \frac{1}{2}\left( 1+ 
     \frac {y / \sqrt{2} }{ \sqrt{ 1 +y^2 / 2 } } \right) 
  \ , \ \ 
     y \sim \frac{ \sqrt{2} (u -1/2) }{ \sqrt{ u(1-u) } } \ .  
  \] 
  Next, let us consider the envelope variate with density and scale 
  factor defined as 
  \[ 
    \frac{1}{\kappa_c \, g_c(x)}= \Gamma(c) \left( 1+ \frac{1}{2}\left( 
    \frac{x- (c-1)}{\sqrt{ 3c/2\, -3/8 } } \right)^2 \; \right)^{3/2} 
    \ , 
  \] 
  \[ 
     \kappa_c =  \ 
     \frac{ 2 }{ \Gamma(c)  } \sqrt{ 3c -\frac{3}{4} } 
     \left( \frac{c-1}{e} \right)^{c-1}   
     \ \leq \  \sqrt{\frac{6}{\pi}} \; e^{1/c}  \ .  
  \]  
  The envelope variate can be generated from the 
  auxiliar variate as 
  \[ 
    x \sim (c-1) + y\sqrt{3c\, -3/4} \ . 
  \] 

  (iv) It is easy to check that if $y$ is a gamma variate with parameter 
  $c+1$ and $u$ is uniform in $[0,1[$, than $x=yu^{1/c}$ is a gamma 
  variate with parameter $c$. This property can be used to use a 
  gamma generator in the domain $c<1$ to generate a gamma variate 
  with parameter $c>1$, and vice-versa.  

  Appendix B presents characterizations of the Beta an Dirichlet 
 distributions by the Gamma, hence providing implicit generation 
 mechanisms for those distributions.  
  For more non-uniform random generation methods see 
 De\'{a}k (1990), Gentle (1998), Lange (2000), Ripley (1987), and the    
 encyclopedic work of Fishman (1996).



 \section{MCMC - Markov Chain Monte Carlo} 

 This section uses the matrix notation and the basic facts about 
 homogeneous Markov chains reviewd in section H.1.

 Markov Chain Monte Carlo, Conditional Monte Carlo, etc. 
 are common names for methods that generate indirect random sampling 
 for a discrete target density $g$. 
 MCMC sampling is based on a Markov chain that has the target density 
 as its limit distribution.
 Our presentation follows ch.1 of Gilks et al. (1996).  
 For the original papers, see Geman and Geman (1984), Hastings (1970),  
 Metropolis and Ulam (1949), and Metropolis et al. (1953).

 The basic idea of the MCMC algorithms is to adapt 
 a general (irreducible and aperiodic) sampling kernel, $Q$, 
 to the desired target distribution, $g>0$.  
 Starting form state $i$, the MCMC algorithm proceeds 
 as follows: \\ 
 (1) A candidate for the next state, $j$, is proposed with 
     probability $Q_i^j$. \\  
 (2a) The chain moves to the candidate $j$ with 
     {\em acceptance probability} $\alpha(i,j)$. \\   
 (2b) Otherwise, candidate $j$ is rejected, and the chain remains 
     at state $i$. \\  
 (3) Go to step 1. \\ 
 Formally, the MCMC transition kernel, $P$, 
 has the form 
 \[ 
    P_i^j = Q_i^j \alpha(i,j) +I(j=i) \left( 1 - 
    \ssum_j Q_i^j \alpha(i,j) \right) \ ,  
 \]  
 where the first term corresponds to the acceptance of new state $j$,   
 while the second term corresponds to the rejection of the proposed 
 candidate, indicating that the chain remains at state $i$.

 In order to obtain the target distribution, $g$, as the limit 
 distribution of the MCMC, we want to choose an acceptance probability, 
 $\alpha(i,j)$, that enforces the detailed balance equation,  
 \[ 
   g^i P_i^j = g^i Q_i^j \alpha(i,j) = 
   g^j Q_j^i \alpha(j,i) = g^j P_j^i \ .  
 \] 
 It is easy to check that the acceptance probabilities suggested by 
 Metropolis-Hastings and Barker accomplish the goal. 
 They are, respectively, 
 \[ 
    \alpha(i,j) = \min \left( 1 \, , \    
      \frac{g^j Q_j^i}{g^i Q_i^j}  \right) 
     \ \ \mbox{and} \ \ \    
    \alpha(i,j) =    
      \frac{g^j Q_j^i}{g^i Q_i^j +g^jQ_j^i} \ ,     
 \]

 In Bayesian statistics, MCMC methods are typically used to compute 
 $\overha{f}$, the expected value of a function, $f(\theta)$, 
 on a specific region of the parameter space, $T\subseteq \Theta$, 
 with respect to the posterior density, $p_n(\theta)$. 
 In standard Bayesian models, 
 $p_n(\theta)=c(y)^{-1} L(\theta \g y) p_0(\theta)$,  where 
 $p_0(\theta)$ is the prior distribution of the parameter $\theta$, 
 $L(\theta \g y)$ is the likelihood of $\theta$ given the observed 
 data $y$, and $c(y)$ is the posterior normalization constant.  
 Hence, 
 \[   
     \overha{f} =  
     \frac{1}{c(y)} \int_T f(\theta) g(\theta \g y) d\theta 
      \ , \            
      g(\theta)= L(\theta \g y) p_0(\theta) 
      \ , \  
      c(y) = \int_\Theta  g(\theta \g y) d\theta 
      \ . 
 \]  
 Notice that $\alpha(i,j)$, the acceptance probabilities   
 defined above, can be computed from posterior ratios 
 $p_n(\theta^j)/p_n(\theta^i)=g^j/g^i$.  
 Hence, actual implementations of these MCMC algorithms  
 do not require the explicit knowledge of the target 
 distribution normalization constant, $c(y)$. 
 It suffices to have an un-normalized function that is proportional 
 to the target distribution, $g(\theta) \propto p_n(\theta)$, 
 as it is the case for the likelihood-prior product.

 The original Metropolis algorithm uses a symmetric sampling 
 kernel, $Q_i^j=Q_j^i$, see Metropolis et al. (1954). 
 In this caese, Metropolis-Hastings acceptance probability can be 
 simplified to the form  $\alpha(i,j) = \min ( 1 , g^j / g^i )$. 
   In statistical physics, the density of interest, $g^i$, 
 often takes the form of the Boltzmann distribution, 
 $g^i=\exp(-\beta H(i))$, where the Hamiltonian function, $H(i)$,  
 gives the energy of the corresponding state. 
  In this case, a new state of lower energy,   
 $j \g \Delta H = H(j) - H(i) \leq 0$, 
 is accepted for sure, while a state of higher 
 energy is accepted with probability $\exp(-\beta \Delta H)$. 
 In section H.1, the same acceptance rejection mechanism 
 reappears in  Metropolis version of Simulated Annealing. 

  Random Walk Metropolis algorithms use a symmetric kernel 
 that is a function only of the random walk step, $z=y-x$, 
 that is, $Q(x,y)=Q(z)=Q(-z)$.  
 A common option in practical implementations is to chose 
 the random walk step from a multivariate Normal distribution, 
 $z\sim N(0,\Sigma)$. 
  The covariance matrix, $\Sigma$, scales the random walk steps. 
  If the steps are too large, the proposed steps would often 
 result in sharp decrease of the traget density, so the acceptance 
 rate is low, making the MCMC inefficient. 
  If the steps are too small, the acceptance rate may be high, 
 but too many steps are required to move effectively across the 
 integration region and, again, the MCMC is inefficient.     

  A practical solution is to take the covariance matrix, $\Sigma$,  
 proportional to the inverse Hessian matrix,  
 $(-\del^2 \log g(x) / \del x' \del x )^{-1}$, computed at the 
 estimated mode, $\overha{x}= \arg \max g(x)$.    
  Alternatively, one can take $\Sigma$ proportional to  
 a convex combination of the diagonal matrix $D$, a prior estimate 
 of marginal variances, and the current estimate of the 
 sampled covariance matrix.      
 \[ 
    \Sigma \propto (1-\lambda) S +\lambda D \ , \ \  
  S = \frac{1}{n}\sum_{j=1}^n (x^j -\bar{x}) (x^j -\bar{x})' 
     = \frac{1}{n} (X -\bar{x}) (X -\bar{x})'  \ . 
 \]   
 In both cases, the proportionality constant is interactively adapted 
 in order to obtain an acceptance rate in a specified range.   
 If the target distribution has heavy tails, this sampling 
 kernel may be modified to a multivariate student's t-distribution.  
 For furthe details, see Gilks et al. (1996).

  Cyclic MCMC schemes use a ``composit kernel'' that updates, one by one, 
  the individual components of a $k$-dimensional vector state, $x$. 
  That is, a cyclic MCMC goes from the current state, $x$ 
  to the next state, $y$, by $k$ intermediate steps,   
  $x=[x_1, x_2, \ldots x_k]$, $[y_1, x_2, \ldots x_k]$,   
  $[y_1, y_2, \ldots x_k]$, \ldots $[y_1, y_2, \ldots y_k]=y$. 
  Cyclic schemes include the Gibbs sampler, popularized by  
  Geman and Geman (1984), and many useful variations.

 \section{Estimation of Ratios}

 This section presents the derivation of the Monte Carlo procedure for
 the numerical integrations required to implement the FBST. 
 The symbol $X$ represents the observed data or 
 some sufficient statistics.  
 The best approach to the numerical integration step required by the 
 FBST is approximation by Monte Carlo (MC) simulation, 
 see Appendix A for the FBST definition, and  
 Evans and Swartz (2000) and Zacks and Stern (2003) 
 for the Monte Carlo approach to this integration problem.  
 We want an estimate of the ratio 
 \begin{eqnarray*} 
 \ev(H;X) &=& 
 \frac{ \int_{\Tb} f(\theta;X) d\theta }     
      { \int_{\Theta}   f(\theta;X) d\theta }
 \end{eqnarray*} 
  \[  
     \Tb = \Tb(s^*) \ , \ \ 
     \Tb(v) =  \{ \theta \in \Theta \g s(\theta) > v \} \ .  
  \]

 Since the space $\Theta$ is unbounded, we randomly chose the values 
of $\theta$ according to an ``importance sampling'' density $g(\theta)$, 
which is positive on $\Theta$. 
 The evidence function is equivalent to  
 \begin{eqnarray*} 
 \ev(H;X) &=& 
 \frac{ \int_{\Theta} Z_g^*(\theta;X) g(\theta) d\theta }     
      { \int_{\Theta}   Z_g(\theta;X) g(\theta) d\theta } 
 \end{eqnarray*} 
 where 
 \begin{eqnarray*}  
 Z_g(\theta;X) &=& \frac{f(\theta;X)}{ g(\theta) } 
   \ \ \ \ \ \mbox{and} \\ 
 Z_g^*(\theta;X) &=& 
           I^*(\theta;X) Z_g(\theta;X) \\ 
           I^*(\theta;X) &=& 1(\theta \in \Tb )      
 \end{eqnarray*} 
 
 Thus, a Monte Carlo estimate of the evidence is 
 \begin{eqnarray*} 
 \hat{Ev}_{g,m}(X) &=& 
 \frac{ \sum_{j=1}^m Z_g^*(\theta^j;X) }     
      { \sum_{j=1}^m Z_g(\theta^j;X) }  
 \end{eqnarray*} 
 where $\theta^j, j=1\ldots m$ are iid and independently 
chosen in $\Theta$ according to the importance sampling density 
$g(\theta)$. Thus, 
 \begin{eqnarray*} 
  \hat{Ev}_{g,m}(X)
  \stackrel{m\rightarrow \infty}{\longrightarrow}    
  \ev(H;X) \ \ \mbox{a.s.} [g]     
 \end{eqnarray*} 

  The goodness of the MC estimation depends on the choice of 
 $g$ and $m$. Standard statistical software libraries have 
 univariate random generators for most common distributions. 
 These univariate generators can also be used to build 
 vector variates from multivariate distributions. 
  Appendix D describes how to generate a Dirichlet vector variate 
 from univariate Gammas. 
   
  Johnson (1980) describes a simple procedure to generate the Cholesky 
 factor of a Wishart variate $W=U'U$ with $n$ degrees of freedom, from 
 the Cholesky factorization of the covariance parameter $V=R^{-1}=C'C$:  
  \begin{eqnarray*} 
  L_i^j &=& N(0,1)\ , \ i>j \\   
  L_i^i &=& \sqrt{ \chi^2(n-i+1) } \ ; \ \ U= L' \, C  
 \end{eqnarray*}    
 At the integration step it is important to perform all matrix 
computations directly from Cholesky factors, 
Golub and van Loan (1989), Jones (1985).   
  In this problem we can therefore use ``exact sampling'', 
 what simplifies substantially the integration step, i.e., 
 $Z_g(\theta;X)=1$.

 \subsection*{Precision of the MC Simulation} 

 In order to control the number of points, $m$, used at each MC 
simulation, we need an estimate of  MC precision for 
evidence estimation. 
 For a fixed large value $m$, the asymptotic distribution of 
 $\hat{Ev}_{g,m}(X)$ is normal with mean $\ev(H;X)$ and 
asymptotic variance $V_g(X)$. 
 According to the delta method, Bickel and Doksum (2001), 
 we obtain that 
 \begin{eqnarray*} 
  V_g(X) &=&  \frac{1}{m} \left( 
    \frac{{\sigma^*_g}^2}{{\mu_g}^2} 
   +\frac{{\sigma_g}^2{\mu^*_g}^2}{{\mu_g}^4} 
   -2\frac{{\mu^*_g}}{{\mu_g}^3}\gamma_g \right) 
 \end{eqnarray*} 
 where  
 \begin{eqnarray*} 
  \mu_g &=& 
   \int_\Theta Z_g(\theta;X)g(\theta)d(\theta) \\    
  \mu^*_g &=& 
   \int_\Theta Z^*_g(\theta;X)g(\theta)d(\theta) \\    
  {\sigma_g}^2 &=& 
   \int_\Theta \left( Z_g(\theta;X) -\mu_g \right)^2  
   g(\theta)d(\theta) \\    
  {\sigma^*_g}^2 &=& 
   \int_\Theta \left( Z^*_g(\theta;X) -\mu^*_g \right)^2  
   g(\theta)d(\theta) \\    
  {\gamma_g}^2 &=& \int_\Theta 
  \left( Z_g(\theta;X) -\mu_g \right) 
  \left( Z^*_g(\theta;X) -\mu^*_g \right) 
   g(\theta)d(\theta) \\    
  \end{eqnarray*} 
  are the expected values, variances and covariance of 
  $Z(\theta;X)$ and $Z^*(\theta;X)$ with respect to 
  $g(\theta)$. 

  Define the coefficients 
  \begin{eqnarray*} 
  {\xi_g} &=& \frac{{\sigma_g} }{{\mu_g} } \ \ , \ \ 
  {\xi^*_g} \ \ = \ \ \frac{{\sigma^*_g} }{{\mu_g} }   
  \end{eqnarray*}   

 For abbreviation, let $\eta=\ev(H;X)$. 
 Also note that $\eta=\mu^*_g/\mu_g$. 
 Then the asymptotic variance is 
 \begin{eqnarray*} 
  V_g(X) &=& \frac{1}{m} \left( 
   {\xi^*_g}^2 +\eta^2{\xi_g}^2 
    -2\, \frac{{\eta\, \gamma_g}}{{{\mu_g}^2}} \right)  
 \end{eqnarray*}

 Let us define the complementary variables 
 \begin{eqnarray*} 
  Z_g^c(\theta;X) &=& 
    I^c(\theta;X) Z_g(\theta;X) \\   
  I^c(\theta;X) &=& 1-I^*(\theta;X) \\ 
  {\sigma^c_g}^2 &=& V_g \left( Z^c(\theta;X) \right) \\ 
  {\xi^c_g} &=& \frac{{\sigma^c_g} }{{\mu_g} }       
 \end{eqnarray*} 
 
 Some algebraic manipulation give us $V_g(X)$ in terms of 
 $\xi^*_g$ and $\xi^c_g$, namely 
 \begin{eqnarray*} 
  V_g(X) &=& \frac{1}{m} \left( 
   {\xi^*_g}^2 (1-\eta)^2 
  +{\xi^c_g}^2 {\eta}^2 
  +2\,{\eta}^2 (1-\eta)^2 \right) 
 \end{eqnarray*}   

 For large values of $m$, the asymptotic $(1-\beta)$ level 
 confidence level confidence interval for $\eta$ is 
 $\hat{Ev}_{g,m}(X) \pm \Delta_{g,m,\beta}$, where   
 \begin{eqnarray*} 
  \Delta_{g,m,\beta}^2 &=& 
   \frac{ F_{1-\beta}(1,m) }{m} 
   \left( {\hat{\xi^*_g}}^2 (1-\hat{\eta})^2 
   +{\hat{\xi^c_g}}^2 {\hat{\eta}}^2 
   +2\,{\hat{\eta}}^2 (1-\hat{\eta})^2 \right)  
 \end{eqnarray*} 
 where $F_{1-\beta}(1,m)$ is the $1-\beta$ quantile 
 of the $F(1,m)$ distribution, and  
 $\hat{\eta}$, ${\hat{\xi^*_g}}$ and ${\hat{\xi^c_g}}$ 
 are consistent estimators of the respective quantities. 

 For large $m$, we can also use the approximation, 
 \begin{eqnarray*} 
  \Delta_{g,m,\beta}^2 &=& 
   \frac{ \chi^2_{1-\beta}(1) }{m} 
   \left( {\hat{\xi^*_g}}^2 (1-\hat{\eta})^2 
   +{\hat{\xi^c_g}}^2 {\hat{\eta}}^2 
   +2\,{\hat{\eta}}^2 (1-\hat{\eta})^2 \right)  
 \end{eqnarray*} 
 since $F(1,m)$ converges in distribution to the 
chi-square distribution with 1 degree of freedom, 
as $m\rightarrow \infty$.  

 If we wish to have $\Delta_{g,m,\beta}\leq \delta$, 
 for a prescribed value of $\delta$, then $m$ should 
 be such that  
 \begin{eqnarray*} 
  m &\geq&    
   \frac{ \chi^2_{1-\beta}(m) }{\delta^2} 
   \left( {\hat{\xi^*_g}}^2 (1-\hat{\eta})^2 
   +{\hat{\xi^c_g}}^2 {\hat{\eta}}^2 
   +2\,{\hat{\eta}}^2 (1-\hat{\eta})^2 \right)  
 \end{eqnarray*} 

 \vfill 


 \section{Monte Carlo for Linear Systems} 
 
 Want to solve the simultaneous matrix equation, 
 \[ 
    x= Hx +b \ , \ \ H \ n\times n  
 \] 
   
 The (Direct) Monte Carlo methods of von Neumann and Ulam (NU) and of
Wasow (WS)  are based on probability transitions, $P_i^j$, and
multipliers or  weights, $V_i^j$, satisfying the following conditions: 
 \[ 
    V_i^j = ( H_i^j / P_i^j ) I(P_i^j >0) \ \g \ \  
    H_i^j \neq 0 \Rightarrow P_i^j >0  \ \wedge \ 
    P_{1:n}^{1:n} \uno < \uno 
 \]  
 We also define the extended Stochastic matrix,  
 \[ 
    P = \left[ \begin{array}{cccc} 
        P_1^1 & \cdots & P_1^n & P_1^{n+1} \\  
        \vdots & & \vdots & \vdots \\ 
        P_n^1 & \cdots & P_n^n & P_n^{n+1} \\  
        0 & \cdots & 0 & 1 
        \end{array} \right] 
        \ , \ \  P_i^{n+1}= 1- \ssum_{j=1}^n P_i^j 
  \]

 $P$ defines a Markov chain in a space of $n+1$ states, 
 $\left\{ 1, 2, \ldots n, n+1 \right\}$, where the last state, 
 $n+1$, is an absorbing state. 

 We want to consider a random path or trajectory, $T$, 
 of a ``particle'' starting at state $i$, until the particle is 
 absorbed at step $m+1$, that is, 
 \[ 
    T= \left[ T(1)=i, T(2), \ldots, T(m), T(m+1)=n+1 \right] 
 \] 

 We define a random variable, $X(T)$, associated to each trajectory. 

 First we define the multipliers products  
 \[ 
    v_1 = 1 \ \ \mbox{and} \ \ 
    v_k = v_{k-1} V_{T(k-1)}^{T(k)} \ , \  2\leq k \leq m \ . 
 \]  
 Von Neumann - Ulam's and Wasow's versions of the Monte Carlo Algorithm, 
 use $X(T)$ equal to, respectively, 
 \[ 
    NU(T) = v_m b_{T(m)} / P_{T(m)}^{n+1} \ \ \mbox{and} \ \ 
    WS(T) = \ssum_{k=1}^m v_k b_{T(k)} \ .  
 \] 

 The key to these Monte Carlo algorithm is that the expected value of 
the variable $X(T)$, over all trajectories starting at state $i$, 
is the solution of the simultaneous equation, provided these expected 
values are well defined, that is, if 
 \[ 
    \mbox{if} \  \   e_i = E( X(T) \g T(1)=i ) 
    \ \  \mbox{then} \ \  e= He +b 
 \] 
   
 Let us prove the statement above for Wasow's version. 
 By definition, 

 \[ 
    \Pr(T) = \prod_{k=1}^{m} P_{T(k)}^{T(k+1)} \ \ \mbox{and} 
 \]    
 \[ 
    e_i = \sum_{\begin{array}{c}
                T=[T(1)=i,T(2)=j,\ldots T(m+1)=n+1] \ , \\   
                m=1,2,\ldots \infty \ . \end{array} } 
           X(T) \Pr(T) \ . 
 \] 
 Given a trajectory $T$, we can separate the terms in 
 $X(T)$ with index $1$, that is, 
 \[ 
   X(T) =  b_{T(1)} +V_{T(1)}^{T(2)} X(T(2:m\ma1 )) 
   \ , \ \ \mbox{hence,}  
 \] 
 \[ e_i = \sum_{j=1}^{n+1} P_i^j \sum_{S=[j,\ldots n+1]} 
    \left( b_i +V_i^j X(S) \right) \Pr(S) 
 \] 
 \[ 
   = \sum_{j=1}^{n} P_i^j  \sum_{S=[j,\ldots n+1]}  
    \left( b_i +V_i^j X(S) \right) \Pr(S)
    +P_i^{n+1} b_i 
 \]
 \[ 
   = \sum_{j=1}^{n} P_i^j V_i^j \sum_{S=[j,\ldots n+1]} X(S) \Pr(S)
    +b_i \left( P_i^{n+1} +\sum_{j=1}^n P_i^j \sum_S \Pr(S) \right)  
 \]
 \[ 
    = \sum_{j=1}^n H_i^j e_j + b_i  \ , \ \ \mbox{Q.E.D.} 
 \]

 The Reverse or Adjoint Monte Carlo methods of von Neumann and Ulam (NU)
and of Wasow (WS)  are based on probability transitions, $Q_i^j$, and 
multipliers or  weights, $W_i^j$, satisfying the following
conditions: 
 \[ 
    W_i^j = ( H_j^i / Q_i^j ) I(Q_i^j>0) \ \g \ \  
    H_j^i \neq 0 \Rightarrow Q_i^j >0  \ \wedge \ 
    Q_{1:n}^{1:n} \uno < \uno 
 \]  
 We also define the extended Stochastic matrix,  
 \[ 
    Q = \left[ \begin{array}{cccc} 
        Q_1^1 & \cdots & Q_1^n & Q_1^{n+1} \\  
        \vdots & & \vdots & \vdots \\ 
        Q_n^1 & \cdots & Q_n^n & Q_n^{n+1} \\  
        0 & \cdots & 0 & 1 
        \end{array} \right] 
        \ , \ \  Q_i^{n+1}= 1- \ssum_{j=1}^n Q_i^j 
  \]   

 We want to consider a random path or trajectory, $T$, of a ``particle'' 
 starting at state $i$, chosen at random with probability $r_i$, 
 until the particle is absorbed at step $m+1$, 
 just after visiting state $T(m)$ at step $m$ , that is, 
 \[ 
    T= \left[ T(1)=i, T(2), \ldots, T(m), T(m+1)=n+1 \right] 
 \] 

 We define random variables  associated to each trajectory. 
 First we define the multiplier products  
 \[ 
    w_1 = b_i/r_i \ \ \mbox{and} \ \ 
    w_k = w_{k-1} W_{T(k-1)}^{T(k)} \ , \  2\leq k \leq m \ . 
 \]  
 Von Neumann - Ulam's and Wasow's versions of the reverse or adjoint
 Monte Carlo Algorithm, use $X_j(T)$ equal to, respectively, 
 \[ 
    NU_j(T) = w_m \delta_{T(m)}^j / Q_{T(m)}^{n+1} \ \ \mbox{and} \ \ 
    WS_j(T) = \ssum_{k=1}^m w_k \delta_{T(k)}^j \ .  
 \] 

 Again, the key to these Monte Carlo algorithm is that the expected
value of  the variable $X_j(T)$, over all trajectories ending at state
$j$,  is the solution of the simultaneous equation, provided these
expected  values are well defined.
 The proof for the reverse method is similar to the direct case. 

 \vfill 

 \mbox{} 

%% file: CAPSOPT.TEX
 \chapter{Stochastic Evolution and Optimization}

 \mbox{}

 {\flushright

 {\it  
 ``God does not play dice (with the universe).'' 
 } 


 Albert Einstein (1879 - 1955).  

 \mbox{} 

 {\it 
 ``Einstein, stop telling God what to do (with his dice).'' 
 }

 Niels Bohr (1885 - 1962). 

 \mbox{} 

 {\it 
 ``God not only plays dice, He also sometimes \\ 
   throws the dice where they are not seen.'' 
 }

 Stephen Hawking (1942 - \ \mbox{} ). 


 }



  This section gives a condensed introduction to inhomogeneous Markov 
 chains, the theory that is needed to formalize Simulated Annealing  
 (SA) and related algorithms presented in chapter 5. 
  We follow the presentations in Jetschke (1989) and Pflug (1996, ch.2),   
 and assume some familiarity with homogeneous Markov chains, as presented 
 in Feller (1957, ch.15) or H\"{a}ggstr\"{o}m (2002).

 \section{Inhomogeneous Markov Chains}

 We begin by introducing some notation for this chapter. First, a
notational idiosyncrasy: In almost all areas of mathematics it is usual
to write a $d$-dimensional vector as a $d\times 1$ {\it column matrix},
$x$, and a linear transformation as the left multiplication of $x$ by a
$d\times d$ square matrix $A$, that is, $Ax$. However, in the literature
of Markov chains, it is usual to write a a $d$-dimensional vector as a
$1\times d$ or {\it row matrix}, $v$, and a linear transformation as the
right multiplication of $v$ by a $d\times d$ square matrix $P$, that is,
$vP$. Herein, we make use of the two forms, according to the context. 

 $d$-Dimensional vectors are written in lower case format, $v$. A
density or probability vector is a vector in the simplex support, $v>0$
and $\lVert v \rVert =1$. $d$-Dimensional (square) matrices, on the
other hand, are written in upper case format, $P$. In particular, 
 $I$ is reserved to denote the $d$-dimensional identity matrix. 
 A $d$-dimensional kernel or transition probability matrix has its 
rows in the simplex support. Right subscripts and superscripts will
index matrices rows and columns. For instance, $P_i$, $P^j$ and $P_i^j$
will indicate the $i$-th row, the $j$-th column, and the element or
entry $i,j$ of matrix $P$, respectively. In the same way, $x_i$ and
$v^j$ denote, respectively, the $i$-th element of the column vector $x$,
and the $j$-th element of the row vector $v$.
  
 Braces are used to index a sequence of objects, such as
 $P\{1\},P\{2\},\ldots P\{t\}$. The symbol $P\{s\atee t\}$ will denote
the product of the objects indexed from $s$ to $t$, that is,
 \[ 
    P\{s\atee t\} \equiv \pprod_{k=s}^t \, P\{k\} \ .   
 \] 

 Finaly, given scalars, $\alpha$ and $\beta$, we have, as usual,
 $\alpha \wedge \beta= \min(\alpha,\beta)$, 
 $\alpha \vee \beta= \max(\alpha,\beta)$, 
 $\alpha^+ = 0 \vee \alpha$, 
 $\alpha^- = 0 \vee -\alpha$.

 \subsubsection*{Homogeneous Markov Chains}  
 
 In a Markov chain with kernel or transition matrix $P$, 
 $P_i^j\geq0$ such that $P_i\uno =1$,  
 $P_i^j$ represents transition probability 
 from state $x\{i\}$ to state $x\{j\}$ in a finite state space, 
 $S=\{x\{ 1\}, x\{ 2\}, \ldots x\{ d\} \}$.  
 For the sake of simplicity, we often write the index, $i$, 
 instead of the indexed state, $x\{i\}$, that is, we identify the 
 state space with its index set, $S=\{1,2,\ldots d\}$. 

 A trajectory or path of length $t$ from an initial state $i$ to 
 a final state $j$ is given by 
 $\tau=[\tau(1)=i, k(2),\ldots \tau(t), \tau(t+1)=j]$.  
 If a Markov chain is initially at state $i$, the probability that 
 it will follow the trajectory $\tau$ is 
 \[ 
    \Pr (\tau) = \pprod_{k=1}^{t} P_{\tau(k)}^{\tau(k+1)} 
 \] 

 If we select the initial state state, $i$, from distribution  
 $v$, $v\geq 0$, $v\uno=1$, the probability that the chain 
 is at state $j$ after $t$ transitions, following any possible 
 trajectories trough intermediate states, is given by $w^j$, 
 where   
 \[ 
    w = v \, \pprod_{k=1}^{t} P  \ . 
 \]

 A trajectory $\tau$ is possible if it has non-zero probability. 
 A Markov chain is irreducible if there is a possible trajectory 
 connecting any initial state, $i$, to any final state, $j$. 
 A cycle is a trajectory with the same initial and final states. 
 State $i$ has period $k>1$ if any possible cycle starting 
 at $i$ has length multiple of $k$. 
 Otherwise, state $i$ is aperiodic.  
 A Markov chain is aperiodic if has no periodic states.

 The probability distribution $g$ is invariant by kernel 
 $P$ if $g=gP$. 
 An invariant distribution is also known as eigen-solution,  
 equilibrium or stable distribution for $P$.   
 It can be shown that an irreducible and aperiodic Markov chain 
 has a unique  invariant distribution, see Feller (1957). 
 Under the same regularity conditions, it can also be shown that 
 the invariant distribution is the chain's limiting distribution, 
 that is, 
 \[ 
    \lim_{ t\rightarrow \infty } \pprod_{k=1}^{t} P 
    \ = \  
    \left[ \begin{array}{c} g \\ g \\ \vdots \\ g \end{array} \right] 
    \ = \ 
    \left[ \begin{array}{cccc} 
     g^1 & g^2 & \ldots & g^d \\ g^1 & g^2 & \ldots & g^d \\ 
     \vdots & \vdots & & \vdots \\ g^1 & g^2 & \ldots & g^d 
     \end{array} \right]  \ .    
 \]  
 Hence, for any initial distribution, $v$,  
 \[ 
    v \left( \lim_{ t\rightarrow \infty } \pprod_{k=1}^{t} P \right)  
    \ = \ g \ .   
 \]

 Given the irreducible and aperiodic kernel $P$, having the stable 
 distribution $g$, the {\em reverse} kernel, $R$, is defined as 
 $R_i^j= g^jP_j^i/g^i$. 
 The reverse kernel can be interpreted, using Bayes theorem, as 
 the kernel of the Markov chain $P$ going backwards in time, 
 that is, 
 \[ 
   R_i^j = \Pr ( x\{t\}=j \g x\{t+1\}=i ) = 
   \frac{ \Pr ( x\{t+1\}=i \g x\{t\}=j ) \Pr( x\{t\}=j ) } 
        { \Pr ( x\{t+1\}=i ) } = 
   \frac{ P_j^i g^j}{g^i } \ . 
 \] 
 Kernel $P$ is  {\em reversible} if there is a distribution $g$ 
 statisfying the {\em detailed balance equation}, 
 $g^i P_i^j = g^j P_j^i$. 
 Summing both sides of the detailed balance equation over index $i$, 
 we obtain $g^j = \ssum_i g^i P_i^j$, showing that this is a 
 sufficient condition for $g$ to be an invariant distribution.  
 Hence, for a reversible chain, the forward and backward 
 kernels are identical, $R_i^j=P_i^j$.

 \subsubsection*{Vector and Matrix Norms}

 A {\bf norm}, in a vector space $E$, is a function
 \[ 
    \lVert\ .\ \rVert \ :\ E \Rightarrow {\bf R}  \mid 
   \forall x, y \in E \ \ \mbox{and} \ \ \alpha \in {\bf R} \ , 
 \] 
 \begin{enumerate}
 \item $ \lVert x \rVert \geq 0,\ \mbox{and}\ \lVert x \rVert = 0 
    \Leftrightarrow x = 0 $. 
 \item $ \lVert \alpha x \rVert = | \alpha | \ \lVert x \rVert $.
 \item $ \lVert x + y \rVert \leq \lVert x \rVert + \lVert y \rVert $, 
  the triangular inequality.  
 \end{enumerate}
 In particular, for $x \in {\bf R}^n$ and $p>0$,
 \[ 
    \lVert x \rVert_p = ( \ssum_1^n |x_i|^p )^{1/p} 
    \ \ , \ \ 
    \lVert x \rVert_\infty = \max_{i=1}^n |x_i| \ . 
 \]  
 defines the standard $L_p$ norms in ${\bf R}^n$.

 Given a normed vector space, $(E,\ \lVert\ \rVert )$, 
 \[ \lVert T \rVert \equiv \max_{x\neq 0} ( \ \lVert T(x) \rVert \ /\ 
    \lVert x \rVert \ ) \ . 
 \]  
 defines the {\bf induced norm} on the vector space of linear
transformations, $T:E \rightarrow E$, for which 
 $\exists \alpha \in {\bf R} \ \mid \ 
  \lVert T(x) \rVert \leq \alpha \lVert x \rVert, \ \forall x \in E$, 
 that is, the vector space of bounded linear transformations on $E$. 
 By linearity,
 \[ \lVert T \rVert \equiv \max_{x\mid \ \lVert x \rVert =1} \lVert T(x) 
    \rVert \ .
 \] 

 In $( {\bf R}^n,\ \lVert \ \rVert )$ the induced norm on set of bounded
linear transformations, $T:{\bf R}^n \rightarrow {\bf R}^n$, defines the
matrix norm in $( {\bf R}^n,\ \lVert \ \rVert )$. 
 Speciffically, for an $n\times n$ matrix $A$, 
 $\lVert A \rVert = \lVert T \rVert$, where $T(x) = Ax$. 

 {\bf Lemma 1:} The matrix norm in $( {\bf R}^n,\ \lVert \ \rVert )$, 
 has the following properties: If $A$ and $B$ are $n\times n$ matrices,
 \begin{enumerate}
 \item $\lVert A \rVert \geq 0 \ \mbox{and}\ \lVert A \rVert = 0 
   \Leftrightarrow A=0 $
 \item $ \lVert A+B \rVert \leq \lVert A \rVert + \lVert B \rVert $
 \item $ \lVert AB \rVert \leq \lVert A \rVert \ \lVert B \rVert $ 
 \end{enumerate}

 {\bf Lemma 2:} ($L_1$ and $L_\infty$ explicit expressions).  
 \begin{eqnarray*}
  \lVert A \rVert_1 &=& \max_{j=1}^n \ssum_{i=1}^n  | A_i^j |  \\ 
 \lVert A \rVert_\infty &=& \max_{i=1}^n \ssum_{j=1}^n | A_i^j | 
 \end{eqnarray*} 

 {\bf Proof:} To check the expression for $L_1$ and $L_\infty$ 
 observe that 
 \begin{eqnarray*} 
  \lVert Ax \rVert_1 &=& 
  \ssum_{i=1}^n \lvert \; \ssum_{j=1}^n A_i^j x_j \rvert  
       \ \ \leq \ \  \ssum_{i=1}^n \ssum_{j=1}^n |A_i^j | \ |x_j | \\ 
 &\leq & \ssum_{j=1}^n |x_j | \max_{j=1}^n \ssum_{i=1}^n |A_i^j | 
       \ \ = \ \  \lVert A \rVert_1 \ \lVert x \rVert_1 
 \end{eqnarray*} 
 \begin{eqnarray*} 
 \lVert Ax \rVert_{\infty } &=& 
  \max_{i=1}^n \mid \ssum_{j=1}^n A_i^j x_j \mid  
      \ \ \leq \ \  \max_{i=1}^n \ssum_{j=1}^n |A_i^j | \ |x_j | \\ 
 &\leq & \max_{j=1}^n |x_j | \max_{i=1}^n \ssum_{j=1}^n |A_i^j | 
      \ \ = \ \  ||x||_\infty \ ||A||_\infty  
 \end{eqnarray*} 
 and that, if $k$ is the index that realizes the maximum in the norm
 definition, then the equality is realized 
 by the vector  $x = I^k$ for $L_1$, and 
 by the vector $x \mid  x_j = sig(A_i^j )$ for $L_\infty$.

 One can check that    
  $\lVert x \rVert_\infty \leq \lVert x \rVert_1 
    \leq  n\lVert x \rVert_\infty$ and 
  $\lVert x \rVert _\infty \leq \lVert x \rVert_2 
    \leq  n^{1/2}\lVert x \rVert_\infty $. 
 In fact, any given $p$ norn can provide a bound to another $q$ norm and, 
in this sense, they are equivalent. 
 In the remaining of this section the $L_1$ norm will be used
throughout, so we will write 
 $\lVert x \rVert$ for $\lVert x \rVert_1$.

 \subsubsection*{Dobroushin's Contraction Coefficient} 

 {\bf Lemma 3} (Total Variation). Given two probability (non-negative,
unitary, row) vectors, $v$ and $w$, their Total Variation or $L_1$
difference has the alternative expressions: 
 \[ 
   \lVert v -w \rVert = 2 \left( 1 -\ssum_k v^k \wedge w^k \right) 
   = 2 \ssum_k \left( v^k -w^k \right)^+ 
 \]    

 {\bf Proof:} Just notice that 
 \[ 
   2 -2\ssum_k v^k \wedge w^k = 
   \ssum_k v^k + \ssum_k w^k -2 \ssum_k v^k \wedge w^k = 
   \ssum_k \lvert v^k - w^k \rvert  \ , \ \ \mbox{and}   
 \]    
 \[ 
    \left( v^k -w^k \right)^+ = \left( v^k - v^k \wedge w^k \right) 
    \ \mbox{hence} \ \ 
    \ssum_k  \left( v^k -w^k \right)^+ = 1 -\ssum_k v^k \wedge w^k \ .  
 \] 
   
 The Dobroushin Contraction Coefficient or Ergodicity Coefficient of a
transition probability matrix, $P$, is defined as 
 \[ 
   \rho(P) \equiv  
   \frac{1}{2} \max_{i,j} \ssum_k \lvert P_i^k -P_j^k \rvert = 
   \frac{1}{2} \max_{i,j} \lVert I_i P - I_j P \rVert \ . 
 \] 
 It is clear from the definition that $\rho(P)$ measures the maximum
$L_1$ distance between the rows of $P$. If a sequence of kernels,
$P\{k\}$, is clear from the context, we shall also write
 \[ 
   \rho\{k\} \equiv \rho (\, P\{k\} ) \ , \ \ \mbox{and} \ \  
   \rho\{m\atee n\}   \equiv \pprod_{k=s}^t \rho\{k\}  \ . 
 \]  

 {\bf Lemma 4} (Vector Contraction). Two probability vectors, $v$ and
$w$, are contracted by the transition matrix $P$ in the sense that: 
 \[ 
   \lVert vP -wP \rVert \leq \rho (P) \lVert v-w \rVert \ . 
 \] 

 {\bf Proof:} If $v=w$ or if $v=I_i$ and $w=I_j$, the result is trivial. 
 Otherwise, let $v\neq w$ and $m= v \wedge w$. Defining
 \[ 
    G_i^j = 2 \frac{ (v_i - m_i ) (w_j -m_j) }{ \lVert v-w \rVert } \ ,  
 \] 
 it is easy to check that: 

 (a) $G_i^j \geq 0$ , 

 (b) $v-w = \ssum_{i,j} G_i^j ( I_i -I_j )$ , and 

 (c) $\frac{1}{2} \lVert v-w \rVert = \ssum_{i,j} G_i^j$ .  

 Hence, 
 \[ 
    \lVert vP -wP \rVert = 
    \lVert \ssum_{i,j} G_i^j ( I_i -I_j ) P \rVert \leq 
 \] 
 \[ 
    \left( \ssum_{i,j} G_i^j \right) 
      \max_{i,j} \lVert ( I_i -I_j ) P \rVert = 
    \frac{1}{2} \lVert v-w \rVert 2 \rho (P) = \rho (P) \lVert v-w \rVert \ . 
 \]  

 {\bf Lemma 5} (Matrix Contraction). Two transition matrices, $P$ and $Q$, 
  are contracted in the sense that:
 \[ 
    \rho ( PQ ) \leq \rho (P) \, \rho (Q)  \ . 
 \] 

 {\bf Proof:} 
 \[ 
  \rho (PQ) \ = 
  \frac{1}{2} \max_{i,j} \lVert ( I_i -I_j ) PQ \rVert \leq 
  \rho (Q) \frac{1}{2} \max_{i,j} \lVert  ( I_i -I_j ) P \rVert =  
  \rho (P) \, \rho (Q) \ . 
 \] 

 {\bf Theorem 6} (Weak Ergodicity (loss of memory)).
 \[ 
    \lim_{t\rightarrow \infty} 
     \rho \{1\atee t\} =0   
     \Rightarrow  
    \lim_{t\rightarrow \infty} 
    \lVert (v-w) \, P\{1\atee t\} \rVert =0 \ . 
 \]  

 {\bf Proof:} Immediate, from Lemma 2.

 {\bf Lemma 7} (Strong Ergodicity). Assume that the following conditions
hold: 

 \noindent 
  (a) Each $P\{k\}$ has a unique invariant distribution, 
    $v\{k\}= \, v\{k\} \, P\{k\}$ , such that \\  
  $\ssum_{k=1}^\infty \lVert \, v\{k+1\} - v\{k\} \rVert < \infty$ \ ; \ \
     
 \noindent 
 (b) $\rho\{k\} > 0$  \ ;  

 \noindent 
 (c) $\rho\{1\atee \infty\} = 0$  \ .  
 
 Then, there is a limiting distribution, $v\{\infty\}$, such that, for
any distribution $w$, 
 \[ 
   \lim_{t\rightarrow \infty} \lVert \, w  
     P\{1\atee t\} -v\{\infty\}  \rVert =0 
 \] 

 {\bf Proof.} Condition 7a ensures that, with respect to the $L_1$ norm,
$v\{k\}$ is a Cauchy sequence in the compact simplex support. 
 Hence, the sequence has a unique accumulation point, 
 $v\{\infty\} = \lim_{k\rightarrow \infty} v\{k\}$.

 Since for $1<s<t<\infty$     
 \[ 
    v\{\infty\} \, P\{s\atee t\} - v\{\infty\} = 
    ( v\{\infty\}  - v\{s\} ) \, P\{s\atee t\} + v\{s\} \, 
    P\{s\atee t\} - v\{\infty\} = 
 \] 
 \[ 
     ( v\{\infty\}  - v\{s\} ) \, P\{s\atee t\} 
      +\ssum_{k=s}^{t-1} ( v\{k\}  - v\{k+1\} ) \, P\{k+1\atee t\}  
      + v\{t\} - v\{\infty\}  \ ,
 \]    
 we have the inequality 
 \[ 
   \lVert w  P\{1\atee t\} - v\{\infty\} \rVert \leq 
   \lVert ( w P\{1\atee t-1\} - v\{\infty\} ) \, P\{s\atee t\} \rVert 
   + \lVert \, v\{\infty\} \, P\{s\atee t\} - v\{\infty\} \rVert \leq 
 \] 
 \[ 
    2 \rho\{s\atee t\} +\lVert v\{\infty\} -v\{s\} \rVert  
    + \ssum_{k=s}^{n-1}  \lVert v\{k\} -v\{k+1\} \rVert 
     +\lVert v\{\infty\} -v\{t\} \rVert \leq 
 \] 
 \[ 
    2 \rho\{s\atee t\} +2 \sup_{k\geq s} \lVert \, v\{\infty\} -v\{k\} \rVert  
    + \ssum_{k=s}^{t-1}  \lVert v\{k\} -v\{k+1\} \rVert \ .  
\] 

 Letting $t\rightarrow \infty$, all terms in the right hand side can be
made arbitrarily small for an appropriate large value of $s$.
Consequently, the left hand side converges to zero, Q.E.D. 

 {\bf Theorem 8} (Small Perturbations). It is possible to use a perturbed 
  sequence of kernels, $Q\{k\}$, instead of $P\{k\}$, and still obtain 
  convergence to the same invariant distribution provided that
 \[ 
    \ssum_{k=1}^\infty \lVert \, P\{k\} - Q\{k\} \rVert < \infty \ . 
 \] 
 
 {\bf Proof.} The result follows from the inequality
 \[ 
    \lVert \, P\{s\atee t\} - Q\{s\atee t\} \rVert \leq 
    \ssum_{k=s}^t \lVert \, P\{k\} - Q\{k\} \rVert
 \]   

 The Small Perturbations theorem plays an important role in the design
of efficient algorithms based on heuristic perturbations, a technique
that can greatly expedite the annealing process, see Stern (1991) and   
Pflug (1996, ch.2).

 \section{Simulated Annealing}

 \subsubsection*{The Metropolis Algorithm} 

 Consider a system, $X$, where the system state
is parameterized by a $d$-dimentional coordinate vector 
 $x=[x_1,\ldots x_d] \in X$. 
 The neighborhood $N(x)$ is defined as the set of states $y$ that are 
adjacent to $x$, that is, the set of states that can be 
reached dirrectly from $x$, taht is, with one move, or in a single step. 
 The neighborhood size is $n(x)=|N(x)|\leq n=\max_x n(x)$. 
 We assume that the neighborhood structure is symmetric, that is, 
 $y\in N(x) \Rightarrow x\in N(y)$, and that any two states, $x$ and $y$, 
 are linked by a path with at most $m$ steps.  
 Our aim is to minimize a finite and positive objective function,
 $H(x)$, with an unique global minimum attained at $x^*$.
 The system's Lipschitz constant, $\Delta$, is the maximum difference in 
the value of $H$, for adjacent states, that is, 
 \[
   \Delta = \max_x \ \max_{y\in N(x)}  
      \lvert H(y) - H(x) \rvert \ . 
 \]

 The Gibbs distribution is defined as 
 \[ 
    g(\theta)_x = \frac{n(x)}{Z(\theta)}  \exp( -\theta H(x) ) \ ,   
      \ \ \mbox{with} \ \ \ 
    Z(\theta) = \ssum_{x} n(x) \, \exp( -\theta H(x) ) \ . 
 \] 
 The Gibbs distribution specifies state probabilities in many systems of   
 Statistical Physics, where the Hamiltonian function, $H$, represents 
 state energies, and the parameter $\theta$ is the system's inverse 
temperature. The normalization constant, $Z(\theta)$, is called the 
partition function.

 The Metropolis kernel is defined by  
 \[
    P(\theta)_x^y = \left\{ \begin{array}{ll} 
    \frac{1}{n(x)} \exp\left( \left( H(x)-H(y) \right)^{+} \right)  
    \ , \ \ 
    & \mbox{if\ } y\in N(x) \\  
    1 -\sum_{y\in N(x)} P(\theta)_x^y &  \mbox{if\ } y=x \\ 
    0      & \mbox{otherwise} 
    \end{array} \right. \ . 
 \]

 {\bf Theorem 9} (Metropolis sampling). The Gibbs distribution 
 $g(\theta)$ is invariant for the metropolis kernel $P(\theta)$. 

 {\bf Proof.} It suffices to prove the detailed balance equation 
 \[
    g(\theta)_x P(\theta)_x^y = g(\theta)_y P(\theta)_y^x 
 \]

 If $y\notin N(x)$, balance is trivial. Otherwise,  we use 
 \[ 
    \frac{1}{n(x)} \exp\left( \left( H(x)-H(y) \right)^{+} \right)  
    \ = \ 
    \frac{1}{n(x)} \left( \frac{g(\theta)_y n(x)}{g(\theta)_x n(y)} 
      \wedge 1 \right) \ . 
 \] 
 Assuming that $(g(\theta)_y n(x))/(g(\theta)_x n(y)) \geq 1$, 
 \[
    g(\theta)_x P(\theta)_x^y = \frac{g(\theta)_x}{n(x)} \ \ \mbox{and} \ 
    g(\theta)_y P(\theta)_y^x = 
      \frac{g(\theta)_y}{n(y)} \frac{g(\theta)_x n(y)}{g(\theta)_y n(x)} = 
    \frac{g(\theta)_x}{n(x)} \ . 
 \]
 The case $(g(\theta)_y n(x))/(g(\theta)_x n(y)) < 1$ follows similarly.

 We will now study an appropriate cooling schedule  
 $\theta_1, \theta_2, \ldots$, for the Simplified Metropolis Algorithm  
 where, at each temperature $1/\theta_t$, we take $m$ steps using 
 the kernel $P\{ t \}=P(\theta_t)$, or a single step using the kernel 
 $Q\{ t \}= P\{ t \}^m$

 {\bf Theorem 10} (Logarithmic Cooling). 
 In the simplified Metropolis algorithm, 
 for any monotone decreasing cooling schedule 
 \[
    \frac{1}{\theta_t} \geq \frac{\Delta m^2 \ln(n)}{\ln(t)} 
 \]
 and any initial distribution $w$,
 \[ 
    \lim_{t\rightarrow \infty} 
    \lVert  w Q\{ 1 \} Q\{ 2 \} \ldots Q\{ t \} 
           -v\{\infty\}  \rVert =0 \ . 
 \] 
 
 {\bf Proof.} From the definition of the system's Lipschitz constant, 
 and from the fact that any two states of the system are 
 conected by a path of lenght at most $m$, it follows that, 
 for any two states, $x$ and $y$,  
 \[ 
    Q\{ t \}_x^y \geq  
   \left( \frac{1}{n} \exp\left( -\Delta \theta_t \right) \right)^m 
   \ = \    
   \frac{1}{n^m} \exp\left( -m\Delta \theta_t \right) \ .   
 \] 
 Hence, 
 \[ 
    \rho\{t\} = \rho ( Q\{ t \} ) =      
     \max_{x,y} \left( 1 -  \sum_z \left( \, 
     Q\{ t \}_x^z  \wedge  Q\{ t \}_y^z   \, \right) \right)  
    \leq 
 \] 
 \[ 
     \max_{x,y} \left( 1 -   \left( \, 
     Q\{ t \}_x^{z*}  \wedge Q\{ t \}_y^{z*} \, \right) \right)  
     \leq 
   1- \frac{1}{n^m} \exp\left( -m \Delta \theta_t \right) =  
 \] 
 \[ 
   1- \frac{1}{n^m} \exp 
     \left( -m \Delta \frac{\ln(t)}{\Delta m^2 \ln(n)} \right) =   
   1 - \frac{1}{ n^m } \frac{ n^m}{ t } = 1 - \frac{1}{t} \ . 
 \] 
 Condition (c) of the strong ergodicity lemma follows from 
 \[   
   \sum_{t=1}^\infty \frac{1}{ t } = \infty  
    \ \Rightarrow \ 
    \rho\{1\atee \infty\} = 
    \prod_{t=1}^\infty \left( 1 -\frac{1}{ t } \right) = 0 \ . 
 \]

 Finally, in orther to check condition (a) the strong ergodicity lemma, 
 we must show that the invariant measures  $v\{t\}=v(\theta_{t})$ 
 constitute a Cauchy sequence. 
 However, as $\theta \rightarrow \infty$, the elements of $v\{t\}$ are 
either increasing for $x=x^*$ or decreasing for $x\neq x^*$ and 
sufficiently large $\theta$. Hence, for $t\geq l$,   
 \[
    \ssum_{t=l}^\infty \lVert v\{t+1\} -v\{t\} \rVert = 
    \ssum_{t=l}^\infty \ssum_{x\in X} 
    \left( v\{t+1\} -v\{t\} \right)^+ \leq 
 \]
 \[
    \ssum_{x\in X} 
    \left( v\{\infty\} -v\{l\} \right)^+ < \infty \ . 
 \]

 There is an implicit choice of scale in the unit taken to measure the
Hamiltonian or objective function, $H(x)$. An adequate scale should
start the annealing process with a good acceptance rate for hill
climbing moves. 
 The step size of the logarithmic cooling schedule is inversely 
proportional to the  {\it cooling constant}, $\Delta m^2 \ln(n)$.  
 An alternative to the simplified Metroplis algorithm, taking $m$ steps
at each temperature $\theta_t$, is to implement the standard Metropolis 
algorithm using the cooling constant $\Delta m^3 \ln(n)$.

 \section{Genetic Programming}

 \subsubsection*{The Intrinsic Parallelism Argument}   

 Consider programs coded as binary (0-1) arrays of length $n$. 
 A pattern or {\it schema} of {\it length} $l$, is a partial specification 
 of a binary array of length $l$,
 \[
   s[i]= 0 , 1 \ \mbox{or} \ * \ \mbox{(don't-care)}  
    \ \ , \ \ 1\leq i \leq l \ .
 \]   
 The number of specified positions or {\it loci}, that is, $l$ minus the 
 number of don't-cares, defines the schema's {\it order}. 
 The program's sub-array $p[j]$ in the {\it window} 
 $k\leq j \leq k+l$, is an {\it instance} of schema $s$ iff they coincide, 
 in the specified loci, that is, iff $p[k+i]=s[i]$, for all $s[i]\neq *$.  

 The intrinsic parallelism argument, presented in chapter 6, requires an
estimate of how many schemata of order $l$ and length $2l$ can be
represented in a program of length $n$.

 Following Reeves (1991), consider the window of lenght $2l$ at the
beginning or leftmost locus, $1\leq j \leq 2l$, and let $B(2l,l)$ be the
number of choices for the specified loci, $l$, among the $2l$ available
positions. This first window can obvously represent $B(2l,l)\; 2^l$
distinct schemata, for once the $l$ loci have been chosen, there are
$2^l$ possible 0-1 attributions to their values.
 
 Now slide the window $2l$ position to the right, so as to span
positions $2l+1\leq j\leq 4l$. This new window has no positions in
common with the previous one and can, therefore, represent the same
number of schemata. If we keep sliding the window $2l$ position to the
right until positions $n-2l\leq j\leq n$ are spanned, it follows from
Stirling's approximation that the total count of possible represented
schemata, satisfies the relation
 \[
    \frac{n}{2l} B(2l,l) 2^l  \approx  2^{3l} \propto m^3 \ ,  
 \]
 where the population size is taken as $m= c\; 2^l$. 
 The constant $c$ is interpreted as the expected number of instances of
any given schema (of order $l$ and length $2l$) present in this
population. Hence, under all the conditions above, we can (under)
estimate the number of schemata present in the population as
proportional to $m^3$. 
 For generazations of the implicit parallelism theorem, 
see Bertoni, M.Dorigo (1993).   

 \subsubsection*{Stirling's Approximation}

  For large $n$,
 \[
 \ln n! = \ssum_{j=1}^n \ln j \approx \int_{j=1}^{n} \ln j dj = 
    \left[ j \ln j -j \right]_1^n = n \ln n -n +1 \ .  
 \]   
 A more detailed analysis of the remainder gives us
 \[ 
   \ln n! \approx n \ln n -n +O(\ln n ) \ . 
 \]

 From Stirling's approximation, the following Binomial approximations
hold:
 \[
   \ln \left( \begin{array}{c} n \\ np \end{array} \right) 
   \approx   n H(p) 
    \ \ \mbox{where} \ \ 
    H(p) = -p \ln p -(1-p) \ln (1-p) \ . 
 \]
 \[
    \mbox{and} \ \ 
   \ln \left( \begin{array}{c} 2l \\ l \end{array} \right) 
   \approx   2l H(1/2) \ . 
 \]

 \section{Ontogenic Development}

 Autopoietic and alopoietic systems, living organisms and artificial
machines, have both to be built up and have their basic components
maintained. However, there are profound differences in their development
processes. In this section we examine the structural similarities and
differences between such systems, and how such structures can explain
some properties of systemic development. 

 Herein, the adult or after construction systemic feature known as
aging, receives special attention. Elementary or simple components have
no structure, no internal states, and hence no memory. They can,
therefore, exhibit no aging. Complex systems, however, exhibit some form
of aging. We will see how the aging process of complex system can
reflect systemic structure. We will contrast, in particular, bottom-up
and top-down system construction, and their respective aging processes.
Our analysis will follow Gavrilov (1981, 2001, 2006).  

 {\it ``The first fundamental feature of biosystems is that, in contrast
to technical (artificial) devices which are constructed out of
previously manufactured and tested components, organisms form themselves
in ontogenesis through a process of self-assembly out of {\em de novo}
forming and externally untested elements (cells). The second property of
organisms is the extraordinary degree of miniaturization of their
components (the microscopic dimensions of cells, as well as the
molecular dimensions of information carriers like DNA and RNA),
permitting the creation of a huge redundancy in the number of elements.
Thus, we expect that for living organisms, in distinction to many
technical (manufactured) devices, the reliability of the system is
achieved not  by the high initial quality of all the elements but by
their huge numbers (redundancy).''} 
 Gavrilov (2001,p.531.)

 \subsection*{Aging Processes}

 In this section we follow Gavrilov (1981, 2001, 2006) to analyse the
aging process of some redundant series / parallel  reliability systems. 

 As usual in reliability theory, $t$ will denote failure time, $f(t)$
and $F(t)$ the density and cumulative distribution functions of the
failure time, $S(t)=1-F(t)$ the survival function and
 \[
   h(t)= \frac{d\; S(t)}{S(t) \; dt} = 
           \frac{d\; \ln S(t) }{dt} 
 \]
 the hazard function, failure rate, or mortality force, see Barlow and
Prochan (1981). 

 Simple, memoryless or non-aging components are characterized by
exponentially distributed failure time. In this case, the failure time
has constant hazard rate, $h(t)=\kappa$, and $S(t))=\exp(-\kappa t)$,
$\kappa,t\geq 0$. Complex systems are characterized by different aging
regimes which, in turn, reflect their structural characteristics. Two
aging regimes are of special interest to us: 

 1- The Weibull or power law regime, with $h(t)= \kappa t^\alpha$,
 $\kappa,\alpha>0$, characteristic of complex top-down, external assembly
 or alopoietic systems, and 

 2- The Gompertz-Mekeham regime, with $h(t)= A+ R\exp(\alpha t)$,
 $A,R,\alpha>0$, characteristic of complex bottom-up, self assembly or
autopoietc systems. In biological models, the Mekeham parameter, $A$,
indicates an external mortality force, whereas the pure Gompertz regime,
for $A=0$, models the internal or systemic hazard function.
 
 In what follows, we will see some structural models that explain these
two regimes and test them on some engineering and biological systems. 
 
 The two basic structures in reliability theory are parallel and series
compositions. Complex systems in general, are recursive compositions of
series and parallel blocks. A parallel block fails if all its components
fail, whereas a series block fails if any one of its components fail,
alternatively, a parallel block fails with its last failing component,
whereas a series block fails with its first failing component. Hence the
series-parallel reliability compositional rules:  

 - The cumulative distribution function of a parallel system with
independent components equals the product of its components' cumulative
distribution functions. 

 - The hazard function of a series system with independent components
equals the sum of its components' hazard functions. 

 Let us now consider the ``simplest complex system'' modeling an
organism or machine with multiple, $m$, functions, where each function
is performed by an independent block of redundant simple components.
That is, a system is assembled as a series of $m$ blocks, 
 $b_j, j=1\ldots m$, such that block $j$ is assembled as a parallel 
 (sub) system with $n_j$ simple components. 

 Top-down projects typically use a small number of redundant units, in
order to optimize production costs as well as to meet other project
constraints such as maximum space or weight. Hence, components have to
comply with strict standards, achieved by several forms of quality
control tasks in the manufacturing process. In such systems all
components are initially alive, operational or working, since they would
have been otherwise rejected by quality control. They are typically
depicted in block diagram of such as shown in Figure 1A. In this example
each block has the same number, $n_j=i$, of redundant components.

 Since each simple component has an exponential failure distribution,
the reliability compositional rules lead to the following systemic
hazard functions for each block,  
 \[
    F_j = \left( 1-e^{-\kappa t} \right)^i \ , \ \ 
    h_j(t) = \frac{ i\kappa e^{-\kappa t} 
                    \left( 1-e^{-\kappa t} \right)^{i-1} } 
                  {1- \left( 1-e^{-\kappa t} \right)^{i} } \ ;  
 \]
 and to the following systemic hazard function for the whole system,  
 \[
    h(t) = \ssum_{j=1}^m h_j (t) = 
   \frac{ m i \kappa e^{-\kappa t} 
           \left( 1-e^{-\kappa t} \right)^{i-1} } 
        {1- \left( 1-e^{-\kappa t} \right)^{i} } \ .   
 \]

 Using the early-life and late-life asymptotic approximations, 
 $1-\exp(-\kappa t)\approx \kappa t$, for 
 $t<< 1/\kappa$, and $1-\exp(-\kappa t)\approx 1$, for 
 $t>> 1/\kappa$, the $i$ elements parallel block and systemic hazard 
 functions can be approximated as 
 \[
   h_i(t) \approx \left\{ \begin{array}{ll} 
   i \kappa^i t^{i-1} & \mbox{if} \ t << 1/\kappa \ \ \mbox{and} \\ 
   \kappa & \mbox{if} \ t >> 1/\kappa \ ; 
   \end{array} \right. 
   \ , \ \ 
   h(t) \approx \left\{ \begin{array}{ll} 
   m i \kappa^i t^{i-1} & \mbox{if} \ t << 1/\kappa \ \ \mbox{and} \\ 
   m \kappa & \mbox{if} \ t >> 1/\kappa \ ; 
   \end{array} \right. 
   \ .  
 \]

 Let us now consider self-assembled blocks where the number $i$ of
initially working elements follows a Poisson distribution with parameter
$\lambda=nq$, \ $P(i)= \exp(-\lambda){\lambda^i}/{i!}$. We should also
truncate the Poisson distribution, to account for the facts that the
organism is initially alive, implying the exclusion of the $i=0$ case,
and that the organism is finite, implying a cut-off $\Pr(i>n)=0$. 
 The corrected normalization constant for this truncated Poisson is 
 $c^{-1}= 1 -\exp(-\lambda) -\exp(-\lambda) 
  \ssum_{i=n+1}^\infty \lambda^i / i!\ .$    

 As in the previous model, the systemic hazard function is the sum of
those of its blocks', where each block begins with $i$, Poisson
distributed, working elements. Hence, the expected systemic hazard
function can be written as:  
 \[
    h(t) = \ssum_{j=1}^m h_j(t) = m \ssum_{i=1}^n P(i) h_i(t) \ . 
 \]

 Substitution of $h_i(t)$ yields the following systemic hazard rate and
approximations:
 \[
   h(t) = c m \kappa \lambda e^{-\lambda} e^{-\kappa t} 
    \ssum_{i=1}^n 
    \frac{ \lambda^{i-1} \left( 1-e^{-\kappa t} \right)^{i-1} } 
    { (i-1)! \left( 1- \left( 1-e^{-\kappa t} \right)^{i} \right) } 
    \ ,  
 \]
 \[
  h(t) \approx \  \left\{ \begin{array}{ll} 
    c m \kappa \lambda e^{-\lambda} \ssum_{i=1}^{n} 
    \frac{ (\kappa \lambda t )^{i-1} }{ (i-1)! } = 
    R\left( e^{\alpha t} -\epsilon(t) \right) 
    & \mbox{if} \ t << 1/\kappa \ \ \mbox{and} \\ 
     m \kappa & \mbox{if} \ t >> 1/\kappa \ ; 
   \end{array} \right. 
   \ .    
 \]

 In the last expression, $R= c m \kappa \lambda \exp(-\lambda)$,
 $\alpha= \kappa \lambda$, and 
 $\epsilon(t)= \ssum_{i=n+1}^\infty (\kappa \lambda t)^{i-1}/(i-1)!.$ 
 For for fixed $\kappa$ and $\lambda$ and sufficiently small $t$,
 $\epsilon(t)$ is close to zero. Hence, in early life, 
 $h(t) \approx R \exp(\alpha t)$, as in the pure Gompertz regime.

  \vfill 

 \pagebreak 

%% file: CAPRES.TEX
\chapter{Research Projects}

 In the last courses we have had classes of very diverse students. 
 As expected, we had students coming from the courses of Applied
Mathematics, Physics and, of course, Statistics, but we also had some
students with quite different backgrounds, such as Computer Science,
Economics, Law, Logic and Philosophy. This appendix proposes some
research projects that may be specially interesting to some of these
students. I do believe, of course, that most of them will also be
interesting to the student of Statistics. If you are interested in one
of these projects, send me an e-mail, or stop by at my office, an let us
talk about how to proceed.

 \subsubsection*{Bayesian and other Credal Networks}  

 The sparse factorization techniques described in Appendix F can be
transposed to Bayesian Networks and other belief propagation networks as
well. 

 1- Symbolic phase: Implement the algorithms used to find a good
elimination order, like the Gibbs heuristics, the Bayes-ball algorithm,
and the other graph algorithms mentioned in appendix F. A language such
as $C$ or $C++$, providing good support for dynamic data structures, is
recommended.   
  
 2- Numeric phase: Once the elimination order, requisite variables, etc.
are determined, implement the numerical elimination process using static
data structures. A language such as Fortran, providing good support for  
automatic parallelization, is recommended. 

 3- Investigate the potential for parallelization of the sequential
codes implemented in steps 1 and 2. Discuss the possibility,
difficulties and advantages of developing tailor made parallel code
versus the use of automatic parallelization tools.    
 
 4- Implement efficient MC or MCMC processes for computing the evidence
supporting the existence of a given causal link, that is, the existence
of a given arrow in (a) a given Bayes network topology (b) all or a given 
subset of topolgies.

 \subsubsection*{Mixture of Factor Analyzers}  
  
 Extend the theory and methods for Mixtures of Multivariate Gaussians,
as described in appendix B, to Mixtures of FA's. The geometric
interpretation of these models is very similar, but whereas all Mixtures
of Gaussians lie in the same $d$-dimensional space, each Mixture of FA's
lies in a different hyperplane of the full $d$-dimensional space. In
particular: \\ 
 a) Test the existence of a given component in the mixture. \\ 
 b) Test the existence of the least significative dimension of a given
component.

 \subsubsection*{Polynomial Networks}  

 1- Discuss the use of edge annotations and heuristic merit functions 
  in the synthesis of sub-networks, that is, the use of 
  heuristic ``recombinative guidance'', in the terminology of 
  Nikolaev and Iba (2001, 2006).   

 2- Discuss the use of time dependent objective functions, 
    as is section 5.2,  to guide the synthesis of the entire network. 

 3- Discuss how to test (sub) topologies of a given network.

 \subsubsection*{(De)Coupling, (De)Composition, Complementarity}  

 1- Discuss the possibility of using complementary models in contexts 
other than Quantum Mechanics. Give examples of such applications. 

 2- Discuss the possibility of extending the results of Borges and Stern
(2007) to models with limited dependence using, for example, the
formalism of Copulas.

 3- Investigate the meaning and interpretation of  decoupling or 
separation schemes generated by alternative sparse and/or structured 
matrix factorizations. 

 4- Using wavelet or other self-similar representations, it is possible 
to overcome the strict version of Heisenberg uncertainty relation, see
Vidakovic (1999, p.xxx). 
 However, these representations may introduce non-local, delayed,
integral, long-rage, long-memory or other forms of coupling or dependence. 
 Investigate how to obtain generalized Heisenberg type relations for such
cases.

 5- Give suitable interpretations and implement statistical models 
 for the ``necessary or consequential randomness'' implied in the 
 following examples:  

 5a- Morgenstern and von Neumann (1947) and Nash (1951), proved the 
 existence of equilibrium strategies for non-cooperative games. 
 However, in general, these equilibria are not at deterministic or 
 {\it pure} strategies, but at randomized or {\it mixed} strategies.  

 5b- The concept of Impossible (or Inconsistent, or Unholy) Trinity,  
 also known as the Mundell-Fleming trilemma, 
 is an hypothesis of international economics, stating the 
 impossibility of achieving simultaneously the follong goals: 
 1- fixed exchange rate; 2- free capital movement; and  
 3- independent monetary policy.

 \subsubsection*{Economics} 

 The economic system may be characterized by eigen-solutions, equilibria
or fixed points resulting from the collective interaction of many
economic agents. Some of the most important of such eigen-values are
prices, see for example Ingrao and Israel (1990). 
 
 1- Give concrete examples of such situations, that are well suited for
experimental research. 

 2- Discuss how to measure the epistemic value of such an economic or
finacial eigen-value. 

 3- Discuss how to assess the consistency of such eigen-values, for
example, by means of sensitivity analyses.   
   
 4- Discuss the need for regulatory mechanisms protecting such
eigen-solutions such as, for example, anti-trust laws.

 5- Discuss the consequences of Zangwill's global convergence theorem 
 to the design of good regulatory policies; see, for example, 
 Border (1989), Ingrao and Israel (1990) and Zangwill (1964).

 \subsubsection*{Law} 
 
 The Objective / Subjective dichotomy manifests itself in the legal
arena via the notion of responsibility. Responsibility may require
either two or three conditions, namely. 

 a) Damage: A loss suffered by the victim (or offended party). 
 
 b) Causal relation: A causal nexus linking an action (or lack thereof)
of the accused (or defendant, offending party, perpetrator) to the
damage suffered by the victim.  

 c) Illicitness: An explanation why the action (or lack thereof) of the
accused was illegal or unlawful.

 While the programs and codes (in Luhmann's sense) needed for checking
condition (c) are internal ones, that is, programs and codes within the
legal system itself, the programs and codes needed for checking
conditions (a) and (b) are often external, that is, programs and codes
of another systems, such as science or economics, for example. 

 Hence, it is not surprising that a responsibility entailed by
conditions (a) and (b) alone is called ``objective'', while one
requiring conditions (a), (b) and (c) is called ``subjective'', see
Stern (2007). 

 R.B.Stern (2007) suggests the following principle, hereby named
``Transference of Objectivity'' (TrOb), for systems characterized by the
existence of eigen-solutions resulting from complex collective
interactions: 

 If an individual agent (or a small group of agents) in the system
disrupts such an eigen-solution, hence destroying its objective
character, then this agent becomes, in the same measure, objectively
responsible for consequential damages caused by the disruption. 

 1- Discuss the plausibility of the TrOb principle.  

 2- Discuss possible justifications for the TrOb principle.  

 3- Discuss the applicability of the TrOb principle in: \\ 
 a) Economic law;  b) Environmental law.

 5- Discuss the applicability of TrOb for state actions.  

 5- Discuss the applicability of TrOb for lost revenues. 

 6- Discuss the applicability of TrOb for the loss of a chance.

 \subsubsection*{Experiment Design and Philosophy} 

 1) Discuss the possibility of conciliating the objective inference
entailed by randomization methods with biased allocation and selection
procedures. 

 2) Discuss the possibility of using optimal selections or allocations
obtained by Multi-Objective or Goal Programming, where some (fake or
artificial) explaining variables have randomly generated values. 

 3) Discuss the possibility of using low-discrepancy selections or 
allocations obtained by quasi-random or hybrid (scrambled quasi-random) 
lattices.   

 4) How can we corroborate the objective character of such inference
procedures? For example, what is the importance of sensitivity analyses
in these allocation?   

 5) What kind of protocols are appropriate for such inference
procedures? 

 6) What criteria can be used in balancing the epistemic value of a
clinical study versus the well being of the participants? 
 What kind of moral, ethical and legal arguments can be used to 
support these criteria?

 \subsubsection*{Art} 

 Make your contribution to the Art Gallery.


%% file: CAPARTA.TEX
 
 \chapter{Image and Art Gallery} 
 
 \mbox{} \vspace{-1.3cm} \mbox{}  



  The images in this gallery are somehow related to topics discussed  
 in the main text -- 
  {\it Cognitive Constructivism and the Epistemic Significance of 
        Sharp Statistical Hypotheses.}
  These images are provided with no fixed or definite interpretation,  
 and are only meant as a stimulus to imagination and creativity. 
  Paraphrasing aphorisms of Fernando Pessoa: \vspace{1mm} \\  
  {\it \mbox{} -- There is no good science that is vague, 
       nor good art or poetry that is not. }

   and also: \vspace{1mm} \\ 
   {\it \mbox{} -- Good navigation is precise, 
                   living in never so  precise.}


 \begin{figure}[!h]
  \centerline{\includegraphics*[height=4.0in]{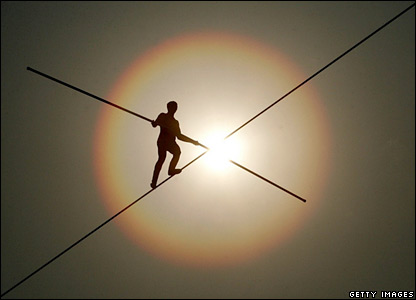}} 
  \centerline{Figure JA.1: Wire Walking.} 
  \centerline{The most important thing is not to fear at all.} 
 \end{figure}

 \pagebreak 


 \begin{figure}[!h]
  \centerline{\includegraphics*[height=3.5in]{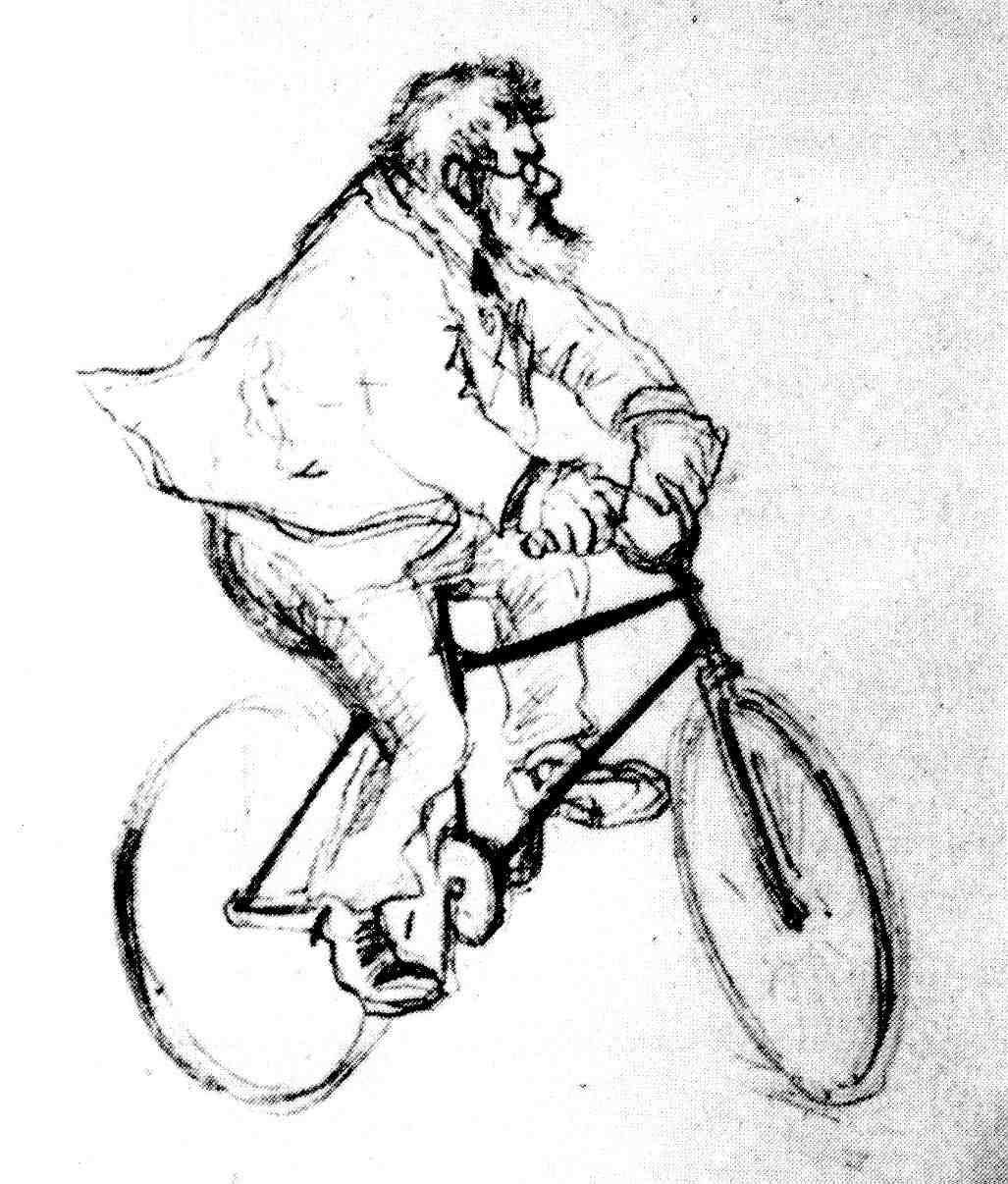}} 
  \centerline{Figure JA.2: Ludwig Boltzmann, Cartoon by K.Przibram.} 
  \centerline{Moving ahead, no matter what.} 
 \end{figure}

 \mbox{} 

 \begin{figure}[!b]
  \centerline{\includegraphics*[height=3.5in]{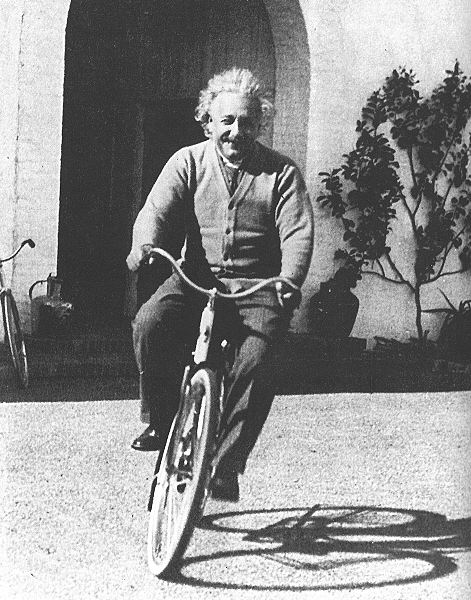}} 
  \centerline{Figure JA.3: Albert Einstein in his Bicycle.} 
  \centerline{Following the gentle curvature of the garden's geometry.}  
 \end{figure}

 \pagebreak 


 \begin{figure}[!h]
  \centerline{\includegraphics*[height=3.5in,  width=2.7in,  
           viewport=1 70 240 360 , clip]{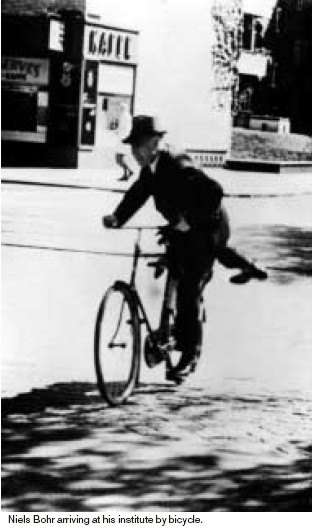}} 
  \centerline{Figure JA.4: Niels Bohr in his Bicycle.} 
  \centerline{Complementary pedals must be pushed one at a time.}  
 \end{figure}

 \mbox{} 

 \begin{figure}[!b]
  \centerline{\includegraphics*[height=3.5in]{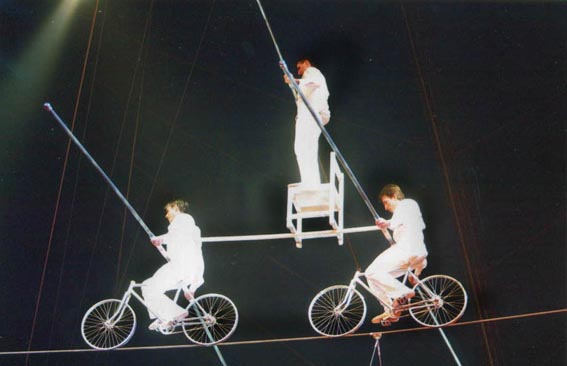}} 
  \centerline{Figure JA.5: Empirical Science: All at Once!} 
  \centerline{Caution: Do this only at a fully equipped laboratory.} 
 \end{figure}

  \pagebreak 

 \begin{figure}[!t]
 \centerline{\includegraphics*[angle=0, height=7.5in]{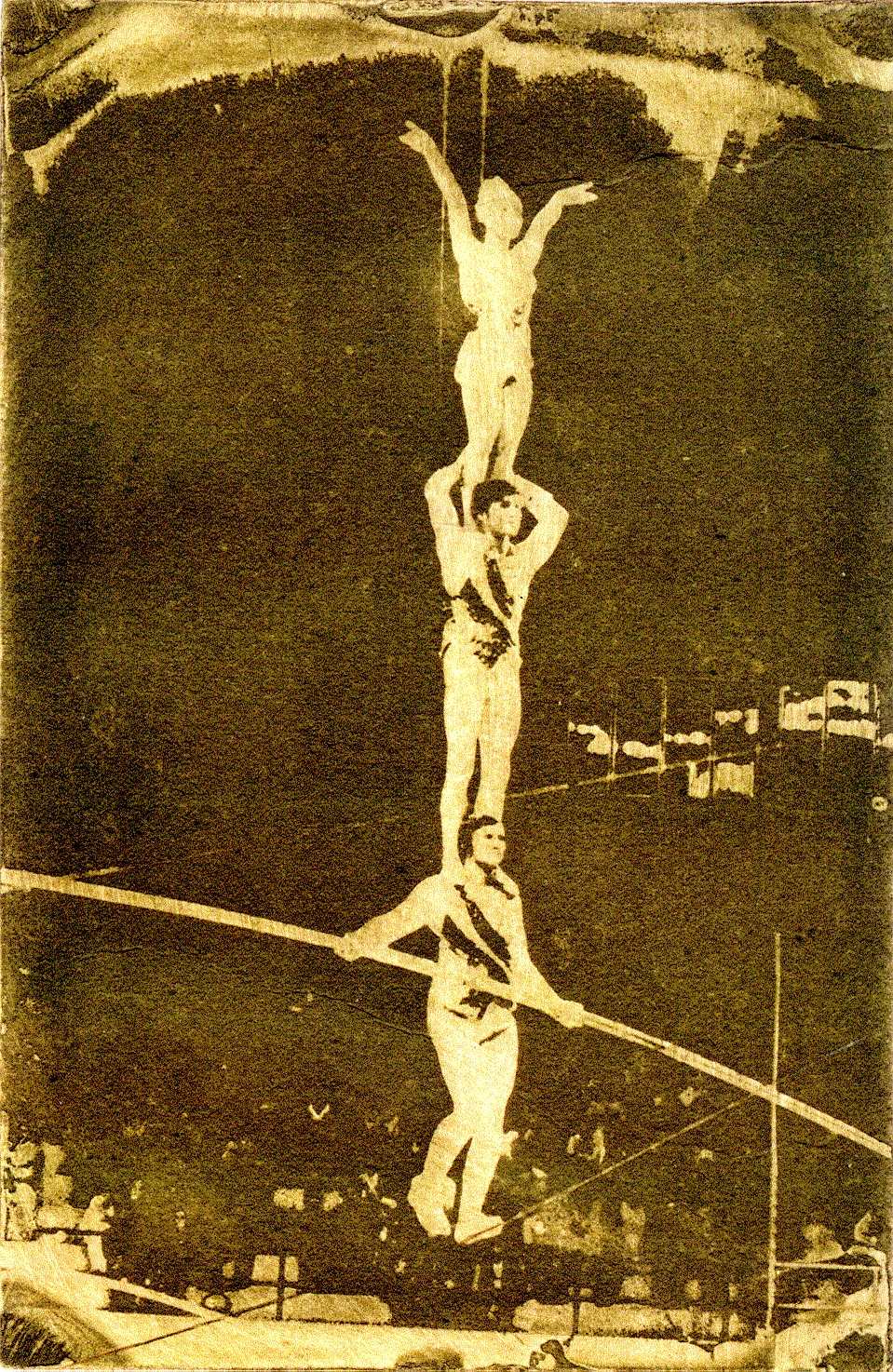}} 
 \centerline{Figure JA.6: Triadic (or Semiotic) Wire Walking.} 
 \centerline{Etching by Alex Flemming (untitled, 1979, PA III/X),} 
 \centerline{based on a photo of the Moscow Circus at S\~{a}o Paulo.}  
 \centerline{Private collection of Marisa Bassi Stern.} 
 \end{figure}


 \vfill 

 \mbox{} 

 \pagebreak

 \subsection*{Additional Contributions to the Art Gallery}

 \begin{figure}[!h]
  \centerline{\includegraphics*[height=2.7in,  width=4.0in,  
           viewport=10 1 180 115 , clip]{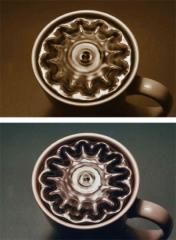}} 
  \centerline{Figure JB.1: Coffee a la MODE.}  
  \centerline{Vibration Eigen-Solutions on a Liquid Surface.}  
  \centerline{Published at The Science Creative Quarterly, Sept. 2006.}
 \end{figure}

 \mbox{} 

 \begin{figure}[!h]
  \centerline{\includegraphics*[height=3.8in]{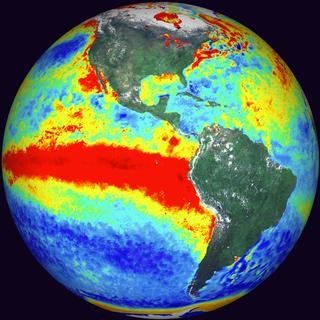}} 
  \center{Figure JB.2: Planet a la MODE.} 
  \centerline{Sea Surface Temperature at the 1988  El Ni\~{n}o Event.}  
  \centerline{NASA - Scientific Visualization Studio.} 
 \end{figure}

  \pagebreak 






 \input{ARVIDA3.TEX} 



 \pagebreak

 \begin{figure}[!h]
  \centerline{\includegraphics*[width=2.5in]{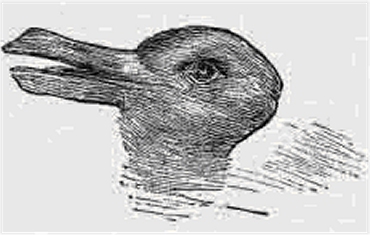}
   \mbox{} \includegraphics*[width=2.5in]{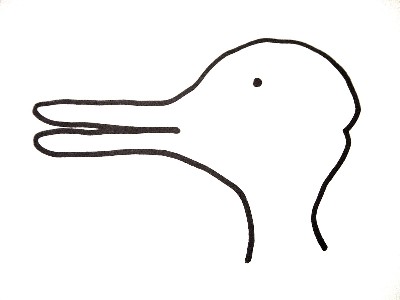} }  
  \centerline{\mbox{}} 
  \centerline{\includegraphics*[width=2.3in, height=2.0in]{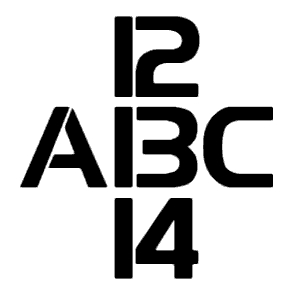}
   \mbox{} \ \ \includegraphics*[width=2.5in, height=2.0in]{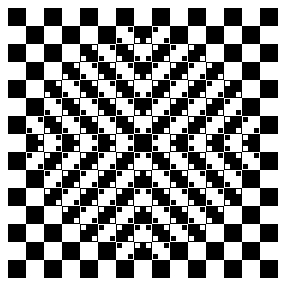} }  
  \centerline{\mbox{}} 
  \centerline{\includegraphics*[width=1.0in]{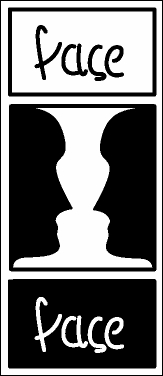}
   \mbox{} \ \ \includegraphics*[width=3.5in, height=2.3in]{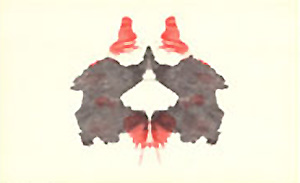} }  
  \centerline{\mbox{} \vspace{0.5cm} \mbox{} } 
   \centerline{\Huge \mbox{} \hspace{2cm} X \hspace{6cm} $\bullet$ \hspace{2cm} \mbox{} } 
  \centerline{\mbox{} \vspace{0.5cm} \mbox{} } 
  \centerline{Figures JB.4: Ambigrams, Optical Illusions, and all that,} 
  \centerline{Including the Duck-Rabbit, by Jastrow and Wittgenstein,}
  \centerline{an Inkblot of the Rorschach Personality Test,}
  \centerline{and von Foerster's Blind Spot Finder.}
 \end{figure} 

 \pagebreak

 \begin{figure}[!h]
  \centerline{
   \includegraphics*[height=3.0in, width=3.5in]{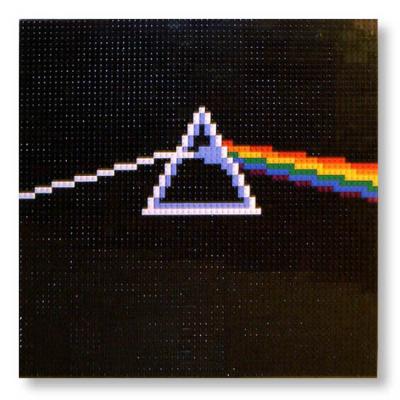}}
  \centerline{Figure JB.5: A Spectrum from LeGogh.com Gallery.} 
 \end{figure}

 \begin{figure}[!h]
  \centerline{
    \includegraphics*[height=2.0in, width=1.6in]{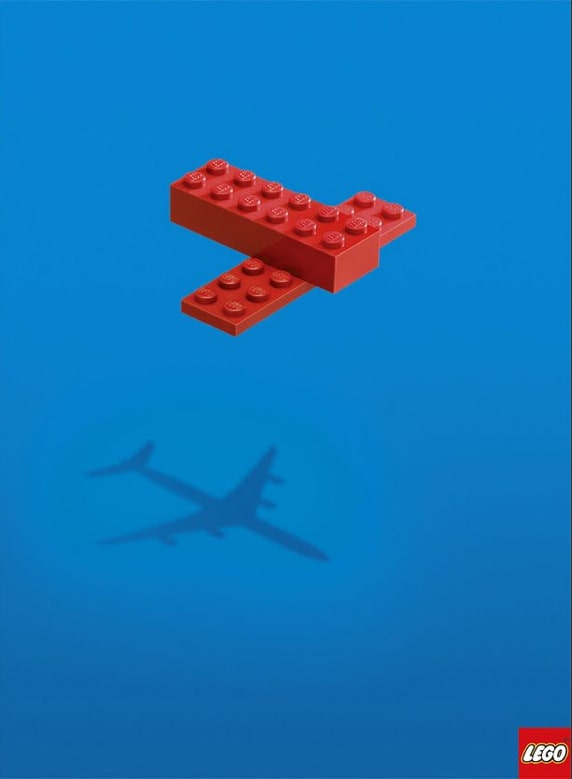}
    \mbox{} 
    \includegraphics*[height=2.0in, width=1.6in]{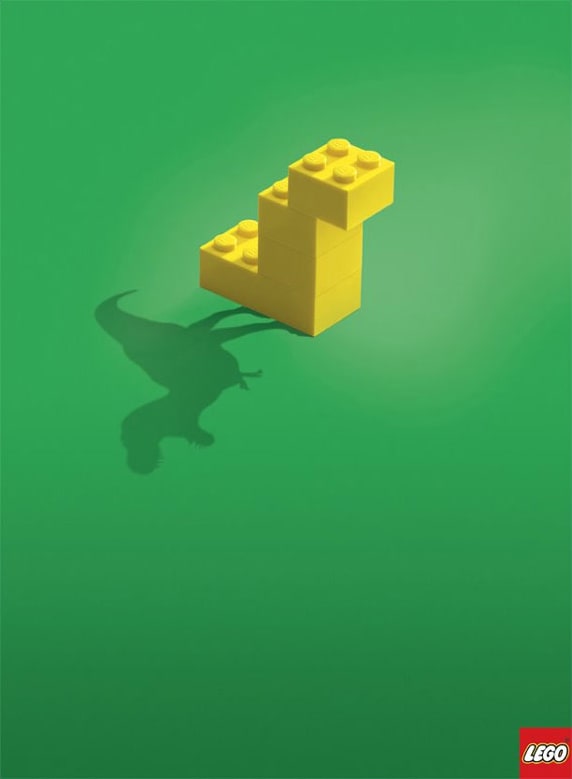}     
   }
 \centerline{\mbox{}}  
 \centerline{\includegraphics*[height=2.0in, width=3.2in]{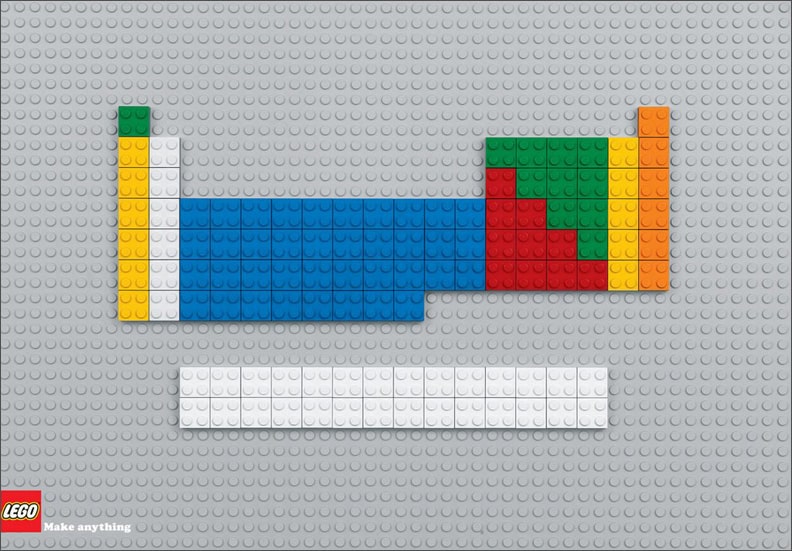}}  
 \centerline{\mbox{}}  
 \centerline{Figure JB.6: Lego Ads by FCB Johannesburg.}  
 \end{figure}

 \pagebreak

 \begin{figure}[!h]
  \centerline{
    \includegraphics*[height=3.5in, width=3.0in]{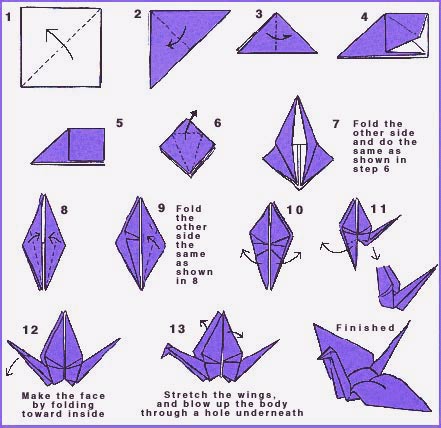}
    \mbox{} 
    \includegraphics*[height=3.5in, width=3.0in]{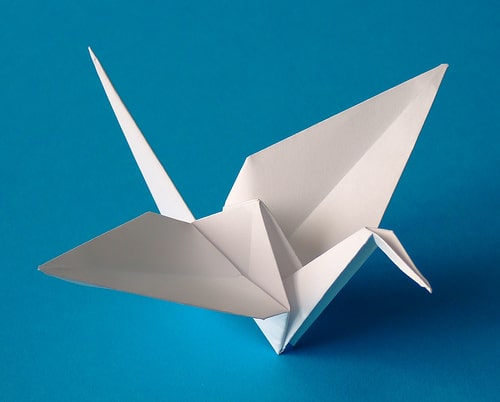}     
   }
 \centerline{\mbox{}}  
  \centerline{
    \includegraphics*[height=3.5in, width=3.0in]{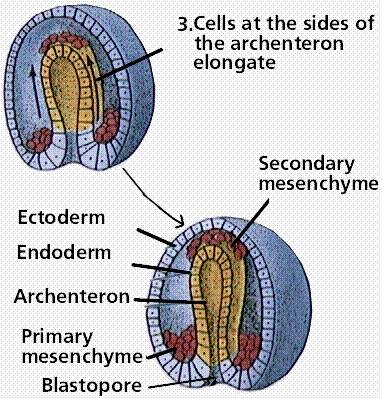}
    \mbox{} 
    \includegraphics*[height=3.5in, width=3.0in]{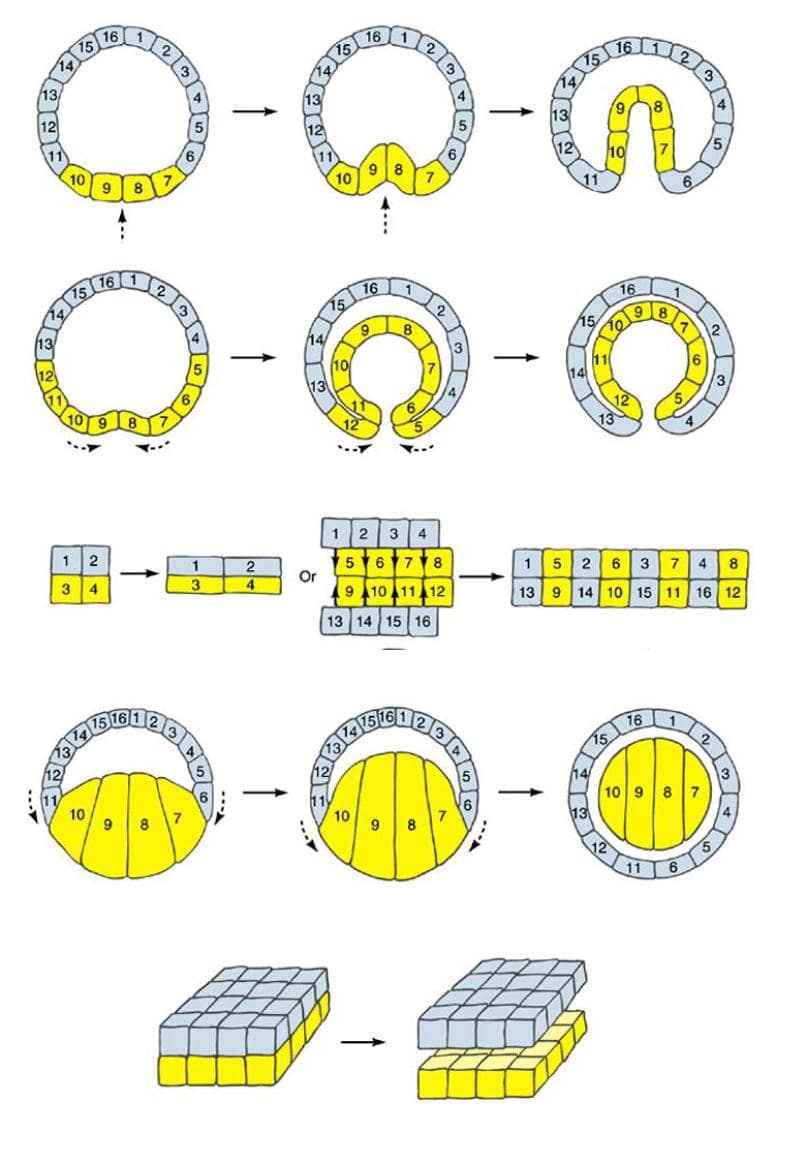}     
   }
 \centerline{\mbox{}}  
 \centerline{Figure JB.7: Foldings. A,B: Origami crane.  
  Organic morphogenesis. C: Gastrulation;} 
   \centerline{D: Tissue movements: Invagination, involution, 
  convergent extension, epiboly, delamination.}  
 \end{figure}

 \pagebreak

 \begin{figure}[!h]
  \centerline{
    \includegraphics*[height=2.0in, width=2.0in]{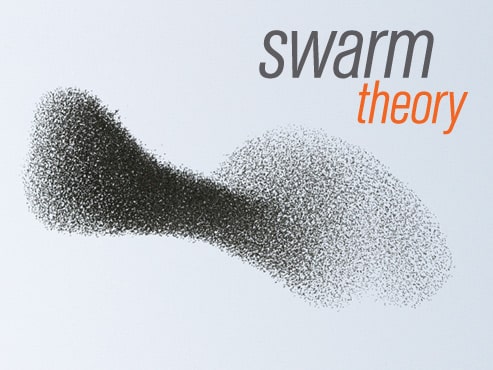}
    \mbox{} 
    \includegraphics*[height=2.0in, width=2.0in]{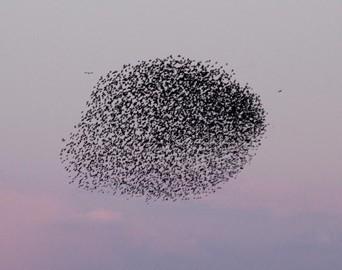}     
   }
 \centerline{\mbox{}}  
  \centerline{
    \includegraphics*[height=2.0in, width=2.0in]{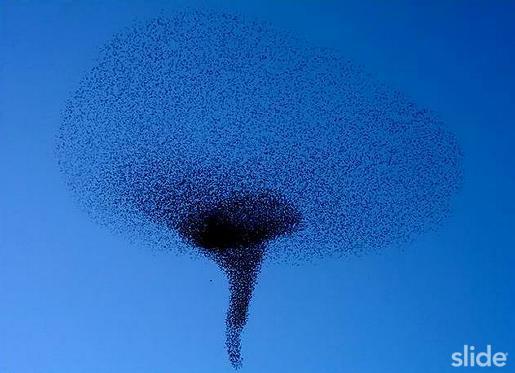}
    \mbox{} 
    \includegraphics*[height=2.0in, width=2.0in]{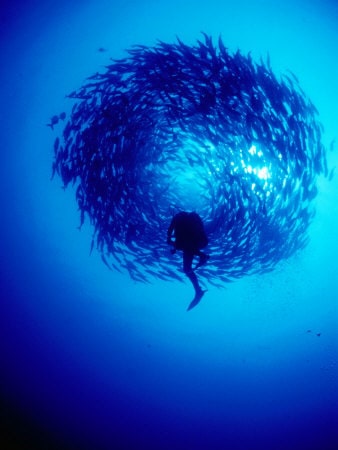}     
   }
 \centerline{\mbox{}}  
  \centerline{
    \includegraphics*[height=2.0in, width=2.0in]{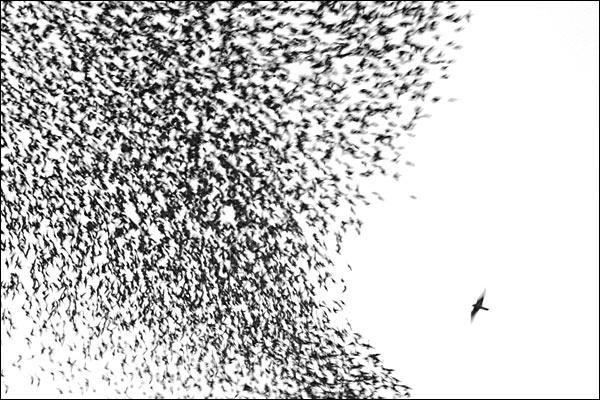}
    \mbox{} 
    \includegraphics*[height=2.0in, width=2.0in]{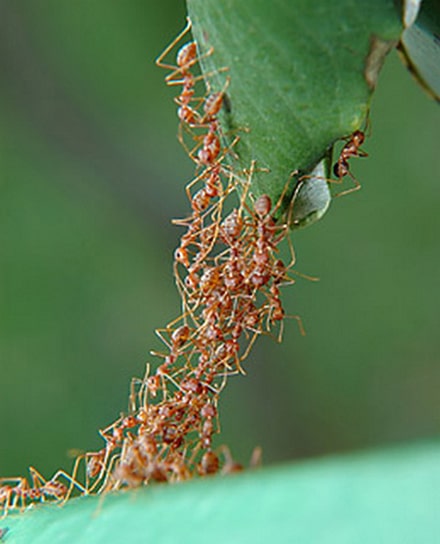}     
   }
 \centerline{\mbox{}}  
 \centerline{Figure JB.8: Flocks, Schools and Swarms.}  
 \end{figure}

 \pagebreak

 \centerline{
    \includegraphics*[height=2.7in, width=3.0in]{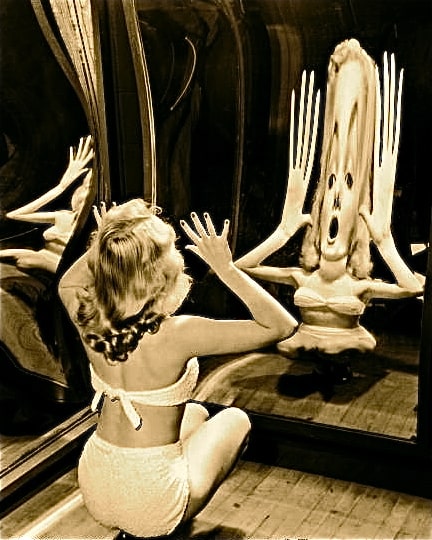}}
 \centerline{\mbox{}}  
  \centerline{
    \includegraphics*[height=3.5in, width=4.5in, 
      viewport=0 150 1280 960, clip]{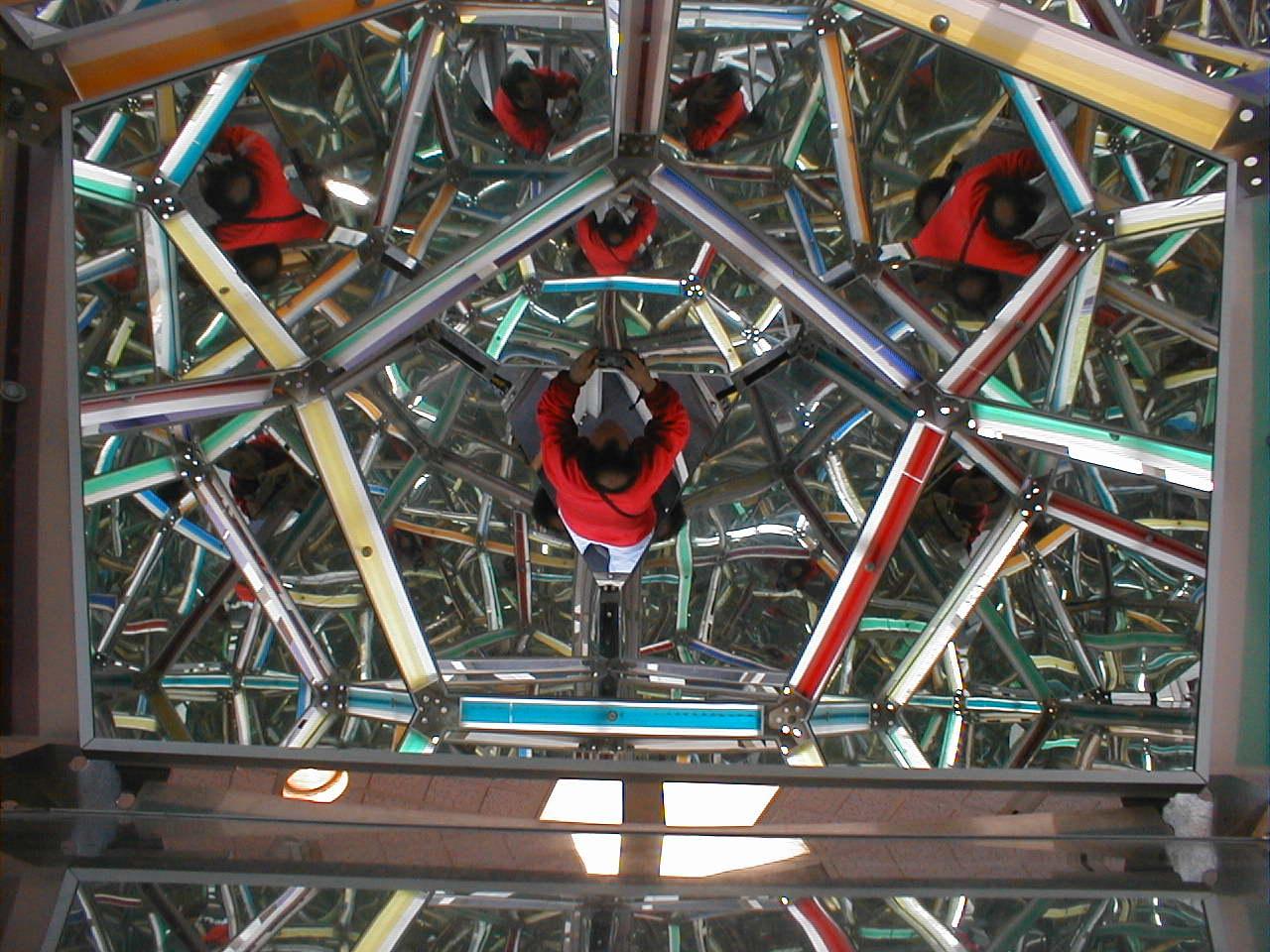}}  
 \centerline{\mbox{}}  
 \centerline{ \includegraphics*[height=1.3in,  width=5.0in,  
           viewport=0 0 640 160 , clip]{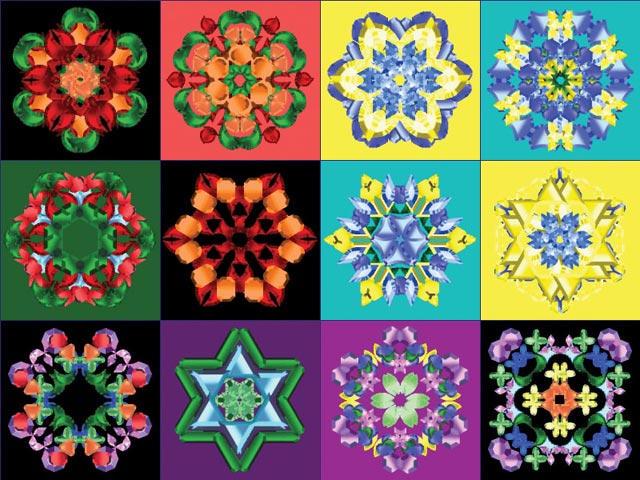}} 
 \centerline{\mbox{}}  
 \centerline{Figure JB.9: Kaleidoscopes and Mirror Houses.}

 \pagebreak

 \begin{figure}[!h]
  \centerline{\includegraphics*[width=2.5in, height=2.5in]{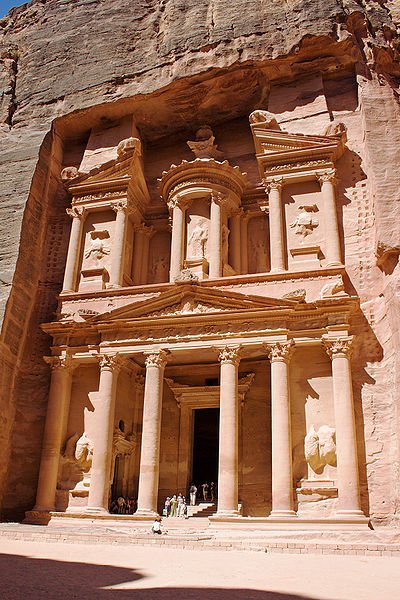}
   \mbox{} \includegraphics*[width=2.5in, height=2.5in]{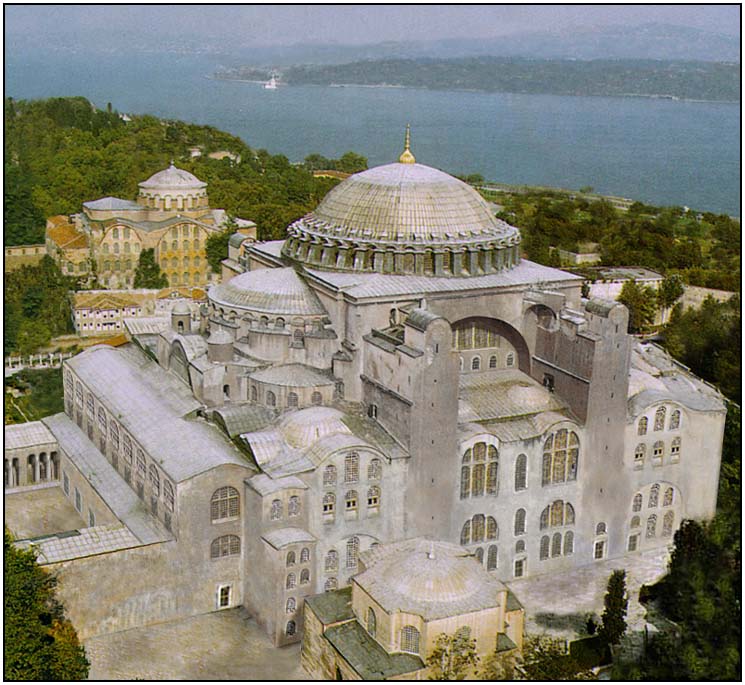} }  
  \centerline{\mbox{}} 
  \centerline{\includegraphics*[width=2.7in, height=2.0in]{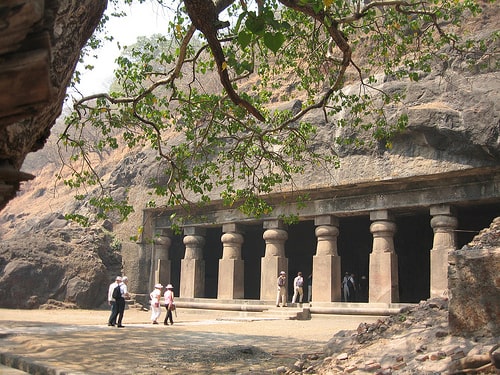} 
   \mbox{} \ \ \includegraphics*[width=2.7in, height=2.0in]{GIZA2.JPG} } 
  \centerline{\mbox{}}  
  \centerline{\includegraphics*[width=3.0in, height=2.0in]{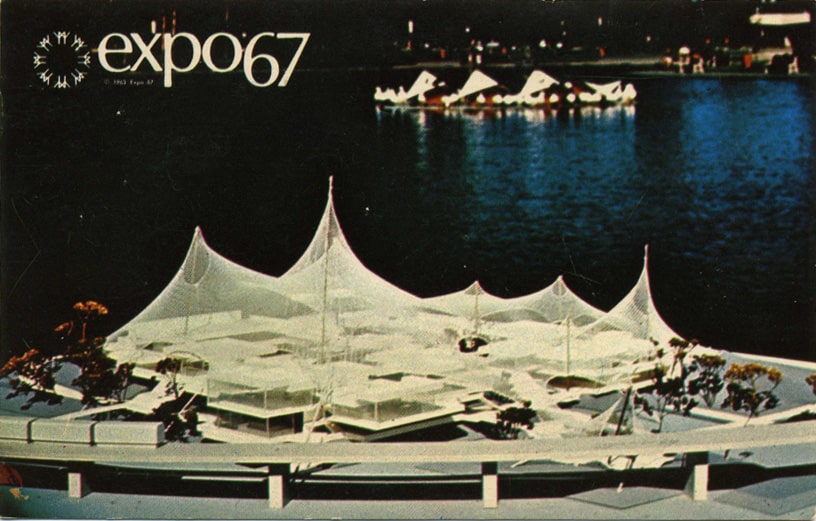}
   \mbox{} \ \ \includegraphics*[width=3.0in, height=2.0in]{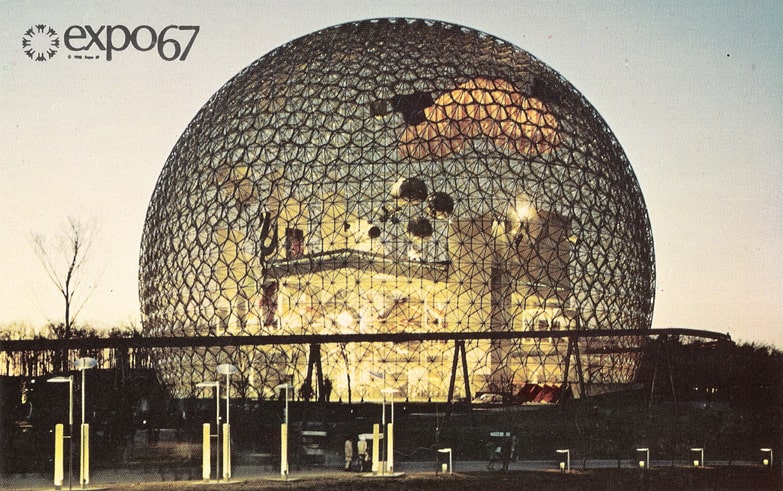} }  
  \centerline{\mbox{}} 
  \centerline{Figures JB.10: Architectural masterpieces.} 
  \centerline{Petra and Elephanta: Carved in solid rock.}
  \centerline{Giza and Hagia-Sophia: Stone building blocks.} 
  \centerline{expo67: New tensile constructive elements.} 

 \end{figure}

 \pagebreak

 \begin{figure}[!h]
  \centerline{ 
   \includegraphics*[height=2.5in, width=3.5in,  
         viewport=60 50 740 550 , clip]{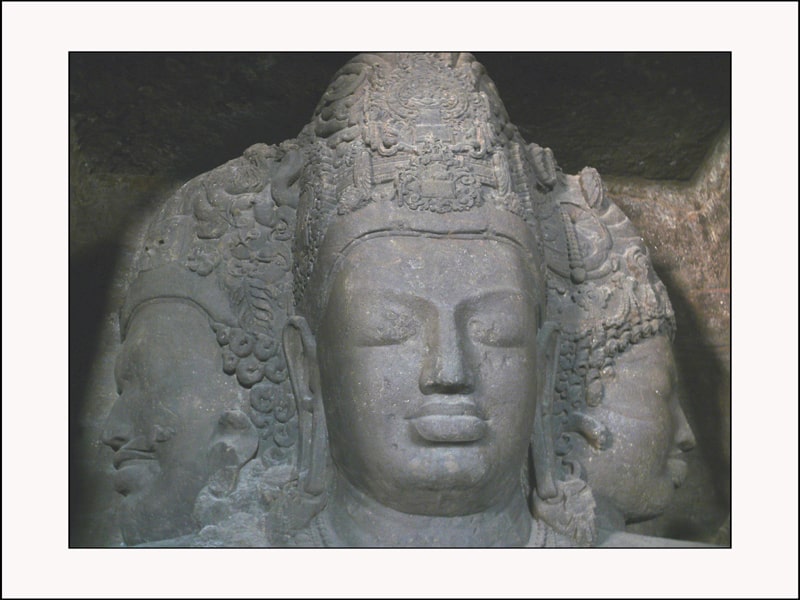}} 
  \centerline{Figure JB.11: Main sculpture at the temple of Elephanta.}       
 \centerline{The mask of eternity, in space and time; The mask of existence,}
 \centerline{as wave and particles; The mask of ethics, between good and evil;} 
 \centerline{The mask of evolution, continuous but modular,  
             from chaos to cosmos.} 
  \centerline{\mbox{}}  
  \centerline{ 
    \includegraphics*[height=4.7in, width=4.5in]{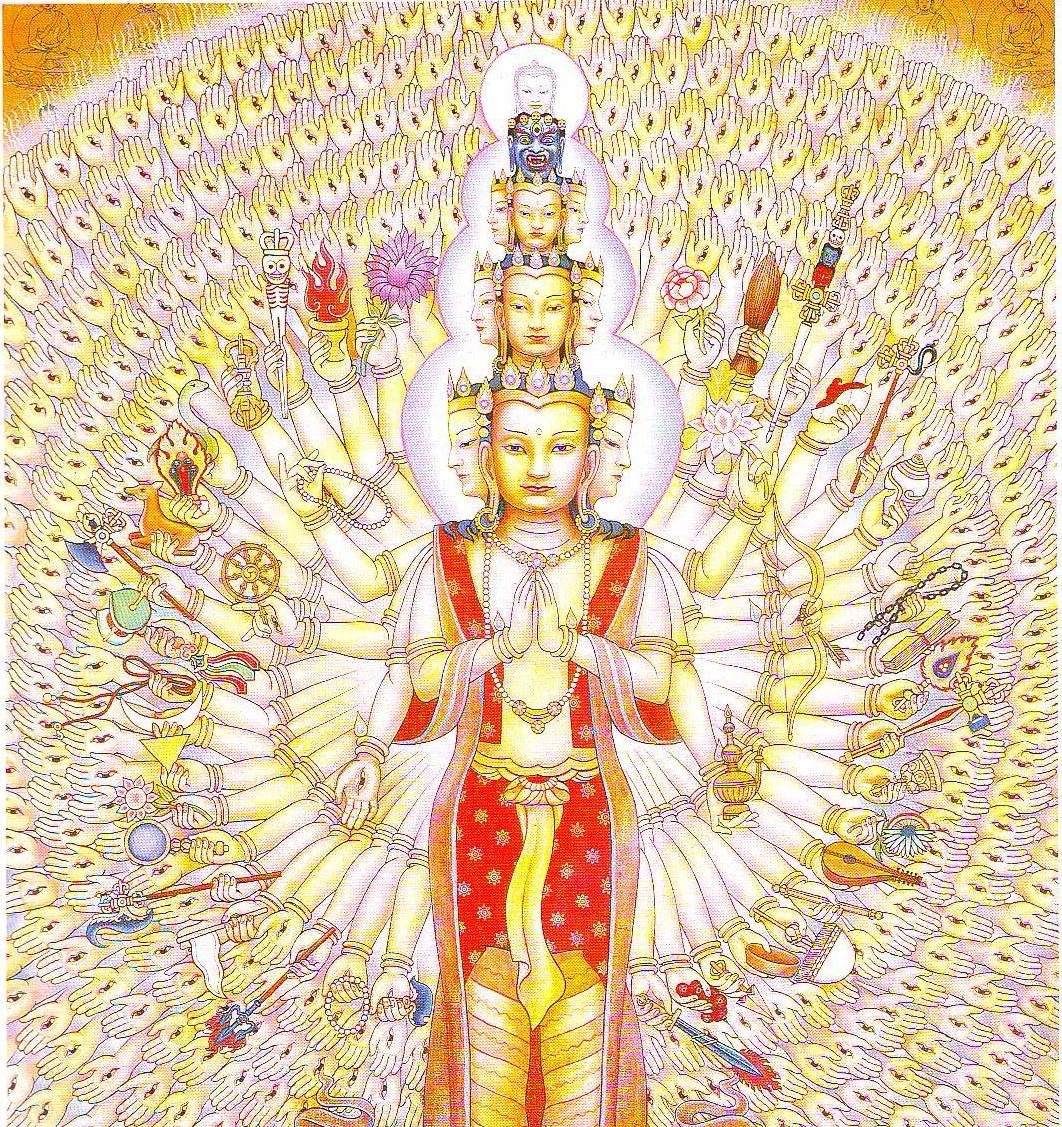}} 
  \centerline{Figure JB.12: Avalokitesvara by Alex Gray.}    
  \centerline{Making and overcomming conceptual distinctions.} 
 \end{figure} 

 \pagebreak


 %


 \subsubsection*{Cartoons} 

 \begin{figure}[!h]
  \centerline{\mbox{}} 
  \centerline{\includegraphics*[width=5.5in]{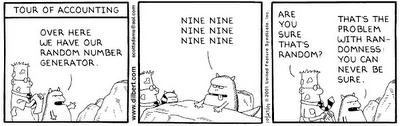}}  
  \centerline{\mbox{}} 
   \centerline{\includegraphics*[height=3.0in]{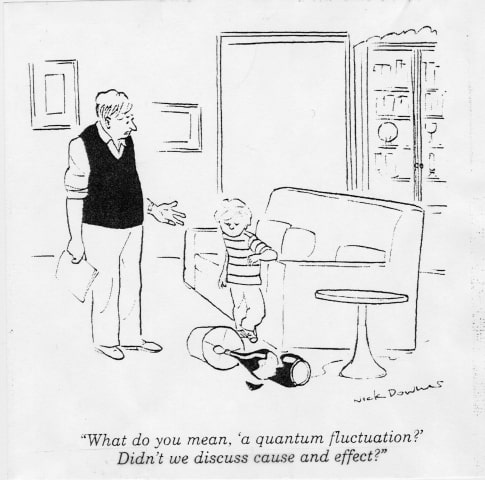} }  
  \centerline{\mbox{}} 
  \centerline{
  \includegraphics*[height=1.5in, width=1.5in]{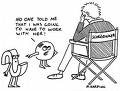}
   \mbox{}  
  \includegraphics*[height=2.0in, width=1.5in, 
      viewport= 10 5 310 350, clip]{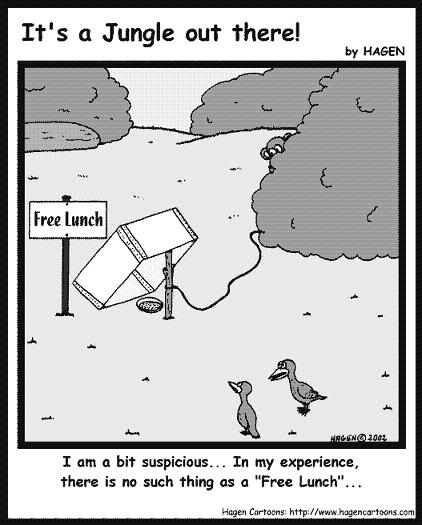}  
  \includegraphics*[height=2.0in, width=2.0in, 
      viewport= 10 0 400 280, clip]{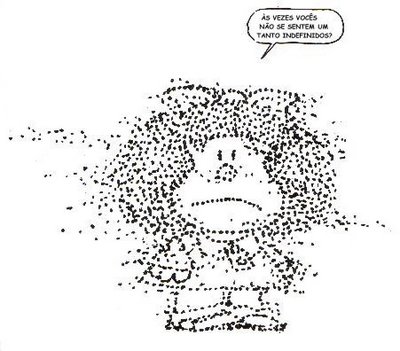} }  
  \centerline{\mbox{}}  
  \centerline{Figure JB.13: Probabilistic Cartoons.} 
 \end{figure}

  \pagebreak 
 
 \begin{figure}[!h]
   \centerline{\mbox{}} 
   \centerline{\includegraphics*[width=5.5in]{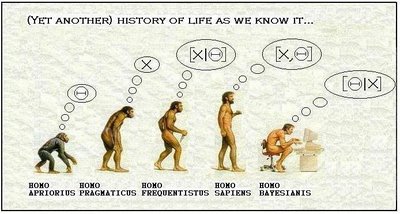}}  
   \centerline{\mbox{}} 
   \centerline{\includegraphics*[width=5.5in]{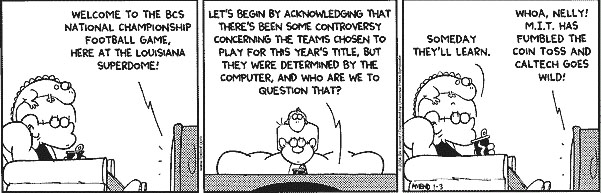}}  
   \centerline{\mbox{}} 
  \centerline{\includegraphics*[width=5.5in]{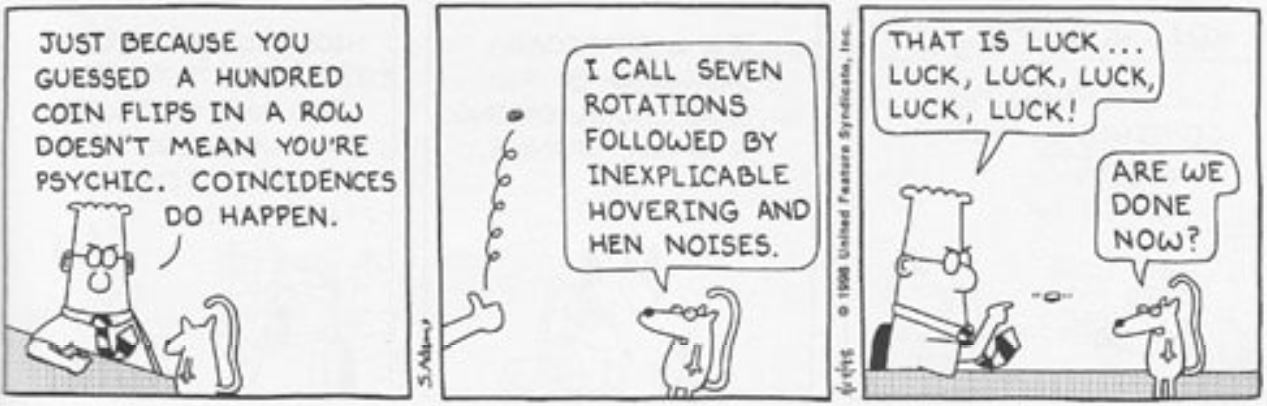}}  
   \centerline{\mbox{}} 
  \centerline{Figure JB.14: Statistical Cartoons.} 
 \end{figure}

  \pagebreak 

   \centerline{\mbox{}} 
   \centerline{\includegraphics*[height=2.5in,width=5.5in]{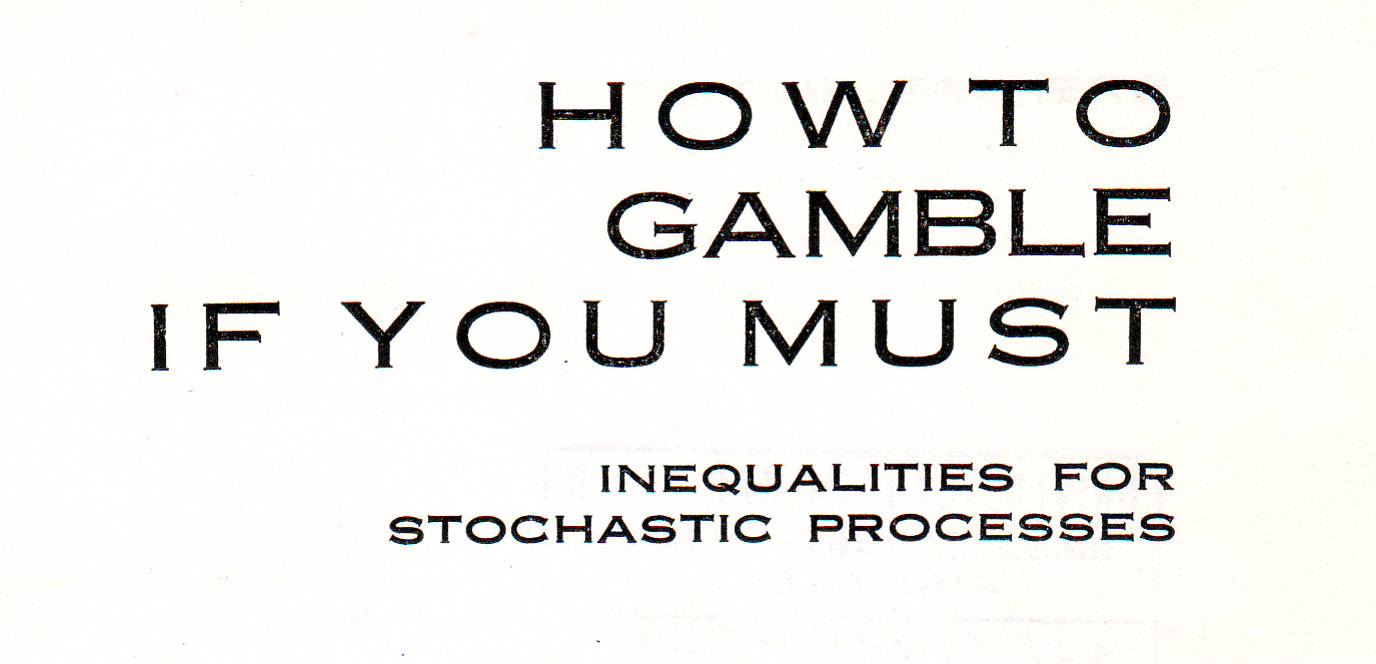}}  
   \centerline{\mbox{}} 
   \centerline{\includegraphics*[height=2.5in,width=5.5in]{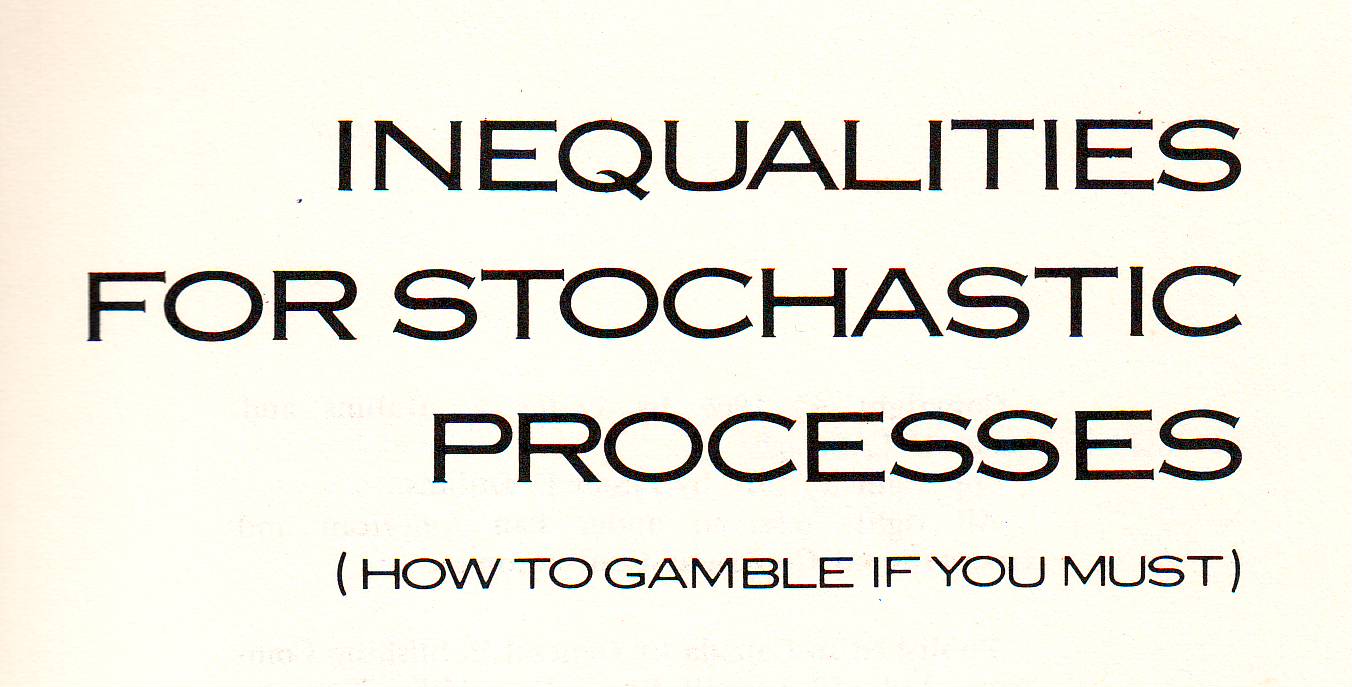}}  
   \centerline{\mbox{}} 
   \centerline{Figure JB.15: Covers of Dubins and Savage book, 1965, 1976.}   
 
 \vspace{0cm} 

 \begin{quote} 

 \mbox{} \ ``In order to alleviate the present publisher's concerns 
 about possible misunderstandings as to the nature of the book, its 
 title and subtitle have been permuted for this edition.'' (p.v). \\ 
 \mbox{} \ ``Gambling problems in which the distributions of various
 quantities are prominent in the description of the gambler's fortune
 seem to embrace the whole of theoretical statistics according to one
 view (which might be called the decision-theoretic Bayesian view) of
 the subject...  \\ 
 \mbox{} \  From the point of view of decision-theoretic statistics, 
 the gambler in this problem is a person who must ultimately act in 
 one of two ways (the two guesses), one of which would be appropriate 
 under one hypothesis and the other under its negation.'' 
 (sec.12.8, p.229,230). 

 \end{quote} 

  \pagebreak

 \begin{figure}[!h]
   \centerline{\mbox{}} 
   \centerline{
     \includegraphics*[width=2.5in]{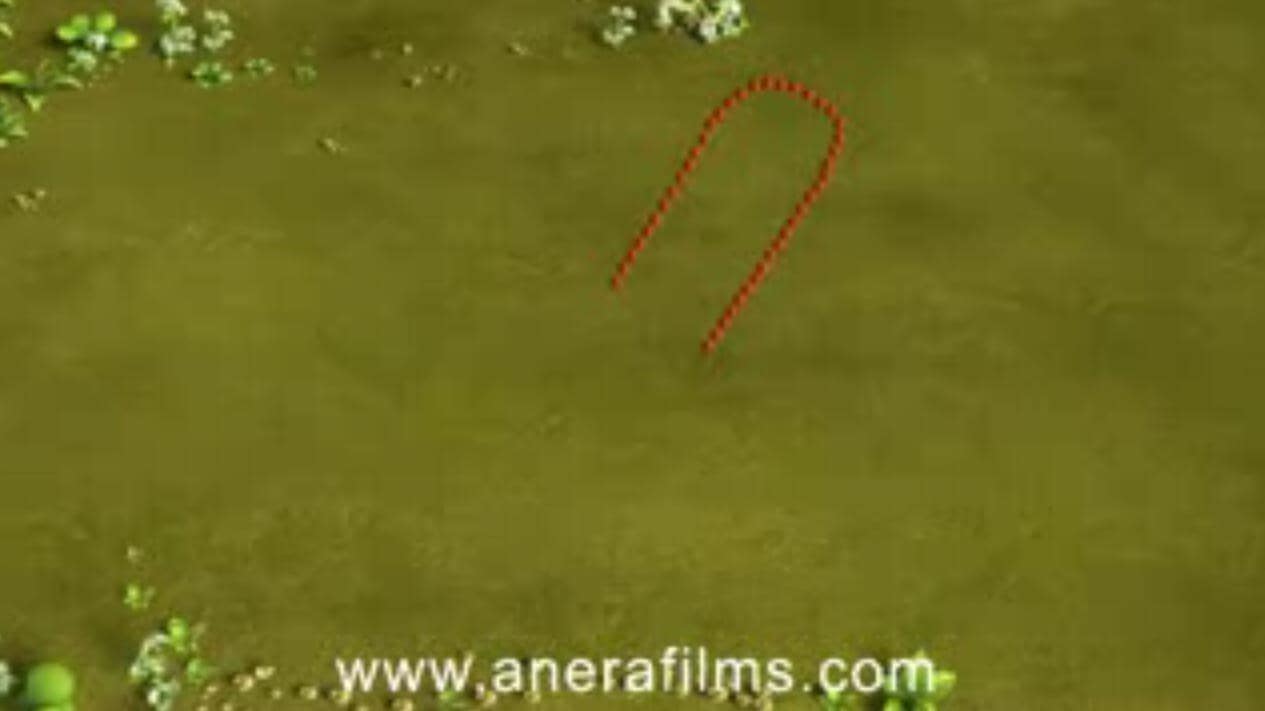} 
     \ \mbox{} \ \mbox{} \ \mbox{} \   
     \includegraphics*[width=2.5in]{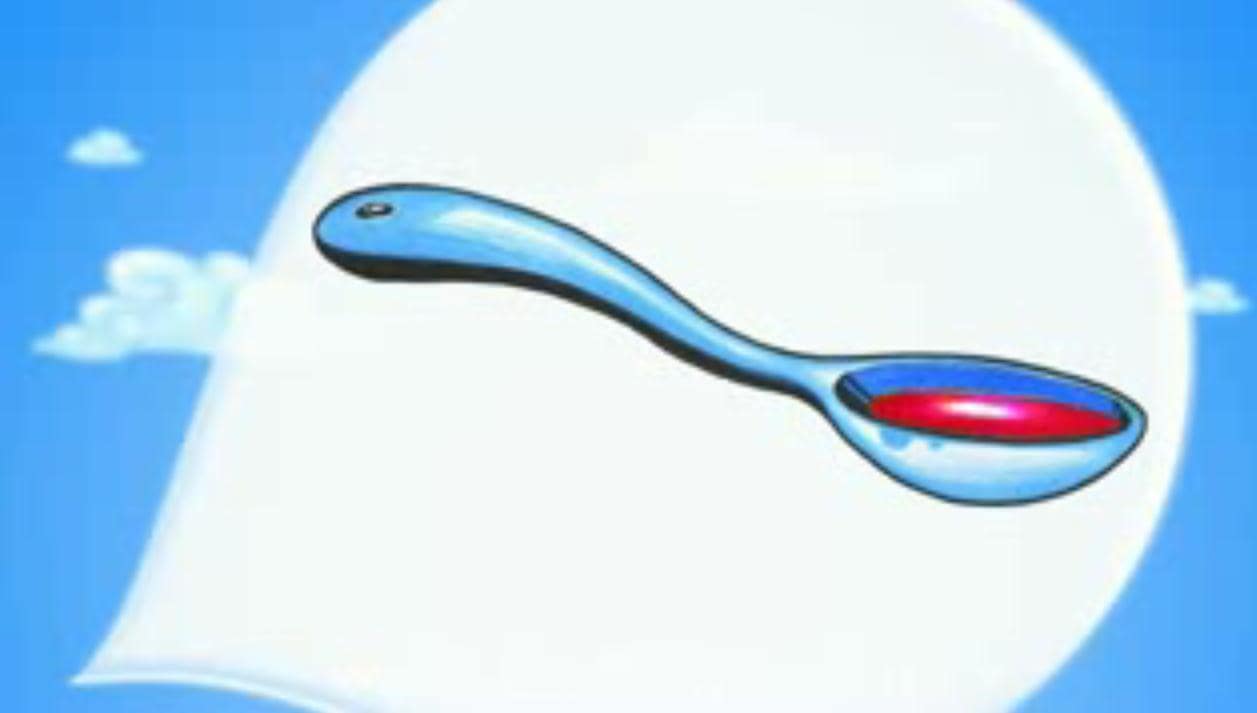}  }    
   \centerline{\mbox{}} 
   \centerline{
     \includegraphics*[width=2.5in]{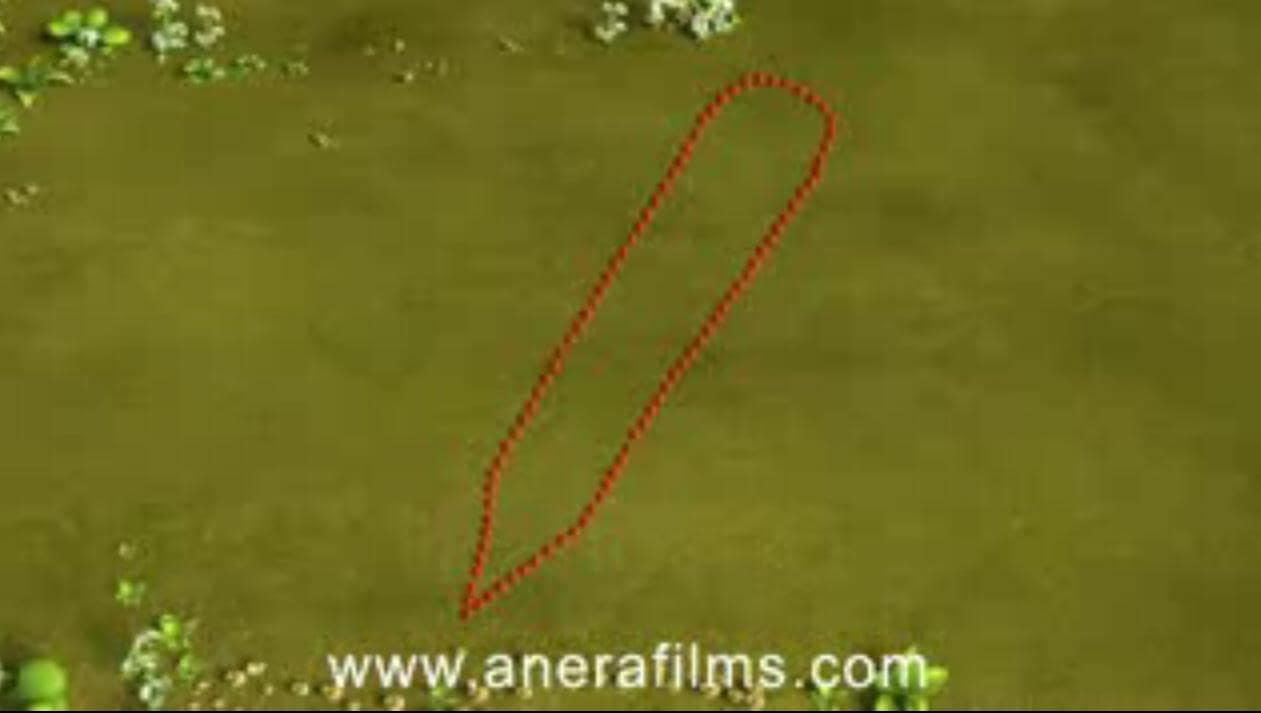} 
     \ \mbox{} \ \mbox{} \ \mbox{} \   
     \includegraphics*[width=2.5in]{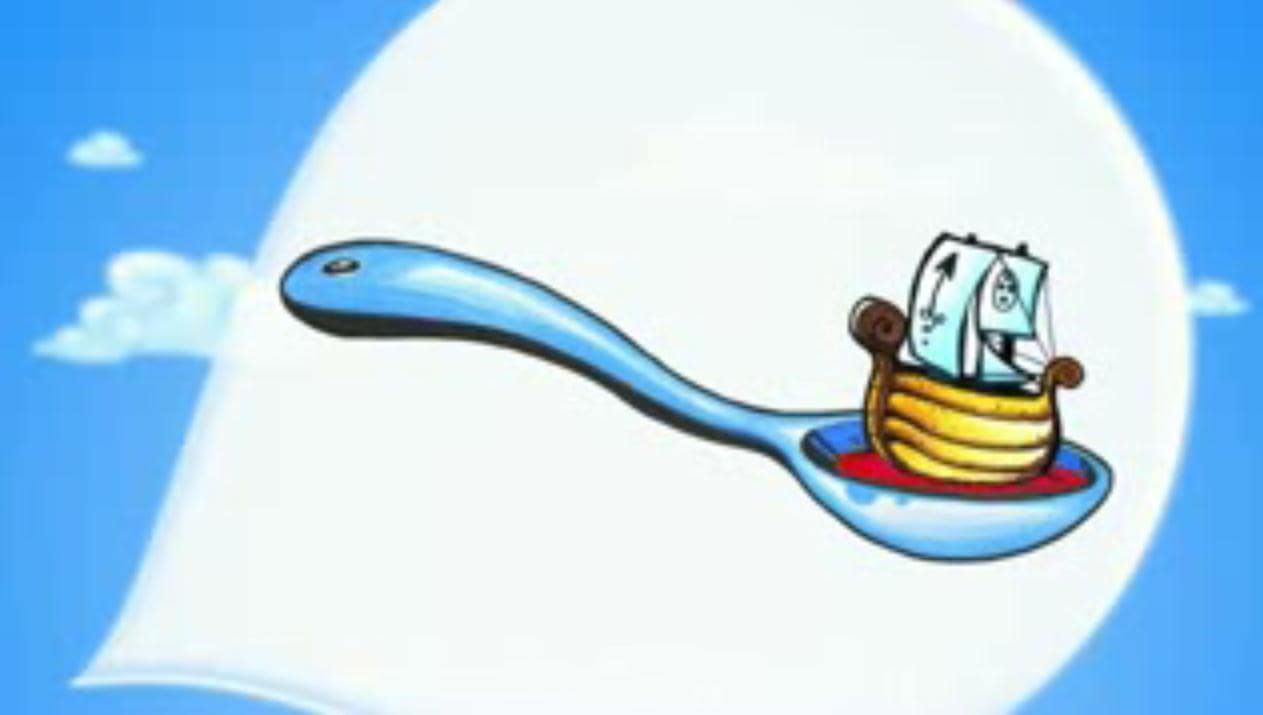}  } 
   \centerline{\mbox{}} 
   \centerline{
     \includegraphics*[width=2.5in]{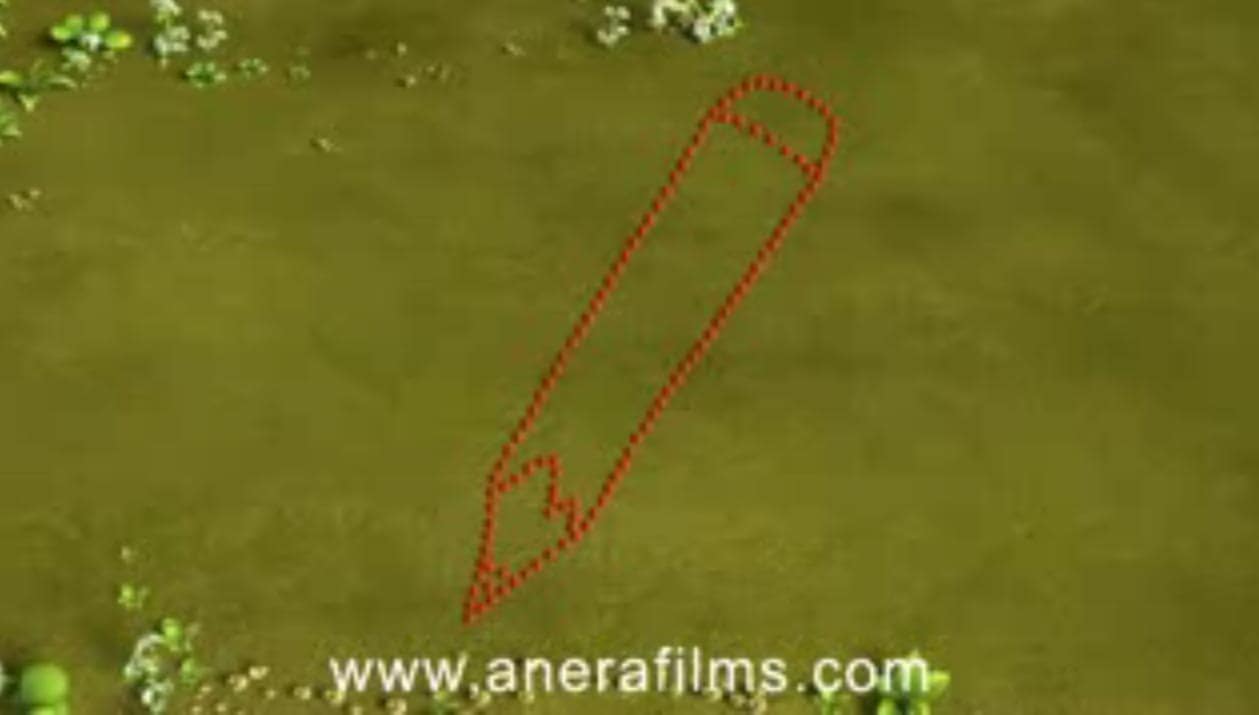} 
     \ \mbox{} \ \mbox{} \ \mbox{} \   
     \includegraphics*[width=2.5in]{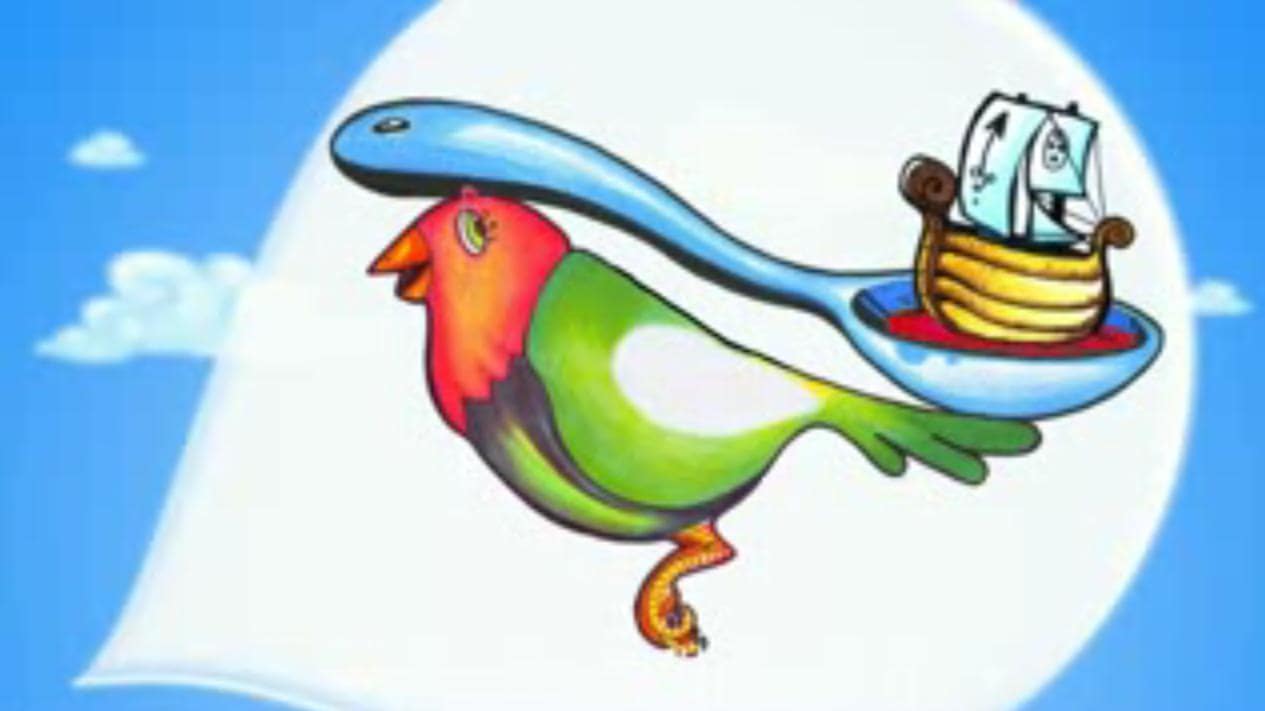}  }    
   \centerline{\mbox{}} 
   \centerline{
     \includegraphics*[width=2.5in]{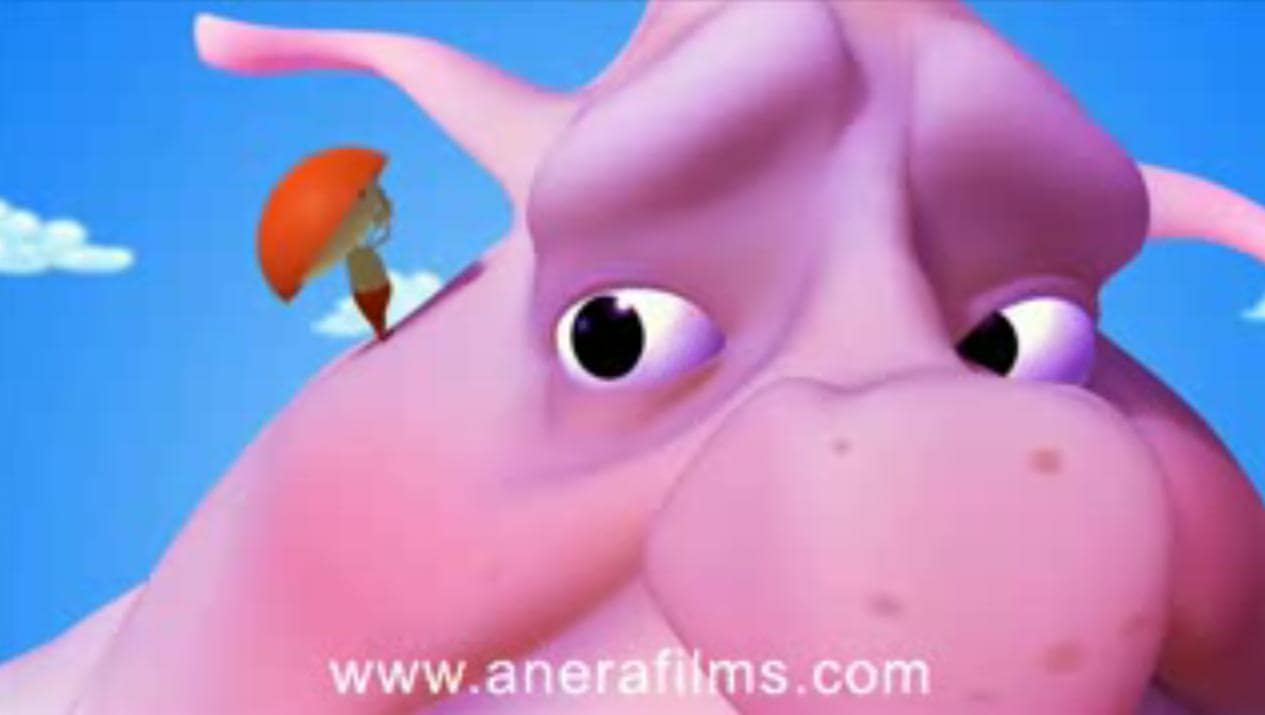} 
     \ \mbox{} \ \mbox{} \ \mbox{} \   
     \includegraphics*[width=2.5in]{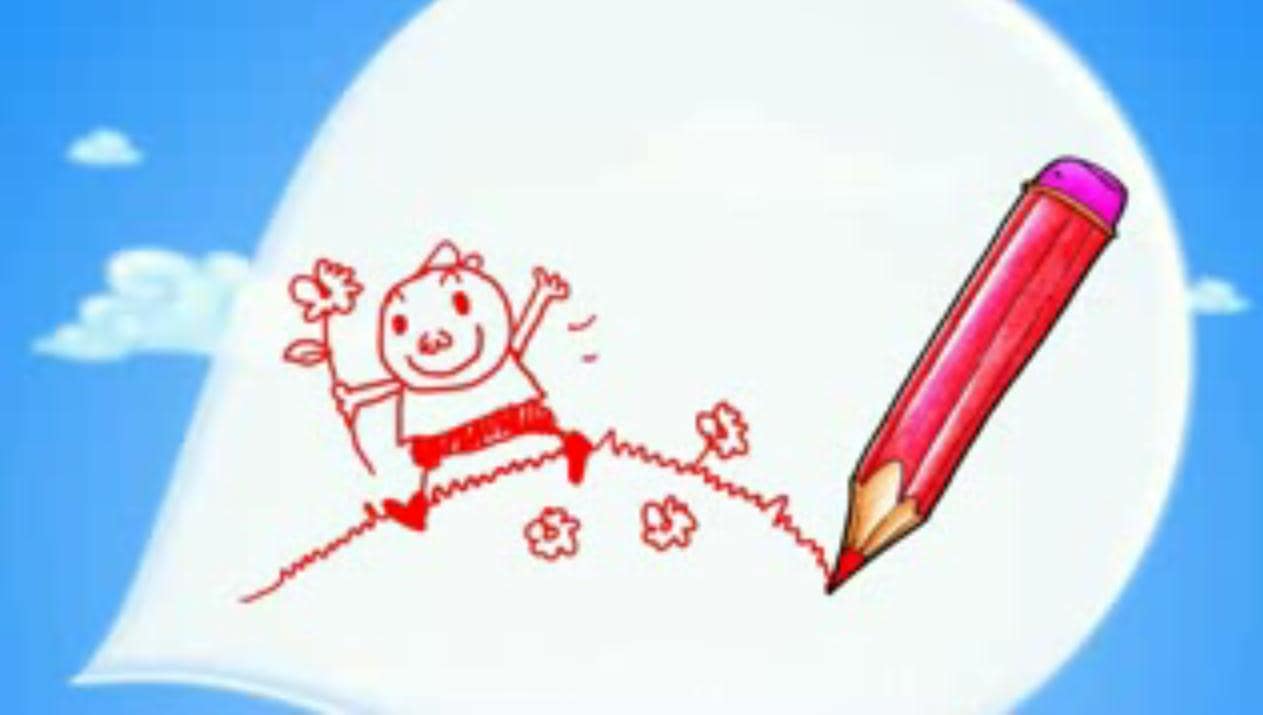}  }        
 \end{figure}
 
 \vspace{-2cm} 

 \begin{quote} 
 Figure JB.16: Boom and Reds, a cartoon by Anera films, features a 
friendly purple ogre, Boom, who, in every episode, will try to  find out
what a big crowd of peppy little mushroom-headed creatures, the Reds,
are drawing in the floor. 
 Boom gets stuck with fixed idea that he desperately tries to use to 
make sense of the drawing, for the great amusement of the Reds 
(and the viewers alike).    
 Finally, with some hints offered by a merciful Red, Boom is able to 
correctly ``see'' (with his mind's eye) what is going on.   
 \end{quote} 

  \pagebreak


 \centerline{
  \includegraphics*[height=2.0in,width=3.0in]{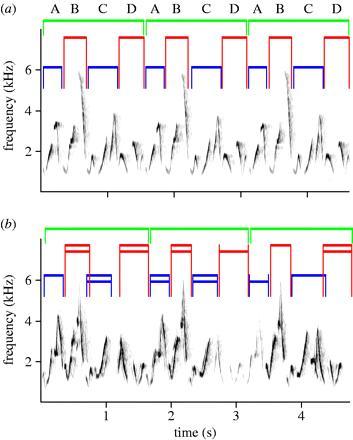}
   } 

 \mbox{} 
 
 \centerline{
  \includegraphics*[height=1.0in,width=4.0in]{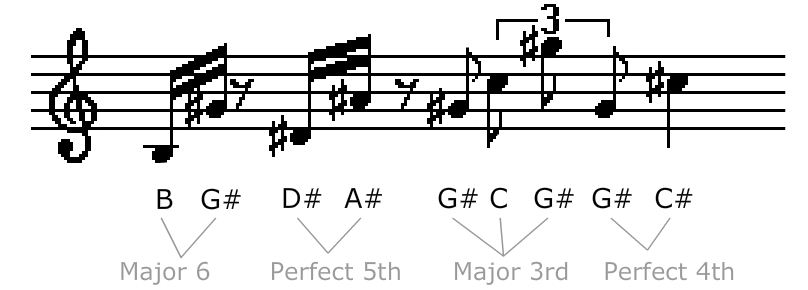}
  } 
 
 \mbox{} 

 \centerline{
  \includegraphics*[width=4.6in, angle=0, 
    trim=0in 1.5in 0.0in 0.0in, clip]{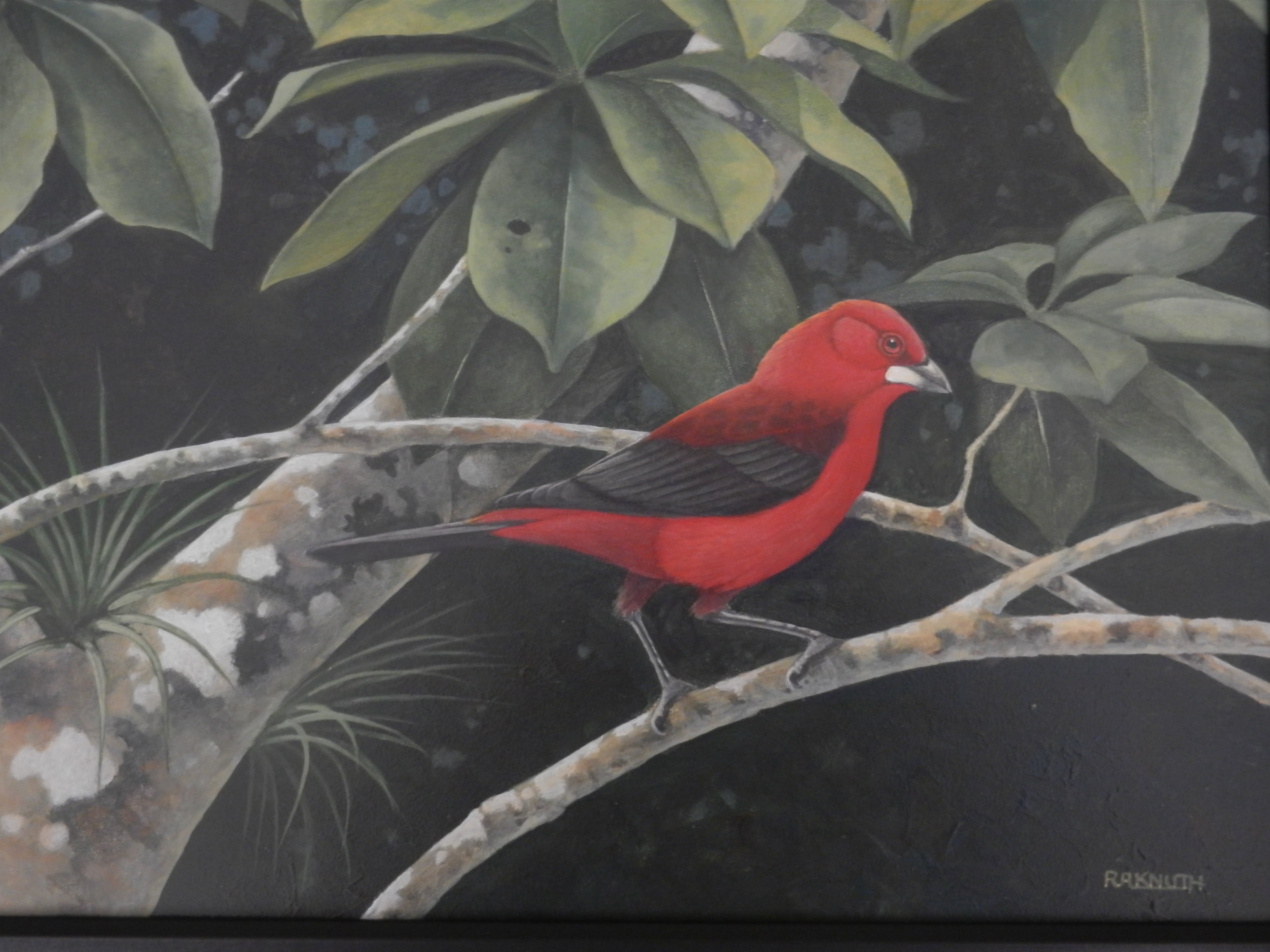} 
  \mbox{} 
  \includegraphics*[
  width=1.5in]{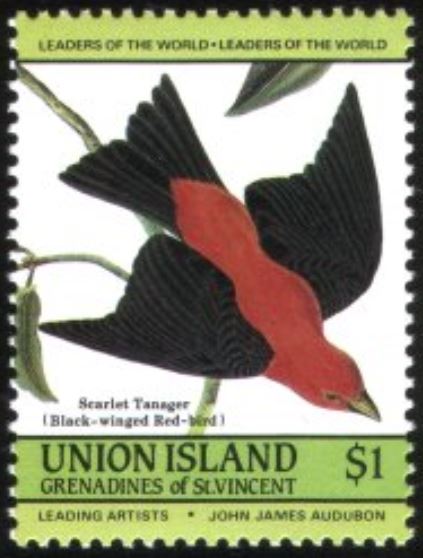} 
   } 

 {\center Figure JB.17: Speciation is an evolutionary 
 process in which sharp, stable, separable and composable 
 biological eigen-solutions may emerge in the form of  
 musical or visual patterns. For example,  
 ``hear'' the harmonic duet of the Neotropic Wren (Thryothorus euophrys)  
 from the sonogram in Mann et al. (2006)  
 or from the musical score at LosDoggies (2010);  
 or see the beautiful plumage of the 
 Scarlet Tanager (Ramphocelus bresilius)  
 in paintings by Rockne Knuth and John James Audubon. 
 The Ti\^{e} Sangue is in Brazil a symbol of freedom, 
 since the vivid colors that a healthy bird displays in the wild 
 quickly fade away when it is held in captivity. 
 }     
 
  \pagebreak 


 \centerline{
  \includegraphics*[height=2.2in,width=2.0in]{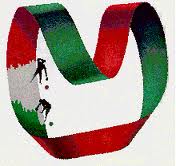}
  \mbox{} 
  \includegraphics*[height=2.2in,width=2.0in]{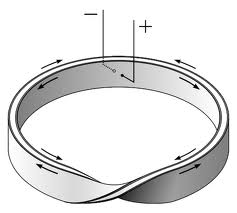} 
  \mbox{} 
  \includegraphics*[height=2.2in,width=2.0in]{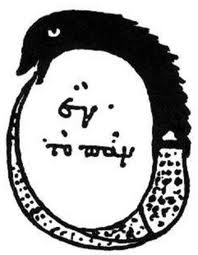}
   } 

 \mbox{} 
 
 \centerline{
  \includegraphics*[height=2.2in,width=2.5in]{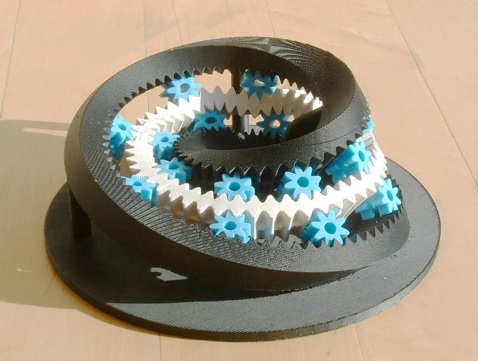}
  \mbox{} 
  \includegraphics*[height=2.2in,width=3.5in]{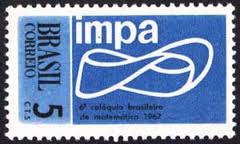}
   } 
 
 \mbox{} 

 \centerline{
  \includegraphics*[height=2.2in,width=2.0in]{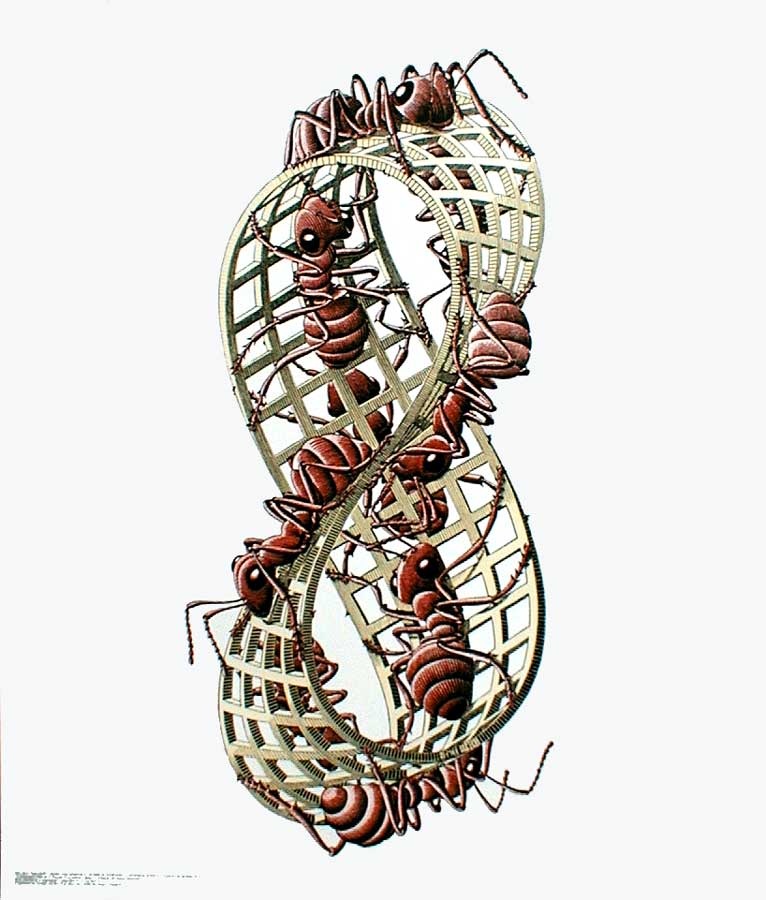}
  \mbox{} 
  \includegraphics*[height=2.2in,width=2.0in]{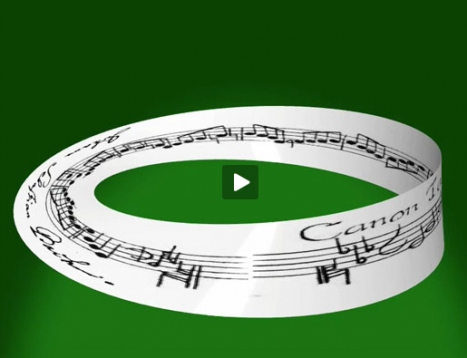}
  \mbox{} 
  \includegraphics*[height=2.2in,width=2.0in]{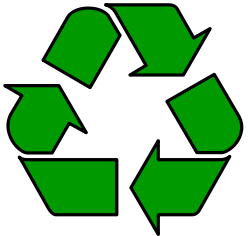}
   } 

 {\center Figure JB.18: The Moebius band in Technology and Art,  
 including R.Davis' noninductive and nonreactive resistor, 
 the `all-seeing' Ouroboros in the Greek manuscript Chrysopoeia of 
 Cleopatra (circa 100 AD), A.M.Hoover's M\"{o}bius Gear, 
 Brazilian postal stamp showing IMPA's symbol, 
 M.C.Escher's M\"{o}bius Strip II woodcut, 
 J.Leys' animation of J.S.Bach's Musikalisches Opfers Krebskanon,  
 and G.Anderson's universal recycling symbol.}

 \pagebreak

 \centerline{
  \includegraphics*[height=1.2in,width=1.7in]{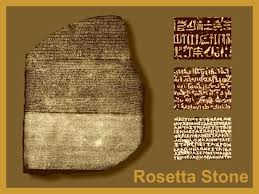}
  \mbox{} \ \ \mbox{} 
  \includegraphics*[height=1.2in,width=1.3in]{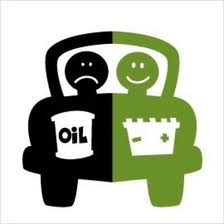}
  \mbox{} \ \ \mbox{} 
  \includegraphics*[height=1.2in,width=1.3in]{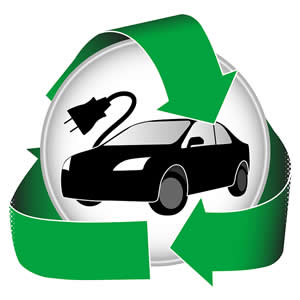}
   } 

 \centerline{
  \includegraphics*[height=1.3in,width=2.6in]{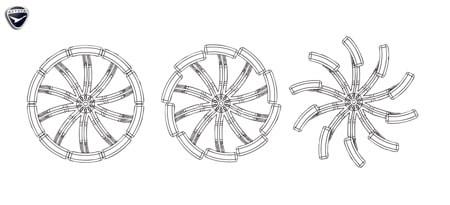}
  \mbox{} \ \ \mbox{} 
  \includegraphics*[height=1.3in,width=2.4in]{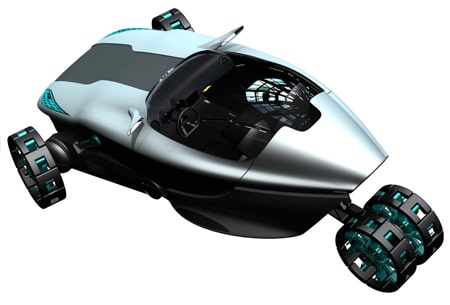}
   } 

 \centerline{Figure JB.19: Hybrid / Amphibious Vehicles.}  
 \noindent 
  {The Rosetta stone (196 BC) is a hybrid vehicle that provides 
 synchronic and diachronic alignments. 
  The text at the top, written in ancient Egyptian hieroglyphs, 
 is translated to Demotic script at the middle, and to  
 ancient Greek at the bottom part of the stela.} \\ 
  {Gas-electric hybrid cars are an engineering nightmare, 
 requiring a thorough understanding of all the details of each 
 of the individual systems plus the necessary coupling mechanisms. 
  However, they may be the only way to achieve a sustainable 
 future. 
  Notice that the last logo is ill-conceived,  wrapping a  
 non-orientable m\"{o}bius bad around a orientable sphere.}  \\   
  {Amphibious car by Aythya employing an ingenious 
   adaptable tire / paddle-weel.} 
 
  \mbox{} 

 \centerline{
  \includegraphics*[height=1.6in,width=3.7in]{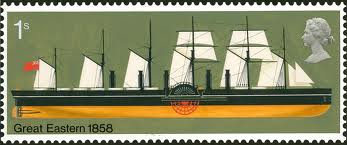}
  \mbox{} \ \ \mbox{} 
  \includegraphics*[height=1.6in,width=2.2in]{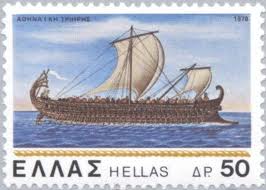}
   } 

 \centerline{Figure JB.20: Famous ships with hybrid propulsion systems.}  
 \noindent 
 {Argo ($\alpha \rho \gamma \omega \varsigma$, bright, swift): 
 A ship planned or constructed with the help of 
 Pallas Athena  on which Jason  
 ($\iota \alpha \sigma \theta \alpha \iota$, cure, heal) or 
 Diomedes  
 ($\delta \iota o$, god;  $\mu \eta \delta o \varsigma$, cousel) 
 sailed  to retrieve his beloved Medeia and their sons Pheres 
 ($\phi \epsilon \rho \omega$, bring, bear, endure, suffer) 
 and Mermeros 
 ($\mu \epsilon \rho \iota \mu \nu \alpha$,  
  care, thought, anxiety).  
  Medeia was a powerful sorcerer, daughter of Aietes 
 ($\alpha \iota \epsilon \tau o \varsigma$, eagle),   
 son of Helios 
 ($\eta \lambda \iota o \varsigma$, sun),  
 and sister of  Aegialeus 
 ($\alpha \iota \gamma \iota \alpha \lambda o \varsigma$, sea-shore). 
 Greek galleys used a combination of rows and sails. 
  %
  }  \\ 
 {The Leviathan or Great Eastern:  
 Her 6MW (8Khp) steam engines drove side paddle-wheels 17m in diameter 
 and a 7.3m four-bladed screw-propeller.  
 Additionaly, she could use 5,400m2 of sails distributed on 6 masts. 
 Converted into a cable-laying ship, in 1866 she helped to establish the 
 first lasting transatlantic (telegraph) communication line.}

 \pagebreak 

 \subsubsection*{Tokens for Economic Eigen-Values.} 

 \mbox{} 
 
 \begin{figure}[!h]
  \centerline{\includegraphics*[height=1.7in]{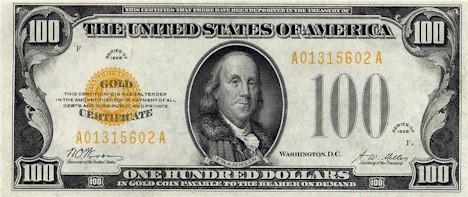}} 
   \centerline{\mbox{}}    
  \centerline{\includegraphics*[height=1.7in, width=2.5in]{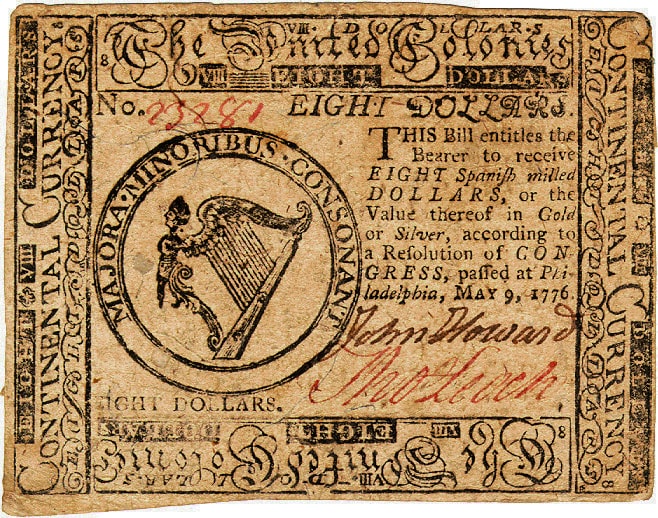}}
   \centerline{\mbox{}}    
  \centerline{\includegraphics*[height=3.4in]{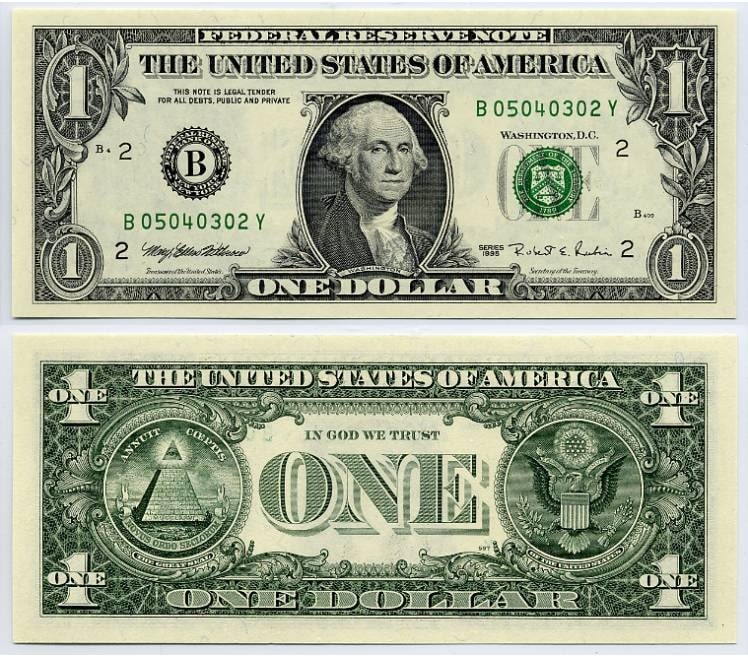}} 
  \centerline{\mbox{}}    
  \centerline{Figure JC.1: Economic token backed by another token (gold),} 
  \centerline{by two (gold or silver), or by no other token (fiat money).}
 \end{figure} 


 \pagebreak 
 
  \centerline{\includegraphics*[height=2.4in, width=4.5in]{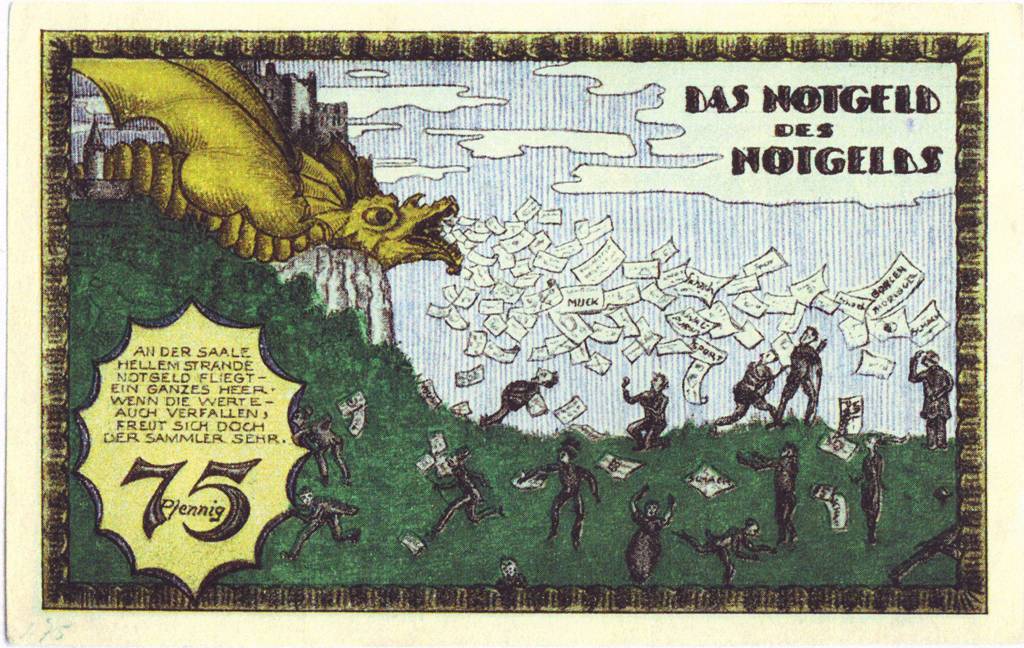}} 
  \vspace{3mm} 
  \centerline{\includegraphics*[height=2.4in, width=4.5in]{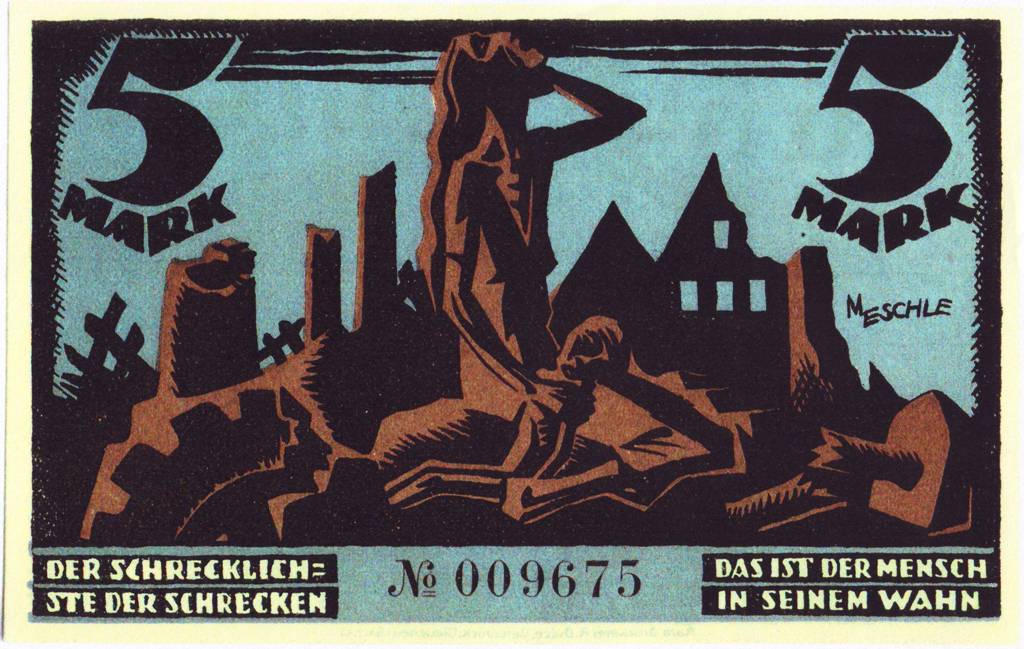}} 
  \vspace{3mm} 
  \centerline{\includegraphics*[height=2.4in, width=4.5in]{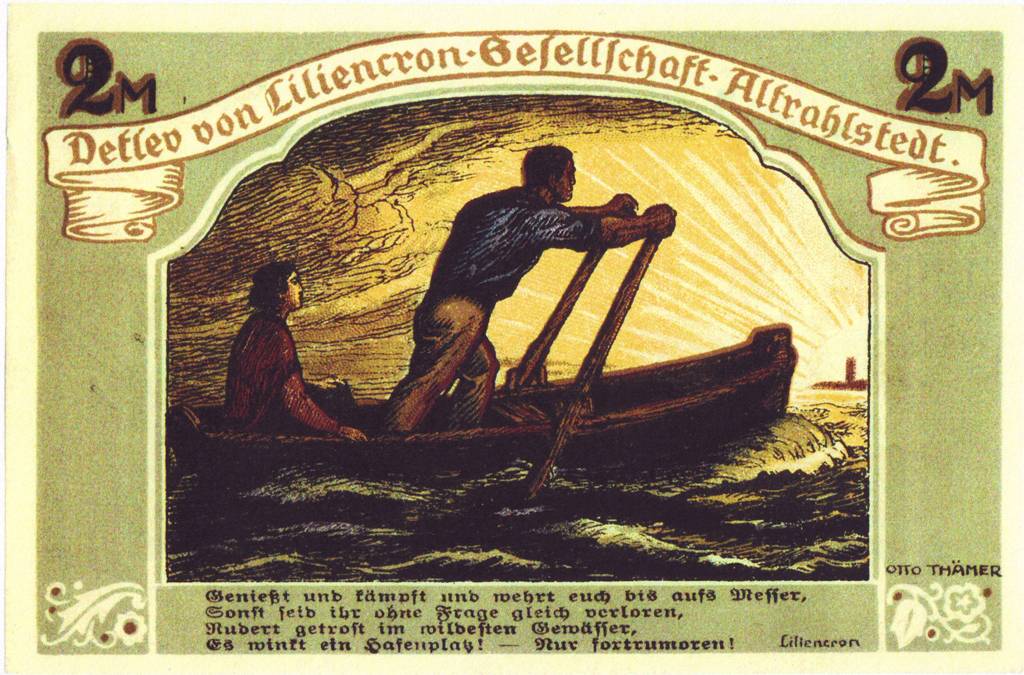}} 
  \centerline{\mbox{}}  
  \centerline{Figure JC.2: Emergency Money (Notgeld),}
  \centerline{from the private collection of Heinz Stern.} 
  \centerline{Weakly Objective Tokens used at WW-I / Hyperinflation Germany.}
  \centerline{5M Legend: The most frightful of frights is man in his delusion,} 
  \centerline{from The Song of the Bell, by Friedrich Schiller, 1800. } 
  \centerline{Knowledge comes with responsibility!} 

 \pagebreak 

  \centerline{\includegraphics*[width=5.2in]{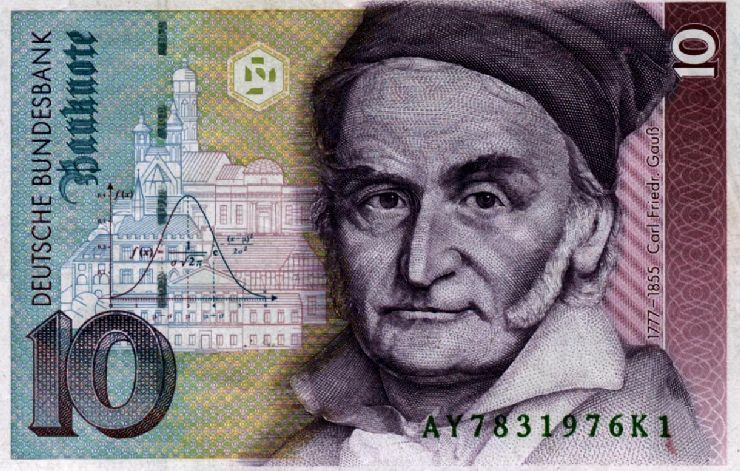}}  
  \centerline{\mbox{}} 
  \centerline{\includegraphics*[width=3.2in]{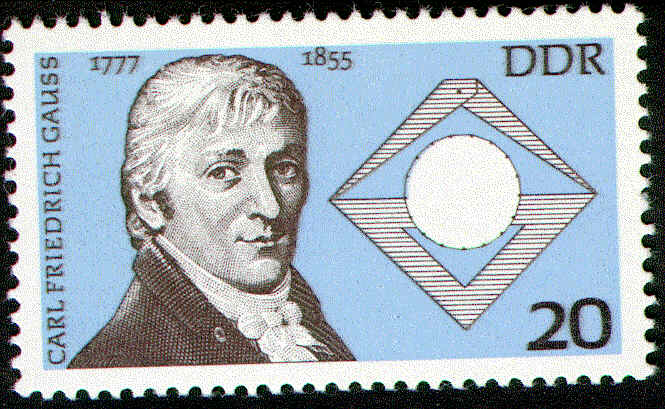}
   \mbox{} \includegraphics*[width=1.2in, 
       viewport=1 1 70 113 , clip]{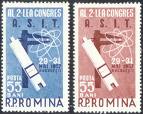} }  
  \centerline{} 
  \centerline{\includegraphics*[width=2.2in]{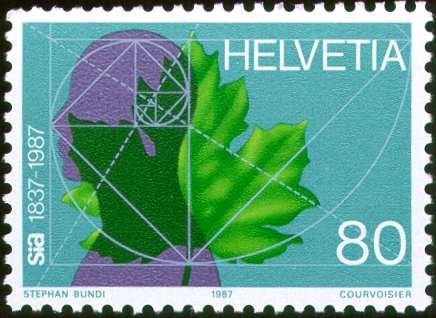}
   \mbox{} \includegraphics*[width=2.2in]{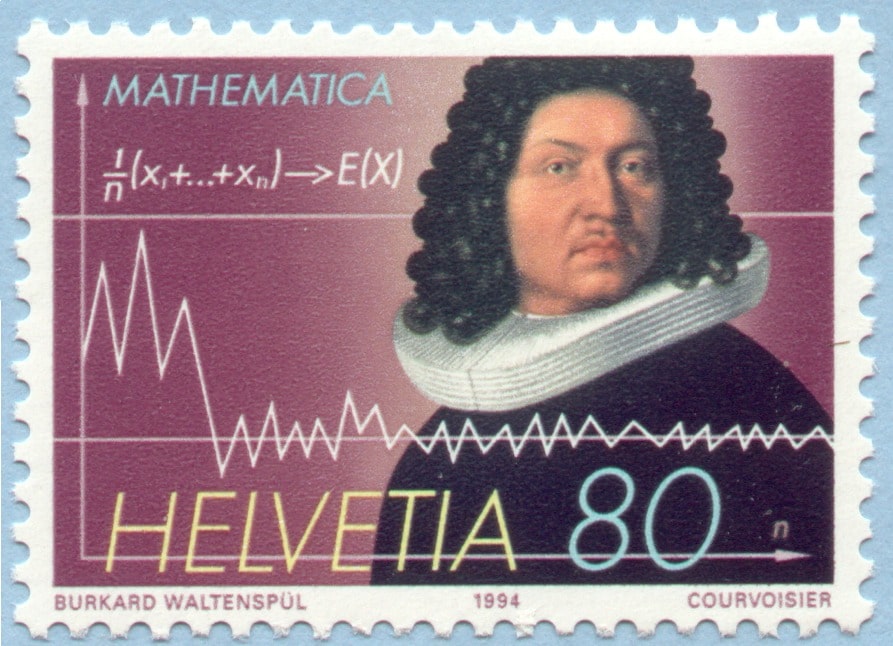} }  
  \centerline{} 
  \centerline{Figures JC.3: Symmetry and Asymptotic Characterizations of 
              Distributions.}
  \centerline{The Standard Gaussian: Center of Symmetry and 
              Orthogonality Relations } 
  \centerline{(Mean and Covariance Matrix) and Limit of Binomial. 
               Poisson / Exponential:}    
  \centerline{Log-Linear Probabilities of 
              Event Counts / Waiting Times.}

 \pagebreak

 \begin{figure}[!h]
  \centerline{\includegraphics*[width=1.5in]{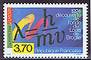} 
  \mbox{} \includegraphics*[width=1.8in]{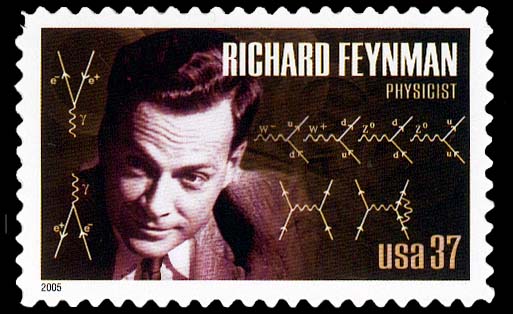} }  
  \centerline{\mbox{}} 
  \centerline{\includegraphics*[width=3.5in]{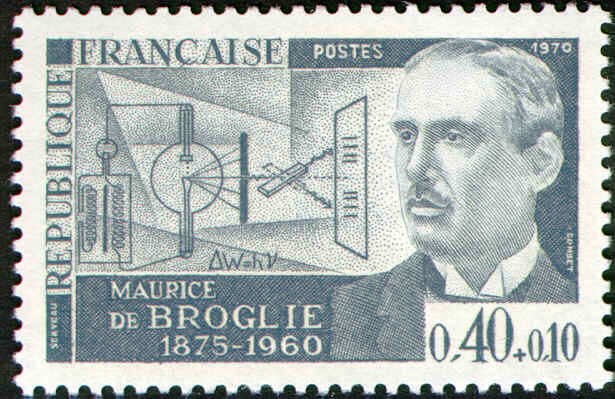}}  
  \centerline{\mbox{}} 
  \centerline{\includegraphics*[width=3.5in]{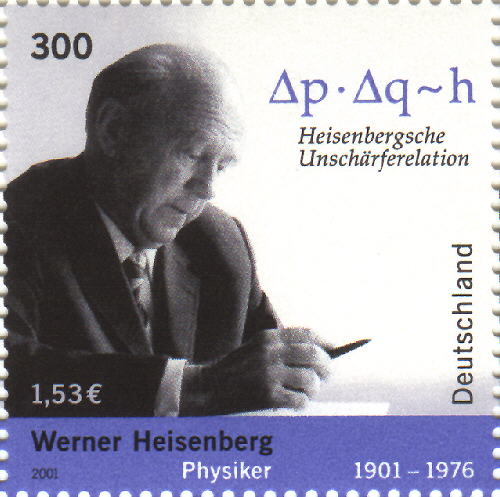}}  
  \centerline{\mbox{}} 
  \centerline{Figures JC.4: More Probabilistic Eigen-Solutions.}
 \end{figure} 

 \pagebreak

 \begin{figure}[!h] 
  \centerline{\includegraphics*[width=3.1in]{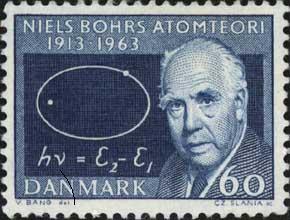} \mbox{} 
         \includegraphics*[width=2.0in, 
         clip]{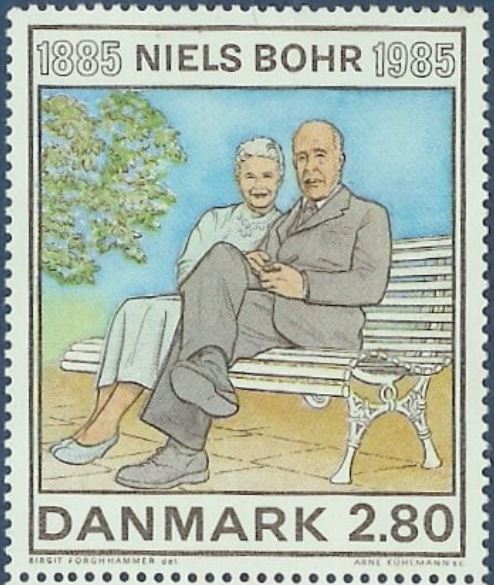} }
  \centerline{\mbox{}} 
  \centerline{\includegraphics*[width=2.65in]{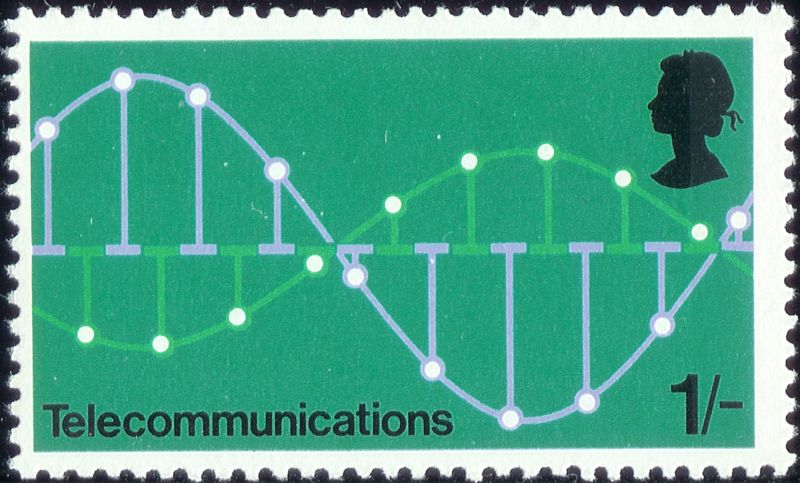}  
     \mbox{}   \includegraphics*[width=2.35in]{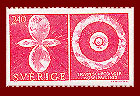} } 
  \centerline{\mbox{}} 
  \centerline{\includegraphics*[width=5.0in]{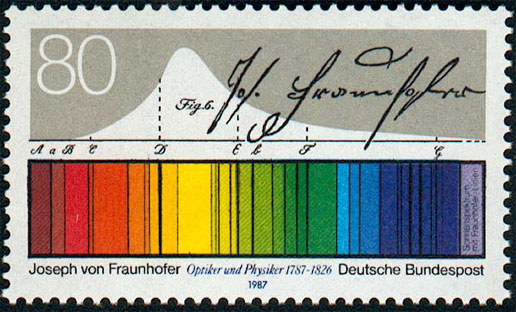}} 
  \centerline{\mbox{}}    
  \centerline{Figures JC.5: Atomic Orbitals, as Eigen-Solutions of} 
 \centerline{Bohr's Complementarity Model and Schr\"{o}dinger's Wave Equation,} 
 \centerline{also the corresponding spectral lines, as observed by Fraunhofer.} 
 \end{figure} 

 \pagebreak

 \begin{figure}[!h]
  \centerline{\includegraphics*[width=4.0in]{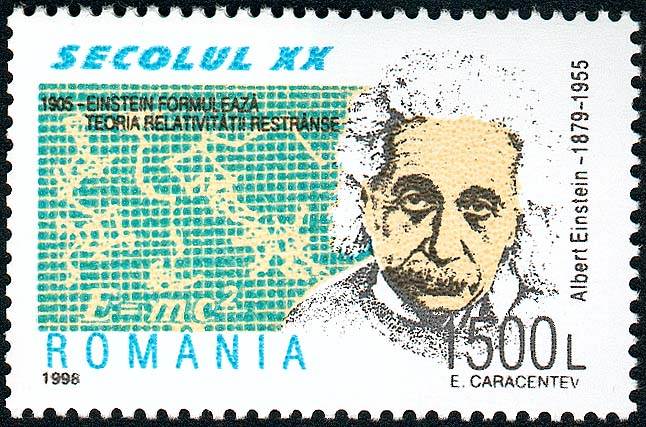}}  
  \centerline{\mbox{}} 
  \centerline{\includegraphics*[width=4.0in]{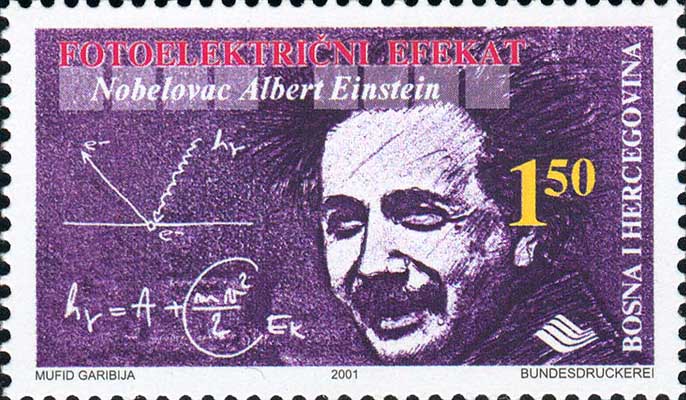}} 
  \centerline{\mbox{}} 
  \centerline{\includegraphics*[width=4.0in]{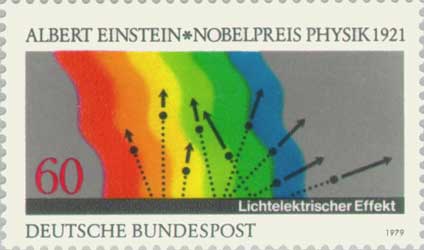}} 
  \centerline{\mbox{}} 
  \centerline{Figures JC.6: Einstein on Brownian Motion and Quantization.}
 \end{figure}

 \pagebreak 

 \begin{figure}[!h]
  \centerline{\includegraphics*[width=2.2in]{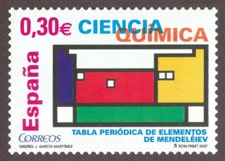}    
   \mbox{} \includegraphics*[width=2.2in]{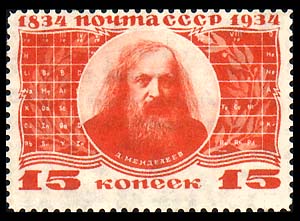}  }  
  \centerline{\mbox{}}  
  \centerline{\includegraphics*[width=2.2in]{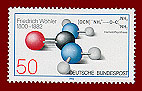}     
   \mbox{} \includegraphics*[width=2.2in]{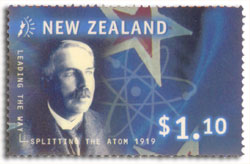}  }    
  \centerline{\mbox{}}  
  \centerline{\includegraphics*[width=1.5in]{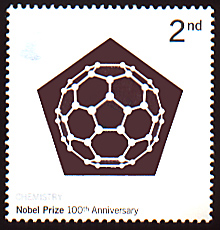} 
   \mbox{} \includegraphics*[width=2.3in]{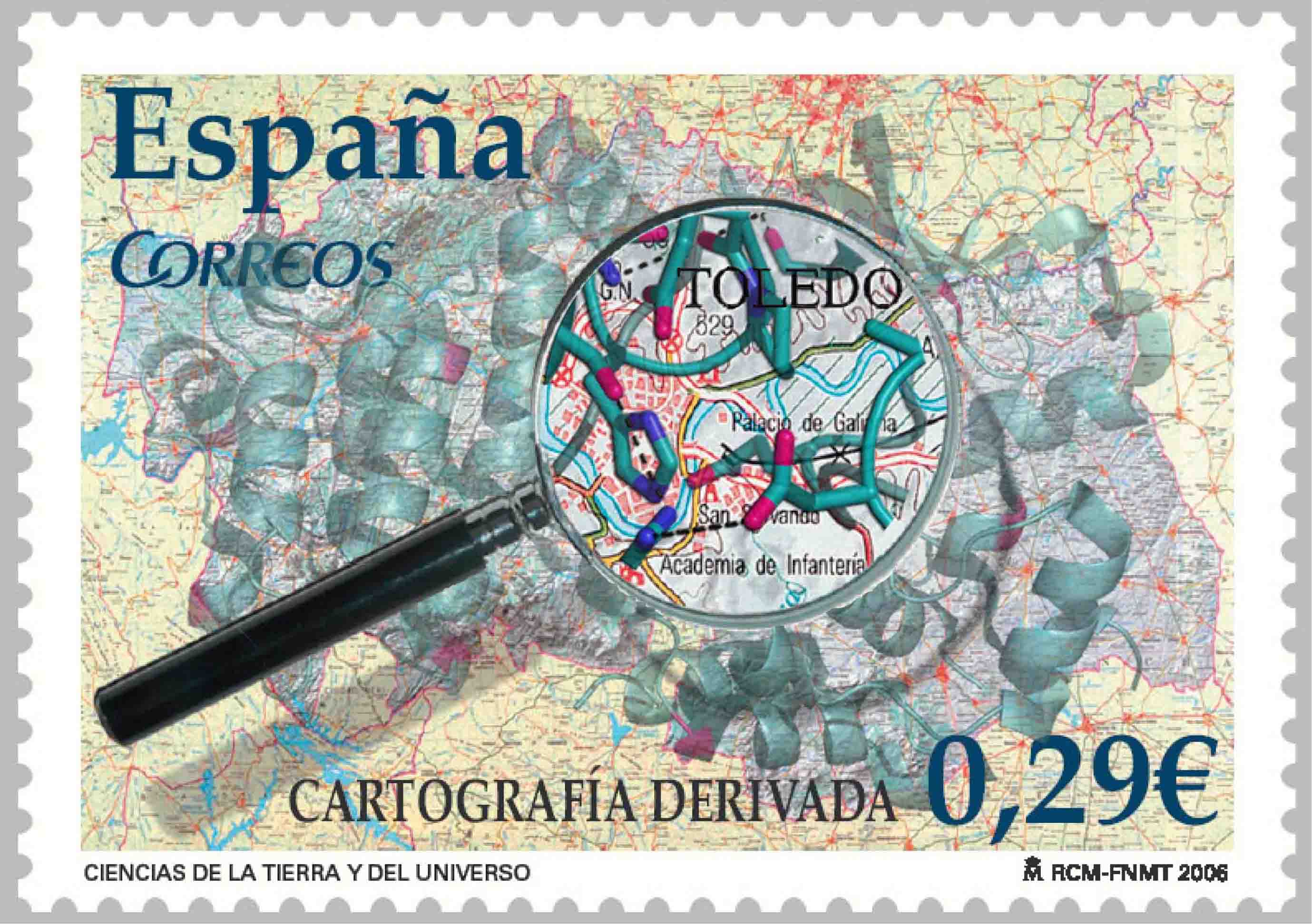}  }  
  \centerline{\mbox{}}  
  \centerline{\includegraphics*[height=2.45in]{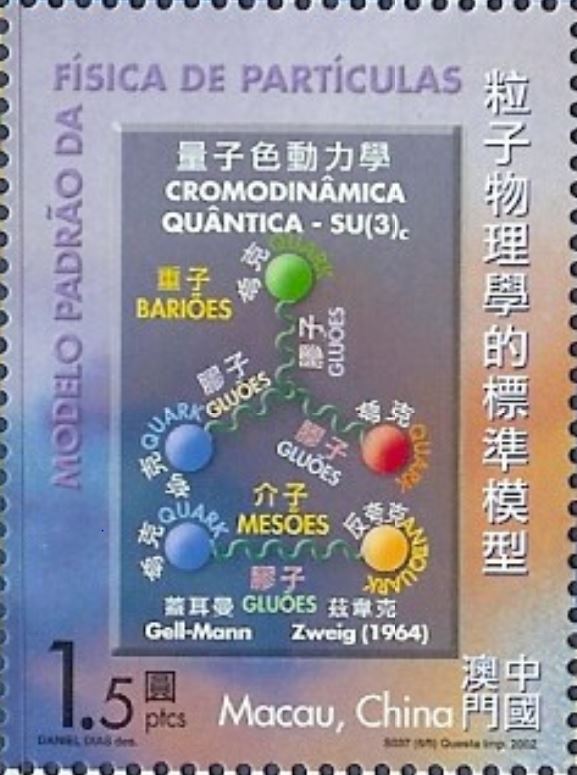}    
   \mbox{} \includegraphics*[width=2.2in, height=2.5in]{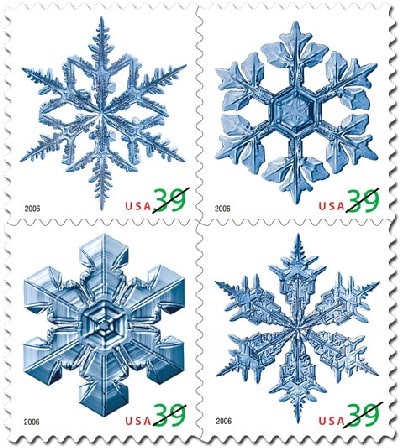}  }  
  \centerline{\mbox{}}  
 \centerline{Figure JC.7: Natural Modular Systems, in several scales.}  
 \end{figure}

 \pagebreak

 \begin{figure}[!h]
  \centerline{\includegraphics*[width=3.5in]{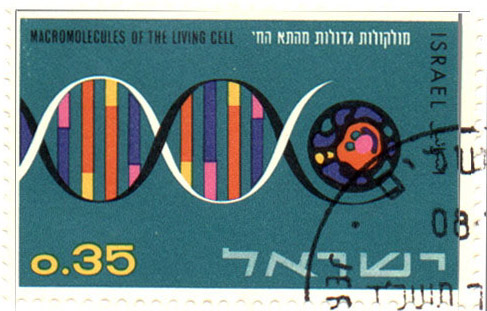}}  
  \centerline{\mbox{}}  
  \centerline{\includegraphics*[height=3.0in]{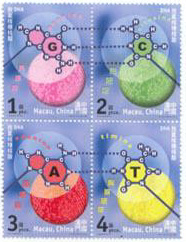}    
   \mbox{} \includegraphics*[height=3.0in]{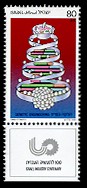}  }  
  \centerline{\mbox{}}  
  \centerline{\includegraphics*[width=3.5in]{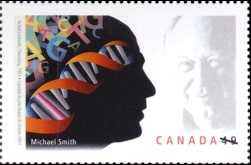}}  
  \centerline{\mbox{}} 
  \centerline{Figure JC.8: The DNA Genetic Code.}
 \end{figure} 

 \pagebreak

 \begin{figure}[!h]
  \centerline{\mbox{}} 
  \centerline{\includegraphics*[width=5.5in]{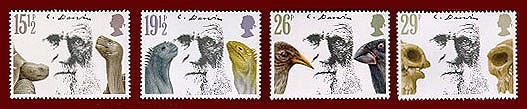}}  
  \centerline{\mbox{}} 
  \centerline{\includegraphics*[width=5.5in]{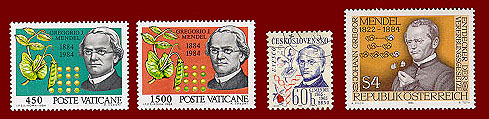}}  
  \centerline{\mbox{}} 
  \centerline{\includegraphics*[width=3.0in]{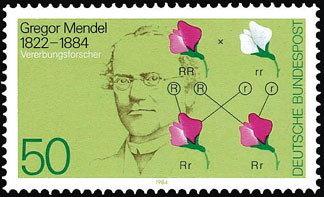}}  
  \centerline{\mbox{}}  
  \centerline{\includegraphics*[height=3.0in]{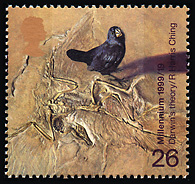} 
   \mbox{} \includegraphics*[height=3.0in]{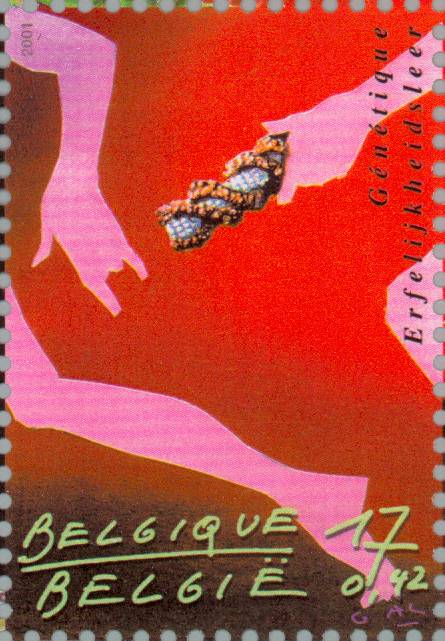}  }  
  \centerline{\mbox{}} 
  \centerline{Figure JC.9: Evolution, Continuous and Discrete Views.} 
 \end{figure}





%% file: ARVIDA3.TEX


\definecolor{mycolorR}{rgb}{0.95,0.00,0.00} 
\definecolor{mycolorY}{rgb}{0.95,0.80,0.00} 
\definecolor{mycolorG}{rgb}{0.00,0.80,0.00} 
\definecolor{mycolorB}{rgb}{0.00,0.10,0.99} 
\definecolor{mycolorZ}{rgb}{0.70,0.70,0.70} 
\definecolor{mycolorC}{rgb}{0.00,0.75,0.90} 
\definecolor{mycolorM}{rgb}{0.70,0.00,0.75} 

\pgfdeclarelayer{bottom}
\pgfsetlayers{bottom,main}


\mbox{} \\[2cm] 

\mbox{} \hspace{0.7cm} 
\begin{tikzpicture}[
    vertexnum/.style={font=\sffamily\bfseries\Huge},
    pathnum/.style={font=\sffamily\bfseries\normalsize, text=black,  inner sep=1pt, circle},
    connector/.style={draw=black!60, line width=1.2pt},
    heavyconnector/.style={draw=black!90, line width=2.8pt},
    ball/.style={circle, draw=black, line width=0.8pt, minimum size=1.8cm, inner sep=2pt},
    colorNode/.style={ball, text=black}
    ]

    \begin{pgfonlayer}{bottom}
        \node[colorNode, fill=mycolorB!40] (0) at (0, 6.0) {\Huge 0};
    \end{pgfonlayer}

    \node[circle, minimum size=1.8cm, draw=black, line width=0.8pt] (1) at (0, 9) {};
    \begin{scope}
        \clip (1.center) circle (0.9cm);
        \fill[black] (1.west) rectangle ($(1.east)+(0,1)$);
        \fill[white] (1.west) rectangle ($(1.east)+(0,-1)$);
    \end{scope}
    \node at (1) [vertexnum, text=white, yshift=0.3cm] {1};
    
    \node[colorNode, fill=mycolorZ!40, drop shadow] (6) at (0, 3.0) {\Huge 6};
    \node[colorNode, fill=mycolorY!60] (9) at (0, 0) {\Huge 9};
    \node[circle, minimum size=1.8cm, draw=black, line width=0.8pt] (10) at (0, -3.0) {};
    \begin{scope}
        \clip (10.center) circle (0.9cm);
        \fill[white] (10.west) rectangle ($(10.east)+(0,1)$);
        \fill[black] (10.west) rectangle ($(10.east)+(0,-1)$);
    \end{scope}
    \node at (10) [vertexnum, text=white, yshift=-0.3cm] {10};

    \node[ball, fill=white, line width=5pt, minimum size=1.62cm] (3) at (-3, 7.5) {\Huge 3};
    \node[colorNode, fill=mycolorM!40] (5) at (-3, 4.5) {\Huge 5};
    \node[colorNode, fill=mycolorR!50] (8) at (-3, 1.5) {\Huge 8};

    \node[ball, fill=white] (2) at (3, 7.5) {};
    \fill[black] (2.center) circle (0.5cm);
    \node at (2.center) [vertexnum, text=white] {2};
    \node[colorNode, fill=mycolorC!40] (4) at (3, 4.5) {\Huge 4};
    \node[colorNode, fill=mycolorG!40] (7) at (3, 1.5) {\Huge 7};

    \draw[connector] (1) -- (2) node[pos=0.5, above right, pathnum] {5};
    \draw[connector] (1) -- (3) node[pos=0.5, above left, pathnum] {6};
    \draw[connector] (3) -- (2) node[pos=0.3, above, pathnum] {21};
    \draw[connector] (1) -- (6) node[pos=0.78, right, pathnum] {4}; 
    \draw[connector] (3) -- (5) node[pos=0.5, left, pathnum] {3};
    \draw[connector] (2) -- (4) node[pos=0.5, right, pathnum] {2};

    \draw[heavyconnector] (5) -- (4) node[pos=0.35, above, pathnum] {1};
    \draw[heavyconnector] (5) -- (8) node[pos=0.5, left, pathnum] {17};
    \draw[heavyconnector] (4) -- (7) node[pos=0.5, right, pathnum] {11};
    \draw[heavyconnector] (8) -- (7) node[pos=0.27, above, pathnum] {13};

    \draw[connector] (3) -- (4) node[pos=0.15, above right, pathnum] {19};
    \draw[connector] (2) -- (5) node[pos=0.15, above left, pathnum] {7};
    \draw[connector] (3) -- (6) node[pos=0.35, left, pathnum] {16};
    \draw[connector] (2) -- (6) node[pos=0.35, right, pathnum] {9};
    \draw[connector] (5) -- (6) node[pos=0.40, below left, pathnum] {18};
    \draw[connector] (4) -- (6) node[pos=0.40, below right, pathnum] {8};
    \draw[connector] (6) -- (8) node[pos=0.4, above left, pathnum] {15};
    \draw[connector] (6) -- (7) node[pos=0.4, above right, pathnum] {10};
    \draw[connector] (6) -- (9) node[pos=0.77, right, pathnum] {20};
    \draw[connector] (8) -- (9) node[pos=0.5, below left, pathnum] {12};
    \draw[connector] (7) -- (9) node[pos=0.5, below right, pathnum] {14};
    \draw[connector] (9) -- (10) node[pos=0.5, right, pathnum] {22};

    \node[
 anchor=north west, 
 text width=4.6cm, 
 align=flush left, 
 font=\sffamily\fontsize{14pt}{14.5pt}\selectfont, 
 inner sep=0pt        
    ] at (5.5, 9.85) {
\mbox{} \\[-6mm] 
\textit{The Tree of Life:} \\[0.3em]    
A lattice diagram of \\ 
11 key concepts and \\ 
22 inter-connections. \\[0.5em]
\textit{Free interpretation:} \\[0.3em]
1- Kosmos noetos/ \\  
 Immanent intellect, \\ 
 Harmonia mundi; \\ 
2- Symmetry, \\ 
   Convexification; \\
3- Quantization, \\ 
   Discretization; \\
0- Language, Onto\\ 
   -logy, Metaphysics; \\
4- Precision, Exact
   Pr(H)=0 statement; \\
5- Stability; \\
6- Eigen-Solution; \\
7- Composition, \\ 
   Associatn, Connectn;\\ 
8- Separation; \\
9- Recombination, \\ 
   in sex, technology..; \\ 
10- Singular object, \\  
 construct, solution,\\ 
Individual organism. 
 };

\end{tikzpicture}

\mbox{} \\[1cm] 

\centerline{Figure JB.3: The Tree of Life,}  
\centerline{Free allegoric interpretation.}  